\documentclass[a4paper,10pt]{book}

\usepackage[utf8]{inputenc}
\usepackage{graphicx}
\usepackage{amsmath,amsthm,amscd,amsfonts,amssymb}
\usepackage[english]{babel}
\usepackage{mathrsfs}
\usepackage{url}
\usepackage[usenames]{xcolor}
\usepackage{soul}

\usepackage{xy}
\xyoption{all}

\usepackage{array}

\usepackage{anysize}
\marginsize{2.5cm}{2.5cm}{2.5cm}{2.5cm}

\usepackage{fancyhdr}
\usepackage{hyperref}
\hypersetup{pdftitle={Geometrical structures of higher-order dynamical systems and field theories}, pdfsubject={Ph.D. Thesis}, pdfauthor={Pedro Daniel Prieto Mart\'{\i}nez},pdfkeywords={Higher-order autonomous and non-autonomous dynamical systems.Higher-order field theory. Hamilton-Jacobi equations. Variational principles. Skinner-Rusk formalism. Lagrangian and Hamiltonian formalisms. Symplectic, presymplectic and multisymplectic manifolds.}}

\newtheorem{theorem}{Theorem}[chapter]
\newtheorem{proposition}[theorem]{Proposition}
\newtheorem{corollary}[theorem]{Corollary}
\newtheorem{lemma}[theorem]{Lemma}
\newtheorem{definition}{Definition}[chapter]
\newtheorem*{theorem*}{Theorem}
\newtheorem*{corollary*}{Corollary}
\newtheorem*{proposition*}{Proposition}
\newtheorem*{definition*}{Definition}

\theoremstyle{definition}
\newtheorem{example}{Example}[chapter]
\newtheorem*{remarkth}{Remark}
\newtheorem*{remarks}{Remarks}

\newenvironment{remark}{\begin{remarkth}}{\hfill$\lozenge$\end{remarkth}}

\newcommand{\derpar}[2]{\frac{\partial #1}{\partial #2}}
\newcommand{\derpars}[3]{\frac{\partial^2 #1}{\partial #2 \partial #3}}

\newcommand{\restric}[2]{\left. #1 \right| _{#2}}

\newcommand{\N}{\mathbb{N}}

\newcommand{\R}{\mathbb{R}}
\newcommand{\C}{\mathcal{C}}
\newcommand{\W}{\mathcal{W}}
\newcommand{\Lag}{\mathcal{L}}
\newcommand{\Leg}{\mathcal{FL}}
\renewcommand{\P}{\mathcal{P}}
\newcommand{\Tan}{\textnormal{T}}
\newcommand{\vf}{\mathfrak{X}}
\newcommand{\df}{\Omega}
\renewcommand{\d}{{\rm d}}
\newcommand{\Lie}{{\mathop{\rm L}\nolimits}}
\newcommand{\Cinfty}{C^\infty}
\newcommand{\X}{\mathcal{X}}
\newcommand{\D}{\mathcal{D}}
\newcommand{\V}{\mathcal{V}}
\newcommand{\p}{\textnormal{p}}

\newcommand{\pr}{\operatorname{pr}}
\newcommand{\Id}{\operatorname{Id}}
\renewcommand{\Im}{\operatorname{Im}}
\renewcommand{\graph}{\operatorname{graph}}
\newcommand{\rank}{\operatorname{rank}}
\newcommand{\inn}{\mathop{i}\nolimits}
\newcommand{\supp}{\operatorname{supp}}

\renewcommand\maketitle{
\thispagestyle{empty}
\noindent \rule{\textwidth}{1pt}

\begin{center}
{\large Programa de Doctorat en Matem\`{a}tica Aplicada}

\

{\large Ph.D. Thesis}

\vspace{5mm}

{\LARGE \textsc{Geometrical structures of higher-order dynamical systems and field theories}} \\

\vspace{10mm}

{\large Author \\ \textbf{Pedro Daniel Prieto Mart\'{\i}nez}}

\vspace{5mm}

{\large Thesis Advisor \\ \textbf{Narciso Rom\'{a}n Roy}}

\

Departament de Matem\`{a}tica Aplicada IV
\end{center}

\noindent \rule{\textwidth}{1pt}

\vfill

\begin{center}
\textsf{Facultat de Matem\`{a}tiques i Estad\'{\i}stica \\ \vspace{2mm} Universitat Polit\`{e}cnica de Catalunya}
\end{center}

}

\begin{document}


\pagenumbering{alph}
\maketitle
\clearpage{\pagestyle{empty}\cleardoublepage}


\frontmatter


\cleardoublepage
\chapter{Abstract}

Geometrical physics is a relatively young branch of applied mathematics that was initiated by the
60's and the 70's when A. Lichnerowicz, W.M. Tulczyjew and J.M. Souriau, among many others, began
to study various topics in physics using methods of differential geometry. This ``geometrization''
provides a way to analyze the features of the physical systems from a global viewpoint, thus
obtaining qualitative properties that help us in the integration of the equations that describe them.
Since then, there has been a strong development in the intrinsic treatment of a variety of topics
in theoretical physics, applied mathematics and control theory using methods of differential geometry.

Most of the work done in geometrical physics since its first days has been devoted to study
first-order theories, that is, those theories whose physical information depends on (at most)
first-order derivatives of the generalized coordinates of position (\textsl{velocities}). However,
there are theories in physics in which the physical information depends explicitly on accelerations
or higher-order derivatives of the generalized coordinates of position, and thus more sophisticated
geometrical tools are needed to model them accurately.

In this Ph.D. Thesis we pretend to give a geometrical description of some of these higher-order
theories. In particular, we focus on dynamical systems and field theories whose dynamical information
can be given in terms of a Lagrangian function, or a Hamiltonian that admits Lagrangian counterpart.
More precisely, we will use the Lagrangian-Hamiltonian unified approach in order to develop a geometric
framework for higher-order autonomous and non-autonomous dynamical system, and for second-order field
theories. This geometric framework will be used to study several relevant physical examples and
applications, such as the Hamilton-Jacobi theory for higher-order mechanical systems, relativistic spin
particles and deformation problems in mechanics, and the Korteweg-de Vries equation and other systems
in field theory.

\

\noindent \textbf{Keywords:} Higher-order autonomous and non-autonomous dynamical systems. Higher-order
field theory. Hamilton-Jacobi equations. Variational principles. Skinner-Rusk formalism. Lagrangian and
Hamiltonian formalisms. Symplectic, presymplectic and multisymplectic manifolds.

\

\noindent \textbf{MSC2010:} 70H50, 70H03, 70H05, 70H20, 53D42, 53C15, 53C80, 35A15, 35G99, 37B55, 55R10.
\clearpage{\pagestyle{plain}\cleardoublepage}


\renewcommand{\headrulewidth}{0.4pt}
\tableofcontents


\mainmatter


\cleardoublepage
\phantomsection
\addcontentsline{toc}{chapter}{Introduction}
\chapter*{Introduction}
\renewcommand{\chaptermark}[1]{\markboth{#1}{}}
\renewcommand{\sectionmark}[1]{\markright{#1}{}}
\chaptermark{Introduction}
\sectionmark{Introduction}

\subsection*{Geometric mechanics: motivation and historical note}

Historically, the research in dynamical systems (including mechanical ones) had a major
impact on
other areas of mathematics and physics as well as in the development of various
engineering
technologies. Most of these advances have been based on applied numerical
and analytical methods.
However, in the 60's, more sophisticated and powerful techniques were introduced in this area
when A. Lichnerowicz \cite{art:Lichnerowicz51,art:Lichnerowicz75}, J.M. Souriau \cite{book:Souriau70},
and W.M. Tulczyjew \cite{art:Tulczyjew76_2,art:Tulczyjew76_1,art:Tulczyjew77_1}, among many others,
began to study various topics in physics using methods of modern differential geometry.

This process of ``geometrization'' provides a natural framework where the features of the
physical systems can be analyzed from a global viewpoint. That is, equations, constraints and
solutions are translated to global and intrinsically defined geometric objects, and the
particularities of each structure become well-known properties of the geometric object being
considered. This enables us to establish a correspondence ``physics $\leftrightarrow$ geometry''.
Some examples of this correspondence are the following:
\begin{itemize}
\item Differential equations defining a physical system are vector fields in the phase space of the
system.

\item Symmetries are identified with actions of Lie groups on the manifold that models the phase
space of the system.

\item The regularity or singularity of dynamical systems is characterized by the (non)-degeneracy
of a form in the phase space.

\item Constraints arising in the physical system give rise to submanifolds on the phase space of
the system.

\item Canonical transformations are fiber bundle isomorphisms between two phase spaces that preserve
the geometrical objects (usually, symplectomorphisms).
\end{itemize}

Because of this, in recent decades, a strong development in the intrinsic study of a wide variety
of topics in theoretical physics, control theory and applied mathematics has been done, using methods
of differential geometry \cite{book:Abraham_Marsden78,book:Arnold89,book:DeLeon_Rodrigues89}. Thus, the intrinsic formulation
of Lagrangian and Hamiltonian formalisms has been developed for autonomous and non-autonomous
systems, as well as field theories. This study has been carried out mainly for first-order theories;
that is, those whose Lagrangian or Hamiltonian functions depend on the generalized coordinates of
position and velocity (or momentum). From the geometric point of view, this means that the phase
space of the system is in most cases a tangent (or cotangent) bundle for autonomous dynamical
systems, or a first-order jet bundle (or the corresponding bundle of forms) for non-autonomous
dynamical systems and field theories.

\subsection*{Higher-order dynamical systems and field theories}

Although the geometric study of physical systems has been carried out mainly for first-order theories,
there are a significant number of relevant dynamical systems and field theories in which
the dynamics have explicit dependence on accelerations or higher-order derivatives of the generalized
coordinates of position. These theories, usually called \textsl{higher-order dynamical systems} and
\textsl{higher-order field theories}, respectively, can be modeled geometrically using higher-order
tangent and jet bundles as the main tool \cite{book:DeLeon_Rodrigues85,book:Saunders89}. In recent
years, much work has been devoted  to the development of geometric formalisms for these kind of
theories (see, for instance, \cite{art:Aldaya_Azcarraga78_2,proc:Cantrijn_Crampin_Sarlet86,
art:Carinena_Lopez92,book:DeLeon_Rodrigues85,art:Gracia_Pons_Roman91,art:Gracia_Pons_Roman92,
art:Krupkova00,art:Saunders_Crampin90}, and references therein).

Higher-order dynamical systems play a relevant role in certain branches of theoretical physics,
applied mathematics and numerical analysis. In particular, they appear in theoretical physics,
in the mathematical description of the interaction of relativistic particles with spin, string
theories, from Polyakov and others, Hilbert's Lagrangian for gravitation, Podolsky's generalization
of electromagnetism and others \cite{art:Batlle_Gomis_Pons_Roman88,art:Nesterenko89,art:Plyushchay91,
art:Popescu_Popescu07,art:Zloshchastiev00}, as well as in some problems of fluid mechanics and
classical physics (see, for instance, the regular example in \cite{art:Prieto_Roman12} taken from
\cite{book:Elsgoltz83}), and in numerical models arising from the discretization of first-order
dynamical systems that preserve their inherent geometric structures
\cite{art:DeLeon_Martin_Santamaria04}. In these kinds of systems, the dynamics have explicit
dependence on accelerations or higher-order derivatives of the generalized coordinates of position.

Nevertheless, although the geometrization of both higher-order Lagrangian and Hamiltonian formalisms
was already developed for autonomous mechanical systems \cite{proc:Cantrijn_Crampin_Sarlet86,
book:DeLeon_Rodrigues85,art:Gracia_Pons_Roman91}, a complete generalization to higher-order
non-autonomous dynamical systems had yet to be developed.

For field theories, there have been some works giving a geometric formulation of higher-order field
theories \cite{art:Campos_DeLeon_Martin_Vankerschaver09,art:Vitagliano10} using a Skinner-Rusk
approach (which is described in the following). However, ambiguities in the definition of the
Poincar\'e-Cartan form arise when dealing with higher-order field theories, that is, given a
Lagrangian density, there are non-equivalent Poincar\'e-Cartan forms from which we obtain the same
Euler-Lagrange equations. Thus, due to its definition, these ambiguities in the Poincar\'{e}-Cartan
form are transferred to the Legendre map, thus obtaining ``different'' Legendre maps for the same
field theory. Up to our knowledge, the only unambiguous geometric formulations for higher-order
field theories are those of first and second order field theories, and those on which the base
manifold has dimension $1$ (which corresponds to non-autonomous mechanics), regardless of the order
\cite{art:Campos_DeLeon_Martin_Vankerschaver09}. Another approach is dealing with jet bundles of
infinite order \cite{art:Vitagliano10}.

\subsection*{Skinner-Rusk formalism}

A generalization of the Lagrangian and Hamiltonian formalisms exists that compresses them into a
single formalism. This is the so-called \textsl{Lagrangian-Hamiltonian unified formalism}, or
\textsl{Skinner-Rusk formalism} due to the authors' names of the original paper. It was originally
developed for first-order autonomous mechanical systems \cite{art:Skinner_Rusk83}, and later
generalized to non-autonomous dynamical systems \cite{art:Barbero_Echeverria_Martin_Munoz_Roman08,
art:Cortes_Martinez_Cantrijn02}, control systems \cite{art:Barbero_Echeverria_Martin_Munoz_Roman07},
first-order classical field theories \cite{art:DeLeon_Marrero_Martin03,art:Echeverria_Lopez_Marin_Munoz_Roman04}
and, more recently, to higher-order classical field theories
\cite{art:Campos_DeLeon_Martin_Vankerschaver09,art:Vitagliano10}.

As we show in Section \ref{Chap02_sec:SkinnerRuskAutonomousFirstOrder},
in autonomous first-order dynamical systems, this formulation is based on the use of the Whitney
sum of the tangent and cotangent bundles $\W = \Tan Q \times_Q \Tan^*Q$ (the velocity and momentum phase
spaces of the system). Observe that $\W$ has obviously higher dimension than $\Tan Q$ and $\Tan^*Q$, and
it is endowed with canonical projections over each factor and the configuration manifold.

The bundle $\W$ is endowed with a canonical presymplectic form $\Omega$, which is the pull-back of the
canonical symplectic form in $\Tan^*Q$. Then, given a Lagrangian function $\Lag \in \Cinfty(\Tan Q)$,
a Hamiltonian function $H \in \Cinfty(\W)$ is determined, and we obtain a presymplectic Hamiltonian system
$(\W, \Omega, H)$. Thus, the standard geometric equation for a presymplectic Hamiltonian system,
$\inn(X)\Omega = \d H$, can be stated, and the vector field $X$ solution to this equation gives the
dynamics of the system.

Some advantages of this unified framework are the following:
\begin{enumerate}
\item The equations $p_i = \partial\Lag/\partial v^i$ defining the momenta
(and, thus, the Legendre map $\Leg$) are obtained as constraints from the compatibility
condition.
\item The dynamical equation contains the second-order condition $v^i = dq^i/dt$ for the
Lagrangian vector field without any additional assumption, regardless of the
regularity of the Lagrangian function.
\item The first constraint submanifold $\W_c$ is diffeomorphic to $\Tan Q$,
thus allowing us to recover the Lagrangian formalism
(structures, equations and solutions) from the unified one.
\item The Legendre map, obtained from the dynamical equations, and the canonical projection
on $\Tan^*Q$ allow us to recover the Hamiltonian formalism, including constraints (if $\Lag$
is singular).
\end{enumerate}
Accordingly, the Skinner-Rusk formalism provides a suitable framework when dealing with dynamical
systems described by singular Lagrangian functions, and allows us to obtain both the Lagrangian
and Hamiltonian formalisms in a single geometrical equation, recovering each formalism to our convenience.

On the other hand, the main drawback of this formulation is that the Hamiltonian system $(\W,\Omega,H)$
is always presymplectic, since the $2$-form in $\W$ is defined as the pull-back of the symplectic
form in $\Tan^*Q$ by a submersion with non-zero kernel. Thus, a constraint algorithm is needed in order to obtain
the first constraint submanifold \cite{art:Gotay_Nester79,art:Gotay_Nester80,art:Gotay_Nester_Hinds78}
(see also \cite{art:deLeon_Marin_Marrero96,art:DeLeon_Marin_Marrero_Munoz_Roman05} for formulations
in jet bundles), and the tangency condition for the vector field solution to the dynamical equation
must be checked at least once, even if the Lagrangian function is regular.

As we see in Chapter \ref{Chap:MathPhysBackground}, in a more general situation (higher-order systems,
non-autonomous dynamical systems or field theories), $\Tan Q$ and $\Tan^*Q$ are replaced by the
corresponding Lagrangian and Hamiltonian phase spaces, respectively, and $\W$ is the fiber product
of those. In addition, the presymplectic form $\Omega$ is the pull-back
of the corresponding non-degenerate form in the Hamiltonian phase space (cosymplectic in non-autonomous
mechanics and multisymplectic in field theories). Some technical issues are also needed in order to
obtain the dynamical equations or the field equations.

\subsection*{Geometric Hamilton-Jacobi theory}

The Hamilton-Jacobi theory provides an important physical example of the deep connection between
first-order partial differential equations and systems of first-order ordinary differential
equations. It is well-known \cite{book:Jose_Saletan98,book:Landau_Lifschitz69} that the Hamilton-Jacobi
equation for a first-order autonomous Hamiltonian function $H$, with $n$ degrees of freedom,
can be written as
$$
H \left( q^A, \derpar{W}{q^A} \right) = \mbox{const} \, .
$$
Its complete solution depends on $n$ arbitrary parameters (and one additive constant),
$W \equiv W(q^A,\tilde{p}_A) + c$. The function $W(q^A,\tilde{p}_A)$ can be considered
as a generating function of a canonical transformation
$\Psi \colon (q^A,p_A) \mapsto (\tilde{q}^A,\tilde{p}_A)$; that is,
$$
\derpar{W}{\tilde{p}_A} = \tilde{q}^A \quad ; \quad
\derpar{W}{q^A} = p_A \, .
$$
This transformation leads the system to equilibrium ($\tilde{H} = 0$), and hence the Hamilton
equations for the new coordinates are trivial
$$
\frac{d\tilde{p}_A}{dt} = 0 \quad ; \quad \frac{d\tilde{q}^A}{dt} = 0 \, .
$$
From these equations, and using $W(q^A,\tilde{p}_A)$, the dynamical solution $(q^A(t),p_A(t))$ is
obtained.

From a more geometrical point of view, the transformation $\Psi$ can be associated with a foliation
in the cotangent bundle $\Tan^*Q$, which constitutes the phase space of the system. This foliation
is transversal to the fibers of $\Tan^*Q$, is invariant under the dynamical evolution, and is
Lagrangian with respect to the canonical symplectic structure of $\Tan^*Q$. In some particular
situations (for instance, when dealing with bihamiltonian systems) the second aspect can be ignored
(and then we obtain the so-called ``generalized Hamilton-Jacobi problem''). On each leaf $S_\lambda$
of that foliation, the Hamiltonian dynamical vector field $X_{h}$ defines a vector field
$\restric{X_{h}}{S_{\lambda}}$, and each of these gives rise to a vector field $X_{\lambda}$ on
the base $Q$. The integral curves of $X_\lambda$ provide the integral curves of $\restric{X_h}{S_\lambda}$,
while the integral curves of all the set $\left\{ X_\lambda \right\}$ provide all the integral
curves of $X_h$. All these considerations can be made in the same way for the Lagrangian formalism.

The geometrical setting for Hamilton-Jacobi theory was pointed out in several works
\cite{book:Abraham_Marsden78,proc:Benenti_Tulczyjew80, art:Dominici_Gomis_Longhi_Pons84,book:Esposito_Marmo_Sudarshan04,
proc:Longhi_Dominici_Gomis_Pons82,art:Marmo_Morandi_Mukunda90}, and has been finally established in
\cite{art:Carinena_Gracia_Marmo_Martinez_Munoz_Roman06} for first-order autonomous systems, both in
the regular case and in constrained systems. More recently, it has been generalized to nonholonomic
dynamical systems \cite{art:Carinena_Gracia_Marmo_Martinez_Munoz_Roman10,art:DeLeon_Marrero_Martin10,
art:Iglesias_DeLeon_Martin08,art:Ohsawa_Bloch09},
discrete mechanics \cite{art:Ohsawa_Bloch_Leok11}, Lie affgebroids \cite{art:Marrero_Sosa06} and
field theories \cite{art:DeLeon_Martin_Marrero_Salgado_Vilarino10,art:DeLeon_Vilarino13,
art:Vitagliano10_2,art:Vitagliano12}. Nowadays, a geometrical Hamilton-Jacobi theory for singular
systems has been developed using a Skinner-Rusk approach \cite{art:DeLeon_Martin_Vaquero12}.
However, up to our knowledge, there is not a definitive Hamilton-Jacobi theory for higher-order systems,
not even a non-geometrical one.

\subsection*{Structure of the dissertation}

This dissertation is structured in 6 Chapters. The first two Chapters review the mathematical and
physical backgrounds needed, and fix the notation used along the rest of the dissertation. The last
four Chapters contain the main original contributions.
A reader which is familiar with the geometric tools and formulations, and with the notation, can
skip the first two Chapters and start with the main contributions of the thesis, in Chapter
\ref{Chap:HOAutonomousDynamicalSystems}.

Chapter \ref{Chap:MathBackground} is devoted to introduce the main mathematical tools needed to
give a geometric description of a physical theory. In particular, the concepts of symplectic,
cosymplectic and multisymplectic manifolds are introduced, as well as jet bundles,
higher-order tangent bundles and multivector fields. In addition, a purely geometric description
of the classical constraint algorithm is given.

The second review, now focused on the background in mathematical physics, is found in Chapter
\ref{Chap:MathPhysBackground}, where the geometric formulations of several different theories
are reviewed in detail. In particular, we give the geometric Lagrangian and Hamiltonian formalisms
of first-order dynamical systems, both in the autonomous and non-autonomous cases; higher-order
autonomous systems, first-order Hamilton-Jacobi theory, and first-order field theories.
In addition, we also review the Lagrangian-Hamiltonian formalism of first-order dynamical systems
(again, both in the autonomous and non-autonomous cases) and first-order field theories.

Main contributions of the thesis begin in Chapter \ref{Chap:HOAutonomousDynamicalSystems},
where we give the Lagrangian-Hamiltonian formalism for higher-order autonomous dynamical
systems. In addition, we recover both the Lagrangian and Hamiltonian formalisms for higher-order
autonomous systems from the unified setting, following the patterns of the original work
\cite{art:Skinner_Rusk83} by R. Skinner and R. Rusk, which is reviewed in the previous Chapter.
In this way we prove that our results are consistent with the Lagrangian and Hamiltonian formalisms
for higher-order autonomous systems described by M. de Le\'on and P.R. Rodrigues in
\cite{book:DeLeon_Rodrigues85}, which have been reviewed in the previous Chapter.
Finally, two physical models are analyzed to show the application of the formalism:
the \textsl{Pais-Uhlenbeck oscillator} and a \textsl{second-order relativistic particle}.

Next, Chapter \ref{Chap:HOHamiltonJacobi} is devoted to generalize the geometric formulation
of the Hamilton-Jacobi theory \cite{art:Carinena_Gracia_Marmo_Martinez_Munoz_Roman06} to
higher-order autonomous dynamical systems. More particularly, starting from the Lagrangian and
Hamiltonian formalisms for higher-order dynamical systems \cite{book:DeLeon_Rodrigues85}, we
generalize the construction in \cite{art:Carinena_Gracia_Marmo_Martinez_Munoz_Roman06} to
higher-order autonomous systems described by regular Lagrangian functions. In addition,
using the results of Chapter \ref{Chap:HOAutonomousDynamicalSystems}, we also establish the
unified formulation of the Hamilton-Jacobi problem for these systems, as a first-step to study
the case of singular Lagrangian functions in further research. Finally, two regular examples are
analyzed to illustrate the features of all the three formulations: the \textsl{end of a thrown javelin}
and the \textsl{shape of a homogeneous deformed elastic cylindrical beam with fixed ends}.

In Chapter \ref{Chap:HONonAutonomousDynamicalSystems} we combine the geometric Lagrangian-Hamiltonian
unified formalism for higher-order non-autonomous systems \cite{art:Barbero_Echeverria_Martin_Munoz_Roman08}
with the geometric formulations for higher-order autonomous systems in \cite{book:DeLeon_Rodrigues85}
and Chapter \ref{Chap:HOAutonomousDynamicalSystems} to state the Lagrangian-Hamiltonian formalism
for higher-order non-autonomous systems. From this unified setting, and following the patterns in
Chapter \ref{Chap:HOAutonomousDynamicalSystems}, we derive a complete description of both the
Lagrangian and Hamiltonian formalisms for these systems. In addition, two physical models are studied
using our formulations: the \textsl{shape of a non-homogeneous deformed elastic cylindrical beam with
fixed ends} and a \textsl{second-order relativistic particle subjected to a time dependent potential}.

Finally, Chapter \ref{Chap:HOClassicalFieldTheories} focuses on giving an unambiguous geometric
formulation of second-order field theories using a similar approach as in Chapter
\ref{Chap:HONonAutonomousDynamicalSystems}: we first state the Lagrangian-Hamiltonian unified
formalism for these theories, and then we derive both the Lagrangian and Hamiltonian formalisms from
the unified setting. This formulation removes all the usual ambiguities of a second-order field
theory introducing a relation of symmetry among the highest-order multimomentum coordinates. As a
consequence of this, a unique Legendre map is obtained from the constraint algorithm and, therefore,
a unique Poincar\'{e}-Cartan is obtained to state the Lagrangian formalism for second-order field
theories. In addition, some comments on the general higher-order case are given. Finally, two physical
models are studied with these formulations: the \textsl{bending of a clamped plate under a uniform load}
and the classic \textsl{Korteweg--de Vries equation}.

Observe that, except for Chapter \ref{Chap:HOHamiltonJacobi}, where only the regular case is analyzed, there
is a regular and a singular example in every Chapter.

All the manifolds are connected, second countable and $\Cinfty$.
The maps and the structures are assumed to be $\Cinfty$.
Summation over crossed repeated indexes is understood, although on some occasions
the symbol of summation is written explicitly in order to avoid confusion.


\clearpage
\thispagestyle{empty}



\renewcommand{\chaptermark}[1]{\markboth{\chaptername\ \thechapter.\ \ #1}{}}
\renewcommand{\sectionmark}[1]{\markright{\thesection.\ \ #1}{}}


\clearpage
\chapter{Mathematical background}
\label{Chap:MathBackground}


In this first Chapter we review the main mathematical tools used along this dissertation: definitions,
main results and, in some cases, fundamental examples that will be used further. This Chapter will also
be useful to fix the common notation along different Chapters. Since this is a review Chapter, no proofs
are given: several references containing proofs and details are included at the beginning of each Section.
Note that only the results used in further Chapters are given, and therefore this Chapter should not be
considered as a thorough introduction to any of the topics given.

The structure of this Chapter is the following: Section \ref{Chap01_sec:SymplecticGeom} introduces the
concept of \textsl{symplectic forms}, that is, nondegenerate closed $2$-forms on a manifold, as well as
consequences of having such a form on a manifold, and several geometric structures derived from it.
Sections \ref{Chap01_sec:CosymplecticGeom} and \ref{Chap01_sec:MultisymplecticGeom} introduce
generalizations of the concept of symplectic form to odd-dimensional manifolds and forms of degree greater
than $2$, respectively. Sections \ref{Chap01_sec:HOJetBundles} and \ref{Chap01_sec:HOTanBundle} generalize
the tangent bundle of a manifold to consider derivatives of higher-order and with respect to more than one
independent variable. In Section \ref{Chap01_sec:MultivectorFields} we introduce the concept of multivector
fields as skew-symmetric contravariant tensors of arbitrary degree on a manifold, in an analogous way to
differential forms of higher degree. Finally, Section \ref{Chap01_sec:ConstraintAlgorithm} is devoted to
study the problem of solving a geometric equation of the type $\inn(X)\omega = \alpha$, when the form $\omega$
is degenerate.

\section{Symplectic geometry}
\label{Chap01_sec:SymplecticGeom}

In this first section we introduce the basic concepts on symplectic manifolds. Many references introduce
the foundations on symplectic manifolds. For details and proofs, see, for example,
\cite{book:Abraham_Marsden78,book:Cannas01,book:Libermann_Marle87,book:Weinstein77,art:Weinstein81}
(among others).

Throughout this Section, $M$ will denote a finite-dimensional smooth manifold.

\subsection{Symplectic forms. Darboux's Theorem. Symplectomorphisms}
\label{Chap01_sec:SymplecticFormsDef}

\begin{definition}
A \textnormal{symplectic form} in $M$ is a closed $2$-form $\omega \in \df^{2}(M)$ which is nondegenerate,
that is, for every $p \in M$, $\inn(X_p)\,\omega_p = 0$ if, and only if, $X_p = 0$. If $\omega$ is closed
and degenerate, it is called a \textnormal{presymplectic form}.
A \textnormal{symplectic manifold} (resp., a \textnormal{presymplectic manifold})
is a couple $(M,\omega)$, where $M$ is a smooth manifold and $\omega$
is a symplectic form (resp., a presymplectic form).
\end{definition}

\begin{remark}
If $(M,\omega)$ is a symplectic manifold, then the nondegeneracy of $\omega$ implies that $M$ has even
dimension, that is, $\dim M = 2n$.
\end{remark}

The fundamental result in symplectic geometry is Darboux's Theorem, which gives a local model for every
symplectic manifold.

\begin{theorem}[Darboux]
Let $(M,\omega)$ be a $2n$-dimensional symplectic manifold. Then for every $p \in M$ there exists a local
chart $(U;(x^i,y_i))$ on $p$, with $1 \leqslant i \leqslant n$, such that the coordinate expression of
$\omega$ in this local chart is 
\begin{equation*}
\restric{\omega}{U} = \d x^i \wedge \d y_i \, .
\end{equation*}
Such a local chart is called \textnormal{Darboux}, \textnormal{symplectic} or \textnormal{canonical chart},
and its coordinates are called \textnormal{Darboux}, \textnormal{symplectic} or \textnormal{canonical coordinates}.
\end{theorem}

\begin{remark}
There is a similar result for presymplectic manifolds. Indeed, if $(M,\omega)$ is a $(2n+k)$-dimensional
presymplectic manifold and $\rank(\omega) = 2n$, then for every $p \in M$ there exists a local chart
$(U;(x^i,y_i,z^j))$ on $p$, with $1 \leqslant i \leqslant n$ and $1 \leqslant j \leqslant k$, such that
the coordinate expression of $\omega$ in this local chart is
\begin{equation*}
\restric{\omega}{U} = \d x^i \wedge \d y_i \, .
\end{equation*}
\end{remark}

Finally, we define morphisms between symplectic manifolds.

\begin{definition}
Let $(M_1,\omega_1)$ and $(M_2,\omega_2)$ be two symplectic manifolds. A \textnormal{symplectic map}
is a smooth map $\Phi \colon M_1 \to M_2$ such that $\Phi^*\omega_2 = \omega_1$.
If, in addition, $\Phi$ is a diffeomorphism then it is called a \textnormal{symplectomorphism}.
\end{definition}

\begin{example}[The cotangent bundle]\label{Chap01_exa:CotangentBundle}
Let $Q$ be a $n$-dimensional smooth manifold, and consider its cotangent bundle $\Tan^*Q$. We define a
$1$-form $\theta \in \df^{1}(\Tan^*Q)$ by
\begin{equation*}
\theta_\alpha(X_\alpha) = \alpha((\Tan_\alpha\pi_Q)(X_\alpha)) \, ,
\end{equation*}
where $X_\alpha \in \Tan_\alpha(\Tan^*Q)$ and $\alpha \in \Tan^*Q$. This $1$-form is called the
\textsl{Liouville $1$-form},
or also \textsl{canonical} or \textsl{tautological} $1$-form in $\Tan^*Q$.
We now define on $\Tan^*Q$ the canonical $2$-form
\begin{equation}
\omega = - \d\theta \, ,
\end{equation}
which is nondegenerate, and thus symplectic. It is called the \textsl{Liouville $2$-form}, or also the
\textsl{canonical symplectic form} of the cotangent bundle.

In coordinates, if $(q^A)$, $1 \leqslant A \leqslant n$ are local coordinates in $Q$, then the induced
local coordinates in $\Tan^*Q$ are $(q^A,p_A)$. Then the local expression of the tautological form is
\begin{equation}\label{Chap01_eqn:CanonLiouvilleFormCotanBundle}
\theta = p_A\d q^A \, ,
\end{equation}
from where the coordinate expression of the canonical symplectic form of $\Tan^*Q$ is
\begin{equation}\label{Chap01_eqn:CanonSympFormCotanBundle}
\omega = \d q^A \wedge \d p_A \, .
\end{equation}
Observe that the natural coordinates of the cotangent bundle coincide with the Darboux coordinates.
\end{example}

\begin{remark}
Notice that, from Darboux's Theorem, every symplectic manifold is locally symplectomorphic to a cotangent bundle.
\end{remark}

\subsection{Canonical isomorphism. Hamiltonian vector fields}
\label{Chap01_sec:CanonicalIsomorphismSymplectic}

Given a $2$-form $\omega \in \df^{2}(M)$, we can define a linear bundle morphism between the tangent
and cotangent bundles of $M$ as follows
\begin{equation*}
\begin{array}{rcl}
\omega^\flat \colon \Tan M & \longrightarrow & \Tan^*M \\
(p,v_p) & \longmapsto & (p,\inn(v_p)\omega_p)
\end{array} \, .
\end{equation*}
This bundle morphism is extended to the modules of vector fields and $1$-forms in a natural way, obtaining
the following morphism of $\Cinfty(M)$-modules (which, in an abuse of notation, we also denote by $\omega^\flat$)
\begin{equation*}
\begin{array}{rcl}
\omega^\flat \colon \vf(M) & \longrightarrow & \df^{1}(M) \\
X & \longmapsto & \inn(X)\omega
\end{array} \, .
\end{equation*}
Now, given a $2n$-dimensional smooth manifold and a closed form $\omega \in \df^{2}(M)$, it is clear that
$\omega$ is nondegenerate (that is, symplectic) if, and only if, the map $\omega^\flat$ is an isomorphism.

\begin{remark}
If $M$ has infinite dimension, then the map $\omega^\flat$ can be injective but not bijective. In this case,
it is said that $\omega$ is \textsl{weakly nondegenerate} (resp., \textsl{strong nondegenerate}) if, and
only if, $\omega^\flat$ is injective (resp., bijective), and therefore we have weak and strong symplectic forms.
\end{remark}

\begin{definition}
If $(M,\omega)$ is a symplectic manifold, the map $\omega^\flat \colon \vf(M) \to \df^{1}(M)$ defined above
is the \textnormal{canonical isomorphism}, or also \textnormal{musical} or \textnormal{flat isomorphism}.
Its inverse is denoted $\omega^\sharp \colon \df^{1}(M) \to \vf(M)$, and is called \textnormal{sharp isomorphism}.
\end{definition}

Now, given a symplectic manifold $(M,\omega)$, every function $f \in \Cinfty(M)$ has a unique vector field
$X_f \in \vf(M)$ associated to it using the map $\omega^\sharp \circ \d \colon \Cinfty(M) \to \vf(M)$, that
is, $X_f$ is defined explicitly by $X_f = \omega^\sharp(\d f)$, or implicitly as the solution to the equation
\begin{equation}\label{Chap01_eqn:GeneralHamVF}
\inn(X_f)\omega = \d f \, .
\end{equation}

\begin{remark}
Notice that the map $\omega^\sharp \circ \d$ is not injective neither surjective. It is clear that it is
not injective since two functions differing in a constant have the same exterior derivative, and therefore
the same associated vector field. On the other hand, it is not surjective since, even if the sharp isomorphism
gives a one-to-one correspondence between $1$-forms and vector fields, the $1$-form obtained may not be exact
(in general, not even closed).
\end{remark}

As a consequence of this remark, we can give the following definition.

\begin{definition}\label{Chap01_def:HamiltonianVF}
Let $(M,\omega)$ be a symplectic manifold. A vector field $X \in \vf(M)$ is a \textnormal{(global) Hamiltonian
vector field}
if the $1$-form $\inn(X)\,\omega$ is exact. In this case, the function $f \in \Cinfty(M)$
satisfying $\inn(X)\,\omega = \d f$ is the \textnormal{(global) Hamiltonian function}
of the vector field $X$.
\end{definition}

\begin{remark}
Usually, the function $f \in \Cinfty(M)$ is given, and we must look for the vector field $X_f \in \vf(M)$
solution to \eqref{Chap01_eqn:GeneralHamVF}.
In this cases, we refer to $X_f$ as the ``Hamiltonian vector field associated to $f$''.
\end{remark}

\begin{remark}
There is a less restrictive definition of Hamiltonian vector fields, which comes from lessening the condition
of $\inn(X)\omega$ being an exact $1$-form to just a closed $1$-form. These vector fields are called
\textsl{local Hamiltonian vector fields}, and the local function $f$ satisfying locally
\eqref{Chap01_eqn:GeneralHamVF}, which exists due to Poincar\'{e}'s Lemma, is the \textsl{local Hamiltonian function}.
\end{remark}

In coordinates, let $(U;(x^i,y_i))$, $1 \leqslant i \leqslant n$, be a symplectic chart on $M$.
In these coordinates, a generic vector field and the exterior derivative of any function are given by
\begin{equation*}
X = A^i\derpar{}{x^i} + B_i\derpar{}{y_i} \quad ; \quad \d f = \derpar{f}{x^i}\d x^i + \derpar{f}{y_i} \d y_i \, .
\end{equation*}
Then, the vector field $X$ is a Hamiltonian vector field for $f$ if the following system of $2n$ equations holds
\begin{equation*}
A^i = \derpar{f}{y_i} \quad ; \quad B_i = - \derpar{f}{x^i} \, ,
\end{equation*}
that is, if $X$ is given by
\begin{equation}\label{Chap01_eqn:GenericHamVFLocal}
X = \derpar{f}{y_i}\derpar{}{x^i} - \derpar{f}{x^i}\derpar{}{y_i} \, .
\end{equation}
Finally, if $\gamma(t) = (x^i(t),y_i(t))$ is an integral curve of $X$, then its component functions must
satisfy the following system of $2n$ ordinary differential equations
\begin{equation*}
\dot{x}^i = \derpar{f}{y_i} \circ \gamma \quad ; \quad
\dot{y}_i = - \derpar{f}{x^i} \circ \gamma \, ,
\end{equation*}
which are called \textsl{Hamilton equations} of the Hamiltonian vector field.

\begin{remark}
Observe that if $\gamma \colon \R \to M$ is an integral curve of a Hamiltonian vector field $X_f$ associated
to a function $f$, that is, we have $\dot{\gamma} = X_f \circ \gamma$, then the curve $\gamma$ must satisfy
the following geometric equation
\begin{equation*}
\inn(\dot{\gamma})(\omega \circ \gamma) = \d (f \circ \gamma) \, ,
\end{equation*}
which is the analogous to equation \eqref{Chap01_eqn:GeneralHamVF} for curves.
\end{remark}

\subsection{Isotropic, coisotropic and Lagrangian submanifolds}
\label{Chap01_sec:LagrangianSubmanifolds}

The existence of a nondegenerate $2$-form on symplectic manifolds enables us to define some particular
submanifolds since, in some sense, the symplectic form may be seen as a skew-symmetric ``inner product''
on the tangent bundle of the manifold, in an analogous way to the case of a metric tensor.
This allows us to give the following definition.

\begin{definition}\label{Chap01_def:SymplecticOrthogonal}
Let $(M,\omega)$ be a symplectic manifold, and $F \subseteq \Tan M$ a vector subbundle. The
\textnormal{$\omega$-orthogonal} of $F$, or \textnormal{symplectic orthogonal},
is the subbundle $F^\bot \subseteq \Tan M$ defined as
\begin{equation*}
F^\bot = \left\{ (p,u_p) \in \Tan M \mid \omega_p(u_p,v_p) = 0 \text{ for every } (p,v_p) \in F \right\} \, .
\end{equation*}
\end{definition}

Once the symplectic orthogonal of a subbundle is defined, we can ``classify'' the subbundles of
$\Tan M$ in three classes, depending whether they contain their symplectic orthogonal,
they are contained in it, or they are exactly the same subbundle.

\begin{definition}
Let $(M,\omega)$ be a symplectic manifold, and $F \subseteq \Tan M$ a vector subbundle of $\Tan M$.
\begin{enumerate}
\item $F$ is an \textnormal{isotropic subbundle} if $F \subseteq F^\bot$, that is,
$\omega_p(u_p,v_p) = 0$ for all $(p,u_p),(p,v_p) \in F$.
\item $F$ is a \textnormal{coisotropic subbundle} if $F \supseteq F^\bot$, that is,
$\omega_p(u_p,v_p) = 0$ for every $(p,v_p) \in F$ implies $(p,u_p) \in F$.
\item $F$ is a \textnormal{Lagrangian subbundle} if $F = F^\bot$, that is,
if $F$ is both an isotropic and coisotropic subbundle.
\end{enumerate}
\end{definition}

Finally, the definition of isotropic, coisotropic and Lagrangian subbundles is generalized to
immersed submanifolds as follows.

\begin{definition}
Let $(M,\omega)$ be a symplectic manifold, and $N \hookrightarrow M$ a submanifold with canonical embedding
$i \colon N \hookrightarrow M$. Let us consider the subbundle $\Tan i(\Tan N) \subseteq \Tan M$.
\begin{enumerate}
\item $N$ is an \textnormal{isotropic immersed submanifold} if $\Tan i(\Tan N)$ is an isotropic subbundle.
\item $N$ is a \textnormal{coisotropic immersed submanifold} if $\Tan i(\Tan N)$ is a coisotropic subbundle.
\item $N$ is a \textnormal{Lagrangian immersed submanifold} if $\Tan i(\Tan N)$ is a Lagrangian subbundle.
\end{enumerate}
\end{definition}

\begin{remark}
In the following, we will call an isotropic (resp., coisotropic, Lagrangian) immersed submanifold simply
as an \textsl{isotropic} (resp., \textsl{coisotropic, Lagrangian}) \textsl{submanifold}.
\end{remark}

\begin{remark}
If the $2$-form $\omega$ is presymplectic (that is, closed and degenerate), we are still able to define the
(pre)symplectic orthogonal with respect to $\omega$, exactly in the same way, and the notions of isotropic,
coisotropic and Lagrangian submanifolds are defined analogously. See \cite{phd:Gotay} for details.
\end{remark}

Finally, some characterizations of both isotropic and Lagrangian submanifolds are the following.

\begin{lemma}
A submanifold $i \colon N \hookrightarrow M$ is isotropic if, and only if, $i^*\omega = 0$.
\end{lemma}

\begin{proposition}
Let $(M,\omega)$ be a symplectic manifold, and $N \hookrightarrow M$ an embedded submanifold.
Then, the following assertions are equivalent:
\begin{enumerate}
\item $N$ is a Lagrangian submanifold of $(M,\omega)$.
\item $N$ is an isotropic submanifold with $\dim N = \frac{1}{2}\dim M$.
\item $N$ is an isotropic submanifold and $\Tan N$ admits an isotropic complement, that is,
there exists an isotropic subbundle $E \subseteq \restric{\Tan M}{N}$ such that $\restric{\Tan M}{N} = \Tan N \oplus E$.
\end{enumerate}
\end{proposition}

\begin{example}[The graph of a closed $1$-form]\label{Chap01_exa:Closed1Form}
Let $Q$ be a $n$-dimensional smooth manifold, and let us consider the cotangent bundle of $Q$, endowed with
the canonical symplectic form $\omega \in \df^{2}(\Tan^*Q)$, as we have seen in Example
\ref{Chap01_exa:CotangentBundle}. Let $\alpha \in \df^{1}(Q)$ be a $1$-form on $Q$. Then, the submanifold
$\Im(\alpha) \hookrightarrow \Tan^*Q$ is a Lagrangian submanifold of $(\Tan^*Q,\omega)$ if, and only if, $\alpha$ is closed.

In order to prove this, first observe that the canonical embedding $\Im(\alpha) \hookrightarrow \Tan^*Q$ may be identified
with the $1$-form $\alpha$ itself. Then, from the definition of the Liouville $1$-form $\theta \in \df^{1}(\Tan^*Q)$,
we have $\alpha^*\theta = \alpha$, and therefore
\begin{equation*}
\alpha^*\omega = \alpha^*(-\d\theta) = -\d\alpha^*\theta = -\d\alpha \, ,
\end{equation*}
which proves that $\Im(\alpha)$ is an isotropic submanifold of $\Tan^*Q$ if, and only if, $\alpha$ is closed. However,
since $\dim\Im(\alpha) = n = \frac{1}{2}\dim \Tan^*Q$, this is equivalent to $\Im(\alpha)$ being a Lagrangian submanifold
of $\Tan^*Q$.
\end{example}

\subsection{Poisson bracket}

In a symplectic manifold, the symplectic form induces in a natural way some well-known operations in
analytical mechanics.

\begin{definition}\label{Chap01_def:PoissonBracketDef}
Let $(M,\omega)$ be a symplectic manifold. The \textnormal{Poisson bracket}
(induced by $\omega$)
of two functions $f,g \in \Cinfty(M)$ is the bilinear map defined as
\begin{equation}\label{Chap01_eqn:PoissonBracketDef}
\begin{array}{rcl}
\left\{ \cdot \, , \cdot \right\} \colon \Cinfty(M) \times \Cinfty(M) & \longrightarrow & \Cinfty(M) \\
(f,g) & \longmapsto & \left\{ f,g \right\}
\end{array}
\end{equation}
where $\{f,g\} = \omega(X_f,X_g) = \inn(X_g)\inn(X_f)\,\omega$, and $X_f,X_g \in \vf(M)$ are the Hamiltonian
vector fields associated to $f$ and $g$, respectively.
\end{definition}

The Poisson bracket satisfies the following properties:
\begin{enumerate}
\item Skew-symmetric: $\{f,g\} = -\{g,f\}$.
\item Jacobi identity: $\{f,\{g,h\}\} + \{g,\{h,f\}\} + \{h,\{f,g\}\} = 0$.
\item $\{f,g\} = \Lie(X_g)f = -\Lie(X_f)g$.
\item $X_{\{f,g\}} = [X_g,X_f]$, where $[ \cdot\,,\cdot] \colon \vf(M) \times \vf(M) \to \vf(M)$
is the usual Lie bracket of vector fields.
\end{enumerate}

\begin{remarks} \ \vspace{-5pt}
\begin{itemize}
\item From properties $1$ and $2$ we conclude that $(\Cinfty(M),\{ \cdot \, , \cdot \})$ is a Lie algebra.
On the other hand, from the fourth property there exists an antihomomorphism of Lie algebras between
$(\vf(M),[ \cdot \, , \cdot ])$ and $(\Cinfty(M),\{ \cdot \, , \cdot \})$. \hfill$\lozenge$
\item The Poisson bracket can be extended to the set of differential $1$-forms using the canonical isomorphisms.
In particular, the Poisson bracket of $1$-forms is the bilinear map
\begin{equation*}
\begin{array}{rcl}
\left\{ \cdot \, , \cdot \right\} \colon \df^{1}(M) \times \df^{1}(M) & \longrightarrow & \df^{1}(M) \\
(\alpha,\beta) & \longmapsto & \left\{ \alpha,\beta \right\}
\end{array}
\end{equation*}
defined by $\left\{ \alpha,\beta \right\} = \omega^\flat([\omega^\sharp(\alpha),\omega^\sharp(\beta)])$.
\hfill$\lozenge$
\end{itemize}
\end{remarks}

In coordinates, let $(U;(x^i,y_i))$, $1 \leqslant i \leqslant n$, be a symplectic chart on $M$.
Bearing in mind the local expression on a Darboux chart \eqref{Chap01_eqn:GenericHamVFLocal}
of the Hamiltonian vector field associated to a function $f$, we have that the local expression
of the Poisson bracket of two functions $f$ and $g$ is
\begin{equation}\label{Chap01_eqn:PoissonBracketLocal}
\left\{ f,g \right\} = \derpar{f}{x^i}\derpar{g}{y_i} - \derpar{f}{y_i}\derpar{g}{x^i} \, .
\end{equation}


\section{Cosymplectic geometry}
\label{Chap01_sec:CosymplecticGeom}

Cosymplectic geometry is the natural extension of symplectic geometry to odd-dimensional manifolds. For
details and proofs, see, for example,
\cite{art:Cappelletti_deNicola_Yudin,art:Chinea_DeLeon_Marrero91,art:DeLeon_Tuynman96,art:Vaisman85}.

Through this Section, $M$ will denote an odd-dimensional smooth manifold, that is, $\dim M = 2n+1$.

\subsection{Cosymplectic structures. Darboux's Theorem}

\begin{definition}
A \textnormal{cosymplectic structure} on an odd-dimensional smooth
manifold $M$ is a pair $(\omega,\eta)$, where $\omega \in \df^{2}(M)$ and $\eta \in \df^{1}(M)$ are
both closed forms, such that the exterior product
$(\Lambda^n \omega) \wedge \eta \equiv \omega^n \wedge \eta$ is a volume form on $M$. If this last condition
fails, then $(\omega,\eta)$ is a \textnormal{precosymplectic structure}. A \textnormal{(pre)cosymplectic manifold}
is an odd-dimensional smooth manifold endowed with a (pre)cosymplectic structure, that is, a triple $(M,\omega,\eta)$
where $\dim M = 2n+1$ and $(\omega,\eta)$ is a (pre)cosymplectic structure.
\end{definition}

As in the symplectic geometry, a fundamental result in cosymplectic geometry is an analogous to Darboux's
Theorem, which is also called ``Darboux's Theorem'', and that gives a local model for every cosymplectic manifold.

\begin{theorem}[Darboux]
Let $(M,\omega,\eta)$ be a $(2n+1)$-dimensional cosymplectic manifold. Then for every $p \in M$ there exists
a local chart $(U;(t,x^i,y_i))$ on $p$, with $1 \leqslant i \leqslant n$, such that the coordinate expressions
of $\omega$ and $\eta$ in this local chart are
\begin{equation*}
\omega = \d x^i \wedge \d y_i \quad ; \quad \eta = \d t \, .
\end{equation*}
Such a local chart is called a \textnormal{Darboux}, \textnormal{canonical} or \textnormal{cosymplectic chart},
and its coordinates are called \textnormal{Darboux}, \textnormal{canonical} or \textnormal{cosymplectic coordinates}.
\end{theorem}

\begin{example}
Let $Q$ be a $n$-dimensional smooth manifold, and let us consider its cotangent bundle $\Tan^*Q$. As we have
seen in the example in Section \ref{Chap01_sec:SymplecticGeom}, the cotangent bundle is endowed with a canonical
symplectic form $\omega \in \df^{2}(\Tan^*Q)$. Now let us consider the product of the real line with the cotangent
bundle of $Q$, that is, $\R \times \Tan^*Q$. This manifold is endowed with a canonical projection over each factor,
namely $\pr_1 \colon \R \times \Tan^*Q \to \R$ and $\pr_2 \colon \R \times \Tan^*Q \to \Tan^*Q$. Since $\R$ is
an oriented manifold, let $\eta \in \df^{1}(\R)$ be the canonical volume form. Then, the pair
$(\pr_2^*\omega,\pr_1^*\eta)$ is a cosymplectic structure on $\R \times \Tan^*Q$.

In coordinates, let $(t)$ be the global coordinate on $\R$ such that $\eta = \d t$ and $(q^A,p_A)$ the induced
local coordinates on $\Tan^*Q$. Then, the induced coordinates in $\R \times \Tan^*Q$ adapted to the bundle
structure are $(t,q^A,p_A)$, and they coincide with the Darboux coordinates of the cosymplectic manifold,
since the forms $\pr_2^*\omega$ and $\pr_1^*\eta$ have the following coordinate expression
\begin{equation*}
\pr_2^*\omega = \d q^A \wedge \d p_A \quad ; \quad \pr_1^*\eta = \d t \, .
\end{equation*}
\end{example}

\subsection{Canonical isomorphism. Reeb vector fields}
\label{Chap01_sec:CosymplecticGeomReebVF}

As in the case of symplectic geometry, given a $2$-form and a $1$-form on a manifold $M$, we can define
a linear bundle morphism between the tangent and cotangent bundles of $M$ as follows
\begin{equation*}
\begin{array}{rcl}
\flat \colon \Tan M & \longrightarrow & \Tan^*M \\
(p,v_p) & \longmapsto & (p,\inn(v_p)\omega_p + \langle \eta_p,v_p \rangle \eta_p)
\end{array} \, .
\end{equation*}
where $\langle \cdot\,,\cdot \rangle \colon \Tan^*_pM \times \Tan_pM \to \R$ is the canonical pairing between
elements of the vector space $\Tan_pM$ and its dual $\Tan^*_pM$. This bundle morphism can be extended to the
modules of vector fields and $1$-forms in a natural way, obtaining the following morphism of
$\Cinfty(M)$-modules (which, in an abuse of notation, we also denote by $\flat$)
\begin{equation*}
\begin{array}{rcl}
\flat \colon \vf(M) & \longrightarrow & \df^{1}(M) \\
X & \longmapsto & \inn(X)\omega + (\inn(X)\eta)\eta
\end{array} \, .
\end{equation*}
Now, if $M$ has $\dim M = 2n + 1$ and both $\omega$ and $\eta$ are closed forms, then it is clear that
the pair $(\omega,\eta)$ is a cosymplectic structure on $M$ if, and only if, the map $\flat$ is an
isomorphism of $\Cinfty(M)$-modules.

\begin{definition}
If $(M,\omega,\eta)$ is a cosymplectic manifold, the map $\flat \colon \vf(M) \to \df^{1}(M)$ defined above
is the \textnormal{canonical isomorphism}, or also \textnormal{musical} or \textnormal{flat isomorphism}.
Its inverse is denoted $\sharp \colon \df^{1}(M) \to \vf(M)$ and is called \textnormal{sharp isomorphism}.
\end{definition}

\begin{remark}
Even if we denote them differently, we named this canonical isomorphism in the same way as we did for symplectic
manifolds. In most cases, it will be clear to which isomorphism we refer to, but we will clarify it to avoid
confusion in subsequent Chapters.
\end{remark}

Since $\flat$ is an isomorphism between the modules of vector fields and $1$-forms in the cosymplectic
manifold $(M,\omega,\eta)$, we can take the pre-image of any $1$-form in $M$ to obtain a unique vector
field. In particular, we can take the pre-image of the closed $1$-form $\eta$. The unique vector field
$R \in \vf(M)$ satisfying $R = \flat^{-1}(\eta) = \sharp(\eta)$ is called the \textsl{Reeb vector field}
of the cosymplectic manifold $(M,\omega,\eta)$. Note that the Reeb vector field is characterized by the equations
\begin{equation}\label{Chap01_eqn:GeneralReebVF}
\inn(R)\omega = 0 \quad ; \quad \inn(R)\eta = 1 \, .
\end{equation}

In coordinates, let $(U;(t,x^i,y_i))$, $1 \leqslant i \leqslant n$, be a cosymplectic chart on $(M,\omega,\eta)$.
In these coordinates, a generic vector field is given by
\begin{equation*}
X = C\derpar{}{t} + A^i\derpar{}{x^i} + B_i\derpar{}{y_i} \, .
\end{equation*}
Then, the vector field $X$ is the Reeb vector field of the cosymplectic manifold $(M,\omega,\eta)$ if the
following system of $2n+1$ equations holds
\begin{equation*}
A^i = 0 \quad ; \quad B_i = 0 \quad ; \quad C = 1 \, , 
\end{equation*}
that is, if $X$ is given by
\begin{equation*}
X = \derpar{}{t} \, .
\end{equation*}
Finally, if $\gamma(s) = (t(s),x^i(s),y_i(s))$ is an integral curve of $X$, then its component functions
must satisfy the following system of $2n+1$ ordinary differential equations
\begin{equation*}
\dot{x}^i = 0 \quad ; \quad \dot{y}_i = 0 \quad ; \quad \dot{t} = 1 \, .
\end{equation*}

\begin{remark}
From a physical point of view, if we consider the coordinate $t$ on a cosymplectic manifold as the ``time''
of a time-dependent dynamical system, then the Reeb vector field is the vector field that fixes the
progression of time to its ``standard'' value, that is, the Reeb vector field fixes the gauge among all the
reparametrizations of the time coordinate on a time-dependent dynamical system.
\end{remark}


\section{Multisymplectic geometry}
\label{Chap01_sec:MultisymplecticGeom}

Multisymplectic forms are a natural generalization of the concept of symplectic forms to forms of degree
greater than $2$. That is, while a symplectic form is a closed $2$-form which is nondegenerate, a
multisymplectic form will be a closed $k$-form which is ``nondegenerate'' in some sense. For details
and proofs, we refer to
\cite{art:Cantrijn_Ibort_DeLeon96,art:Cantrijn_Ibort_DeLeon99,art:Echeverria_Ibort_Munoz_Roman12}

Along this section, $M$ will denote a smooth manifold with $\dim M = m$.

\begin{definition}
A \textnormal{multisymplectic $k$-form}
in $M$ is a closed $k$-form $\omega \in \df^{k}(M)$ which, in addition, is
$1$-nondegenerate, that is, for every $p \in M$, $\inn(X_p)\omega_p = 0$ if, and only if, $X_p = 0$.
If $\omega$ is closed and $1$-degenerate, it is called a \textnormal{premultisymplectic $k$-form}.
A manifold endowed with a multisymplectic $k$-form (resp., a premultisymplectic $k$-form) is called a
\textnormal{multisymplectic manifold of order $k$}
(resp., a \textnormal{premultisymplectic manifold of order $k$}).
\end{definition}

Observe that a necessary condition for a $k$-form to be $1$-nondegenerate is $1 < k \leqslant \dim M$.
The nondegeneracy condition is sometimes written in terms of the analog of the canonical isomorphism
for symplectic manifolds given in Section \ref{Chap01_sec:CanonicalIsomorphismSymplectic}. In particular,
given a $k$-form $\omega \in \df^{k}(M)$ we define the following morphism of $\Cinfty(M)$-modules
\begin{equation*}
\begin{array}{rcl}
\omega^\flat \colon \vf(M) & \longrightarrow & \df^{k-1}(M) \\
X & \longmapsto & \inn(X) \omega
\end{array} \, .
\end{equation*}
Then, a closed $k$-form $\omega \in \df^{k}(M)$ is $1$-nondegenerate (that is, multisymplectic),
and only if, the morphism $\omega^\flat$ defined above is injective.

\begin{remark}
Multisymplectic $2$-forms are just symplectic forms, as defined in Section \ref{Chap01_sec:SymplecticFormsDef}.
\end{remark}

\begin{example}[The multicotangent bundle]\label{Chap01_exa:MulticotangentBundle}
Let $Q$ be a $n$-dimensional smooth manifold, and let us consider the bundle of $k$-forms on $Q$, which is
$\Lambda^k(\Tan^*Q)$, that is, the $k$th exterior power of the cotangent bundle of $Q$. This bundle is
called the \textsl{multicotangent bundle of order $k$} of $Q$, and is sometimes denoted by $\Lambda^k(Q)$,
for short. Following the patterns in Example \ref{Chap01_exa:CotangentBundle}, we define a $k$-form
$\theta \in \df^{k}(\Lambda^k(\Tan^*Q))$ by
\begin{equation*}
\theta_\alpha(X_1(\alpha),\ldots,X_k(\alpha)) =
\alpha((\Tan_\alpha\pi_Q^k)(X_1(\alpha)),\ldots,(\Tan_\alpha\pi_Q^k)(X_k(\alpha)))
\end{equation*}
where $X_i(\alpha) \in \Tan_\alpha(\Lambda^k(\Tan^*Q))$, $1 \leqslant i \leqslant k$, and
$\alpha \in \Lambda^k(\Tan^*Q)$. This $k$-form is called the \textsl{tautological $k$-form} of $\Lambda^k(\Tan^*Q)$.
Then, taking its exterior derivative, we define the following $(k+1)$-form on $\Lambda^{k}(\Tan^*Q)$
\begin{equation*}
\omega = -\d \theta \, .
\end{equation*}
which, as we will see in the coordinate expression, is $1$-nondegenerate, and thus multisymplectic.
This $(k+1)$-form is called the \textsl{canonical multisymplectic form} on $\Lambda^k(\Tan^*Q)$.

In coordinates, if $(q^i)$, $1 \leqslant i \leqslant n$, are local coordinates in $Q$, then the induced natural
coordinates on $\Lambda^k\Tan^*Q$ are $(q^i,p_{i_1\ldots i_k})$, $1 \leqslant i_1 < \ldots < i_k \leqslant n$.
In these coordinates, the local expression of the tautological $k$-form $\theta$ is
\begin{equation*}
\theta = \sum_{1\leqslant i_1 < \ldots < i_k \leqslant n} p_{i_1\ldots i_k} \d q^{i_1} \wedge \ldots \wedge \d q^{i_k} \, ,
\end{equation*}
from where the coordinate expression of the canonical multisymplectic form of $\Lambda^{k}\Tan^*Q$ is
\begin{equation}\label{Chap01_eqn:CanonMultiSympFormMultiCotanBundle}
\omega = \sum_{1\leqslant i_1 < \ldots < i_k \leqslant n} -\d p_{i_1\ldots i_k} \wedge \d q^{i_1} \wedge \ldots \wedge \d q^{i_k} \, .
\end{equation}
\end{example}

\begin{example}\label{Chap01_exa:MulticotangentBundleVertical}
Let $\pi \colon E \to M$ be a fiber bundle. Let us consider the bundle of $k$-forms
on $E$ which are annihilated by the action of $r$ $\pi$-vertical vector fields, that is
\begin{equation*}
\Lambda^k_r(\Tan^*E) \equiv \Lambda^k_r E = \left\{ \alpha \in \Lambda^k(\Tan^*E) \mid
\inn(V_r)\ldots\inn(V_1)\alpha = 0 \, , \, \forall \, V_1,\ldots,V_r \in \vf^{V(\pi)}(E)\right\} \, .
\end{equation*}
Then, the restriction of the canonical multisymplectic $(k+1)$-form of the multicotangent bundle
$\Lambda^k(\Tan^*E)$ to this subbundle $\Lambda^k_r(\Tan^*E)$ is also a multisymplectic $(k+1)$-form.
Note that, if $M = E$, then we recover the whole $\Lambda^{k}(\Tan^*E)$.
\end{example}


\section{Geometry of higher-order jet bundles}
\label{Chap01_sec:HOJetBundles}

In this Section we generalize the definition of the tangent bundle of a manifold to consider derivatives
with respect to several independent variables $x^1,\ldots,x^m$, instead of derivatives with respect to a
single variable $t$, that is, ``partial derivatives''. The case of derivatives of higher-order is also introduced.

Along this Section, $M$ will denote a $m$-dimensional smooth manifold with no additional structure,
$\pi \colon E \to M$ (or $(E,\pi,M)$) will denote a smooth fiber bundle over $M$ with $\dim E = m + n$,
and $\Gamma(\pi)$ the set of sections of $\pi$, that is, maps $\phi \colon M \to E$ satisfying
$\pi \circ \phi = \Id_M$. Finally, $k \geqslant 1$ will be a fixed, but arbitrary, integer. We refer to
\cite{book:Saunders89} for details and proofs.

\subsection{Multi-index notation}
\label{Chap01_sec:HOJetBundlesMultiIndex}

(See \cite{phd:Campos} (Appendix A) and \cite{book:Saunders89} (\S $6.1$) for details).

Given a function $f \colon \R^m \to \R$, it is usual to denote its partial derivatives as
\begin{equation*}
f_{i_1,i_2,\ldots,i_k} = \frac{\partial^kf}{\partial x_{i_1} \partial x_{i_2} \ldots \partial x_{i_k}} \, .
\end{equation*}
Nevertheless, when smooth functions are considered, their cross derivatives coincide. In particular,
the order in which the derivatives are taken is no longer relevant, but only the number of times
with respect to each variable.

An alternative notation to denote partial derivatives is defined through ``symmetric'' multi-indexes.
A \textsl{multi-index} $I$ is an $m$-tuple of non-negative integers. The components of $I$ are
$I(i)$, with $1 \leqslant i \leqslant m$. Addition and subtraction of multi-indexes are defined
component-wise (although the result of a subtraction may not be a multi-index), that is,
$(I \pm J)(i) = I(i) \pm J(i)$. Given a fixed $1 \leqslant k \leqslant m$, the symbol ``$1_k$''
denotes the multi-index defined as $I(i) = \delta_i^k$, $1 \leqslant j \leqslant m$, that is,
all of its components are zero but the $k$th, which takes the value $1$.
The \textsl{length} and the \textsl{factorial} of a multi-index $I$ is
\begin{equation*}
|I| = \sum_{i=1}^{m} I(i) \quad ; \quad
I! = \prod_{i=1}^{m} (I(i))! \, .
\end{equation*}
With these notations, the symbol $\partial^{|I|}/ \partial x^I$ is defined as
\begin{equation*}
\frac{\partial^{|I|}}{\partial x^I} = \prod_{i=1}^{m} \left( \derpar{}{x_i} \right)^{I(i)} \, ,
\end{equation*}
where we adopt the convention that if $|I|=0$, then we have the identity operator.

As an example, let $f \colon \R^3 \to \R$ be a smooth function. Then, the partial derivatives
of $f$ with the multi-index notation are denoted by
\begin{equation*}
f_I = \frac{\partial^{|I|}f}{\partial x^I} =
\frac{\partial^{I(1)+I(2)+\ldots+I(m)}f}{\partial x_1^{I(1)} \partial x_2^{I(2)} \ldots \partial x_m^{I(m)}}
\end{equation*}
Then, first-order derivatives are denoted by
\begin{equation*}
f_{(1,0,0)} \quad ; \quad f_{(0,1,0)} \quad ; \quad
f_{(0,0,1)} \, ,
\end{equation*}
second-order derivatives are
\begin{equation*}
f_{(2,0,0)} \quad ; \quad f_{(0,2,0)} \quad ; \quad f_{(0,0,2)} \quad ; \quad
f_{(1,1,0)} \quad ; \quad f_{(1,0,1)} \quad ; \quad f_{(0,1,1)} \, ,
\end{equation*}
and so on.

Along this dissertation we will usually mix both notations. In particular, first-order partial
derivatives of a smooth function $f \colon \R^m \to \R$ will still be denoted by $f_i$,
$1 \leqslant i \leqslant m$, and multi-index notation will be kept for partial derivatives of
order greater than $1$.

Finally, sum over repeated multi-indexes will be understood, and expressions of the type
``for every $|I| = k$'' and ``$\sum_{|I|=k}$'' mean that the expression or the sum is taken for every
multi-index of length $k$. The same applies for inequalities.

\subsection{Definition and fiber bundle structures. Natural coordinates}
\label{Chap01_sec:HOJetBundlesDef&Coord}

Let $x \in M$ be a point, and $\Gamma_x(\pi)$ the set of sections of $\pi$ defined in a neighborhood
of $x$. The first we need is to define an equivalence relation in the set $\Gamma_x(\pi)$, which will be
introduced in coordinates. Thus, let $(x^i)$, $1 \leqslant i \leqslant m$, be a system of coordinates in
$M$, and $(x^i,u^\alpha)$, $1 \leqslant \alpha \leqslant n$, local coordinates in $E$ adapted to the
bundle structure. If $\phi,\psi \in \Gamma_x(\pi)$, we denote $\phi^\alpha = u^\alpha \circ \phi$ and
$\psi^\alpha = u^\alpha \circ \psi$, so that $\phi(x^i) = (x^i,\phi^\alpha(x^i))$ and
$\psi(x^i) = (x^i,\psi^\alpha(x^i))$. With these notations, we have:

\begin{definition}
The local sections $\phi,\psi \in \Gamma_x(\pi)$ are \textnormal{$k$-equivalent at $x$} if
\begin{enumerate}
\item $\phi(x) = \psi(x)$.
\item In some adapted coordinate system $(x^i,u^\alpha)$ around $\phi(x)$ (or $\psi(x)$) we have
\begin{equation*}\label{Chap01_eqn:HOJetBundlesEquivRel}
\restric{\frac{\partial^{|I|}\phi^\alpha}{\partial x^I}}{x} = \restric{\frac{\partial^{|I|}\psi^\alpha}{\partial x^I}}{x} \, ,
\end{equation*}
for $1 \leqslant |I| \leqslant k$ and $1 \leqslant \alpha \leqslant n$.
\end{enumerate}
\end{definition}

This relation does not depend on the chosen coordinate system, and then we have a well-defined relation
in the set of local sections $\Gamma_x(\pi)$.

\begin{lemma}
Let $x \in M$ be a point. The $k$-equivalence relation in the set of local sections $\Gamma_x(\pi)$
is independent of the chosen coordinate system.
\end{lemma}

In particular, the $k$-equivalence relation is a well-defined relation in the whole set $\Gamma_x(\pi)$,
and it is easy to prove that it is an equivalence relation. The equivalence class containing $\phi$ is
called the \textsl{$k$-jet of $\phi$ at $x$}, and is denoted $j^k_x\phi$.

\begin{definition}
The \textnormal{$k$-jet manifold of $\pi$} is the set
\begin{equation*}\label{Chap01_eqn:HOJetBundlesDefinition}
J^k\pi = \left\{ j_x^k\phi \mid x \in M \, , \, \phi \in \Gamma_x(\pi) \right\} \, .
\end{equation*}
\end{definition}

The $k$-jet manifold of $\pi$, $J^k\pi$, has a natural structure of smooth manifold. In addition,
it is endowed with the following natural projections: if $r\leqslant k$, then
\begin{equation*}
\begin{array}{rcl}
\pi^k_r \colon J^k\pi & \longrightarrow & J^r\pi \\
j^k_x\phi & \longmapsto & j^r_x\phi
\end{array}
\quad ; \quad
\begin{array}{rcl}
\pi^k \colon J^k\pi & \longrightarrow & E \\
j^k_x\phi & \longmapsto & \phi(x)
\end{array}
\quad ; \quad
\begin{array}{rcl}
\bar{\pi}^k \colon J^k\pi & \longrightarrow & M \\
j^k_x\phi & \longmapsto & x
\end{array}
\end{equation*}
which are called the \textsl{$r$-jet}, \textsl{target} and \textsl{source} projections, respectively, and
all of them are smooth surjective submersions. Observe that $\pi^s_r\circ\pi^k_s = \pi^k_r$, $\pi^k_0 = \pi^k$ 
(where $J^0\pi$ is canonically identified with $E$), $\pi^k_k = \Id_{J^k\pi}$, and $\bar{\pi}^k = \pi \circ \pi^k$.

\begin{proposition}\label{Chap01_prop:HOJetBundlesBundleStruct}
Let $(E,\pi,M)$ be a fibered manifold. Then the triples $(J^k\pi,\pi^{k}_{r},J^r\pi)$ and $(J^k\pi,\pi^{k},E)$
are fiber bundles, and $(J^k\pi,\bar{\pi}^k,M)$ is a fibered manifold. If $(E,\pi,M)$ is a fiber bundle, then
the triple $(J^k\pi,\bar{\pi}^k,M)$ is also a fiber bundle.
\end{proposition}

\begin{remark}
If $x \in M$, then the fiber $(\bar{\pi}^k)^{-1}(x) \hookrightarrow J^k\pi$ will be denoted $J_x^k\pi$
rather than $(J^k\pi)_x$. Observe that $J^k_x\pi$ is a $m$-codimensional submanifold of $J^k\pi$.
\end{remark}

In particular, the bundle $(J^{k}\pi,\pi^{k}_{k-1},J^{k-1}\pi)$ is canonically endowed with additional structure.

\begin{theorem}
The triple $(J^{k}\pi,\pi^{k}_{k-1},J^{k-1}\pi)$ is an affine bundle modeled on the vector bundle
\begin{equation*}
(\bar{\pi}^{k-1})^*(S^k\Tan^*M) \otimes_{J^{k-1}\pi}(\pi^{k-1})^*(V(\pi)) \, ,
\end{equation*}
where $S^k\Tan^*M$ is the space of symmetric covariant tensors of order $k$ over $M$ and $V(\pi)$ is the
vertical bundle of $\pi$.
\end{theorem}

Local coordinates in $J^k\pi$ are defined as follows: let $(x^i)$, $1 \leqslant i \leqslant m$, be local
coordinates in $M$, and $(x^i,u^\alpha)$, $1 \leqslant \alpha \leqslant n$, a set of local coordinates
in $E$ adapted to the bundle structure. Let $\phi \in \Gamma(\pi)$ be a section with coordinate expression
$\phi(x^i) = (x^i,\phi^\alpha(x^i))$. Then, local coordinates in $J^k\pi$ are $(x^i,u^\alpha,u_I^\alpha)$, where
\begin{equation*}
u^\alpha = \phi^\alpha \quad ; \quad u_I^\alpha = \frac{\partial^{|I|}\phi^\alpha}{\partial x^I} \, ,
\end{equation*}
with $1 \leqslant |I| \leqslant k$. We usually write $u_{\mathbf{0}}^\alpha$, with $|\mathbf{0}| = 0$,
instead of $u^\alpha$, and so the coordinates in $J^k\pi$ are $(x^i,u_I^\alpha)$, where now
$0 \leqslant |I| \leqslant k$. Observe that the dimension of $J^k\pi$ is
\begin{equation*}
\dim J^k\pi = m + n \binom{m+k}{k} = m + n \sum_{r=0}^{k}\binom{m+r-1}{r} \, .
\end{equation*}

Using these coordinates, the local expressions of the natural projections are
\begin{equation*}
\pi^k_r(x^i,u_I^\alpha) = (x^i,u_J^\alpha) \quad ; \quad
\pi^k(x^i,u_I^\alpha) = (x^i,u^\alpha) \quad ;\quad
\bar{\pi}^k(x^i,u_I^\alpha) = (x^i) \, ,
\end{equation*}
where $(x^i,u_J^\alpha)$, with $0 \leqslant |J| \leqslant r \leqslant k$ are the corresponding natural
coordinates in $J^r\pi$.

\subsection{Prolongation of sections. Holonomic sections}
\label{Chap01_sec:HOJetBundlesProlongationHolonomicSections}

\begin{definition}
Let $\phi \in \Gamma(\pi)$ a (local) section of $\pi$ with domain $U \subseteq M$. The
\textnormal{$k$th prolongation of $\phi$} is the (local) section $j^k\phi \in \Gamma(\bar{\pi}^{k})$ defined by
\begin{equation*}
j^k\phi(x) = j_x^k\phi \, ,
\end{equation*}
for every $x \in M$.
\end{definition}

\begin{definition}\label{Chap01_def:HOJetBundlesHolonomicSect}
A section $\psi \in \Gamma(\bar{\pi}^k)$ is \textnormal{holonomic of type $r$},
$1 \leqslant r \leqslant k$,
if $j^{k-r+1}\phi = \pi^{k}_{k-r+1} \circ \psi$, where $\phi = \pi^k \circ \psi \in \Gamma(\pi)$; that is,
the section $\pi^{k}_{k-r+1} \circ \psi$ is the prolongation of a section $\phi \in \Gamma(\pi)$ up to the
jet bundle $J^{k-r+1}\pi$.
\begin{equation*}
\xymatrix{
\ & \ & J^k\pi \ar[d]_{\pi^k_{k-r+1}} \ar@/^2.5pc/[ddd]^{\pi^k} \\
M \ar@/^1.5pc/[urr]^{\psi} \ar@/_1.5pc/[ddrr]_{\phi = \pi^k\circ\psi}
\ar[rr]^-{\pi^k_{k-r+1}\circ\psi} \ar[drr]_{j^{k-r+1}\phi} 
& \ & J^{k-r+1}\pi \ar[d]_{\textnormal{Id}} \\
\ & \ & J^{k-r+1}\pi \ar[d]_{\pi^{k-r+1}} \\
\ & \ & E
}
\end{equation*}
In particular, a section $\psi$ is \textnormal{holonomic of type $1$} (or simply \textnormal{holonomic})
if $j^k(\pi^{k} \circ \psi) = \psi$; that is, $\psi$ is the canonical $k$th prolongation of a section
$\phi = \pi^{k} \circ \psi \in \Gamma(\pi)$.
\end{definition}

In coordinates, the $k$th prolongation of a section $\phi(x^i) = (x^i,\phi^\alpha(x^i))$ is locally given by
\begin{equation*}\label{Chap01_eqn:HOJetBundlesProlongationSect}
j^k\phi(x^i) = \left( x^i,\phi^\alpha,\frac{\partial^{|I|}\phi^\alpha}{\partial x^{I}} \right) \, ,
\end{equation*}
with $1 \leqslant |I| \leqslant k$. On the other hand, let $\psi \in \Gamma(\bar{\pi}^k)$ be given by
$\psi(x^i) = (x^i,\psi^\alpha,\psi_I^\alpha)$, where $1 \leqslant |I| \leqslant k$, and let
$1 \leqslant r \leqslant k$ be a fixed, but arbitrary, integer. Then the condition for $\psi$ to be holonomic
of type $r$ gives the following system of partial differential equations
\begin{equation}\label{Chap01_eqn:HOJetBundlesHolonomyConditionSect1}
\psi_{I}^\alpha = \frac{\partial^{|I|} \psi^\alpha}{\partial x^{I}} \, ,
\qquad 1 \leqslant |I| \leqslant k-r+1 \, , \ 1 \leqslant \alpha \leqslant n \, ,
\end{equation}
or, equivalently,
\begin{equation}\label{Chap01_eqn:HOJetBundlesHolonomyConditionSect2}
\psi_{I+1_i}^\alpha = \derpar{\psi_I^\alpha}{x^i} \, ,
\qquad 0 \leqslant |I| \leqslant k-r \, , \ 1 \leqslant i \leqslant m \, ,
\ 1 \leqslant \alpha \leqslant n \, .
\end{equation}

\subsection{Contact forms. Cartan distribution}

\begin{definition}
Let $\phi \in \Gamma(\pi)$ be a section, $x \in M$ a point and $\bar{u} \equiv j^{k-1}_x\phi \in J^{k-1}\pi$.
The \textnormal{vertical differential} of $\phi$ at $\bar{u} \in J^{k-1}\pi$ is the map
$d_{\bar{u}}^{\rm v} \phi \colon \Tan_{\bar{u}}J^{k-1}\pi \to \Tan_{\bar{u}}J^{k-1}\pi$ defined as
\begin{equation*}
d_{\bar{u}}^{\rm v}\phi = \Id_{\bar{u}} - \Tan_{\bar{u}}(j^{k-1}\phi \circ \bar{\pi}^{k-1}) \, .
\end{equation*}
\end{definition}

Observe that $\Tan_{\bar{u}}\bar{\pi}^{k-1} \circ d_{\bar{u}}\phi = 0$, and therefore $d^{\rm v}_{\bar{u}}$
takes values in $V_{\bar{u}}(\bar{\pi}^{k-1})$. In the natural coordinates $(x^i,u_I^\alpha)$ of $J^{k-1}\pi$,
the vertical differential has the following coordinate expression
\begin{equation}\label{Chap01_eqn:HOJetBundlesVerticalDifferential}
d^{\rm v}_{\bar{u}}\phi = \left( \d u_I^\alpha - \frac{\partial^{|I|+1}\phi^\alpha}{\partial x^{I+1_i}} \, \d x^i \right)
\otimes \derpar{}{u_I^\alpha} \, ,
\end{equation}
from where it is clear that $d_{\bar{u}}^{\rm v}\phi$ depends only on $j^{k}_x\phi$.

\begin{definition}
The \textnormal{canonical structure form} of $J^k\pi$ is the $1$-form $\theta$ in $J^k\pi$
with values in $V(\bar{\pi}^{k-1})$ defined by
\begin{equation*}
\theta_{j^k_x\phi}(v) = (d_{j^{k-1}_x\phi}^{\rm v}\phi)(\Tan_{j^k_x\phi}\pi^{k}_{k-1}(v))
\end{equation*}
where $v \in \Tan_{j^k_x\phi}J^k\pi$. 
\end{definition}

The contraction of covectors in $(V(\bar{\pi}^{k-1}))^*$ with $\theta$ defines a ``distribution'' in
$\Tan^*J^k\pi$, which is called the \textsl{contact module} or \textsl{Cartan codistribution of order $k$},
and it is denoted $\C^k$. The annihilator of $\C^k$ is the \textsl{Cartan distribution of order $k$}.

In the natural coordinates of $J^k\pi$, and bearing in mind the coordinate expression
\eqref{Chap01_eqn:HOJetBundlesVerticalDifferential} of the vertical differential,
the canonical structure form is given by
\begin{equation}\label{Chap01_eqn:HOJetBundlesCanonicalStructureForm}
\theta = \left( \d u_I^\alpha - u_{I+1_i}^\alpha \d x^i \right) \otimes \derpar{}{u_I^\alpha} \, , \quad
0 \leqslant |I| \leqslant k-1 \, .
\end{equation}
The forms $\theta_I^\alpha = \d u_I^\alpha - u_{I+1_i}\d x^i \in \C^k$ are the \textsl{coordinate contact forms}.

\begin{proposition}
Let $(x^i,u_I^\alpha)$ be adapted coordinates in $J^k\pi$. A basis of the Cartan codistribution
is given by the coordinate contact forms $\theta_I^\alpha = \d u_I^\alpha - u_{I+1_i}\d x^i$.
\end{proposition}

Contact forms may be distinguished from the rest of $1$-forms defined on $J^k\pi$ by their relation with
prolongations of sections $\phi \in \Gamma(\pi)$, as it is shown in the following result.

\begin{proposition}
Let $\omega \in \df^{1}(J^k\pi)$ be a $1$-form. Then, $\omega$ is a contact form
if, and only if, $(j^k\phi)^*\omega = 0$ for every $\phi \in \Gamma(\pi)$.
\end{proposition}

Finally, the following result relates this Section with the previous one.

\begin{proposition}
Let $\psi \in \Gamma(\bar{\pi}^{k})$ be a section. The following assertions are equivalent:
\begin{enumerate}
\item $\psi$ is holonomic.
\item $\psi^*\theta = 0$, where $\theta \in \Gamma(\Tan^*J^k\pi \otimes_{J^k\pi} V(\bar{\pi}^{k-1}))$
is the canonical structure form.
\item $\psi^*\omega = 0$ for every $\omega \in \C^k$.
\end{enumerate}
\end{proposition}

\subsection{The vertical endomorphisms}
\label{Chap01_sec:HOJetBundlesVerticalEndomorphisms}

In this Section we will assume that $k=1$, since the vertical endomorphism cannot be generalized to higher-order
jet bundles in a unique way. In fact, this is the main issue when we want to give a geometric formulation
of higher-order field theories, since both the Cartan $m$-form and the Legendre map depend on the vertical endomorphism.

\begin{definition}
The \textnormal{vertical lift} is a morphism of vector bundles
$\mathcal{S} \colon \Tan^*M \otimes_{J^1\pi} V(\pi) \to V(\pi^1)$
over the identity of $J^1\pi$ defined as follows: given $j^1_x\phi \in J^1\pi$
and $\beta \in \Tan^*_xM \otimes V_{\phi(x)}(\pi)$, we have
\begin{equation*}
\mathcal{S}_{j^1_x\phi}(\beta)(f) = \restric{\frac{d}{dt}}{t=0} f(j^1_x\phi + t\beta) \, ,
\quad \mbox{for every } f \in \Cinfty(J^1_{\phi(x)}\pi) \, .
\end{equation*}
\end{definition}

Observe that $\Tan \pi^1 \circ \mathcal{S} = 0$, and therefore the image of $\mathcal{S}$ is certainly
in $V(\pi^1)$. In addition, for every $j^1_x\phi \in J^1\pi$, the vertical lift at $j_x^1\phi$,
$S_{j^1_x\phi} \colon \Tan^*_xM \otimes V_{\phi(x)}(\pi) \to V_{j^1_x\phi}(\pi)$,
is a linear isomorphism.

\begin{definition}
The \textnormal{canonical vertical isomorphism} $\V$ arises from the natural contraction between the factors
in $V(\pi)$ of the structure canonical form $\theta$ and the factors in $(V(\pi))^*$ of the vertical lift
$\mathcal{S}$, that is,
\begin{equation*}
\V = \inn(\mathcal{S})\theta \in \Gamma(\Tan^*J^1\pi \otimes_{J^1\pi} \Tan M \otimes_{J^1\pi} V(\pi^1)) \, .
\end{equation*}
\end{definition}

In the natural coordinates $(x^i,u^\alpha,u_i^\alpha)$ if $J^1\pi$, the vertical lift is given by
\begin{equation*}
\mathcal{S} = \d u^\alpha \otimes \derpar{}{x^i} \otimes \derpar{}{u^\alpha_i} \, .
\end{equation*}
From here, and bearing in mind the coordinate expression
\eqref{Chap01_eqn:HOJetBundlesCanonicalStructureForm} of the canonical
structure form $\theta$ (taking $k=1$), the local expression of the vertical endomorphism is
\begin{equation}\label{Chap01_eqn:HOJetBundleVertEnd}
\V = \left( \d u^\alpha - u_j^\alpha \d x^j \right) \otimes \derpar{}{x^i} \otimes \derpar{}{u_i^\alpha}
= \theta^\alpha \otimes \derpar{}{x^i} \otimes \derpar{}{u_i^\alpha} \, ,
\end{equation}
where $\theta^\alpha = \d u^\alpha - u_j^\alpha \d x^j$ are the coordinate contact forms.

\subsection{Iterated jet bundles}

From Proposition \ref{Chap01_prop:HOJetBundlesBundleStruct} we know that $(J^k\pi,\bar{\pi}^k,M)$ is a
fiber bundle. Hence, we can consider the $r$th-order jet bundle of $\bar{\pi}^k$, that is, repeated
(or iterated) jet bundles. The $r$-jet manifold of $\bar{\pi}^k$, which will be denoted $J^r\bar{\pi}^{k}$,
will contain $r$-jets of all the local sections of $\bar{\pi}^{k}$, that is, it is the manifold
\begin{equation*}
J^r\bar{\pi}^{k} = \left\{ j^r_x\psi \mid
x \in M \, , \, \psi \in \Gamma_{x}(\bar{\pi}^{k}) \right\} \, .
\end{equation*}
There is a distinguished subset in $J^r\bar{\pi}^k$ containing those elements $j^r_x\psi$, where the local
section $\psi$ is holonomic, that is, $\psi$ itself is the $k$th prolongation $j^k\phi$ of a local section
$\phi \in \Gamma(\pi)$. This set is a submanifold of $J^r\bar{\pi}^k$ which can be identified with the
image of a highest-order jet bundle, $J^{r+k}\pi$, by the following embedding.

\begin{definition}
The \textnormal{canonical embedding} is the map
\begin{equation}\label{Chap01_eqn:JetBundlesCanonicalEmbeddings}
\begin{array}{rcl}
\iota_{r,k} \colon J^{k+r}\pi & \longrightarrow & J^r\bar{\pi}^k \\
j^{k+r}_x\phi & \longmapsto & j^r_x(j^k\phi)
\end{array} \, .
\end{equation}
The elements in the image of $\iota_{r,k}$ are called \textnormal{holonomic}.
\end{definition}

\begin{remark}
It is important not to confuse this notion of holonomy with the one given in Definition
\ref{Chap01_def:HOJetBundlesHolonomicSect}. The former refers to the holonomy when considering iterated
jets, the latter to the holonomy of a jet section itself.
\end{remark}

Local coordinates in $J^{r}\bar{\pi}^k$ are constructed in an analogous way to $J^k\pi$. Let $(x^i)$ be
local coordinates in $M$, and $(x^i,u_I^\alpha)$, $0 \leqslant |I| \leqslant k$, the induced natural
coordinates in $J^k\pi$. Let $\psi \in \Gamma(\bar{\pi}^{k})$ be a section locally given by
$\psi(x^i) = (x^i,\psi_I^\alpha(x^i))$, where we take $\psi_I^\alpha = u_I^\alpha \circ \psi$.
Then, local coordinates in $J^{r}\bar{\pi}^{k}$ are $(x^i,u_{I;J}^\alpha)$, where
\begin{equation*}
u_{I;J}^\alpha = \frac{\partial^{|J|}\psi_I^\alpha}{\partial x^J} \, .
\end{equation*}
with $0 \leqslant |J| \leqslant r$. In these coordinates, the canonical embedding $\iota_{r,k}$ is given by
\begin{equation*}
\iota_{r,k}^*u_{I;J}^\alpha = u_{I+J}^\alpha \, .
\end{equation*}
It follows from this coordinate expression that $J^{k+r}\pi$ may be identified with the submanifold
of $J^r\bar{\pi}^k$ given locally by
\begin{equation*}
J^{k+r}\pi = \left\{ j^r_x\psi \in J^{r}\bar{\pi}^k \mid u_{I_1;J_1}^\alpha = u_{I_2;J_2}^\alpha
\text{ whenever } I_1 + J_1 = I_2 + J_2 \right\} \, .
\end{equation*}

\subsection{Coordinate total derivatives}
\label{Chap01_sec:HOJetBundlesTotalDeriv}

\begin{definition}
Let $x \in M$, $\phi \in \Gamma_x(\pi)$, and $v \in \Tan_xM$. The \textnormal{$k$th holonomic lift}
of $v$ by $\phi$ is defined as
\begin{equation*}\label{Chap01_eqn:HOJetBundlesHolonomicLiftVector}
((j^k\phi)_*(v),j^{k+1}_x\phi) \in (\pi^{k+1}_{k})^*\Tan J^k\pi \, .
\end{equation*}
\end{definition}

From this definition, observe that we can split $(\pi^{k+1}_{k})^*(\Tan J^{k}\pi)_{j^{k+1}_x\phi}$ to
distinguish the vectors which are $k$th holonomic lifts of vectors in the base manifold from those which are not.

\begin{theorem}\label{Chap01_thm:HOJetBundlesTotalDerivSplittingVectorSpace}
Let $\pi \colon E \to M$ be a fiber bundle, and let $j_x^{k+1}\phi \in J^{k+1}\pi$. Then the vector space
$(\pi^{k+1}_{k})^*(\Tan J^k\pi)_{j^{k+1}_x\pi}$ has a canonical decomposition as a direct sum of two subspaces
\begin{equation*}
(\pi^{k+1}_{k})^*(\Tan J^k\pi)_{j^{k+1}_x\phi} =
(\pi^{k+1}_{k})^*(V(\bar{\pi}^k))_{j^{k+1}_x\phi}
\oplus (j^k\phi)_*(\Tan_xM) \, ,
\end{equation*}
where $(j^k\phi)_*(\Tan_xM)$ denotes the set of $k$th holonomic lifts of tangent vectors in $\Tan_xM$ by $\phi$.
\end{theorem}

Since Theorem \ref{Chap01_thm:HOJetBundlesTotalDerivSplittingVectorSpace}
gives a pointwise decomposition, we have the following result straightforwardly.

\begin{corollary}\label{Chap01_corol:HOJetBundlesTotalDerivSplittingBundles}
The vector bundle $(\pi^{k+1}_{k})^*\tau_{J^k\pi} \colon (\pi^{k+1}_{k})^*\Tan J^k\pi \to J^k\pi$ has a
canonical splitting in the direct sum of two subbundles
\begin{equation*}
\xymatrix{
(\pi^{k+1}_{k})^*\Tan J^k\pi = (\pi^{k+1}_{k})^*V(\bar{\pi}^k) \oplus H(\pi^{k+1}_{k}) \ar[rr]^-{(\pi^{k+1}_{k})^*\tau_{J^k\pi}}
& \ & J^k\pi
} \, ,
\end{equation*}
where $H(\pi^{k+1}_k)$ is the reunion of the fibers $(j^k\phi)_*(\Tan_xM)$, for $x \in M$.
\end{corollary}

In local coordinates, if $v \in \Tan_xM$ is given by
\begin{equation*}
v = v^i\restric{\derpar{}{x^i}}{x} \, ,
\end{equation*}
the $k$th holonomic lift of $v$ is then given by
\begin{equation}\label{Chap01_eqn:HOJetBundlesHolonomicLiftVectorLocal}
(j^k\phi)_*(v) = v^i\left( \restric{\derpar{}{x^i}}{j_x^{k}\phi}
+ \sum_{|I|=0}^{k}\restric{u_{I+1_i}^\alpha(j_x^{k+1}\phi)\derpar{}{u_{I}^\alpha}}{j_x^k\phi} \right) \, .
\end{equation}

Now, if $\vf(\pi^{k+1}_{k})$ denotes the module of vector fields along the projection $\pi^{k+1}_{k}$,
the submodule corresponding to sections of $\restric{(\pi^{k+1}_{k})^*\tau_{J^{k}\pi}}{(\pi^{k+1}_{k})^*V(\bar{\pi}^k)}$
is denoted by $\vf^v(\pi^{k+1}_{k})$, and the submodule corresponding to sections of
$\restric{(\pi^{k+1}_{k})^*\tau_{J^{k}\pi}}{H(\pi^{k+1}_{k})}$ is denoted by $\vf^h(\pi^{k+1}_{k})$.

\begin{definition}
An element of the submodule $\vf^h(\pi^{k+1}_{k})$ is called a \textnormal{total derivative}.
\end{definition}

\begin{remark}
Total derivatives may be defined in the following equivalent way. Since the contact forms are
$\pi^{k+1}_{k}$-semibasic, they may be thought as forms along $\pi^{k+1}_{k}$ rather than in
$J^{k+1}\pi$. Then, a total derivative is a vector field along $\pi^{k+1}_{k}$ which is annihilated
by the Cartan codistribution (as forms along $\pi^{k+1}_{k}$).
\end{remark}

The splitting given in Corollary \ref{Chap01_corol:HOJetBundlesTotalDerivSplittingBundles} induces
the following canonical splitting for the module $\vf(\pi^{k+1}_{k})$:
\begin{equation*}
\vf(\pi^{k+1}_{k}) = \vf^v(\pi^{k+1}_{k}) \oplus \vf^h(\pi^{k+1}_{k}) \, .
\end{equation*}

\begin{definition}
Given a vector field $X \in \vf(M)$, a section $\phi \in \Gamma(\pi)$ and a point $x \in M$, the
\textnormal{$k$th holonomic lift} of $X$ by $\phi$,
$j^kX \in \vf^{h}(\pi^{k+1}_{k})$, is defined as
\begin{equation*}
(j^kX)_{j^{k+1}_x\phi} = (j^k\phi)_*(X_{x}) \, .
\end{equation*}
\end{definition}

In local coordinates, if $X \in \vf(M)$ is given by
\begin{equation*}
X = X^i\derpar{}{x^i} \, ,
\end{equation*}
then, bearing in mind the local expression \eqref{Chap01_eqn:HOJetBundlesHolonomicLiftVectorLocal}
of the $k$th holonomic lift for tangent vectors, the $k$th holonomic lift of $X$ is
\begin{equation*}
j^kX = X^i\left( \derpar{}{x^i} + \sum_{|I|=0}^{k}u_{I+1_i}^\alpha \derpar{}{u_I^\alpha} \right) \, .
\end{equation*}

Finally, the \textsl{coordinate total derivatives} are the holonomic lifts of the local vector fields
$\partial / \partial x^i \in \vf(M)$, which are denoted by $d / dx^i \in \vf(\pi^{k+1}_{k})$, and whose
coordinate expressions are
\begin{equation*}
\frac{d}{dx^i} = \derpar{}{x^i} + \sum_{|I|=0}^{k} u_{I+1_i}^\alpha \derpar{}{u_I^\alpha}
\quad (1 \leqslant i \leqslant m) \, .
\end{equation*}

\subsection{Dual jet bundles}
\label{Chap01_sec:HOJetBundlesDualBundles}

Let us consider the iterated jet bundle $J^1\bar{\pi}^{k-1}$, and its dual space as an affine bundle
over $J^{k-1}\pi$, which we denote by $(J^1\bar{\pi}^{k-1})^*$. Since $J^{k}\pi$ is affinely embedded
into $J^1\bar{\pi}^{k-1}$, we can restrict the elements of $(J^1\bar{\pi}^{k-1})^*$ to the points of $J^k\pi$.

\begin{definition}
The \textnormal{$k$th-order extended dual jet bundle of $\pi$}, denoted $J^k\pi^\circ$, is the reunion
of the affine maps from $J^1_u\bar{\pi}^{k-1}$ to $(\Lambda^m\Tan^*M)_{\bar{\pi}^{k-1}(u)}$, where
$u \in J^{k-1}\pi$, that is,
\begin{equation*}
J^k\pi^\circ = \bigcup_{u \in J^{k-1}\pi} \textnormal{Aff}(J^1_u\bar{\pi}^{k-1},(\Lambda^m\Tan^*M)_{\bar{\pi}^{k-1}(u)})
\end{equation*}
\end{definition}

The $k$th-order extended dual jet bundle admits a structure of smooth manifold.
Furthermore, it may be endowed with a fiber bundle structure, as shown in the following results.

\begin{proposition}\label{Chap01_prop:HOJetBundlesDualBundleDiffeomorphicMulticotangentBundle}
The $k$th-order extended dual jet bundle, $J^{k}\pi^\circ$, is diffeomorphic to the bundle of $\pi$-semibasic
$m$-forms over $J^{k-1}\pi$, $\Lambda^{m}_{2}(\Tan^*J^{k-1}\pi)$, that is, the manifold
\begin{equation*}
\Lambda^{m}_{2}(\Tan^*J^{k-1}\pi) = \left\{ \alpha \in \Lambda^m(\Tan^*J^{k-1}\pi) \mid
\inn(V_2)\inn(V_1)\alpha = 0 \, , \, \forall \, V_1,V_2 \in \vf^{V(\bar{\pi}^{k-1})}(J^{k-1}\pi)\right\} \, .
\end{equation*}
\end{proposition}

\begin{remark}
In the following we denote $J^{k}\pi^\circ$ by $\Lambda^{m}_{2}(\Tan^*J^{k-1}\pi)$, or
$\Lambda^{m}_{2}(J^{k-1}\pi)$ for short.
\end{remark}

In addition to Proposition \ref{Chap01_prop:HOJetBundlesDualBundleDiffeomorphicMulticotangentBundle},
the following result gives the precise structure of the extended dual jet bundle.

\begin{proposition}\label{Chap01_prop:HOJetBundlesDualBundleVectorBundle}
The triple $(\Lambda_2^m(J^{k-1}\pi),\pi_{J^{k-1}\pi},J^{k-1}\pi)$ is a smooth vector bundle.
\end{proposition}

Now, from Proposition \ref{Chap01_prop:HOJetBundlesDualBundleDiffeomorphicMulticotangentBundle}
the $k$th-order extended dual jet bundle is endowed with the following canonical projections:
\begin{equation*}
\begin{array}{rcl}
\pi_{J^{k-1}\pi} \colon \Lambda^m_2(J^{k-1}\pi) & \longrightarrow & J^{k-1}\pi \\
(u,\omega_u) & \longmapsto & u
\end{array}
\quad ; \quad
\begin{array}{rcl}
\bar{\pi}_{J^{k-1}\pi} \colon \Lambda^m_2(J^{k-1}\pi) & \longrightarrow & M \\
(u,\omega_u) & \longmapsto & \bar{\pi}^{k-1}(u)
\end{array} \, .
\end{equation*}

The bundle $\Lambda^m_2(J^{k-1}\pi)$ is endowed with some canonical structures. First, as we have seen in
Examples \ref{Chap01_exa:MulticotangentBundle} and \ref{Chap01_exa:MulticotangentBundleVertical}, we can
define a couple of forms in $\Lambda^{m}_2(J^{k-1}\pi)$ as follows.

\begin{definition}
The \textnormal{Liouville $m$-form}, or \textnormal{tautological} or \textnormal{canonical $m$-form},
on $\Lambda^{m}_{2}(J^{k-1}\pi)$ is the form $\Theta_{k-1} \in \df^{m}(\Lambda^m_2(J^{k-1}\pi))$
defined as
\begin{equation*}
\Theta_{k-1}(\omega)(X_1,\ldots,X_m) =
\omega(\Tan\pi_{J^{k-1}\pi}(X_1),\ldots,\Tan\pi_{J^{k-1}\pi}(X_m)) \, ,
\end{equation*}
where $\omega \in \Lambda_2^m(J^{k-1}\pi)$, and
$X_1,\ldots,X_m \in \Tan_{\omega}\Lambda_2^m(J^{k-1}\pi)$.
The \textnormal{Liouville $(m+1)$-form}, or \textnormal{canonical multisymplectic $(m+1)$-form},
is the form $\Omega_{k-1} \in \df^{m}(\Lambda^m_2(J^{k-1}\pi))$ given by
\begin{equation*}
\Omega_{k-1} = -\d\Theta_{k-1} \, .
\end{equation*}
\end{definition}

The second canonical structure is the pairing due to the duality between $J^1\bar{\pi}^{k-1}$
and $\Lambda^{m}_2(J^{k-1}\pi)$, and the fact that $J^{k}\pi$ is embedded in the former.

\begin{definition}
The \textnormal{canonical pairing} between the elements of $J^k\pi$ and the elements of
$\Lambda_2^m(J^{k-1}\pi)$ is the fibered map over $J^{k-1}\pi$ defined as follows
\begin{equation*}\label{Chap01_eqn:HOJetBundlesDualBundlesCanonicalPairingDef}
\begin{array}{rcl}
\C \colon J^{k}\pi \times_{J^{k-1}\pi} \Lambda_2^m(J^{k-1}\pi) & \longrightarrow & \Lambda_1^m(J^{k-1}\pi) \\
(j^{k}_x\phi,\omega) & \longmapsto & (j^{k-1}\phi)^*_{j^{k-1}_x\phi}\omega
\end{array}
\end{equation*}
\end{definition}

Local coordinates in $\Lambda^m_2(J^{k-1}\pi)$ are constructed as follows: let $(x^i)$ be a system
of coordinates in $M$, and $(x^i,u_I^\alpha)$ the induced coordinates in $J^{k-1}\pi$, with
$0 \leqslant |I| \leqslant k-1$. Then, local coordinates in $\Lambda^m_2(J^{k-1}\pi)$ are
$(x^i,u_I^\alpha,p,p_\alpha^{I,i})$, where $1 \leqslant i  \leqslant m$, $1 \leqslant \alpha \leqslant n$
and $0 \leqslant |I| \leqslant k-1$.
In these coordinates, the canonical projections have the following local expressions
\begin{equation*}
\pi_{J^{k-1}\pi}(x^i,u_I^\alpha,p,p_\alpha^I) = (x^i,u_I^\alpha) \quad ; \quad
\bar{\pi}_{J^{k-1}\pi}(x^i,u_I^\alpha,p,p_\alpha^I) = (x^i) \, .
\end{equation*}
On the other hand, the Liouville $m$ and $(m+1)$-forms have the following local expressions
\begin{equation}\label{Chap01_eqn:HOJetBundlesDualBundlesCanonicalFormsLocal}
\begin{array}{l}
\Theta_{k-1} = p\d^mx + p^i_\alpha \d u^\alpha \wedge \d^{m-1}x_i + p^{Ii}_\alpha \d u_I^\alpha \wedge \d^{m-1}x_i \, , \\[10pt]
\Omega_{k-1} = -\d p \wedge \d^mx - \d p^i_\alpha \wedge \d u^\alpha \wedge \d^{m-1}x_i - \d p^{Ii}_\alpha \wedge \d u_I^\alpha \wedge \d^{m-1}x_i \, ,
\end{array}
\end{equation}
where $\d^mx = \d x^1 \wedge \ldots \wedge \d x^m$ and $\d^{m-1}x_i = \inn(\partial/\partial x^i)\d^m x$.
Finally, the canonical pairing $\C$ has the following coordinate expression
\begin{equation}\label{Chap01_eqn:HOJetBundlesDualBundlesCanonicalPairingLocal}
\C(x^i,u^\alpha,u_I^\alpha,p,p_\alpha^i,p_\alpha^{Ii}) = (p + p_\alpha^{Ii}u_{I+1_i}^\alpha)\d^mx \, .
\end{equation}

Using Proposition \ref{Chap01_prop:HOJetBundlesDualBundleVectorBundle}, we can now give the following definition.

\begin{definition}
The \textnormal{$k$th-order reduced dual jet bundle of $\pi$}, denoted $J^{k-1}\pi^*$, is the quotient of the
$k$th-order extended dual jet bundle, $\Lambda^{m}_{2}(J^{k-1}\pi)$, by constant affine transformations along the
fibers of $\pi^{k}$, and is diffeomorphic to $\Lambda^{m}_{2}(J^{k-1}\pi)/\Lambda^{m}_{1}(J^{k-1}\pi)$.
The quotient map is $\mu \colon \Lambda^{m}_{2}(J^{k-1}\pi) \to J^{k-1}\pi^*$.
\end{definition}

It can be proved that $J^{k-1}\pi^*$ may be endowed with the structure of a smooth manifold and, moreover,
$(\Lambda^{m}_{2}(J^{k-1}\pi),\mu,J^{k-1}\pi^*)$ is a smooth vector bundle of rank $1$. In addition, using
the universal property of the quotient, from the canonical projections of the extended dual jet bundle and
he natural quotient map $\mu \colon \Lambda^{m}_{2}(J^{k-1}\pi) \to J^{k-1}\pi^*$, we obtain the canonical
projections of the restricted dual jet bundle
\begin{equation*}
\begin{array}{rcl}
\pi_{J^{k-1}\pi}^r \colon J^{k-1}\pi^* & \longrightarrow & J^{k-1}\pi \\
(u,[\omega_u]) & \longmapsto & u
\end{array}
\quad ; \quad
\begin{array}{rcl}
\bar{\pi}_{J^{k-1}\pi}^r \colon J^{k-1}\pi^* & \longrightarrow & M \\
(u,[\omega_u]) & \longmapsto & \bar{\pi}^{k-1}(u)
\end{array} \, .
\end{equation*}
Finally, adapted coordinates $(x^i,u^\alpha)$ in $E$ induce coordinates $(x^i,u_I^\alpha,p_\alpha^{Ii})$
in $J^{k-1}\pi^*$ such that the coordinate expression of the natural quotient map is
\begin{equation*}
\mu(x^i,u_I^\alpha,p,p_\alpha^{Ii}) = (x^i,u_I^\alpha,p_\alpha^{Ii}) \, ,
\end{equation*}
where $(x^i,u_I^\alpha,p,p_\alpha^{Ii})$ are the induced coordinates in $\Lambda^{m}_2(J^{k-1}\pi)$.
In these coordinates, the natural projections are given by
\begin{equation*}
\pi_{J^{k-1}\pi}^r(x^i,u_I^\alpha,p_\alpha^{Ii}) = (x^i,u_I^\alpha) \quad ; \quad
\bar{\pi}_{J^{k-1}\pi}^r(x^i,u_I^\alpha,p_\alpha^{Ii}) = (x^i) \, .
\end{equation*}


\section{Geometry of higher-order tangent bundles}
\label{Chap01_sec:HOTanBundle}

In this Section we generalize the definition of the tangent bundle of a manifold to consider not only
first-order derivatives of the coordinates in the base $M$, but also derivatives of higher-order.

As we will see, all the canonical structures of the tangent bundle, namely the vertical endomorphism
and the Liouville vector field, can be defined in higher-order tangent bundles, with some minor differences.
In addition, new structures arise when considering derivatives of order greater than $1$.

Along this Section, $M$ will denote a $m$-dimensional smooth manifold with no additional structure,
and $k \geqslant 1$ will be a fixed, but arbitrary, integer. We refer to \cite{book:DeLeon_Rodrigues85}
for details and proofs.

\subsection{Definition and fiber bundle structures. Natural coordinates}

Let $p \in M$ be a point, and let $C(M,p)$ be the set of curves on $M$ passing through $p$ at $t=0$, that is,
\begin{equation*}
C(M,p) = \left\{ \phi \colon \R \to M \mid \phi(0) = p \right\}
\end{equation*}
We define in $C(M,p)$ the following relation: given $\phi_1,\phi_2 \in C(M,p)$, then $\phi_1 \sim_k \phi_2$
if, and only if, $D^i\widehat{\phi}_1(0) = D^i\widehat{\phi}_2(0)$ for every $i = 0,\ldots,k$, where
$\widehat{\phi}$ denotes the local expression of $\phi$ and $D^i$ is the $i$th derivative. That is,
$\phi_1$ and $\phi_2$ must pass through $p$ at $t=0$, and all of their derivatives up to order $k$ must
coincide in $p$. It is easy to check that this defines an equivalence relation in $C(M,p)$.

\begin{definition}
The quotient set $\Tan_p^kM = C(M,p) / \sim_k$ is the \textnormal{$k$th-order tangent space} of $M$ at $p$,
which has dimension $km$. The \textnormal{$k$th-order tangent bundle} of $M$, denoted $\Tan^kM$ is the
disjoint union, indexed by $p \in M$, of every $k$th-order tangent space of $M$ at the point $p$, that is,
\begin{equation*}
\Tan^kM = \bigsqcup_{p\in M} \Tan_p^kM \, ,
\end{equation*}
which has dimension $(k+1)m$.
\end{definition}

\begin{remark}
Taking $k=1$, we recover one of the usual definitions of the tangent bundle of $M$.
\end{remark}

Bearing in mind the results stated in the previous Section on higher-order jet bundles, an alternative,
but equivalent, definition of the $k$th-order tangent bundle of $M$ is the following: the
\textsl{$k$th-order tangent bundle} of $M$ is the $(k+1)m$-dimensional manifold made of $k$-jets of the
trivial bundle $\pi \colon \R \times M \to \R$ with source point $0 \in \R$, that is, $\Tan^{k}M = J^k_0\pi$.
It is a $1$-codimensional submanifold of $J^k\pi$.

\begin{remark}
Observe that if $\pi \colon \R \times M \to \R$, then $J^k\pi \simeq \R \times \Tan^kM$.
\end{remark}

Hence, a point in $\Tan^kM$ will be denoted by $j^k_0\phi$, that is, the equivalence class of a curve
$\phi \colon \R \to M$ by the $k$-jet relation at $t = 0$. In addition, the canonical projections introduced
in Section \ref{Chap01_sec:HOJetBundlesDef&Coord} restrict to the $k$th-order tangent bundle, and we have
for $r \leqslant k$:
\begin{equation*}
\begin{array}{rcl}
\rho_r^k \colon \Tan^kM & \longrightarrow & \Tan^rM \\
j^k_0\phi & \longmapsto & j^r_0\phi
\end{array} \qquad
\begin{array}{rcl}
\beta^k \colon \Tan^kM & \longrightarrow & M \\
j^k_0\phi & \longmapsto & \phi(0)
\end{array}
\end{equation*}
Observe that $\rho^s_r \circ \rho^k_s = \rho^k_r$ for every $0 \leqslant r\leqslant s \leqslant k$,
$\rho^k_0 = \beta^k$, and $\rho^k_k = \Id_{\Tan^kM}$.

From the results in Section \ref{Chap01_sec:HOJetBundlesDef&Coord}, the natural projections
$\rho^r_s \colon \Tan^rM \to \Tan^sM$ are surjective submersions for every
$0 \leqslant s \leqslant r \leqslant k$. Furthermore, the triple $(\Tan^rM,\rho^r_s,\Tan^sM)$ is a fiber
bundle with fiber $\R^{(r-s)n}$. In particular, $(\Tan^kM,\rho^k_r,\Tan^rM)$ is a fiber bundle with fiber
$\R^{(k-r)n}$, for $0\leqslant r \leqslant k$; that is, $\Tan^kM$ is canonically endowed with $k+1$ different
fiber bundle structures given by the projections $\rho^k_0,\rho^k_1,\ldots,\rho^k_k$. In the sequel, we refer
to this fiber bundle structure as the \textsl{$\rho^k_r$-bundle structure} of $\Tan^kM$.

\begin{remark}
Notice that $\rho^k_k = \Id_{\Tan^kM}$ and hence $(\Tan^kM,\Id_{\Tan^kM},\Tan^kM)$ is nor a relevant,
neither interesting, fiber bundle. In the following, we restrict to the fiber bundle structures of
$\Tan^kM$ given by the projections $\rho^k_0,\rho^k_1,\ldots,\rho^k_{k-1}$, and consider that
$\Tan^kM$ is canonically endowed with $k$ different bundle structures.
\end{remark}

\begin{remark}
The notation is changed with respect to Section \ref{Chap01_sec:HOJetBundlesDef&Coord}
to keep in mind that we are considering the equivalence class in a fixed point of the
base manifold (the ``autonomous'' case), but also to take into account that
higher-order tangent bundles can be defined independently of higher-order jet bundles.
\end{remark}

If $\phi \colon \R \to M$ is a curve in $M$, the \textsl{$k$th-order lift} of $\phi$ to $\Tan^kM$ is the
curve $j^k_0\phi \colon \R \to \Tan^kM$ defined as $j^k_0\phi(t) = j^k_{t}\phi(0)$, that is, the $k$th
prolongation of $\phi$ evaluated in $t=0$.

Local coordinates in $\Tan^kM$ are constructed in a similar way to the local coordinates in the
higher-order jet bundles. Let $(U,\varphi)$ be a local chart of $M$, with $\varphi = (\varphi^i)$,
$1 \leqslant i \leqslant m$, and $\phi \colon \R \to M$ a curve in $M$ such that $\phi(0) \in U$.
Then, by writing $\phi^i = \varphi^i \circ \phi$, the equivalence class $j^k_0\phi$ of $\phi$ is
given in $(\beta^k)^{-1}(U) = \Tan^kU$ by $(x^i,x^i_1,\ldots,x^i_k)$, where
\begin{equation*}
x^i = \phi^i(0) \quad \mbox{and} \quad x^i_j = \restric{\frac{d^j\phi^i}{dt^j}}{t=0} \, ,
\end{equation*}
with $1 \leqslant j \leqslant k$. Usually we write $x^i_0$ instead of $x^i$, and so we have the local
chart $(\beta^k)^{-1}(U)$ in $\Tan^{k}M$ with local coordinates $(x^i_0,x^i_1,\ldots,x^i_k) \equiv (x^i_j)$,
where $1 \leqslant i \leqslant m$ and $0 \leqslant j \leqslant k$. When dealing with tangent bundles over
higher-order tangent bundles, that is, the manifold $\Tan(\Tan^{k}M)$, we will only consider the natural
coordinates of the tangent bundle structure, which will be denoted
$(x^i_0,\ldots,x^i_k,v^i_0,\ldots,v^i_k) \equiv (x^i_j,v^i_j)$,
with $1 \leqslant i \leqslant m$ and $0 \leqslant j \leqslant k$.

Using these coordinates, the local expression of the canonical projections are
\begin{equation*}
\rho_r^k(x^i_0,\ldots,x^i_k) = (x^i_0,\ldots,x^i_r) \quad ; \quad
\beta^k(x^i_0,\ldots,x^i_k) = (x^i_0) \, .
\end{equation*}
Then, their tangent maps are given by
\begin{equation*}
\Tan\rho_r^k(x^i_0,\ldots,x^i_k,v^i_0,\ldots,v^i_k) = (x^i_0,\ldots,x^i_r,v^i_0,\ldots,v^i_r) \quad ; \quad
\Tan\beta^k(x^i_0,\ldots,x^i_k,v^i_0,\ldots,v^i_k) = (x^i_0,v^i_0) \, .
\end{equation*}

\subsection{Geometric structures of higher-order tangent bundles}

In order to define the canonical structures of the higher-order tangent bundles, we first need an auxiliary
tool: the \textsl{fundamental sequences}. As we will see, every bundle structure of $\Tan^{k}M$ over $\Tan^rM$
defines an exact sequence of vector bundles over $\Tan^kM$.

Let $V(\rho^{k}_{r-1})$ be the vertical bundle of the projection $\rho^k_{r-1}$, that is,
$V(\rho^{k}_{r-1}) = \ker\Tan\rho^{k}_{r-1}$. In the natural coordinates of $\Tan(\Tan^{k}M)$ introduced
in the previous Section, for every $p \in M$ and $u_p \in V_p(\rho^{k}_{r-1})$, we have that the components
of $u_p$ are $u_p = (0,\ldots,0,v_r^{i},\ldots,v_k^{i})$. Furthermore, if
$i_{k-r+1} \colon V(\rho^{k}_{r-1}) \hookrightarrow \Tan(\Tan^{k}M)$ is the canonical embedding, then
\begin{equation}\label{Chap01_eqn:FundSeqInjectiveMapLocal}
i_{k-r+1}\left(x_0^i,\ldots,x_k^i,v_r^i,\ldots,v_k^i\right) = \left(x_0^i,\ldots,x_k^i,0,\ldots,0,v_r^i,\ldots,v_k^i\right) \ .
\end{equation}

Consider now the induced bundle of $\tau_{\Tan^{r-1}M} \colon \Tan(\Tan^{r-1}M) \to \Tan^{r-1}M$
by the canonical projection $\rho^k_{r-1}$, denoted by $\Tan^kM \times_{\Tan^{r-1}M}\Tan(\Tan^{r-1}M)$,
which is a vector bundle over $\Tan^kM$. Recall that $\Tan^kM \times_{\Tan^{r-1}M}\Tan(\Tan^{r-1}M)$
is the set of points $(p,u) \in \Tan^kM \times \Tan(\Tan^{r-1}M)$ such that
$\rho^{k}_{r-1}(p) = \tau_{\Tan^{r-1}M}(u)$. Then we have the following commutative diagrams
\begin{equation*}
\xymatrix{
\Tan^kM \times_{\Tan^{r-1}M}\Tan(\Tan^{r-1}M) \ar@{-->}[r] \ar@{-->}[d] & \Tan(\Tan^{r-1}M) \ar[d]^{\tau_{\Tan^{r-1}M}} \\
\Tan^kM \ar[r]^{\rho_{r-1}^k} & \Tan^{r-1}M
}
\qquad
\xymatrix{
\Tan(\Tan^kM) \ar[rr]^{\Tan\rho_{r-1}^k} \ar[d]^{\tau_{\Tan^kM}} & \ & \Tan(\Tan^{r-1}M) \ar[d]^{\tau_{\Tan^{r-1}M}} \\
\Tan^kM \ar[rr]^{\rho^k_{r-1}} & \ & \Tan^{r-1}M
}
\end{equation*}
where the dashed lines in the first diagram correspond to the canonical projections of the direct product
$\Tan^kM \times \Tan(\Tan^{r-1}M)$ restricted to $\Tan^kM \times_{\Tan^{r-1}M}\Tan(\Tan^{r-1}M)$. Then,
we have the following result.

\begin{proposition}
There exists a unique vector bundle morphism
$s_{k-r+1} \colon \Tan(\Tan^kM) \to \Tan^kM \times_{\Tan^{r-1}M}\Tan(\Tan^{r-1}M)$
such that the following diagram is commutative:
\begin{equation*}
\xymatrix{
\Tan(\Tan^kM) \ar@/_/[ddr]_{\tau_{\Tan^kM}} \ar[dr]^{s_{k-r+1}} \ar@/^/[drr]^{\Tan\rho^k_{r-1}} & \ & \ \\
\ & \Tan^kM \times_{\Tan^{r-1}M}\Tan(\Tan^{r-1}M) \ar@{-->}[r] \ar@{-->}[d] & \Tan(\Tan^{r-1}M) \ar[d]^{\tau_{\Tan^{r-1}M}} \\
\ & \Tan^kM \ar[r]^{\rho^k_{r-1}} & \Tan^{r-1}M \ . \\
}
\end{equation*}
\end{proposition}

This vector bundle morphism is defined as follows: if $u \in \Tan(\Tan^kM)$, we have
\begin{equation}\label{Chap01_eqn:FundSeqSurjectiveMapLocal}
s_{k-r+1}(u) = \left(\tau_{\Tan^kM}(u), \Tan\rho^k_{r-1}(u)\right) \, .
\end{equation}
In the natural coordinates of $\Tan^kM$ introduced in the previous Section,
its coordinate expression is
\begin{equation*}
s_{k-r+1}(x_0^i,\ldots,x_k^i,v_0^i,\ldots,v_k^i) = (x_0^i,\ldots,x_{r-1}^i,x_{r}^i,\ldots,x_{k}^i,v_0^i,\ldots,v_{r-1}^i)\ .
\end{equation*}
From its coordinate expression, it is clear that $s_{k-r+1}$ is a surjective map.
On the other hand, $i_{k-r+1}$ is an injective map, and in addition we have
$\Im i_{k-r+1} = \ker s_{k-r+1}$. Therefore, we have constructed the following
exact sequence of vector bundles over $\Tan^kM$:
\begin{equation*}
\xymatrix{
0 \ar[r] & V(\rho^k_{r-1}) \ar[rr]^-{i_{k-r+1}} & \ & \Tan(\Tan^kM) \ar[rr]^-{s_{k-r+1}}
& \ & \Tan^kM \times_{\Tan^{r-1}M}\Tan(\Tan^{r-1}M) \ar[r] & 0 \ ,
}
\end{equation*}
which is called the \textsl{$(k-r+1)$-fundamental exact sequence}.
In local coordinates, it is given by
\begin{align*}
& \xymatrix{
0 \ar@{|->}[r] & (x_0^i,\ldots,x_k^i,v_r^i,\ldots,v_k^i) \ar@{|->}[rr]^-{i_{k-r+1}} & \ & (x_0^i,\ldots,x_k^i,0,\ldots,0,v_r^i,\ldots,v_k^i)
}
\\
& \xymatrix{
\ & (x_0^i,\ldots,x_k^i,v_0^i,\ldots,v_k^i) \ar@{|->}[rr]^-{s_{k-r+1}} & \ & (x_0^i,\ldots,x_k^i;x_0^i,\ldots,x_{r-1}^i,v_0^i,\ldots,v_{r-1}^i) \ar@{|->}[r] & 0
}
\end{align*}
Thus, we have $k$ exact sequences of vector bundles given by
\begin{align*}
& 1st \colon
\xymatrix{
0 \ar[r] & V(\rho^k_{k-1}) \ar[r]^-{i_{1}} & \Tan(\Tan^kM) \ar[r]^-{s_{1}}
& \Tan^kM \times_{\Tan^{k-1}M}\Tan(\Tan^{k-1}M) \ar[r] & 0 \ ,
}
\\ & \qquad \vdots \\
& rth \colon
\xymatrix{
0 \ar[r] & V(\rho^k_{k-r}) \ar[r]^-{i_{r}} & \Tan(\Tan^kM) \ar[r]^-{s_{r}}
& \Tan^kM \times_{\Tan^{k-r}M}\Tan(\Tan^{k-r}M) \ar[r] & 0 \ ,
}
\\ & \qquad \vdots \\
& kth \colon
\xymatrix{
0 \ar[r] & V(\beta^k) \ar[r]^-{i_{k}} & \Tan(\Tan^kM) \ar[r]^-{s_{k}}
& \Tan^kM \times_{M}\Tan M \ar[r] & 0 \ ,
}
\end{align*}
These sequences can be connected by means of the following connecting maps
\begin{equation*}\label{Chap01_eqn:FundSeqConnectingMapDef}
h_{k-r+1} \colon \Tan^kM \times_{\Tan^{k-r}M}\Tan(\Tan^{k-r}M) \longrightarrow V(\rho^k_{r-1}) \, ,
\end{equation*}
locally defined as
\begin{equation}\label{Chap01_eqn:FundSeqConnectingMapLocal}
h_{k-r+1}\left(x_0^i,\ldots,x_{k}^i,v_0^i,\ldots,v_{k-r}^i\right) = 
\left(x_0^i,\ldots,x_k^i,0,\ldots,0,\frac{r!}{0!}v_0^i,\frac{(r+1)!}{1!}v_1^i,\ldots,\frac{k!}{(k-r)!}v_{k-r}^i\right) \ .
\end{equation}
It can be easily proved that these maps are globally well-defined and are vector bundle isomorphisms
over $\Tan^kM$. Then we have the following connection between two fundamental exact sequences:
\begin{equation*}
\xymatrix{
0 \ar[r] & V(\rho^k_{k-r}) \ar[rr]^{i_r} & \ & \Tan(\Tan^kM) \ar[rr]^-{s_r} & \ & \Tan^kM \times_{\Tan^{k-r}M} \Tan(\Tan^{k-r}M) \ar[dllll]^(.25){h_{k-r+1}}|(.5){\hole} \ar[r] & 0 \\
0 \ar[r] & V(\rho^k_{r-1}) \ar[rr]_{i_{k-r+1}} & \ & \Tan(\Tan^kM) \ar[rr]_-{s_{k-r+1}} & \ & \Tan^kM \times_{\Tan^{r-1}M} \Tan(\Tan^{r-1}M) \ar[ullll]^(.75){h_r} \ar[r] & 0
}
\end{equation*}

\begin{remark}
The connecting maps $h_{k-r+1}$ defined above are just the generalization of the vertical lift
of tangent vectors in higher-order tangent bundles.
\end{remark}

\subsubsection{Canonical vector fields. Liouville vector field}

The canonical embeddings defined in \eqref{Chap01_eqn:JetBundlesCanonicalEmbeddings} restrict
to the higher-order tangent bundles and enable us to define the following maps:
\begin{equation} \label{Chap01_eqn:HOTanBundleCanonicalImmersionDef}
\begin{array}{rcl}
j_{r} \colon \Tan^{k}M & \longrightarrow & \Tan(\Tan^{r-1}M) \\
j^{k}_0\phi & \longmapsto & j^1_0(j^{r-1}_0\phi)
\end{array} \, .
\end{equation}
where $1 \leqslant r \leqslant k$. In the natural coordinates of $\Tan^kM$ we have
\begin{equation}\label{Chap01_eqn:HOTanBundleCanonicalImmersionLocal}
j_r(x_0^i,\ldots,x_k^i) = (x_0^i,\ldots,x_{r-1}^i;x_1^i,x_2^i,\ldots,x_r^i) \ .
\end{equation}

Then, the following composition defines a vector field $\Delta_r \in \vf(\Tan^kM)$
\begin{equation*}
\xymatrix{
\Tan^kM \ar[rr]^-{\Id \times j_{k-r+1}} \ar@/_1.5pc/[rrrrrr]_{\Delta_r} & \ & 
\Tan^kM \times_{\Tan^{k-r}M} \Tan(\Tan^{k-r}M) \ar[rr]^-{h_{k-r+1}} & \ &
 V(\rho^k_{r-1}) \ar[rr]^-{i_{k-r+1}} & \ & \Tan(\Tan^kM) \, ,
}
\end{equation*}
that is, $\Delta_r = i_{k-r+1} \circ h_{k-r+1} \circ \left({\rm Id} \times j_{k-r+1}\right)$.
From the local expressions of $i_{k-r+1}$, $h_{k-r+1}$ and $j_{k-r+1}$ given by
\eqref{Chap01_eqn:FundSeqInjectiveMapLocal}, \eqref{Chap01_eqn:FundSeqConnectingMapLocal}
and \eqref{Chap01_eqn:HOTanBundleCanonicalImmersionLocal}, respectively we obtain that
\begin{equation*}
\Delta_r\left( x_0^i,\ldots,x_k^i \right) = 
\left( x_0^i,\ldots,x_k^i,0,\ldots,0,r!\,x_1^i,(r+1)!\,x_2^i,\ldots,\frac{k!}{(k-r)!}x_{k-r+1}^i \right) \, ,
\end{equation*}
or, equivalently,
\begin{equation}\label{Chap01_eqn:HOTanBundleCanonicalVF}
\Delta_r = \sum_{j=0}^{k-r} \frac{(r+j)!}{j!} x_{j+1}^i \derpar{}{x_{r+j}^i} = r!\,x_1^i\derpar{}{x_r^i} + 
(r+1)!\,x_2^i\derpar{}{x_{r+1}^i} + \ldots + \frac{k!}{(k-r)!}\,x_{k-r+1}^i\derpar{}{x_k^i} \, .
\end{equation}
In particular, taking $r = 1$, we obtain
\begin{equation}\label{Chap01_eqn:HOTanBundleLiouvilleVF}
\Delta_1 = \sum_{j=1}^{k} j x_j^i \derpar{}{x_{j}^i} = \sum_{j=0}^{k-1} (j+1) x_{j+1}^i \derpar{}{x_{j+1}^i} = 
x_1^i\derpar{}{x_1^i} + 2x_2^i\derpar{}{x_2^i} + \ldots + kx_{k}^i\derpar{}{x_k^i} \ .
\end{equation}

\begin{definition}
The vector field $\Delta_r \in \vf(\Tan^kM)$ is the \textnormal{$r$th-canonical vector field}.
In particular, $\Delta_1$ is called the \textnormal{Liouville vector field} in $\Tan^kM$.
\end{definition}

In the particular case $k = 1$, we obtain a single vector field $\Delta \in \vf(\Tan M)$ locally given by
\begin{equation}\label{Chap01_eqn:FOTanBundleLiouvilleVF}
\Delta = v^i \derpar{}{v^i} \, ,
\end{equation}
that is, the usual Liouville vector field of the tangent bundle.

\subsubsection{Almost-tangent structures. Vertical endomorphisms}

\begin{definition}
A \textnormal{$k$th-order almost-tangent structure} on a $(k+1)n$-dimensional manifold $N$
is an endomorphism $J \colon \Tan N \to \Tan N$ satisfying:
\begin{enumerate}
\item $J^{k+1} = 0$.
\item $\rank(J) = kn$.
\end{enumerate}
\end{definition}

Observe that a first-order almost-tangent structure is an endomorphism $J \colon \Tan N \to \Tan N$,
where $\dim N = 2n$, such that $J^2 = 0$ and $\rank(J) = n$. In particular, the tangent bundle of every
manifold $M$ is endowed with a canonical first-order almost-tangent structure given by the vertical
endomorphism $J \colon \Tan(\Tan M) \to \Tan(\Tan M)$. This endomorphism is given in coordinates by
\begin{equation}\label{Chap01_eqn:FOTanBundleVertEnd}
J = \d x^i \otimes \derpar{}{v^i} \, .
\end{equation}
In this Section we show that the $k$th-order tangent bundle of $M$ is endowed with a canonical $k$th-order
almost-tangent structure.

\begin{definition}
For $1 \leqslant r \leqslant k$, let $i_{k-r+1}$, $h_{k-r+1}$, $s_r$ be the morphisms
of the fundamental exact sequences introduced previously. The map
\begin{equation*}
J_r = i_{k-r+1} \circ h_{k-r+1} \circ s_r \colon \Tan(\Tan^kM) \longrightarrow \Tan(\Tan^kM) \, ,
\end{equation*}
defined by the composition
\begin{equation*}
\xymatrix{
\Tan(\Tan^kM) \ar[rr]^-{s_r} \ar@/_1.5pc/[rrrrrr]_{J_r} & \ & 
\Tan^kM \times_{\Tan^{k-r}M} \Tan(\Tan^{k-r}M) \ar[rr]^-{h_{k-r+1}} & \ & 
V(\rho^k_{r-1}) \ar[rr]^-{i_{k-r+1}} & \ & \Tan(\Tan^kM)
} \, ,
\end{equation*}
is called the \textnormal{$r$th vertical endomorphism} of $\Tan(\Tan^kM)$.
\end{definition}

From the local expressions of $i_{k-r+1}$, $s_r$, $h_{k-r+1}$ given by
\eqref{Chap01_eqn:FundSeqInjectiveMapLocal}, \eqref{Chap01_eqn:FundSeqSurjectiveMapLocal}
and \eqref{Chap01_eqn:FundSeqConnectingMapLocal}, respectively, we obtain the coordinate
expression of the $r$th vertical endomorphism, which is
\begin{equation*}
J_r(x_0^i,\ldots,x_k^i,v_0^i,\ldots,v_k^i) =
\left(x_0^i,\ldots,x_k^i,0,\ldots,0,r!\,v_0^i,(r+1)!\,v_1^i,\ldots,\frac{k!}{(k-r)!}\,v_{k-r}^i\right) \, ,
\end{equation*}
that is,
\begin{equation}\label{Chap01_eqn:HoTanBundleVertEndR}
J_r = \sum_{j=0}^{k-r} \frac{(r+j)!}{j!} \, \d x_j^i \otimes \derpar{}{x_{r+j}^i} \, .
\end{equation}
In particular, for $r = 1$, we have
\begin{equation}\label{Chap01_eqn:HoTanBundleVertEnd1}
J_1 = \sum_{j=0}^{k-1} (j+1) \d x_j^i \otimes \derpar{}{x_{j+1}^i} \, .
\end{equation}

\begin{proposition}
The $r$th-vertical endomorphism $J_r$ has constant rank $(k-r+1)n$ and satisfies that
\begin{equation*}
\left(J_r\right)^s =
\begin{cases}
J_{rs} & \mbox{ if } \ rs \leqslant k \\
0 & \mbox{ if } \ rs > k
\end{cases}
\end{equation*}
\end{proposition}

As a consequence of this last result, the $1$st-vertical endomorphism $J_1$ defines a $k$th-order
almost-tangent structure in $\Tan^kM$, which is called the \textsl{canonical almost-tangent structure}
of $\Tan^kM$. In addition, every other vertical endomorphism $J_r$ is obtained by composing $J_1$ with
itself $r$ times. Furthermore, we have the following result relating the canonical vector fields $\Delta_s$
with the vertical endomorphisms $J_r$.

\begin{proposition} 
Let $1 \leqslant r,s \leqslant k$ be two integers. Then,
\begin{enumerate}
\item $J_r\circ \Delta_s = \begin{cases} \Delta_{r+s} & \mbox{ if } \ r+s \leqslant k \\
0 & \mbox{ if } \ r+s > k \end{cases}$
\item $\left[ \Delta_r,J_s \right] =  \begin{cases} -sJ_{r+s-1} & \mbox{ if } \ r+s-1 \leqslant k \\
 0 & \mbox{ if } \ r+s-1 > k \end{cases}$
\item $\left[ J_r,J_s \right] = 0$, with $1\leqslant rs \leqslant k$.
\end{enumerate}
\end{proposition}

As a consequence, starting from the Liouville vector field and the vertical endomorphisms,
we can recover all the canonical vector fields. However, since all the vertical endomorphisms
are obtained from $J_1$, we conclude that all the canonical structures in $\Tan^kM$ are
obtained from the Liouville vector field and the canonical almost-tangent structure.

Consider now the dual maps $J_r^*$ of $J_r$, $1 \leqslant r \leqslant k$;
that is, the maps $J_r^* \colon \Tan^*(\Tan^kM) \to \Tan^*(\Tan^kM)$, 
and their natural extensions to the exterior algebra $\Lambda\Tan^*(\Tan^kM)$ 
(also denoted by $J_r^*$). Their action on the set of differential forms is given by
\begin{equation*}
(J_r^*\omega)(X_1,\ldots,X_p) = \omega(J_r(X_1),\ldots,J_r(X_p)) \, ,
\end{equation*}
for $\omega \in \df^p(\Tan^kM)$ and $X_1,\ldots,X_p \in \vf(\Tan^kM)$,
and for every $f \in \Cinfty(\Tan^kM)$ we write $J_r^*(f) = f$.

\begin{definition}
The endomorphism $J_r^* \colon \df(\Tan^kM) \to \df(\Tan^kM)$, $1\leqslant r \leqslant k$, 
is called the \textsl{$r$th vertical operator}, and it is locally given by
\begin{equation*}
J_r^*(f) = f \, , \mbox{ for every } \ f \in \Cinfty(\Tan^kM) \quad ; \quad
J_r^*(\d x_j^i) =
\begin{cases}
\displaystyle \frac{j!}{(j-r)!}\, \d x_{j-r}^i & \mbox{ if } \ j \geqslant r \\[10pt]
\displaystyle 0 & \mbox{ if } \ j < r
\end{cases} \ .
\end{equation*}
\end{definition}

\subsection{Tulczyjew's derivation}

In the set $\displaystyle \bigoplus_{k\geqslant 0}\df(\Tan^kM)$
we define an equivalence relation as follows:
for $\alpha \in \df(\Tan^kM)$ and $\beta \in \df(\Tan^{k'}M)$,
\begin{equation*}
\alpha \sim \beta \Longleftrightarrow
\begin{cases} \alpha = (\rho^k_{k'})^*(\beta) & \mbox{ if }k'\leqslant k \\ 
\beta = (\rho^{k'}_k)^*(\alpha) & \mbox{ if }k' \geqslant k
\end{cases} \ .
\end{equation*}
Then we consider the quotient set
\begin{equation*}
{\mit\Omega} = \bigoplus_{k \geqslant 0}\df(\Tan^kM)/ \sim \, ,
\end{equation*}
which is a commutative graded algebra. In this set we can define the
\textsl{Tulczyjew's derivation}, denoted by $d_T$, as follows:
for every $f \in \Cinfty(\Tan^kM)$ we construct the function $d_Tf \in \Cinfty(\Tan^{k+1}M)$
given by
\begin{equation*}\label{Chap01_eqn:HOTanBundleTulczyjewDerDef}
(d_Tf)(j^{k+1}_0\phi) = (\d_{j^{k}_0\phi}f)(j_{k+1}(j^{k+1}_0\phi)) \, ,
\end{equation*}
where $j_{k+1} \colon \Tan^{k+1}M \to \Tan(\Tan^kM)$ is the canonical injection introduced in
\eqref{Chap01_eqn:HOTanBundleCanonicalImmersionDef}, and $d_{j^{k}_0\phi}f$ is the exterior
derivative of $f$ in $j^{k}_0\phi \in \Tan^kM$.
From the coordinate expression \eqref{Chap01_eqn:HOTanBundleCanonicalImmersionLocal} for $j_{k+1}$, we obtain that
\begin{equation}\label{Chap01_eqn:HOTanBundleTulczyjewDerLocal}
d_Tf\left(x_0^i,\ldots,x_{k+1}^i\right) = \sum_{j=0}^{k}x_{j+1}^i \derpar{f}{x_j^i}(x_0^i,\ldots,x_{k}^i) \, .
\end{equation}
This map $d_T$ extends to a derivation of degree $0$ in $\mit\Omega$ and,
as $d_T\d = \d d_T$, it is determined by its action on functions
and by the property $d_T(\d x_j^i) = \d x_{j+1}^i$.

\begin{remark}
Bearing in mind the results in Section \ref{Chap01_sec:HOJetBundlesTotalDeriv}, the Tulczyjew's derivation
can be defined in the following equivalent way: let us consider the fiber bundle $\pi \colon \R \times M \to \R$,
$J^k\pi$ the $k$th-order jet bundle of $\pi$, and $d/dt \in \vf(\pi^{k+1}_{k})$ the total time derivative
associated to the canonical vector field in $\R$. Let us consider the following commutative diagram
\begin{equation*}
\xymatrix{
\R \times \Tan^{k+1}M \ar[rr]^-{\pr_2} \ar[dd]_{\pi^{k+1}_{k}} & \ & \Tan^{k+1}M \ar[dd]^{\rho^{k+1}_{k}} \\
\ & \ & \ \\
\R \times \Tan^{k}M \ar[rr]^-{\pr_2} & \ & \Tan^{k}M
}
\end{equation*}
where we have used the identification $J^{k}\pi \simeq \R \times \Tan^{k}M$,
and $\pr_2 \colon \R \times \Tan^{k}M \to \Tan^{k}M$ is the canonical projection
on the second factor. Then, the total time derivative $d/dt$ induces an operator
$T \in \vf(\rho^{k+1}_{k})$ which satisfies
\begin{equation*}
T_{\pr_2(j^{k+1}_p\phi)} = \Tan\pr_2 \left( \restric{\frac{d}{dt}}{j^{k+1}_p\phi} \right) \, .
\end{equation*}
In coordinates,
\begin{equation*}
T = \sum_{j=0}^{k}x_{j+1}^i \derpar{}{x_j^i}
\end{equation*}
The derivation corresponding to $T$ is denoted $d_T$ and coincides with the
Tulczyjew's derivation.
\end{remark}

\subsection{Higher-order semisprays}

\begin{definition}\label{Chap01_def:HOTanBundleHolonomicCurve}
A curve $\psi \colon \R \to \Tan^{k}M$ is \textnormal{holonomic of type $r$}, $1 \leqslant r \leqslant k$,
if, denoting $\phi = \beta^k \circ \psi$, then $j^{k-r+1}_0\phi = \rho^k_{k-r+1} \circ \psi$, where
$j^{k-r+1}_0\phi$ is the canonical lifting of $\phi$ to $\Tan^{k-r+1}M$. That is, the following
diagram is commutative
\begin{equation*}
\xymatrix{
\ & \ & \Tan^kM \ar[d]_{\rho^k_{k-r+1}} \ar@/^2.5pc/[ddd]^{\beta^k} \\
\R \ar@/^1.5pc/[urr]^{\psi} \ar@/_1.5pc/[ddrr]_{\phi = \beta^k\circ\psi}
 \ar[rr]^-{\rho^k_{k-r+1}\circ\psi} \ar[drr]_{j^{k-r+1}_0\phi} 
 & \ & \Tan^{k-r+1}M \ar[d]_{\Id} \\
\ & \ & \Tan^{k-r+1}M \ar[d]_{\beta^{k-r+1}} \\
\ & \ & M
}
\end{equation*}
In particular, a curve $\psi \colon \R \to \Tan^{k}M$ is \textnormal{holonomic of type $1$},
or simply \textnormal{holonomic} if it is the canonical lifting of a curve $\phi \colon \R \to M$,
that is, $j^k_0\phi = \psi$.
\end{definition}

\begin{definition}
A vector field $X \in \vf(\Tan^kM)$ is a \textnormal{semispray of type $r$},
$1 \leqslant r \leqslant k$, if every integral curve $\psi$ of $X$ is
holonomic of type $r$. If $r = 1$, the vector field is said to be
a \textnormal{semispray of type $1$}, a \textnormal{$k$th-order differential equation} ($k$-O.D.E.),
or a \textnormal{holonomic vector field}.
\end{definition}

In coordinates, let $\psi(t) = (\psi^i_0(t),\ldots,\psi^i_k(t))$ be a curve in $\Tan^kM$.
Then $\psi$ is holonomic of type $r$ if its component functions satisfy the following
systems of differential equations
\begin{equation*}
\psi^i_{j} = \frac{d^j\psi^i_0}{dt^j} \, , \quad 1 \leqslant j \leqslant k-r+1 \, , \ 1 \leqslant i \leqslant m \, ,
\end{equation*}
or, equivalently
\begin{equation*}
\psi^i_{j+1} = \frac{d\psi^i_j}{dt} \, , \quad 0 \leqslant j \leqslant k-r \, , \ 1 \leqslant i \leqslant m \, .
\end{equation*}
From this, the local expression of a semispray of type $r$ is
\begin{equation*}\label{Chap01_eqn:HoTanBundleSemisprayTypeR}
X = x_1^i\derpar{}{x_0^i} + x_2^i\derpar{}{x_1^i} + \ldots + x_{k-r+1}^i\derpar{}{x_{k-r}^i} +
X_{k-r+1}^i\derpar{}{x_{k-r+1}^i} + \ldots + X_k^i\derpar{}{x_k^i} \, ,
\end{equation*}
and, in particular, the coordinate expression of a semispray of type $1$ is
\begin{equation*}\label{Chap01_eqn:HoTanBundleSemisprayType1}
X = x_1^i\derpar{}{x_0^i} + x_2^i\derpar{}{x_1^i} + \ldots + x_{k}^i\derpar{}{x_{k-1}^i} + X_k^i\derpar{}{x_k^i} \, ,
\end{equation*}

From this coordinate expressions it is clear that every semispray of type $r$
is a semispray of type $s$, for $s \geqslant r$. In addition, we can state
the following result.

\begin{proposition}\label{Chap01_prop:HOTanBundleEquivalenceSemisprays}
Let $X \in \vf(\Tan^{k}M)$ be a vector field. The following assertions are equivalent:
\begin{enumerate}
\item $X$ is a semispray of type $r$.
\item $\Tan\rho^k_{k-r} \circ X = j_{k-r+1}$, that is, the following diagram commutes
\begin{equation*}
\xymatrix{
\Tan(\Tan^kM) \ar[drr]^{\Tan\rho^k_{k-r}} \\
\Tan^kM \ar[u]^X \ar[rr]^-{j_{k-r+1}} & \ & \Tan(\Tan^{k-r}M) \ .
}
\end{equation*}
\item $J_r \circ X = \Delta_r$.
\end{enumerate}
\end{proposition}

\begin{remark}
Taking $k = 1$ in the previous Definitions we recover the definitions of
holonomic curve in $\Tan M$ and the S.O.D.E. vector fields.
Hence, holonomic curves of type $1$ and semisprays of type $1$ are the natural
generalization of these concepts to the higher-order tangent bundles.
\end{remark}

\begin{definition}\label{Chap01_def:HoTanBundleSemisprayPath}
Let $X\in\vf(\Tan^kM)$ be a semispray of type $r$. A curve $\phi \colon \R \to M$
is said to be a \textnormal{path} or \textnormal{solution} of $X$ if
$j^k_0\phi \colon \R \to \Tan^kM$ is an integral curve of $X$.
\end{definition}

In coordinates, $\phi(t) = (\phi^i(t))$ verifies the following
system of differential equations of order $k+1$
\begin{align*}
\frac{d^{k-r+2}\phi^i}{dt^{k-r+2}} = X_{k-r+1}^i\left(\phi,\frac{d\phi}{dt},\ldots,\frac{d^k\phi}{dt^k}\right) \ , \ \ldots \ , \
\frac{d^{k+1}\phi^i}{dt^{k+1}} = X_k^i\left(\phi,\frac{d\phi}{dt},\ldots,\frac{d^k\phi}{dt^k}\right) \, .
\end{align*}


\section{Multivector fields}
\label{Chap01_sec:MultivectorFields}

In this Section we introduce the analog to differential forms of arbitrary degree
for vector fields:
the multivector fields, which are just contravariant skew-symmetric tensors of
arbitrary degree on a manifold $M$. We will study their relation with distributions.
(See \cite{art:Echeverria_Munoz_Roman98} for details).

Along this Section, $M$ will denote a $m$-dimensional smooth manifold.

\subsection{(Locally) Decomposable multivector fields. Integrability conditions}

\begin{definition}
A \textnormal{multivector field of degree $k$}, or \textnormal{$k$-multivector field},
is a section of the bundle $\Lambda^{k}\Tan M$.
The set of all multivector fields of degree $k$ in $M$ is denoted $\vf^{k}(M)$.
\end{definition}

In general, given a $k$-multivector field $\X \in \vf^k(M)$, for every $p \in M$
there exists an open neighborhood $U_p \subseteq M$ and $X_1,\ldots,X_r \in \vf(U_p)$
such that the multivector field $\X$ can be written in $U_p$ as
\begin{equation}\label{Chap01_eqn:GenericMultiVF}
\X = \sum_{1 \leqslant i_1 < \ldots < i_k \leqslant r} f^{i_1\dots i_k} X_{i_1} \wedge \ldots \wedge X_{i_k} \, ,
\end{equation}
with $f^{i_1\dots i_k} \in \Cinfty(U_p)$ and $k \leqslant r \leqslant \dim M$.
Now, if for every $p$ we have $r = k$, we have the following definition.

\begin{definition}
A $k$-multivector field $\X \in \vf^{k}(M)$ is \textnormal{decomposable}
if there are $X_1,\ldots,X_k \in \vf(M)$ such that
$\X = X_1 \wedge \ldots \wedge X_k$.
The multivector field $\X \in \vf^{k}(M)$ is \textnormal{locally decomposable}
if for every $p \in M$ there exists an open neighborhood $U_p \subseteq M$ and
$X_1,\ldots,X_k \in \vf(U_p)$ such that $\X = X_1 \wedge \ldots \wedge X_k$ on $U_p$.
\end{definition}

Every multivector field $\X \in \vf^{k}(M)$ defines an operation $\inn(\X)$ of degree
$-k$ in the algebra of differential forms $\df(M)$. In particular, if $\omega \in \df^{n}(M)$
is a $n$-form in $M$, then bearing in mind \eqref{Chap01_eqn:GenericMultiVF} we have
$$
\inn(\X)\,\omega
= \sum_{1 \leqslant i_1 < \ldots < i_k \leqslant r} f^{i_1\dots i_k} \inn(X_{i_1} \wedge \ldots \wedge X_{i_k}) \, \omega
= \sum_{1 \leqslant i_1 < \ldots < i_k \leqslant r} f^{i_1\dots i_k} \inn(X_{i_1}) \ldots \inn(X_{i_k}) \, \omega
$$
if $n \geqslant k$, and it vanishes if $n < k$. A $n$-form $\omega$ is said to be
\textsl{$j$-nondegenerate}, $1 \leqslant j \leqslant n-1$, if for every $p \in M$
and $\X \in \vf^{j}(M)$, $\inn(\X_p)\omega_p = 0$ if, and only if, $\X_p = 0$.

Let $\D$ be a $k$-dimensional distribution in $M$, that is, a $k$-dimensional
subbundle of $\Tan M$. It is clear that sections of $\Lambda^k\D$ are $k$-multivector
fields in $M$, and that the existence of a non-vanishing global section of $\Lambda^k\D$
is equivalent to the orientability of the distribution $\D$. Then, we want to
study the relation between non-vanishing $k$-multivector fields in $M$ and
$k$-dimensional distributions in $\Tan M$.

\begin{definition}
A non-vanishing multivector field $\X \in \vf^{k}(M)$ and a $k$-dimensional
distribution $\D \subset \Tan M$ are \textnormal{locally associated} if there
exists a connected open set $U \subseteq M$ such that $\restric{\X}{U}$ is a section of
$\restric{\Lambda^{k}\D}{U}$.
\end{definition}

As a consequence of this last Definition we can introduce an equivalence relation
on the set of non-vanishing $k$-multivector fields in $M$ as follows:
two $k$-multivector fields $\X,\X^\prime \in \vf^{k}(M)$ are related if, and only if,
they are both locally associated, on the same connected open set $U \subseteq M$,
with the same distribution $\D$. In addition, in this case there exists a non-vanishing
function $f \in \Cinfty(U)$ such that $\X^\prime = f\X$ on $U$. The equivalence classes
of this quotient set will be denoted by $\{ \X \}_U$.

\begin{theorem}\label{Chap01_thm:CorrespondenceMultiVFOrientableDist}
There is a bijective correspondence between the set of $k$-dimensional orientable
distributions $\D \subseteq \Tan M$ and set of equivalence classes $\{ \X \}_M$
of non-vanishing, locally decomposable $k$-multivector fields in $M$.
\end{theorem}

\begin{remark}
If $\D \subseteq \Tan M$ is a non-orientable $k$-dimensional distribution, then for every
$p \in M$ there exists an open neighborhood $U_p \subseteq M$ and a non-vanishing
$k$-multivector field $\X \in \vf^{k}(U)$ such that $\restric{\D}{U_p} = \D_U(\X)$.
\end{remark}

If $\X \in \vf^{k}(M)$ is a non-vanishing, locally decomposable $k$-multivector field and
$U \subseteq M$ is a connected open set, then the distribution associated
to the equivalence class $\{\X\}_U$ will be denoted by $\D_U(\X)$. If $U = M$,
then we write simply $\D(\X)$.

\begin{definition}
Let $\X \in \vf^{k}(M)$ be a multivector field.
\begin{itemize}
\item A submanifold $N \hookrightarrow M$ with $\dim N = k$ is an
\textnormal{integral manifold} of $\X$ if for every $p \in N$,
$\X_p$ spans $\Lambda^{k}\Tan_pN$.
\item Given an open subset $U \subseteq M$, $\X$ is \textnormal{integrable on $U$}
if for every $p \in U$ there exists an integral manifold $N \hookrightarrow U$ of $\X$
containing $p$.
\item $\X$ is \textnormal{integrable} if it is integrable in $M$.
\end{itemize}
\end{definition}

It is clear from the definition that every integrable multivector field is
non-vanishing. Now, using Theorem \ref{Chap01_thm:CorrespondenceMultiVFOrientableDist},
we can give the following definitions.

\begin{definition}
Let $\X \in \vf^{k}(M)$ be a multivector field.
\begin{itemize}
\item Given a connected open set $U \subseteq M$, $\X$ is \textnormal{involutive on $U$}
if it is locally decomposable in $U$ and its associated distribution $\D_U(\X)$ is involutive.
\item $\X$ is \textnormal{locally involutive around $p \in M$} if there exists
a connected open neighborhood $U_p \ni p$ such that $\X$ is involutive on $U_p$.
\item $\X$ is \textnormal{involutive} if it is involutive on $M$ or, equivalently,
if it is locally involutive around every $p \in M$.
\end{itemize}
\end{definition}

These definitions enable us to reformulate the classical Frobenius' Theorem in
the setting of multivector fields.

\begin{theorem}[Frobenius]
A non-vanishing and locally decomposable multivector field $\X \in \vf^{k}(M)$
is integrable on a connected open set $U \subseteq M$ if, and only if, it is involutive
on $U$.
\end{theorem}

\begin{remark}
If a multivector field $\X \in \vf^{k}(M)$ is integrable, then so is every other multivector
field in it equivalence class $\{\X\}$, and all of them have the same integral manifolds.
\end{remark}

Recall that a $k$-dimensional distribution $\D \subseteq \Tan M$ is integrable if, and only if,
it is locally spanned by a set of vector fields $X_1,\ldots,X_k \in \vf(M)$ such that
$[X_i,X_j] = 0$ for every pair $X_i,X_j$. Then, a multivector field $\X \in \vf^{k}(M)$ is
integrable if, and only if, for every $p \in M$ there exists an open neighborhood $U_p \subseteq M$
and $X_1,\ldots,X_k \in \vf(U_p)$ such that
\begin{enumerate}
\item $X_1,\ldots,X_k$ span $\D_{U_p}(\X)$.
\item $[X_i,X_j] = 0$ for every pair $X_i,X_j$.
\end{enumerate}
Then there exists a non-vanishing function $f \in \Cinfty(U_p)$
such that $\X = f X_1 \wedge \ldots \wedge X_k$.

\begin{remark}
In many applications we have locally decomposable multivector fields $\X \in \vf^k(M)$ 
which are not integrable in $M$, but integrable in a submanifold of $M$.
A (local) algorithm for finding this submanifold as been developed \cite{art:Echeverria_Munoz_Roman98}.
\end{remark}

\begin{definition}
A multivector field $\X \in \vf^{k}(M)$ is a \textnormal{dynamical multivector field} if
\begin{enumerate}
\item $\X$ is integrable.
\item For every $p \in M$ there exists an open neighborhood $U_p \subseteq M$
and $X_1,\ldots,X_k \in \vf(M)$ such that $[X_i,X_j] = 0$ for every pair $X_i,X_j$,
and $\X = X_1 \wedge \ldots \wedge X_k$ on $U_p$.
\end{enumerate}
\end{definition}

\begin{proposition}\label{Chap01_eqn:MultiVFDynamicalExistence}
Let $\{\X\} \subseteq \vf^{k}(M)$ be a class of integrable $k$-multivector fields.
Then there is a representative $\X$ of the class which is a dynamical multivector field.
\end{proposition}

\subsection{Multivector fields in fiber bundles and jet bundles. Holonomy condition}
\label{Chap01_sec:MultivectorFieldsHolonomic}

We are interested in the particular situation of a fiber bundle and, more
precisely, of jet bundles.

First, let $\pi \colon E \to M$ be a fiber bundle, with $\dim M = m$ and $\dim E = m + n$.
We are interested in the case where the integral manifolds of multivector fields are
sections of the projection $\pi$.

\begin{definition}
A multivector field $\X \in \vf^{m}(M)$ is \textnormal{transverse to the projection $\pi$},
or \textnormal{$\pi$-transverse}, if at every point $y \in E$ we have
$$
(\inn(\X)(\pi^*\omega))_y \neq 0 \, ,
$$
for every $\omega \in \df^{m}(M)$ satisfying $\omega(\pi(y)) = 0$.
\end{definition}

Observe that if $\X \in \vf^{m}(E)$ is a locally decomposable multivector field,
then $\X$ is $\pi$-transverse if, and only if, $\Tan_y\pi(\D(\X)) = \Tan_{\pi(y)}M$
for every $y \in E$.

\begin{theorem}
Let $\X \in \vf^{m}(E)$ be an integrable multivector field. Then $\X$ is $\pi$-transverse
if, and only if, its integral manifolds are local sections of $\pi$.
\end{theorem}

In this case, if $\phi \colon U \subseteq M \to E$ is a local section with $\phi(x) = y$
and $\phi(U)$ is the integral manifold of $\X$, then $\Tan_y(\Im\phi) = \D_y(\X)$.
Now, let us consider the following diagram
$$
\xymatrix{
\Lambda^{m}\Tan(\pi^{-1}(U)) \ar[dd]^{\Lambda^m\tau_E} \ar[rrr]^-{\Lambda^m\Tan\pi} & \ & \ & \Lambda^m\Tan U \ar@/^1.2pc/[lll]^-{\Lambda^m\Tan\phi} \ar[dd]^{\Lambda^m\tau_U} \\
\ & \ & \ & \ \\
\pi^{-1}(U) \ar[rrr]^-{\pi} \ar@/^1pc/[uu]^{\X} & \ & \ & U \ar@/^1.2pc/[lll]^-{\phi}
}
$$
where $U \subseteq M$ is an open set. Then we have

\begin{proposition}
A multivector field $\X \in \vf^{m}(E)$ is integrable and $\pi$-transverse if, and only if,
for every $y \in E$ there exists a local section $\phi \in \Gamma_U(\pi)$ such that $\phi(\pi(y)) = y$
and a non-vanishing function $f \in \Cinfty(E)$ such that
$\Lambda^{m}\Tan\phi = f\X \circ \phi \circ \Lambda^{m}\tau_U$.
\end{proposition}

Now, let us consider the $k$th-order jet bundle of $\pi$, $J^k\pi$.
We are interested in the case when the integral sections of the multivector
field $\X \in \vf^{m}(J^{k}\pi)$ are the $k$th prolongations of sections of $\pi$.

\begin{definition}
A multivector field $\X \in \vf^m(J^{k}\pi)$ is \textnormal{holonomic of type $r$},
$1 \leqslant r \leqslant k$, if
\begin{enumerate}
\item $\X$ is integrable.
\item $\X$ is $\bar{\pi}^k$-transverse.
\item The integral sections $\psi \in \Gamma(\bar{\pi}^k)$ of $\X$ are holonomic of type $r$.
\end{enumerate}
In particular, a multivector field $\X \in \vf^{m}(J^{k}\pi)$ is
\textnormal{holonomic of type $1$} (or simply \textnormal{holonomic}) if
it is integrable, $\bar{\pi}^{k}$-transverse and its integral
sections $\psi \in \Gamma(\bar{\pi}^k)$ are the $k$th prolongations of
sections $\phi \in \Gamma(\pi)$.
\end{definition}

In natural coordinates, let $\X \in \vf^{m}(J^k\pi)$ be a locally decomposable
and $\bar{\pi}^k$-transverse multivector field. From Proposition
\ref{Chap01_eqn:MultiVFDynamicalExistence}, this multivector field $\X$ may be chosen to
have the following coordinate expression
\begin{equation*}
\X = \bigwedge_{i=1}^{m} f_i
\left(  \derpar{}{x^i} + F_i^\alpha\derpar{}{u^\alpha} + F_{I,i}^\alpha\derpar{}{u_I^\alpha} \right) \, ,
\qquad (1 \leqslant |I| \leqslant k) \ ,
\end{equation*}
with $f_i$ non-vanishing local functions. Then, the condition for $\X$
to be holonomic of type $r$ gives the following equations:
\begin{equation}\label{Chap01_eqn:MultiVFHolonomyLocal}
F_i^\alpha = u_i^\alpha \quad ; \quad
F_{I,i}^\alpha = u_{I+1_i}^\alpha \, , \qquad 1 \leqslant |I| \leqslant k-r \, , \
1 \leqslant i \leqslant m \, , \ 1 \leqslant \alpha \leqslant n \, .
\end{equation}
Hence, the local expression of a locally decomposable holonomic of type $r$
multivector field is
\begin{equation*}
\X = \bigwedge_{i=1}^{m} f_i
\left(  \derpar{}{x^i} + u_i^\alpha\derpar{}{u^\alpha} + \sum_{|I|=1}^{k-r} u_{I+1_i}^\alpha\derpar{}{u_I^\alpha}
+ \sum_{|I|=k-r+1}^{k} F_{I,i}^\alpha\derpar{}{u_I^\alpha}\right) \, ,
\end{equation*}
In the particular case $r=1$, that is, $\X$ being a locally decomposable holonomic multivector
field, its local expression is
\begin{equation*}
\X = \bigwedge_{i=1}^{m} f_i
\left(  \derpar{}{x^i} + u_i^\alpha\derpar{}{u^\alpha} + \sum_{|I|=1}^{k-1} u_{I+1_i}^\alpha\derpar{}{u_I^\alpha}
+ F_{K,i}^\alpha\derpar{}{u_K^\alpha}\right) \, .
\end{equation*}
where $|K| = k$.

\begin{remark}
It is important to point out that a locally decomposable and $\bar{\pi}^k$-transverse multivector field
$\X$ satisfying the local equations \eqref{Chap01_eqn:MultiVFHolonomyLocal} may not be holonomic of type
$r$, since these local equations are not a sufficient nor necessary condition for the multivector field
to be integrable. However, we can assure that if such a multivector field admits integral sections, then
its integral sections are holonomic of type $r$. In first-order theories using jet bundles, these equations
are equivalent to the so-called \textsl{semi-holonomy condition} (or \textsl{S.O.P.D.E condition}, from
\textsl{Second Order Partial Differential Equation}) \cite{art:Echeverria_Munoz_Roman98}.
In the general setting, a locally decomposable and $\bar{\pi}^k$-transverse multivector field
satisfying equations \eqref{Chap01_eqn:MultiVFHolonomyLocal} is called \textsl{semi-holonomic}.
\end{remark}


\section{The constraint algorithm}
\label{Chap01_sec:ConstraintAlgorithm}

There are many works devoted to the study of a constraint algorithm for implicit differential equations.
See, for example,
\cite{art:Gotay_Nester79,art:Gotay_Nester80,art:Gotay_Nester_Hinds78,art:Mendella_Marmo_Tulczyjew95,art:Munoz_Roman92}.
Moreover, this algorithm has been generalized to many different situations, including time-dependent mechanical systems,
both for a trivial bundle and a general bundle over $\R$ endowed with a cosymplectic structure (see
\cite{art:Chinea_deLeon_Marrero94,art:deLeon_Marin_Marrero96,art:DeLeon_Marin_Marrero_Munoz_Roman02}),
and field theories in the $k$-symplectic and multisymplectic settings (see
\cite{art:DeLeon_Marin_Marrero_Munoz_Roman05,art:Gracia_Martin_Roman09}).

In this Section we briefly review the constraint algorithm for \textsl{presymplectic} systems, which is a purely
geometric algorithm based on the Dirac--Bergmann algorithm.

Let $M$ be a $m$-dimensional smooth manifold, $\omega \in \df^{2}(M)$ a closed $2$-form on $M$, and
$\alpha \in \df^{1}(M)$ a $1$-form on $M$. If $\omega$ is nondegenerate (that is, symplectic), then the equation
\begin{equation}\label{Chap01_eqn:GeneralDynEq}
\inn(X)\Omega = \alpha \, ,
\end{equation}
has a unique solution $X \in \vf(M)$ for every $1$-form $\alpha$ that we consider. In particular, the
vector field solution to equation \eqref{Chap01_eqn:GeneralDynEq} is given by
\begin{equation*}
X = \omega^\sharp(\alpha) = (\omega^\flat)^{-1}(\alpha) \, ,
\end{equation*}
where $\omega^\flat \colon \vf(M) \to \df^{1}(M)$ is the canonical isomorphism introduced in Section
\ref{Chap01_sec:SymplecticGeom}. Nevertheless, if $\omega$ is degenerate (that is, presymplectic),
then equation \eqref{Chap01_eqn:GeneralDynEq} may not have a solution defined on the whole manifold
$M$, but only on some points of $M$. The triple $(M,\omega,\alpha)$ is said to be a presymplectic system.
The aim of the \textsl{constraint algorithm}, is to find a submanifold $N \hookrightarrow M$ such that
the equation \eqref{Chap01_eqn:GeneralDynEq} has solutions in $N$ (if such a submanifold exists).
More precisely, the constraint algorithm returns the maximal submanifold $N$ of $M$ such that there exists
a vector field $X \in \vf(M)$ satisfying equation \eqref{Chap01_eqn:GeneralDynEq} with support on $N$.

The algorithm proceeds as follows. Since $\omega$ is degenerate, equation \eqref{Chap01_eqn:GeneralDynEq}
has no solution in general, or the solutions are not defined everywhere. In the most favorable cases,
equation \eqref{Chap01_eqn:GeneralDynEq} admits a global (but not unique) solution $X \in \vf(M)$.
Otherwise, we consider the subset of points in $M$ where such a solution exists, that is, we define
\begin{align*}
M_1 &= \left\{ p \in M \mid \mbox{ there exists } X_p \in \Tan_pM \mbox{ satisfying } \inn(X_p)\omega_p = \alpha_p \right\} \\
&= \left\{ p \in M \mid (\inn(Y)\alpha)(p) = 0 \mbox{ for every } Y \in \ker\omega \right\} \, ,
\end{align*}
and we assume that it is a submanifold of $M$. The submanifold $M_1 \hookrightarrow M$ is the
\textsl{compatibility submanifold}, or the \textsl{first constraint submanifold}, of the system.
Then, equation \eqref{Chap01_eqn:GeneralDynEq} admits a solution $X$ defined everywhere in $M_1$, but
$X$ is not necessarily tangent to $M_1$, and thus it does not necessarily induce a dynamics on $M_1$.
So we impose a tangency condition along $M_1$, and we obtain a new submanifold
\begin{equation*}
M_2 = \left\{ p \in M_1 \mid \mbox{ there exists } X_p \in \Tan_pM_1 \mbox{ satisfying } \inn(X_p)\omega_p = \alpha_p \right\} \, .
\end{equation*}
A solution $X$ to equation \eqref{Chap01_eqn:GeneralDynEq} does exist in $M_2$ but, again, such an $X$ is
not necessarily tangent to $M_2$, and this condition must be required. Following this process we obtain a
sequence of submanifolds
\begin{equation*}
\ldots \hookrightarrow M_{i} \hookrightarrow \ldots \hookrightarrow M_2 \hookrightarrow M_1 \hookrightarrow M \, ,
\end{equation*}
where the general description of $M_i$ is
\begin{equation*}
M_i = \left\{ p \in M_{i-1} \mid \mbox{ there exists } X_p \in \Tan_pM_{i-1} \mbox{ satisfying } \inn(X_p)\omega_p = \alpha_p \right\}
\end{equation*}

If the algorithm terminates at a nonempty set, in the sense that at some $s \geqslant 1$ we have
$M_{i+1} = M_{i}$ for every $i \geqslant s$, then we say that $M_s$ is the
\textsl{final constraint submanifold}, which is denoted by $M_f$. It may still happen that
$\dim M_f = 0$, that is, $M_f$ is a discrete set of points, and in this case the system
does not admit a proper dynamics. But if $\dim M_f > 0$, by construction, there exists a well-defined
solution $X$ of equation \eqref{Chap01_eqn:GeneralDynEq} along $M_f$.


\clearpage
\thispagestyle{empty}


\clearpage
\chapter{Mathematical physics background}
\label{Chap:MathPhysBackground}


In this second Chapter we review the geometric formulations of several kinds of physical systems.
In particular, we focus on dynamical systems and field theories whose dynamical information is
given in terms of a Lagrangian function or density. As for Chapter \ref{Chap:MathBackground}, this
is a review Chapter. Hence, no proofs or detailed calculations are given: several references
containing proofs, calculations and details are included within each Section.

The structure of the Chapter is the following. In Section \ref{Chap02_sec:AutonomousFirstOrder},
we review the Lagrangian, Hamiltonian and unified formalisms for first-order autonomous dynamical
systems. Using this geometric setting, Section \ref{Chap02_sec:HamiltonJacobi} is devoted to
introduce the geometric version of the Hamilton-Jacobi problem for these kinds of systems, both in
the Lagrangian and Hamiltonian formulations. Section \ref{Chap02_sec:AutonomousHigherOrder} is
devoted to give the geometric setting of both the Lagrangian and Hamiltonian formalisms for
higher-order dynamical systems. Finally, in Sections \ref{Chap02_sec:NonAutonomous} and
\ref{Chap02_sec:FieldTheories} we review the Lagrangian, Hamiltonian and Skinner-Rusk formulations
of non-autonomous first-order dynamical systems and first-order field theories, respectively.

We point out that the reader will find obvious similarities between the different Sections within
this Chapter. In fact, since we assume that all our physical systems are given in terms of a
Lagrangian, the geometrization of these kind of systems gives rise to similar geometric models that
may be adapted from one particular situation to another.

\section{First-order autonomous dynamical systems}
\label{Chap02_sec:AutonomousFirstOrder}

Consider a first-order autonomous Lagrangian dynamical system with $n$ degrees of freedom. Let $Q$
be a $n$-dimensional smooth manifold modeling the configuration space of this first-order dynamical
system, and $\Lag \in \Cinfty(\Tan Q)$ a Lagrangian function describing the dynamics of the system.

\subsection{Lagrangian formalism}
\label{Chap02_sec:LagrangianAutonomousFirstOrder}

(See \cite{book:Arnold89} and \cite{art:Crampin83,art:Munoz_Roman92} for details).

\subsubsection{Geometric and dynamical structures}

From the Lagrangian function $\Lag \in \Cinfty(\Tan Q)$ and the canonical
structures of the tangent bundle, namely the vertical endomorphism
$J \colon \vf(\Tan Q) \to \vf^{V(\tau_Q)}(\Tan Q)$ and the
Liouville vector field $\Delta \in \vf(\Tan Q)$, we construct the following
structures.

\begin{definition}
The \textnormal{Poincar\'{e}-Cartan $1$-form} associated to $\Lag \in \Cinfty(\Tan Q)$,
or \textnormal{Lagrangian $1$-form}, is the form $\theta_\Lag \in \df^{1}(\Tan Q)$
defined as
\begin{equation*}\label{Chap02_eqn:LagFO1FormDef}
\theta_\Lag = \inn(J)\d\Lag = \d\Lag \circ J \, .
\end{equation*}
From this, the \textnormal{Poincar\'{e}-Cartan $2$-form} associated to $\Lag$,
or \textnormal{Lagrangian $2$-form}, is the form $\omega_\Lag \in \df^{2}(\Tan Q)$
defined as
\begin{equation*}\label{Chap02_eqn:LagFO2FormDef}
\omega_\Lag = -\d\theta_\Lag  \, .
\end{equation*}
\end{definition}

It is important to point out that, given an arbitrary Lagrangian function
$\Lag \in \Cinfty(\Tan Q)$, the $2$-form $\omega_\Lag$ may not have constant rank
at every point in $\Tan Q$. If $\rank\omega_\Lag(p,v_p) = \text{const.}$
for every $(p,v_p) \in \Tan Q$, then $\Lag \in \Cinfty(\Tan Q)$ is said to be a
\textsl{geometrically admissible Lagrangian}. We will only consider Lagrangian
functions satisfying this property.

\begin{definition}
The \textnormal{Lagrangian energy} associated to $\Lag \in \Cinfty(\Tan Q)$
is the function $E_\Lag \in \Cinfty(\Tan Q)$ defined as
\begin{equation*}\label{Chap02_eqn:LagFOEnergyDef}
E_\Lag = \Delta(\Lag) - \Lag \, .
\end{equation*}
\end{definition}

\begin{remark}
In some references, the first term of the Lagrangian energy is referred to as
\textsl{Lagrangian action} associated to $\Lag$, and it is denoted by $A_\Lag$.
Then we have $E_\Lag = A_\Lag - \Lag$.
\end{remark}

It is clear from these definitions that the phase space of a first-order autonomous Lagrangian
dynamical system is the tangent bundle of the configuration manifold $Q$.

\begin{definition}
A \textnormal{first-order autonomous Lagrangian system} is a pair $(\Tan Q,\Lag)$, where $Q$
represents the configuration space and $\Lag \in \Cinfty(\Tan Q)$ is the Lagrangian function.
\end{definition}

In coordinates, bearing in mind the local expressions of the vertical endomorphism
of the tangent bundle given by \eqref{Chap01_eqn:FOTanBundleVertEnd}
and the Liouville vector field given by \eqref{Chap01_eqn:FOTanBundleLiouvilleVF},
the coordinate expression of the Poincar\'{e}-Cartan $1$-form is
\begin{equation}\label{Chap02_eqn:LagAFO1FormLocal}
\theta_\Lag =
\derpar{\Lag}{v^A} \, \d q^A \, ,
\end{equation}
from which the coordinate expression of the Poincar\'{e}-Cartan $2$-form is
\begin{equation}\label{Chap02_eqn:LagAFO2FormLocal}
\omega_\Lag
= \derpars{\Lag}{v^A}{q^B}\d q^A \wedge \d q^B + \derpars{\Lag}{v^A}{v^B} \d q^A \wedge \d v^B \, .
\end{equation}
On the other hand, the Lagrangian energy has the following coordinate expression
\begin{equation}\label{Chap02_eqn:LagAFOEnergyLocal}
E_\Lag
= v^A\derpar{\Lag}{v^A} - \Lag(q^A,v^A) \, .
\end{equation}

\begin{remark}
From the coordinate expression \eqref{Chap02_eqn:LagAFO1FormLocal}, it is clear that
the Lagrangian $1$-form $\theta_\Lag$ is a semibasic form on $\Tan Q$ since
$\theta_\Lag \in \Im(J^*)$, where $J^*\colon \Tan^*\Tan Q \to \Tan^*\Tan Q$
is the conjugate of the vertical endomorphism.
\end{remark}

Observe that, given an arbitrary Lagrangian function $\Lag \in \Cinfty(\Tan Q)$,
the Poincar\'{e}-Cartan $2$-form $\omega_\Lag \in \df^{2}(\Tan Q)$ is always
closed by definition, but not necessarily nondegenerate. That is, the Lagrangian $2$-form
is always presymplectic, but not necessarily symplectic. This leads to the
following definition.

\begin{definition}\label{Chap02_def:LagAFORegularLagrangian}
A Lagrangian function $\Lag \in \Cinfty(\Tan Q)$ is \textnormal{regular}
(and thus $(\Tan Q,\Lag)$ is a \textnormal{regular system}) if the Poincar\'{e}-Cartan
$2$-form $\omega_\Lag \in \df^{2}(\Tan Q)$ associated to $\Lag$ is symplectic.
Otherwise, the Lagrangian is said to be \textnormal{singular}
(and thus $(\Tan Q, \Lag)$ is a \textnormal{singular system}).
\end{definition}

From the coordinate expression \eqref{Chap02_eqn:LagAFO2FormLocal} of the
Poincar\'{e}-Cartan $2$-form it is clear that the nondegeneracy of $\omega_\Lag$
is locally equivalent to
\begin{equation*}
\det\left( \derpars{\Lag}{v^A}{v^B} \right)(p,v_p) \neq 0 \ , \,
\mbox{for every } (p,v_p) \in \Tan Q \, .
\end{equation*}
That is, a Lagrangian function is regular if, and only if, its Hessian matrix
with respect to the velocities is invertible at every point of $\Tan Q$.

\subsubsection{Dynamical vector field}

The dynamical trajectories of the system are given by the integral curves of
a holonomic vector field $X_\Lag \in \vf(\Tan Q)$ satisfying
\begin{equation}\label{Chap02_eqn:LagAFODynEq}
\inn(X_\Lag)\omega_\Lag = \d E_\Lag \, .
\end{equation}
This equation is the \textsl{Lagrangian equation}, and a vector field $X_\Lag$
solution to \eqref{Chap02_eqn:LagAFODynEq} (if such a vector field exists)
is called a \textsl{Lagrangian vector field}. If, in addition, $X_\Lag$ is
holonomic, then it is called the \textsl{Euler-Lagrange vector field},
and its integral curves are solutions to the \textsl{Euler-Lagrange equations}.

\begin{remark}
Notice that, following the terminology introduced in Section
\ref{Chap01_sec:CanonicalIsomorphismSymplectic}, the vector field
$X_\Lag$ solution to equation \eqref{Chap02_eqn:LagAFODynEq} is nothing but the
Hamiltonian vector field associated to the Lagrangian energy $E_\Lag$.
\end{remark}

In the natural coordinates of $\Tan Q$, a generic vector field $X_\Lag \in \vf(\Tan Q)$
is given by
\begin{equation*}
X_\Lag = f^A \derpar{}{q^A} + F^A \derpar{}{v^A} \, .
\end{equation*}
Then, bearing in mind the local expression \eqref{Chap02_eqn:LagAFO2FormLocal} of the
Lagrangian $2$-form $\omega_\Lag$, and
requiring equation \eqref{Chap02_eqn:LagAFODynEq} to hold, we obtain the following system
of $2n$ equations for the component functions of $X_\Lag$
\begin{align}
F^A\derpars{\Lag}{v^A}{v^B} = f^A \left( \derpars{\Lag}{v^A}{q^B} - \derpars{\Lag}{v^B}{q^A} \right)
- v^A\derpars{\Lag}{v^A}{q^B} + \derpar{\Lag}{q^B} \, , \label{Chap02_eqn:LagAFODynEqLocal} \\
(f^A - v^A) \derpars{\Lag}{v^A}{v^B} = 0 \, . \label{Chap02_eqn:LagAFODynEqHolonomyLocal}
\end{align}
Observe that equations \eqref{Chap02_eqn:LagAFODynEqHolonomyLocal} are the local equations
for the holonomy condition of the vector field $X_\Lag$, while
equations \eqref{Chap02_eqn:LagAFODynEqLocal} are the dynamical equations.
Observe that if the condition to be holonomic is required from the beginning, then
equations \eqref{Chap02_eqn:LagAFODynEqHolonomyLocal} are an identity, and equations
\eqref{Chap02_eqn:LagAFODynEqLocal} reduce to
\begin{equation}\label{Chap02_eqn:LagAFODynEqWithHolonomyLocal}
F^A\derpars{\Lag}{v^A}{v^B} = \derpar{\Lag}{q^B} - v^A\derpars{\Lag}{q^A}{v^B} \, .
\end{equation}
In all of these equations the Hessian matrix of $\Lag$ with respect to the velocities
appears alongside the coefficients to be determined. Therefore, we have the following result.

\begin{proposition}
If the Lagrangian function $\Lag \in \Cinfty(\Tan Q)$ is regular, then there exists a unique
vector field $X_\Lag \in \vf(\Tan Q)$ solution to the equation \eqref{Chap02_eqn:LagAFODynEq}
which, in addition, is holonomic.
\end{proposition}

\begin{remark}
Notice that the existence and uniqueness of the solution to equation \eqref{Chap02_eqn:LagAFODynEq}
is also a straightforward consequence to the fact that $\Lag \in \Cinfty(\Tan Q)$
is regular if, and only if, the $2$-form $\omega_\Lag \in \df^{2}(\Tan Q)$ is symplectic.
\end{remark}

If the Lagrangian function provided is not regular, then the $2$-form $\omega_\Lag$
is merely presymplectic, so the existence of solutions to the equation
\eqref{Chap02_eqn:LagAFODynEq} is not assured, except in special cases or requiring some
additional conditions to the Lagrangian function. In general, we must use the constraint
algorithm described in Section \ref{Chap01_sec:ConstraintAlgorithm} and, in the most
favorable cases, there exists a submanifold $S_f \hookrightarrow \Tan Q$ where the
equation
\begin{equation}\label{Chap02_eqn:LagAFODynEqSingular}
\restric{[\inn(X_\Lag) \omega_\Lag - \d E_\Lag]}{S_f} = 0 \, .
\end{equation}
admits a well-defined solution $X_\Lag$, which is tangent to $S_f$. Nevertheless,
these vector fields solution are not necessarily holonomic on $S_f$, but only in
the points of another submanifold $S_f^h \hookrightarrow S_f$.

\subsubsection{Integral curves}

Let $X_\Lag \in \vf(\Tan Q)$ be a holonomic vector field solution to the equation
\eqref{Chap02_eqn:LagAFODynEq}, and let $\psi_\Lag \colon \R \to \Tan Q$ be an integral
curve of $X_\Lag$. Since $X_\Lag$ is a holonomic vector field, there exists a curve
$\phi_\Lag \colon \R \to Q$ such that $\dot{\phi}_\Lag = \psi_\Lag$.
Since $\dot{\psi}_\Lag = X_\Lag \circ \psi_\Lag$, the geometric
equation for the dynamical trajectories of the system is
\begin{equation*}
\inn(\dot{\psi}_\Lag)(\omega_\Lag \circ \psi_\Lag) = \d E_\Lag \circ \psi_\Lag \, ,
\end{equation*}
or, equivalently,
\begin{equation*}
\inn(\ddot{\phi}_\Lag)(\omega_\Lag \circ \dot{\phi}_\Lag) = \d E_\Lag \circ \dot{\phi}_\Lag \, .
\end{equation*}

In coordinates, the curve $\phi_\Lag \colon \R \to Q$ must satisfy the following system of $n$
second-order ordinary differential equations
\begin{equation}\label{Chap02_eqn:FOEulerLagrange}
\restric{\derpar{\Lag}{q^A}}{\dot\phi_\Lag} - \restric{\frac{d}{dt}\derpar{\Lag}{v^A}}{\dot\phi_\Lag} = 0 \, .
\end{equation}
These are the \textsl{Euler-Lagrange equations} for this dynamical system.

\subsection{Hamiltonian formalism associated to a Lagrangian system}
\label{Chap02_sec:HamiltonianAutonomousFirstOrder}

(See \cite{book:Abraham_Marsden78,book:Arnold89} for details).

\subsubsection{The Legendre map}

The Legendre map in first-order dynamical systems can be introduced in several
equivalent ways. In this dissertation we define this transformation using the
fact that the Poincar\'{e}-Cartan $1$-form is semibasic on $\Tan Q$,
since it will be the easiest way when we generalize it to the higher-order setting
in Section \ref{Chap02_sec:LegendreOstrogradskyMap}. For alternative definitions, see
\cite{book:Abraham_Marsden78,book:Arnold89,book:Cannas01,book:Libermann_Marle87}.

\begin{definition}
The \textnormal{Legendre map} associated to the Lagrangian function $\Lag \in \Cinfty(\Tan Q)$
is the bundle morphism $\Leg \colon \Tan Q \to \Tan^*Q$ over $Q$ defined as follows: for every
$u \in \Tan\Tan Q$,
\begin{equation*}\label{Chap02_eqn:HamAFOLegendreMapDef}
\theta_\Lag(u) = \left\langle \Tan \tau_Q(u) \, , \, \Leg(\tau_{\Tan Q}(u)) \right\rangle \, .
\end{equation*}
\end{definition}

It is clear from the definition that this map satisfies $\pi_Q \circ \Leg = \tau_Q$.
Furthermore, if $\theta \in \df^{1}(\Tan^*Q)$ and $\omega = -\d\theta \in \df^{2}(\Tan^*Q)$
are the canonical $1$ and $2$ forms of the cotangent bundle, we have that
$\Leg^*\theta = \theta_\Lag$ and $\Leg^*\omega = \omega_\Lag$.

In coordinates, bearing in mind the local expressions of the Liouville $1$-form
$\theta \in \df^{1}(\Tan^*Q)$ given by \eqref{Chap01_eqn:CanonLiouvilleFormCotanBundle}
and the Poincar\'{e}-Cartan $1$-form $\theta_\Lag \in \df^{1}(\Tan Q)$ given by
\eqref{Chap02_eqn:LagAFO1FormLocal}, we obtain the coordinate expression of the
Legendre map, which is
\begin{equation}\label{Chap02_eqn:HamAFOLegendreMapLocal}
\Leg^*q^A = q^A \quad ; \quad \Leg^*p_A = \derpar{\Lag}{v^A} \, .
\end{equation}

Observe that from this coordinate expression, the rank of the tangent map of $\Leg$ depends
only on the rank of the Hessian matrix of $\Lag$ with respect to the velocities.
Hence, $\Lag \in \Cinfty(\Tan Q)$ is a regular Lagrangian function if, and only if,
the Legendre map $\Leg \colon \Tan Q \to \Tan^*Q$ is a local diffeomorphism.

\begin{definition}
A Lagrangian function $\Lag \in \Cinfty(\Tan Q)$ is \textnormal{hyperregular} if
the Legendre map $\Leg \colon \Tan Q \to \Tan^*Q$ is a global diffeomorphism.
\end{definition}

\begin{remark}
If the Lagrangian function is hyperregular, then the Legendre map is a symplectomorphism
between the symplectic manifolds $(\Tan Q,\omega_\Lag)$ and $(\Tan^*Q,\omega)$.
\end{remark}

In order to describe the dynamical trajectories in the canonical Hamiltonian formalism
of a Lagrangian system, we distinguish between the regular and non-regular cases.
In fact, the only singular kind of systems that we will consider are the \textsl{almost-regular} ones,
since we must require the Lagrangian to satisfy some minimal regularity conditions in order to
give a general description of these systems.

\begin{definition}
A Lagrangian function $\Lag \in \Cinfty(\Tan Q)$ is \textnormal{almost-regular} if
\begin{enumerate}
\item $\Leg(\Tan Q) \hookrightarrow \Tan^*Q$ is a closed submanifold.
\item $\Leg$ is a surjective submersion onto its image.
\item For every $(p,u_p) \in \Tan Q$ the fibers $\Leg^{-1}(\Leg(p,u_p))$ are connected submanifolds
of $\Tan Q$.
\end{enumerate}
\end{definition}

\subsubsection{Regular and hyperregular Lagrangian functions}

Let us assume that the Lagrangian function $\Lag \in \Cinfty(\Tan Q)$ is hyperregular,
since the regular case is recovered from this one by restriction on the open sets
where the Legendre map is a local diffeomorphism.

Since $\Leg \colon \Tan Q \to \Tan^*Q$ is a global diffeomorphism, there exists a unique
function $h \in \Cinfty(\Tan^*Q)$ such that $\Leg^*h = E_\Lag$.

\begin{definition}
The \textnormal{canonical Hamiltonian function} is the unique function
$h \in \Cinfty(\Tan^*Q)$ satisfying $\Leg^*h = E_\Lag$.
\end{definition}

The dynamical trajectories of the system are given by the integral curves of a vector field
$X_h \in \vf(\Tan^*Q)$ satisfying
\begin{equation}\label{Chap02_eqn:HamAFODynEq}
\inn(X_h) \omega = \d h \, .
\end{equation}
This equation is the \textsl{Hamiltonian equation}, and the unique vector field solution
to this equation is called the \textsl{Hamiltonian vector field}.

\begin{remark}
This concept of Hamiltonian vector field should not be confused with the one introduced
in Section \ref{Chap01_sec:CanonicalIsomorphismSymplectic}, although the vector field
solution to equation \eqref{Chap02_eqn:HamAFODynEq} is obviously the Hamiltonian vector field
(in the sense of Definition \ref{Chap01_def:HamiltonianVF}) of the function $h \in \Cinfty(\Tan^*Q)$.
\end{remark}

Let us compute the coordinate expression of the Hamiltonian function $h \in \Cinfty(\Tan^*Q)$.
Bearing in mind the local expressions \eqref{Chap02_eqn:HamAFOLegendreMapLocal} of the
Legendre map $\Leg$ and \eqref{Chap02_eqn:LagAFOEnergyLocal} of the Lagrangian energy, we have
\begin{equation*}\label{Chap02_eqn:HamFOHamiltonianFunctionLocal}
h = (\Leg^{-1})^*v^Ap_A - (\Leg^{-1})^*\Lag \, .
\end{equation*}
Now, for the equation \eqref{Chap02_eqn:HamAFODynEq}, let $X_h \in \Cinfty(\Tan^*Q)$ be
a generic vector field given by
\begin{equation*}
X_h = f^A \derpar{}{q^A} + G_A\derpar{}{p_A} \, .
\end{equation*}
Bearing in mind the coordinate expression \eqref{Chap01_eqn:CanonSympFormCotanBundle}
of the canonical symplectic form of $\Tan^*Q$ we have that
the equation \eqref{Chap02_eqn:HamAFODynEq} gives locally the following system of $2n$ equations
\begin{equation}\label{Chap02_eqn:HamAFODynEqLocal}
f^A = \derpar{h}{p_A} \quad ; \quad G_A = - \derpar{h}{q^A} \, .
\end{equation}

Finally, we establish the relation between the vector fields solution to the dynamical equation
\eqref{Chap02_eqn:LagAFODynEq} in the Lagrangian formalism and the vector fields solution to the
dynamical equation \eqref{Chap02_eqn:HamAFODynEq} in the Hamiltonian formalism associated to a
hyperregular Lagrangian system.

\begin{theorem}\label{Chap02_thm:HamAFORelationLagHamRegular}
Let $\Lag \in \Cinfty(\Tan Q)$ be a hyperregular Lagrangian function.
Then we have:
\begin{enumerate}
\item Let $X_\Lag \in \vf(\Tan Q)$ be the unique holonomic vector field solution
to equation \eqref{Chap02_eqn:LagAFODynEq}. Then the vector field
$X_h = \Leg_*X_\Lag \in \vf(\Tan^*Q)$ is a solution to equation \eqref{Chap02_eqn:HamAFODynEq}.

\item Conversely, let $X_h \in \vf(\Tan^*Q)$ be the unique vector field solution
to equation \eqref{Chap02_eqn:HamAFODynEq}. Then the vector field
$X_\Lag = (\Leg^{-1})_*X_h \in \vf(\Tan Q)$ is holonomic, and is a solution to
equation \eqref{Chap02_eqn:LagAFODynEq}.
\end{enumerate}
\end{theorem}

Now, if $\psi_h \colon \R \to \Tan^*Q$ is an integral curve of $X_h$, the geometric
equation for the dynamical trajectories of the system is 
\begin{equation*}
\inn(\dot{\psi}_h)(\omega \circ \psi_h) = \d h \circ \psi_h \, .
\end{equation*}

In coordinates, if the curve $\psi_h$ is given by $\psi_h(t) = (q^A(t),p_A(t))$, then its
component functions must satisfy the following system of $2n$ first-order ordinary
differential equations
\begin{equation}\label{Chap02_eqn:AFOHamiltonEq}
\dot{q}^A = \restric{\derpar{h}{p_A}}{\psi_h} \quad ; \quad \dot{p}_A = - \restric{\derpar{h}{q^A}}{\psi_h} \, .
\end{equation}
These are the \textsl{Hamilton equations} for this dynamical system.

\subsubsection{Singular (almost-regular) Lagrangian functions}

In the case of almost-regular Lagrangian systems, the Legendre map is no longer
a diffeomorphism, and therefore the image of $\Leg$ is a proper submanifold of $\Tan^*Q$.
Let $\P = \Im(\Leg) \hookrightarrow \Tan^*Q$ be the image set of the Legendre map,
with natural embedding $\jmath \colon \P \hookrightarrow \Tan^*Q$, and we denote by
$\Leg_o \colon \Tan Q \to \P$ the map defined by $\Leg = \jmath \circ \Leg_o$.
With these notations, we have the following result.

\begin{proposition}
The Lagrangian energy $E_\Lag \in \Cinfty(\Tan Q)$ is $\Leg_o$-projectable.
\end{proposition}

As a consequence of this last result, we can define a Hamiltonian function
in $\P$ as follows.

\begin{definition}
The \textnormal{canonical Hamiltonian function} is the unique function
$h_o \in \Cinfty(\P)$ such that $\Leg_o^*h_o = E_\Lag$.
\end{definition}

Then, taking $\omega_o = \jmath^*\omega \in \df^{2}(\P)$, we can state the Hamilton equation
for this system: we look for a vector field $X_{h_o} \in \vf(\P)$ satisfying
\begin{equation*}\label{Chap02_eqn:HamFODynEqSingular1}
\inn(X_{h_o})\omega_o = \d h_o \, .
\end{equation*}
Since the $2$-form $\omega_o \in \df^{2}(\P)$ is, in general, presymplectic,
we must apply the constraint algorithm described in Section
\ref{Chap01_sec:ConstraintAlgorithm}. In the most favorable cases, this equation
admits a vector field solution only on the points of some submanifold
$\P_f \hookrightarrow \P \hookrightarrow \Tan^*Q$, and is tangent to it,
so the following equation holds
\begin{equation}\label{Chap02_eqn:HamAFODynEqSingular2}
\restric{[\inn(X_{h_o}) \omega_o - \d h_o]}{\P_f} = 0 \, .
\end{equation}
This vector field is not unique, in general.

In this situation, we have an analogous result to Theorem \ref{Chap02_thm:HamAFORelationLagHamRegular}.

\begin{theorem}\label{Chap02_thm:HamAFORelationLagHamSingular}
Let $\Lag \in \vf(\Tan Q)$ be an almost-regular Lagrangian function.
Then we have:
\begin{enumerate}
\item Let $X_\Lag \in \vf(\Tan Q)$ be a holonomic vector field solution
to equation \eqref{Chap02_eqn:LagAFODynEqSingular} in the points of a submanifold
$S_f \hookrightarrow \Tan Q$. Then there exists a vector field
$X_{h_o} \in \vf(\P)$ which is $\Leg_o$-related to $X_\Lag$
and is a solution to equation \eqref{Chap02_eqn:HamAFODynEqSingular2},
where $\P_f = \Leg_o(S_f) \hookrightarrow \P$.

\item Conversely, let $X_{h_o} \in \vf(\P)$ be a vector field solution
to equation \eqref{Chap02_eqn:HamAFODynEqSingular2} on the points of some
submanifold $\P_f \hookrightarrow \P$. Then there exist vector fields
$X_\Lag \in \vf(\Tan Q)$ which are $\Leg_o$-related to $X_{h_o}$, and are solutions to
equation \eqref{Chap02_eqn:LagAFODynEqSingular}, where $S_f = \Leg^{-1}(\P_f)$.
\end{enumerate}
\end{theorem}

Observe that the vector fields $X_\Lag \in \vf(\Tan Q)$ which are
$\Leg_o$-related to $X_{h_o}$ may not be holonomic, since this condition
can not be assured in the singular case.
These $\Leg_o$-projectable holonomic vector fields could be defined only on the
points of another submanifold $M_f\hookrightarrow S_f$.

\subsection{Lagrangian-Hamiltonian unified formalism}
\label{Chap02_sec:SkinnerRuskAutonomousFirstOrder}

(See \cite{art:Skinner_Rusk83} for details).

\subsubsection{Unified phase space. Geometric and dynamical structures}

Let us consider the bundle
\begin{equation*}
\W = \Tan Q \times_{Q} \Tan^{*}Q \, ,
\end{equation*}
that is, the product over $Q$ of the Lagrangian and Hamiltonian phase spaces.
This bundle is endowed with canonical projections over each factor,
namely $\rho_1 \colon \W \to \Tan Q$ and $\rho_2 \colon \W \to \Tan^*Q$.
Using these projections, and the canonical projections of the tangent and
cotangent bundles of $Q$, we introduce the following commutative diagram
\begin{equation*}
\xymatrix{
\ & \Tan Q \times_{Q} \Tan^{*}Q \ar[dl]_-{\rho_1} \ar[dr]^-{\rho_2} & \ \\
\Tan Q \ar[dr]_-{\tau_{Q}} & \ & \Tan^{*}Q \ar[dl]^-{\pi_{Q}} \\
\ & Q & \
}
\end{equation*}
Local coordinates in $\W$ are constructed as follows: if $(U,\varphi)$ is a local
chart of $Q$ with $\varphi = (q^A)$, $1 \leqslant A \leqslant n$, then the induced
local charts in $\Tan Q$ and $\Tan^{*}Q$ are $(\tau_{Q}^{-1}(U),(q^A,v^A))$
and $(\pi_{Q}^{-1}(U),(q^A,p_A))$, respectively.
Therefore, natural coordinates in  the open set
$(\tau_{Q} \circ \rho_1)^{-1}(U) = (\pi_{Q} \circ \rho_2)^{-1}(U) \subseteq \W$ are $(q^A,v^A,p_A)$.
Observe that $\dim\W = 3\dim Q = 3n$. Using these coordinates, the above
projections have the following local expressions
\begin{equation*}
\rho_1(q^A,v^A,p_A) = (q^A,v^A) \quad ; \quad
\rho_2(q^A,v^A,p_A) = (q^A,p_A) \, .
\end{equation*}

The bundle $\W$ is endowed with some canonical geometric structures.
First, let $\theta \in \df^{1}(\Tan^{*}Q)$ be the Liouville $1$-form
on the cotangent bundle, and $\omega = -\d\theta \in \df^{2}(\Tan^{*}Q)$
the canonical symplectic form on $\Tan^{*}Q$. From this we can define
a $2$-form $\Omega$ in $\W$ as
\begin{equation*}
\Omega = \rho_{2}^{*}\,\omega \in \df^{2}(\W) \, .
\end{equation*}
It is clear that $\Omega$ is a closed $2$-form, since
\begin{equation*}
\Omega = \rho_{2}^{*}\,\omega = \rho_{2}^{*}\,(-\d\theta) = -\d\rho_{2}^{*}\,\theta \, .
\end{equation*}
Nevertheless, this form is degenerate, and therefore it is a presymplectic form.
Indeed, let $X \in \vf^{V(\rho_{2})}(\W)$. Then we have
\begin{equation*}
\inn(X)\Omega = \inn(X)\rho_{2}^*\,\omega = \rho_{2}^*(\inn(Y)\omega) \, ,
\end{equation*}
where $Y \in \vf(\Tan^{*}Q)$ is a vector field $\rho_{2}$-related to $X$.
However, since $X$ is vertical with respect to $\rho_{2}$, we have
$Y = 0$, and therefore
\begin{equation*}
\rho_{2}^{*}(\inn(Y)\omega) = \rho_{2}^{*}(\inn(0)\omega) = 0 \, .
\end{equation*}
In particular, $\{ 0 \} \varsubsetneq X^{V(\rho_2)}(\W) \subseteq \ker\Omega$,
and thus $\Omega$ is a degenerate $2$-form.

In coordinates, bearing in mind the local expression of the canonical symplectic
form of the cotangent bundle given by \eqref{Chap01_eqn:CanonSympFormCotanBundle}
and the local expression of the projection $\rho_2$ given above,
the coordinate expression of the presymplectic form $\Omega$ is
\begin{equation}\label{Chap02_eqn:UnifAPresympFormLocal}
\Omega
= \d q^A \wedge \d p_A \, .
\end{equation}
From this local expression it is clear that $\Omega$ is closed and that its kernel
is locally given by
\begin{equation*}
\ker\Omega = \left\langle \derpar{}{v^A} \right\rangle = \vf^{V(\rho_2)}(\W) \, .
\end{equation*}

The second relevant geometric structure in $\W$ is the following.

\begin{definition}
Let $p \in Q$ be a point, $v_p \in \Tan_pQ$ a tangent vector at $p$ and $\alpha_p \in \Tan_p^*Q$
a covector on $p$. Then we define the \textnormal{coupling function} $\C \in \Cinfty(\W)$ as
\begin{equation*}
\C(p,v_p,\alpha_p) = \langle \alpha_p,v_p \rangle \, ,
\end{equation*}
where $\langle \alpha_p,v_p \rangle \equiv \alpha_p(v_p)$ is the canonical pairing
between the elements of the vector space $\Tan_pQ$ and its dual $\Tan_p^*Q$.
\end{definition}

In local coordinates, if we consider a local chart on $p \in Q$ such that
$\alpha_p = \restric{p_A\d q^A}{p}$ and $\displaystyle v_p = \restric{v^A\derpar{}{q^A}}{p}$,
then the local expression of the coupling function is
\begin{equation*}
\C(p,v_p,\alpha_p) = \langle \alpha_p, v_p \rangle
= \left\langle \restric{p_A\d q^A}{p}, \restric{v^A\derpar{}{q^A}}{p} \right\rangle
= \restric{p_Av^A}{p} \, .
\end{equation*}

Using the coupling function defined above and the given Lagrangian function
$\Lag \in \Cinfty(\Tan Q)$, we define the \textsl{Hamiltonian function}
$H \in \Cinfty(\W)$ by
\begin{equation}\label{Chap02_eqn:UnifAHamFunctDef}
H = \C - \rho_1^*\Lag \, ,
\end{equation}
whose local expression is
\begin{equation}\label{Chap02_eqn:UnifAHamFunctLocal}
H(q^A,v^A,p_A) = p_Av^A - \Lag(q^A,v^A) \, .
\end{equation}

\subsubsection{Dynamical vector fields in $\W$}

Hence, we have constructed a presymplectic Hamiltonian system $(\W,\Omega,H)$.
The dynamical equation for this kind of systems is
\begin{equation}\label{Chap02_eqn:UnifADynEq}
\inn(X)\Omega = \d H \ , \ X \in \vf(\W) \, .
\end{equation}
Observe that, since the system is presymplectic, the above equation
may not admit a global solution $X \in \vf(\W)$, and we have to use
the constraint algorithm given in Section \ref{Chap01_sec:ConstraintAlgorithm}.
From the algorithm given in the aforementioned Section, we can state
the following result.

\begin{proposition}\label{Chap02_prop:UnifAFirstConstSubm}
Given the presymplectic Hamiltonian system $(\W,\Omega,H)$,
a solution $X \in \vf(\W)$ to equation \eqref{Chap02_eqn:UnifADynEq}
exists only on the points of the submanifold $\W_\Lag \hookrightarrow \W$
defined by
\begin{equation*}\label{Chap02_eqn:UnifAFirstConstSubmDef}
\W_\Lag = \left\{ w \in \W \mid (\inn(Y)\d H)(w) = 0 \ , \, \forall \, Y \in \ker\Omega \right\} \, ,
\end{equation*}
with natural embedding $j_\Lag \colon \W_\Lag \hookrightarrow \W$.
\end{proposition}

We have the following characterization of the first constraint submanifold $\W_\Lag$.

\begin{proposition}\label{Chap02_prop:UnifAGraphLegendreMap}
The submanifold $\W_\Lag \hookrightarrow \W$ is the graph of the Legendre map
defined by $\Lag$, that is, $\W_\Lag = \graph\Leg$.
\end{proposition}

\begin{remark}
If we denote by $\vf_{\W_\Lag}(\W)$ the set of vector fields in $\W$ at support
on $\W_\Lag$, then the dynamical equation for the presymplectic system $(\W,\Omega,H)$
can be stated as follows: we look for vector fields $X \in \vf_{\W_\Lag}(\W)$ which
are solutions to the equation
\begin{equation*}
\restric{\left[\inn(X)\Omega - \d H\right]}{\W_\Lag} = 0 \, .
\end{equation*}
Nevertheless, since we do not have a distinguished system of coordinates
in $\W_\Lag$, we will stick to the general setting: we consider a vector field
$X \in \vf(\W)$ and the equation \eqref{Chap02_eqn:UnifADynEq}, and at the end
we require the vector field $X$ to be tangent to the submanifold $\W_\Lag$.
\end{remark}

In the natural coordinates of $\W$,
a generic vector field $X \in \vf(\W)$ is given by
\begin{equation*}
X = f^A\derpar{}{q^A} + F^A\derpar{}{v^A} + G_A\derpar{}{p_A} \, .
\end{equation*}
Then, bearing in mind the local expressions \eqref{Chap02_eqn:UnifAPresympFormLocal}
of the presymplectic form $\Omega$ and \eqref{Chap02_eqn:UnifAHamFunctLocal}
of the Hamiltonian function $H$, the equation \eqref{Chap02_eqn:UnifADynEq}
gives the following system of $3n$ equations
\begin{align}
f^A = v^A \, , \label{Chap02_eqn:UnifADynEqHolonomyLocal} \\
G_A = \derpar{\Lag}{q^A} \, , \label{Chap02_eqn:UnifADynEqLocal} \\
p_A - \derpar{\Lag}{v^A} = 0 \, . \label{Chap02_eqn:UnifADynEqLegendreLocal}
\end{align}
where $1 \leqslant A \leqslant n$. Then a vector field $X \in \vf(\W)$
solution to equation \eqref{Chap02_eqn:UnifADynEq} has the following
coordinate expression
\begin{equation*}
X = v^A \derpar{}{q^A} + F^A \derpar{}{v^A} + \derpar{\Lag}{q^A}\derpar{}{p_A} \, .
\end{equation*}
Observe that equations \eqref{Chap02_eqn:UnifADynEqHolonomyLocal} are the holonomy
condition for a vector field in the Lagrangian formalism, as we have seen in Section
\ref{Chap02_sec:LagrangianAutonomousFirstOrder}. On the other hand, equations
\eqref{Chap02_eqn:UnifADynEqLegendreLocal} are a compatibility condition that
state that the vector fields $X$ exist only with support on the submanifold
defined as the graph of the Legendre map. So we recover, in coordinates,
the result stated in Propositions \ref{Chap02_prop:UnifAFirstConstSubm} and
\ref{Chap02_prop:UnifAGraphLegendreMap}. Finally, equations \eqref{Chap02_eqn:UnifADynEqLocal}
are the dynamical equations of the system.

\begin{remark}
It is important to point out that the holonomy of the vector field $X \in \vf(\W)$ is obtained
regardless of the regularity of the Lagrangian function $\Lag \in \Cinfty(\Tan Q)$ provided.
\end{remark}

Notice that the component functions $F^A$ of the vector field remain undetermined.
This is due to the fact that these functions are the components of the $\rho_2$-vertical
part of the vector field $X$, and therefore they are annihilated by the presymplectic form
$\Omega$. Nevertheless, since $X$ is defined at support on the submanifold $\W_\Lag$,
we must study the tangency of $X$ along this submanifold. That is, we must require that
$\restric{\Lie(X)\xi}{\W_\Lag} \equiv \restric{X(\xi)}{\W_\Lag} = 0$ for every constraint
function $\xi$ defining $\W_\Lag$. From Proposition \ref{Chap02_prop:UnifAGraphLegendreMap}
we have that the submanifold $\W_\Lag$ is defined by the $n$ constraints
\begin{equation*}
\xi^B \equiv p_B - \derpar{\Lag}{v^B} = 0 \ , \ B = 1,\ldots,n \, ,
\end{equation*}
and therefore the tangency condition for $X$ along $\W_\Lag$ leads to
the following $n$ equations
\begin{equation*}
X(\xi^B) =
\left( v^A \derpar{}{q^A} + F^A \derpar{}{v^A} + \derpar{\Lag}{q^A}\derpar{}{p_A} \right)
\left( p_B - \derpar{\Lag}{v^B} \right)
= \derpar{\Lag}{q^B} - v^A\derpars{\Lag}{q^A}{v^B} - F^A\derpars{\Lag}{v^A}{v^B} = 0 \, .
\end{equation*}
Notice that these are the Lagrangian equations for the components of a vector field
once the holonomy condition is satisfied,
as we have seen in \eqref{Chap02_eqn:LagAFODynEqWithHolonomyLocal}. These equations can be
compatible or not, and a sufficient condition to ensure compatibility is the
regularity of the Lagrangian function. In particular, we have

\begin{proposition}\label{Chap02_prop:UnifARegLagUniqueVF}
If $\Lag \in \Cinfty(\Tan Q)$ is a regular Lagrangian function, then there exists a unique
vector field $X \in \vf(\W)$ which is a solution to equation \eqref{Chap02_eqn:UnifADynEq}
and is tangent to $\W_\Lag$.
\end{proposition}

If the Lagrangian $\Lag$ is not regular, the above equations can be compatible or not,
and new constraints may arise from them, thus requiring the constraint algorithm
to continue. In the most favorable cases, there exists a submanifold $\W_f \hookrightarrow \W_\Lag$
(it could be $\W_f = \W_\Lag)$ such that there exist vector fields $X \in \vf_{\W_f}(\W)$,
tangent to $\W_f$, which are solutions to the equation
\begin{equation}\label{Chap02_eqn:UnifADynEqSingular}
\restric{\left[ \inn(X)\Omega - \d H \right]}{\W_f} = 0 \, .
\end{equation}

\subsubsection{Lagrangian dynamics}

Now we study how to recover the Lagrangian vector fields from
the dynamical vector fields in the unified setting.
In fact, we will show that there exists a bijective correspondence between
the set of vector fields solution to the dynamical equation in the unified setting
and the set of vector fields solution to the dynamical equation in the Lagrangian formalism.

The first fundamental result is the following.

\begin{proposition}
The map $\rho_1^\Lag = \rho_1 \circ j_\Lag \colon \W_\Lag \to \Tan Q$
is a diffeomorphism.
\end{proposition}

This result allows us to recover the geometric and dynamical structures of
the Lagrangian formalism from the unified setting. In particular,
we have the following results.

\begin{lemma}
If $\omega \in \df^{2}(\Tan^*Q)$ is the canonical symplectic form in $\Tan^*Q$,
and $\omega_\Lag = \Leg^*\omega \in \df^{2}(\Tan Q)$ is the Poincar\'{e}--Cartan
$2$-form, then $\Omega = \rho_1^*\omega_\Lag$.
\end{lemma}

\begin{lemma}
There exists a unique function $E_\Lag \in \Cinfty(\Tan Q)$ such that
$\rho_1^*E_\Lag = H$.
\end{lemma}

The function obtained in this last result is the \textsl{Lagrangian energy}
from Section \ref{Chap02_sec:LagrangianAutonomousFirstOrder}.
With all these results, we can now state the equivalence theorem.

\begin{theorem}\label{Chap02_thm:EquivUnifLagAVF}
Let $X \in \vf(\W)$ be a vector field solution to equation \eqref{Chap02_eqn:UnifADynEq}
and tangent to $\W_\Lag$ (at least on the points of a submanifold $\W_f \hookrightarrow \W_\Lag$).
Then there exists a unique holonomic vector field $X_\Lag \in \vf(\Tan Q)$ which is a solution
to equation \eqref{Chap02_eqn:LagAFODynEq} (at least on the points of $S_f = \rho_1^\Lag(\W_f)$).

\noindent Conversely, if $X_\Lag \in \vf(\W)$ is a holonomic vector field solution to equation
\eqref{Chap02_eqn:LagAFODynEq} (at least on the points of a submanifold
$S_f \hookrightarrow \Tan Q$), then there exists a unique vector field $X \in \vf(\W)$
which is a solution to equation \eqref{Chap02_eqn:UnifADynEq} and tangent to $\W_\Lag$
(at least on the points of the submanifold $\W_f = (\rho_1^\Lag)^{-1}(S_f)$).
\end{theorem}

Note that Theorem \ref{Chap02_thm:EquivUnifLagAVF} states that there is a one-to-one
correspondence between vector fields $X \in \vf(\W)$ which are solutions to equation
\eqref{Chap02_eqn:UnifADynEq} and vector fields $X_\Lag \in \vf(\Tan Q)$ which are solution to
\eqref{Chap02_eqn:LagAFODynEq}, but it does not state the uniqueness of any of them.
In fact, the uniqueness can not be assured in the general case, but only when the
Lagrangian function is regular, as it is stated in Proposition
\ref{Chap02_prop:UnifARegLagUniqueVF}.

\subsubsection{Hamiltonian dynamics}

As in the usual formulation of the Hamiltonian formalism described in Section
\ref{Chap02_sec:HamiltonianAutonomousFirstOrder}, in order to recover the Hamiltonian
dynamics from the unified setting we must distinguish between regular and
singular (almost-regular) Lagrangian functions.

\paragraph{Hyperregular and regular Lagrangians.}
Assume that the Lagrangian function $\Lag \in \Cinfty(\Tan Q)$ is hyperregular
(the regular case is recovered from this by restriction on the corresponding
open sets where the Legendre map is a local diffeomorphism),
and let $\rho_2^\Lag = \rho_2 \circ j_\Lag \colon \W_\Lag \to \Tan^*Q$.
Then, we have the following commutative diagram
\begin{equation*}
\xymatrix{
\ & \ & \W \ar@/_1pc/[ddll]_{\rho_1} \ar@/^1pc/[ddrr]^{\rho_2} & \ & \ \\
\ & \ & \W_\Lag \ar[dll]_{\rho_1^\Lag} \ar[drr]^{\rho_2^\Lag} \ar@{^{(}->}[u]^-{j_\Lag}  & \ & \ \\
\Tan Q \ar[drr]_{\tau_Q} \ar[rrrr]^{\Leg} & \ & \ & \ & \Tan^*Q \ar[dll]^{\pi_Q} \\
\ & \ & Q & \ & \ \\
}
\end{equation*}
In particular, $\rho_2^\Lag = \Leg \circ \rho_1^\Lag$ is a diffeomorphism, since
both $\Leg$ and $\rho_1^\Lag$ are diffeomorphisms. Therefore, we can state
the following result.

\begin{lemma}
There exists a unique function $h \in \Cinfty(\Tan^*Q)$ such that $\rho_2^*h = H$.
\end{lemma}

The function obtained in this last result is the \textsl{Hamiltonian function} from
Section \ref{Chap02_sec:HamiltonianAutonomousFirstOrder}. Now we can state the equivalence
theorem in this case.

\begin{theorem}\label{Chap02_thm:EquivUnifHamAVF}
Let $X \in \vf(\W)$ be a vector field solution to equation \eqref{Chap02_eqn:UnifADynEq}
and tangent to $\W_\Lag$. Then there exists a unique vector field $X_h \in \vf(\Tan^*Q)$
which is a solution to equation \eqref{Chap02_eqn:HamAFODynEq}.

\noindent Conversely, if $X_h \in \vf(\Tan^*Q)$ is a vector field solution to equation
\eqref{Chap02_eqn:HamAFODynEq}, then there exists a unique vector field $X \in \vf(\W)$
which is a solution to equation \eqref{Chap02_eqn:UnifADynEq} and tangent to $\W_\Lag$.
\end{theorem}

\paragraph{Singular (almost-regular) Lagrangian functions.}
If the Lagrangian function is not regular, then we can not recover the Hamiltonian
dynamics straightforwardly from the unified setting, but rather passing through the
Lagrangian formalism.

Remember that, for almost-regular Lagrangian functions, only in the most favorable cases
we can assure the existence of a submanifold $\W_f \hookrightarrow \W_\Lag$ and
vector fields $X \in \vf(\W)$ tangent to $\W_f$ which are solutions to equation
\eqref{Chap02_eqn:UnifADynEqSingular}. Thus, we can consider the submanifold
$S_f = \rho_1(\W_f) \hookrightarrow \Tan Q$ and then, using Theorem
\ref{Chap02_thm:EquivUnifLagAVF}, from the vector fields $X \in \vf(\W)$
we obtain the corresponding holonomic vector fields $X_\Lag \in \vf(\Tan Q)$
solutions to \eqref{Chap02_eqn:LagAFODynEqSingular}. With these elements, we can
apply the procedure described in Section \ref{Chap02_sec:HamiltonianAutonomousFirstOrder}
for singular (almost-regular) Lagrangian functions, and recover the Hamiltonian
dynamics from the Lagrangian formalism.

\subsubsection{Integral curves}

After studying the vector fields which are solutions to the dynamical equations,
we analyze their integral curves, showing how to recover the Lagrangian and
Hamiltonian dynamical trajectories from the dynamical trajectories in the unified
formalism.

Let $X \in \vf(\W)$ be a vector field tangent to $\W_\Lag$ which is a solution
to the equation \eqref{Chap02_eqn:UnifADynEq}, and let
$\psi \colon I \subseteq \R \to \W$ be an integral curve of $X$.
Since $\dot{\psi} = X \circ \psi$, the geometric equation for the dynamical
trajectories of the system is
\begin{equation*}
\inn(\dot{\psi})(\Omega \circ \psi) = \d H \circ \psi \, .
\end{equation*}

In coordinates, if $\psi(t) = (q^A(t),v^A(t),p_A(t))$, the condition
$\dot{\psi} = X \circ \psi$ gives the following system of differential equations
for the component functions of $\psi$
\begin{equation*}
\dot{q}^A(t) = v^A \circ \psi \quad ; \quad \dot{p}_A(t) = \derpar{\Lag}{q^A} \circ \psi \, ,
\end{equation*}
together with the equations defining locally the Legendre map.

Now, for the Lagrangian dynamical trajectories we have the following result:

\begin{proposition}\label{Chap02_thm:EquivUnifLagAIntCur}
Let $\psi \colon I \subseteq \R \to \W$ be an integral curve of a vector field $X$
solution to equation \eqref{Chap02_eqn:UnifADynEq} on $\W_\Lag$. Then the curve
$\psi_\Lag = \rho_1 \circ \psi \colon I \to \Tan Q$ is holonomic and is an integral curve
of $X_\Lag$.
\end{proposition}

\begin{remark}
Since $\psi_\Lag$ is holonomic, there is a curve $\phi_\Lag \colon \R \to Q$
such that $\dot{\phi}_\Lag = \psi_\Lag$.
\end{remark}

And for the Hamiltonian trajectories, we have:

\begin{proposition}\label{Chap02_thm:EquivUnifHamAIntCur}
Let $\psi \colon I \subseteq \R \to \W$ be an integral curve of a vector field $X$
solution to equation \eqref{Chap02_eqn:UnifADynEq} on $\W_\Lag$. Then the curve
$\psi_h = \rho_2 \circ \psi \colon I \to \Tan^*Q$ is an integral curve of $X_h$.
\end{proposition}

The relation among all these integral curves is summarized in the following diagram.
\begin{equation*}
\xymatrix{
\ & \ & \W \ar[dll]_{\rho_1} \ar[drr]^{\rho_2} & \ & \ \\
\Tan Q \ar[ddrr]_{\tau_Q} \ar[rrrr]^(.65){\Leg}|(.495){\hole} & \ & \ & \ & \Tan^*Q \ar[ddll]^{\pi_Q} \\
\ & \ & \R \ar[uu]^(.35){\psi} \ar@{-->}[ull]_{\psi_\Lag} \ar@{-->}[urr]^{\psi_h} \ar[d]^-{\phi} & \ & \ \\
\ & \ & Q & \ & \ \\
}
\end{equation*}

\begin{remark}
Observe that in Propositions \ref{Chap02_thm:EquivUnifLagAIntCur} and \ref{Chap02_thm:EquivUnifHamAIntCur}
we make no assumptions on the regularity of the system. In fact, the only considerations in the
almost-regular case are that, in general, the curves lie in some submanifolds which are
determined by the constraint algorithm described in Section \ref{Chap01_sec:ConstraintAlgorithm}.
\end{remark}


\section{Hamilton-Jacobi theory for first-order autonomous systems}
\label{Chap02_sec:HamiltonJacobi}

(See \cite{book:Abraham_Marsden78,book:Arnold89} and
\cite{art:Carinena_Gracia_Marmo_Martinez_Munoz_Roman06} for details).

Let us consider a first-order autonomous Lagrangian dynamical system with
$n$ degrees of freedom.
Let $Q$ be a $n$-dimensional smooth manifold modeling the configuration space
of this first-order dynamical system, and $\Lag \in \Cinfty(\Tan Q)$ a
Lagrangian function describing the dynamics of the system.
Along this Section, we assume that the Lagrangian function $\Lag$ is regular
(see Definition \ref{Chap02_def:LagAFORegularLagrangian}).

\begin{remark}
The geometric Hamilton-Jacobi problem has been established also in the unified
Lagrangian-Hamiltonian formalism in a recent paper \cite{art:DeLeon_Martin_Vaquero12}.
The goal of the paper is to give a geometric formulation of the Hamilton-Jacobi theory
for dynamical systems described by singular Lagrangian functions, and therefore
no distinction between the generalized and standard Hamilton-Jacobi problems is made,
but rather between the cases of regular and singular Lagrangian functions.
In this Section we do not review this paper, which is beyond the scope of this dissertation.
\end{remark}

\subsection{Lagrangian formulation of the Hamilton-Jacobi problem}
\label{Chap02_sec:HamiltonJacobiLag}

Since $\Lag \in \Cinfty(\Tan Q)$ is a regular Lagrangian function,
the Poincar\'e-Cartan $2$-form $\omega_\Lag \in \df^{2}(\Tan Q)$ is symplectic, and
hence the equation \eqref{Chap02_eqn:LagAFODynEq} admits a unique solution, which
in addition is holonomic. Thus, let $X_\Lag \in \Cinfty(\Tan Q)$ be the unique holonomic vector field
solution to equation \eqref{Chap02_eqn:LagAFODynEq}.

\subsubsection{Generalized Lagrangian Hamilton-Jacobi problem}

\begin{definition}
The \textnormal{generalized Lagrangian Hamilton-Jacobi problem}
consists in finding a vector field $X \in \vf(Q)$ such that
if $\gamma \colon \R \to Q$ is an integral curve of $X$, then
its canonical lifting $\dot{\gamma} \colon \R \to \Tan Q$ is an integral curve of $X_\Lag$;
that is,
\begin{equation*}
X \circ \gamma = \dot{\gamma} \Rightarrow X_\Lag \circ \dot{\gamma} = \ddot{\gamma} \, .
\end{equation*}
$X$ is said to be a \textnormal{solution to the generalized Lagrangian Hamilton-Jacobi problem}.
\end{definition}

\begin{theorem}\label{Chap02_thm:HJLagFOEquivalenceGeneralized}
Let $X \in \vf(Q)$. The following assertions are equivalent:
\begin{enumerate}
\item $X$ is a solution to the generalized Lagrangian Hamilton-Jacobi problem.
\item $X$ and $X_\Lag$ are $X$-related, that is, $X_\Lag \circ X = \Tan X \circ X$.
\item The submanifold $\Im(X) \hookrightarrow \Tan Q$ is invariant by the Euler-Lagrange
vector field $X_\Lag$ (that is, $X_\Lag$ is tangent to the submanifold $X(Q)$).
\item $X$ satisfies the equation
\begin{equation*}
\inn(X)(X^*\omega_\Lag) = \d(X^*E_\Lag) \, .
\end{equation*}
\end{enumerate}
\end{theorem}

In coordinates, let $(U;(q^A))$, $1 \leqslant A \leqslant n$, be a local chart in $Q$,
and let $(q^A,v^A)$ be the induced natural coordinates in $\tau_Q^{-1}(U) \subseteq \Tan Q$.
In these coordinates, a generic vector field $X \in \vf(Q)$ has the following coordinate expression
\begin{equation*}
X = X^A\derpar{}{q^A} \, ,
\end{equation*}
and the Euler-Lagrange vector field is given locally by
\begin{equation*}
X_\Lag = v^A\derpar{}{q^A} + F^A\derpar{}{v^A} \, ,
\end{equation*}
where the functions $F^A$ are the unique solutions to the system of $n$
equations \eqref{Chap02_eqn:LagAFODynEqWithHolonomyLocal}. Then, bearing
in mind that the submanifold $\Im(X) \hookrightarrow \Tan Q$ is locally
defined by the constraints $v^A - X^A = 0$, and the third item in Theorem
\ref{Chap02_thm:HJLagFOEquivalenceGeneralized}, the condition for $X$
to be a solution to the generalized Lagrangian Hamilton-Jacobi problem
gives the following system of $n$ partial differential equations for the
component functions of $X$
\begin{equation}\label{Chap02_eqn:HJLagFOLocalEqGeneralized}
\restric{F^A - X^B\derpar{X^A}{q^B}}{\Im(X)} = 0 \, .
\end{equation}

\subsubsection{Lagrangian Hamilton-Jacobi problem}

In general, to solve the generalized Lagrangian Hamilton-Jacobi problem is a
difficult task, since it amounts to finding $n$-codimensional $X_\Lag$-invariant
submanifolds of $\Tan Q$. Thus, it is convenient to consider a less general problem.

\begin{definition}
The \textnormal{Lagrangian Hamilton-Jacobi problem} consists in finding
solutions $X \in \vf(Q)$ to the generalized Lagrangian Hamilton-Jacobi problem
satisfying that $X^*\omega_\Lag = 0$. Such a vector field is called a
\textnormal{solution to the Lagrangian Hamilton-Jacobi problem}.
\end{definition}

Observe that the condition required to the vector field $X$ is equivalent to require
an isotropy condition to the submanifold $\Im(X)$, which in addition satisfies
$\dim\Im(X) = n = \frac{1}{2}2n = \frac{1}{2} \dim\Tan Q$.
Hence, we have the following result as a straightforward consequence of Theorem
\ref{Chap02_thm:HJLagFOEquivalenceGeneralized}.

\begin{theorem}
Let $X \in \vf(Q)$ be a vector field satisfying $X^*\omega_\Lag = 0$.
Then, the following assertions are equivalent:
\begin{enumerate}
\item $X$ is a solution to the Lagrangian Hamilton-Jacobi problem.
\item $\d(X^*E_\Lag) = 0$.
\item $\Im(X)$ is a Lagrangian submanifold of $\Tan Q$ invariant by $X_\Lag$.
\item The integral curves of $X_\Lag$ with initial conditions in $\Im(X)$ project
onto the integral curves of $X$.
\end{enumerate}
\end{theorem}

In the induced natural coordinates $(q^A,v^A)$ of $\Tan Q$, bearing in mind the
coordinate expression \eqref{Chap02_eqn:LagAFO2FormLocal} of the Poincar\'e-Cartan
$2$-form $\omega_\Lag \in \df^{2}(\Tan Q)$, the local expression of equation
$X^*\omega_\Lag = 0$ is
\begin{equation}\label{Chap02_eqn:HJLagFOLocalEqStandard1}
\restric{\derpars{\Lag}{v^A}{q^B} + \derpars{\Lag}{v^A}{v^C}\derpar{X^C}{q^B}}{\Im(X)} = 0 \, ,
\end{equation}
or, equivalently, bearing in mind the coordinate expression
\eqref{Chap02_eqn:LagAFOEnergyLocal} of the Lagrangian energy
$E_\Lag \in \Cinfty(\Tan Q)$, the local expression of equation
$\d(X^*E_\Lag) = X^*(\d E_\Lag) = 0$ is
\begin{equation}\label{Chap02_eqn:HJLagFOLocalEqStandard2}
\restric{\derpars{\Lag}{v^A}{v^B}\derpar{X^B}{q^C}X^C + \derpars{\Lag}{v^A}{q^B}X^B - \derpar{\Lag}{q^A}}{\Im(X)} = 0 \, .
\end{equation}
Therefore, a vector field $X \in \vf(Q)$ is a solution to the Lagrangian Hamilton-Jacobi
problem if, and only if, its component functions satisfy the system of partial differential
equations given by \eqref{Chap02_eqn:HJLagFOLocalEqGeneralized} and
\eqref{Chap02_eqn:HJLagFOLocalEqStandard1}, or, equivalently,
the system of partial differential equations given by
\eqref{Chap02_eqn:HJLagFOLocalEqGeneralized} and \eqref{Chap02_eqn:HJLagFOLocalEqStandard2}.

In addition, in this case we can obtain the classical Hamilton-Jacobi
equation in the Lagrangian formalism as follows.
As $\omega_\Lag = -\d\theta_\Lag$, we have
$0 = X^*\omega_\Lag = -X^*\d\theta_\Lag = -\d(X^*\theta_\Lag)$, that is,
$X^*\theta_\Lag \in \df^{1}(Q)$ is a closed form. In particular, using Poincar\'e's Lemma,
we have that every point in $Q$ has an open neighborhood $U \subseteq Q$ where there
exists a local function $W \in \Cinfty(U)$ such that $X^*\theta_\Lag = \d W$ (in $U$).
Then, bearing in mind the coordinate expression \eqref{Chap02_eqn:LagAFO1FormLocal} of
the Poincar\'{e}-Cartan $1$-form, we have
\begin{equation*}
X^*\theta_\Lag = \restric{\derpar{\Lag}{v^A}}{\Im(X)} \d q^A \, ,
\end{equation*}
from which the equation $X^*\theta_\Lag = \d W$ gives the following system of $n$ partial
differential equations
\begin{equation}\label{Chap02_eqn:HJLagFOHamiltonJacobiEq}
\derpar{W}{q^A} = \restric{\derpar{\Lag}{v^A}}{\Im(X)} \, .
\end{equation}
which are the standard Hamilton-Jacobi equations in the Lagrangian formalism.

\subsubsection{Complete solutions}

Observe that, in the previous Sections, we established the general setting
to obtain a \textit{particular} solution of the system, since only the
integral curves of $X_\Lag$ whose initial conditions lie in the
submanifold $\Im(X)$ can be recovered. Hence, in order to obtain a complete
solution to the problem, it is clear that we need to foliate $\Tan Q$
with Lagrangian submanifolds invariant by the Euler-Lagrange vector field
$X_\Lag \in \vf(\Tan Q)$.

\begin{definition}
A \textnormal{complete solution to the Lagrangian Hamilton-Jacobi problem}
is a local diffeomorphism $\Phi \colon Q \times U \to \Tan Q$, with $U \subseteq \R^n$
an open set, such that
for every $\lambda \in U$, the vector field $\Phi(\bullet,\lambda) \equiv X_\lambda \in \vf(Q)$
is a solution to the Lagrangian Hamilton-Jacobi problem.
\end{definition}

\begin{remark}
Usually, it is the set of vector fields $\{ X_\lambda \mid \lambda \in U \}$
which is called a complete solution of the Lagrangian Hamilton-Jacobi problem,
instead of the map $\Phi$. Both definitions are clearly equivalent.
\end{remark}

It is clear from the definition that a complete solution endows $\Tan Q$ with
a foliation transverse to the fibers, and such that every leaf is Lagrangian
and invariant by the Euler-Lagrange vector field $X_\Lag$.

If $\{ X_\lambda \mid \lambda \in U \}$ is a complete solution, the integral
curves of $X_\lambda$, for different $\lambda \in U$, will provide all the integral
curves of the Euler-Lagrange vector field $X_\Lag$. That is, if $(q_o,v_o) \in \Tan Q$,
then there exists $\lambda_o \in U$ such that $X_{\lambda_o}(q_o) = (q_o,v_o)$, and the
integral curves of $X_{\lambda_o}$ through $q_o$ lifted to $\Tan Q$ by $X_{\lambda_o}$
give the integral curves of $X_\Lag$ through $(q_o,v_o)$.

\subsection{Hamiltonian formulation of the Hamilton-Jacobi problem}
\label{Chap02_sec:HamiltonJacobiHam}

Since the Lagrangian function $\Lag \in \Cinfty(\Tan Q)$ is regular, the
associated Legendre map $\Leg \colon \Tan Q \to \Tan^*Q$ is, at least,
a local diffeomorphism. For simplicity, we will assume that the Lagrangian
function is hyperregular, so the associated Legendre map will be a
symplectomorphism between $(\Tan Q,\omega_\Lag)$ and $(\Tan^*Q,\omega)$
(see Section \ref{Chap02_sec:HamiltonianAutonomousFirstOrder}).
In particular, let $h \in \Cinfty(\Tan^*Q)$ be the canonical Hamiltonian
function and $X_h \in \vf(\Tan^*Q)$ the Hamiltonian vector field solution
to equation \eqref{Chap02_eqn:HamAFODynEq}.

(The regular, but not hyperregular case, is recovered by restriction
in the open sets where $\Leg$ is a local diffeomorphism).

\subsubsection{Generalized Hamiltonian Hamilton-Jacobi problem}

\begin{definition}
The \textnormal{generalized Hamiltonian Hamilton-Jacobi problem} consists
in finding a vector field $X \in \vf(Q)$ and a $1$-form $\alpha \in \df^{1}(Q)$
such that, if $\gamma \colon \R \to Q$ is an integral curve of $X$, then
$\alpha \circ \gamma \colon \R \to \Tan^*Q$ is an integral curve of $X_h$;
that is,
\begin{equation}\label{Chap02_eqn:HJHamFOGeneralizedProblemDef}
X \circ \gamma = \dot{\gamma} \Rightarrow X_h \circ (\alpha \circ \gamma) = \dot{\overline{\alpha \circ \gamma}} \, .
\end{equation}
\end{definition}

It is clear from the definition that the vector field $X \in \vf(Q)$ and the
$1$-form $\alpha \in \df^{1}(Q)$ can not be chosen independently.
In particular, we have the following result.

\begin{proposition}\label{Chap02_prop:HJHamFORelatedVF}
The pair $(\alpha,X) \in \df^{1}(Q) \times \vf(Q)$ satisfies the condition
\eqref{Chap02_eqn:HJHamFOGeneralizedProblemDef} if, and only if, $X$ and $X_h$
are $\alpha$-related, that is, $X_h \circ \alpha = \Tan\alpha \circ X$.
\end{proposition}

Now, from Proposition \ref{Chap02_prop:HJHamFORelatedVF}, composing both sides of the
equality with $\Tan\pi_{Q}$, and bearing in mind that $\alpha \in \df^{1}(Q) = \Gamma(\pi_Q)$,
we obtain the following result.

\begin{corollary}
If the pair $(\alpha,X) \in \df^{1}(Q) \times \vf(Q)$ satisfies the condition
\eqref{Chap02_eqn:HJHamFOGeneralizedProblemDef}, then $X = \Tan\pi_Q \circ X_h \circ \alpha$.
\end{corollary}

Hence, $X$ is determined by $\alpha$. This enables us to introduce the following definition.

\begin{definition}
A \textnormal{solution to the generalized Hamiltonian Hamilton-Jacobi problem}
is a $1$-form $\alpha \in \df^{1}(Q)$ such that, if $\gamma \colon \R \to Q$ is
an integral curve of $X = \Tan\pi_Q \circ X_h \circ \alpha \in \vf(Q)$, then
$\alpha \circ \gamma \colon \R \to \Tan^*Q$ is an integral curve of $X_h$;
that is,
\begin{equation}
\Tan\pi_Q \circ X_h \circ \alpha \circ \gamma = \dot{\gamma} \Rightarrow
X_h \circ (\alpha \circ \gamma) = \dot{\overline{\alpha \circ \gamma}} \, .
\end{equation}
The vector field $X = \Tan\pi_Q \circ X_h \circ \alpha \in \vf(Q)$ is said to be
the \textnormal{vector field associated with $\alpha$}.
\end{definition}

\begin{theorem}\label{Chap02_thm:HJHamFOEquivalenceGeneralized}
Let $\alpha \in \df^{1}(Q)$. The following assertions are equivalent:
\begin{enumerate}
\item $\alpha$ is a solution to the generalized Hamiltonian Hamilton-Jacobi problem.
\item The submanifold $\Im(\alpha) \hookrightarrow \Tan^*Q$ is invariant by the
Hamiltonian vector field $X_h$ (that is, $X_h$ is tangent to the submanifold $\Im(\alpha)$).
\item $\alpha$ satisfies the equation
\begin{equation*}
\inn(X)\d\alpha = -\d(\alpha^*h) \, ,
\end{equation*}
where $X = \Tan\pi_Q \circ X_h \circ \alpha$ is the vector field associated with $\alpha$.
\end{enumerate}
\end{theorem}

In coordinates, let $(U;q^A)$, $1 \leqslant A \leqslant n$, be a local chart in $Q$,
and let $(q^A,v^A)$ be the induced natural coordinates in $\Tan^*Q$. In these coordinates,
a generic $1$-form $\alpha \in \df^{1}(Q)$ has the following coordinate expression
\begin{equation*}
\alpha = \alpha_A\d q^A \, ,
\end{equation*}
and the Hamiltonian vector field is given by
\begin{equation*}
X_h = \derpar{h}{p_A}\derpar{}{q^A} - \derpar{h}{q^A}\derpar{}{p_A} \, .
\end{equation*}
Then, taking into account that the submanifold $\Im(\alpha) \hookrightarrow \Tan^*Q$
is locally defined by the constraints $p_A - \alpha_A = 0$, and the second item in
Theorem \ref{Chap02_thm:HJHamFOEquivalenceGeneralized}, the condition for $\alpha$ to
be a solution to the generalized Hamiltonian Hamilton-Jacobi problem gives the
following system of $n$ partial differential equations for the component functions
$\alpha_A$ of $\alpha$
\begin{equation}\label{Chap02_eqn:HJHamFOLocalEqGeneralized}
\restric{\derpar{h}{q^A} + \derpar{h}{p_B}\derpar{\alpha_A}{q^B}}{\Im(\alpha)} = 0 \, .
\end{equation}

\subsubsection{Hamiltonian Hamilton-Jacobi problem}

As in the Lagrangian setting, to solve the generalized Hamiltonian Hamilton-Jacobi
problem is, in general, a difficult task, and thus we require some additional conditions
to the $1$-form $\alpha$ to consider a less general problem.

\begin{definition}
The \textnormal{Hamiltonian Hamilton-Jacobi problem} consists in finding $1$-forms
$\alpha \in \df^{1}(Q)$ solution to the generalized Hamiltonian Hamilton-Jacobi
problem, which are, moreover, closed, that is, $\d\alpha = 0$.
Such a $1$-form is called a \textnormal{solution to the Hamiltonian Hamilton-Jacobi problem}.
\end{definition}

Observe that, since $\theta \in \df^{1}(\Tan^*Q)$ is the tautological form
of the cotangent bundle, we have $\alpha^*\theta = \alpha$ for every
$\alpha \in \df^{1}(Q)$, and thus
\begin{equation*}
\alpha^*\omega = -\alpha^*\d\theta = -\d(\alpha^*\theta) = -\d\alpha \, .
\end{equation*}
In particular, $\alpha$ is a closed $1$-form if, and only if, $\alpha^*\omega=0$,
that is, the submanifold $\Im(\alpha) \hookrightarrow \Tan^*Q$ is a Lagrangian
submanifold. Hence, we have the following result as a straightforward
consequence of Theorem \ref{Chap02_thm:HJHamFOEquivalenceGeneralized}.

\begin{theorem}
Let $\alpha \in \df^{1}(Q)$ be a closed $1$-form. Then, the following assertions
are equivalent:
\begin{enumerate}
\item $\alpha$ is a solution to the Hamiltonian Hamilton-Jacobi problem.
\item $\d(\alpha^*h) = 0$.
\item $\Im(\alpha) \hookrightarrow \Tan^*Q$ is a Lagrangian submanifold invariant
by $X_h$.
\item The integral curves of $X_h$ with initial conditions in $\Im(\alpha)$
project onto the integral curves of $X = \Tan\pi_Q \circ X_h \circ \alpha$.
\end{enumerate}
\end{theorem}

In the induced natural coordinates $(q^A,p_A)$ of $\Tan^*Q$, if the
$1$-form $\alpha \in \df^{1}(Q)$ is locally given by
$\alpha = \alpha_A\d q^A$, then its differential is
\begin{equation*}
\d\alpha = \derpar{\alpha_A}{q^B}\d q^B \wedge \d q^A \, ,
\end{equation*}
and thus the condition for $\alpha$ to be
closed gives the following system of partial differential equations
\begin{equation}\label{Chap02_eqn:HJHamLocalClosedForm}
\derpar{\alpha_A}{q^B} - \derpar{\alpha_B}{q^A} = 0 \, , \mbox{ for } A \neq B \, .
\end{equation}
Therefore, a $1$-form $\alpha \in \df^{1}(Q)$ is a solution to the Hamiltonian
Hamilton-Jacobi problem if, and only if, its component functions satisfy the system
of partial differential equations given by \eqref{Chap02_eqn:HJHamFOLocalEqGeneralized}
and \eqref{Chap02_eqn:HJHamLocalClosedForm}.

In addition, in this case we can obtain the classical Hamilton-Jacobi equation.
Since $\alpha \in \df^{1}(Q)$ is a closed form, using Poincar\'{e}'s Lemma,
every point in $Q$ admits an open neighborhood $U \subseteq Q$ where there exists
a local function $W \in \Cinfty(U)$ such that $\alpha = \d W$ (in $U$).
Then, we obtain the following system of partial differential equations
\begin{equation*}
\derpar{W}{q^A} = \alpha_A \, .
\end{equation*}
Finally, taking into account that $\d(\alpha^*h) = 0$, then we
have that $h \circ \d W$ must be constant, and therefore
\begin{equation}\label{Chap02_eqn:HJHamFOHamiltonJacobiEq}
h \left( q^A,\derpar{W}{q^A} \right) = \text{const.} \, ,
\end{equation}
which is the standard Hamilton-Jacobi equation.

\subsubsection{Complete solutions}

As in the Lagrangian setting, up to this point we have stated the equation
to obtain a \textit{particular} solution of the dynamical system,
rather than a complete solution, since only the integral curves of $X_h$
with initial conditions lying in the submanifold $\Im(\alpha)$ can be recovered.
In order to obtain a complete solution of the system, we proceed in an analogous way
than we did for the Lagrangian formalism.

\begin{definition}
A \textnormal{complete solution to the Hamiltonian Hamilton-Jacobi problem}
is a local diffeomorphism $\Phi \colon Q \times U \to \Tan^*Q$, with
$U \subseteq \R^n$ an open set, such that for every $\lambda \in U$, the
$1$-form $\Phi(\bullet,\lambda) = \alpha_\lambda \in \df^{1}(Q)$ is a solution
to the Hamiltonian Hamilton-Jacobi problem.
\end{definition}

\begin{remark}
It is usually the set of $1$-forms $\{ \alpha_\lambda \mid \lambda \in U \}$
which is called a complete solution of the Hamiltonian Hamilton-Jacobi problem,
instead of the map $\Phi$. Both definitions are clearly equivalent.
\end{remark}

We deduce from the definition that a complete solution provides $\Tan^*Q$ with
a foliation transverse to the fibers, and such that every leaf is Lagrangian
and invariant by the Hamiltonian vector field $X_h$.

If $\{ \alpha_\lambda \mid \lambda \in U \}$ is a complete solution, the integral
curves of the vector fields associated with $\alpha_\lambda$, for different $\lambda \in U$,
will provide all the integral curves of the Hamiltonian vector field $X_h$.
That is, if $(q_o,p_o) \in \Tan^*Q$,
then there exists $\lambda_o \in U$ such that $\alpha_{\lambda_o}(q_o) = (q_o,p_o)$, and the
integral curves of $\Tan\pi_Q \circ X_h \circ \alpha_{\lambda_o}$ through $q_o$ lifted
to $\Tan^*Q$ by $\alpha_{\lambda_o}$ give the integral curves of $X_h$ through $(q_o,p_o)$.


\section{Higher-order autonomous dynamical systems}
\label{Chap02_sec:AutonomousHigherOrder}

Let us consider a $k$th-order autonomous Lagrangian dynamical system
with $n$ degrees of freedom. Let $Q$ be a $n$-dimensional smooth manifold
modeling the configuration space of this $k$th-order dynamical system, and
$\Lag \in \Cinfty(\Tan^{k}Q)$ a $k$th-order Lagrangian function describing the dynamics
of the system.

We refer to \cite{art:Batlle_Gomis_Pons_Roman88,proc:Cantrijn_Crampin_Sarlet86,book:DeLeon_Rodrigues85,
proc:Garcia_Munoz83,art:Gracia_Pons_Roman91,art:Gracia_Pons_Roman92,phd:Martinez,art:Nesterenko89}
for details and proofs.

\subsection{Lagrangian formalism}
\label{Chap02_sec:LagrangianAutonomousHigherOrder}

\subsubsection{Geometric and dynamical structures}

From the Lagrangian function $\Lag \in \Cinfty(\Tan^{k}Q)$ and the
canonical structures of the $k$th-order tangent bundle, namely the
vertical endomorphisms $J_r \colon \vf(\Tan^{k}Q) \to \vf^{V(\rho^{k}_{r})}(\Tan^kQ)$
and the canonical vector fields $\Delta_r \in \vf(\Tan^{k}Q)$, we construct the following
structures.

\begin{definition}
The \textnormal{$k$th-order Poincar\'{e}-Cartan $1$-form} associated to
$\Lag \in \Cinfty(\Tan^{k}Q)$, or \textnormal{$k$th-order Lagrangian $1$-form},
is the form $\theta_\Lag \in \df^{1}(\Tan^{2k-1}Q)$ defined as
\begin{equation*}\label{Chap02_eqn:LagHO1FormDef}
\theta_\Lag = \sum_{i=1}^{k} (-1)^{i-1} \frac{1}{i!} \, d^{i-1}_T\left(\inn(J_i)\d\Lag \right) \, .
\end{equation*}
From this, the \textnormal{$k$th-order Poincar\'{e}-Cartan $2$-form} associated to
$\Lag \in \Cinfty(\Tan^{k}Q)$, or \textnormal{$k$th-order Lagrangian $2$-form}, is the
form $\omega_\Lag \in \df^{2}(\Tan^{2k-1}Q)$ defined as
\begin{equation*}\label{Chap02_eqn:LagHO2FormDef}
\omega_\Lag = -\d\theta_\Lag \, .
\end{equation*}
\end{definition}

As in the first-order formalism described in Section
\ref{Chap02_sec:LagrangianAutonomousFirstOrder}, the $2$-form $\omega_\Lag$
may not have constant rank at every point of $\Tan^{2k-1}Q$ for an arbitrary
$k$th-order Lagrangian function.
If $\rank(\omega_\Lag(j^{2k-1}_0\phi)) = \text{const.}$
for every $j^{2k-1}_0\phi \in \Tan^{2k-1}Q$,
then $\Lag \in \Cinfty(\Tan^{k}Q)$ is said to be a
\textsl{geometrically admissible $k$th-order Lagrangian}.
Again, we will only consider $k$th-order Lagrangian
functions satisfying this property.

\begin{definition}
The \textnormal{$k$th-order Lagrangian energy} associated to $\Lag$
is the function $E_\Lag \in \Cinfty(\Tan^{2k-1}Q)$ defined as
\begin{equation*}\label{Chap02_eqn:LagHOEnergyDef}
E_\Lag = \left( \sum_{i=1}^{k} (-1)^{i-1} \frac{1}{i!} \, d_T^{i-1}\left( \Delta_{i}(\Lag)) \right) \right)
- (\rho^{2k-1}_{k})^*\Lag \, .
\end{equation*}
(In an abuse of notation, in the following we write simply $\Lag$ instead of $(\rho^{2k-1}_{k})^*\Lag$.)
\end{definition}

From these definitions, it is clear that the phase space of a $k$th-order Lagrangian dynamical system
is the $(2k-1)$th-order tangent bundle of the configuration manifold $Q$.

\begin{definition}
A \textnormal{Lagrangian system of order $k$} is a pair $(\Tan^{2k-1}Q,\Lag)$, where $Q$
represents the configuration space and $\Lag \in \Cinfty(\Tan^{k}Q)$ is the $k$th-order Lagrangian
function.
\end{definition}

In coordinates, bearing in mind the coordinate expressions
\eqref{Chap01_eqn:HoTanBundleVertEndR} of the vertical
endomorphisms and \eqref{Chap01_eqn:HOTanBundleTulczyjewDerLocal}
of the Tulczyjew's derivation, we obtain, after a long but straightforward calculation,
the local expression of the $k$th-order Lagrangian $1$-form, which is
\begin{equation}\label{Chap02_eqn:LagHO1FormLocal}
\theta_\Lag = \sum_{r=1}^k \sum_{i=0}^{k-r}(-1)^i d_T^i\left(\derpar{\Lag}{q_{r+i}^A}\right) \d q_{r-1}^A \, ,
\end{equation}
where the terms $d_T(\bullet)$ are not expanded to avoid a long expression involving higher derivatives of
the $k$th-order Lagrangian function. The coordinate expression of the $k$th-order Lagrangian $2$-form
is omitted for the same reason. Now, for the $k$th-order Lagrangian energy, bearing in mind the local
expressions \eqref{Chap01_eqn:HOTanBundleCanonicalVF} of the canonical vector fields and
\eqref{Chap01_eqn:HOTanBundleTulczyjewDerLocal} of the Tulczyjew's derivation, we have
\begin{equation}\label{Chap02_eqn:LagHOEnergyLocal}
E_\Lag = \sum_{r=1}^{k} q_{r}^A \sum_{i=0}^{k-r} (-1)^i d_T^i\left( \derpar{\Lag}{q_{r+i}^A} \right) - \Lag(q_0^A,\ldots,q_k^A) \, .
\end{equation}

\begin{remark}
From the coordinate expression \eqref{Chap02_eqn:LagHO1FormLocal} we deduce that the $k$th-order
Lagrangian $1$-form is a $\rho^{2k-1}_{k-1}$-semibasic $1$-form in $\Tan^{2k-1}Q$, sinc
$\theta_\Lag \in \Im(J_k^{*})$.
\end{remark}

Observe that, given an arbitrary $k$th-order Lagrangian function $\Lag \in \Cinfty(\Tan^{k}Q)$,
the $k$th-order Lagrangian $2$-form is always closed by definition. In addition, note that
$\dim\Tan^{2k-1}Q = (2k-1+1)n = 2kn$, that is, $\Tan^{2k-1}Q$ has even dimension for every $k \in \N$.
In particular, $\omega_\Lag \in \df^{2}(\Tan^{2k-1}Q)$ may be nondegenerate, and therefore a symplectic
form on $\Tan^{2k-1}Q$. This leads to the following definition.

\begin{definition}\label{Chap02_def:LagHORegularLagrangian}
A $k$th-order Lagrangian function $\Lag \in \Cinfty(\Tan^{k}Q)$ is \textnormal{regular}
(and thus $(\Tan^{2k-1}Q,\Lag)$ is a \textnormal{$k$th-order regular system})
if the $k$th-order Lagrangian $2$-form $\omega_\Lag \in \df^{2}(\Tan^{2k-1}Q)$
associated to $\Lag$ is symplectic. Otherwise, the $k$th-order Lagrangian is said
to be \textnormal{singular} (and thus $(\Tan^{2k-1}Q,\Lag)$ is a
\textnormal{$k$th-order singular system}).
\end{definition}

After a long and tedious calculation in coordinates,
we can prove that the nondegeneracy of the $2$-form $\omega_\Lag$
is locally equivalent to
\begin{equation*}
\det\left( \derpars{\Lag}{q_k^A}{q_k^B} \right)(j^{2k-1}_0\phi) \neq 0 \ , \,
\mbox{for every } j^{2k-1}_0\phi \in \Tan^{2k-1}Q \, .
\end{equation*}
That is, a $k$th-order Lagrangian function is regular if, and only if, the Hessian matrix of $\Lag$
with respect to the ``velocities'' of highest order is invertible at every point of $\Tan^{2k-1}Q$.

\subsubsection{Dynamical vector field}

The dynamical trajectories of this $k$th-order system are given by the integral curves of
a semispray of type $1$, $X_\Lag \in \vf(\Tan^{2k-1}Q)$, satisfying
\begin{equation}\label{Chap02_eqn:LagHODynEq}
\inn(X_\Lag)\omega_\Lag = \d E_\Lag \, .
\end{equation}
This equation is the \textsl{$k$th-order Lagrangian equation}, and a vector field $X_\Lag$
solution to \eqref{Chap02_eqn:LagHODynEq} (if such a vector field exists)
is called a \textsl{$k$th-order Lagrangian vector field}. If, in addition, $X_\Lag$ is
a semispray of type $1$, then it is called the \textsl{$k$th-order Euler-Lagrange vector field},
and its integral curves are solutions to the \textsl{$k$th-order Euler-Lagrange equations}.

In the natural coordinates of $\Tan^{2k-1}Q$, let $X_\Lag \in \vf(\Tan^{2k-1}Q)$ be a generic
vector field locally given by
\begin{equation*}
X_\Lag = f_i^A \derpar{}{q_i^A} = f_0^A \derpar{}{q_0^A} + \ldots + f_{2k-1}^A \derpar{}{q_{2k-1}^A} \, .
\end{equation*}
Taking into account the coordinate expression
\eqref{Chap02_eqn:LagHOEnergyLocal} of the $k$th-order Lagrangian energy
and after a long and tedious calculation, from equation \eqref{Chap02_eqn:LagHODynEq} we obtain the
following system of $2kn$ equations for the component functions of $X_\Lag$
\begin{align}
&&\left.\begin{array}{r}
\displaystyle \left(f_0^B-q_1^B\right) \derpars{\Lag}{q_k^B}{q_k^A} = 0 \\[10pt]
\displaystyle \left(f_{1}^B - q_{2}^B\right)\derpars{\Lag}{q_k^B}{q_k^A} - \left(f_0^B-q_{1}^B \right)(\,\cdots\cdots) = 0 \\
\vdots \qquad \qquad \qquad \vdots \qquad \qquad \\
\displaystyle \left(f_{2k-2}^B - q_{2k-1}^B\right)\derpars{\Lag}{q_k^B}{q_k^A} - \sum_{i=0}^{2k-3} \left(f_{i}^B-q_{i+1}^B \right) (\,\cdots\cdots) = 0
\end{array} \right\}
\label{Chap02_eqn:LagHODynEqHolonomyLocal}
\end{align}
\begin{equation}
(-1)^k\left(f_{2k-1}^B - d_T\left(q_{2k-1}^B\right)\right) \derpars{\Lag}{q_k^B}{q_k^A} +
\sum_{i=0}^{k} (-1)^id_T^i\left( \derpar{L}{q_i^A} \right) - \sum_{i=0}^{2k-2} \left(f_{i}^B-q_{i+1}^B \right) (\,\cdots\cdots) = 0 \, , \ \ \,
\label{Chap02_eqn:LagHODynEqLocal}
\end{equation}
where the terms in brackets $(\cdots\cdots)$ contain relations involving partial derivatives
of the $k$th-order Lagrangian and applications of $d_T$, which for simplicity are not written.
These are the local equations arising from the $k$th-order Lagrangian equation for $X_\Lag$.

Observe that the $(2k-1)n$ equations \eqref{Chap02_eqn:LagHODynEqHolonomyLocal} are the local
equations for the condition of semispray of type $1$ that we require to the vector field $X_\Lag$.
If this condition is required from the beginning, then equations
\eqref{Chap02_eqn:LagHODynEqHolonomyLocal} are an identity, and equations
\eqref{Chap02_eqn:LagHODynEqLocal} reduce to
\begin{equation}\label{Chap02_eqn:LagHODynEqWithHolonomyLocal}
(-1)^k\left(f_{2k-1}^B - d_T\left(q_{2k-1}^B\right)\right) \derpars{\Lag}{q_k^B}{q_k^A} + \sum_{i=0}^{k} (-1)^id_T^i\left( \derpar{L}{q_i^A} \right) = 0 \, ,
\end{equation}
which are clearly a generalization of equations \eqref{Chap02_eqn:LagAFODynEqWithHolonomyLocal}
to higher-order systems.

On the other hand, notice that in all of the equations \eqref{Chap02_eqn:LagHODynEqHolonomyLocal}
and \eqref{Chap02_eqn:LagHODynEqLocal}
the Hessian matrix of $\Lag$ with respect to the highest order ``velocities'' appears alongside
the coefficients to be determined. Therefore, we have the following result.

\begin{proposition}\label{Chap02_prop:LagHORegLagUniqueVF}
If the $k$th-order Lagrangian function $\Lag \in \Cinfty(\Tan^{k}Q)$ is regular,
then there exists a unique vector field $X_\Lag \in \vf(\Tan^{2k-1}Q)$ solution to
equation \eqref{Chap02_eqn:LagHODynEq} which, in addition, is a semispray of type
$1$ in $\Tan^{2k-1}Q$.
\end{proposition}

\begin{remark}
As in the first-order formalism described in Section \ref{Chap02_sec:LagrangianAutonomousFirstOrder},
the existence and uniqueness of the solution to equation \eqref{Chap02_eqn:LagHODynEq}
is a direct consequence to the fact that $\Lag \in \Cinfty(\Tan^{k}Q)$ is regular
if, and only if, the $k$th-order Lagrangian $2$-form $\omega_\Lag \in \df^{2}(\Tan^{2k-1}Q)$
is symplectic.
\end{remark}

If the $k$th-order Lagrangian function $\Lag \in \Cinfty(\Tan^{k}Q)$ is not regular,
then the $2$-form $\omega_\Lag \in \df^{2}(\Tan^{2k-1}Q)$ is degenerate, and hence
the existence of solutions to the equation \eqref{Chap02_eqn:LagHODynEq} can not be assured
in general, but only in some special cases or requiring some
additional conditions to the $k$th-order Lagrangian function. In general, we must use the constraint
algorithm described in Section \ref{Chap01_sec:ConstraintAlgorithm} and, in the most
favorable cases, there exists a submanifold $S_f \hookrightarrow \Tan^{2k-1}Q$ where the
equation
\begin{equation}\label{Chap02_eqn:LagHODynEqSingular}
\restric{[\inn(X_\Lag) \omega_\Lag - \d E_\Lag]}{S_f} = 0 \, ,
\end{equation}
admits a well-defined solution $X_\Lag$, which is tangent to $S_f$. Nevertheless,
these vector fields solution are not necessarily semisprays of type $1$ on $S_f$, but only in
the points of another submanifold $S_f^h \hookrightarrow S_f$.

\subsubsection{Integral curves}

Let $X_\Lag$ be a semispray of type $1$ in $\Tan^{2k-1}Q$ solution to the equation
\eqref{Chap02_eqn:LagHODynEq}, and let $\psi_\Lag \colon \R \to \Tan^{2k-1}Q$ be an integral
curve of $X_\Lag$. Since $X_\Lag$ is a semispray of type $1$, the curve $\psi_\Lag$ is holonomic,
and therefore there exists a curve $\phi_\Lag \colon \R \to Q$ such that $j^{2k-1}_0\phi_\Lag = \psi_\Lag$; that is,
$\phi$ is a path of $X_\Lag$, in the sense of Definition \ref{Chap01_def:HoTanBundleSemisprayPath}.
From the condition of being an integral curve, we deduce the following geometric equation for
the curve $\psi_\Lag$
\begin{equation}\label{Chap02_eqn:LagHODynEqIC}
\inn(\dot{\psi}_\Lag)(\omega_\Lag \circ \psi_\Lag) = \d E_\Lag \circ \psi_\Lag \, ,
\end{equation}
or, equivalently, the following geometric equation for $\phi_\Lag$
\begin{equation*}
\inn(\dot{\overline{j^{2k-1}_0\phi_\Lag}})(\omega_\Lag \circ j^{2k-1}_0\phi_\Lag) = \d E_\Lag \circ j^{2k-1}_0\phi_\Lag \, .
\end{equation*}

In coordinates, the curve $\phi_\Lag \colon \R \to Q$ must satisfy the
following system of $n$ ordinary differential equations of order $2k$
\begin{equation}\label{Chap02_eqn:HOEulerLagrange}
\sum_{i=0}^{k} \restric{(-1)^i\derpar{\Lag}{q_i^A}}{j^{2k-1}_0\phi_\Lag} = 0 \, .
\end{equation}
These are the \textsl{$k$th-order Euler-Lagrange equations} for this dynamical system.

\subsection{Hamiltonian formalism associated to a Lagrangian system}
\label{Chap02_sec:HamiltonianAutonomousHigherOrder}

\subsubsection{The Legendre-Ostrogradsky map}
\label{Chap02_sec:LegendreOstrogradskyMap}

\begin{definition}
The \textnormal{Legendre-Ostrogradsky map} associated to the $k$th-order
Lagrangian function $\Lag$ is the fiber bundle morphism
$\Leg \colon \Tan^{2k-1}Q \to \Tan^*(\Tan^{k-1}Q)$ over $\Tan^{k-1}Q$ defined as
follows: for every $u \in \Tan(\Tan^{2k-1}Q)$,
\begin{equation}\label{Chap02_eqn:HamHOLegendreMapDef}
\theta_\Lag(u) = \left\langle \Tan\rho^{2k-1}_{k-1}(u) \, , \, \Leg(\tau_{\Tan^{2k-1}Q}(u) \right\rangle \, .
\end{equation}
\end{definition}

From the definition we have that $\pi_{\Tan^{k-1}Q} \circ \Leg = \rho^{2k-1}_{k-1}$.
In addition, if $\theta_{k-1} \in \df^{1}(\Tan^*(\Tan^{k-1}Q))$ and
$\omega_{k-1} = -\d\theta_{k-1} \in \df^{2}(\Tan^*(\Tan^{k-1}Q))$ are the canonical $1$ and $2$ forms
of the cotangent bundle $\Tan^*(\Tan^{k-1}Q)$, then $\Leg^*\theta_{k-1} = \theta_\Lag$ and
$\Leg^*\omega_{k-1} = \omega_\Lag$.

In the natural coordinates of $\Tan^{2k-1}Q$, we define the following local functions
\begin{equation*}\label{Chap02_eqn:HamHOMomentumCoord}
\hat p^{r-1}_A = \sum_{i=0}^{k-r}(-1)^i d_T^i\left(\derpar{\Lag}{q_{r+i}^A}\right) \, .
\end{equation*}
Observe that we have the following relation between $\hat{p}^r_A$ and $\hat{p}^{r-1}_A$
\begin{equation}\label{Chap02_eqn:HamHOMomentumCoordRelation}
\hat p^{r-1}_A = \derpar{\Lag}{q_r^A} - d_T(\hat p^r_A) \quad , \quad \mbox{for } 1 \leqslant r \leqslant k-1 \, .
\end{equation}
Thus, bearing in mind the local expression \eqref{Chap02_eqn:LagHO1FormLocal}
of the form $\theta_\Lag$, we can write
\begin{equation}\label{Chap02_eqn:LagHO1FormLocalMomenta}
\theta_\Lag = \sum_{r=1}^k \hat p^{r-1}_A \d q_{r-1}^A \, ,
\end{equation}
On the other hand, let $(q^A)$, $1 \leqslant A \leqslant n$, be a set of local coordinates
in an open set $U \subseteq Q$, and $(q_i^A)$, $0 \leqslant i \leqslant k-1$, the
induced natural coordinates in $\Tan^{k-1}Q$. Then, natural coordinates in $\Tan^*(\Tan^{k-1}Q)$
are $(q_i^A,p_A^i)$, with $1 \leqslant A \leqslant n$ and $0 \leqslant i \leqslant k-1$.
In these coordinates, the canonical $1$ and $2$ forms of $\Tan^*(\Tan^{k-1}Q)$ have the
following coordinate expressions
\begin{equation}\label{Chap02_eqn:HamHOCanonicalForms}
\theta_{k-1} = p_A^i \d q_i^A \quad ; \quad \omega_{k-1} = -\d\theta_{k-1} = \d q_i^A \wedge \d p_A^i \, .
\end{equation}
Finally, taking into account the local expressions
\eqref{Chap02_eqn:LagHO1FormLocalMomenta} of $\theta_\Lag$ and
\eqref{Chap02_eqn:HamHOCanonicalForms} of $\theta_{k-1}$, the coordinate
expression of the Legendre-Ostrogradsky map associated to $\Lag$ is
\begin{equation}\label{Chap02_eqn:HamHOLegendreMapLocal}
\Leg^*q_{r-1}^A = q_{r-1}^A \quad ; \quad
\Leg^*p_A^{r-1} = \hat p^{r-1}_A = \sum_{i=0}^{k-r}(-1)^i d_T^i\left(\derpar{\Lag}{q_{r+i}^A}\right) \, ,
\end{equation}
where $1 \leqslant r \leqslant k$.

From the local expression \eqref{Chap02_eqn:HamHOLegendreMapLocal},
the rank of the tangent map $\Tan\Leg \colon \Tan(\Tan^{2k-1}Q) \to \Tan(\Tan^*(\Tan^{k-1}Q))$
depends only of the rank of the Hessian matrix of $\Lag$ with respect to the highest order
``velocities''. Therefore, $\Lag \in \Cinfty(\Tan^{k}Q)$ is a regular $k$th-order
Lagrangian function if, and only if, the Legendre-Ostrogradsky map
$\Leg \colon \Tan^{2k-1}Q \to \Tan^*(\Tan^{k-1}Q)$ is a local diffeomorphism.
As a consequence of this, we have that if $\Lag$ is a $k$th-order regular Lagrangian then the set
$(q_i^A,\hat p^i_A)$, $0\leqslant i\leqslant k-1$, is a set of local coordinates in $\Tan^{2k-1}Q$,
and $(\hat p^i_A)$ are called the \textsl{Jacobi-Ostrogradsky momentum coordinates}.

\begin{remark}
The relation \eqref{Chap02_eqn:HamHOMomentumCoordRelation} means that we can recover
all the Jacobi-Ostrogradsky momentum coordinates from the set of highest order momenta
$(\hat p^{k-1}_A)$.
\end{remark}

\begin{definition}
A $k$th-order Lagrangian function $\Lag \in \Cinfty(\Tan^{k}Q)$ is \textnormal{hyperregular} if the
Legendre-Ostrogradsky map $\Leg \colon \Tan^{2k-1}Q \to \Tan^*(\Tan^{k-1}Q)$ is a global
diffeomorphism.
\end{definition}

\begin{remark}
If the $k$th-order Lagrangian function is hyperregular, then the Legendre-Ostrogradsky map is a
symplectomorphism between the symplectic manifolds $(\Tan^{2k-1}Q,\omega_\Lag)$ and
$(\Tan^*(\Tan^{k-1}Q),\omega_{k-1})$.
\end{remark}

As in the first-order Hamiltonian formalism described in Section \ref{Chap02_sec:HamiltonianAutonomousFirstOrder},
we will distinguish between the regular and non-regular cases to describe the dynamical trajectories
of the system. Nevertheless, as in the first-order setting,
the only singular systems that we will consider are the almost-regular ones.

\begin{definition}
A $k$th-order Lagrangian function $\Lag \in \Cinfty(\Tan^{k}Q)$ is \textnormal{almost-regular} if
\begin{enumerate}
\item $\Leg(\Tan^{2k-1}Q) \hookrightarrow \Tan^*(\Tan^{k-1}Q)$ is a closed submanifold.
\item $\Leg$ is a surjective submersion onto its image.
\item For every $j^{2k-1}_0\phi \in \Tan^{2k-1}Q$ the fibers $\Leg^{-1}(\Leg(j^{2k-1}_0\phi))$
are connected submanifolds of $\Tan^{2k-1}Q$.
\end{enumerate}
\end{definition}

\subsubsection{Regular and hyperregular Lagrangian functions}

Suppose that the $k$th-order Lagrangian function $\Lag \in \Cinfty(\Tan^{k}Q)$
is hyperregular (the regular case is recovered from this one by restriction on the open sets
where the Legendre-Ostrogradsky map is a local diffeomorphism).
Since $\Leg \colon \Tan^{2k-1}Q \to \Tan^*(\Tan^{k-1}Q)$ is a global diffeomorphism,
there exists a unique function $h \in \Cinfty(\Tan^*(\Tan^{k-1}Q))$ such that $\Leg^*h = E_\Lag$.

\begin{definition}\label{Chap02_def:HamHOHamiltonianFunctionDef}
The \textnormal{canonical $k$th-order Hamiltonian function} $h \in \Cinfty(\Tan^*(\Tan^{k-1}Q))$
is the unique function satisfying $\Leg^*h = E_\Lag$.
\end{definition}

The dynamical trajectories of the system are given by the integral curves of a vector field
$X_h \in \vf(\Tan^*(\Tan^{k-1}Q))$ satisfying
\begin{equation}\label{Chap02_eqn:HamHODynEq}
\inn(X_h) \omega_{k-1} = \d h \, .
\end{equation}
This equation is the \textsl{$k$th-order Hamiltonian equation}, and the unique vector field solution
to this equation is called the \textsl{$k$th-order Hamiltonian vector field}.

In coordinates, bearing in mind the local expressions
\eqref{Chap02_eqn:HamHOLegendreMapLocal} of the
Legendre-Ostrogradsky map $\Leg$ and \eqref{Chap02_eqn:LagHOEnergyLocal}
of the $k$th-order Lagrangian energy, we have
\begin{equation*}\label{Chap02_eqn:HamHOHamiltonianFunctionLocal}
h = \sum_{i=0}^{k-2} q_{i+1}^{A}p_{A}^{i} + (\Leg^{-1})^*q_k^Ap^{k-1}_A - (\rho^{2k-1}_{k} \circ \Leg^{-1})^*\Lag \, .
\end{equation*}
Now, for the equation \eqref{Chap02_eqn:HamHODynEq}, let $X_h \in \Cinfty(\Tan^*(\Tan^{k-1}Q))$ be
a generic vector field given by
\begin{equation*}
X_h = f_i^A \derpar{}{q_i^A} + G^i_A\derpar{}{p^i_A} \, .
\end{equation*}
Then, bearing in mind the coordinate expression \eqref{Chap02_eqn:HamHOCanonicalForms}
of the canonical symplectic form of $\Tan^*(\Tan^{k-1}Q)$,
equation \eqref{Chap02_eqn:HamHODynEq} gives the following system of $2kn$ equations
\begin{equation}\label{Chap02_eqn:HamHODynEqLocal}
f_i^A = \derpar{h}{p^i_A} \quad ; \quad G^i_A = - \derpar{h}{q_i^A} \, .
\end{equation}

Finally, as in the first-order setting, we establish the relation between the solutions to the dynamical
equation \eqref{Chap02_eqn:LagHODynEq} in the Lagrangian formalism and the solutions to the dynamical
equation \eqref{Chap02_eqn:HamHODynEq} in the Hamiltonian formalism associated to a hyperregular Lagrangian system.

\begin{theorem}\label{Chap02_thm:HamHORelationLagHamRegular}
Let $\Lag \in \Cinfty(\Tan^{k}Q)$ be a hyperregular $k$th-order Lagrangian function.
Then we have:
\begin{enumerate}
\item Let $X_\Lag \in \vf(\Tan^{2k-1}Q)$ be the unique semispray of type $1$ solution
to equation \eqref{Chap02_eqn:LagHODynEq}. Then the vector field
$X_h = \Leg_*X_\Lag \in \vf(\Tan^*(\Tan^{k-1}Q))$ is a solution to equation \eqref{Chap02_eqn:HamHODynEq}.

\item Conversely, let $X_h \in \vf(\Tan^*(\Tan^{k-1}Q))$ be the unique vector field solution
to equation \eqref{Chap02_eqn:HamHODynEq}. Then the vector field
$X_\Lag = (\Leg^{-1})_*X_h \in \vf(\Tan^{2k-1}Q)$ is a semispray of type $1$, and is a solution to
equation \eqref{Chap02_eqn:LagHODynEq}.
\end{enumerate}
\end{theorem}

Now, if $\psi_h \colon \R \to \Tan^*(\Tan^{k-1}Q)$ is an integral curve of $X_h$, the geometric
equation for the dynamical trajectories of the system is
\begin{equation}\label{Chap02_eqn:HamHODynEqIC}
\inn(\dot{\psi}_h)(\omega_{k-1} \circ \psi_h) = \d h \circ \psi_h \, .
\end{equation}

In coordinates, if the curve $\psi_h$ is given by $\psi_h(t) = (q_i^A(t),p^i_A(t))$, then its
component functions must satisfy the following system of $2kn$ first-order
differential equations
\begin{equation}\label{Chap02_eqn:HOHamiltonEq}
\dot{q}_i^A = \restric{\derpar{h}{p^i_A}}{\psi_h} \quad ; \quad \dot{p}^i_A = - \restric{\derpar{h}{q_i^A}}{\psi_h} \, .
\end{equation}
These are the \textsl{$k$th-order Hamilton equations} for this dynamical system.

\subsubsection{Singular (almost-regular) Lagrangian functions}

Suppose now that the $k$th-order Lagrangian function $\Lag \in \Cinfty(\Tan^{k}Q)$
is almost-regular. This implies that the Legendre-Ostrogradsky map
$\Leg \colon \Tan^{2k-1}Q \to \Tan^*(\Tan^{k-1}Q)$ is no longer
a diffeomorphism, and therefore its image set is a proper submanifold of $\Tan^*(\Tan^{k-1}Q)$.
Let $\P = \Im(\Leg) \hookrightarrow \Tan^*(\Tan^{k-1}Q)$ be the image set of the Legendre-Ostrogradsky map,
with natural embedding $\jmath \colon \P \hookrightarrow \Tan^*(\Tan^{k-1}Q)$, and we denote by
$\Leg_o \colon \Tan^{2k-1} Q \to \P$ the map defined by $\Leg = \jmath \circ \Leg_o$.
With these notations, we have the following result.

\begin{proposition}\label{Chap02_prop:HamHOLagrangianEnergyProjectable}
Let $\Lag \in \Cinfty(\Tan^{k}Q)$ be a $k$th-order almost-regular Lagrangian function.
Then the $k$th-order Lagrangian energy $E_\Lag \in \Cinfty(\Tan^{2k-1}Q)$ is $\Leg_o$-projectable.
\end{proposition}

As a consequence of this last result, we can define a Hamiltonian function
in $\P$ as follows.

\begin{definition}\label{Chap02_def:HamHOHamiltonianFunctionDefSingular}
The \textnormal{canonical Hamiltonian function} is the unique function
$h_o \in \Cinfty(\P)$ such that $\Leg_o^*h_o = E_\Lag$.
\end{definition}

Then, taking $\omega_o = \jmath^*\omega \in \df^{2}(\P)$, we can state the $k$th-order
Hamilton equation for this system: we look for a vector field $X_{h_o} \in \vf(\P)$ satisfying
\begin{equation*}\label{Chap02_eqn:HamHODynEqSingular1}
\inn(X_{h_o})\omega_o = \d h_o \, .
\end{equation*}
Since the form $\omega_o$ is, in general, presymplectic,
we must apply the constraint algorithm described in Section
\ref{Chap01_sec:ConstraintAlgorithm}. In the most favorable cases, this equation
admits a solution only on the points of some submanifold
$\P_f \hookrightarrow \P \hookrightarrow \Tan^*(\Tan^{k-1}Q)$, and is tangent to it,
so the following equation holds
\begin{equation}\label{Chap02_eqn:HamHODynEqSingular2}
\restric{[\inn(X_{h_o}) \omega_o - \d h_o]}{\P_f} = 0 \, .
\end{equation}
This vector field is not unique, in general.

In this situation, we have an analogous result to Theorem \ref{Chap02_thm:HamHORelationLagHamRegular}.

\begin{theorem}\label{Chap02_thm:HamHORelationLagHamSingular}
Let $\Lag \in \Cinfty(\Tan^{k}Q)$ be an almost-regular Lagrangian function.
Then we have:
\begin{enumerate}
\item Let $X_\Lag \in \vf(\Tan^{2k-1} Q)$ be a semispray of type $1$ solution
to equation \eqref{Chap02_eqn:LagHODynEqSingular} in the points of a submanifold
$S_f \hookrightarrow \Tan^{2k-1}Q$. Then there exists a vector field
$X_{h_o} \in \vf(\P)$ which is $\Leg_o$-related to $X_\Lag$
and is a solution to equation \eqref{Chap02_eqn:HamHODynEqSingular2},
where $\P_f = \Leg_o(S_f) \hookrightarrow \P$

\item Conversely, let $X_{h_o} \in \vf(\P)$ be a vector field solution
to equation \eqref{Chap02_eqn:HamHODynEqSingular2} on the points of some
submanifold $\P_f \hookrightarrow \P$. Then there exist vector fields
$X_\Lag \in \vf(\Tan^{2k-1}Q)$ which are $\Leg_o$-related to $X_{h_o}$, and are solutions to
equation \eqref{Chap02_eqn:LagHODynEqSingular}, where $S_f = \Leg_o^{-1}(\P_f)$.
\end{enumerate}
\end{theorem}

Observe that the vector fields $X_\Lag \in \vf(\Tan^{2k-1}Q)$ which are
$\Leg_o$-related to $X_{h_o}$ may not be semisprays of type $1$, since this condition
can not be assured in the singular case.
These $\Leg_o$-projectable semisprays of type $1$ could be defined only on the
points of another submanifold $S^h_f\hookrightarrow S_f$.
(See \cite{art:Gracia_Pons_Roman91,art:Gracia_Pons_Roman92} for a detailed exposition of all these topics).


\section{First-order non-autonomous dynamical systems}
\label{Chap02_sec:NonAutonomous}

Let us consider a first-order non-autonomous Lagrangian dynamical system
with $n$ degrees of freedom. The configuration space for this system
is a bundle $\pi \colon E \to \R$, with $\dim E = n+1$. The dynamical
information is given in terms of a Lagrangian density $\Lag \in \df^{1}(J^1\pi)$,
which is a $\bar{\pi}^1$-semibasic $1$-form. Because of this, we usually write
$\Lag = L\cdot(\bar{\pi}^1)^*\eta$, where $\eta \in \df^{1}(\R)$ is the canonical
volume form in $\R$ and $L \in \Cinfty(J^1\pi)$ is the Lagrangian function
associated to $\Lag$ and $\eta$.

\subsection{Lagrangian formalism}
\label{Chap02_sec:LagrangianNonAutonomousFirstOrder}

(See \cite{art:Cappelletti_deNicola_Yudin,art:Chinea_deLeon_Marrero94,art:deLeon_Marin_Marrero96,
art:DeLeon_Marin_Marrero_Munoz_Roman02,art:Echeverria_Munoz_Roman91,book:Mangiarotti_Sardanashvily98} for details).

\subsubsection{Geometric and dynamical structures}

From the Lagrangian density $\Lag$ and the vertical endomorphism
$\V \in \Gamma(\Tan^*J^1\pi \otimes_{J^1\pi} \Tan M \otimes_{J^1\pi} V(\pi^1))$
of the jet bundle $J^1\pi$, we construct the following structures.

\begin{definition}
The \textnormal{Poincar\'{e}-Cartan $1$-form} associated to $\Lag \in \df^{1}(J^1\pi)$ and $\eta \in \df^{1}(\R)$
is the $1$-form $\Theta_\Lag \in \df^{1}(J^1\pi)$ defined as
\begin{equation*}
\Theta_\Lag = \inn(\V)\d\Lag + \Lag \, .
\end{equation*}
From this, the \textnormal{Poincar\'{e}-Cartan $2$-form} associated to $\Lag$ and $\eta$ is the $2$-form
$\Omega_\Lag \in \df^{2}(J^1\pi)$ defined as
\begin{equation*}
\Omega_\Lag = -\d\Theta_\Lag \, .
\end{equation*}
\end{definition}

As in the autonomous setting, given an arbitrary Lagrangian density $\Lag \in \df^{1}(J^1\pi)$,
the Poincar\'{e}-Cartan $2$-form may not have constant rank at every point in $J^1\pi$.
If $\rank(\Omega_\Lag(j^1_t\phi)) = \text{const.}$ for every $j^1_t\phi \in J^1\pi$,
then the Lagrangian density $\Lag$ is said to be a \textsl{geometrically admissible Lagrangian}.
We will only consider Lagrangian densities satisfying this property.

It is clear from the previous definitions that the phase space of a first-order
non-autonomous Lagrangian dynamical system is the first-order jet bundle of the configuration
bundle $\pi \colon E \to \R$.

\begin{definition}
A \textnormal{first-order non-autonomous Lagrangian system} is a pair $(J^1\pi,\Lag)$,
where $(E,\pi,\R)$ is the configuration bundle, and $\Lag \in \df^{1}(J^1\pi)$ the Lagrangian density.
\end{definition}

In the natural coordinates $(t,q^A,v^A)$ of $J^1\pi$,
bearing in mind the local expression \eqref{Chap01_eqn:HOJetBundleVertEnd}
of the vertical endomorphism $\V$, the coordinate expression of the Poincar\'{e}-Cartan $1$-form is
\begin{equation}\label{Chap02_eqn:LagNAFO1FormLocal}
\Theta_\Lag
= \derpar{L}{v^A}\d q^A - \left( v^A \derpar{L}{v^A} - L \right) \d t \, ,
\end{equation}
from which the coordinate expression of the Poincar\'{e}-Cartan $2$-form is
\begin{equation}\label{Chap02_eqn:LagNAFO2FormLocal}
\begin{array}{l}
\displaystyle \Omega_\Lag = \derpars{L}{v^A}{q^B}\d q^A \wedge \d q^B + \derpars{L}{v^A}{v^B} \d q^A \wedge \d v^B \\[12pt]
\displaystyle \qquad\qquad + \left( v^A\derpars{L}{v^A}{q^B} - \derpar{L}{q^B} \right) \d q^B \wedge \d t + v^A\derpars{L}{v^A}{v^B} \d v^B \wedge \d t \, .
\end{array}
\end{equation}

\begin{remark}
As in the autonomous formulation, it is clear from the coordinate expression
\eqref{Chap02_eqn:LagNAFO1FormLocal} that the Poincar\'{e}-Cartan $1$-form $\Theta_\Lag$
is $\pi^1$-semibasic.
\end{remark}

Notice that, given an arbitrary Lagrangian density $\Lag \in \df^{1}(J^1\pi)$,
the Poincar\'{e}-Cartan $2$-form is always a closed form.

\begin{definition}\label{Chap02_def:LagNAFORegularLagrangian}
A Lagrangian density $\Lag \in \df^{1}(J^1\pi)$ is \textnormal{regular}
(and thus $(J^1\pi,\Lag)$ is a \textnormal{regular system}) if the pair
$(\Omega_\Lag,(\bar{\pi}^1)^*\eta)$ is a cosymplectic structure on $J^1\pi$.
Otherwise, the Lagrangian density is said to be \textnormal{singular}
(and thus $(J^1\pi,\Lag)$ is a \textnormal{singular system}).
\end{definition}

Bearing in mind the coordinate expression \eqref{Chap02_eqn:LagNAFO2FormLocal}
of the Poincar\'{e}-Cartan $2$-form, it is clear that the regularity condition
for the Lagrangian density $\Lag$ is locally equivalent to
\begin{equation*}
\det\left( \derpars{L}{v^A}{v^B} \right) (j^1_t\phi) \neq 0 \ , \,
\mbox{for every } j^1_t\phi \in J^1\pi \, ,
\end{equation*}
where $L \in \Cinfty(J^1\pi)$ is the Lagrangian function associated to $\Lag$ and $\eta$.
That is, a Lagrangian density is regular if, and only it, the Hessian matrix of its associated
Lagrangian function with respect to the velocities is invertible at every point of $J^1\pi$.
Observe also that this condition is equivalent to require $\Omega_\Lag$ to have maximal
rank $2n$ in $J^1\pi$.

\subsubsection{Dynamical equations for sections}

The \textsl{first-order Lagrangian problem for sections} associated with the system
$(J^1\pi,\Lag)$ consists in finding sections $\phi \in \Gamma(\pi)$ characterized
by the condition
\begin{equation}\label{Chap02_eqn:LagNAFODynEqSect}
(j^1\phi)^*\inn(X)\Omega_\Lag = 0 \, , \ \mbox{for every } X \in \vf(J^1\pi) \, ,
\end{equation}
where $j^1\phi \in \Gamma(\bar{\pi}^1)$ is the first prolongation of $\phi$ to $J^1\pi$.

In the natural coordinates of $J^1\pi$, the section $\phi \in \Gamma(\pi)$ must
satisfy the following system $n$ second-order differential equations
\begin{equation*}
\restric{\derpar{L}{q^A}}{j^1\phi} - \restric{\frac{d}{dt}\derpar{L}{v^A}}{j^1\phi} = 0 \, .
\end{equation*}
These equations are the \textsl{non-autonomous Euler-Lagrange equations}.

\subsubsection{Dynamical equations for vector fields}

If we assume that the first prolongations of the sections $\phi \in \Gamma(\pi)$
which are solutions to the equation \eqref{Chap02_eqn:LagNAFODynEqSect} are the
integral curves of some vector fields in $J^1\pi$, then we can state the problem
in terms of vector fields. The \textsl{first-order Lagrangian problem for vector fields}
consists in finding holonomic vector fields $X_\Lag \in \vf(J^1\pi)$ satisfying the
following equations
\begin{equation}\label{Chap02_eqn:LagNAFODynEqVF}
\inn(X_\Lag)\Omega_\Lag = 0 \quad ; \quad \inn(X_\Lag)(\bar{\pi}^1)^*\eta \neq 0 \, .
\end{equation}

\begin{remark}
The second equation in \eqref{Chap02_eqn:LagNAFODynEqVF} is just a transverse condition
for the vector field $X_\Lag$ with respect to the projection onto the base manifold. This equation
is usually considered with a fixed non-zero value, which is equivalent in physics to
fixing the Gauge of the system. In most cases we take $1$, which gives the
following equations
\begin{equation}\label{Chap02_eqn:LagNAFODynEqVFGaugeOne}
\inn(X_\Lag)\Omega_\Lag = 0 \quad ; \quad \inn(X_\Lag)(\bar{\pi}^1)^*\eta = 1 \, .
\end{equation}
Observe that, in this case, $X_\Lag$ is nothing but the Reeb vector field of the
(pre)cosymplectic structure $(\Omega_\Lag,(\bar{\pi}^1)^*\eta)$ (see Section
\ref{Chap01_sec:CosymplecticGeomReebVF}).
\end{remark}

In the natural coordinates $(t,q^A,v^A)$ of $J^1\pi$, let $X_\Lag \in \vf(J^1\pi)$
be a generic vector field given by
\begin{equation*}
X_\Lag = f\derpar{}{t} + f^A \derpar{}{q^A} + F^A\derpar{}{v^A} \, .
\end{equation*}
Then, bearing in mind the coordinate expression \eqref{Chap02_eqn:LagNAFO2FormLocal}
of the Poincar\'{e}-Cartan $2$-form, and requiring equations \eqref{Chap02_eqn:LagNAFODynEqVF}
to hold, we obtain the following system of $(2n+1)$ equations
\begin{align}
F^A\derpars{L}{v^A}{v^B} =  - \derpars{L}{t}{v^B} + f^A \left( \derpars{L}{v^A}{q^B} - \derpars{L}{v^B}{q^A} \right)
- f\left( v^A\derpars{L}{v^A}{q^B} - \derpar{L}{q^B}\right) \, , \label{Chap02_eqn:LagNAFODynEqLocal1} \\
(f^A - fv^A) \derpars{L}{v^A}{v^B} = 0 \, , \label{Chap02_eqn:LagNAFODynEqHolonomyLocal1} \\
f \neq 0 \, , \label{Chap02_eqn:LagNAFODynEqLocalGaugeNonZero}
\end{align}
where equation \eqref{Chap02_eqn:LagNAFODynEqLocalGaugeNonZero} arises from the second equation
in \eqref{Chap02_eqn:LagNAFODynEqVF}. The extra equation alongside the form $\d t$ has been omitted,
since it is a combination of the others and gives no additional information.
Fixing the non-zero value to $1$ by the equation \eqref{Chap02_eqn:LagNAFODynEqVFGaugeOne}, we obtain the system
\begin{align}
F^A\derpars{L}{v^A}{v^B} =  - \derpars{\hat{L}}{t}{v^B} + f^A \left( \derpars{L}{v^A}{q^B} - \derpars{L}{v^B}{q^A} \right)
- v^A\derpars{L}{v^A}{q^B} + \derpar{L}{q^B} \, , \label{Chap02_eqn:LagNAFODynEqLocal2} \\
(f^A - v^A) \derpars{L}{v^A}{v^B} = 0 \, , \label{Chap02_eqn:LagNAFODynEqHolonomyLocal2} \\
f = 1 \, . \label{Chap02_eqn:LagNAFODynEqLocalGaugeOne}
\end{align}
Observe that equations \eqref{Chap02_eqn:LagNAFODynEqLocal2} and \eqref{Chap02_eqn:LagNAFODynEqHolonomyLocal2}
are exactly equations \eqref{Chap02_eqn:LagAFODynEqLocal} and \eqref{Chap02_eqn:LagAFODynEqHolonomyLocal}
in the autonomous setting, and the same comments apply in this case. In particular, equations
\eqref{Chap02_eqn:LagNAFODynEqHolonomyLocal2} are the local equations for the holonomy condition
required to the vector field $X_\Lag$, while equations \eqref{Chap02_eqn:LagNAFODynEqLocal2}
are the dynamical equations. Notice that, as in the autonomous case, if the holonomy condition
is required from the beginning, then equations \eqref{Chap02_eqn:LagNAFODynEqHolonomyLocal2}
are an identity, and equations \eqref{Chap02_eqn:LagNAFODynEqLocal2} reduce to
\begin{equation}\label{Chap02_eqn:LagNAFODynEqWithHolonomyLocal}
F^A\derpars{L}{v^A}{v^B} = \derpar{L}{q^B}  - \derpars{L}{t}{v^B} - v^A\derpars{L}{q^A}{v^B} \, .
\end{equation}
Note that in equations \eqref{Chap02_eqn:LagNAFODynEqLocal2} and
\eqref{Chap02_eqn:LagNAFODynEqHolonomyLocal2} the Hessian of the Lagrangian function $L$
associated to the Lagrangian density $\Lag$ and the volume form $\eta$ appears alongside
the coefficients to be determined. Therefore, we have the following result.

\begin{proposition}
If the Lagrangian density $\Lag \in \df^{1}(J^1\pi)$ is regular, then there exists a unique
vector field $X_\Lag$ solution to the equations \eqref{Chap02_eqn:LagNAFODynEqVFGaugeOne} which,
in addition, is holonomic.
\end{proposition}

If the Lagrangian density is not regular, then the pair $(\Omega_\Lag,(\bar{\pi}^1)^*\eta)$
is just a precosymplectic structure on $J^1\pi$, and so the existence of solutions to equations
\eqref{Chap02_eqn:LagNAFODynEqVFGaugeOne} (or \eqref{Chap02_eqn:LagNAFODynEqVF}) can not be assured
in the general case. Hence, an adapted version of the constraint algorithm described in Section
\ref{Chap01_sec:ConstraintAlgorithm} for time-dependent Lagrangian systems or jet bundle formulations must be used
(see \cite{art:Chinea_deLeon_Marrero94,art:deLeon_Marin_Marrero96,art:DeLeon_Marin_Marrero_Munoz_Roman02}),
and, in the most favorable cases, there exists a submanifold $S_f \hookrightarrow J^1\pi$ where the equations
\begin{equation}\label{Chap02_eqn:LagNAFODynEqSingular}
\restric{[\inn(X_\Lag) \Omega_\Lag]}{S_f} = 0 \quad ; \quad
\restric{[\inn(X_\Lag) (\bar{\pi}^1)^*\eta - 1]}{S_f} = 0 \, ,
\end{equation}
admit a well-defined solution $X_\Lag$, which is tangent to $S_f$. Nevertheless,
these vector fields solution are not necessarily holonomic on $S_f$, but only in
the points of another submanifold $S_f^h \hookrightarrow S_f$.

\begin{remark}
Notice that the second equation in \eqref{Chap02_eqn:LagNAFODynEqSingular} is
redundant, since a vector field which is $\bar{\pi}^1$-transverse in $J^1\pi$
is also $\bar{\pi}^1$-transverse in every submanifold of $J^1\pi$.
\end{remark}

\subsection{Extended Hamiltonian formalism}
\label{Chap02_sec:NonAutonomousHamiltonian}

In the extended Hamiltonian formalism associated to a Lagrangian system
$(J^1\pi,\Lag)$, two phase spaces are considered: the
\textsl{extended momentum bundle} and the \textsl{restricted momentum bundle}.
The former is exactly the extended dual jet bundle $\Lambda^1_{2}(J^0\pi)$
of $J^1\pi$ described in Section \ref{Chap01_sec:HOJetBundlesDualBundles},
which in this case reduces simply to $\Tan^*E$, while the latter is the
reduced dual jet bundle $\Lambda^1_{2}(J^0\pi)/\Lambda^1_{1}(J^0\pi)$.
To avoid confusion with the notation, we denote the restricted momentum bundle
by $J^0\pi^*$, instead of $E^*$. The quotient map is denoted by
$\mu \colon \Tan^*E \to J^0\pi^*$.
Natural coordinates in $\Tan^*E$ are $(t,q^A,p,p_A)$,
and the induced natural coordinates in $J^0\pi^*$ are $(t,q^A,p_A)$.

(See \cite{art:Chinea_DeLeon_Marrero91,art:DeLeon_Marrero_Martin96,
art:Echeverria_Munoz_Roman91,book:Mangiarotti_Sardanashvily98,
art:Ranada92} for details).

\subsubsection{The extended and restricted Legendre maps}

As in the autonomous setting described in Section
\ref{Chap02_sec:HamiltonianAutonomousFirstOrder}, we begin by introducing
the Legendre map. Since the Poincar\'{e}-Cartan $1$-form $\Theta_\Lag \in \df^{1}(J^1\pi)$
is $\pi^1$-semibasic, we can give the following definition.

\begin{definition}
The \textnormal{extended Legendre map} associated to the Lagrangian density $\Lag \in \df^{1}(J^1\pi)$
is the bundle morphism $\widetilde{\Leg} \colon J^1\pi \to \Tan^*E$ over $E$ defined as follows:
for every $u \in \Tan J^1\pi$,
\begin{equation*}
\Theta_\Lag(u) = \left\langle \Tan\pi^1(u) \, , \, \widetilde{\Leg}(\tau_{J^1\pi}(u)) \right\rangle \, ,
\end{equation*}
where $\tau_{J^1\pi} \colon \Tan(J^1\pi) \to J^1\pi$ is the canonical submersion.
\end{definition}

It is clear from the definition that $\pi_E \circ \widetilde{\Leg} = \pi^1$,
where $\pi_E \colon \Tan^*E \to E$ is the canonical submersion.
Furthermore, let $\Theta \in \df^{1}(\Tan^*E)$ be the tautological form
of $\Tan^*E$, and $\Omega = -\d\Theta \in \df^{2}(\Tan^*E)$
the canonical symplectic form. Then, we have $\widetilde{\Leg}^*\Theta = \Theta_\Lag$
and $\widetilde{\Leg}^*\Omega = \Omega_\Lag$.
From Example \ref{Chap01_exa:CotangentBundle},
the coordinate expression of $\Theta$ in this case is
\begin{equation*}
\Theta = p_A\d q^A + p\d t \, .
\end{equation*}
Thus, the canonical symplectic form $\Omega$ in $\Tan^*E$ is given in
coordinates by
\begin{equation}\label{Chap02_eqn:HamNAFOCanonicalSymplecticForm}
\Omega = \d q^A \wedge \d p_A + \d t \wedge \d p \, .
\end{equation}
Then, bearing in mind the coordinate expression \eqref{Chap02_eqn:LagNAFO1FormLocal}
of the Poincar\'{e}-Cartan $1$-form $\Theta_\Lag \in \df^{1}(J^1\pi)$, the
coordinate expression of the extended Legendre map is
\begin{equation}\label{Chap02_eqn:HamNAFOExtendedLegendreMapLocal}
\widetilde{\Leg}^*t = t \quad ; \quad \widetilde{\Leg}^*q^A = q^A \quad ; \quad
\widetilde{\Leg}^*p_A = \derpar{L}{v^A} \quad ; \quad
\widetilde{\Leg}^*p = L - v^A \derpar{L}{v^A} \, .
\end{equation}

Now, composing the extended Legendre map with the quotient map $\mu \colon \Tan^*E \to J^0\pi^*$,
we can give the following definition.

\begin{definition}
The \textnormal{restricted Legendre map} associated to the Lagrangian density $\Lag \in \df^{1}(J^1\pi)$
is the map $\Leg \colon J^1\pi \to J^0\pi^*$ defined as $\Leg = \mu \circ \widetilde{\Leg}$.
\end{definition}

In the natural coordinates of $J^0\pi^*$, the coordinate expression of the
restricted Legendre map is
\begin{equation*}
\Leg^*t = t \quad ; \quad \Leg^*q^A = q^A \quad ; \quad
\Leg^*p_A = \derpar{L}{v^A} \, .
\end{equation*}

A fundamental result relating both Legendre maps is the following.

\begin{proposition}\label{Chap02_prop:HamNAFOLegendreMapsEqualRank}
For every $j^1_t\phi \in J^1\pi$ we have that
$\rank(\widetilde{\Leg}(j^1_t\phi)) = \rank(\Leg(j^1_t\phi))$.
\end{proposition}

We refer to \cite{art:deLeon_Marin_Marrero96} for the proof of this result.
As a consequence of Proposition \ref{Chap02_prop:HamNAFOLegendreMapsEqualRank},
and bearing in mind the coordinate expressions of both Legendre maps
and the results stated in Section \ref{Chap02_sec:LagrangianNonAutonomousFirstOrder},
we have the following result.

\begin{proposition}
Let $\Lag \in \df^{1}(J^1\pi)$ be a Lagrangian density. The following
statements are equivalent:
\begin{enumerate}
\item $\Omega_\Lag$ has maximal rank $2n$ on $J^1\pi$.
\item The pair $(\Omega_\Lag,(\bar{\pi}^1)^*\eta)$ is a cosymplectic
structure on $J^1\pi$.
\item In the natural coordinates of $J^1\pi$, we have
\begin{equation*}
\det\left( \derpars{L}{v^B}{v^A} \right) (j^1_t\phi) \neq 0 \, ,
\end{equation*}
for every $j^1_t\phi \in J^1\pi$, where $L \in \Cinfty(J^1\pi)$
is the Lagrangian function associated with $\Lag$ and $\eta$.
\item The restricted Legendre map $\Leg \colon J^1\pi \to J^0\pi^*$ is a local
diffeomorphism.
\item The extended Legendre map $\widetilde{\Leg} \colon J^1\pi \to \Tan^*E$
is an immersion.
\end{enumerate}
In this case, $\Lag$ is a \textnormal{regular} Lagrangian density.
\end{proposition}

\begin{definition}
A Lagrangian density $\Lag \in \df^{1}(J^1\pi)$ is \textnormal{hyperregular} if the restricted
Legendre map $\Leg \colon J^1\pi \to J^0\pi^*$ is a global diffeomorphism.
\end{definition}

Now, let $\widetilde{\P} = \Im(\widetilde{\Leg}) \hookrightarrow \Tan^*E$ be the image of
the extended Legendre map, with natural embedding
$\tilde{\jmath} \colon \widetilde{\P} \hookrightarrow \Tan^*E$,
and $\P = \Im(\Leg) \hookrightarrow J^0\pi^*$ the image of the restricted Legendre map,
with canonical embedding $\jmath \colon \P \hookrightarrow J^0\pi^*$.
Let $\bar{\pi}_\P = \bar{\pi}_{E}^r \circ \jmath \colon \P \to \R$ be the canonical projection,
and $\Leg_o \colon J^1\pi \to \P$ the map defined by $\Leg = \jmath \circ \Leg_o$.
We can now give the following definition.

\begin{definition}
A Lagrangian density $\Lag \in \df^{1}(J^1\pi)$ is \textnormal{almost-regular} if
\begin{enumerate}
\item $\P$ is a closed submanifold of $J^0\pi^*$.
\item $\Leg$ is a submersion onto its image.
\item For every $j^1_t\phi \in J^1\pi$, the fibers $\Leg^{-1}(\Leg(j^1_t\phi))$
are connected submanifolds of $J^1\pi$.
\end{enumerate}
\end{definition}

Observe that, as a consequence of Proposition \ref{Chap02_prop:HamNAFOLegendreMapsEqualRank},
we have that $\widetilde{\P}$ is diffeomorphic to $\P$. This diffeomorphism is just $\mu$ restricted
on the image set $\widetilde{\P}$, and we denote it by $\widetilde{\mu}$. Then, we have the following
definition.

\begin{definition}
The \textnormal{canonical Hamiltonian section} $h \in \Gamma(\widetilde{\mu})$ is
defined as the map $h = \widetilde{\mu}^{-1} \colon \P \to \widetilde{\P}$.
\end{definition}

\begin{remark}
Observe that the Hamiltonian section $h \in \Gamma(\widetilde{\mu})$ depends only
on the Lagrangian density $\Lag \in \df^{1}(J^1\pi)$, since both $\widetilde{\P}$
and $\P$ depend only on the Legendre maps and, more particularly, on the Lagrangian density.
\end{remark}

From the Hamiltonian section $h \in \Gamma(\widetilde{\mu})$ we can define the following forms
on $\P$.

\begin{definition}
The \textnormal{Hamilton-Cartan forms} are the forms $\Theta_h \in \df^{1}(\P)$
and $\Omega_h \in \df^{2}(\P)$ defined as
\begin{equation*}
\Theta_h = (\tilde{\jmath} \circ h)^*\Theta \quad ; \quad
\Omega_h = (\tilde{\jmath} \circ h)^*\Omega = -\d\Theta_h \, ,
\end{equation*}
where $\Theta$ and $\Omega$ are the canonical Liouville forms of the cotangent bundle $\Tan^*E$.
\end{definition}

The triple $(\P,\Omega_h,\bar{\pi}_\P^*\eta)$ is called the \textsl{Hamiltonian system}
associated with the Lagrangian system $(J^1\pi,\Lag)$.

\subsubsection{Regular and hyperregular Lagrangian densities}

Suppose that the Lagrangian density $\Lag \in \df^{1}(J^1\pi)$ is hyperregular, since
the regular case can be recovered from this by restriction on the open sets where
the restricted Legendre map is a local diffeomorphism.

In the hyperregular case we have $\P = J^0\pi^*$, and that $\widetilde{\P}$
is a $1$-codimensional and $\mu$-transverse submanifold of $\Tan^*E$ which
is diffeomorphic to $J^0\pi^*$. In addition, in this case the Hamiltonian
section may be defined equivalently as $h = \widetilde{\Leg} \circ \Leg^{-1}$.

In the natural coordinates of $J^0\pi^*$, the Hamiltonian section is specified
by a Hamiltonian function $H \in \Cinfty(J^0\pi^*)$ as
\begin{equation*}
h(t,q^A,p_A) = (t,q^A,-H(t,q^A,p_A),p_A) \, ,
\end{equation*}
with the Hamiltonian function $H$ being locally given by
\begin{equation*}
H(t,q^A,p_A) = p_A(\Leg^{-1})^*v^A - (\Leg^{-1})^*L(t,q^A,v^A) \, ,
\end{equation*}
where $(t,q^A,v^A)$ are the natural coordinates in $J^1\pi$.
From this, and bearing in mind the coordinate expressions of the
canonical Liouville forms of the cotangent bundle given in Example
\ref{Chap01_exa:CotangentBundle}, the local expressions of the
Hamilton-Cartan forms are
\begin{equation}\label{Chap02_eqn:HamNAFOHamiltonCartanFormsLocal}
\Theta_h = p_A\d q^A - H\d t \quad ; \quad
\Omega_h = \d q^A \wedge \d p_A + \d H \wedge \d t \, .
\end{equation}

Then, the \textsl{first-order Hamiltonian problem for sections} associated with
the system $(J^0\pi^*,\Omega_h,(\bar{\pi}_E^r)^*\eta)$ consists in finding sections
$\psi \in \Gamma(\bar{\pi}_{E}^r)$ satisfying the equation
\begin{equation}\label{Chap02_eqn:HamNAFODynEqSectRegular}
\psi^*\inn(X)\Omega_h = 0 \, , \ \mbox{for every } X \in \vf(J^0\pi^*) \, .
\end{equation}

In the natural coordinates of $J^0\pi^*$, the section $\psi(t) = (t,q^A(t),p_A(t))
\in \Gamma(\bar{\pi}_{E}^{r})$
must satisfy the following system of $2n$ first-order differential equations
\begin{equation*}
\dot{q}^A = \restric{\derpar{H}{p_A}}{\psi} \quad ; \quad
\dot{p}_A = -\restric{\derpar{H}{q^A}}{\psi} \, .
\end{equation*}
These are the \textsl{non-autonomous Hamilton equations}.

Now, if we assume that the sections $\psi \in \Gamma(\bar{\pi}_E^r)$ solution to the
equation \eqref{Chap02_eqn:HamNAFODynEqSectRegular} are the integral curves of some
vector fields in $J^0\pi^*$, we can state the problem in terms of vector fields.
The \textsl{first-order Hamiltonian problem for vector fields} consists in finding
vector fields $X_h \in \vf(J^0\pi^*)$ satisfying the equations
\begin{equation}\label{Chap02_eqn:HamNAFODynEqVFRegular}
\inn(X_h) \Omega_h = 0 \quad ; \quad
\inn(X_h)(\bar{\pi}_E^r)^*\eta \neq 0 \, .
\end{equation}

\begin{remark}
As in the Lagrangian formalism, the second equation in
\eqref{Chap02_eqn:HamNAFODynEqVFRegular} is just a transverse condition
for the vector field $X_h$ with respect to the projection onto $\R$,
and it is usual to take the non-zero value equal to $1$, thus giving
the following equations for $X_h$
\begin{equation}\label{Chap02_eqn:HamRegNAFODynEqVFGaugeOne}
\inn(X_h) \Omega_h = 0 \quad ; \quad
\inn(X_h)(\bar{\pi}_E^r)^*\eta = 1 \, .
\end{equation}
In this case, $X_h$ is the Reeb vector field of the cosymplectic
structure $(\Omega_h,(\bar{\pi}_E^r)^*\eta)$.
\end{remark}

In the natural coordinates $(t,q^A,p_A)$ of $J^0\pi^*$, let $X_h$ in $\vf(J^0\pi^*)$
be a generic vector field given by
\begin{equation*}
X_h = f\derpar{}{t} + f^A \derpar{}{q^A} + G_A\derpar{}{p_A} \, .
\end{equation*}
Now, bearing in mind the coordinate expression \eqref{Chap02_eqn:HamNAFOHamiltonCartanFormsLocal}
of the Hamilton-Cartan $2$-form, and requiring equations
\eqref{Chap02_eqn:HamNAFODynEqVFRegular} to hold, we obtain the following system
of $(2n+1)$ equations
\begin{align}
f^A = f\derpar{H}{p_A} \quad ; \quad G_A = - f\derpar{H}{q^A} \, , \label{Chap02_eqn:HamNAFODynEqLocal1} \\
f \neq 0 \, , \label{Chap02_eqn:HamNAFODynEqLocalGaugeNonZero}
\end{align}
where equation \eqref{Chap02_eqn:HamNAFODynEqLocalGaugeNonZero} arises from the second equation
in \eqref{Chap02_eqn:HamNAFODynEqVFRegular}. As in the Lagrangian formalism, the extra equation
alongside the form $\d t$ is a combination from the others, and we omit it. Fixing
the non-zero value of $f$ to $1$ by the second equation in \eqref{Chap02_eqn:HamRegNAFODynEqVFGaugeOne},
we obtain the system
\begin{align}
f^A = \derpar{H}{p_A} \quad ; \quad G_A = - \derpar{H}{q^A} \, , \label{Chap02_eqn:HamNAFODynEqLocal2} \\
f = 1 \, , \label{Chap02_eqn:HamNAFODynEqLocalGaugeOne}
\end{align}
Observe that equations \eqref{Chap02_eqn:HamNAFODynEqLocal2} are exactly equations
\eqref{Chap02_eqn:HamAFODynEqLocal} in the autonomous setting.

Finally, we establish the a one-to-one correspondence between the vector fields solution
to equations \eqref{Chap02_eqn:LagNAFODynEqVF}
and the vector fields solution to equations \eqref{Chap02_eqn:HamNAFODynEqVFRegular}.

\begin{theorem}\label{Chap02_thm:HamNAFORelationLagHamRegular}
Let $\Lag \in \df^{1}(J^1\pi)$ be a hyperregular Lagrangian density.
Then we have:
\begin{enumerate}
\item Let $X_\Lag \in \vf(J^1\pi)$ be the unique holonomic vector field solution
to equations \eqref{Chap02_eqn:LagNAFODynEqVF}. Then the vector field
$X_h = \Leg_*X_\Lag \in \vf(J^0\pi^*)$ is a solution to equations \eqref{Chap02_eqn:HamNAFODynEqVFRegular}.

\item Conversely, let $X_h \in \vf(J^0\pi^*)$ be the unique vector field solution
to equations \eqref{Chap02_eqn:HamNAFODynEqVFRegular}. Then the vector field
$X_\Lag = (\Leg^{-1})_*X_h \in \vf(J^1\pi)$ is holonomic, and is a solution to
equations \eqref{Chap02_eqn:LagNAFODynEqVF}.
\end{enumerate}
\end{theorem}

\begin{remark}
There is an analogous result to Theorem \ref{Chap02_thm:HamNAFORelationLagHamRegular}
for sections, which is a straightforward consequence of this last theorem bearing in mind that the
sections solution to equations \eqref{Chap02_eqn:LagNAFODynEqSect} (respectively, to equations
\eqref{Chap02_eqn:HamNAFODynEqSectRegular}) are integral curves of the vector fields
solution to equations \eqref{Chap02_eqn:LagNAFODynEqVF} (respectively, to equations
\eqref{Chap02_eqn:HamNAFODynEqVFRegular}).
\end{remark}

\subsubsection{Singular (almost-regular) Lagrangian densities}

For almost-regular Lagrangian densities, the restricted Legendre map is no longer a (local) diffeomorphism,
and therefore the image of $\Leg$ is a proper submanifold of
$J^0\pi^*$. Nevertheless, we can still state a Hamiltonian formulation for an almost-regular Lagrangian
system $(J^1\pi,\Lag)$.

The \textsl{first-order Hamiltonian problem for sections} in this case consists
in finding sections $\psi_o \in \Gamma(\bar{\pi}_\P)$ characterized by the condition
\begin{equation*}
\psi_o^*\inn(X_o)\Omega_h = 0 \, , \, \mbox{for every } X_o \in \vf(\P) \, .
\end{equation*}

On the other hand, the \textsl{first-order Hamiltonian problem for vector fields}
consists in finding vector fields $X_h \in \vf(\P)$ satisfying the equations
\begin{equation}\label{Chap02_eqn:HamNAFODynEqVFSingular1}
\inn(X_h) \Omega_h = 0 \quad ; \quad
\inn(X_h)\bar{\pi}_\P^*\eta \neq 0 \, .
\end{equation}

Since the pair $(\Omega_h,\bar{\pi}_\P)$ is, in general, a precosymplectic structure in
$\P$, we must apply an adaptation of the constraint algorithm given in Section
\ref{Chap01_sec:ConstraintAlgorithm} for precosymplectic structures. In the most
favorable cases, equations \eqref{Chap02_eqn:HamNAFODynEqVFSingular1} admit a solution
only on the points of some submanifold $\P_f \hookrightarrow \P$, and is tangent to
it. In this case, the following equations hold
\begin{equation}\label{Chap02_eqn:HamNAFODynEqVFSingular2}
\restric{[\inn(X_h) \Omega_h]}{\P_f} = 0 \quad ; \quad
\restric{[\inn(X_h)\bar{\pi}_\P^*\eta]}{\P_f} \neq 0 \, .
\end{equation}
Note that this vector field is not unique, in general.

As in the autonomous setting, we have an analogous result to Theorem
\ref{Chap02_thm:HamNAFORelationLagHamRegular}.

\begin{theorem}\label{Chap02_thm:HamNAFORelationLagHamSingular}
Let $\Lag \in \vf(J^1\pi)$ be an almost-regular Lagrangian density.
Then we have:
\begin{enumerate}
\item Let $X_\Lag \in \vf(J^1\pi)$ be a holonomic vector field solution
to equations \eqref{Chap02_eqn:LagNAFODynEqSingular} in the points of a submanifold
$S_f \hookrightarrow J^1\pi$. Then there exists a vector field
$X_{h} \in \vf(\P)$ which is $\Leg_o$-related to $X_\Lag$
and is a solution to equations \eqref{Chap02_eqn:HamNAFODynEqVFSingular2},
where $\P_f = \Leg_o(S_f) \hookrightarrow \P$.

\item Conversely, let $X_{h} \in \vf(\P)$ be a vector field solution
to equations \eqref{Chap02_eqn:HamNAFODynEqVFSingular2} on the points of some
submanifold $\P_f \hookrightarrow \P$. Then there exist vector fields
$X_\Lag \in \vf(J^1\pi)$ which are $\Leg_o$-related to $X_{h}$, and are solutions to
equations \eqref{Chap02_eqn:LagNAFODynEqSingular}, where $S_f = \Leg^{-1}(\P_f)$.
\end{enumerate}
\end{theorem}

Notice that the vector fields $X_\Lag \in \vf(J^1\pi)$ which are
$\Leg_o$-related to $X_{h}$ are not necessarily holonomic, since this condition
can not be assured in the singular case. These $\Leg_o$-projectable holonomic
vector fields could be defined only on the points of another submanifold $S^h_f\hookrightarrow S_f$.

\subsection{Lagrangian-Hamiltonian unified formalism}
\label{Chap02_sec:NonAutonomousUnified}

(See \cite{art:Barbero_Echeverria_Martin_Munoz_Roman08,art:Cortes_Martinez_Cantrijn02} for details).

\subsubsection{Unified phase spaces. Geometric and dynamical structures}

As in the extended Hamiltonian formalism stated in the previous Section,
we consider two phase spaces in this formulation, which are the bundles
\begin{equation*}
\W = J^1\pi \times_E \Tan^*E \quad ; \quad \W_r = J^1\pi \times_E J^0\pi^* \, ,
\end{equation*}
known as the \textsl{extended jet-momentum bundle} and the \textsl{restricted jet-momentum bundle},
respectively. These bundles are endowed with the canonical projections
\begin{equation*}
\rho_1 \colon \W \to J^1\pi \quad ; \quad \rho_2 \colon \W \to \Tan^*E \quad ; \quad
\rho_E \colon \W \to E \quad ; \quad \rho_\R \colon \W \to \R \, ,
\end{equation*}
\begin{equation*}
\rho_1^r \colon \W_r \to J^1\pi \quad ; \quad \rho_2^r \colon \W_r \to J^0\pi^* \quad ; \quad
\rho_E^r \colon \W_r \to E \quad ; \quad \rho_\R^r \colon \W_r \to \R \, .
\end{equation*}

In addition, the natural quotient map $\mu \colon \Tan^*E \to J^0\pi^*$ induces a surjective submersion
$\mu_\W \colon \W \to \W_r$. Hence, we have the following diagram
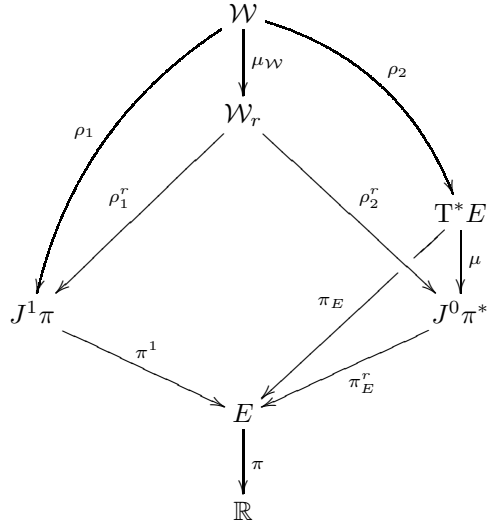
\begin{equation}\label{Chap02_fig:UnifNABundleDiagram1}
\xymatrix{
\ & \ & \W \ar@/_1.3pc/[llddd]_{\rho_1} \ar[d]^-{\mu_\W} \ar@/^1.3pc/[rrdd]^{\rho_2} & \ & \ \\
\ & \ & \W_r \ar[lldd]_{\rho_1^r} \ar[rrdd]^{\rho_2^r} & \ & \ \\
\ & \ & \ & \ & \Tan^*E \ar[d]^-{\mu} \ar[lldd]_{\pi_E}|(.25){\hole} \\
J^{1}\pi \ar[rrd]^{\pi^{1}} & \ & \ & \ & J^0\pi^* \ar[dll]^{\pi_{E}^r} \\
\ & \ & E \ar[d]^{\pi} & \ & \ \\
\ & \ & \R & \ & \
} \nonumber
\end{equation}
Local coordinates in $\W$ and $\W_r$ are constructed in an analogous way to the autonomous setting.
Let $(t,q^A)$, $1 \leqslant A \leqslant n$, be a set of local coordinates in $E$ adapted
to the bundle structure and such that the canonical volume form in $\R$ is given locally
by $\eta = \d t$. Then, the induced natural coordinates in $J^1\pi$, $\Tan^*E$ and $J^0\pi^*$
are $(t,q^A,v^A)$, $(t,q^A,p,p_A)$ and $(t,q^A,p_A)$, respectively.
Therefore, the natural coordinates in $\W$ and $\W_r$ are $(t,q^A,v^A,p,p_A)$ and
$(t,q^A,v^A,p_A)$, respectively. Observe that $\dim\W = 3n+2$ and $\dim\W_r = 3n+1$.
In these coordinates, the above projections have the following coordinate expressions
\begin{equation*}
\rho_1(t,q^A,v^A,p,p_A) = (t,q^A,v^A) \quad ; \quad \rho_2(t,q^A,v^A,p,p_A) = (t,q^A,p,p_A) \quad ; \quad
\rho_E(t,q^A,v^A,p,p_A) = (t,q^A) \, ,
\end{equation*}
\begin{equation*}
\rho_1^r(t,q^A,v^A,p_A) = (t,q^A,v^A) \quad ; \quad \rho_2^r(t,q^A,v^A,p_A) = (t,q^A,p_A) \quad ; \quad
\rho_E^r(t,q^A,v^A,p_A) = (t,q^A) \, ,
\end{equation*}
\begin{equation*}
\rho_\R(t,q^A,v^A,p,p_A) = t \quad ; \quad \rho_\R^r(t,q^A,v^A,p_A) = t \, .
\end{equation*}

Let us introduce some canonical structures on the extended jet-momentum bundle $\W$.
First, let $\Theta \in \df^{1}(\Tan^*E)$ and $\Omega = -\d\Theta \in \df^{2}(\Tan^*E)$
be the canonical Liouville forms of the cotangent bundle. Then, we define the following forms in $\W$
\begin{equation*}
\Theta_\W = \rho_2^*\Theta \in \df^{1}(\W) \quad ; \quad
\Omega_\W = \rho_2^*\Omega = -\d\Theta_\W \in \df^{2}(\W) \, .
\end{equation*}
Bearing in mind the coordinate expressions of the Liouville forms of the cotangent bundle
given in Example \ref{Chap01_exa:CotangentBundle} and the local expression of the projection
$\rho_2$ given above, the forms $\Theta_\W$ and $\Omega_\W$ are given locally by
\begin{equation*}
\Theta_\W = p_A\d q^A + p\d t \quad ; \quad
\Omega_\W = \d q^A \wedge \d p_A - \d p \wedge \d t \, .
\end{equation*}
It is clear from these coordinate expressions that $\Omega_\W$ is a closed $2$-form,
and that the $2$-form $\Omega_\W$ is degenerate, since we have
\begin{equation*}
\inn(\partial/\partial v^A)\Omega_\W = 0 \, , \quad \mbox{for every } 1 \leqslant A \leqslant n \, .
\end{equation*}
In particular, a local basis of $\ker\Omega_\W$ is given by
\begin{equation*}
\ker\Omega_\W = \left\langle \derpar{}{v^A} \right\rangle = \vf^{V(\rho_2)}(\W) \, .
\end{equation*}
Thus, the pair $(\Omega_\W,\rho_\R^*\eta)$ is a precosymplectic structure in $\W$.

The second canonical structure in $\W$ is the following.

\begin{definition}
The \textnormal{coupling form} in $\W$ is the $\rho_\R$-semibasic $1$-form
$\hat{\C} \in \df^{1}(\W)$ defined as follows: for every $w = (j_t^1\phi,\alpha) \in \W$
(that is, $\alpha \in \Tan^*_{\rho_E(w)}E$) and $v \in \Tan_w\W$, then
\begin{equation*}
\langle \hat{\C}(w), v \rangle = \langle \alpha, \Tan_w(\phi \circ \rho_\R)v \rangle \, .
\end{equation*}
\end{definition}

Since $\hat{\C}$ is a $\rho_\R$-semibasic form, there exists a function
$\hat{C} \in \Cinfty(\W)$ such that $\hat{\C} = \hat{C}\rho_\R^*\eta = \hat{C}\d t$.
A straightforward computation in coordinates gives the following local expression
for the coupling form
\begin{equation}\label{Chap02_eqn:UnifNACouplingFormLocal}
\hat{\C} = \left( p + p_Av^A \right) \d t \, .
\end{equation}

Given a Lagrangian density $\Lag \in \df^{1}(J^1\pi)$, we denote
$\hat{\Lag} = \rho_1^*\Lag \in \df^{1}(\W)$. As the Lagrangian density
is a $\bar{\pi}^1$-semibasic form, we have that $\hat{\Lag}$ is a
$\rho_\R$-semibasic form, and thus we can write $\hat{\Lag} = \hat{L}\rho_\R^*\eta$,
where $\hat{L} = \rho_2^*L \in \Cinfty(\W)$, $L \in \Cinfty(J^1\pi)$ being the
Lagrangian function associated to $\Lag$ and $\eta$. Then we define a
\textsl{Hamiltonian submanifold}
\begin{equation*}
\W_o = \left\{ w \in \W \mid \hat{\Lag}(w) = \hat{\C}(w) \right\} \stackrel{j_o}{\hookrightarrow} \W \, .
\end{equation*}
Since both $\hat{\C}$ and $\hat{\Lag}$ are $\rho_\R$-semibasic $1$-forms, the submanifold
$\W_o$ is defined by the regular constraint $\hat{C} - \hat{L} = 0$. In the natural
coordinates of $\W$, bearing in mind the local expression \eqref{Chap02_eqn:UnifNACouplingFormLocal}
of $\hat{\C}$, the constraint function is locally given by
\begin{equation*}
\hat{C} - \hat{L} = p + p_Av^A - \hat{L}(t,q^A,v^A) = 0 \, .
\end{equation*}

\begin{proposition}
The submanifold $\W_o \hookrightarrow \W$ is $1$-codimensional, $\mu_\W$-transverse
and diffeomorphic to $\W_r$ via the map $\mu_\W \circ j_o \colon \W_o \to \W_r$.
\end{proposition}

As a consequence of this last Proposition, the submanifold $\W_o$ induces a section
$\hat{h} \in \Gamma(\mu_\W)$ defined as
$\hat{h} = j_o \circ (\mu_\W \circ j_o)^{-1} \colon \W_r \to \W$. This section
is called the \textsl{Hamiltonian $\mu_\W$-section}, and is specified by giving the
\textsl{local Hamiltonian function}
\begin{equation*}
\hat{H} = -\hat{L} + p_Av^A \, ,
\end{equation*}
that is, $\hat{h}(t,q^A,v^A,p_A) = (t,q^A,v^A,-\hat{H},p_A)$.

\begin{remark}
If the Lagrangian density $\Lag \in \df^{1}(J^1\pi)$ is, at least, almost-regular, then from the
Hamiltonian $\mu_\W$-section $\hat{h} \in \Gamma(\mu_\W)$ in the unified formalism
we can recover the Hamiltonian $\mu$-section $h \in \Gamma(\mu)$ in the extended Hamiltonian
formalism. In fact, given $[\alpha] \in J^0\pi^*$, the section $\hat{h}$ maps every point
$(j^1_t\phi,[\alpha]) \in (\rho_2^r)^{-1}([\alpha])$ into $\rho_2^{-1}(\rho_2(\hat{h}(j^1_t\phi,[\alpha])))$.
Hence, the crucial point is the $\rho_2$-projectability of the local function $\hat{H}$.
However, since $\partial / \partial v^A$ is a local basis for $\ker\Tan\rho_2$, the local function
$\hat{H}$ is $\rho_2$-projectable if, and only if, $p_A = \partial\hat{L} / \partial v^A$, and this
condition is fulfilled when $[\alpha] \in \P = \Im\Leg \hookrightarrow J^0\pi^*$, which implies
that $\rho_2(\hat{h}((\rho_2^r)^{-1}([\alpha]))) \in \widetilde{\P} = \Im\widetilde{\Leg} \hookrightarrow \Tan^*E$.
Then, the Hamiltonian section $h$ is defined as
\begin{equation*}
h([\alpha]) = (\rho_2 \circ \hat{h})((\rho_2^r)^{-1}(\jmath([\alpha])))
= (\tilde{\jmath} \circ \widetilde{\mu}^{-1})([\alpha]) \, ,
\end{equation*}
for every $[\alpha] \in \P$.
\end{remark}

Finally, we can define the forms
\begin{equation*}
\Theta_r = \hat{h}^*\Theta_\W \in \df^{1}(\W_r) \quad ; \quad
\Omega_r = \hat{h}^*\Omega_\W \in \df^{2}(\W_r) \, ,
\end{equation*}
with local expressions
\begin{equation}\label{Chap02_eqn:UnifNAPrecosymplecticFormsLocal}
\Theta_r = p_A\d q^A + (\hat{L} - p_Av^A)\d t \quad ; \quad
\Omega_r = \d q^A \wedge \d p_A + \d(p_A v^A - \hat{L}) \wedge \d t \, .
\end{equation}
Then, the triple $(\W_r,\Omega_r,(\rho_\R^r)^*\eta)$ is a precosymplectic Hamiltonian system.

\subsubsection{Dynamical equations for sections}

The \textsl{first-order Lagrangian-Hamiltonian problem for sections} associated with the system
$(\W_r,\Omega_r,(\rho_\R^r)^*\eta)$ consists in finding sections $\psi \in \Gamma(\rho_\R^r)$
characterized by the condition
\begin{equation}\label{Chap02_eqn:UnifNADynEqSect}
\psi^*\inn(Y)\Omega_r = 0 \, , \quad \mbox{for every } Y \in \vf(\W_r) \, .
\end{equation}

In the natural coordinates of $\W_r$, if the section $\psi$
is locally given by $\psi(t) = (t,q^A(t),v^A(t),p_A(t))$, then, bearing in mind
the coordinate expression \eqref{Chap02_eqn:UnifNAPrecosymplecticFormsLocal}
of the $2$-form $\Omega_r$, the equation \eqref{Chap02_eqn:UnifNADynEqSect}
gives the following system of $3n$ equations
\begin{align}
\dot{q}^A = v^A \, , \label{Chap02_eqn:UnifNADynEqSectHolonomyLocal} \\
\dot{p}_A = \derpar{\hat{L}}{q^A} \, , \label{Chap02_eqn:UnifNADynEqSectLocal} \\
p_A - \derpar{\hat{L}}{v^A} = 0 \, . \label{Chap02_eqn:UnifNADynEqSectLegendreLocal}
\end{align}
Observe that equations \eqref{Chap02_eqn:UnifNADynEqSectHolonomyLocal} and
\eqref{Chap02_eqn:UnifNADynEqSectLocal} are differential equations whose solutions
are the component functions of the section $\psi$. More particularly, equations
\eqref{Chap02_eqn:UnifNADynEqSectHolonomyLocal} give the holonomy condition
for the section $\psi$ that must be satisfied once it is projected to $J^1\pi$,
while equations \eqref{Chap02_eqn:UnifNADynEqSectLocal} are the real dynamical equations
of the system. On the other hand, equations \eqref{Chap02_eqn:UnifNADynEqSectLegendreLocal}
do not involve any derivative of the component functions: they are point-wise algebraic equations
that must satisfy every section $\psi \in \Gamma(\rho_\R^r)$ to be a solution to equation
\eqref{Chap02_eqn:UnifNADynEqSect}. These equations arise from the $\rho_2^r$-vertical component
of the vector fields $Y$. In particular, we have the following result.

\begin{lemma}
If $Y \in \vf^{V(\rho_2^r)}(\W_r)$, then $\inn(Y)\Omega_r$ is a $\rho_\R^r$-semibasic
$1$-form.
\end{lemma}

As a consequence of this result, we can define the submanifold
\begin{equation*}
\W_\Lag = \left\{ [w] \in \W_r \mid (\inn(Y)\Omega_r)([w]) = 0 \mbox{ for every } Y \in \vf^{V(\rho_2^r)}(\W_r) \right\}
\stackrel{j_\Lag}{\hookrightarrow} \W_r \, ,
\end{equation*}
where every section solution to equation \eqref{Chap02_eqn:UnifNADynEqSect} must take values.
Locally, the submanifold $\W_\Lag$ is defined by the constraint $p_A - \partial \hat{L} / \partial v^A = 0$.
Moreover, we have the following characterization of $\W_\Lag$.

\begin{proposition}
$\W_\Lag \hookrightarrow \W_r$ is the graph of the restricted Legendre map $\Leg \colon J^1\pi \to J^0\pi^*$.
\end{proposition}

As a consequence of this result, since $\W_\Lag$ is the graph of the restricted Legendre map, then
it is diffeomorphic to $J^1\pi$. In addition, every section $\psi \in \Gamma(\rho_\R^r)$ is of the
form $\psi = (\psi_\Lag,\psi_h)$, with $\psi_\Lag = \rho_1^r \circ \psi \in \Gamma(\bar{\pi}^1)$ and
$\psi_h = \Leg \circ \psi_\Lag \in \Gamma(\bar{\pi}_E^r)$. In this way, every constraint, differential
equation, etc., in the unified formalism can be translated to the Lagrangian and Hamiltonian formalisms
by projection to the first factor of the product bundle or using the Legendre map.
Hence, we have the following result.

\begin{theorem}
Let $\psi \in \Gamma(\rho_\R^r)$ be a section solution to equation \eqref{Chap02_eqn:UnifNADynEqSect}.
Then we have
\begin{enumerate}
\item The section $\psi_\Lag = \rho_1^r \circ \psi \in \Gamma(\bar{\pi}^{1})$ is holonomic, and is a solution
to equation \eqref{Chap02_eqn:LagNAFODynEqSect}.
\item The section $\psi_h = \Leg \circ \psi_\Lag \in \Gamma(\bar{\pi}_E^{r})$ is a solution
to equation \eqref{Chap02_eqn:HamNAFODynEqSectRegular}.
\end{enumerate}
\end{theorem}

\subsubsection{Dynamical equations for vector fields}

As in the Lagrangian and Hamiltonian formalisms, if we assume that the sections
$\psi \in \Gamma(\rho_\R^r)$ solutions to equation \eqref{Chap02_eqn:UnifNADynEqSect}
are the integral curves of some vector fields in $\W_r$, then we can state
the problem in terms of vector fields.
The \textsl{first-order Lagrangian-Hamiltonian problem for vector fields} consists in finding
vector fields $X \in \vf(\W_r)$ satisfying the following equations
\begin{equation}\label{Chap02_eqn:UnifNADynEqVF}
\inn(X)\Omega_r = 0 \quad ; \quad
\inn(X)(\rho_\R^r)^*\eta \neq 0 \, .
\end{equation}

\begin{remark}
As in previous sections, the second equation in \eqref{Chap02_eqn:UnifNADynEqVF} is just
a transverse condition for the vector field $X$ with respect to the projection onto $\R$,
and the non-zero value is usually fixed to $1$, thus giving the following equations
\begin{equation*}
\inn(X)\Omega_r = 0 \quad ; \quad
\inn(X)(\rho_\R^r)^*\eta = 1 \, .
\end{equation*}
\end{remark}

Recall that que pair $(\Omega_r,(\rho_\R^r)^*\eta)$ is a precosymplectic structure on $\W_r$.
Hence, equations \eqref{Chap02_eqn:UnifNADynEqVF} may not admit a global solution $X \in \vf(\W_r)$,
and an adapted version of the constraint algorithm to precosymplectic structures must be used.
From the algorithm, we can state the following result.

\begin{proposition}\label{Chap02_prop:UnifNAFirstConstSubm}
Given the precosymplectic Hamiltonian system $(\W_r,\Omega_r,(\rho_\R^r)^*\eta)$,
a solution $X \in \vf(\W_r)$ to equations \eqref{Chap02_eqn:UnifNADynEqVF} exists only
on the points of the submanifold $\mathcal{S}_\Lag \hookrightarrow \W_r$ defined by
\begin{equation*}
\mathcal{S}_\Lag = \left\{ [w] \in \W_r \mid (\inn(Y)\d\hat{H})([w]) = 0
\mbox{ for every } Y \in \ker\Omega_\W \right\} \, .
\end{equation*}
\end{proposition}

As in the autonomous setting described in Section \ref{Chap02_sec:SkinnerRuskAutonomousFirstOrder},
we have the following characterization of the submanifold $\mathcal{S}_\Lag \hookrightarrow \W_r$.

\begin{proposition}\label{Chap02_prop:UnifNAGraphLegendreMap}
The submanifold $\mathcal{S}_\Lag$ is the graph of the restricted Legendre map $\Leg \colon J^1\pi \to J^0\pi^*$,
and therefore $\mathcal{S}_\Lag = \W_\Lag$.
\end{proposition}

In the natural coordinates of $\W_r$, let $X \in \vf(\W_r)$ be a generic vector field
locally given by
\begin{equation*}
X = f\derpar{}{t} + f^A \derpar{}{q^A} + F^A\derpar{}{v^A} + G_A\derpar{}{p_A} \, .
\end{equation*}
Then, bearing in mind the coordinate expression \eqref{Chap02_eqn:UnifNAPrecosymplecticFormsLocal}
of the $2$-form $\Omega_r$, the equation \eqref{Chap02_eqn:UnifNADynEqVF} gives in coordinates
the following system of $3n+1$  equations for the component functions of $X$
\begin{align}
f^A = fv^A \, , \label{Chap02_eqn:UnifNADynEqVFHolonomyLocal1} \\
G_A = f\derpar{\hat{L}}{q^A} \, , \label{Chap02_eqn:UnifNADynEqVFLocal1} \\
f\left( p_A - \derpar{\hat{L}}{v^A}\right) = 0 \, , \label{Chap02_eqn:UnifNADynEqVFLegendreLocal1} \\
f \neq 0 \label{Chap02_eqn:UnifNADynEqVFLocalGaugeNonZero} \, .
\end{align}
where equation \eqref{Chap02_eqn:UnifNADynEqVFLocalGaugeNonZero} arises from the second equation
in \eqref{Chap02_eqn:UnifNADynEqVF}. Fixing the non-zero value of $f$ to $1$, the above equations
become
\begin{align}
f^A = v^A \, , \label{Chap02_eqn:UnifNADynEqVFHolonomyLocal2} \\
G_A = \derpar{\hat{L}}{q^A} \, , \label{Chap02_eqn:UnifNADynEqVFLocal2} \\
p_A - \derpar{\hat{L}}{v^A} = 0 \, , \label{Chap02_eqn:UnifNADynEqVFLegendreLocal2} \\
f = 1 \label{Chap02_eqn:UnifNADynEqVFLocalGaugeOne} \, ,
\end{align}
Note that equations \eqref{Chap02_eqn:UnifNADynEqVFHolonomyLocal2} are the holonomy
condition for a vector field in the Lagrangian formalism, as we have seen in Section
\ref{Chap02_sec:LagrangianNonAutonomousFirstOrder}, while equations 
\eqref{Chap02_eqn:UnifNADynEqVFLocal2} are the dynamical equations of the system.
On the other hand, equations \eqref{Chap02_eqn:UnifNADynEqVFLegendreLocal2} are a
compatibility condition stating that the vector fields solution to equations \eqref{Chap02_eqn:UnifNADynEqVF}
exist only with support on the submanifold defined as the graph of the restricted Legendre map.
Thus we recover, in coordinates, the result stated in Propositions \ref{Chap02_prop:UnifNAFirstConstSubm} and
\ref{Chap02_prop:UnifNAGraphLegendreMap}.

\begin{remark}
As in the autonomous setting, the holonomy of the vector field $X \in \vf(\W_r)$
is obtained regardless of the regularity of the Lagrangian density $\Lag \in \df^{1}(J^1\pi)$ provided.
\end{remark}

Therefore, a vector field $X \in \vf(\W_r)$ solution to equation \eqref{Chap02_eqn:UnifNADynEqVF}
is locally given by
\begin{equation*}
X = \derpar{}{t} + v^A\derpar{}{q^A} + F^A\derpar{}{v^A} + \derpar{\hat{L}}{q^A}\derpar{}{p_A} \, .
\end{equation*}

Observe that the component functions $F^A$ of the vector field $X$ remain undetermined.
However, since the vector field $X$ is defined at support on the submanifold $\W_\Lag$, we must
study the tangency of $X$ along this submanifold. That is, we must require that
$\restric{\Lie(X)\xi}{\W_\Lag} = 0$ for every constraint function defining $\W_\Lag$.
From Proposition \ref{Chap02_prop:UnifNAGraphLegendreMap} the submanifold $\W_\Lag$
is the graph of the restricted Legendre map, and so it is defined by the $n$
constraints
\begin{equation*}
\xi^B \equiv p_B - \derpar{\hat{L}}{v^B} = 0 \ , \ B = 1,\ldots,n \, .
\end{equation*}
Therefore, the tangency condition for $X$ along $\W_\Lag$ gives the following $n$ equations
\begin{align*}
\Lie(X)\xi_B &=
\left( \derpar{}{t} + v^A \derpar{}{q^A} + F^A \derpar{}{v^A} + \derpar{\hat{L}}{q^A}\derpar{}{p_A} \right)
\left( p_B - \derpar{\hat{L}}{v^B} \right) \\
&= \derpar{\hat{L}}{q^B} - \derpars{\hat{L}}{t}{v^B} - v^A\derpars{\hat{L}}{q^A}{v^B} - F^A\derpars{\hat{L}}{v^A}{v^B} = 0 \, .
\end{align*}
Observe that these are the Lagrangian equations for a vector field once the holonomy condition is satisfied,
as we have seen in \eqref{Chap02_eqn:LagNAFODynEqWithHolonomyLocal}. These equations can be
compatible or not, and a sufficient condition to ensure compatibility is the
regularity of the Lagrangian density. In particular, we have the following analogous result
to Proposition \ref{Chap02_prop:UnifARegLagUniqueVF} in Section \ref{Chap02_sec:SkinnerRuskAutonomousFirstOrder}.

\begin{proposition}\label{Chap02_prop:UnifNARegLagUniqueVF}
If $\Lag \in \df^{1}(J^1\pi)$ is a regular Lagrangian density, then there exists a unique
vector field $X \in \vf(\W_r)$ which is a solution to equation \eqref{Chap02_eqn:UnifNADynEqVF}
and is tangent to $\W_\Lag$.
\end{proposition}

If the Lagrangian density $\Lag$ is singular, then the above equations can be compatible or not,
and new constraints may arise from them, thus requiring the constraint algorithm
to continue. In the most favorable cases, there exists a submanifold $\W_f \hookrightarrow \W_\Lag$
(it could be $\W_f = \W_\Lag)$ such that there exist vector fields $X \in \vf_{\W_f}(\W_r)$,
tangent to $\W_f$, which are solutions to the equation
\begin{equation*}\label{Chap02_eqn:UnifNADynEqSingular}
\restric{\left[ \inn(X)\Omega_r \right]}{\W_f} = 0 \, .
\end{equation*}

Now, the equivalence of the unified formalism with the Lagrangian and Hamiltonian formalisms can
be recovered as follows.

\begin{theorem}
Let $X \in \vf(\W_r)$ be a vector field solution to equations \eqref{Chap02_eqn:UnifNADynEqVF}
and tangent to $\W_\Lag$ (at least on the points of a submanifold $\W_f \hookrightarrow \W_\Lag$).
Then the vector field $X_\Lag \in \vf(J^1\pi)$ defined by $X_\Lag \circ \rho_1^r = \Tan\rho_1^r \circ X$
is holonomic, and is a solution to equations \eqref{Chap02_eqn:LagNAFODynEqVF} (at least on the
points of a submanifold $\mathcal{S}_f = \rho_1^r(\W_f) \hookrightarrow J^1\pi$).

\noindent Moreover, every holonomic vector field $X_\Lag \in \vf^{1}(J^1\pi)$ solution to equations
\eqref{Chap02_eqn:LagNAFODynEqVF} (at least on the points of a submanifold $\mathcal{S}_f \hookrightarrow J^1\pi$)
can be recovered in this way from a vector field $X \in \vf(\W_r)$ solution to equations
\eqref{Chap02_eqn:UnifNADynEqVF} and tangent to $\W_\Lag$ (at least on the points of a submanifold
$\W_f \hookrightarrow \W_\Lag$).
\end{theorem}

Finally, the Hamiltonian formalism is recovered from the Lagrangian one using Theorem
\ref{Chap02_thm:HamNAFORelationLagHamRegular} in the regular case and Theorem
\ref{Chap02_thm:HamNAFORelationLagHamSingular} in the singular case.


\section{First-order classical field theories}
\label{Chap02_sec:FieldTheories}

Let us consider a first-order classical field theory with $n$ fields depending on $m$
independent variables. The configuration space for this theory is a bundle $\pi \colon E \to M$,
where $M$ is a $m$-dimensional orientable smooth manifold with fixed volume form $\eta \in \df^{m}(M)$,
and $\dim E = m + n$. The physical information is given in terms of a Lagrangian density $\Lag \in \df^{m}(J^1\pi)$,
which is a $\bar{\pi}^1$-semibasic $m$-form. Because of this, we can write $\Lag = L \cdot (\bar{\pi}^1)^*\eta$,
where $L \in \Cinfty(J^1\pi)$ is the Lagrangian function associated to $\Lag$ and $\eta$.

\subsection{Lagrangian formalism}
\label{Chap02_sec:LagFieldTheories}

(See \cite{art:Aldaya_Azcarraga80,proc:Carinena_Crampin_Ibort91,
art:Echeverria_Munoz_Roman96,art:Echeverria_Munoz_Roman98,proc:Garcia_Munoz83,
book:Giachetta_Mangiarotti_Sardanashvily97,art:Goldschmidt_Sternberg73,art:Paufler_Romer02,art:Roman09} for details).

\subsubsection{Geometrical setting}

Using the Lagrangian density $\Lag$ and the vertical endomorphism 
$\V \in \Gamma(\Tan^*J^1\pi \otimes_{J^1\pi} \Tan M \otimes_{J^1\pi} V(\pi^1))$
described in Section \ref{Chap01_sec:HOJetBundlesVerticalEndomorphisms} we can construct
the following forms in $J^1\pi$.

\begin{definition}
The \textnormal{Poincar\'{e}-Cartan $m$-form} associated to $\Lag$ and
$\eta$ is the form $\Theta_\Lag \in \df^{m}(J^1\pi)$ defined as
\begin{equation*}
\Theta_\Lag = \inn(\V)\d\Lag + \Lag \, .
\end{equation*}
Then, the \textnormal{Poincar\'{e}-Cartan $(m+1)$-form} associated to $\Lag$ and $\eta$
is the form $\Omega_\Lag \in \df^{m+1}(J^1\pi)$ defined as
\begin{equation*}
\Omega_\Lag = -\d\Theta_\Lag \, .
\end{equation*}
\end{definition}

From the previous definitions it is clear that the phase space of a first-order
classical field theory described by a Lagrangian density is the first-order jet bundle
of the configuration bundle $\pi \colon E \to M$.

\begin{definition}
A \textnormal{first-order Lagrangian field theory} is a pair $(J^1\pi,\Lag)$,
where $\pi \colon E \to M$ is the configuration bundle and $\Lag \in \df^{m}(J^1\pi)$
the Lagrangian density.
\end{definition}

In the natural coordinates $(x^i,u^\alpha,u_i^\alpha)$ of $J^1\pi$, and bearing in mind
the local expression \eqref{Chap01_eqn:HOJetBundleVertEnd} of the vertical endomorphism
$\V$, the coordinate expression of the Poincar\'{e}-Cartan $m$-form is
\begin{equation}\label{Chap02_eqn:LagCFTmFormLocal}
\Theta_\Lag = \derpar{L}{u_i^\alpha}\d u^\alpha \wedge \d^{m-1}x_i -
\left( \derpar{L}{u_i^\alpha}u_i^\alpha - L \right)\d^{m}x \, .
\end{equation}
From this, the coordinate expression of the Poincar\'{e}-Cartan $(m+1)$-form is
\begin{equation}\label{Chap02_eqn:LagCFTm+1FormLocal}
\begin{array}{c}
\displaystyle \Omega_{\Lag} =
\derpars{L}{u_i^\alpha}{u^\beta} \, \d u^\alpha \wedge \d u^\beta \wedge \d^{m-1}x_i
+ \derpars{L}{u_i^\alpha}{u_j^\beta} \, \d u^\alpha \wedge \d u_j^\beta \wedge \d^{m-1}x_i \\[12pt]
\displaystyle \qquad\qquad + \left( u_i^\alpha \derpars{L}{u_i^\alpha}{u^\beta} - \derpar{L}{u^\beta} \right) \d u^\beta \wedge \d^mx
+ u_i^\alpha\derpars{L}{u_i^\alpha}{u_j^\beta} \, \d u_j^\beta \wedge \d^mx \, .
\end{array}
\end{equation}

\begin{remark}
As in the autonomous and non-autonomous formulations described in previous Sections,
it is clear from the coordinate expression \eqref{Chap02_eqn:LagCFTmFormLocal}
that the Poincar\'{e}-Cartan $m$-form $\Theta_\Lag$ is $\pi^1$-semibasic.
\end{remark}

Notice that, given an arbitrary Lagrangian density $\Lag \in \df^{m}(J^1\pi)$,
the Poincar\'{e}-Cartan $(m+1)$-form is always a closed form (in fact, it
is an exact form). Nevertheless, we can not assure the $1$-nondegeneracy of $\Omega_\Lag$.
Hence, we introduce the following definition.

\begin{definition}
A Lagrangian density $\Lag \in \df^{m}(J^1\pi)$ is \textnormal{regular}
(and thus $(J^1\pi,\Lag)$ is a \textnormal{regular field theory}) if
$\Omega_\Lag$ is a multisymplectic $(m+1)$-form on $J^1\pi$.
Otherwise, the Lagrangian density is said to be \textnormal{singular}
(and thus $(J^1\pi,\Lag)$ is a \textnormal{singular field theory}).
\end{definition}

From the coordinate expression \eqref{Chap02_eqn:LagCFTm+1FormLocal}
of the Poincar\'{e}-Cartan $(m+1)$-form, it is clear that the regularity
of the Lagrangian density $\Lag$ is locally equivalent to
\begin{equation*}
\det\left( \derpars{L}{u_i^\alpha}{u_j^\beta} \right) (j^1_x\phi) \neq 0 \ , \,
\mbox{for every } j^1_x\phi \in J^1\pi \, ,
\end{equation*}
where $L \in \Cinfty(J^1\pi)$ is the Lagrangian function associated to $\Lag$ and $\eta$.

\subsubsection{Lagrangian field equations}

The Lagrangian problem for first-order field theories consists in finding a distribution $\D$
in $J^1\pi$ satisfying
\begin{enumerate}
\item $\D$ is $m$-dimensional.
\item $\D$ is $\bar{\pi}^1$-transverse.
\item $\D$ is integrable.
\item The integral manifolds of $\D$ are the first prolongations
of the critical sections of the Hamilton principle.
\end{enumerate}

As we have seen in Section \ref{Chap01_sec:MultivectorFields}, these kinds of integrable
distributions are associated with classes of integrable (that is, non-vanishing, locally decomposable
and involutive) $\bar{\pi}^1$-transverse multivector fields $\X_\Lag \in \vf^{m}(J^1\pi)$.
In the natural coordinates of $J^1\pi$, the local expression of an element of one of these classes is
\begin{equation*}
\X_\Lag = f \bigwedge_{i=1}^{m}
\left(  \derpar{}{x^i} + f_i^\alpha\derpar{}{u^\alpha} + F_{j,i}^\alpha\derpar{}{u_j^\alpha} \right) \, ,
\end{equation*}
where $f \in \Cinfty(J^1\pi)$ is a non-vanishing function. If, in addition, the integral sections of the
distribution $\D$ are holonomic, then the associated classes of multivector fields are also holonomic
(see Section \ref{Chap01_sec:MultivectorFieldsHolonomic}). Then, we have the following result.

\begin{theorem}
Let $(J^1\pi,\Lag)$ be a Lagrangian field theory. The following assertions on a section
$\phi \in \Gamma(\pi)$ are equivalent.
\begin{enumerate}
\item The section $\phi$ is a solution to the equation
\begin{equation}\label{Chap02_eqn:LagCFTFieldEqSect}
(j^1\phi)^*\inn(X)\Omega_\Lag = 0\, , \quad \mbox{for every } X \in \vf(J^1\pi) \, .
\end{equation}
\item In the natural coordinates $(x^i,u^\alpha,u_i^\alpha)$ of $J^1\pi$, the first prolongation
of the section $\phi \in \Gamma(\pi)$, $j^1\phi(x^i) = (x^i,u^\alpha(x^i),\derpar{u^\alpha}{x^j}(x^i))$,
satisfies the Euler-Lagrange equations
\begin{equation}\label{Chap02_eqn:CFTEulerLagrange}
\restric{\derpar{L}{u^\alpha}}{j^1\phi} - \restric{\frac{d}{dx^i}\derpar{L}{u_i^\alpha}}{j^1\phi} = 0 \, .
\end{equation}
\item $j^1\phi$ is an integral section of a class of holonomic multivector fields
$\{ \X_\Lag \} \subseteq \vf^{m}(J^1\pi)$ satisfying
\begin{equation}\label{Chap02_eqn:LagCFTFieldEqMultiVF}
\inn(\X_\Lag)\Omega_\Lag = 0 \, , \quad \mbox{for every } \X_\Lag \in \{ \X_\Lag \} \, .
\end{equation}
\end{enumerate}
\end{theorem}

Semi-holonomic (but not necessarily integrable) multivector fields which are solutions to
equation \eqref{Chap02_eqn:LagCFTFieldEqMultiVF} are called \textsl{Euler-Lagrange multivector fields}.

Let us compute in coordinates the equation \eqref{Chap02_eqn:LagCFTFieldEqMultiVF}.
Let $\X_\Lag \in \vf^{m}(J^1\pi)$ be a locally decomposable and $\bar{\pi}^1$-transverse
multivector field locally given by
\begin{equation*}
\X_\Lag = f \bigwedge_{i=1}^{m}
\left(  \derpar{}{x^i} + f_i^\alpha\derpar{}{u^\alpha} + F_{j,i}^\alpha\derpar{}{u_j^\alpha} \right) \, ,
\end{equation*}
where $f$ is a non-vanishing function.
Then, taking $f = 1$ as a representative of the equivalence class, and bearing in mind the coordinate
expression \eqref{Chap02_eqn:LagCFTm+1FormLocal} of the Poincar\'{e}-Cartan $(m+1)$-form $\Omega_\Lag$,
the equation \eqref{Chap02_eqn:LagCFTFieldEqMultiVF} gives the following system of $n(m+1)$ equations
\begin{align}
\derpar{L}{u^\alpha} - \derpars{L}{u_i^\alpha}{x^i} - f_i^\beta\derpars{L}{u_i^\alpha}{u^\beta}
-F_{j,i}^\beta\derpars{L}{u_i^\alpha}{u_j^\beta} + (f_i^\beta - u_i^\beta)\derpars{L}{u_i^\beta}{u^\alpha} = 0 \, ,
\label{Chap02_eqn:LagCFTFieldEqMultiVFLocal} \\
(f_j^\beta - u_j^\beta) \derpars{L}{u_i^\alpha}{u_j^\beta} = 0 \label{Chap02_eqn:LagCFTFieldEqHolonomyLocal} \, .
\end{align}
Observe that equations \eqref{Chap02_eqn:LagCFTFieldEqMultiVFLocal} and
\eqref{Chap02_eqn:LagCFTFieldEqHolonomyLocal} correspond exactly to equations
\eqref{Chap02_eqn:LagNAFODynEqLocal2} and \eqref{Chap02_eqn:LagNAFODynEqHolonomyLocal2}
in the formulation of non-autonomous dynamical systems, respectively, and to equations
\eqref{Chap02_eqn:LagAFODynEqLocal} and \eqref{Chap02_eqn:LagAFODynEqHolonomyLocal}
in the formulation of autonomous dynamical systems, respectively. In particular, equations
\eqref{Chap02_eqn:LagCFTFieldEqHolonomyLocal} are the local equations for the semi-holonomy
condition required to the multivector field $\X_\Lag$. If the holonomy condition is required
from the beginning, then $\X_\Lag$ is also semi-holonomic, and therefore equations
\eqref{Chap02_eqn:LagCFTFieldEqHolonomyLocal} are an identity. Then, equations
\eqref{Chap02_eqn:LagCFTFieldEqMultiVFLocal} become
\begin{equation}\label{Chap02_eqn:LagCFTFieldEqWithHolonomyLocal}
\derpar{L}{u^\alpha} - \derpars{L}{u_i^\alpha}{x^i} - u_i^\beta\derpars{L}{u_i^\alpha}{u^\beta}
-F_{j,i}^\beta\derpars{L}{u_i^\alpha}{u_j^\beta} = 0 \, .
\end{equation}

From the coordinate expression of equation \eqref{Chap02_eqn:LagCFTFieldEqMultiVF} we observe that
if $\Lag \in \df^{m}(J^1\pi)$ is a regular Lagrangian density then Euler-Lagrange
multivector fields do exist in $J^1\pi$, although they are not necessarily integrable. Otherwise, if the
Lagrangian density is singular, in the most favorable cases a semi-holonomic multivector field solution
to equation \eqref{Chap02_eqn:LagCFTFieldEqMultiVF} exists only on the points of a submanifold
$S_f \hookrightarrow J^1\pi$, which can be obtained after applying a suitable adapted
version of the constraint algorithm described in Section \ref{Chap01_sec:ConstraintAlgorithm}
(see \cite{art:DeLeon_Marin_Marrero_Munoz_Roman05}).
In these cases, the following equation holds
\begin{equation}\label{Chap02_eqn:LagCFTFieldEqSingular}
\restric{[\inn(\X_\Lag)\Omega_\Lag]}{S_f} = 0 \, .
\end{equation}

\subsection{Extended Hamiltonian formalism}

As in the Hamiltonian formulation for non-autonomous dynamical systems, the extended Hamiltonian
formalism associated to a Lagrangian field theory $(J^1\pi,\Lag)$ makes use of two phase spaces: the
\textsl{extended multimomentum bundle} and the \textsl{restricted multimomentum bundle}.
The former is exactly the extended dual jet bundle $\Lambda^m_{2}(\Tan^*J^0\pi) = \Lambda^m_2(\Tan^*E)$
of $J^1\pi$ described in Section \ref{Chap01_sec:HOJetBundlesDualBundles},
while the latter is the reduced dual jet bundle $\Lambda^m_{2}(J^0\pi)/\Lambda^m_{1}(J^0\pi)$.
As in Section \ref{Chap02_sec:NonAutonomousHamiltonian}, we denote the restricted multimomentum bundle
by $J^0\pi^*$, instead of $E^*$, to avoid confusion. The quotient map is denoted by
$\mu \colon \Lambda^m_2(\Tan^*E) \to J^0\pi^*$.
Natural coordinates in $\Lambda^m_2(\Tan^*E)$ are $(x^i,u^\alpha,p,p_\alpha^i)$,
and the induced natural coordinates in $J^0\pi^*$ are $(x^i,u^\alpha,p_\alpha^i)$,
$1 \leqslant i \leqslant m$, $1 \leqslant \alpha \leqslant n$.

(See \cite{art:Cantrijn_Ibort_DeLeon96,art:Echeverria_DeLeon_Munoz_Roman07,
art:Echeverria_Munoz92,art:Echeverria_Munoz_Roman00_JMP,art:Echeverria_Munoz_Roman00_RMP,
proc:Garcia_Munoz83,art:Goldschmidt_Sternberg73,art:Helein_Kouneiher02,art:Paufler_Romer02} for details).

\subsubsection{The extended and restricted Legendre maps}

As in previous Sections, we begin by introducing the Legendre map that
relates the Lagrangian formulation with the Hamiltonian one. Recall that,
from the coordinate expression \eqref{Chap02_eqn:LagCFTmFormLocal}, it
is clear that the Poincar\'{e}-Cartan $m$-form $\Theta_\Lag \in \df^{m}(J^1\pi)$
is $\pi^1$-semibasic. Then, we can give the following definition.

\begin{definition}
The \textnormal{extended Legendre map} associated with the Lagrangian density $\Lag \in \df^{m}(J^1\pi)$
is the bundle morphism $\widetilde{\Leg} \colon J^1\pi \to \Lambda^m_2(\Tan^*E)$ over $E$ defined as
follows:
\begin{equation*}
(\Theta_\Lag(j^1_x\phi))(Y_1(j^1_x\phi),\ldots,Y_m(j^1_x\phi)) =
(\widetilde{\Leg}(j^1_x\phi))((\Tan_{j^1_x\phi}\pi^1Y_1)(\phi(x)),\ldots,(\Tan_{j^1_x\phi}\pi^1Y_m)(\phi(x))) \, ,
\end{equation*}
where $Y_i \in \vf(J^1\pi)$, and therefore $\Tan\pi^1Y_i \in \vf(E)$.
\end{definition}

It is clear from the definition that $\pi_E \circ \widetilde{\Leg} = \pi^1$, where
$\pi_E \colon \Lambda^m_2(\Tan^*E) \to E$ is the canonical submersion. In addition,
if $\Theta \in \df^{m}(\Lambda^m_2(\Tan^*E))$ is the canonical $m$-form of $\Lambda_2^m(\Tan^*E)$
and $\Omega = -\d\Theta \in \df^{m+1}(\Lambda^m_2(\Tan^*E))$ the canonical multisymplectic $(m+1)$-form,
then we have $\widetilde{\Leg}^*\Theta = \Theta_\Lag$ and $\widetilde{\Leg}^*\Omega = \Omega_\Lag$.
From Examples \ref{Chap01_exa:MulticotangentBundle} and \ref{Chap01_exa:MulticotangentBundleVertical},
the coordinate expression of $\Theta$ in this case is
\begin{equation*}
\Theta = p_\alpha^i \d u^\alpha \wedge \d^{m-1}x_i + p\d^mx \, ,
\end{equation*}
from where the coordinate expression of the multisymplectic $(m+1)$-form is
\begin{equation}\label{Chap02_eqn:HamCFTCanonicalMultisymplecticForm}
\Omega = -\d p_\alpha^i \wedge \d u^\alpha \wedge \d^{m-1}x_i - \d p \wedge \d^mx \, .
\end{equation}
Hence, bearing in mind the coordinate expression \eqref{Chap02_eqn:LagCFTmFormLocal} of the
Poincar\'{e}-Cartan $m$-form $\Theta_\Lag \in \df^{m}(J^1\pi)$, the coordinate expression
of the extended Legendre map is
\begin{equation}\label{Chap02_eqn:HamCFTExtendedLegendreMapLocal}
\widetilde{\Leg}^*x^i = x^i \quad ; \quad \widetilde{\Leg}^*u^\alpha = u^\alpha \quad ; \quad
\widetilde{\Leg}^*p_\alpha^i = \derpar{L}{u_i^\alpha} \quad ; \quad
\widetilde{\Leg}^*p = L - u_i^\alpha \derpar{L}{u_i^\alpha} \, .
\end{equation}

Now, if we compose the extended Legendre map with the natural quotient map
$\mu \colon \Lambda^m_2(\Tan^*E) \to J^0\pi^*$, we obtain a bundle morphism
$\mu \circ \widetilde{\Leg} \colon J^1\pi \to J^0\pi^*$ which leads to the
following definition.

\begin{definition}
The \textnormal{restricted Legendre map} associated to the Lagrangian density
$\Lag \in \df^{m}(J^1\pi)$ is the bundle morphism $\Leg \colon J^1\pi \to J^0\pi^*$
over $E$ defined as $\Leg = \mu \circ \widetilde{\Leg}$.
\end{definition}

In the natural coordinates of $J^0\pi^*$, the local expression of the restricted Legendre map is
\begin{equation*}
\Leg^*x^i = x^i \quad ; \quad \Leg^*u^\alpha = u^\alpha \quad ; \quad
\Leg^*p_\alpha^i = \derpar{L}{u_i^\alpha} \, .
\end{equation*}

As in the Hamiltonian formulation for first-order non-autonomous systems, a fundamental
result relating both Legendre maps is the following.

\begin{proposition}\label{Chap02_prop:HamCFTLegendreMapsEqualRank}
For every $j^1_x\phi \in J^1\pi$ we have that $\rank(\widetilde{\Leg}(j^1_x\phi)) = \rank(\Leg(j^1_x\phi))$.
\end{proposition}

The proof of this result follows the patterns in \cite{art:deLeon_Marin_Marrero96} for
non-autonomous dynamical systems. As a consequence of Proposition \ref{Chap02_prop:HamCFTLegendreMapsEqualRank},
from the coordinate expressions of both Legendre maps, and bearing in mind the results of Section
\ref{Chap02_sec:LagFieldTheories}, we have the following result.

\begin{proposition}
Let $\Lag \in \df^{m}(J^1\pi)$ be a Lagrangian density. The following statements are equivalent:
\begin{enumerate}
\item $\Omega_\Lag$ is $1$-nondegenerate, that is, it is a multisymplectic $(m+1)$-form in $J^1\pi$.
\item In the natural coordinates of $J^1\pi$, we have
\begin{equation*}
\det\left( \derpars{L}{u_j^\beta}{u_i^\alpha} \right)(j^1_x\phi) \neq 0 \, ,
\end{equation*}
for every $j^1_x\phi \in J^1\pi$, where $L \in \Cinfty(J^1\pi)$ is the Lagrangian function associated
with $\Lag$ and $\eta$.
\item The restricted Legendre map $\Leg \colon J^1\pi \to J^0\pi^*$ is a local diffeomorphism.
\item The extended Legendre map $\widetilde{\Leg} \colon J^1\pi \to \Lambda^m_2(\Tan^*E)$ is an immersion.
\end{enumerate}
In this case, $\Lag$ is a \textnormal{regular} Lagrangian density.
\end{proposition}

\begin{definition}
A Lagrangian density $\Lag \in \df^{m}(J^1\pi)$ is \textnormal{hyperregular} if the restricted Legendre map
$\Leg \colon J^1\pi \to J^0\pi^*$ is a global diffeomorphism.
\end{definition}

Now, let $\widetilde{\P} = \Im(\widetilde{\Leg}) \hookrightarrow \Lambda^m_2(\Tan^*E)$ be the image of
the extended Legendre map, with natural embedding
$\tilde{\jmath} \colon \widetilde{\P} \hookrightarrow \Lambda^m_2(\Tan^*E)$,
and $\P = \Im(\Leg) \hookrightarrow J^0\pi^*$ the image of the restricted Legendre map,
with canonical embedding $\jmath \colon \P \hookrightarrow J^0\pi^*$.
Let $\bar{\pi}_\P = \bar{\pi}_{E}^r \circ \jmath \colon \P \to \R$ be the canonical projection,
and $\Leg_o \colon J^1\pi \to \P$ the map defined by $\Leg = \jmath \circ \Leg_o$.
We can now give the following definition.

\begin{definition}
A Lagrangian density $\Lag \in \df^{m}(J^1\pi)$ is \textnormal{almost-regular} if
\begin{enumerate}
\item $\P$ is a closed submanifold of $J^0\pi^*$.
\item $\Leg$ is a submersion onto its image.
\item For every $j^1_x\phi \in J^1\pi$, the fibers $\Leg^{-1}(\Leg(j^1_x\phi))$
are connected submanifolds of $J^1\pi$.
\end{enumerate}
\end{definition}

As in the Hamiltonian formalism for non-autonomous dynamical systems, from
Proposition \ref{Chap02_prop:HamCFTLegendreMapsEqualRank}
we have that the map $\widetilde{\mu} \colon \widetilde{\P} \to \P$ is a diffeomorphism,
where $\widetilde{\mu}$ is the restriction of $\mu$ on the image set $\widetilde{\P}$.
Then, we have the following definition.

\begin{definition}
The \textnormal{canonical Hamiltonian section} $h \in \Gamma(\widetilde{\mu})$ is
defined as the map $h = \widetilde{\mu}^{-1} \colon \P \to \widetilde{\P}$.
\end{definition}

From the Hamiltonian section $h \in \Gamma(\widetilde{\mu})$ we can define the following forms
on $\P$.

\begin{definition}
The \textnormal{Hamilton-Cartan forms} $\Theta_h \in \df^{m}(\P)$
and $\Omega_h \in \df^{m+1}(\P)$ are defined as
\begin{equation*}
\Theta_h = (\tilde{\jmath} \circ h)^*\Theta \quad ; \quad
\Omega_h = (\tilde{\jmath} \circ h)^*\Omega = -\d\Theta_h \, ,
\end{equation*}
where $\Theta$ and $\Omega$ are the canonical $m$ and $(m+1)$-forms of $\Lambda^{m}_2(\Tan^*E)$.
\end{definition}

The pair $(\P,\Omega_h)$ is the \textsl{Hamiltonian field theory}
associated with the Lagrangian field theory $(J^1\pi,\Lag)$.

\subsubsection{Hamiltonian field equations}

Let us suppose first that the Lagrangian density $\Lag \in \df^{m}(J^1\pi)$ is hyperregular.
The regular, but not hyperregular, case can be recovered by restriction on the open sets
where the restricted Legendre map is a local diffeomorphism.

In the hyperregular case we have $\P = J^0\pi^*$, and $\widetilde{\P}$
is a $1$-codimensional and $\mu$-transverse submanifold of $\Lambda_2^{m}(\Tan^*E)$ which
is diffeomorphic to $J^0\pi^*$. In addition, in this case the Hamiltonian
section may be defined equivalently as $h = \widetilde{\Leg} \circ \Leg^{-1}$.

In the natural coordinates of $J^0\pi^*$, the Hamiltonian section is specified
by a Hamiltonian function $H \in \Cinfty(J^0\pi^*)$ as
\begin{equation*}
h(x^i,u^\alpha,p_\alpha^i) = (x^i,u^\alpha,-H(x^i,u^\alpha,p_\alpha^i),p_\alpha^i) \, ,
\end{equation*}
with the Hamiltonian function $H$ being locally given by
\begin{equation*}
H(x^i,u^\alpha,p_\alpha^i) = p_\alpha^i(\Leg^{-1})^*u_i^\alpha - (\Leg^{-1})^*L(x^i,u^\alpha,u_i^\alpha) \, .
\end{equation*}
From this, and bearing in mind the coordinate expressions of the
canonical forms of $\Lambda^m_2(\Tan^*E)$ given in Section
\ref{Chap01_sec:HOJetBundlesDualBundles}, the local expressions of the
Hamilton-Cartan forms are
\begin{equation}\label{Chap02_eqn:HamCFTHamiltonCartanFormsLocal}
\Theta_h = p_\alpha^i\d u^\alpha \wedge \d^{m-1}x_i - H \d^mx \quad ; \quad
\Omega_h = -\d p_\alpha^i \wedge \d u^\alpha \wedge \d^{m-1}x_i + \d H \wedge \d^{m}x \, .
\end{equation}

As in the Lagrangian formalism, the Hamiltonian problem for first-order field theories
consists in finding a distribution $\D$ in $J^0\pi^*$ such that
\begin{enumerate}
\item $\D$ is $m$-dimensional.
\item $\D$ is $\bar{\pi}^r_E$-transverse.
\item $\D$ is integrable.
\item The integral manifolds of $\D$ are the critical sections of the Hamilton-Jacobi principle.
\end{enumerate}

As in the Lagrangian formalism, these kinds of integrable distributions are associated with classes
of integrable and $\bar{\pi}^r_E$-transverse multivector fields $\X_h \in \vf^m(J^0\pi^*)$. In the
natural coordinates of $J^0\pi^*$, the local expression of an element of one of those classes is
\begin{equation*}
\X_h = f \bigwedge_{i=1}^{m} \left( \derpar{}{x^i} + f_i^\alpha \derpar{}{u^\alpha} + G_{\alpha,i}^{j} \derpar{}{p_\alpha^j} \right) \, ,
\end{equation*}
where $f \in \Cinfty(J^0\pi^*)$ is a non-vanishing function. Then, we have the following result.

\begin{theorem}\label{Chap02_thm:HamCFTFieldEqEquivalentWays}
Let $(J^0\pi^*,\Omega_h)$ be a Hamiltonian field theory. The following assertions on a section
$\psi \in \Gamma(\bar{\pi}^r_E)$ are equivalent.
\begin{enumerate}
\item The section $\psi$ is a solution to the equation
\begin{equation}\label{Chap02_eqn:HamCFTFieldEqSect}
\psi^*\inn(X)\Omega_h = 0\, , \quad \mbox{for every } X \in \vf(J^0\pi^*) \, .
\end{equation}
\item In the natural coordinates $(x^i,u^\alpha,p^i_\alpha)$ of $J^0\pi^*$,
the section $\psi \in \Gamma(\bar{\pi}^r_E)$ given locally by $\psi(x^i) = (x^i,u^\alpha(x^i),p_\alpha^i(x^i))$,
satisfies the Hamilton-De Donder-Weyl equations
\begin{equation}\label{Chap02_eqn:CFTHamiltonDeDonderWeylEquations}
\restric{\derpar{u^\alpha}{x^i}}{\psi} = \restric{\derpar{H}{p_\alpha^i}}{\psi} \quad ; \quad
\sum_{i=1}^{m}\restric{\derpar{p_\alpha^i}{x^i}}{\psi} = -\restric{\derpar{H}{u^\alpha}}{\psi}  \, .
\end{equation}
\item $\psi$ is an integral section of a class of locally decomposable, integrable and $\bar{\pi}^r_E$-transverse
multivector fields $\{ \X_h \} \subseteq \vf^{m}(J^0\pi^*)$ satisfying
\begin{equation}\label{Chap02_eqn:HamCFTFieldEqMultiVF}
\inn(\X_h)\Omega_h = 0 \, , \quad \mbox{for every } \X_h \in \{ \X_h \} \, .
\end{equation}
\end{enumerate}
\end{theorem}

The $\bar{\pi}^r_E$-transverse and locally decomposable multivector fields which are solutions to
equation \eqref{Chap02_eqn:HamCFTFieldEqMultiVF}, but are not necessarily integrable, are called
\textsl{Hamilton-De Donder-Weyl multivector fields}.

Let us compute in coordinates the equation \eqref{Chap02_eqn:HamCFTFieldEqMultiVF}. Let $\X_h \in \vf^{m}(J^0\pi^*)$
be a locally decomposable and $\bar{\pi}^r_E$-transverse multivector field locally given by
\begin{equation*}
\X_h = f \bigwedge_{i=1}^{m} \left( \derpar{}{x^i} + f_i^\alpha \derpar{}{u^\alpha} + G_{\alpha,i}^{j} \derpar{}{p_\alpha^j} \right) \, ,
\end{equation*}
where $f$ is a non-vanishing function. Then, taking $f = 1$ as a representative of the equivalence class,
and bearing in mind the coordinate expression \eqref{Chap02_eqn:HamCFTHamiltonCartanFormsLocal}
of the Hamilton-Cartan $(m+1)$-form $\Omega_h$, the equation
\eqref{Chap02_eqn:HamCFTFieldEqMultiVF} gives the following system of $(m+1)n$ equations
\begin{equation}\label{Chap02_eqn:HamCFTFieldEqMultiVFLocal}
f_i^\alpha = \derpar{H}{p_\alpha^i} \quad ; \quad
\sum_{i=1}^{m} G_{\alpha,i}^{i} = -\derpar{H}{u^\alpha} \, .
\end{equation}
From this coordinate expression we can assure the local existence of classes of locally decomposable
and $\bar{\pi}^r_E$-transverse multivector fields $\X_h \in \vf^m(J^0\pi^*)$ solution to equation
\eqref{Chap02_eqn:HamCFTFieldEqMultiVF}. The corresponding global solutions are then obtained using
a partition of unity subordinated to a covering of $J^0\pi^*$ made of local natural charts.

Finally, we can establish the equivalence between the Lagrangian and Hamiltonian formalisms in
the hyperregular case.

\begin{theorem}\label{Chap02_thm:HamCFTEquivalenceLagrangianHamiltonian}
Let $(J^1\pi,\Lag)$ be a hyperregular Lagrangian field theory, and $(J^0\pi^*,\Omega_h)$ the associated
Hamiltonian field theory.
\begin{enumerate}
\item If $\phi \in \Gamma(\pi)$ is a solution to equation \eqref{Chap02_eqn:LagCFTFieldEqSect}, then the section
$\psi = \Leg \circ j^1\phi \in \Gamma(\bar{\pi}^r_E)$ is a solution to equation \eqref{Chap02_eqn:HamCFTFieldEqSect}.
\item Conversely, if $\psi \in \Gamma(\bar{\pi}^r_E)$ is a solution to equation \eqref{Chap02_eqn:HamCFTFieldEqSect},
then the section $\phi = \pi^r_E \circ \psi \in \Gamma(\pi)$ is a solution to equation \eqref{Chap02_eqn:LagCFTFieldEqSect}.
\end{enumerate}
\end{theorem}

\begin{remark}
This last Theorem can be stated also in terms of multivector fields.
\end{remark}

Now, let us assume that the Lagrangian density $\Lag \in \df^{m}(J^1\pi)$ is almost-regular.
In this setting, the Legendre map is no longer a local diffeomorphism and, in particular,
$\P \hookrightarrow J^0\pi^*$ is a proper submanifold. Nevertheless, the Hamiltonian section
$h \in \Gamma(\widetilde{\mu})$ and the Hamilton-Cartan forms can still be defined. Therefore, the
field equations can be stated as in Theorem \ref{Chap02_thm:HamCFTFieldEqEquivalentWays}.

If $\Omega_h \in \df^{m+1}(\P)$ is a premultisymplectic form, Hamilton-De Donder-Weyl multivector
fields exist only, in the most favorable cases, in some submanifold $\P_f \hookrightarrow \P$, and
they are not necessarily integrable. As in the Lagrangian formulation, this submanifold $\P_f$
can be obtained using a suitable adapted version of the constraint algorithm described in Section
\ref{Chap01_sec:ConstraintAlgorithm} \cite{art:DeLeon_Marin_Marrero_Munoz_Roman05}.
Then, the analogous result to Theorem
\ref{Chap02_thm:HamCFTEquivalenceLagrangianHamiltonian} in the almost-regular case can be obtained.

\newpage

\subsection{Lagrangian-Hamiltonian unified formalism}
\label{Chap02_sec:FieldTheoriesUnified}

(See \cite{art:Echeverria_Lopez_Marin_Munoz_Roman04} for details).

\subsubsection{Unified phase spaces. Geometric structures}

As in the extended Hamiltonian formalism stated in the previous Section and the unified framework
for non-autonomous first-order dynamical systems stated in Section \ref{Chap02_sec:NonAutonomousUnified},
in the Skinner-Rusk formulation for first-order classical field theories we consider two phase spaces,
which are the bundles
\begin{equation*}
\W = J^1\pi \times_E \Lambda^m_2(\Tan^*E) \quad ; \quad
\W_r = J^1\pi \times_E J^0\pi^* \, .
\end{equation*}
These bundles are called the \textsl{extended jet-multimomentum bundle} and the \textsl{restricted jet-multimomentum bundle},
respectively. These bundles are endowed with the canonical projections
\begin{equation*}
\rho_1 \colon \W \to J^1\pi \quad ; \quad \rho_2 \colon \W \to \Lambda^m_2(\Tan^*E) \quad ; \quad
\rho_E \colon \W \to E \quad ; \quad \rho_M \colon \W \to M \, ,
\end{equation*}
\begin{equation*}
\rho_1^r \colon \W_r \to J^1\pi \quad ; \quad \rho_2^r \colon \W_r \to J^0\pi^* \quad ; \quad
\rho_E^r \colon \W_r \to E \quad ; \quad \rho_M^r \colon \W_r \to M \, .
\end{equation*}

In addition, the natural quotient map $\mu \colon \Lambda^m_2(\Tan^*E) \to J^0\pi^*$ induces a projection, that is,
a surjective submersion $\mu_\W \colon \W \to \W_r$. Hence, we have the same diagram that we have in Section
\ref{Chap02_sec:NonAutonomousUnified}, on page \pageref{Chap02_fig:UnifNABundleDiagram1}, replacing
$\Tan^*E$ by $\Lambda^m_2(\Tan^*E)$ and $\R$ by $M$.

Local coordinates in $\W$ and $\W_r$ are constructed in an analogous way to the autonomous
and non-autonomous formulations of dynamical systems.
Let $(x^i,u^\alpha)$, $1 \leqslant i \leqslant m$, $1 \leqslant \alpha \leqslant n$,
be a set of local coordinates in $E$ adapted
to the bundle structure and such that the fixed volume form $\eta \in \df^{m}(M)$ is given locally
by $\eta = \d^mx$. Then, the induced natural coordinates in $J^1\pi$, $\Lambda^m_2(\Tan^*E)$ and $J^0\pi^*$
are $(x^i,u^\alpha,u_i^\alpha)$, $(x^i,u^\alpha,p,p_\alpha^i)$ and $(x^i,u^\alpha,p_\alpha^i)$, respectively.
Therefore, the natural coordinates in $\W$ and $\W_r$ are $(x^i,u^\alpha,u_i^\alpha,p,p_\alpha^i)$ and
$(x^i,u^\alpha,u_i^\alpha,p_\alpha^i)$, respectively. Observe that $\dim\W = m+n+2mn+1$ and $\dim\W_r = m+n+2mn$.
  
In these coordinates, the above projections have the following coordinate expressions
\begin{equation*}
\rho_1(x^i,u^\alpha,u_i^\alpha,p,p_\alpha^i) = (x^i,u^\alpha,u_i^\alpha) \ ; \
\rho_2(x^i,u^\alpha,u_i^\alpha,p,p_\alpha^i) = (x^i,u^\alpha,p,p_\alpha^i) \ ; \
\rho_E(x^i,u^\alpha,u_i^\alpha,p,p_\alpha^i) = (x^i,u^\alpha) \, ,
\end{equation*}
\begin{equation*}
\rho_1^r(x^i,u^\alpha,u_i^\alpha,p_\alpha^i) = (x^i,u^\alpha,u_i^\alpha) \quad ; \quad
\rho_2^r(x^i,u^\alpha,u_i^\alpha,p_\alpha^i) = (x^i,u^\alpha,p_\alpha^i) \quad ; \quad
\rho_E^r(x^i,u^\alpha,u_i^\alpha,p_\alpha^i) = (x^i,u^\alpha) \, ,
\end{equation*}
\begin{equation*}
\rho_M(x^i,u^\alpha,u_i^\alpha,p,p_\alpha^i) = (x^i) \quad ; \quad
\rho_M^r(x^i,u^\alpha,u_i^\alpha,p,p_\alpha^i) = (x^i) \, .
\end{equation*}

The extended jet-multimomentum bundle is endowed with some canonical structures. First,
let $\Theta \in \df^{m}(\Lambda^m_2(\Tan^*E))$ and $\Omega = -\d\Theta \in \df^{m+1}(\Lambda^m_2(\Tan^*E))$
be the canonical forms of $\Lambda^m_2(\Tan^*E)$. Then, we define the following forms in $\W$
\begin{equation*}
\Theta_\W = \rho_2^*\Theta \in \df^{m}(\W) \quad ; \quad
\Omega_\W = \rho_2^*\Omega = -\d\Theta_\W \in \df^{m+1}(\W) \, .
\end{equation*}
Then, bearing in mind the coordinate expressions of the forms $\Theta$ and $\Omega$
given in Examples \ref{Chap01_exa:MulticotangentBundle} and \ref{Chap01_exa:MulticotangentBundleVertical},
and also in \eqref{Chap02_eqn:HamCFTCanonicalMultisymplecticForm}, and the local expression
of the projection $\rho_2$ given above, the forms $\Theta_\W$ and $\Omega_\W$ are given locally by
\begin{equation*}
\Theta_\W = p_\alpha^i \d u^\alpha \wedge \d^{m-1}x_i + p\d^mx \quad ; \quad
\Omega_\W = -\d p_\alpha^i \wedge \d u^\alpha \wedge \d^{m-1}x_i - \d p \wedge \d^mx \, .
\end{equation*}
It is clear from these coordinate expressions that $\Omega_\W$ is a closed $(m+1)$-form,
and that it is $1$-degenerate, since we have
\begin{equation*}
\inn(\partial/\partial u_i^\alpha)\Omega_\W = 0 \, , \quad \mbox{for every } 1 \leqslant i \leqslant m \, , 1 \leqslant \alpha \leqslant n \, .
\end{equation*}
In particular, a local basis of $\ker\Omega_\W$ is given by
\begin{equation*}
\ker\Omega_\W = \left\langle \derpar{}{u_i^\alpha} \right\rangle = \vf^{V(\rho_2)}(\W) \, .
\end{equation*}
Thus, the form $\Omega_\W$ is a premultisymplectic $(m+1)$-form in $\W$.

The second canonical structure in $\W$ is the following.

\begin{definition}
The \textnormal{coupling form} in $\W$ is the $\rho_M$-semibasic $m$-form
$\hat{\C} \in \df^{m}(\W)$ defined as follows: for every $w = (j_x^1\phi,\alpha) \in \W$
(that is, $\alpha \in \Lambda^m_2(\Tan^*_{\rho_E(w)}E)$) and $v_1,\ldots,v_m \in \Tan_w\W$, we have
\begin{equation*}
\hat{\C}(w)(v_1,\ldots,v_m) = \alpha(\Tan_w(\phi \circ \rho_M)v_1,\ldots,\Tan_w(\phi \circ \rho_M)v_m) \, .
\end{equation*}
\end{definition}

Since $\hat{\C} \in \df^{m}(\W)$ is $\rho_M$-semibasic, there exists a function
$\hat{C} \in \Cinfty(\W)$ such that $\hat{\C} = \hat{C}\rho_M^*\eta = \hat{C}\d^mx$.
An easy computation in coordinates gives the following local expression
for the coupling form
\begin{equation}\label{Chap02_eqn:UnifCFTCouplingFormLocal}
\hat{\C} = \left( p + p_\alpha^iu_i^\alpha \right) \d^mx \, .
\end{equation}

Given a Lagrangian density $\Lag \in \df^{m}(J^1\pi)$, we denote
$\hat{\Lag} = \rho_1^*\Lag \in \df^{m}(\W)$. Since the Lagrangian density
is $\bar{\pi}^1$-semibasic, then $\hat{\Lag}$ is $\rho_M$-semibasic,
and hence we can write $\hat{\Lag} = \hat{L}\rho_M^*\eta$,
where $\hat{L} = \rho_2^*L \in \Cinfty(\W)$, $L \in \Cinfty(J^1\pi)$ being the
Lagrangian function associated to $\Lag$ and $\eta$. Then we define a
\textsl{Hamiltonian submanifold}
\begin{equation*}
\W_o = \left\{ w \in \W \mid \hat{\Lag}(w) = \hat{\C}(w) \right\} \stackrel{j_o}{\hookrightarrow} \W \, .
\end{equation*}
Since both $\hat{\C}$ and $\hat{\Lag}$ are $\rho_M$-semibasic forms, the submanifold
$\W_o$ is defined by the regular constraint function $\hat{C} - \hat{L} \in \Cinfty(\W)$. In the natural
coordinates of $\W$, bearing in mind the local expression \eqref{Chap02_eqn:UnifCFTCouplingFormLocal}
of $\hat{\C}$, the constraint function is locally given by
\begin{equation*}
\hat{C} - \hat{L} = p + p_A^iu_i^\alpha - \hat{L}(x^i,u^\alpha,u_i^\alpha) \, .
\end{equation*}

\begin{proposition}
The submanifold $\W_o \hookrightarrow \W$ is $1$-codimensional, $\mu_\W$-transverse
and diffeomorphic to $\W_r$. This diffeomorphism is given by the map
$\mu_\W \circ j_o \colon \W_o \to \W_r$.
\end{proposition}

As a consequence of this last Proposition, the submanifold $\W_o$ induces a section
$\hat{h} \in \Gamma(\mu_\W)$ defined as
$\hat{h} = j_o \circ (\mu_\W \circ j_o)^{-1} \colon \W_r \to \W$. This section
is called the \textsl{Hamiltonian $\mu_\W$-section}, and is specified by giving the
\textsl{local Hamiltonian function}
\begin{equation*}
\hat{H}(x^i,u^\alpha,u_i^\alpha,p^i_\alpha) = -\hat{L}(x^i,u^\alpha,u_i^\alpha) + p_\alpha^iu_i^\alpha \, ,
\end{equation*}
that is, $\hat{h}(x^i,u^\alpha,u^\alpha_i,p_\alpha^i) = (x^i,u^\alpha,u_i^\alpha,-\hat{H}(x^i,u^\alpha,u_i^\alpha,p_\alpha^i),p_\alpha^i)$.

\begin{remark}
As in the unified formalism for non-autonomous dynamical systems described in Section \ref{Chap02_sec:NonAutonomousUnified},
if the Lagrangian density $\Lag \in \df^{m}(J^1\pi)$ is, at least, almost-regular, then from the
Hamiltonian $\mu_\W$-section $\hat{h} \in \Gamma(\mu_\W)$ in the unified formalism
we can recover the Hamiltonian $\mu$-section $h \in \Gamma(\mu)$ in the extended Hamiltonian
formalism. In fact, given $[\alpha] \in J^0\pi^*$, the section $\hat{h}$ maps every point
$(j^1_x\phi,[\alpha]) \in (\rho_2^r)^{-1}([\alpha])$ into $\rho_2^{-1}(\rho_2(\hat{h}(j^1_x\phi,[\alpha])))$.
Hence, the crucial point is the $\rho_2$-projectability of the local function $\hat{H}$.
However, since $\partial / \partial u_i^\alpha$ is a local basis for $\ker\Tan\rho_2$, the local function
$\hat{H}$ is $\rho_2$-projectable if, and only if, $p_\alpha^i = \partial\hat{L} / \partial u_i^\alpha$, and this
condition is fulfilled when $[\alpha] \in \P = \Im\Leg \hookrightarrow J^0\pi^*$, which implies
that $\rho_2(\hat{h}((\rho_2^r)^{-1}([\alpha]))) \in \widetilde{\P} = \Im\widetilde{\Leg} \hookrightarrow \Tan^*E$.
Then, the Hamiltonian section $h$ is defined as
\begin{equation*}
h([\alpha]) = (\rho_2 \circ \hat{h})((\rho_2^r)^{-1}(\jmath([\alpha])))
= (\tilde{\jmath} \circ \widetilde{\mu}^{-1})([\alpha]) \, .
\end{equation*}
for every $[\alpha] \in \P$.
\end{remark}

Finally, we can define the forms
\begin{equation*}
\Theta_r = \hat{h}^*\Theta_\W \in \df^{m}(\W_r) \quad ; \quad
\Omega_r = \hat{h}^*\Omega_\W \in \df^{m+1}(\W_r) \, ,
\end{equation*}
with local expressions
\begin{equation}\label{Chap02_eqn:UnifCFTPremultisymplecticFormsLocal}
\Theta_r = p^i_\alpha\d u^\alpha \wedge \d^{m-1} x_i + (\hat{L} - p_\alpha^iu_i^\alpha)\d^{m}x \ ; \
\Omega_r = - \d p^i_\alpha \wedge \d u^\alpha \wedge \d^{m-1} x_i + \d(p_\alpha^iu_i^\alpha - \hat{L}) \wedge \d^{m}x \, .
\end{equation}
Then, the pair $(\W_r,\Omega_r)$ is a premultisymplectic Hamiltonian field theory.

\subsubsection{Field equations for sections}

The \textsl{Lagrangian-Hamiltonian problem for sections} associated
with the field theory $(\W_r,\Omega_r)$ consists in finding sections
$\psi \in \Gamma(\rho_M^r)$ characterized by the condition
\begin{equation}\label{Chap02_eqn:UnifCFTFieldEqSect}
\psi^*\inn(X)\Omega_r = 0 \, , \quad \mbox{for every } X \in \vf(\W_r) \, .
\end{equation}

As in the Lagrangian and Hamiltonian formalisms described in previous Sections,
the Lagrangian-Hamiltonian problem for sections stated above is equivalent to
find a distribution $\D$ in $\W_r$ such that
\begin{enumerate}
\item $\D$ is $m$-dimensional.
\item $\D$ is $\rho_M^r$-transverse.
\item $\D$ is integrable.
\item The integral manifolds of $\D$ are the sections solution to equation
\eqref{Chap02_eqn:UnifCFTFieldEqSect}.
\end{enumerate}

In the natural coordinates of $\W_r$, if the section $\psi$
is locally given by $\psi(x^i) = (x^i,u^\alpha(x^i),u_i^\alpha(x^i),p_\alpha^i(x^i))$,
then, taking into account the local expression \eqref{Chap02_eqn:UnifCFTPremultisymplecticFormsLocal}
of the $(m+1)$-form $\Omega_r$, the equation \eqref{Chap02_eqn:UnifCFTFieldEqSect}
gives the following system of $(2m+1)n$ equations
\begin{align}
\derpar{u^\alpha}{x^i} = u_i^\alpha \, , \label{Chap02_eqn:UnifCFTFieldEqSectHolonomyLocal} \\
\sum_{i=1}^{m}\derpar{p_\alpha^i}{x^i} = \derpar{\hat{L}}{u^\alpha} \, , \label{Chap02_eqn:UnifCFTFieldEqSectLocal} \\
p_\alpha^i - \derpar{\hat{L}}{u_i^\alpha} = 0 \, . \label{Chap02_eqn:UnifCFTFieldEqSectLegendreLocal}
\end{align}
In an analogous way as in the unified formalism for non-autonomous systems described in Section
\ref{Chap02_sec:NonAutonomousUnified}, equations \eqref{Chap02_eqn:UnifCFTFieldEqSectHolonomyLocal} and
\eqref{Chap02_eqn:UnifCFTFieldEqSectLocal} are partial differential equations whose solutions
are the component functions of the section $\psi$. More particularly, equations
\eqref{Chap02_eqn:UnifCFTFieldEqSectHolonomyLocal} give the holonomy condition
for the section $\psi$ that must be satisfied once it is projected to $J^1\pi$,
while equations \eqref{Chap02_eqn:UnifCFTFieldEqSectLocal} are the actual field equations.
On the other hand, equations \eqref{Chap02_eqn:UnifCFTFieldEqSectLegendreLocal}
do not involve any derivative of the component functions: they are point-wise algebraic equations
that must satisfy every section $\psi \in \Gamma(\rho_M^r)$ to be a solution to equation
\eqref{Chap02_eqn:UnifCFTFieldEqSect}. These equations arise from the $\rho_2^r$-vertical component
of the vector fields $X$. In particular, we have the following result.

\begin{lemma}
If $X \in \vf^{V(\rho_2^r)}(\W_r)$, then $\inn(X)\Omega_r$ is a $\rho_M^r$-semibasic $m$-form.
\end{lemma}

As a consequence of this result, we can define the submanifold
\begin{equation*}
\W_\Lag = \left\{ [w] \in \W_r \mid (\inn(X)\Omega_r)([w]) = 0 \mbox{ for every } X \in \vf^{V(\rho_2^r)}(\W_r) \right\}
\stackrel{j_\Lag}{\hookrightarrow} \W_r \, ,
\end{equation*}
where every section solution to equation \eqref{Chap02_eqn:UnifCFTFieldEqSect} must take values.
Locally, the submanifold $\W_\Lag$ is defined by the constraints
$p_\alpha^i - \partial \hat{L} / \partial u_i^\alpha = 0$.
Moreover, we have the following characterization of $\W_\Lag$.

\begin{proposition}
$\W_\Lag \hookrightarrow \W_r$ is the graph of the restricted Legendre map $\Leg \colon J^1\pi \to J^0\pi^*$.
\end{proposition}

As a consequence of this result, since $\W_\Lag$ is the graph of the restricted Legendre map, then
it is diffeomorphic to $J^1\pi$. In addition, every section $\psi \in \Gamma(\rho_M^r)$ is of the
form $\psi = (\psi_\Lag,\psi_h)$, with $\psi_\Lag = \rho_1^r \circ \psi \in \Gamma(\bar{\pi}^1)$ and
$\psi_h = \Leg \circ \psi_\Lag \in \Gamma(\bar{\pi}_E^r)$. In this way, every constraint, differential
equation, etc., in the unified formalism can be translated to the Lagrangian and Hamiltonian formalisms
by projection to the first factor of the product bundle or using the Legendre map.
Hence, we have the following result.

\begin{theorem}
Let $\psi \in \Gamma(\rho_M^r)$ be a section solution to equation \eqref{Chap02_eqn:UnifCFTFieldEqSect}.
Then we have
\begin{enumerate}
\item The section $\psi_\Lag = \rho_1^r \circ \psi \in \Gamma(\bar{\pi}^{1})$ is holonomic, and is a solution
to equation \eqref{Chap02_eqn:LagCFTFieldEqSect}.
\item The section $\psi_h = \Leg \circ \psi_\Lag \in \Gamma(\bar{\pi}_E^{r})$ is a solution
to equation \eqref{Chap02_eqn:HamCFTFieldEqSect}.
\end{enumerate}
\end{theorem}

\subsubsection{Field equations for multivector fields}

As in the Lagrangian and Hamiltonian formalisms, if we assume that the sections
$\psi \in \Gamma(\rho_M^r)$ solutions to equation \eqref{Chap02_eqn:UnifCFTFieldEqSect}
are the integral sections of a class of locally decomposable, integrable and $\rho_M^r$-transverse
multivector fields, then we can state the problem in terms of multivector fields defined in $\W_r$.
The \textsl{Lagrangian-Hamiltonian problem for multivector fields} consists in finding
$m$-vector fields $\X \in \vf^{m}(\W_r)$ satisfying the above conditions and the following equation
\begin{equation}\label{Chap02_eqn:UnifCFTFieldEqMultiVF}
\inn(\X)\Omega_r = 0 \, .
\end{equation}

Since the $(m+1)$-form $\Omega_r$ is premultisymplectic on $\W_r$, we must use a suitable adaptation
of the constraint algorithm described in Section \ref{Chap01_sec:ConstraintAlgorithm} in order to find
a submanifold of $\W_r$ where the equation \eqref{Chap02_eqn:UnifCFTFieldEqMultiVF} is compatible.
From the algorithm, we can state the following result.

\begin{proposition}\label{Chap02_prop:UnifCFTFirstConstSubm}
Given the premultisymplectic Hamiltonian field theory $(\W_r,\Omega_r)$,
a $m$-vector field $\X$ solution to equation \eqref{Chap02_eqn:UnifCFTFieldEqMultiVF} exists only
on the points of the submanifold $\mathcal{S}_\Lag \hookrightarrow \W_r$ defined by
\begin{equation*}
\mathcal{S}_\Lag = \left\{ [w] \in \W_r \mid (\inn(X)\d\hat{H})([w]) = 0
\mbox{ for every } X \in \ker\Omega_\W \right\} \, .
\end{equation*}
\end{proposition}

As in the unified formulations for autonomous and non-autonomous dynamical systems described in
previous Sections we have the following characterization of the submanifold $\mathcal{S}_\Lag \hookrightarrow \W_r$.

\begin{proposition}\label{Chap02_prop:UnifCFTGraphLegendreMap}
The submanifold $\mathcal{S}_\Lag$ is the graph of the restricted Legendre map $\Leg \colon J^1\pi \to J^0\pi^*$,
and therefore $\mathcal{S}_\Lag = \W_\Lag$.
\end{proposition}

In the natural coordinates of $\W_r$, let $\X \in \vf^{m}(\W_r)$ be a locally decomposable and
$\rho_M^r$-transverse multivector field locally given by
\begin{equation*}
\X = f \bigwedge_{i=1}^{m}\left( \derpar{}{x^i} + f_i^\alpha \derpar{}{u^\alpha}
+ F_{j,i}^\alpha\derpar{}{u_j^\alpha} + G_{\alpha,i}^j\derpar{}{p^j_\alpha} \right) \, .
\end{equation*}
Now, taking $f=1$ as a representative of the equivalence class and
using the local expression \eqref{Chap02_eqn:UnifCFTPremultisymplecticFormsLocal}
of the $(m+1)$-form $\Omega_r$, the equation \eqref{Chap02_eqn:UnifCFTFieldEqMultiVF} gives
rise to the following system of $(2m+1)n$ equations for the component functions of $\X$
\begin{align}
f_i^\alpha = u_i^\alpha \, , \label{Chap02_eqn:UnifCFTFieldEqMultiVFHolonomyLocal} \\
\sum_{i=1}^{m}G_{\alpha,i}^i = \derpar{\hat{L}}{u^\alpha} \, , \label{Chap02_eqn:UnifCFTFieldEqMultiVFLocal} \\
p_\alpha^i - \derpar{\hat{L}}{u_i^\alpha} = 0 \, , \label{Chap02_eqn:UnifCFTFieldEqMultiVFLegendreLocal} \, .
\end{align}
Note that equations \eqref{Chap02_eqn:UnifCFTFieldEqMultiVFHolonomyLocal} are the holonomy
condition for a multivector field in the Lagrangian formalism, as we have seen in Section
\ref{Chap02_sec:LagFieldTheories}, while equations equations \eqref{Chap02_eqn:UnifCFTFieldEqMultiVFLocal}
are the field equations of the system.
On the other hand, equations \eqref{Chap02_eqn:UnifCFTFieldEqMultiVFLegendreLocal} are a
compatibility condition stating that the multivector fields solution to equations \eqref{Chap02_eqn:UnifCFTFieldEqMultiVF}
exist only with support on the submanifold defined as the graph of the restricted Legendre map.
Thus we recover, in coordinates, the result stated in Propositions \ref{Chap02_prop:UnifCFTFirstConstSubm} and
\ref{Chap02_prop:UnifCFTGraphLegendreMap}.

\begin{remark}
As in the previous unified formulations described in Sections \ref{Chap02_sec:SkinnerRuskAutonomousFirstOrder}
and \ref{Chap02_sec:NonAutonomousUnified}, the holonomy condition
is obtained regardless of the regularity of the Lagrangian density $\Lag \in \df^{m}(J^1\pi)$ provided.
\end{remark}

Therefore, a multivector field $\X \in \vf^{m}(\W_r)$ solution to equation \eqref{Chap02_eqn:UnifCFTFieldEqMultiVF}
is locally given by
\begin{equation*}
\X = f \bigwedge_{i=1}^{m}\left( \derpar{}{x^i} + u_i^\alpha \derpar{}{u^\alpha}
+ F_{j,i}^\alpha\derpar{}{u_j^\alpha} + G_{\alpha,i}^j\derpar{}{p^j_\alpha} \right) \, .
\end{equation*}

Observe that most of the component functions of the multivector field $\X$ remain undetermined,
even if we take into account equations \eqref{Chap02_eqn:UnifCFTFieldEqMultiVFLocal} relating
the component functions $G_{\alpha,i}^j$.
However, since the multivector field $\X$ is defined at support on the submanifold $\W_\Lag$, we must
study the tangency of $\X$ along this submanifold. Since $\X$ is locally decomposable, that is,
we have $\X = X_1 \wedge \ldots \wedge X_m$ on an open neighborhood around every point,
the tangency of $\X$ is equivalent to the tangency
of every $X_i$ along $\W_\Lag$. That is, we must require that
$\restric{\Lie(X_k)\xi}{\W_\Lag} = 0$ for every constraint function $\xi$ defining $\W_\Lag$ and
for every $1 \leqslant k \leqslant m$.
From Proposition \ref{Chap02_prop:UnifCFTGraphLegendreMap} the submanifold $\W_\Lag$
is the graph of the restricted Legendre map, and so it is defined by the $mn$
constraints
\begin{equation*}
\xi^\beta_j \equiv p_\beta^j - \derpar{\hat{L}}{u_j^\beta} = 0 \ , \ j = 1,\ldots,m \, , \, \beta = 1,\ldots,n \, .
\end{equation*}
Therefore, the tangency condition for $\X$ along $\W_\Lag$ gives the following $n$ equations
\begin{align*}
\Lie(X_k)\xi^\beta_j &=
\left(  \derpar{}{x^k} + u_k^\alpha \derpar{}{u^\alpha} + F_{i,k}^\alpha\derpar{}{u_i^\alpha} + G_{\alpha,k}^i\derpar{}{p^i_\alpha}  \right)
\left( p^j_\beta - \derpar{\hat{L}}{u_j^\beta} \right) \\
&= G_{\beta,k}^{j} - \derpars{\hat{L}}{x^k}{u_j^\beta} - u_k^\alpha\derpars{\hat{L}}{u^\alpha}{u_j^\beta}
- F_{i,k}^\alpha\derpars{\hat{L}}{u_i^\alpha}{u_j^\beta} = 0 \, .
\end{align*}
This first step enables us to determine the $m^2n$ functions $G_{\alpha,k}^j$ in terms of the
functions $F_{k,i}^\alpha$. Now, replacing this last expression in \eqref{Chap02_eqn:UnifCFTFieldEqMultiVFLocal},
we obtain
\begin{equation*}
\derpar{\hat{L}}{u^\beta} - \derpars{\hat{L}}{x^i}{u_i^\beta} - u_i^\alpha\derpars{\hat{L}}{u^\alpha}{u_i^\beta}
- F_{i,j}^\alpha\derpars{\hat{L}}{u_i^\alpha}{u_j^\beta} = 0 \, .
\end{equation*}
Observe that these are the Lagrangian equations for a multivector field once the holonomy condition is satisfied,
as we have seen in \eqref{Chap02_eqn:LagCFTFieldEqWithHolonomyLocal}. These equations can be
compatible or not, and a sufficient condition to ensure compatibility is the
regularity of the Lagrangian density, as we have seen in Section \ref{Chap02_sec:LagFieldTheories}.
If the Lagrangian density $\Lag$ is singular, then in the most favorable cases,
there exists a submanifold $\W_f \hookrightarrow \W_\Lag$
(it could be $\W_f = \W_\Lag)$ such that there exist multivector fields $\X \in \vf_{\W_f}^{m}(\W_r)$,
tangent to $\W_f$, which are solutions to the equation
\begin{equation*}\label{Chap02_eqn:UnifCFTFieldEqMultiVFSingular}
\restric{\left[ \inn(\X)\Omega_r \right]}{\W_f} = 0 \, .
\end{equation*}

Now, the equivalence of the unified formalism with the Lagrangian and Hamiltonian formalisms can
be recovered as follows.

\begin{theorem}
Let $\X \in \vf^{m}(\W_r)$ be a $\rho_M^r$-transverse and integrable
multivector field solution to equations \eqref{Chap02_eqn:UnifCFTFieldEqMultiVF}
and tangent to $\W_\Lag$ (at least on the points of a submanifold $\W_f \hookrightarrow \W_\Lag$).
Then the multivector field $\X_\Lag \in \vf^{m}(J^1\pi)$ defined by $\X_\Lag \circ \rho_1^r = \Lambda^{m}\Tan\rho_1^r \circ \X$
is holonomic, and is a solution to equation \eqref{Chap02_eqn:LagCFTFieldEqMultiVF} (at least on the
points of a submanifold $\mathcal{S}_f = \rho_1^r(\W_f) \hookrightarrow J^1\pi$).

\noindent Moreover, every holonomic multivector field $X_\Lag \in \vf^{1}(J^1\pi)$ solution to equation
\eqref{Chap02_eqn:LagCFTFieldEqMultiVF} (at least on the points of a submanifold $\mathcal{S}_f \hookrightarrow J^1\pi$)
can be recovered in this way from a $\rho_M^r$-transverse and integrable multivector field
$\X \in \vf^{m}(\W_r)$ solution to the equation \eqref{Chap02_eqn:UnifCFTFieldEqMultiVF}
and tangent to $\W_\Lag$ (at least on the points of a submanifold $\W_f \hookrightarrow \W_\Lag$).
\end{theorem}

Finally, the Hamiltonian formalism is recovered from the Lagrangian one using Theorem
\ref{Chap02_thm:HamCFTEquivalenceLagrangianHamiltonian}.


\clearpage
\chapter{Unified formalism for higher-order autonomous dynamical systems}
\label{Chap:HOAutonomousDynamicalSystems}


In this Chapter we state the Lagrangian-Hamiltonian unified formalism for higher-order
autonomous dynamical systems. That is, we generalize the unified formulation described
in Section \ref{Chap02_sec:SkinnerRuskAutonomousFirstOrder} to the higher-order
case described in Section \ref{Chap02_sec:AutonomousHigherOrder}.

The structure of the Chapter is the following. In Section \ref{Chap03_sec:GeometricalSetting}
we introduce the phase space where the formulation takes place, we construct the local
coordinates in this phase space and we define some canonical structures used in the
formulation. The dynamical equations are then stated and analyzed in Section
\ref{Chap03_sec:DynamicalEquations}, both in terms of vector fields and integral curves.
Then, Sections \ref{Chap03_sec:UnifiedToLagrangian} and \ref{Chap03_sec:UnifiedToHamiltonian}
are devoted to recover the Lagrangian and Hamiltonian formalisms for higher-order autonomous
dynamical systems described in Sections \ref{Chap02_sec:LagrangianAutonomousHigherOrder}
and \ref{Chap02_sec:HamiltonianAutonomousHigherOrder}, respectively. Finally, two physical models are studied
using this formulation in Section \ref{Chap03_sec:Examples}: the Pais-Uhlenbeck oscillator
and a second-order relativistic particle.

\section{Geometrical setting}
\label{Chap03_sec:GeometricalSetting}

Let us consider a $k$th-order autonomous Lagrangian dynamical system
with $n$ degrees of freedom. Let $Q$ be a $n$-dimensional smooth manifold
modeling the configuration space of this $k$th-order dynamical system, and
$\Lag \in \Cinfty(\Tan^{k}Q)$ a $k$th-order Lagrangian function describing
the dynamics of the system.

\subsection{Unified phase space and bundle structures. Local coordinates}
\label{Chap03_sec:GeometricalSettingPhaseSpace}

As we have seen in Section \ref{Chap02_sec:AutonomousHigherOrder}, the Lagrangian and
Hamiltonian phase spaces for a $k$th-order autonomous system are $\Tan^{2k-1}Q$ and
$\Tan^*(\Tan^{k-1}Q)$, respectively. Hence, following the patterns in Section
\ref{Chap02_sec:SkinnerRuskAutonomousFirstOrder}, let us consider the bundle
\begin{equation*}
\W = \Tan^{2k-1}Q \times_{\Tan^{k-1}Q} \Tan^*(\Tan^{k-1}Q) \, ,
\end{equation*}
that is, the fiber product over $\Tan^{k-1}Q$ of the Lagrangian and Hamiltonian phase spaces.

\begin{remark}
There is an alternative approach to this formulation, which consists in considering
the bundle $\W^\prime = \Tan^{k}Q \times_{\Tan^{k-1}Q} \Tan^*(\Tan^{k-1}Q)$ as the phase
space of the dynamical system, instead of the bundle $\W$ given above (see 
\cite{phd:Campos,art:Campos_deLeon_Martin10,art:Campos_DeLeon_Martin_Vankerschaver09,
master:Colombo,art:Colombo_Jimenez_Martin12,art:Colombo_Jimenez_Martin12_2,
art:Colombo_Martin11,art:Colombo_Martin_Zuccalli10} for several formulations
on different situations and systems). The similarities
and differences between these two approaches are pointed out along the Chapter.
\end{remark}

The bundle $\W$ is endowed with the canonical projections
\begin{equation*}
\rho_1 \colon \W \to \Tan^{2k-1}Q \quad ; \quad
\rho_2 \colon \W \to \Tan^*(\Tan^{k-1}Q) \, .
\end{equation*}
With these projections and the canonical projections of $\Tan^{2k-1}Q$ and $\Tan^*(\Tan^{k-1}Q)$
over $\Tan^{k-1}Q$, we have the following commutative diagram
\begin{equation*}
\xymatrix{
\ & \W \ar[dl]_-{\rho_1} \ar[dr]^-{\rho_2} & \ \\
\Tan^{2k-1}Q \ar[dr]_-{\rho^{2k-1}_{k-1}} \ar@/_1.3pc/[ddr]_-{\beta^{2k-1}} & \ &
 \Tan^*(\Tan^{k-1}Q) \ar[dl]^-{\pi_{\Tan^{k-1}Q}} \\
\ & \Tan^{k-1}Q \ar[d]_-{\beta^{k-1}} & \ \\
\ & Q & \
}
\end{equation*}

Local coordinates in $\W$ are constructed as follows.
Let $(U,q^A)$, $1 \leqslant A \leqslant n$, be a local chart in $Q$,
and $((\beta^{2k-1})^{-1}(U);(q_i^A,q_j^A))$ and
$((\beta^{k-1} \circ \pi_{\Tan^{k-1}Q})^{-1}(U);(q_i^A,p^i_A))$,
$1 \leqslant A \leqslant n$, $0 \leqslant i \leqslant k-1$, $k \leqslant j \leqslant 2k-1$,
the induced natural charts in $\Tan^{2k-1}Q$ and $\Tan^*(\Tan^{k-1}Q)$, respectively.
Then, the natural coordinates in the open set
$(\beta^{2k-1} \circ \rho_1)^{-1}(U) = (\beta^{k-1} \circ \pi_{\Tan^{k-1}Q} \circ \rho_2)^{-1}(U) \subseteq \W$
are $(q_i^A,q_j^A,p_A^i)$. Note that $\dim\W = 3kn$.

In these coordinates, the above projections have the following local expressions
\begin{equation*}
\rho_1(q_i^A,q_j^A,p_A^i) = (q_i^A,q_j^A) \quad ; \quad
\rho_2(q_i^A,q_j^A,p_A^i) = (q_i^A,p_A^i) \, .
\end{equation*}

\subsection{Canonical geometric structures}
\label{Chap03_sec:GeometricalSettingStructures}

The bundle $\W$ is endowed with some canonical geometric structures. In particular,
we generalize to the higher-order setting the definitions of the presymplectic $2$-form
and the coupling function given in Section \ref{Chap02_sec:SkinnerRuskAutonomousFirstOrder}.

Let $\theta_{k-1} \in \df^{1}(\Tan^*(\Tan^{k-1}Q))$ be the tautological form,
and $\omega_{k-1} = -\d\theta_{k-1} \in \df^{2}(\Tan^*(\Tan^{k-1}Q))$ the canonical
symplectic form of the cotangent bundle. Then, we define a $2$-form $\Omega$ in $\W$ as
\begin{equation*}
\Omega = \rho_2^*\,\omega_{k-1} \in \df^{2}(\W) \, .
\end{equation*}
It is clear from the definition that $\Omega$ is closed, since we have
\begin{equation*}
\Omega = \rho_{2}^{*}\,\omega_{k-1} = \rho_{2}^{*}\,(-\d\theta_{k-1}) = -\d\rho_{2}^{*}\,\theta_{k-1} \, .
\end{equation*}
Nevertheless, this form is degenerate, and therefore it is a presymplectic form.
Indeed, let $X \in \vf^{V(\rho_{2})}(\W)$. Then we have
\begin{equation*}
\inn(X)\Omega = \inn(X)\rho_{2}^*\,\omega_{k-1} = \rho_{2}^*(\inn(Y)\omega_{k-1}) \, ,
\end{equation*}
where $Y \in \vf(\Tan^{*}(\Tan^{k-1}Q))$ is a vector field $\rho_{2}$-related to $X$.
However, since $X$ is vertical with respect to $\rho_{2}$, we have $Y = 0$, and therefore
\begin{equation*}
\rho_{2}^{*}(\inn(Y)\omega_{k-1}) = \rho_{2}^{*}(\inn(0)\omega_{k-1}) = 0 \, .
\end{equation*}
In particular, $\{ 0 \} \varsubsetneq \vf^{V(\rho_2)}(\W) \subseteq \ker\Omega$,
and thus $\Omega$ is a degenerate $2$-form.

Bearing in mind the local expression of the projection $\rho_2$ given in the previous Section
and the coordinate expression \eqref{Chap02_eqn:HamHOCanonicalForms} of the symplectic form
$\omega_{k-1}$, we have that the $2$-form $\Omega$ is given locally by
\begin{equation}\label{Chap03_eqn:UnifPresymplecticFormLocal}
\Omega = \rho_2^*\,\omega_{k-1} = \rho_2^*\left(\d q_i^A \wedge \d p^i_A\right)
= \d\rho_2^*\left(q_i^A\right) \wedge \d\rho_2^*\left(p_A^i\right)  = \d q_i^A \wedge \d p^i_A \, .
\end{equation}
From this local expression it is clear that a local base for the kernel of $\Omega$ is
\begin{equation}\label{Chap03_eqn:UnifPresymplecticFormLocalKernel}
\ker\,\Omega = \left\langle \derpar{}{q_k^A},\ldots,\derpar{}{q_{2k-1}^A} \right\rangle = \vf^{V(\rho_2)}(\W) \ .
\end{equation}

The second geometric structure in $\W$ is the following.

\begin{definition}
Let $p = j^{2k-1}_0\phi \in \Tan^{2k-1}Q$, $q = j^{k-1}_0\phi = \rho^{2k-1}_{k-1}(p) \in \Tan^{k-1}Q$, and
$\alpha_q \in \Tan_{q}^*(\Tan^{k-1}Q)$. 
The \textnormal{$k$th-order coupling function} $\C \in \Cinfty(\W)$ is defined as follows:
\begin{equation*}\label{Chap03_eqn:UnifCouplingFunction}
\begin{array}{rcl}
\C \colon \Tan^{2k-1}Q \times_{\Tan^{k-1}Q} \Tan^*(\Tan^{k-1}Q) & \longrightarrow & \R \\
(p,\alpha_q) & \longmapsto & \langle \alpha_q , j_k(p)_q \rangle_{k-1}
\end{array} \, ,
\end{equation*}
where $j_k \colon \Tan^{2k-1}Q \to \Tan(\Tan^{k-1}Q)$ is the map defined
in \eqref{Chap01_eqn:HOTanBundleCanonicalImmersionDef},
$j_k(p)_q \in \Tan_{q}(\Tan^{k-1}Q)$ is the corresponding tangent vector to $\Tan^{k-1}Q$
in $q = j^{k-1}_0\phi$, and $\langle \bullet , \bullet \rangle_{k-1}$ denotes the canonical
pairing between the elements of the vector space  $\Tan_{q}(\Tan^{k-1}Q)$ and
the elements of its dual $\Tan_{q}^*(\Tan^{k-1}Q)$.
\end{definition}

In the natural coordinates of the bundle $\W$, if $p = j^{2k-1}_0\phi = (q_0^A,\ldots,q_{k-1}^A,q_k^A,\ldots,q_{2k-1}^A)$,
then $q = \rho^{2k-1}_{k-1}(p) = j^{k-1}_0\phi = (q_0^A,\ldots,q_{k-1}^A)$ and, bearing in mind the coordinate
expression \eqref{Chap01_eqn:HOTanBundleCanonicalImmersionLocal} of the map $j_k$, we
have $j_k(p) = (q_0^A,\ldots,q_{k-1}^A,q_1^A,\ldots,q_k^A)$. Therefore, if $j_k(p)_q$ and $\alpha_q$
are locally given by
\begin{equation*}
j_k(p)_q = q_{i+1}^A \restric{\derpar{}{q_i^A}}{q} \quad ; \quad
\alpha_q = p_A^i\restric{\d q_i^A}{q} \, ,
\end{equation*}
then we obtain the following coordinate expression for the $k$th-order coupling function $\C$
\begin{equation}\label{Chap03_eqn:UnifCouplingFunctionLocal}
\C(q_i^A,q_j^A,p_A^i)
= \left\langle p_A^i\restric{\d q_i^A}{q} , q_{i+1}^A \restric{\derpar{}{q_i^A}}{q} \right\rangle_{k-1}
= p_A^iq_i^A \, .
\end{equation}

\begin{remark}
Taking $k = 1$, the map $j_1 \colon \Tan Q \to \Tan Q$ is the identity on the tangent bundle, and hence
we recover the coupling function defined in Section \ref{Chap02_sec:SkinnerRuskAutonomousFirstOrder}
for first-order autonomous systems.
\end{remark}

From the $k$th-order coupling function $\C \in \Cinfty(\W)$ and the $k$th-order
Lagrangian function $\Lag \in \Cinfty(\Tan^{k}Q)$ provided, we define a
\textsl{Hamiltonian function} $H \in \Cinfty(\W)$ as
\begin{equation}\label{Chap03_eqn:UnifHamiltonianFunctionDef}
H = \C - (\rho^{2k-1}_k \circ \rho_1)^*\Lag \, .
\end{equation}
Doing an abuse of notation, in the following we denote $(\rho^{2k-1}_k \circ \rho_1)^*\Lag \in \Cinfty(\W)$
simply by $\Lag$. Bearing in mind the coordinate expression \eqref{Chap03_eqn:UnifCouplingFunctionLocal}
of the $k$th-order coupling function $\C$, we deduce that the Hamiltonian function $H$ is given locally by
\begin{equation}\label{Chap03_eqn:UnifHamiltonianFunctionLocal}
H(q_i^A,q_j^A,p_A^i) = p^i_Aq_{i+1}^A - \Lag(q_0^A,\ldots,q_k^A) \, .
\end{equation}

Therefore, we have constructed a presymplectic Hamiltonian system  $(\W,\Omega,H)$.

Finally, in order to give a complete description of the dynamics of higher-order Lagrangian systems
in terms of the unified formalism, we need to introduce the following concepts.

\begin{definition}
A curve $\psi \colon I \subset \R \to \W$ is \textnormal{holonomic of type $r$} in $\W$,
$1 \leqslant r \leqslant 2k-1$, if the curve $\psi_1 = \rho_1 \circ \psi \colon I \to \Tan^{2k-1}Q$
is holonomic of type $r$ in $\Tan^{2k-1}Q$, in the sense of Definition \ref{Chap01_def:HOTanBundleHolonomicCurve}.
\end{definition}

\begin{definition}
A vector field $X\in\vf(\W)$ is said to be a \textnormal{semispray of type $r$} in $\W$,
$1 \leqslant r \leqslant 2k-1$, if every integral curve $\psi \colon I \subset\R \to \W$
of $X$ is holonomic of type $r$ in $\W$.
\end{definition}

In the natural coordinates of $\W$, the local expression of a semispray of type $r$ in $\W$ is
\begin{equation*}
X = \sum_{i=0}^{2k-1-r}q_{i+1}^A\derpar{}{q_i^A} + \sum_{i=2k-r}^{2k-1}X_i^A\derpar{}{q_i^A}
+\sum_{i=0}^{k-1}G^i_A\derpar{}{p^i_A} \ ,
\end{equation*}
and, in particular, for a semispray of type $1$ in $\W$ we have
\begin{equation*}
X = \sum_{i=0}^{2k-2}q_{i+1}^A\derpar{}{q_i^A} + X_{2k-1}^A\derpar{}{q_{2k-1}^A}
+\sum_{i=0}^{k-1}G^i_A\derpar{}{p^i_A} \, .
\end{equation*}

\section{Dynamical equations}
\label{Chap03_sec:DynamicalEquations}

In this Section we state the dynamical equations for a $k$th-order autonomous dynamical system
in the unified formalism, both for vector fields and integral curves.

\subsection{Dynamical vector fields}
\label{Chap03_sec:DynamicalEquationsVF}

Following the patterns given in Section \ref{Chap02_sec:SkinnerRuskAutonomousFirstOrder},
the dynamical equation of the presymplectic Hamiltonian system $(\W,\Omega,H)$
is geometrically written in terms of vector fields as
\begin{equation}\label{Chap03_eqn:UnifDynEqVF}
\inn(X)\Omega = \d H \ , \ X \in \vf(\W) \, .
\end{equation}
As in the first-order formalism described in the aforementioned Section, the form $\Omega$ is presymplectic
and thus the equation \eqref{Chap03_eqn:UnifDynEqVF} may not admit a global solution $X \in \vf(\W)$.
From the constraint algorithm described in Section \ref{Chap01_sec:ConstraintAlgorithm} we have the
following result, which gives the first constraint submanifold of the system.

\begin{proposition}\label{Chap03_prop:UnifFirstConstraintSubmanifold}
Given the presymplectic Hamiltonian system $(\W,\Omega,H)$, a solution $X \in \vf(\W)$ to equation
\eqref{Chap03_eqn:UnifDynEqVF} exists only on the points of the submanifold $\W_c \hookrightarrow \W$
defined by
\begin{equation*}\label{Chap03_eqn:UnifFirstConstraintSubmanifoldDef}
\W_c = \left\{ w \in \W \mid (\inn(Y)\d H)(w) = 0 \ , \, \forall \, Y \in \ker\Omega \right\} \, .
\end{equation*}
\end{proposition}

In the natural coordinates of $\W$, let us compute the constraint functions defining locally the submanifold
$\W_c$. Taking into account the coordinate expression \eqref{Chap03_eqn:UnifHamiltonianFunctionLocal} of the
Hamiltonian function $H \in \Cinfty(\W)$, then its differential is locally given by
\begin{equation}\label{Chap03_eqn:UnifHamiltonianFunctionDifferentialLocal}
\d H = \sum_{i=0}^{k-1}(q_{i+1}^A\d p^i_A + p^i_A\d q_{i+1}^A) - \sum_{i=0}^{k} \derpar{\Lag}{q_i^A}\d q_i^A \, .
\end{equation}
Then, using the local basis for $\ker\Omega$ given in \eqref{Chap03_eqn:UnifPresymplecticFormLocalKernel},
we obtain
\begin{equation*}
\inn(Y)\d H =
\begin{cases}
\displaystyle p_A^{k-1} - \derpar{\Lag}{q_k^A} & \displaystyle \mbox{if } Y = \derpar{}{q_k^A} \\[10pt]
0 & \displaystyle \mbox{if } Y = \derpar{}{q_j^A} \, , \, j=k+1,\ldots,2k-1
\end{cases}
\end{equation*}
Therefore, $\W_c \hookrightarrow \W$ is a $n$-codimensional submanifold of $\W$ defined locally by the constraints
\begin{equation}\label{Chap03_eqn:UnifFirstConstraintSubmanifoldLocal}
p_A^{k-1} - \derpar{\Lag}{q_k^A} = 0 \, .
\end{equation}

In this setting, we do not have an analogous result to Proposition \ref{Chap02_prop:UnifAGraphLegendreMap},
that is, the submanifold $\W_c$ can not be characterized as the graph of the Legendre-Ostrogradsky map,
since $\W_c$ is $(3k-1)n$-dimensional, but the graph of the Legendre-Ostrogradsky map has dimension $2kn$.
Nevertheless, we can state the following result.

\begin{proposition}\label{Chap03_prop:UnifGraphLegendreOstrogradskyMap}
The submanifold $\W_c \hookrightarrow \W$ contains a submanifold $\W_\Lag \hookrightarrow \W_c$
which is the graph of the Legendre-Ostrogradsky map defined by $\Lag$; that is,
$\W_\Lag = \graph\Leg$.
\end{proposition}
\begin{proof}
We proceed in coordinates.
Since the submanifold $\W_c \hookrightarrow \W$ is defined locally by the constraints
\eqref{Chap03_eqn:UnifFirstConstraintSubmanifoldLocal}, it suffices to prove that
these constraints give rise to those defining locally the graph of the Legendre-Ostrogradsky
map associated to $\Lag$.

Observe that equations \eqref{Chap03_eqn:UnifFirstConstraintSubmanifoldLocal} relate the
highest-order momentum coordinates $p^{k-1}_A$ with the Jacobi-Ostrogradsky functions
$\displaystyle \hat{p}^{k-1}_A = \partial \Lag / \partial q_k^A$ defined in Section
\ref{Chap02_sec:LegendreOstrogradskyMap}, and so we obtain the last group of equations
of the Legendre-Ostrogradsky map. Furthermore, in the aforementioned Section we have seen that
the Jacobi-Ostrogradsky functions satisfy the relations \eqref{Chap02_eqn:HamHOMomentumCoordRelation}.
In particular, from the highest-order Jacobi-Ostrogradsky functions we can recover the full
set of $kn$ functions $\hat{p}_A^i$, and therefore we can consider that $\W_c$ contains a
submanifold $\W_\Lag$ which can be identified with the graph of a map
\begin{equation*}
\begin{array}{rcl}
F \colon \Tan^{2k-1}Q & \longrightarrow & \Tan^*(\Tan^{k-1}Q) \\ 
(q_i^A,q_j^A) & \longmapsto & (q_i^A,p^i_A) 
\end{array}
\end{equation*}
which we identify with the Legendre-Ostrogradsky map by making the identification
$p^{i}_A = \hat{p}^{i}_A$.
\end{proof}

\begin{remark}
If we use the bundle $\W' = \Tan^{k}Q \times_{\Tan^{k-1}Q} \Tan^*(\Tan^{k-1}Q)$, instead of the
bundle $\W$, then Proposition \ref{Chap03_prop:UnifFirstConstraintSubmanifold} remains the same,
but Proposition \ref{Chap03_prop:UnifGraphLegendreOstrogradskyMap} does not hold anymore, due to
dimension restrictions. In fact, $\dim \W' = (k+1)n + 2kn - kn = (2k+1)n$, and since $\W_c$
is a $n$-codimensional submanifold, we have $\dim\W_c = 2kn$ in this setting, which coincides
with the dimension of the submanifold defined by the graph of the Legendre-Ostrogradsky map.
However, it is clear that $\W_c \neq \graph\Leg$. Moreover, the Legendre-Ostrogradsky map
can not be fully recovered in this alternative approach, and this is its main drawback.
\end{remark}

\begin{remark}
It may seem convenient to take the submanifold $\W_\Lag \hookrightarrow \W$ as the initial
phase space of the system, instead of the submanifold $\W_c$, or any other submanifold of $\W_c$.
As we will see in the analysis of the dynamical equations, the submanifold $\W_\Lag$ can be obtained
from $\W_c$ using a constraint algorithm, and hence it is the natural choice as the initial phase space
of the system.
\end{remark}

Let us compute in coordinates the equation \eqref{Chap03_eqn:UnifDynEqVF}. Let $X \in \vf(\W)$
be a generic vector field locally given by
\begin{equation*}
X = f_i^A\derpar{}{q_i^A} + F_j^A\derpar{}{q_j^A} + G^i_A\derpar{}{p^i_A} \, ,
\end{equation*}
where $0 \leqslant i \leqslant k-1$ and $k \leqslant j \leqslant 2k-1$. Then,
bearing in mind the coordinate expression \eqref{Chap03_eqn:UnifPresymplecticFormLocal}
of the presymplectic form $\Omega$, the $1$-form $\inn(X)\Omega$ is locally given by
\begin{equation*}
\inn(X)\Omega = f_i^A\d p_A^i - G_A^i\d q_i^A \, .
\end{equation*}
Now, requiring equation \eqref{Chap03_eqn:UnifDynEqVF} to hold, and taking into account the
coordinate expression \eqref{Chap03_eqn:UnifHamiltonianFunctionDifferentialLocal} of $\d H$,
we obtain the following system of $(2k+1)n$ equations for the component functions of the vector field $X$
\begin{align}
f_i^A  = q_{i+1}^A \, , \label{Chap03_eqn:UnifDynEqVFHolonomyLocalPart1} \\
G^0_A  = \derpar{\Lag}{q_0^A} \quad , \quad G^i_A = \derpar{\Lag}{q_i^A} -p^{i-1}_A \, , \label{Chap03_eqn:UnifDynEqVFLocal} \\
p^{k-1}_A - \derpar{\Lag}{q_k^A} = 0 \, , \label{Chap03_eqn:UnifDynEqVFLegendreLocal}
\end{align}
where $0 \leqslant i \leqslant k-1$ in \eqref{Chap03_eqn:UnifDynEqVFHolonomyLocalPart1}
and $1 \leqslant i \leqslant k-1$ in \eqref{Chap03_eqn:UnifDynEqVFLocal}.
Therefore, the vector field $X$ solution to equation \eqref{Chap03_eqn:UnifDynEqVF} is given
in coordinates by
\begin{equation}\label{Chap03_eqn:UnifDynEqVFSolutionBeforeTangency}
X = q_{i+1}^A\derpar{}{q_i^A} + F_j^A\derpar{}{q_j^A} + \derpar{\Lag}{q_0^A}\derpar{}{p_A^0}
+ \left( \derpar{\Lag}{q_i^A} -p^{i-1}_A \right)\derpar{}{p^i_A} \, .
\end{equation}
Note that equations \eqref{Chap03_eqn:UnifDynEqVFHolonomyLocalPart1} are part of the system
of equations that the vector field $X$ must satisfy to be a semispray of type $1$.
In particular, from equations \eqref{Chap03_eqn:UnifDynEqVFHolonomyLocalPart1} we deduce that
$X$ is a semispray of type $k$, but not necessarily a semispray of type $1$.
On the other hand, equations \eqref{Chap03_eqn:UnifDynEqVFLegendreLocal} are a compatibility
condition stating that the vector fields $X$ solution to equation \eqref{Chap03_eqn:UnifDynEqVF} exist only with support
on the submanifold $\W_c$ given by Proposition \ref{Chap03_prop:UnifFirstConstraintSubmanifold}.
Finally, equations \eqref{Chap03_eqn:UnifDynEqVFLocal} are the dynamical equations of the system.

\begin{remark}
If we take $\W^\prime = \Tan^{k}Q \times_{\Tan^{k-1}Q} \Tan^*(\Tan^{k-1}Q)$ as the phase space of the
formalism, the coordinate expression of the dynamical equation \eqref{Chap03_eqn:UnifDynEqVF}
remains the same. However, in this case equations \eqref{Chap03_eqn:UnifDynEqVFHolonomyLocalPart1}
are exactly the $kn$ equations that enable us to recover the full holonomy condition for the vector field $X$.
That is, using this bundle as the phase space of the system implies that the vector field $X$ is always
a semispray of type $1$.
\end{remark}

Observe that from equations \eqref{Chap03_eqn:UnifDynEqVFHolonomyLocalPart1}
we deduce that the vector field $X$ is a semispray of type $k$ in $\W$, but not
necessarily a semispray of type $1$ in $\W$. Therefore, since this condition
is not recovered from the dynamical equations, in contrast to the first-order formalism
stated in Section \ref{Chap02_sec:SkinnerRuskAutonomousFirstOrder}, we can require the
holonomy condition to be fulfilled from the beginning. If we do so,
the local expression
\eqref{Chap03_eqn:UnifDynEqVFSolutionBeforeTangency} for a vector field
$X \in \vf(\W)$ solution to equation \eqref{Chap03_eqn:UnifDynEqVF} becomes
\begin{equation}\label{Chap03_eqn:UnifDynEqVFSolutionWithHolonomy}
X = \sum_{i=0}^{2k-2} q_{i+1}^A\derpar{}{q_i^A} + F_{2k-1}^A\derpar{}{q_{2k-1}^A} + \derpar{\Lag}{q_0^A}\derpar{}{p_A^0}
+ \left( \derpar{\Lag}{q_i^A} - p^{i-1}_A \right)\derpar{}{p^i_A} \, .
\end{equation}
Observe that the component functions $F_{2k-1}^A$, $k \leqslant j \leqslant 2k-1$, remain undetermined.
Nevertheless, from Proposition \ref{Chap03_prop:UnifFirstConstraintSubmanifold} and the local equations
\eqref{Chap03_eqn:UnifDynEqVFLegendreLocal}, the vector field $X$ is defined at support on the submanifold
$\W_c$. Therefore, we must study the tangency of $X$ along the submanifold $\W_c$; that is,
we have to impose that $\restric{\Lie(X)\xi}{\W_c} = 0$, for every constraint function $\xi$
defining $\W_c$. So, bearing in mind that the submanifold $\W_c$ is defined locally by the $n$
constraints \eqref{Chap03_eqn:UnifDynEqVFLegendreLocal}, we must require
\begin{equation*}
\left(\sum_{i=0}^{2k-2} q_{i+1}^A\derpar{}{q_i^A} + F_{2k-1}^A\derpar{}{q_{2k-1}^A} + \derpar{\Lag}{q_0^A}\derpar{}{p_A^0}
+ \left( \derpar{\Lag}{q_i^A} -p^{i-1}_A \right)\derpar{}{p^i_A}\right)
\left( p_A^{k-1} - \derpar{\Lag}{q_k^A} \right) = 0 \, .
\end{equation*}
Computing, these conditions lead to the following $n$ equations
\begin{equation*}
p^{k-2}_A - \sum_{i=0}^{1}(-1)^i d_T^i\left(\derpar{\Lag}{q_{k-1+i}^A}\right) = 0 \, ,
\end{equation*}
which define a new submanifold $\W_1 \hookrightarrow \W_c$. Then, requiring $X$ to be tangent
to this new submanifold $\W_1$, and iterating this process $k-1$ more times, the constraint
algorithm delivers the submanifold $\W_\Lag$ of Proposition \ref{Chap03_prop:UnifGraphLegendreOstrogradskyMap},
the full Legendre-Ostrogradsky map, and the following system of $n$ equations that must hold
for the vector field $X$ to be tangent to the submanifold $\W_\Lag$
\begin{equation*}
(-1)^k\left(F_{2k-1}^B - d_T\left(q_{2k-1}^B\right)\right) \derpars{\Lag}{q_k^B}{q_k^A} +
\sum_{i=0}^{k} (-1)^id_T^i\left( \derpar{\Lag}{q_i^A} \right) = 0 \, .
\end{equation*}
These are just the Lagrangian equations for the components of $X$ once the condition of being
a semispray of type $1$ is satisfied, as we have seen in \eqref{Chap02_eqn:LagHODynEqWithHolonomyLocal}.
These equations can be compatible or not, and a sufficient condition to ensure compatibility is the regularity of the
Lagrangian function, as we have seen in Proposition \ref{Chap02_prop:LagHORegLagUniqueVF}.

An alternative approach to the study of the tangency condition, without requiring the vector field
$X$ to be a semispray of type $1$ from the beginning, is the following. From the results in
Sections \ref{Chap02_sec:SkinnerRuskAutonomousFirstOrder}, \ref{Chap02_sec:NonAutonomousUnified}
and \ref{Chap02_sec:FieldTheoriesUnified}, we know that the vector field $X$ solution to the dynamical
equation \eqref{Chap03_eqn:UnifDynEqVF} in the unified formalism is defined at support on the submanifold
$\W_\Lag = \graph(\Leg)$, and is tangent to it. Therefore, it is natural to require the tangency
of the vector field $X$ solution to equation \eqref{Chap03_eqn:UnifDynEqVF} along the submanifold $\W_\Lag$,
without further assumptions. If we do so, bearing in mind the coordinate expression
\eqref{Chap02_eqn:HamHOLegendreMapLocal} of the Legendre-Ostrogradsky map
$\Leg \colon \Tan^{2k-1}Q \to \Tan^*(\Tan^{k-1}Q)$, the tangency condition leads to
\begin{align*}
\left(q_{i+1}^A\derpar{}{q_i^A} + F_j^A\derpar{}{q_i^A} + 
\derpar{\Lag}{q_0^A}\derpar{}{p^0_A} + \left( \derpar{\Lag}{q_i^A} -p^{i-1}_A \right)\derpar{}{p^i_A}\right)
\left( p^{k-1}_A - \derpar{\Lag}{q_k^A} \right) = 0 \, , \\
\left(q_{i+1}^A\derpar{}{q_i^A} + F_j^A\derpar{}{q_i^A} + 
\derpar{\Lag}{q_0^A}\derpar{}{p^0_A} + \left( \derpar{\Lag}{q_i^A} -p^{i-1}_A \right)\derpar{}{p^i_A}\right)
\left( p^{k-2}_A - \sum_{i=0}^{1}(-1)^i d_T^i\left(\derpar{\Lag}{q_{k-1+i}^A}\right)  \right) = 0 \, , \\
\vdots \qquad \qquad \qquad \qquad  \\
\left(q_{i+1}^A\derpar{}{q_i^A} + F_j^A\derpar{}{q_i^A} + 
\derpar{\Lag}{q_0^A}\derpar{}{p^0_A} + \left( \derpar{\Lag}{q_i^A} -p^{i-1}_A \right)\derpar{}{p^i_A}\right)
\left( p^{1}_A - \sum_{i=0}^{k-2}(-1)^i d_T^i\left(\derpar{\Lag}{q_{2+i}^A}\right)  \right) = 0 \, , \\
\left(q_{i+1}^A\derpar{}{q_i^A} + F_j^A\derpar{}{q_i^A} + 
\derpar{\Lag}{q_0^A}\derpar{}{p^0_A} + \left( \derpar{\Lag}{q_i^A} -p^{i-1}_A \right)\derpar{}{p^i_A}\right)
\left( p^{0}_A - \sum_{i=0}^{k-1}(-1)^i d_T^i\left(\derpar{\Lag}{q_{1+i}^A}\right)  \right) = 0 \, ,
\end{align*}
and, from here, we obtain the following $kn$ equations
\begin{equation}
\label{Chap03_eqn:UnifDynEqVFTangencyCondition}
\begin{array}{r}
\displaystyle \left(F_k^B-q_{k+1}^B\right)\derpars{\Lag}{q_k^B}{q_k^A} = 0 \, , \\[12pt]
\displaystyle \left(F_{k+1}^B - q_{k+2}^B\right)\derpars{\Lag}{q_k^B}{q_k^A} - 
\left(F_k^B-q_{k+1}^B \right) d_T\left(\derpars{\Lag}{q_k^B}{q_k^A}\right) = 0 \, , \\[10pt]
\vdots \qquad \qquad \qquad \qquad \\
\displaystyle \left(F_{2k-2}^B - q_{2k-1}^B\right)\derpars{\Lag}{q_k^B}{q_k^A} -
 \sum_{i=0}^{k-3} \left(F_{k+i}^B-q_{k+i+1}^B \right) (\cdots) = 0 \, , \\[12pt]
\displaystyle (-1)^k\left(F_{2k-1}^B - d_T\left(q_{2k-1}^B\right)\right) \derpars{\Lag}{q_k^B}{q_k^A} + 
\sum_{i=0}^{k} (-1)^id_T^i\left( \derpar{\Lag}{q_i^A} \right) - \sum_{i=0}^{k-2} \left(F_{k+i}^B-q_{k+i+1}^B \right)
 (\cdots) = 0 \, ,
\end{array}
\end{equation}
where the terms in brackets $(\cdots)$ contain relations involving partial derivatives
of the Lagrangian and applications of $d_T$ which for simplicity are not written.
These $kn$ equations are exactly the Lagrangian equations \eqref{Chap02_eqn:LagHODynEqHolonomyLocal}
and \eqref{Chap02_eqn:LagHODynEqLocal} for a vector field $X$ once the condition of semispray of
type $k$ is required. As in the first approach, these equations can be compatible or not, and a
sufficient condition to ensure compatibility is the regularity of the Lagrangian function.
In particular, we have the following result.

\begin{proposition}\label{Chap03_prop:UnifRegLagUniqueVF}
If $\Lag \in \Cinfty(\Tan^kQ)$ is a $k$th-order regular Lagrangian function, then there exists a unique
vector field $X \in \vf(\W)$ which is a solution to equation \eqref{Chap03_eqn:UnifDynEqVF},
it  is  tangent to $\W_\Lag$, and is a semispray of type $1$ in $\W$.
\end{proposition}
\begin{proof}
Since the $k$th-order Lagrangian function $\Lag$ is regular, the Hessian matrix
of $\Lag$ with respect to the highest-order ``velocities'' is regular at every point.
This enables us to solve the $kn$ equations \eqref{Chap03_eqn:UnifDynEqVFTangencyCondition}
determining all the functions $F_j^A$ uniquely, as follows
\begin{align}
\label{Chap03_eqn:UnifDynEqVFHolonomyLocalPart2}
F_i^A = q_{i+1}^A \quad , \quad  (k \leqslant i \leqslant 2k-2) \, , \\[5pt]
(-1)^k\left(F_{2k-1}^B - d_T\left(q_{2k-1}^B\right)\right) \derpars{\Lag}{q_k^B}{q_k^A} +
\sum_{i=0}^{k} (-1)^id_T^i\left( \derpar{\Lag}{q_i^A} \right) = 0 \, .
\label{Chap03_eqn:UnifDynEqVFTangencyConditionWithHolonomy}
\end{align}
Therefore, from the $(k-1)n$ equation \eqref{Chap03_eqn:UnifDynEqVFHolonomyLocalPart2}
we deduce that the vector field $X$ is a semispray of type $1$ in $\W$. On the other
hand, equations \eqref{Chap03_eqn:UnifDynEqVFTangencyConditionWithHolonomy} are exactly
equations \eqref{Chap02_eqn:LagHODynEqWithHolonomyLocal}, which are compatible and have
a unique solution when the $k$th-order Lagrangian function is regular. Therefore,
$X$ is a semispray of type $1$, it is tangent to $\W_\Lag$ and it is unique.
\end{proof}

If the $k$th-order Lagrangian function $\Lag$ is not regular, then equations
\eqref{Chap03_eqn:UnifDynEqVFTangencyCondition} can be compatible or not.
In the most favorable cases, there is a submanifold $\W_f \hookrightarrow \W_\Lag$
(it could be $\W_f = \W_\Lag$) such that there exist vector fields $X \in \vf(\W)$,
tangent to $\W_f$, which are solutions to the equation
\begin{equation}\label{Chap03_eqn:UnifDynEqVFSingular}
\restric{\left[\inn(X)\Omega - \d H\right]}{\W_f} = 0 \ .
\end{equation}
In this case, the equations \eqref{Chap03_eqn:UnifDynEqVFTangencyCondition}
are not compatible, and the compatibility condition gives rise to new constraints,
and the constraint algorithm continues.

\begin{remark}
If we take $\W^\prime = \Tan^{k}Q \times_{\Tan^{k-1}Q} \Tan^*(\Tan^{k-1}Q)$ as the phase space of the
formalism, there are only $n$ component functions of the vector field $X$ to be determined, since
the coordinate expression of $X$ is
\begin{equation*}
X = q_{i+1}^A\derpar{}{q_i^A} + F_k^A\derpar{}{q_k^A} + \derpar{\Lag}{q_0^A}\derpar{}{p_A^0}
+ \left( \derpar{\Lag}{q_i^A} -p^{i-1}_A \right)\derpar{}{p^i_A} \, .
\end{equation*}
Then, requiring $X$ to be tangent to the submanifold $\W_c$ gives the last $n$ equations
in \eqref{Chap03_eqn:UnifDynEqVFTangencyCondition}, that is,
\begin{equation*}
(-1)^k\left(F_{2k-1}^B - d_T\left(q_{2k-1}^B\right)\right) \derpars{\Lag}{q_k^B}{q_k^A} +
\sum_{i=0}^{k} (-1)^id_T^i\left( \derpar{\Lag}{q_i^A} \right) = 0 \, .
\end{equation*}
In this case, Proposition \ref{Chap03_prop:UnifRegLagUniqueVF} remains almost the same:
the only difference is that the vector field $X$ is already a semispray of type $1$,
regardless of the regularity of the $k$th-order Lagrangian function.
\end{remark}

\subsection{Integral curves}

After studying the vector fields which are solutions to the dynamical equations,
we analyze their integral curves, which are the dynamical trajectories of the system.

Let $X \in \vf(\W)$ be a semispray of type $1$, tangent to $\W_\Lag$,
which is a solution to equation \eqref{Chap03_eqn:UnifDynEqVF}, 
and let $\psi \colon I \subseteq \R \to \W$ be an integral curve of $X$.
Since $\dot{\psi} = X \circ \psi$, the geometric equation for the dynamical trajectories
of the system is
\begin{equation}\label{Chap03_eqn:UnifDynEqIC}
\inn({\dot{\psi}})(\Omega \circ \psi) = \d H \circ \psi \, .
\end{equation}

In local coordinates, if $\psi(t) = (q_i^A(t),q_j^A(t),p^i_A(t))$, we have that
$\dot{\psi}(t) = (\dot{q}_i^A(t),\dot{q}_j^A(t),\dot{p}_A^i(t))$. Then the condition for $\psi$
to be an integral curve of $X$ gives the following system of $3kn$ differential equations
\begin{align*}
\dot{q}_i^A(t) = q_{i+1}^A \circ \psi \, , \\[5pt]
\dot{q}_j^A(t) = F_j^A\circ\psi \, , \\[5pt]
\dot{p}^0_A(t) = \derpar{\Lag}{q_0^A} \circ \psi  \quad ; \quad
\dot{p}^i_A(t) = \derpar{\Lag}{q_i^A} \circ \psi - p_A^{i-1}(t) \, ,
\end{align*}
where the functions $F_j^A$ are solutions to equations \eqref{Chap03_eqn:UnifDynEqVFTangencyCondition}.

\section{The Lagrangian formalism}
\label{Chap03_sec:UnifiedToLagrangian}

Now we study how to recover the Lagrangian formalism described in Section
\ref{Chap02_sec:LagrangianAutonomousHigherOrder} from the unified setting.
In order to do this, we proceed in an analogous way to Section
\ref{Chap02_sec:SkinnerRuskAutonomousFirstOrder}: we first recover the
geometric and dynamical structures from the unified setting, and then we show how
to define a solution of the Lagrangian formalism from a solution in the unified setting.

\subsection{Geometric and dynamical structures}

The first step to recover the Lagrangian formalism from the unified setting described
in previous Sections is the recover the geometric and dynamical structures of the
Lagrangian formalism, namely the $k$th-order Poincar\'{e}-Cartan forms $\theta_\Lag$ and
$\omega_\Lag$, and the $k$th-order Lagrangian energy $E_\Lag$.

The first fundamental result is the following.

\begin{proposition}\label{Chap03_prop:LagRho1LDiffeomorphism}
The map $\rho_1^\Lag = \rho_1 \circ j_\Lag \colon \W_\Lag \to \Tan^{2k-1}Q$ is a diffeomorphism.
\end{proposition}
\begin{proof}
Since $\W_\Lag = \graph(\Leg)$, it is clear that $\Tan^{2k-1}Q$ is diffeomorphic to $\W_\Lag$.
On the other hand, since $\rho_1$ is a surjective submersion by definition, its restriction
to the submanifold $\W_\Lag$ is also a surjective submersion and, due to the fact that
$\dim\W_\Lag = \dim\Tan^{2k-1}Q = 2kn$, the map $\rho_1^\Lag$ is a bijective local diffeomorphism.
In particular, the map $\rho_1^\Lag$ is a global diffeomorphism.
\end{proof}

This result enables us to state a one-to-one correspondence between the solutions of the
unified formalism and the solutions of the Lagrangian formalism in a straightforward way.
Now, the following results enable us to recover the geometric and dynamical structures of
the Lagrangian formalism.

\begin{lemma}\label{Chap03_lemma:LagCartanForms}
If $\omega_{k-1} \in \df^2(\Tan^*(\Tan^{k-1}Q))$ is the canonical symplectic form
of the cotangent bundle over $\Tan^{k-1}Q$, $\Omega = \rho_2^*\,\omega_{k-1} \in \df^{2}(\W)$
the presymplectic form in $\W$, and $\omega_{\Lag} \in \df^{2}(\Tan^{2k-1}Q)$ the
$k$th-order Poincar\'{e}-Cartan $2$-form, then $\Omega = \rho_1^*\,\omega_\Lag$.
\end{lemma}
\begin{proof}
A simple calculation leads to this result. In fact,
\begin{equation*}
\rho_1^*\,\omega_{\Lag} = \rho_1^*(\Leg^*\omega_{k-1})=(\Leg \circ \rho_1)^*\omega_{k-1} = \rho_2^*\,\omega_{k-1} = \Omega \, .\qedhere
\end{equation*}
\end{proof}

\begin{lemma}\label{Chap03_lemma:LagLagrangianEnergy}
There exists a unique function $E_\Lag \in \Cinfty(\Tan^{2k-1} Q)$ such that $\rho_1^*E_\Lag = H$,
and coincides with the $k$th-order Lagrangian energy defined in \eqref{Chap02_eqn:LagHOEnergyDef}.
\end{lemma}
\begin{proof}
Since the map $\rho_1^\Lag \colon \W_\Lag \to \Tan^{2k-1}Q$ is a diffeomorphism, we define the
following function in $\Tan^{2k-1}Q$
\begin{equation*}
E_\Lag = (j_\Lag \circ (\rho_1^{\Lag})^{-1})^*H \in \Cinfty(\Tan^{2k-1}Q) \, .
\end{equation*}
This function is unique because the map $j_\Lag \circ (\rho_1^{\Lag})^{-1} \colon \Tan^{2k-1}Q \to \W$
is an embedding. In addition, $E_\Lag$ verifies that $\rho_1^*E_\Lag = H$, since we have
\begin{equation*}
\rho_1^*E_\Lag = \rho_1^*((j_\Lag \circ (\rho_1^{\Lag})^{-1})^*H)
= (\rho_1 \circ j_\Lag \circ (\rho_1^{\Lag})^{-1})^*H = (\rho_1^\Lag \circ (\rho_1^{\Lag})^{-1})^*H = H \, .
\end{equation*}

Finally, in order to prove that $E_\Lag$ is the $k$th-order Lagrangian energy defined
in \eqref{Chap02_eqn:LagHOEnergyDef}, we compute its coordinate expression.
Thus, from \eqref{Chap03_eqn:UnifHamiltonianFunctionLocal} we have
\begin{equation*}
(\rho_1^\Lag)^*E_\Lag = j_\Lag^*H = j_\Lag^*\left( p^i_Aq_{i+1}^A - \Lag(q_0^A,\ldots,q_k^A) \right) \, ,
\end{equation*}
but since $\W_\Lag \hookrightarrow \W$ is the graph of the Legendre-Ostrogradsky map, the following
relations hold in $\W_\Lag$
\begin{equation*}
p^i_A = \sum_{j=0}^{k-i-1}(-1)^j d_T^j\left(\derpar{\Lag}{q_{i+1+j}^A}\right) \, ,
\end{equation*}
and then, replacing in the previous equation, we obtain
\begin{align*}
(\rho_1^\Lag)^*E_\Lag &= \sum_{i=0}^{k-1}\sum_{j=0}^{k-i-1} q_{i+1}^A(-1)^j d_T^j\left(\derpar{\Lag}{q_{i+1+j}^A}\right) - 
\Lag(q_0^A,\ldots,q_k^A) \\
&= \sum_{i=1}^k \sum_{j=0}^{k-i}q_i^A (-1)^jd_T^j\left(\derpar{\Lag}{q_{i+j}^A}\right) - \Lag(q_0^A,\ldots,q_k^A) \, .
\end{align*}
Now, as $\rho_1^\Lag = \rho_1 \circ j_\Lag$ and $\rho_1^*q_i^A = q_i^A$, we obtain finally
\begin{equation*}
E_\Lag = \sum_{i=1}^k\sum_{j=0}^{k-i} q_i^A (-1)^jd_T^j\left(\derpar{\Lag}{q_{i+j}^A}\right) -\Lag(q_0^A,\ldots,q_k^A) \, ,
\end{equation*}
which is the local expression \eqref{Chap02_eqn:LagHOEnergyLocal} of the $k$th-order Lagrangian energy.
\end{proof}

\subsection{Dynamical equations}

Using the results stated in the previous Section, we can recover an Euler-Lagrange vector field
in $\Tan^{2k-1}Q$ starting from a vector field $X \in \vf(\W)$ solution to equation
\eqref{Chap03_eqn:UnifDynEqVF} and tangent to $\W_\Lag$. First, let us see how to define
a vector field in $\Tan^{2k-1}Q$ from a vector field in $\W$ tangent to $\W_\Lag$.

\begin{lemma}\label{Chap03_lemma:LagLagrangianVF}
Let $X \in \vf(\W)$ be a vector field tangent to $\W_\Lag$. 
Then there exists a unique vector field $X_\Lag \in \vf(\Tan^{2k-1}Q)$ such that
$X_\Lag \circ \rho_1 \circ j_\Lag = \Tan\rho_1 \circ X \circ j_\Lag$.
\end{lemma}
\begin{proof}
Since $X \in \vf(\W)$ is tangent to $\W_\Lag$, there exists a vector field $X_o \in \vf(\W_\Lag)$
which is $j_\Lag$-related to $X$, that is, $\Tan j_\Lag \circ X_o = X \circ j_\Lag$.
Furthermore, as $\rho_1^\Lag$ is a diffeomorphism, there is a unique vector field
$X_\Lag \in \vf(\Tan^{2k-1}Q)$ which is $\rho_1^\Lag$-related with $X_o$; that is,
$X_\Lag \circ \rho_1^\Lag = \Tan\rho_1^\Lag \circ X_o$. Then
\begin{equation*}
X_\Lag \circ \rho_1 \circ j_\Lag = X_\Lag \circ \rho_1^\Lag = \Tan\rho_1^\Lag \circ X_o = 
\Tan\rho_1 \circ \Tan j_\Lag \circ X_o = \Tan\rho_1 \circ X \circ j_\Lag \, . \qedhere
\end{equation*}
\end{proof}

And as a consequence we obtain:

\begin{theorem}\label{Chap03_thm:EquivUnifLagVF}
Let $X \in \vf(\W)$ be a vector field solution to equation \eqref{Chap03_eqn:UnifDynEqVF} and
tangent to $\W_\Lag$ (at least on the points of a submanifold $\W_f \hookrightarrow \W_\Lag$).
Then there exists a unique semispray of type $k$, $X_\Lag \in \vf(\Tan^{2k-1}Q)$,
which is a solution to the equation \eqref{Chap02_eqn:LagHODynEq} (at least on the points
of $S_f = \rho_1(\W_f)$). In addition, if $\Lag \in \Cinfty(\Tan^kQ)$ is a regular $k$th-order
Lagrangian function, then $X_\Lag$ is a semispray of type $1$, and hence it is the Euler-Lagrange vector field.

\noindent Conversely, if $X_\Lag \in \vf(\Tan^{2k-1}Q)$ is a semispray of type $k$ (resp., of type $1$),
which is a solution to equation \eqref{Chap02_eqn:LagHODynEq} (at least on the points of a submanifold
$S_f \hookrightarrow \Tan^{2k-1}Q$), then there exists a unique vector field $X \in \vf(\W)$,
tangent to $\W_\Lag$, which is a solution to equation \eqref{Chap03_eqn:UnifDynEqVF} (at least on
$\W_f = (\rho_1^\Lag)^{-1}(S_f) \hookrightarrow \W_\Lag \hookrightarrow \W$),
and it is a semispray of type $k$ in $\W$ (resp., of type $1$).
\end{theorem}
\begin{proof}
Let $X_\Lag \in \vf(\Tan^{2k-1}Q)$ be the unique vector field given by Lemma
\ref{Chap03_lemma:LagLagrangianVF}. Then, applying Lemmas \ref{Chap03_lemma:LagCartanForms}
and \ref{Chap03_lemma:LagLagrangianEnergy} we have
\begin{equation*}
\inn(X)\Omega - \d H = \inn(X)\rho_1^*\omega_\Lag - \d\rho_1^*E_\Lag =
\rho_1^*\left[\inn(X_\Lag)\omega_\Lag - \d E_\Lag\right] \, ,
\end{equation*}
but, as $\rho_1$ is a surjective submersion, this last equation is equivalent to
\begin{equation*}
\restric{\left[\inn(X_\Lag)\omega_\Lag - \d E_\Lag\right]}{\rho_1(\W)} = 
\restric{\left[\inn(X_\Lag)\omega_\Lag - \d E_\Lag\right]}{\Tan^{2k-1}Q} = 
\inn(X_\Lag)\omega_\Lag - \d E_\Lag \, ,
\end{equation*}
since $\rho_1(\W) = \Tan^{2k-1}Q$.
Hence, we have proved that $X \in \vf(\W)$ is vector field tangent to $\W_\Lag$
and solution to equation \eqref{Chap03_eqn:UnifDynEqVF} if, and only if,
the vector field $X_\Lag \in \vf(\Tan^{2k-1}Q)$ given by Lemma
\ref{Chap03_lemma:LagLagrangianVF} is a solution to equation \eqref{Chap02_eqn:LagHODynEq}.

In order to prove that $X_\Lag$ is a semispray of type $k$, we proceed in coordinates.
From the local expression \eqref{Chap03_eqn:UnifDynEqVFSolutionBeforeTangency} for the vector field $X$
solution to equation \eqref{Chap03_eqn:UnifDynEqVF} (where the functions $F_j^A$ are the solutions of the
system of equations \eqref{Chap03_eqn:UnifDynEqVFTangencyCondition}), and using Lemma
\ref{Chap03_lemma:LagLagrangianVF}, we obtain the local expression of $X_\Lag \in \vf(\Tan^{2k-1}Q)$, which is
\begin{equation*}
X_\Lag = q_{i+1}^A\derpar{}{q_i^A} + F_j^A\derpar{}{q_j^A} \, .
\end{equation*}
Then, composing $X_\Lag$ with the $k$th vertical endomorphism
$J_k \colon \Tan(\Tan^{2k-1}Q) \to \Tan(\Tan^{2k-1}Q)$, and bearing in mind the coordinate
expression of $J_k$ given by \eqref{Chap01_eqn:HoTanBundleVertEndR}, we have
\begin{equation*}
J_k(X_\Lag) = \sum_{i=0}^{k-1}\frac{(k+i)!}{i!}q_{i+1}^A\derpar{}{q_{k+i}^A} = \Delta_k \, ,
\end{equation*}
where $\Delta_k \in \vf(\Tan^{2k-1}Q)$ is the $k$th canonical vector field. Therefore,
since $J_k(X_\Lag) = \Delta_k$, using Proposition \ref{Chap01_prop:HOTanBundleEquivalenceSemisprays}
we conclude that $X_\Lag$ is a semispray of type $k$ in $\Tan^{2k-1}Q$.

Finally, if $\Lag \in \Cinfty(\Tan^kQ)$ is regular, then equations
\eqref{Chap03_eqn:UnifDynEqVFTangencyCondition} become \eqref{Chap03_eqn:UnifDynEqVFHolonomyLocalPart2}
and \eqref{Chap03_eqn:UnifDynEqVFTangencyConditionWithHolonomy} and hence the vector field $X$ is given
locally by \eqref{Chap03_eqn:UnifDynEqVFSolutionWithHolonomy}. Therefore, the vector field $X_\Lag$
has the the following coordinate expression
\begin{equation*}
X_\Lag = \sum_{i=0}^{2k-2}q_{i+1}^A\derpar{}{q_i^A} + F_{2k-1}^A\derpar{}{q_{2k-1}^A} \, .
\end{equation*}
Then, composing $X_\Lag$ with the canonical almost-tangent structure
$J_1 \colon \Tan(\Tan^{2k-1}Q) \to \Tan(\Tan^{2k-1}Q)$ of $\Tan^{2k-1}Q$, and bearing in mind
the coordinate expression \eqref{Chap01_eqn:HoTanBundleVertEnd1} of $J_1$, we have
\begin{equation*}
J_1(X_\Lag) = \sum_{i=0}^{2k-2}(i+1)q_{i+1}^A\derpar{}{q_{i+1}^A} = \Delta_1 \, ,
\end{equation*}
where $\Delta_1 \in \vf(\Tan^{2k-1}Q)$ is the Liouville vector field of $\Tan^{2k-1}Q$.
Hence, using again Proposition \ref{Chap01_prop:HOTanBundleEquivalenceSemisprays},
we conclude that if $\Lag \in \Cinfty(\Tan^{k}Q)$ is a $k$th-order regular Lagrangian function,
then $X_\Lag$ is a semispray of type $1$ in $\Tan^{2k-1}Q$.
\end{proof}

Observe that Theorem \ref{Chap03_thm:EquivUnifLagVF} states that there is a one-to-one
correspondence between vector fields $X \in \vf(\W)$ which are solutions to equation
\eqref{Chap03_eqn:UnifDynEqVF} and vector fields $X_\Lag \in \vf(\Tan^{2k-1}Q)$ solutions to
\eqref{Chap02_eqn:LagHODynEq}, but not uniqueness of any of them. In fact, uniqueness can only
be assured if the $k$th-order Lagrangian function $\Lag$ is regular, as we have seen in
Propositions \ref{Chap02_prop:LagHORegLagUniqueVF} and \ref{Chap03_prop:UnifRegLagUniqueVF}.

\begin{remark}
It is important to point out that, if $\Lag \in \Cinfty(\Tan^{k}Q)$ is not a $k$th-order
regular Lagrangian function, then $X$ is a semispray of type $k$ in $\W$,
but not necessarily a semispray of type $1$.
This means that the vector field $X_\Lag$ given by Theorem \ref{Chap03_thm:EquivUnifLagVF}
is a Lagrangian vector field, but it is not necessarily an Euler-Lagrange
vector field (it is not a semispray of type $1$, but just a semispray of type $k$).
Thus, for singular Lagrangians, this must be imposed as an additional condition in order that the
integral curves of $X_\Lag$ verify the Euler-Lagrange equations.
This is a different situation from the case of first-order dynamical systems described in
Section \ref{Chap02_sec:SkinnerRuskAutonomousFirstOrder}, where the holonomy condition
is obtained straightforwardly in the unified formalism.
\end{remark}

\begin{remark}
In general, only in the most interesting cases have we assured the existence of
a submanifold  $\W_f \hookrightarrow \W_\Lag$ and vector fields $X \in \vf(\W)$ tangent to $\W_f$
which are solutions to equation \eqref{Chap03_eqn:UnifDynEqVFSingular}. Then, considering the submanifold
$S_f = \rho_1(\W_f) \hookrightarrow \Tan^{2k-1}Q$, in the best cases (see
\cite{art:Batlle_Gomis_Pons_Roman88,art:Gracia_Pons_Roman91,art:Gracia_Pons_Roman92} for details), we have that those Euler-Lagrange
vector fields $X_\Lag$ exist, perhaps on another submanifold $S^h_f\hookrightarrow S_f$ where they are tangent,
and are solutions to the equation \eqref{Chap02_eqn:LagHODynEqSingular}
\end{remark}

Finally, let us recover the dynamical trajectories in the Lagrangian formalism from the dynamical
trajectories in the unified formalism. The following result enables us to
project the integral curves of a vector field in $\W$ solution of the dynamical equation in the unified
formalism to the integral curves of a vector field in $\Tan^{2k-1}Q$ solution to the Lagrangian equation.

\begin{proposition}\label{Chap03_prop:EquivUnifLagIC}
Let $\psi \colon I \subseteq \R \to \W$ be an integral curve of a vector field $X \in \vf(\W)$
tangent to $\W_\Lag$ and solution to equation \eqref{Chap03_eqn:UnifDynEqVF}.
Then the curve $\psi_\Lag = \rho_1 \circ \psi \colon I \to \Tan^{2k-1}Q$ is an integral curve
of a vector field solution to equation \eqref{Chap02_eqn:LagHODynEq}.
\end{proposition}
\begin{proof}
Since $X \in \vf(\W)$ is tangent to $\W_\Lag$, there exists a vector field $X_o \in \vf(\W_\Lag)$
which is $j_\Lag$-related to $X$. Moreover, since $\psi$ is an integral curve of $X$, every integral
of $X$ must lie in $\W_\Lag$, and thus we can write $\psi = j_\Lag \circ \psi_o$, where
$\psi_o \colon I \to \W_\Lag$ is a integral curve of $X_o$. Then, using Lemma
\ref{Chap03_lemma:LagLagrangianVF}, we have
\begin{align*}
X_\Lag \circ \psi_\Lag &= X_\Lag \circ \rho_1 \circ \psi = X_\Lag \circ \rho_1 \circ j_\Lag \circ \psi_o
= \Tan\rho_1 \circ X \circ j_\Lag \circ \psi_o \\
&= \Tan\rho_1 \circ X \circ \psi = \Tan\rho_1 \circ \dot{\psi}
= \dot{\overline{\rho_1 \circ \psi}} = \dot{\psi}_\Lag \, .
\end{align*}
Therefore, $\psi_\Lag = \rho_1 \circ \psi$ is an integral curve of $X_\Lag$.

Finally, from Theorem \ref{Chap03_thm:EquivUnifLagVF}, if $X$ is a solution to equation
\eqref{Chap03_eqn:UnifDynEqVF}, then the vector field $X_\Lag$ is a solution to equation
\eqref{Chap02_eqn:LagHODynEq}.
\end{proof}

\begin{remark}
In particular, this last Proposition states that if $\psi \colon I \subseteq \R \to \W$
is a solution to equation \eqref{Chap03_eqn:UnifDynEqIC}, then the curve
$\psi_\Lag = \rho_1 \circ \psi \colon I \subseteq \R \to \Tan^{2k-1}Q$ is a solution to equation
\eqref{Chap02_eqn:LagHODynEqIC}.
\end{remark}

Observe that the curve $\psi_\Lag$ is not necessarily holonomic, since the vector field $X_\Lag$
is not a semispray of type $1$ without further assumptions. This fact leads to the following result.

\begin{corollary}\label{Chap03_corol:HolonomyLagrangianIC}
If $\Lag \in \Cinfty(\Tan^kQ)$ is a $k$th-order regular Lagrangian function, then the curve
$\psi_\Lag = \rho_1 \circ \psi \colon I \to \Tan^{2k-1}Q$ is holonomic, that is,
it is the canonical lifting of a curve on $Q$.
\end{corollary}
\begin{proof}
It is a straighforward consequence of Proposition \ref{Chap03_prop:EquivUnifLagIC}
and Theorem \ref{Chap03_thm:EquivUnifLagVF}.
\end{proof}

\section{The Hamiltonian formalism}
\label{Chap03_sec:UnifiedToHamiltonian}

Now we study how to recover the Hamiltonian formalism described in Section
\ref{Chap02_sec:HamiltonianAutonomousHigherOrder} from the unified setting,
and we will proceed as in Section \ref{Chap02_sec:SkinnerRuskAutonomousFirstOrder}.
As in the usual formulation of the Hamiltonian formalism for higher-order autonomous
dynamical systems, we distinguish between regular and singular (almost-regular) Lagrangian functions.

Note that, since the geometric structures in the unified formalism are constructed
from the canonical forms in the Hamiltonian phase space, we only need to define
a Hamiltonian function $h$ in the Hamiltonian phase space using the Hamiltonian function $H$
defined in \eqref{Chap03_eqn:UnifHamiltonianFunctionDef} for the unified formalism.

\subsection{Hyperregular and regular Lagrangian functions}

Let us suppose that the $k$th-order Lagrangian function $\Lag \in \Cinfty(\Tan^{k}Q)$
is hyperregular. The regular case can be obtained from the hyperregular setting by
restriction on the corresponding open sets where the Legendre-Ostrogradsky map is a
local diffeomorphism.

As in the case of the Lagrangian formalism, the first fundamental result is the following.

\begin{proposition}\label{Chap03_prop:HamRegRho2LDiffeomorphism}
If $\Lag \in \Cinfty(\Tan^{k}Q)$ is a $k$th-order hyperregular Lagrangian function, then
the map $\rho_2^\Lag = \rho_2 \circ j_\Lag \colon \W_\Lag \to \Tan^*(\Tan^{k-1}Q)$ is a diffeomorphism.
\end{proposition}
\begin{proof}
As the $k$th-order Lagrangian function $\Lag \in \Cinfty(\Tan^{k}Q)$ is hyperregular, the Legendre-Ostrogradsky map
$\Leg \colon \Tan^{2k-1}Q \to \Tan^*(\Tan^{k-1}Q)$ is a diffeomorphism. Hence, the have the following
commutative diagram relating the phase spaces of the unified, Lagrangian and Hamiltonian formalisms
\begin{equation*}
\xymatrix{
\ & \W \ar@/_1.3pc/[ddl]_{\rho_1} \ar@/^1.3pc/[ddr]^{\rho_2} & \ \\
\ & \W_\Lag \ar@[red][dl]_{\rho_1^\Lag} \ar[dr]^{\rho_2^\Lag} \ar@{^{(}->}[u]^-{j_\Lag} & \ \\
\Tan^{2k-1}Q \ar[rr]^{\Leg} & \ & \Tan^*(\Tan^{k-1}Q)
}
\end{equation*}
In particular, we have $\rho_2^\Lag = \rho_2 \circ j_\Lag = \Leg \circ \rho_1^\Lag$. Therefore,
as $\rho_1^\Lag \colon \W_\Lag \to \Tan^{2k-1}Q$ is a diffeomorphism by Proposition \ref{Chap03_prop:LagRho1LDiffeomorphism}
and $\Leg$ is a diffeomorphism by hypothesis, we have that $\rho_2^\Lag$ is a composition of diffeomorphisms,
and thus a diffeomorphism itself.
\end{proof}

This last result enables us to recover the Hamiltonian dynamics straightforwardly from the
unified formalism, as we have done for the Lagrangian formalism in the previous Section.
In particular, the following result gives the Hamiltonian function in $\Tan^*(\Tan^{k-1}Q)$
describing the dynamical information of the system.

\begin{lemma}\label{Chap03_lemma:HamRegHamiltonianFunction}
Let $\Lag \in \Cinfty(\Tan^{k}Q)$ be a $k$th-order hyperregular Lagrangian function.
Then there exists a unique function $h \in \Cinfty(\Tan^*(\Tan^{k-1}Q))$ such that $\rho_2^*h = H$,
and it coincides with the canonical $k$th-order Hamiltonian function introduced in Definition
\ref{Chap02_def:HamHOHamiltonianFunctionDef}.
\end{lemma}
\begin{proof}
This proof follows the same patterns that the proof of Lemma
\ref{Chap03_lemma:LagLagrangianEnergy}.
Since by Proposition \ref{Chap03_prop:HamRegRho2LDiffeomorphism} the map
$\rho_2^\Lag \colon \W_\Lag \to \Tan^*(\Tan^{k-1}Q)$ is a diffeomorphism, we define the
following function in $\Tan^*(\Tan^{k-1}Q)$
\begin{equation*}
h = (j_\Lag \circ (\rho_2^{\Lag})^{-1})^*H \in \Cinfty(\Tan^*(\Tan^{k-1}Q)) \, .
\end{equation*}
This function is unique since the map $j_\Lag \circ (\rho_2^{\Lag})^{-1} \colon \Tan^*(\Tan^{k-1}Q) \to \W$
is the composition of a diffeomorphism with an embedding, and thus and embedding itself. Moreover,
this function verifies $\rho_2^*h = H$, since we have
\begin{equation*}
\rho_2^*h = \rho_2^*((j_\Lag \circ (\rho_2^{\Lag})^{-1})^*H)
= (\rho_2 \circ j_\Lag \circ (\rho_2^{\Lag})^{-1})^*H = (\rho_2^\Lag \circ (\rho_2^{\Lag})^{-1})^*H = H \, .
\end{equation*}

Finally, let us prove that this function $h \in \Cinfty(\Tan^*(\Tan^{k-1}Q))$ is the canonical
$k$th-order Hamiltonian function introduced in Definition \ref{Chap02_def:HamHOHamiltonianFunctionDef}.
Using Lemma \ref{Chap03_lemma:LagLagrangianEnergy} and the commutative diagram in the proof
of Proposition \ref{Chap03_prop:HamRegRho2LDiffeomorphism}, we have
\begin{align*}
\Leg^*h &= \Leg^*((j_\Lag \circ (\rho_2^{\Lag})^{-1})^*H) = (j_\Lag \circ (\rho_2^{\Lag})^{-1} \circ \Leg)^*H
= (j_\Lag \circ (\Leg \circ \rho_1^{\Lag})^{-1} \circ \Leg)^*H \\
&= (j_\Lag \circ (\rho_1^{\Lag})^{-1} \circ \Leg^{-1} \circ \Leg)^*H
= (j_\Lag \circ (\rho_1^{\Lag})^{-1})^*H = E_\Lag \, ,
\end{align*}
which is the definition of the canonical $k$th-order Hamiltonian function.
\end{proof}

Now that we have recovered the canonical $k$th-order Hamiltonian function in
$\Tan^*(\Tan^{k-1}Q)$, we want to recover the Hamiltonian vector field from
the vector field solution to the dynamical equation in the unified formalism.
First, let us see how to define a vector field in $\Tan^*(\Tan^{k-1}Q)$ from
a vector field in $\W$ tangent to $\W_\Lag$.

\begin{lemma}\label{Chap03_lemma:HamRegHamiltonianVF}
Let $X \in \vf(\W)$ be a vector field tangent to $\W_\Lag$. 
Then there exists a unique vector field $X_h \in \vf(\Tan^*(\Tan^{k-1}Q))$ such that
$X_h \circ \rho_2 \circ j_\Lag = \Tan\rho_2 \circ X \circ j_\Lag$.
\end{lemma}
\begin{proof}
This proof follows the patterns of the proof of Lemma \ref{Chap03_lemma:LagLagrangianVF}.
Since $X \in \vf(\W)$ is tangent to $\W_\Lag$, there exists a vector field $X_o \in \vf(\W_\Lag)$
which is $j_\Lag$-related to $X$, that is, $\Tan j_\Lag \circ X_o = X \circ j_\Lag$.
Furthermore, as $\rho_2^\Lag$ is a diffeomorphism, there is a unique vector field
$X_h \in \vf(\Tan^*(\Tan^{k-1}Q))$ which is $\rho_2^\Lag$-related with $X_o$; that is,
$X_h \circ \rho_2^\Lag = \Tan\rho_2^\Lag \circ X_o$. Then
\begin{equation*}
X_h \circ \rho_2 \circ j_\Lag = X_h \circ \rho_2^\Lag = \Tan\rho_2^\Lag \circ X_o = 
\Tan\rho_2 \circ \Tan j_\Lag \circ X_o = \Tan\rho_2 \circ X \circ j_\Lag \, . \qedhere
\end{equation*}
\end{proof}

Finally, as a consequence, we can state the equivalence between the vector fields
which are solutions to the dynamical equation \eqref{Chap03_eqn:UnifDynEqVF}
in the unified formalism and the vector fields solutions to the equation
\eqref{Chap02_eqn:HamHODynEq} in the Hamiltonian formalism.

\begin{theorem}\label{Chap03_thm:EquivUnifHamVFReg}
Let $\Lag \in \Cinfty(\Tan^kQ)$ be a $k$th-order hyperregular Lagrangian function,
and $X \in \vf(\W)$ the unique vector field solution to equation
\eqref{Chap03_eqn:UnifDynEqVF} and tangent to $\W_\Lag$. Then, there exists a
unique vector field $X_h \in \vf(\Tan^*(\Tan^{k-1}Q))$ which is a solution to
equation \eqref{Chap02_eqn:HamHODynEq}, where $h \in \Cinfty(\Tan^*(\Tan^{k-1}Q))$
is the Hamiltonian function given by Lemma \ref{Chap03_lemma:HamRegHamiltonianFunction}.

\noindent Conversely, if $X_h \in \vf(\Tan^*(\Tan^{k-1}Q))$ is a solution to equation
\eqref{Chap02_eqn:HamHODynEq}, then there exists a unique vector field $X \in \vf(\W)$,
tangent to $\W_\Lag$, which is a solution to equation \eqref{Chap03_eqn:UnifDynEqVF},
with $H = \rho_2^*h$.
\end{theorem}
\begin{proof}
This proof follows the same patterns that the proof of Theorem \ref{Chap03_thm:EquivUnifLagVF}.
Let $X_h \in \vf(\Tan^*(\Tan^{k-1}Q))$ be the unique vector field given by Lemma
\ref{Chap03_lemma:HamRegHamiltonianVF}. Then, applying Lemma
\ref{Chap03_lemma:HamRegHamiltonianFunction} we have
\begin{equation*}
\inn(X)\Omega - \d H = \inn(X)\rho_2^*\omega_{k-1} - \d\rho_2^*h =
\rho_2^*\left[\inn(X_h)\omega_{k-1} - \d h\right] \, ,
\end{equation*}
but, since $\rho_2$ is a surjective submersion, this last equation is equivalent to
\begin{equation*}
\restric{\left[\inn(X_h)\omega_{k-1} - \d h \right]}{\rho_2(\W)} = 
\restric{\left[\inn(X_h)\omega_{k-1} - \d h \right]}{\Tan^*(\Tan^{k-1}Q)} = 
\inn(X_h)\omega_{k-1} - \d h \, ,
\end{equation*}
since $\rho_2(\W) = \Tan^*(\Tan^{k-1}Q)$.
Hence, we have proved that $X \in \vf(\W)$ is vector field tangent to $\W_\Lag$
and solution to equation \eqref{Chap03_eqn:UnifDynEqVF} if, and only if,
the vector field $X_h \in \vf(\Tan^*(\Tan^{2k-1}Q))$ given by Lemma
\ref{Chap03_lemma:HamRegHamiltonianVF} is a solution to equation \eqref{Chap02_eqn:HamHODynEq}.
\end{proof}

The commutative diagram summarizing the statements and proofs of Theorems \ref{Chap03_thm:EquivUnifLagVF}
and \ref{Chap03_thm:EquivUnifHamVFReg} in the hyperregular case is the following
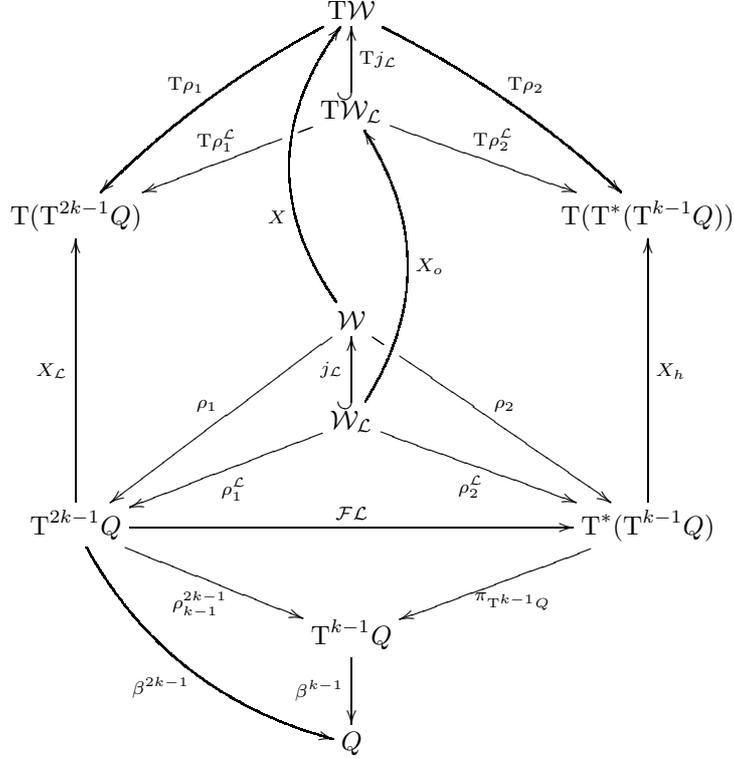
\begin{equation*}
\xymatrix{
\ & \ & \Tan\W \ar@/_/[ddll]_-{\Tan\rho_1} \ar@/^/[ddrr]^-{\Tan\rho_2} & \ & \  \\
\ & \ & \Tan\W_\Lag \ar[dll]_-{\Tan\rho^\Lag_1}|<(.095){\hole} \ar[drr]^-{\Tan\rho^\Lag_2} \ar@{^{(}->}[u]_-{\Tan j_\Lag} & \ & \  \\
\Tan(\Tan^{2k-1} Q) & \ & \ & \ & \Tan(\Tan^*(\Tan^{k-1}Q)) \\
\ & \ & \W \ar[ddll]_-{\rho_1} \ar[ddrr]^-{\rho_2}|<(.07){\hole} \ar@/^1.95pc/[uuu]^(.35){X} & \ & \ \\
\ & \ & \W_\Lag \ar[dll]^-{\rho^\Lag_1} \ar[drr]_-{\rho^\Lag_2} \ar@{^{(}->}[u]^-{j_\Lag} 
\ar@/_1.75pc/[uuu]_-{X_o} & \ & \ \\
\Tan^{2k-1}Q \ar[drr]_-{\rho^{2k-1}_{k-1}} \ar[rrrr]^-{\Leg} \ar[uuu]^-{X_\Lag} \ar@/_1.3pc/[ddrr]_-{\beta^{2k-1}} 
& \ & \ & \ & \Tan^*(\Tan^{k-1}Q) \ar[dll]^-{\pi_{\Tan^{k-1}Q}} \ar[uuu]_-{X_h} \\
\ & \ & \Tan^{k-1}Q \ar[d]_-{\beta^{k-1}} & \ & \ \\
\ & \ & Q & \ & \ \\
}
\end{equation*}

Now, let us recover the dynamical trajectories in the Hamiltonian formalism from the dynamical
trajectories in the unified setting. The following result enables us to
project the integral curves of a vector field in $\W$ solution of the dynamical equation in the unified
formalism to the integral curves of a vector field in $\Tan^*(\Tan^{k-1}Q)$ solution to the Hamiltonian equation.

\begin{proposition}\label{Chap03_prop:EquivUnifHamICReg}
Let $\psi \colon I \subseteq \R \to \W$ be an integral curve of a vector field $X \in \vf(\W)$
tangent to $\W_\Lag$ and solution to equation \eqref{Chap03_eqn:UnifDynEqVF}.
Then the curve $\psi_h = \rho_2 \circ \psi \colon I \to \Tan^*(\Tan^{k-1}Q)$ is the integral curve
of a vector field solution to equation \eqref{Chap02_eqn:HamHODynEq}.
\end{proposition}
\begin{proof}
The proof of this result follows the same patterns that the proof of Proposition
\ref{Chap03_prop:EquivUnifLagIC}.
Since $X \in \vf(\W)$ is tangent to $\W_\Lag$, there exists a vector field $X_o \in \vf(\W_\Lag)$
which is $j_\Lag$-related to $X$. Moreover, since $\psi$ is an integral curve of $X$, every integral
of $X$ must lie in $\W_\Lag$, and thus we can write $\psi = j_\Lag \circ \psi_o$, where
$\psi_o \colon I \to \W_\Lag$ is a integral curve of $X_o$. Then, using Lemma
\ref{Chap03_lemma:HamRegHamiltonianVF}, we have
\begin{align*}
X_h \circ \psi_h &= X_h \circ \rho_2 \circ \psi = X_h \circ \rho_2 \circ j_\Lag \circ \psi_o
= \Tan\rho_2 \circ X \circ j_\Lag \circ \psi_o \\
&= \Tan\rho_2 \circ X \circ \psi = \Tan\rho_2 \circ \dot{\psi}
= \dot{\overline{\rho_2 \circ \psi}} = \dot{\psi}_h \, .
\end{align*}
Therefore, $\psi_h = \rho_2 \circ \psi$ is an integral curve of $X_h$.

Finally, from Theorem \ref{Chap03_thm:EquivUnifHamVFReg}, if $X$ is a solution to equation
\eqref{Chap03_eqn:UnifDynEqVF}, then the vector field $X_h$ is a solution to equation
\eqref{Chap02_eqn:HamHODynEq}.
\end{proof}

\begin{remark}
In particular, this last Proposition states that if $\psi \colon I \to \W$
is a solution to equation \eqref{Chap03_eqn:UnifDynEqIC}, then the curve
$\psi_\Lag = \rho_1 \circ \psi \colon I \to \Tan^{2k-1}Q$ is a solution to equation
\eqref{Chap02_eqn:HamHODynEqIC}.
\end{remark}

Finally, the commutative diagram summarizing the statements and proofs of Propositions
\ref{Chap03_prop:EquivUnifLagIC} and \ref{Chap03_prop:EquivUnifHamICReg}, and of Corollary
\ref{Chap03_corol:HolonomyLagrangianIC}, in the hyperregular case is the following
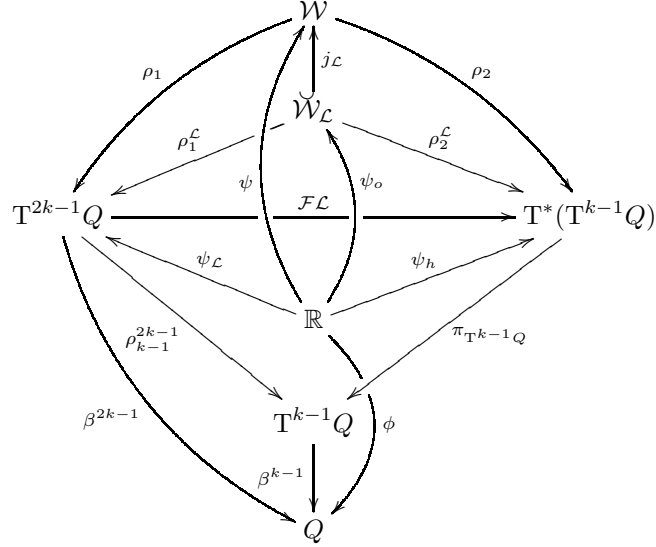
\begin{equation}\label{Chap03_fig:DiagramIC}
\xymatrix{
\ & \ & \W \ar@/_1pc/[ddll]_-{\rho_1} \ar@/^1pc/[ddrr]^-{\rho_2} & \ & \ \\
\ & \ & \W_\Lag \ar[dll]_(.45){\rho^\Lag_1}|(.2){\hole} \ar[drr]^-{\rho^\Lag_2} \ar@{^{(}->}[u]_-{j_\Lag} & \ & \ \\
\Tan^{2k-1}Q \ar[ddrr]_-{\rho^{2k-1}_{k-1}} 
\ar@/_1.5pc/[dddrr]_-{\beta^{2k-1}} \ar[rrrr]^-{\Leg}|(.39){\hole}|(.56){\hole} & \ & \ & \ & \Tan^*(\Tan^{k-1}Q) \ar[ddll]^-{\pi_{\Tan^{k-1}Q}} \\
\ & \ & \R \ar@/^1.9pc/[dd]^-{\phi}|(.31){\hole} \ar@/_1.3pc/[uu]_(.65){\psi_o} 
\ar@/^1.65pc/[uuu]^(.45){\psi} \ar[ull]_-{\psi_\Lag} \ar[urr]^-{\psi_h} & \ & \ \\
\ & \ & \Tan^{k-1}Q \ar[d]_-{\beta^{k-1}} & \ & \ \\
\ & \ & Q & \ & \
}\nonumber
\end{equation}

\subsection{Singular (almost-regular) Lagrangian functions}

Suppose now that the $k$th-order Lagrangian function $\Lag \in \Cinfty(\Tan^kQ)$
is almost-regular. Remember that, for these kinds of Lagrangian functions,
only in the most interesting cases have we assured the existence of a submanifold
$\W_f \hookrightarrow \W_\Lag$  and vector fields $X \in \vf(\W)$, tangent to $\W_f$,
which are solutions to equation \eqref{Chap03_eqn:UnifDynEqVFSingular}. In this case,
the dynamical vector fields in the Hamiltonian formalism cannot be obtained straightforwardly
from the solutions in the unified formalism, but rather by passing through the Lagrangian
formalism and using the Legendre-Ostrogradsky map, which is no longer a (local) diffeomorphism.

As in the Hamiltonian formalism for almost-regular Lagrangian functions described in
Section \ref{Chap02_sec:HamiltonianAutonomousHigherOrder}, let
$\P = \Im(\Leg) \hookrightarrow \Tan^*(\Tan^{k-1}Q)$ be the image set of the Legendre-Ostrogradsky map,
with natural embedding $\jmath \colon \P \hookrightarrow \Tan^*(\Tan^{k-1}Q)$, and we denote by
$\Leg_o \colon \Tan^{2k-1} Q \to \P$ the map defined by $\Leg = \jmath \circ \Leg_o$.
In addition, let $\rho_\P = \Leg_o \circ \rho_1 \colon \W \to \P$ the canonical projection.
Then, we have the following result.

\begin{proposition}
Let $\Lag \in \Cinfty(\Tan^{k}Q)$ be a $k$th-order almost-regular Lagrangian function.
Then the Hamiltonian function $H \in \Cinfty(\W)$ is $\rho_\P$-projectable.
\end{proposition}
\begin{proof}
From Lemma \ref{Chap03_lemma:LagLagrangianEnergy}, the function $H \in \Cinfty(\W)$
is $\rho_1$-projectable to the $k$th-order Lagrangian energy $E_\Lag \in \Cinfty(\Tan^{2k-1}Q)$.
Moreover, if the $k$th-order Lagrangian function is, at least, almost-regular, then
the $k$th-order Lagrangian energy is $\Leg_o$-projectable by Proposition
\ref{Chap02_prop:HamHOLagrangianEnergyProjectable}. Therefore, the Hamiltonian function
$H \in \Cinfty(\W)$ is $(\Leg_o \circ \rho_1)$-projectable, that is, $\rho_\P$-projectable.
\end{proof}

As a consequence of this result, we can define a Hamiltonian function in $\P$ as follows.

\begin{definition}
The \textnormal{canonical Hamiltonian function} is the unique function
$h_o \in \Cinfty(\P)$ such that $\rho_\P^*h_o = H$.
\end{definition}

\begin{remark}
This canonical Hamiltonian function coincides with the canonical Hamiltonian function
introduced in Definition \ref{Chap02_def:HamHOHamiltonianFunctionDefSingular}.
\end{remark}

With this canonical Hamiltonian function, and taking
$\omega_o = \jmath^*\omega_{k-1} \in \df^{2}(\P)$, we can now state the equivalence
Theorem for $k$th-order almost-regular Lagrangian functions.

\begin{theorem}\label{Chap03_thm:EquivUnifHamVFSing}
Let $\Lag \in \Cinfty(\Tan^kQ)$ be a $k$th-order almost-regular Lagrangian function,
and $X \in \vf(\W)$ a vector field tangent to $\W_f$ which is a solution to equation
\eqref{Chap03_eqn:UnifDynEqVFSingular}. Then, there exists a
vector field $X_{h_o} \in \vf(\P)$ tangent to $\P_f = \rho_\P(\W_f)$ which is a solution to
the equation \eqref{Chap02_eqn:HamHODynEqSingular2}, where $h_o \in \Cinfty(\P)$
is the canonical Hamiltonian function defined above.

\noindent Conversely, if $X_{h_o} \in \vf(\P)$ is a solution to equation
\eqref{Chap02_eqn:HamHODynEqSingular2} and tangent to $\P_f$, then there exist
vector fields $X \in \vf(\W)$, tangent to $\W_f = \rho_\P^{-1}(\P_f)$, which are
solutions to equation \eqref{Chap03_eqn:UnifDynEqVFSingular}, with $H = \rho_\P^*h_o$.
\end{theorem}
\begin{proof}
From Theorem \ref{Chap03_thm:EquivUnifLagVF} there is a one-to-one correspondence
between the set of vector fields solution to equation \eqref{Chap03_eqn:UnifDynEqVFSingular}
and tangent to $\W_f$, and the set of vector fields solution to equation
\eqref{Chap02_eqn:LagHODynEqSingular} and tangent to $S_f$. From here,
using Theorem \ref{Chap02_thm:HamHORelationLagHamSingular}, we obtain a non-bijective
correspondence, given by the Legendre-Ostrogradsky map, between these vector fields and
the set of vector fields in $\P$, tangent to $\P_f$, which are solutions to equation
\eqref{Chap02_eqn:HamHODynEqSingular2}, thus proving the statement.
\end{proof}

The diagram summarizing the statements and proofs of Theorems \ref{Chap03_thm:EquivUnifLagVF}
and \ref{Chap03_thm:EquivUnifHamVFSing} in the almost-regular case is the following
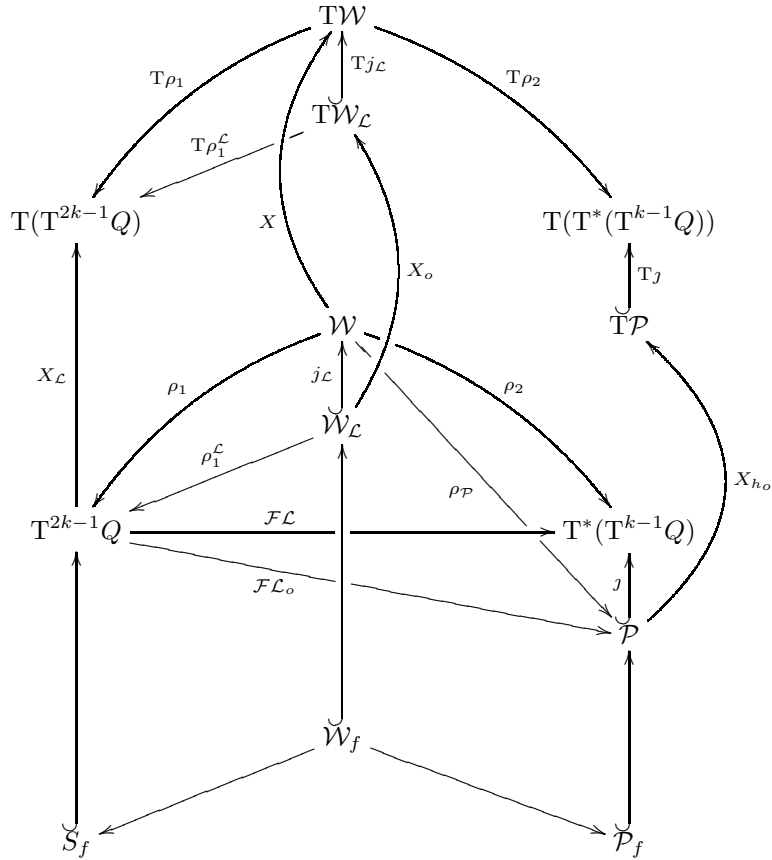
\begin{equation*}
\xymatrix{
\ & \ & \Tan\W \ar@/_1pc/[ddll]_-{\Tan\rho_1} \ar@/^1pc/[ddrr]^-{\Tan\rho_2} & \ & \  \\
\ & \ & \Tan\W_\Lag \ar[dll]_-{\Tan\rho^\Lag_1}|<(.095){\hole} \ar@{^{(}->}[u]_-{\Tan j_\Lag} & \ & \  \\
\Tan(\Tan^{2k-1} Q) & \ & \ & \ & \Tan(\Tan^*(\Tan^{k-1}Q)) \\
\ & \ & \W \ar@/_1pc/[ddll]_-{\rho_1} \ar@/^1pc/[ddrr]^-{\rho_2}|<(.07){\hole} \ar@/^1.95pc/[uuu]^(.35){X}
\ar[dddrr]_-{\rho_\P}|<(.08){\hole}|<(.65){\hole} & \ & \Tan\P \ar@{^{(}->}[u]_-{\Tan\jmath} \\
\ & \ & \W_\Lag \ar[dll]_-{\rho^\Lag_1} \ar@{^{(}->}[u]^-{j_\Lag}
\ar@/_1.75pc/[uuu]_-{X_o} & \ & \ \\
\Tan^{2k-1}Q \ar[rrrr]^<(.3){\Leg}|<(.425){\hole} \ar[drrrr]_<(.3){\Leg_o}|<(.427){\hole} \ar[uuu]^-{X_\Lag} & \ & \ & \ &
\Tan^*(\Tan^{k-1}Q) \\
\ & \ & \  & \ & \P \ar@{^{(}->}[u]^{\jmath} \ar@/_3pc/[uuu]_-{X_{h_o}} \\
\ & \ & \W_f \ar@{^{(}->}[uuu] \ar[dll] \ar[drr] & \ & \ \\
S_f \ar@{^{(}->}[uuu] & \ & \ & \ & \P_f \ar@{^{(}->}[uu]
}
\end{equation*}

Finally, for the dynamical trajectories of the system, we have the following
result, which is the analogous to Proposition \ref{Chap03_prop:EquivUnifHamICReg} in the
almost-regular case.

\begin{proposition}\label{Chap03_prop:EquivUnifHamICSing}
Let $\psi \colon I \subseteq \R \to \W$ be an integral curve of a vector field $X \in \vf(\W)$
tangent to $\W_f$ and solution to equation \eqref{Chap03_eqn:UnifDynEqVFSingular}.
Then the curve $\psi_{h_o} = \rho_\P \circ \psi \colon I \to \P$ is the integral curve
of a vector field solution to equation \eqref{Chap02_eqn:HamHODynEqSingular2} and tangent to
$\P_f = \rho_\P(\W_f)$.
\end{proposition}
\begin{proof}
Bearing in mind Proposition \ref{Chap03_prop:EquivUnifLagIC} and the fact that
$X_\Lag$ and $X_{h_o}$ are $\Leg_o$-related, we have
\begin{align*}
X_{h_o} \circ \psi_{h_o} &= X_{h_o} \circ \rho_\P \circ \psi = X_{h_o} \circ \Leg \circ \rho_1 \circ \psi
= \Tan\Leg_o \circ X_\Lag \circ \psi_\Lag = \Tan\Leg_o \circ \dot{\psi}_\Lag = \dot{\overline{\Leg_o \circ \psi_\Lag}}
= \dot{\psi}_{h_o} \, .
\end{align*}
Therefore, $\psi_{h_o} = \rho_\P \circ \psi$ is an integral curve of $X_{h_o}$.

Finally, from Theorem \ref{Chap03_thm:EquivUnifHamVFSing}, if $X$ is a solution to equation
\eqref{Chap03_eqn:UnifDynEqVFSingular} tangent to $\W_f$, then the vector field $X_{h_o}$ is a solution to equation
\eqref{Chap02_eqn:HamHODynEqSingular2} tangent to $\P_f$.
\end{proof}

Finally, the commutative diagram in the almost-regular case is the same that the diagram in page
\pageref{Chap03_fig:DiagramIC}, replacing $\Tan^{2k-1}Q$, $\Tan^*(\Tan^{k-1}Q)$ and $\W_\Lag$ by $S_f$,
$\P_f$ and $\W_f$, respectively.

\section{Examples}
\label{Chap03_sec:Examples}

In this last Section of the Chapter, two physical models are analyzed as examples to show the application
of the formalism. The first example is a regular system, the so-called \textsl{Pais-Uhlenbeck oscillator},
while the second is a singular one, the \textsl{second-order relativistic particle}.

\subsection{The Pais-Uhlenbeck oscillator}
\label{Chap03_exa:PaisUhlenbeckOscillator}

The Pais-Uhlenbeck oscillator is one of the simplest regular systems that can be used to
explore the features of higher-order dynamical systems, and has been analyzed in detail in
many publications (see \cite{art:Pais_Uhlenbeck50} for the original statement, and
\cite{art:Martinez_Montemayor_Urrutia11} for a more recent analysis).
Here we study it using the unified formalism.

The configuration space for this system is a $1$-dimensional smooth manifold $Q$ with local
coordinate $(q_0)$. Taking natural coordinates $(q_0,q_1,q_2)$ in the second-order tangent
bundle over $Q$, the second-order Lagrangian function $\Lag \in \Cinfty(\Tan^2Q)$
for this system is locally given by
\begin{equation}\label{Chap03_eqn:ExampleRegular_LagrangianFunction}
\Lag(q_0,q_1,q_2) = \frac{1}{2} \left( q_1^2 - \omega^2q_0^2 - \gamma q_2^2 \right) \, ,
\end{equation}
where $\gamma \in \R$ is a nonzero constant, and $\omega \in \R$ is a constant.
Observe that $\Lag$ is regular, since the Hessian matrix of $\Lag$
with respect to $q_2$ is
\begin{equation*}
\left( \derpars{\Lag}{q_2}{q_2} \right) = - \gamma \, ,
\end{equation*}
which has maximum rank, since we assume that $\gamma$ is nonzero. Notice that,
if we take $\gamma = 0$, then $\Lag$ becomes a first-order regular Lagrangian function,
and thus it is a nonsense to study this system using the higher-order unified formalism.

As this is a second-order dynamical system, the phase space that we consider is
\begin{equation*}
\xymatrix{
\ & \W = \Tan^3Q \times_{\Tan Q} \Tan^*(\Tan Q) \ar[dl]_-{\rho_1} \ar[dr]^-{\rho_2} & \ \\
\Tan^3Q \ar[dr]_-{\rho^{3}_{1}} & \ & \Tan^*(\Tan Q) \ar[dl]^-{\pi_{\Tan Q}} \\
\ & \Tan Q & \
}
\end{equation*}

Denoting the canonical symplectic form of $\Tan^*(\Tan Q)$ by
$\omega_1 \in \df^2(\Tan^*(\Tan Q))$, we define the presymplectic form
$\Omega = \rho_2^*\omega_1 \in \df^2(\W)$ with the local expression
\begin{equation*}
\Omega = \d q_0 \wedge \d p^0 + \d q_1 \wedge \d p^1 \, .
\end{equation*}
The Hamiltonian function $H \in \Cinfty(\W)$ in the unified formalism is defined
by \eqref{Chap03_eqn:UnifHamiltonianFunctionDef}, which in this case is
$H = \C - (\rho^3_2 \circ \rho_1)^*\Lag$, where $\C$ is the coupling function,
whose local expression \eqref{Chap03_eqn:UnifCouplingFunctionLocal} in this case is
\begin{equation*}
\C(q_0,q_1,q_2,q_3,p^0,p^1) = p^0q_1 + p^1q_2 \, .
\end{equation*}
Then, bearing in mind the coordinate expression \eqref{Chap03_eqn:ExampleRegular_LagrangianFunction}
of the Lagrangian function for this system, the Hamiltonian function can be written locally
\begin{equation}\label{Chap03_eqn:ExampleRegular_UnifiedHamiltonianFunction}
H(q_0,q_1,q_2,q_3,p^0,p^1) = p^0q_1 + p^1q_2 - \frac{1}{2} \left( q_1^2 - \omega^2q_0^2 - \gamma q_2^2 \right) \, .
\end{equation}

As stated in Section \ref{Chap03_sec:DynamicalEquations}, we can describe the dynamics for this
system in terms of the integral curves of vector fields $X \in \vf(\W)$ which are
solutions to equation \eqref{Chap03_eqn:UnifDynEqVF}. Let $X$ be a generic vector field
in $\W$, given locally by
\begin{equation*}
X = f_0 \derpar{}{q_0} + f_1 \derpar{}{q_1} + F_2 \derpar{}{q_2} + F_3 \derpar{}{q_3} +
 G^0\derpar{}{p^0} + G^1\derpar{}{p^1} \, .
\end{equation*}
Then, from the coordinate expression \eqref{Chap03_eqn:ExampleRegular_UnifiedHamiltonianFunction}
of the Hamiltonian function $H$, we have
\begin{equation*}
\d H = \omega^2q_0\d q_0 + (p^0-q_1)\d q_1 + (p^1 + \gamma q_2)\d q_2 + q_1 \d p^0 + q_2 \d p^1 \, .
\end{equation*}
Now, requiring the dynamical equation $\inn(X)\Omega = \d H$ to hold, we obtain the following system
of $5$ linear equations for the coefficients of the vector field
\begin{align}
f_0 = q_1 \quad ; \quad f_1 = q_2 \, ,\label{Chap03_eqn:ExampleRegular_Semispray2} \\
G^0 = - \omega^2 q_0 \quad ; \quad G^1 = q_1 - p^0 \label{Chap03_eqn:ExampleRegular_VectorFieldG} \, , \\
p^1 + \gamma q_2 = 0 \, . \label{Chap03_eqn:ExampleRegular_LegTransformation}
\end{align}
Equations \eqref{Chap03_eqn:ExampleRegular_Semispray2} give us the condition of semispray of type
$2$ for the vector field $X$. Furthermore, equation \eqref{Chap03_eqn:ExampleRegular_LegTransformation}
is an algebraic relation stating that the vector field $X$ is defined along the submanifold $\W_c$,
as we have seen in Proposition \ref{Chap03_prop:UnifFirstConstraintSubmanifold}. Thus, using equations
\eqref{Chap03_eqn:ExampleRegular_Semispray2} and \eqref{Chap03_eqn:ExampleRegular_VectorFieldG},
the vector field is given locally by
\begin{equation}\label{Chap03_eqn:ExampleRegular_VectorFieldX}
X = q_1 \derpar{}{q_0} + q_2 \derpar{}{q_1} + F_2 \derpar{}{q_2} + F_3 \derpar{}{q_3} -
\omega^2q_0\derpar{}{p^0} + \left(q_1 - p^0\right) \derpar{}{p_1} \, .
\end{equation}

As our goal is to recover the Lagrangian and Hamiltonian solutions from the vector field $X$,
we must require $X$ to be a semispray of type $1$. Nevertheless, as $\Lag$ is a regular Lagrangian
function, this condition is naturally deduced from the formalism when requiring the vector field
$X$ to be tangent to the submanifold $\W_\Lag$, as we have seen in \eqref{Chap03_eqn:UnifDynEqVFTangencyCondition}.

Notice that the functions $F_2$ and $F_3$ in \eqref{Chap03_eqn:ExampleRegular_VectorFieldX}
are not determined until the tangency of the vector field $X$ on $\W_\Lag$ is required.
Hence, let us compute locally the Legendre-Ostrogradsky map associated to the Lagrangian function
\eqref{Chap03_eqn:ExampleRegular_LagrangianFunction}. The Legendre-Ostrogradsky transformation is
the bundle morphism $\Leg \colon \Tan^3Q \longrightarrow \Tan^*(\Tan Q)$ over $\Tan Q$ given in
local coordinates by
\begin{align*}
\Leg^*p^0 = \derpar{\Lag}{q_1} - d_T\left(\derpar{\Lag}{q_2}\right) = q_1 + \gamma q_3 \quad ; \quad
\Leg^*p^1 = \derpar{\Lag}{q_2} = - \gamma q_2 \, ,
\end{align*}
and, as $\gamma \neq 0$, we see that $\Lag$ is a regular Lagrangian since
$\Leg$ is a (local) diffeomorphism. Then, the submanifold
$\W_\Lag = \graph\Leg$ is defined by
\begin{equation*}
\W_\Lag = \left\{ w \in \W \mid \xi_0(w) = \xi_1(w) = 0 \right\} \, ,
\end{equation*}
where $\xi_r = p^r - \Leg^*p^r$, $r=1,2$. The diagram for this situation is
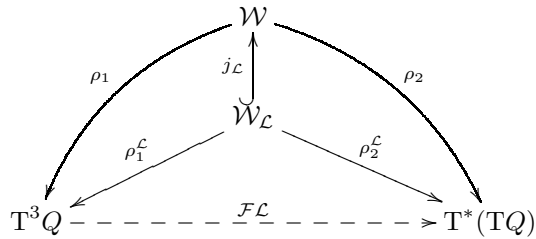
\begin{equation*}
\xymatrix{
\ & \ & \W \ar@/_1.25pc/[ddll]_-{\rho_1} \ar@/^1.25pc/[ddrr]^-{\rho_2} & \ & \ \\
\ & \ & \W_\Lag \ar@{^{(}->}[u]^-{j_\Lag} \ar[dll]_-{\rho^\Lag_{1}} \ar[drr]^-{\rho^\Lag_{2}} & \ & \ \\
\Tan^3Q \ar@{-->}[rrrr]^-{\Leg} & \ & \ & \ & \Tan^*(\Tan Q)
}
\end{equation*}

Now we compute the tangency condition for the vector field $X \in \vf(\W)$ given locally by
\eqref{Chap03_eqn:ExampleRegular_VectorFieldX} along the submanifold $\W_\Lag \hookrightarrow \W$,
by checking if the following identities hold
\begin{equation*}
\restric{\Lie(X)\xi_0}{\W_o} = 0  \quad ; \quad \restric{\Lie(X)\xi_1}{\W_o} = 0 \, .
\end{equation*}
As we have seen in Section \ref{Chap03_sec:DynamicalEquations}, these equations give the
Lagrangian equations for the vector field $X$; that is, on the points of $\W_\Lag$ we obtain
\begin{align}
\Lie(X)\xi_0 = - \omega^2 q_0 - q_2 - \gamma F_3 = 0 \, ,
 \label{Chap03_eqn:ExampleRegular_EulerLagrangeVectFieldEq} \\
\Lie(X)\xi_1 = \gamma\left(F_2 - q_3\right) = 0  \, .
\label{Chap03_eqn:ExampleRegular_Semispray1}
\end{align}
Equation \eqref{Chap03_eqn:ExampleRegular_Semispray1} gives the condition of semispray of type $1$
for the vector field $X$ (recall that $\gamma \neq 0$), and equation
\eqref{Chap03_eqn:ExampleRegular_EulerLagrangeVectFieldEq} is the Euler-Lagrange equation for
the vector field $X$. Notice that, as $\gamma \neq 0$, these equations have a unique solution
for $F_2$ and $F_3$. Thus, there is a unique vector field $X \in \vf(\W)$ solution to the
dynamical equation which is tangent to the submanifold $\W_\Lag \hookrightarrow \W$,
and it is given locally by
\begin{equation*}
X = q_1 \derpar{}{q_0} + q_2 \derpar{}{q_1} + q_3 \derpar{}{q_2} - 
\frac{1}{\gamma}\left(\omega^2q_0 + q_2\right) \derpar{}{q_3} - \omega^2q_0\derpar{}{p^0} +
 \left(q_1 - p^0\right) \derpar{}{p_1} \, .
\end{equation*}

If we require the vector field $X$ to be a semispray of type $1$ from the beginning,
then the coordinate expression \eqref{Chap03_eqn:ExampleRegular_VectorFieldX} becomes
\begin{equation*}
X = q_1 \derpar{}{q_0} + q_2 \derpar{}{q_1} + q_3 \derpar{}{q_2} + F_3 \derpar{}{q_3} -
\omega^2q_0\derpar{}{p^0} + \left(q_1 - p^0\right) \derpar{}{p_1} \, .
\end{equation*}
Then, the tangency condition of $X$ along the submanifold $\W_c$ defined locally by equation
\eqref{Chap03_eqn:ExampleRegular_LegTransformation} gives the following equation on $\W_c$
\begin{equation*}
\Lie(X)(p^1 + \gamma q_2) = q_1 - p^0 + \gamma q_3 = 0 \, ,
\end{equation*}
which gives rise to a new constraint, defining a submanifold $\W_\Lag = \graph\Leg$,
as we have seen in Section \ref{Chap03_sec:DynamicalEquations}. Now, if we require $X$
to be tangent to this new submanifold, we obtain
\begin{equation*}
\Lie(X)(q_1 - p^0 + \gamma q_3) = q_2 + \omega^2 q_0 + \gamma F_3 = 0 \, ,
\end{equation*}
which is exactly equation \eqref{Chap03_eqn:ExampleRegular_EulerLagrangeVectFieldEq}.

Now, if $\psi \colon \R \to \W$ is an integral curve of $X$ locally given by
\begin{equation}
\label{Chap03_eqn:ExampleRegular_LocalCoordSigma}
\psi(t) = \left(q_0(t),q_1(t),q_2(t),q_3(t),p^0(t),p^1(t)\right) \, ,
\end{equation}
then its component functions are solutions to the system
\begin{align}
\dot{q}_0(t) = q_1(t) \quad ; \quad \dot{q}_1(t) = q_2(t) \quad ; \quad \dot{q}_2(t) = q_3(t) \, ,
\label{Chap03_eqn:ExampleRegular_IntCurveX_Semispray1} \\
\dot{q}_3(t) = -\frac{1}{\gamma}\left(\omega^2q_0(t) + q_2(t)\right) \, , \label{Chap03_eqn:ExampleRegular_IntCurveX_EulerLagrange} \\
\dot{p}^0(t) =  - \omega^2q_0(t) \quad ; \quad \dot{p}^1(t) = q_1(t) - p^0(t) \, . \label{Chap03_eqn:ExampleRegular_IntCurveX_Hamiltonian}
\end{align}

Finally we recover the Lagrangian and Hamiltonian solutions for this system.
For the Lagrangian solutions, as we have shown in Lemma \ref{Chap03_lemma:LagLagrangianVF} and
Theorem \ref{Chap03_thm:EquivUnifLagVF}, the Euler-Lagrange vector field is the unique semispray
of type $1$, $X_\Lag \in \vf(\Tan^3Q)$, such that
$X_\Lag \circ \rho_1 \circ j_\Lag = \Tan\rho_1 \circ X \circ j_\Lag$.
Thus this vector field $X_\Lag$ is locally given by
\begin{equation*}
X_\Lag = q_1 \derpar{}{q_0} + q_2 \derpar{}{q_1} + q_3 \derpar{}{q_2} - 
\frac{1}{\gamma}\left(\omega^2q_0 + q_2\right) \derpar{}{q_3} \, .
\end{equation*}
For the integral curves of $X_\Lag$ we know from Proposition \ref{Chap03_prop:EquivUnifLagIC}
that if $\psi \colon \R \to \W$ is an integral curve of $X$, then $\psi_\Lag = \rho_1 \circ \psi$
is an integral curve of $X_\Lag$. Thus, if $\psi$ is given locally by
\eqref{Chap03_eqn:ExampleRegular_LocalCoordSigma}, then $\psi_\Lag$ has the following local expression
\begin{equation}\label{Chap03_eqn:ExampleRegular_LocalCoordSigmaL}
\psi_\Lag(t) = \left(q_0(t),q_1(t),q_2(t),q_3(t)\right) \ ,
\end{equation}
and its components satisfy equations \eqref{Chap03_eqn:ExampleRegular_IntCurveX_Semispray1}
and \eqref{Chap03_eqn:ExampleRegular_IntCurveX_EulerLagrange}. Notice that equations
\eqref{Chap03_eqn:ExampleRegular_IntCurveX_Semispray1} state that $\psi_\Lag$ is the canonical
lifting of a curve in the basis, that is, there exists a curve $\phi \colon \R \to Q$ such that
$j^3_0\phi = \psi_\Lag$. Furthermore, equation \eqref{Chap03_eqn:ExampleRegular_IntCurveX_EulerLagrange}
is the Euler-Lagrange equation for this system, which can be written in the standard form
\begin{equation*}
\frac{d^4q_0}{dt^4} + \frac{1}{\gamma}\left( \omega^2q_0 + \frac{d^2q_0}{dt^2} \right) = 0 \, .
\end{equation*}

Now, for the Hamiltonian solutions, as $\Lag$ is a regular Lagrangian, Theorem
\ref{Chap03_thm:EquivUnifHamVFReg} states that there exists a unique vector field
$X_h \in \vf(\Tan^*(\Tan Q))$ satisfying
$X_h \circ \rho_2 \circ j_\Lag = \Tan\rho_2 \circ X \circ j_\Lag$, and it is a solution to the
Hamiltonian dynamical equation. Hence, it is given locally by
\begin{equation*}
X_h = q_1 \derpar{}{q_0} + q_2 \derpar{}{q_1} - \omega^2q_0\derpar{}{p^0} + \left(q_1 - p^0\right) \derpar{}{p_1} \, .
\end{equation*}
For the integral curves of $X_h$, Proposition \ref{Chap03_prop:EquivUnifHamICReg} states that
if $\psi \colon \R \to \W$ is an integral curve of $X$, then $\psi_h = \rho_2 \circ \psi$ is an
integral curve of the vector field $X_h$. Therefore, if $\psi$ is given locally by
\eqref{Chap03_eqn:ExampleRegular_LocalCoordSigma}, $\psi_h$ can be locally written
\begin{equation*}
\psi_h(t) = \left(q_0(t),q_1(t),p^0(t),p^1(t)\right) \, ,
\end{equation*}
and its component functions must satisfy the first two equations in
\eqref{Chap03_eqn:ExampleRegular_IntCurveX_Semispray1} and equations
\eqref{Chap03_eqn:ExampleRegular_IntCurveX_Hamiltonian}.
Notice that these equations are the standard Hamilton equations for this system,
since the Hamiltonian function $h \in \Cinfty(\Tan^*(\Tan Q))$ of this system is
\begin{equation*}
h(q_0,q_1,p^0,p^1) = p^0q_1 - \frac{1}{2} \left( q_1^2 - \omega^2q_0^2 + \frac{1}{\gamma} (p^1)^2 \right) \, .
\end{equation*}

\subsection{The second-order relativistic particle}
\label{Chap03_exa:2ndOrderRelativisticParticle}

Let us consider a relativistic particle whose action is proportional to its extrinsic curvature. 
This system has been analyzed in several papers \cite{art:Batlle_Gomis_Pons_Roman88,art:Nesterenko89,art:Pisarski86,art:Plyushchay88},
and here we study it using the Lagrangian-Hamiltonian unified  formalism.
  
The configuration space is a $n$-dimensional smooth manifold $Q$ with local coordinates
$(q_0^A)$, $1 \leqslant A \leqslant n$. Then, if we take the natural set of coordinates
on the second-order tangent bundle over $Q$, the second-order Lagrangian function for this system,
$\Lag \in \Cinfty(\Tan^2Q)$, can be written locally as
\begin{equation}\label{Chap03_eqn:ExampleSingular_Lagrangian}
\Lag(q_0^i,q_1^i,q_2^i) = \frac{\alpha}{(q_1^i)^2} \left[ (q_1^i)^2(q_2^i)^2 - (q_1^iq_2^i)^2 \right]^{1/2} 
\equiv \frac{\alpha}{(q_1^i)^2} \sqrt{g}  \, ,
\end{equation}
where $\alpha \in \R$ is some nonzero constant and $g = (q_1^i)^2(q_2^i)^2 - (q_1^iq_2^i)^2$. This
is a singular Lagrangian, as we can see by computing the Hessian matrix of $\Lag$ with respect to
$q_2^A$, which is
\begin{equation*}
\derpars{\Lag}{q_2^B}{q_2^A} =
\begin{cases}
\displaystyle \frac{\alpha}{2(q_1^i)^2\sqrt{g^3}} \left[ \left((q_1^iq_2^i)^2 - 2(q_1^i)^2(q_2^i)^2 \right)q_1^Bq_1^A \right. &  \\
\displaystyle \qquad\qquad\quad \left. + (q_1^i)^2(q_1^iq_2^i)(q_2^Bq_1^A-q_1^Bq_2^A) - (q_1^i)^2(q_2^i)^2q_2^Bq_2^A \right]
& \mbox{ if } B \neq A \, , \\[10pt]
\displaystyle \frac{\alpha}{\sqrt{g^3}}\left[ g - 
(q_2^i)^2q_1^Aq_1^A + 2(q_1^iq_2^i)q_1^Aq_2^A - (q_1^i)^2q_2^Aq_2^A \right] & \mbox{ if } B = A \, .
\end{cases}
\end{equation*}
Then, after a long calculation, we obtain that
\begin{equation*}
\det\left( \derpars{\Lag}{q_2^B}{q_2^A} \right) = 0 \, .
\end{equation*}
In particular, the second-order Lagrangian function $\Lag$ is an almost-regular Lagrangian.

As this is a second-order dynamical system, the phase space that we consider is
\begin{equation*}
\xymatrix{
\ & \W = \Tan^3Q \times_{\Tan Q} \Tan^*(\Tan Q) \ar[dl]_-{\rho_1} \ar[dr]^-{\rho_2} & \ \\
\Tan^3Q \ar[dr]_-{\rho^{3}_{1}} & \ & \Tan^*(\Tan Q) \ar[dl]^-{\pi_{\Tan Q}} \\
\ & \Tan Q & \
}
\end{equation*}

If $\omega_{1} \in \df^2(\Tan^*(\Tan Q))$ denotes the canonical symplectic form, we define
the presymplectic form $\Omega = \rho_2^*\omega_{1}\in \df^2(\W)$, whose local expression is
\begin{equation*}
\Omega = \d q_0^i \wedge \d p_i^0 + \d q_1^i \wedge \d p_i^1 \ .
\end{equation*}
The Hamiltonian function $H \in \Cinfty(\W)$ is defined by \eqref{Chap03_eqn:UnifHamiltonianFunctionDef},
which in this case is $H = \C - (\rho^3_2 \circ \rho_1)^*\Lag$, where $\C$ is the coupling function,
whose local expression \eqref{Chap03_eqn:UnifCouplingFunctionLocal} in this case is
\begin{equation*}
\C\left(q_0^i,q_1^i,q_2^i,q_3^i,p_i^0,p_i^1\right) = p_i^0q_1^i + p_i^1q_2^i \, .
\end{equation*}
Then, the Hamiltonian function can be written locally
\begin{equation*}
H(q_0^i,q_1^i,q_2^i,q_3^i,p_i^0,p_i^1) =
 p_i^0q_1^i + p_i^1q_2^i - \frac{\alpha}{(q_1^i)^2} \left[ (q_1^i)^2(q_2^i)^2 - (q_1^iq_2^i)^2 \right]^{1/2} \, .
\end{equation*}

The dynamics for this system are described as the integral curves of vector fields $X \in \vf(\W)$
which are solutions to equation \eqref{Chap03_eqn:UnifDynEqVF}. If we take a generic vector field
$X\in\vf(\W)$, given locally by
\begin{equation*}
X = f_0^A \derpar{}{q_0^A} + f_1^A \derpar{}{q_1^A} + F_2^A \derpar{}{q_2^A} + 
F_3^A \derpar{}{q_3^A} + G_A^0 \derpar{}{p_A^0} + G_A^1 \derpar{}{p_A^1} \ ,
\end{equation*}
taking into account that
\begin{align*}
\d H &= q_1^A\d p_A^0 + q_2^A\d p_A^1 + 
\left[ p^0_A + \frac{\alpha}{((q_1^i)^2)^2\sqrt{g}}\left[ \left((q_1^i)^2(q_2^i)^2 - 2(q_1^iq_2^i)^2\right)q_1^A + 
(q_1^iq_2^i)(q_1^i)^2q_2^A \right] \right]\d q_1^A \\
&\quad{} + \left[ p_A^1 - \frac{\alpha}{(q_1^i)^2\sqrt{g}}\left( (q^i_1)^2 q_2^A - (q_1^iq_2^i)q_1^A \right) \right]\d q_2^A \, ,
\end{align*}
then, from the dynamical equation we obtain the following linear equations for the coefficients of $X$
\begin{align}
f_0^A = q_1^A \quad ; \quad f_1^A = q_2^A \, , \label{Chap03_eqn:ExampleSingular_Semispray2} \\[8pt]
G_A^0 = 0 \quad ; \quad
G_A^1 = - p^0_A - \frac{\alpha}{((q_1^i)^2)^2\sqrt{g}}\left[ \left((q_1^i)^2(q_2^i)^2 - 2(q_1^iq_2^i)^2\right)q_1^A + 
(q_1^iq_2^i)(q_1^i)^2q_2^A \right] \, , \label{Chap03_eqn:ExampleSingular_VectorFieldG} \\
p_A^1 - \frac{\alpha}{(q_1^i)^2\sqrt{g}}\left( (q^i_1)^2 q_2^A - (q_1^iq_2^i)q_1^A \right) = 0 \, .
\label{Chap03_eqn:ExampleSingular_LegTransformation}
\end{align}
Note that from equations \eqref{Chap03_eqn:ExampleSingular_Semispray2} we obtain the condition
of semispray of type $2$ for $X$. Furthermore, equations
\eqref{Chap03_eqn:ExampleSingular_LegTransformation} are algebraic relations between the coordinates
in $\W$ stating that the vector field $X$ is defined along the submanifold $\W_c$, as it is stated
in Proposition \ref{Chap03_prop:UnifFirstConstraintSubmanifold}. Thus, the vector field $X$ is given locally by
\begin{equation}\label{Chap03_eqn:ExampleSingular_VectorFieldBeforeHolonomy}
X = q_1^A \derpar{}{q_0^A} + q_2^A \derpar{}{q_1^A} + F_2^A \derpar{}{q_2^A} +
F_3^A \derpar{}{q_3^A} + G_A^1 \derpar{}{p_A^1} \, ,
\end{equation}
where the functions $G_A^1$ are determined by \eqref{Chap03_eqn:ExampleSingular_VectorFieldG}.
As we want to recover the Lagrangian solutions from the vector field $X$, we must require $X$
to be a semispray of type $1$. This condition reduces the set of vector fields $X \in \vf(\W)$ given by 
\eqref{Chap03_eqn:ExampleSingular_VectorFieldBeforeHolonomy} to the following ones
\begin{equation}\label{Chap03_eqn:ExampleSingular_VectorFieldHolonomy}
X = q_1^A \derpar{}{q_0^A} + q_2^A \derpar{}{q_1^A} + q_3^A \derpar{}{q_2^A} + 
F_3^A \derpar{}{q_3^A} + G_A^1\derpar{}{p_A^1} \, .
\end{equation}
Notice that the functions $F_3^A$ are not determined until the tangency of the vector field $X$
along $\W_c$ is required. Since this example has a Lagrangian function far more complicated than
the example analyzed in the previous Section, in this case we study directly the tangency of the
vector field along the submanifold $\W_\Lag = \graph(\Leg)$. Hence, we need to compute the coordinate
expression of the Legendre-Ostrogradsky map. From the results in Section
\ref{Chap02_sec:HamiltonianAutonomousHigherOrder}, the Legendre-Ostrogradsky map is the bundle morphism
$\Leg \colon \Tan^3Q \to \Tan^*(\Tan Q)$ over $\Tan Q$ locally given by
\begin{align*}
\Leg^*(p^0_A) &= \frac{\alpha}{(q_1^i)^2\sqrt{g^3}} \left[ \left( (q_2^i)^2g + (q_1^i)^2(q_2^i)^2(q_1^iq_3^i) -
(q_1^i)^2(q_1^iq_2^i)(q_ 2^iq_3^i) \right)q_1^A\right] \\
& \quad{} + \frac{\alpha}{(q_1^i)^2\sqrt{g^3}} \left[ \left( ((q_1^i)^2)^2(q_2^iq_3^i) - (q_1^i)^2(q_1^iq_2^i)(q_1^iq_3^i) - 
(q_1^iq_2^i)g \right)q_2^A - (q_1^i)^2gq_3^A\right] \, , \\ 
\Leg^*(p^1_A) &= \frac{\alpha}{(q^i_1)^2\sqrt{g}} \left[ (q^i_1)^2q_2^A - (q^i_1q^i_2)q^A_1 \right] \, .
\end{align*}
From this coordinate expression we can check that the second-order Lagrangian function of this system is,
in fact, almost-regular. Thus, from the expression in local coordinates of the map $\Leg$, we obtain the
(primary) constraints that define the closed submanifold $\P = \Im(\Leg)$, which are
\begin{equation}
\phi^{(0)}_1 \equiv p^1_iq_1^i = 0 \quad ; \quad \phi^{(0)}_2 \equiv (p_i^1)^2 - \frac{\alpha^2}{(q^i_1)^2} = 0 \, .
\label{Chap03_eqn:ExampleSingular_Constraints0}
\end{equation}
Let $\Leg_o \colon \Tan^{3}Q \to \P$. Then, the submanifold $\W_o = \graph(\Leg_o)$ is defined by
\begin{equation*}
\W_o = \left\{ w \in \W \mid \xi^A_0(w) = \xi^A_1(w) = \phi^{(0)}_1(w) = \phi^{(0)}_2(w) = 0,
\ 1 \leqslant A \leqslant n \right\} \, ,
\end{equation*}
where $\xi_r^A \equiv p_A^r - \Leg^*p_A^r$. Notice that $\W_o$ is a submanifold of $\W_\Lag$,
and that $\W_o$ is the real phase space of the system, where the dynamics take place.

Next we compute the tangency condition for $X \in \vf(\W)$ given locally by
\eqref{Chap03_eqn:ExampleSingular_VectorFieldHolonomy} on the submanifold 
$\W_o \hookrightarrow \W_{\Lag} \hookrightarrow \W$, by checking if the following identities hold
\begin{align}
\restric{\Lie(X)\xi^A_0}{\W_o} = 0 \quad ; \quad \restric{\Lie(X)\xi^A_1}{\W_o} = 0 \, ,
\label{Chap03_eqn:ExampleSingular_LagEquations1} \\[10pt]
\restric{\Lie(X)\phi^{(0)}_1}{\W_o} = 0  \quad ; \quad \restric{\Lie(X)\phi^{(0)}_2}{\W_o} = 0 \, .
\label{Chap03_eqn:ExampleSingular_LieDerConstraints0}
\end{align}
As we have seen in Section \ref{Chap03_sec:DynamicalEquations}, equations
\eqref{Chap03_eqn:ExampleSingular_LagEquations1} give the Lagrangian equations for the vector field $X$.
However, equations (\ref{Chap03_eqn:ExampleSingular_LieDerConstraints0}) do not hold since
\begin{equation*}
\Lie(X)\phi^{(0)}_1 = \Lie(X)(p^1_iq_1^i) = - p_i^0q_1^i \quad ; \quad
\Lie(X)\phi^{(0)}_2 = \Lie(X)((p^1_i)^2 - \alpha^2 / (q_1^i)^2) = - 2p_i^0q_i^1 \ ,
\end{equation*}
and hence we obtain two first-generation secondary constraints
\begin{equation}\label{Chap03_eqn:ExampleSingular_Constraints1}
\phi^{(1)}_1 \equiv p_i^0 q_1^i = 0 \quad ; \quad \phi^{(1)}_2 \equiv p_i^0p_i^1 = 0 
\end{equation}
that define a new submanifold $\W_1 \hookrightarrow \W_o$. Now, checking the tangency
of the vector field $X$ along this new submanifold, we obtain
\begin{equation*}
\Lie(X)\phi^{(1)}_1 = \Lie(X)(p_i^0q_1^i) = 0 \quad ; \quad
\Lie(X)\phi^{(1)}_2 = \Lie(X)(p_i^0p_i^1) = -(p^0_i)^2 \, ,
\end{equation*}
and a second-generation secondary constraint appears
\begin{equation}\label{Chap03_eqn:ExampleSingular_Constraints2}
\phi^{(2)} \equiv (p^0_i)^2 = 0 \, ,
\end{equation}
which defines a new submanifold $\W_2 \hookrightarrow \W_1$.
Finally, the tangency of the vector field $X$ along this submanifold gives no new constraints, since
\begin{equation*}
\Lie(X)\phi^{(2)} = \Lie(X)((p^0_i)^2) = 0 \, .
\end{equation*}
So we have two primary constraints \eqref{Chap03_eqn:ExampleSingular_Constraints0},
two first-generation secondary constraints \eqref{Chap03_eqn:ExampleSingular_Constraints1},
and a single second-generation secondary constraint \eqref{Chap03_eqn:ExampleSingular_Constraints2}.
Notice that these five constraints depend only on $q_1^A$, $p^0_A$ and $p^1_A$, 
and so they are $\rho_2$-projectable. Thus, we have the following diagram
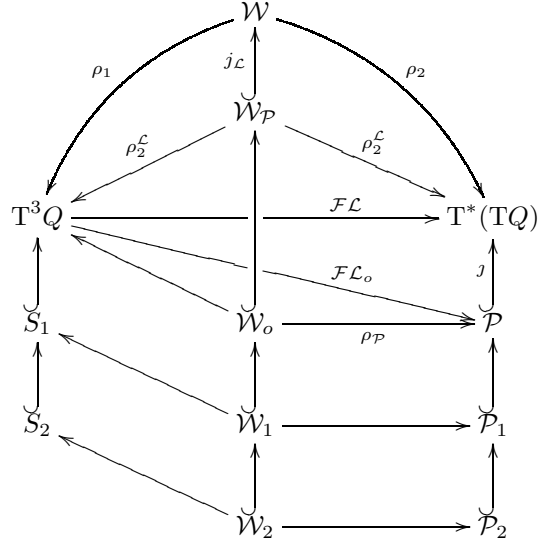
\begin{equation*}
\xymatrix{
\ & \ & \W \ar@/_1.25pc/[ddll]_-{\rho_1} \ar@/^1.25pc/[ddrr]^-{\rho_2} & \ & \ \\
\ & \ & \W_{\P} \ar@{^{(}->}[u]^-{j_\Lag} \ar[dll]_-{\rho_2^\Lag} \ar[drr]^-{\rho_2^\Lag} & \  & \ \\
\Tan^3Q \ar[rrrr]^<(.65){\Leg}|<(.435){\hole} \ar[drrrr]^<(.65){\Leg_o}|<(.435){\hole} & \ & \ & \ & \Tan^*(\Tan Q) \\
S_1 \ar@{^{(}->}[u] & \ & \W_o \ar@{^{(}->}[uu] \ar[ull] \ar[rr]_{\rho_\P} & \ & \P \ar@{^{(}->}[u]^{\jmath} \\
S_2 \ar@{^{(}->}[u] & \ & \W_1 \ar@{^{(}->}[u] \ar[ull] \ar[rr] & \ & \P_1 \ar@{^{(}->}[u] \\
\ & \ & \W_2 \ar@{^{(}->}[u] \ar[ull] \ar[rr] & \ & \P_2 \ar@{^{(}->}[u]
}
\end{equation*}
where
\begin{equation*}
\P_1 = \left\{ p \in \P \mid \phi^{(1)}_1(p) = \phi^{(1)}_2(p) = 0 \right\} = \rho_\P(\W_1) \quad ; \quad
\P_2 = \left\{ p \in \P_1 \mid \phi^{(2)}(p) = 0 \right\} = \rho_\P(\W_2) \, ,
\end{equation*}
\begin{equation*}
S_1 = \Leg_o^{-1}(\P_1) = \rho_1(\W_1) \quad ; \quad S_2 = \Leg_o^{-1}(\P_2) = \rho_1(\W_2) \, .
\end{equation*}
Focusing only on the Legendre-Ostrogradsky map, and ignoring the unified part of the diagram, we have
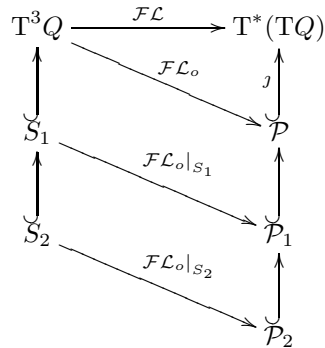
\begin{equation*}
\xymatrix{
\Tan^3 Q \ar[rr]^-{\Leg} \ar[drr]^-{\Leg_o} & \ & \Tan^*(\Tan Q) \\
S_1 \ar@{^{(}->}[u] \ar[drr]^-{\restric{\Leg_o}{S_1}} & \ & \P \ar@{^{(}->}[u]^-{\jmath} \\
S_2 \ar@{^{(}->}[u] \ar[drr]^-{\restric{\Leg_o}{S_2}} & \ & \P_1 \ar@{^{(}->}[u] \\
\ & \ & \P_2 \ar@{^{(}->}[u]
}
\end{equation*}

Notice that we still have to check \eqref{Chap03_eqn:ExampleSingular_LagEquations1}.
As we have seen in Section \ref{Chap03_sec:DynamicalEquations}, we obtain the following
system of equations
\begin{align}
\left(F_3^B - d_T\left(q_3^B\right)\right)\frac{\partial^2\Lag}{\partial q_2^B\partial q_2^A} +
\derpar{\Lag}{q_0^A} - d_T\left(\derpar{\Lag}{q_1^A}\right) + 
d_T^2\left(\derpar{\Lag}{q_2^A}\right) + 
\left(F_2^B - q_3^B\right)d_T\left(\frac{\partial^2\Lag}{\partial q_2^B\partial q_2^A}\right) = 0 \, ,
\label{Chap03_eqn:ExampleSingular_EulerLagrangeInitial} \\
\left(F_2^B - q_3^B\right)\frac{\partial^2\Lag}{\partial q_2^B\partial q_2^A} = 0 \, .
\label{Chap03_eqn:ExampleSingular_Semispray1}
\end{align}
As we have already required the vector field $X$ to be a semispray of type $1$, equations
\eqref{Chap03_eqn:ExampleSingular_Semispray1} are satisfied identically and equations
\eqref{Chap03_eqn:ExampleSingular_EulerLagrangeInitial} become
\begin{equation}\label{Chap03_eqn:ExampleSingular_EulerLagrangeFinal}
\left(F_3^B - d_T\left(q_3^B\right)\right)\frac{\partial^2\Lag}{\partial q_2^B\partial q_2^A} +
\derpar{\Lag}{q_0^A} -
d_T\left(\derpar{\Lag}{q_1^A}\right) + d_T^2\left(\derpar{\Lag}{q_2^A}\right) = 0 \ .
\end{equation}
A long calculation shows that this equation is compatible and so no new constraints arise.
Thus, we have no Lagrangian constraints appearing from the semispray condition.
If some constraint had appeared, it would not be $\Leg_o$-projectable.

Thus, the vector fields $X \in \vf(\W)$ given locally by
\eqref{Chap03_eqn:ExampleSingular_VectorFieldHolonomy} which are solutions to the equation
\begin{equation*}
\restric{\left[\inn(X)\Omega - \d H\right]}{\W_2} = 0 \ ,
\end{equation*}
are tangent to the submanifold $j_2 \colon \W_2 \hookrightarrow \W_\Lag$. Therefore, taking the vector
fields $X_o \in \vf(\W_2)$ such that $\Tan j_2\circ X_o = X \circ j_2$, the form
$\Omega_o = (j_\Lag \circ j_2)^*\Omega$, and the canonical Hamiltonian function
$H_o = (j_\Lag \circ j_2)^*H$, the above equation leads to
\begin{equation*}
\inn(X_o)\Omega_o - \d H_o = 0 \, ,
\end{equation*}
but a simple calculation in local coordinates shows that $H_o=0$, and thus
the last equation becomes simply $\inn(X_o)\Omega_o= 0$.

One can easily check that, if the semispray condition is not required at the beginning
and we perform all this procedure with the vector field given by
\eqref{Chap03_eqn:ExampleSingular_VectorFieldBeforeHolonomy}, the final result is the same.
This means that, in this case, the semispray condition does not give any additional constraint.

As final results, we recover the Lagrangian and Hamiltonian vector fields from the vector field
$X \in \vf(\W)$.
For the Lagrangian vector field, by using Lemma \ref{Chap03_lemma:LagLagrangianVF} and
Theorem \ref{Chap03_thm:EquivUnifLagVF} we obtain a semispray of type $2$, $X_\Lag \in \vf(\Tan^3Q)$,
tangent to $S_2$. Thus, requiring the condition of semispray of type $1$ to be satisfied
(perhaps on another submanifold $S^h_2 \hookrightarrow S_2$), the local expression for the vector field $X_\Lag$ is
\begin{equation*}
X_\Lag = q_1^A \derpar{}{q_0^A} + q_2^A \derpar{}{q_1^A} + q_3^A \derpar{}{q_2^A} + F_3^A \derpar{}{q_3^A} \, ,
\end{equation*}
where the functions $F_3^A$ are determined by \eqref{Chap03_eqn:ExampleSingular_EulerLagrangeFinal}.
For the Hamiltonian vector fields, recall that $\Lag$ is an almost-regular Lagrangian
function. Thus, we know that there are Euler-Lagrange vector fields which are
$\Leg_o$-projectable on $\P_2$, tangent to $\P_2$ and solutions to the Hamilton equation.


\clearpage
\chapter{Geometric Hamilton-Jacobi theory for higher-order autonomous systems}
\label{Chap:HOHamiltonJacobi}


In this Chapter we give the geometric description of the Hamilton-Jacobi problem for higher-order
autonomous dynamical systems. That is, we generalize the results in Section \ref{Chap02_sec:HamiltonJacobi}
to the higher-order setting described in Section \ref{Chap02_sec:AutonomousHigherOrder}. In addition,
using the results in Chapter \ref{Chap:HOAutonomousDynamicalSystems}, we also state the problem in the
Lagrangian-Hamiltonian unified formalism.

The structure of the Chapter is the following. In Sections \ref{Chap04_sec:LagrangianFormalism},
\ref{Chap04_sec:HamiltonianFormalism} and \ref{Chap04_sec:UnifiedFormalism} we introduce the geometric
formulation of the Hamilton-Jacobi problem in the Lagrangian, Hamiltonian and unified formalisms, respectively.
In particular, following the patterns in \cite{art:Carinena_Gracia_Marmo_Martinez_Munoz_Roman06}, we first
introduce the generalized version of the Hamilton-Jacobi problem. Then, the standard Hamilton-Jacobi problem
is stated adding an isotropy condition to the generalized problem. Finally, the concept of complete solutions
is defined in these settings. Relations between these three formulations in terms of the Legendre-Ostrogradsky
map and the natural projections in the unified formalism are also analyzed. Finally, in Section
\ref{Chap04_sec:Examples}, two physical models are analyzed using this formulation: the end of a thrown
javelin and the shape of a homogeneous deformed elastic cylindrical beam with fixed ends.

Throughout this Chapter we consider a $k$th-order autonomous Lagrangian dynamical system with $n$ degrees
of freedom. Let $Q$ be a $n$-dimensional smooth manifold modeling the configuration space of this system,
and $\Lag \in \Cinfty(\Tan^{k}Q)$ a $k$th-order Lagrangian function describing the dynamics of the system.
In addition, we assume that the Lagrangian function is regular (see Definition \ref{Chap02_def:LagHORegularLagrangian}).
Finally, to avoid confusion, points in the higher-order tangent bundles of $Q$ are denoted by $\bar{y}$.

\section{The Lagrangian formalism}
\label{Chap04_sec:LagrangianFormalism}

Recall that, in the Lagrangian formalism for a $k$th-order dynamical system, from the $k$th-order Lagrangian
function $\Lag \in \Cinfty(\Tan^{k}Q)$ and using the canonical structures of the $k$th-order tangent bundle
of $Q$, we construct the $k$th-order Poincar\'{e}-Cartan forms $\theta_{\Lag} \in \df^{1}(\Tan^{2k-1}Q)$ and
$\omega_\Lag = -\d\theta_{\Lag}\in \df^{2}(\Tan^{2k-1}Q)$, as well as the $k$th-order Lagrangian energy
$E_\Lag \in \Cinfty(\Tan^{2k-1}Q)$. Then, using these geometric objects, we can state the geometric equation,
which is the search for a semispray of type $1$, $X_\Lag \in \vf(\Tan^{2k-1}Q)$, satisfying equation
\eqref{Chap02_eqn:LagHODynEq}, that is,
\begin{equation*}
\inn(X_\Lag)\omega_\Lag = \d E_\Lag \, .
\end{equation*}
Since the Lagrangian function is regular, the $k$th-order Poincar\'{e}-Cartan $2$-form $\omega_\Lag$ is symplectic, and
then the above equation has a unique solution $X_\Lag \in \vf(\Tan^{2k-1}Q)$ which is a semispray of type $1$
in $\Tan^{2k-1}Q$ without further assumption. See Section \ref{Chap02_sec:LagrangianAutonomousHigherOrder}
and references therein for a detailed description of the Lagrangian formalism for higher-order dynamical
systems.

\subsection{The generalized Lagrangian Hamilton-Jacobi problem}
\label{Chap04_sec:GenLagHJProb}

Following \cite{book:Abraham_Marsden78} and \cite{art:Carinena_Gracia_Marmo_Martinez_Munoz_Roman06},
and the results in Section \ref{Chap02_sec:HamiltonJacobiLag}, we first state a general version of the
Hamilton-Jacobi problem in the Lagrangian setting, which is the so-called
\textsl{generalized Lagrangian Hamilton-Jacobi problem}. As we have seen in Section
\ref{Chap02_sec:HamiltonJacobiLag}, in the first-order setting this problem consists in finding vector fields
$X \in \vf(Q)$ such that the lifting of every integral curve of $X$ to $\Tan Q$ by $X$ itself is an integral
curve of the Lagrangian vector field $X_\Lag$. For higher-order systems we can state an analogous problem.

\begin{definition}\label{Chap04_def:GenLagHJDef}
The \textnormal{generalized $k$th-order Lagrangian Hamilton-Jacobi problem} consists in finding a section
$s \in \Gamma(\rho^{2k-1}_{k-1})$ and a vector field $X \in \vf(\Tan^{k-1}Q)$ such that, if
$\gamma \colon \R \to \Tan^{k-1}Q$ is an integral curve of $X$, then $s \circ \gamma \colon \R \to \Tan^{2k-1}Q$
is an integral curve of $X_\Lag$; that is,
\begin{equation}\label{Chap04_eqn:GenLagHJDef}
X \circ \gamma = \dot \gamma \Longrightarrow X_\Lag \circ (s \circ \gamma) = \dot{\overline{s \circ \gamma}} \, .
\end{equation}
\end{definition}

\begin{remark}
Observe that, since $X_\Lag$ is a semispray of type $1$, then every integral curve of $X_\Lag$ is the
canonical lifting of a curve in $Q$ to $\Tan^{2k-1}Q$. In particular, this holds for the curve $s \circ \gamma$,
that is, there exists a curve $\phi \colon \R \to Q$ such that
\begin{equation*}
j^{2k-1}_0\phi = s \circ \gamma \, .
\end{equation*}
Then, composing both sides of the equality with $\rho^{2k-1}_{k-1}$ and bearing in mind that
$s \in \Gamma(\rho^{2k-1}_{k-1})$, we obtain
\begin{equation*}
\gamma = j^{k-1}_0\phi \, ,
\end{equation*}
that is, the curve $\gamma$ is the $(k-1)$-jet lifting of a curve in $Q$. This enables us to restate the problem
as follows: \textit{The \textnormal{generalized $k$th-order Lagrangian Hamilton-Jacobi problem} consists
in finding a vector field $X_o \in \vf(Q)$ such that, if $\phi \colon \R \to Q$ is an integral curve of $X_o$,
then $j^{2k-1}_0\phi \colon \R \to \Tan^{2k-1}Q$ is an integral curve of $X_\Lag$; that is,}
\begin{equation*}
X_o \circ \phi = \dot{\phi} \Longrightarrow X_\Lag \circ (j^{2k-1}_0\phi) = \dot{\overline{j^{2k-1}_0\phi}} \, .
\end{equation*}
Nevertheless, we will stick to the previous statement (Definition \ref{Chap04_def:GenLagHJDef}) in order to give several
different characterizations of the problem.
\end{remark}

It is clear from Definition \ref{Chap04_def:GenLagHJDef} that the vector field $X \in \vf(\Tan^{k-1}Q)$ cannot be
chosen independently from the section $s \in \Gamma(\rho^{2k-1}_{k-1})$. In fact, we have the following result.

\begin{proposition}\label{Chap04_prop:GenLagHJRelatedVF}
The pair $(s,X) \in \Gamma(\rho^{2k-1}_{k-1}) \times \vf(\Tan^{k-1}Q)$ satisfies the condition
\eqref{Chap04_eqn:GenLagHJDef} if, and only if, $X$ and $X_\Lag$ are $s$-related; that is,
$X_\Lag \circ s = \Tan s \circ X$.
\end{proposition}
\begin{proof}
The proof of this result follows the patterns of the proof of Proposition 5 in
\cite{art:Carinena_Gracia_Marmo_Martinez_Munoz_Roman06}. In particular, if $(s,X)$ satisfies the condition
\eqref{Chap04_eqn:GenLagHJDef}, then for every integral curve $\gamma$ of $X$, we have
\begin{equation*}
X_\Lag \circ ( s \circ \gamma) = \dot{\overline{s \circ \gamma}} = \Tan s \circ \dot\gamma = \Tan s \circ X \circ \gamma \, ,
\end{equation*}
but, since $X$ has integral curves through every point $\bar{y} \in \Tan^{k-1}Q$, this is equivalent to
$X_\Lag \circ s = \Tan s \circ X$.

Conversely, if $X_{\Lag}$ and $X$ are $s$-related and $\gamma \colon \R \to \Tan^{k-1}Q$ is an
integral curve of $X$, we have
\begin{equation*}
X_\Lag \circ s \circ \gamma = \Tan s\circ X \circ \gamma = \Tan s\circ\dot{\gamma} = \dot{\overline{s \circ \gamma}} \, .
\qedhere
\end{equation*}
\end{proof}

Hence, the vector field $X \in \vf(\Tan^{k-1}Q)$ is related with the Lagrangian vector field $X_\Lag$
and with the section $s \in \Gamma(\rho^{2k-1}_{k-1})$. As a consequence of Proposition
\ref{Chap04_prop:GenLagHJRelatedVF}, we have the following result.

\begin{corollary}\label{Chap04_corol:GenLagHJRelatedVF}
If the pair $(s,X) $ satisfies condition \eqref{Chap04_eqn:GenLagHJDef}, then
$X = \Tan\rho^{2k-1}_{k-1} \circ X_\Lag \circ s$.
\end{corollary}
\begin{proof}
If $(s,X)$ satisfies the condition \eqref{Chap04_eqn:GenLagHJDef}, then from Proposition
\ref{Chap04_prop:GenLagHJRelatedVF} we know that $X$ and $X_\Lag$ are $s$-related, that is, we have
$\Tan s \circ X = X_\Lag \circ s$. Then, composing both sides of the equality with $\Tan\rho^{2k-1}_{k-1}$
and bearing in mind that $\rho^{2k-1}_{k-1} \circ s = \Id_{\Tan^{k-1}Q}$, we obtain
$X = \Tan\rho^{2k-1}_{k-1} \circ X_\Lag \circ s$.
\end{proof}

Thus, the vector field $X$ is completely determined by the section $s \in \Gamma(\rho^{2k-1}_{k-1})$,
and it is called the \textsl{vector field associated to $s$}. The following diagram illustrates the situation
\begin{equation*}
\xymatrix{
\Tan(\Tan^{k-1}Q) \ar@/_1.5pc/[rr]_{\Tan s}  & \ & \ar[ll]_{\Tan\rho^{2k-1}_{k-1}} \Tan(\Tan^{2k-1}Q) \\
\ & \ & \ \\
\Tan^{k-1}Q \ar[uu]^{X} \ar@/_1.5pc/[rr]_{s} & \ & \ar[ll]_{\rho^{2k-1}_{k-1}} \Tan^{2k-1}Q \ar[uu]_{X_\Lag}
}
\end{equation*}

Since the vector field $X \in \vf(\Tan^{k-1}Q)$ is completely determined by the section $s$, the search of
a pair $(s,X) \in \Gamma(\rho^{2k-1}_{k-1}) \times \vf(\Tan^{k-1}Q)$ satisfying condition
\eqref{Chap04_eqn:GenLagHJDef} is equivalent to the search of a section $s \in \Gamma(\rho^{2k-1}_{k-1})$
such that the pair $(s,\Tan\rho^{2k-1}_{k-1} \circ X_\Lag \circ s)$ satisfies the same condition.
Thus, we can give the following definition.

\begin{definition}
A \textnormal{solution to the generalized $k$th-order Lagrangian Hamilton-Jacobi problem} is a section
$s \in \Gamma(\rho^{2k-1}_{k-1})$ such that, if $\gamma \colon \R \to \Tan^{k-1}Q$ is an integral curve of
$\Tan\rho^{2k-1}_{k-1} \circ X_\Lag \circ s \in \vf(\Tan^{k-1}Q)$, then
$s \circ \gamma \colon \R \to \Tan^{2k-1}Q$ is an integral curve of $X_\Lag$, that is,
\begin{equation*}
\Tan\rho^{2k-1}_{k-1} \circ X_\Lag \circ s \circ \gamma = \dot{\gamma}
\Longrightarrow X_\Lag \circ (s \circ \gamma) = \dot{\overline{s \circ \gamma}} \, .
\end{equation*}
\end{definition}

Finally, we have the following result, which gives some equivalent conditions for a section to be a
solution to the generalized $k$th-order Lagrangian Hamilton-Jacobi problem. This Proposition is the
analogous to Theorem \ref{Chap02_thm:HJLagFOEquivalenceGeneralized} in the higher-order setting.

\begin{proposition}\label{Chap04_prop:GenLagHJEquiv}
The following assertions on a section $s \in \Gamma(\rho^{2k-1}_{k-1})$ are equivalent.
\begin{enumerate}
\item The section $s$ is a solution to the generalized $k$th-order Lagrangian Hamilton-Jacobi problem.
\item The submanifold $\Im(s) \hookrightarrow \Tan^{2k-1}Q$ is invariant by the Euler-Lagrange vector field
$X_\Lag$ (that is, $X_\Lag$ is tangent to the submanifold $s(\Tan^{k-1}Q) \hookrightarrow \Tan^{2k-1}Q$).
\item The section $s$ satisfies the equation
\begin{equation*}
\inn(X)(s^*\omega_\Lag) = \d(s^*E_\Lag) \, ,
\end{equation*}
where $X = \Tan\rho^{2k-1}_{k-1} \circ X_\Lag \circ s \in \vf(\Tan^{k-1}Q)$ is the vector field associated to $s$.
\end{enumerate}
\end{proposition}
\begin{proof} This proof follows the same patterns as the proof of Proposition $2$ and Theorem $1$ in
\cite{art:Carinena_Gracia_Marmo_Martinez_Munoz_Roman06}.
\begin{description}
\item[\textnormal{($1 \, \Longleftrightarrow \, 2$)}]
Let $s$ be a solution to the generalized $k$th-order Lagrangian Hamilton-Jacobi problem. Then by
Proposition \ref{Chap04_prop:GenLagHJRelatedVF} the Lagrangian vector field $X_{\Lag} \in \vf(\Tan^{2k-1}Q)$
is $s$-related to the vector field $X = \Tan\rho^{2k-1}_{k-1} \circ X_\Lag \circ s \in \vf(\Tan^{k-1}Q)$
associated to $s$, and thus for every $\bar{y} \in \Tan^{k-1}Q$ we have
\begin{equation*}
X_{\Lag}(s(\bar{y})) = (X_{\Lag}\circ s)(\bar{y}) = (\Tan s \circ X)(\bar{y}) = \Tan s(X(\bar{y})) \, .
\end{equation*}
Hence, $X_{\Lag}(s(\bar{y})) = \Tan s(X(\bar{y}))$ and therefore $X_{\Lag}$ is tangent to the
submanifold $\Im(s) \hookrightarrow \Tan^{2k-1}Q.$

Conversely, if the submanifold $\Im(s)$ is invariant under the flow of $X_{\Lag}$, then
$X_{\Lag}(s(\bar{y})) \in \Tan_{s(\bar{y})}\Im(s)$, for every $\bar{y} \in \Tan^{k-1}Q$;
that is, there exists an element $u_{\bar{y}} \in \Tan_{\bar{y}}\Tan^{k-1}Q$ such that
$X_{\Lag}(s(\bar{y})) = \Tan_{\bar{y}}s(u_{\bar{y}})$. If we define $X \in \vf(\Tan^{k-1}Q)$
as the vector field that satisfies $\Tan_{\bar{y}}s(X_{\bar{y}}) = X_{\Lag}(s(\bar{y}))$,
then $X$ is a vector field in $\Tan^{k-1}Q$, since $X = \Tan\rho^{2k-1}_{k-1} \circ X_\Lag \circ s$,
and it is $s$-related with $X_{\Lag}$. Therefore, by Proposition \ref{Chap04_prop:GenLagHJRelatedVF},
$s$ is a solution to the generalized $k$th-order Lagrangian Hamilton-Jacobi problem.

\item[\textnormal{($1 \, \Longleftrightarrow \, 3$)}]
Let $s$ be a solution to the generalized $k$th-order Lagrangian Hamilton-Jacobi problem. Taking the
pull-back of the Lagrangian dynamical equation \eqref{Chap02_eqn:LagHODynEq} by the section $s$ we have
\begin{equation*}
s^*\inn(X_{\Lag})\omega_{\Lag} = s^*(\d E_{\Lag}) = \d(s^*E_{\Lag}) \, ,
\end{equation*}
but since $X$ and $X_{\Lag}$ are $s$-related by Proposition \ref{Chap04_prop:GenLagHJRelatedVF}, we have
that $s^*\inn(X_{\Lag})\omega_{\Lag} = \inn(X)s^*\omega_{\Lag}$, and hence we obtain
\begin{equation*}
\inn(X)s^*\omega_{\Lag} = \d(s^*E_{\Lag}) \, .
\end{equation*}

Conversely, consider the following vector field defined along the section $s\in\Gamma(\rho_{k-1}^{2k-1})$
\begin{equation*}
D_\Lag = X_{\Lag} \circ s - \Tan s \circ X \colon \Tan^{k-1}Q \to \Tan(\Tan^{2k-1}Q) \, .
\end{equation*}
We want to prove that $D_\Lag = 0$, or equivalently, as $\omega_{\Lag}$ is nondegenerate,
$(\omega_{\Lag})_{s(\bar{y})}(D_\Lag(\bar{y}),Z_{s(\bar{y})}) = 0$ for every tangent
vector $Z_{s(\bar{y})} \in \Tan_{s(\bar{y})}\Tan^{2k-1}Q$. Taking the pull-back of the
Lagrangian dynamical equation \eqref{Chap02_eqn:LagHODynEq}, and using the hypothesis, we have
\begin{equation*}
s^{*}(\inn(X_{\Lag})\omega_{\Lag}) = s^{*}(\d E_{\Lag}) = \d(s^{*}E_{\Lag}) = \inn(X)(s^{*}\omega_{\Lag}) \, ,
\end{equation*}
and then the form $s^{*}(\inn(X_{\Lag})\omega_{\Lag}) - \inn(X)(s^{*}\omega_{\Lag}) \in \df^1(\Tan^{k-1}Q)$
vanishes. Therefore, for every $\bar{y} \in \Tan^{k-1}Q$ and $u_{\bar{y}}\in \Tan_{\bar{y}}\Tan^{k-1}Q$, we have
\begin{align*}
0 &=
(s^*\inn(X_{\Lag})\omega_{\Lag} - \inn(X)s^*\omega_{\Lag})_{\bar{y}}(u_{\bar{y}}) \\
&= (\omega_{\Lag})_{s(\bar{y})}(X_{\Lag}(s(\bar{y})),\Tan_{\bar{y}}s(u_{\bar{y}}))
- (\omega_{\Lag})_{s(\bar{y})}(\Tan_{\bar{y}}s(X_{\bar{y}}),\Tan_{\bar{y}}s(u_{\bar{y}})) \\
&= (\omega_{\Lag})_{s(\bar{y})}(X_{\Lag}(s(\bar{y})) - \Tan_{\bar{y}}s(X_{\bar{y}}),\Tan_{\bar{y}}s(u_{\bar{y}})) \\
&= (\omega_{\Lag})_{s(\bar{y})}(D_\Lag(\bar{y}),\Tan_{\bar{y}}s(u_{\bar{y}})) \, .
\end{align*}
Therefore, $(\omega_{\Lag})_{s(\bar{y})}(D_\Lag(\bar{y}),A_{s(\bar{y})}) = 0$, for every
$A_{s(\bar{y})} \in \Tan_{s(\bar{y})}\Im(s)$. Now recall that every section defines a canonical
splitting of the tangent space of $\Tan^{2k-1}Q$ at every point given by
\begin{equation*}
\Tan_{s(\bar{y})}\Tan^{2k-1}Q = \Tan_{s(\bar{y})}\Im(s) \oplus V_{s(\bar{y})}(\rho^{2k-1}_{k-1}) \, .
\end{equation*}
Thus, we only need to prove that $(\omega_{\Lag})_{s(\bar{y})}(D_\Lag(\bar{y}),B_{s(\bar{y})}) = 0$, for
every vertical tangent vector $B_{s(\bar{y})} \in V_{s(\bar{y})}(\rho^{2k-1}_{k-1})$. Equivalently, as
$\omega_{\Lag}$ is annihilated by the contraction of two $\rho^{2k-1}_{k-1}$-vertical vectors, it suffices
to prove that $D_\Lag$ is vertical with respect to that submersion. Indeed,
\begin{align*}
\Tan\rho^{2k-1}_{k-1} \circ D_\Lag &=
\Tan\rho^{2k-1}_{k-1} \circ (X_{\Lag} \circ s - \Tan s \circ X) \\
&= \Tan\rho^{2k-1}_{k-1} \circ X_{\Lag} \circ s - \Tan\rho^{2k-1}_{k-1} \circ \Tan s \circ X \\
&= \Tan\rho^{2k-1}_{k-1} \circ X_{\Lag} \circ s - \Tan(\rho^{2k-1}_{k-1}\circ s) \circ X \\
&= \Tan\rho^{2k-1}_{k-1} \circ X_{\Lag} \circ s - X = 0 \, .
\end{align*}
Therefore
$(\omega_{\Lag})_{s(\bar{y})}(D_\Lag(\bar{y}),Z_{s(\bar{y})}) = 0$, for every
$Z_{s(\bar{y})} \in \Tan_{s(\bar{y})}\Tan^{2k-1}Q$, and as $\omega_{\Lag}$ is nondegenerate, we have
that $X_{\Lag}$ and $X$ are $s$-related, and, by Proposition \ref{Chap04_prop:GenLagHJRelatedVF}, $s$
is a solution to the generalized $k$th-order Lagrangian Hamilton-Jacobi problem. \qedhere
\end{description}
\end{proof}

Observe that if $s \in \Gamma(\rho^{2k-1}_{k-1})$ is a solution to the generalized $k$th-order Lagrangian
Hamilton-Jacobi problem then, taking into account Corollary \ref{Chap04_corol:GenLagHJRelatedVF}, we can
conclude that the integral curves of the Lagrangian vector field $X_\Lag$ contained in $\Im(s)$ project to
$\Tan^{k-1}Q$ by $\rho^{2k-1}_{k-1}$ to integral curves of $\Tan^{2k-1}_{k-1}\circ X_\Lag \circ s$.
The converse, however, is not true unless we make further assumptions.

\begin{remark}
Notice that, except for the third item in Proposition \ref{Chap04_prop:GenLagHJEquiv}, all the results
stated in this Section hold for every vector field $Z \in \vf(\Tan^{2k-1}Q)$, not only for the Lagrangian
vector field $X_\Lag$. Indeed, the assumption for $X_\Lag$ being the Lagrangian vector field solution to
the equation \eqref{Chap02_eqn:LagHODynEq} is only needed to prove that the section
$s \in \Gamma(\rho^{2k-1}_{k-1})$ and its associated vector field $X \in \vf(\Tan^{k-1}Q)$ satisfy some
kind of dynamical equation in $\Tan^{k-1}Q$.
\end{remark}

Let us compute in coordinates the condition for a section $s \in \Gamma(\rho^{2k-1}_{k-1})$ to be a
solution to the generalized $k$th-order Lagrangian Hamilton-Jacobi problem. Let $(q_0^A)$ be local
coordinates in $Q$, and $(q_0^A,\ldots,q_{2k-1}^A)$ the induced natural coordinates in $\Tan^{2k-1}Q$.
Then, local coordinates in $\Tan^{2k-1}Q$ adapted to the $\rho^{2k-1}_{k-1}$-bundle structure are
$(q_i^A;q_j^A)$, with $0 \leqslant i \leqslant k-1$ and $k \leqslant j \leqslant 2k-1$.  Hence, a section
$s \in \Gamma(\rho^{2k-1}_{k-1})$ is given locally by $s(q_i^A) = (q_i^A,s_{j}^A)$, where $s_j^A$
are local smooth functions in $\Tan^{k-1}Q$.

From Proposition \ref{Chap04_prop:GenLagHJEquiv} we know that $s \in \Gamma(\rho_{k-1}^{2k-1})$ is a
solution to the generalized $k$th-order Lagrangian Hamilton-Jacobi problem if, and only if, the
Euler-Lagrange vector field $X_\Lag \in \vf(\Tan^{2k-1}Q)$ is tangent to the submanifold
$\Im(s) \hookrightarrow \Tan^{2k-1}Q$. As $\Im(s)$ is locally defined by the constraints
$q_j^A - s_j^A = 0$, we must require $\Lie(X_\Lag)(q_j^A-s_j^A) \equiv X_\Lag(q_j^A-s_j^A) = 0$
(on $\Im(s)$), for $k \leqslant j \leqslant 2k-1$, $1 \leqslant A \leqslant n$.
From the results in Section \ref{Chap02_sec:LagrangianAutonomousHigherOrder}, we know that the
Euler-Lagrange vector field $X_\Lag$ has the following local expression
\begin{equation*}
X_\Lag = q_1^A\derpar{}{q_0^A} + q_2^A\derpar{}{q_1^A} + \ldots + q_{2k-1}^A\derpar{}{q_{2k-2}^A}
+ F^A \derpar{}{q_{2k-1}^A} \, ,
\end{equation*}
where $F^A$ are the functions solution to equations \eqref{Chap02_eqn:LagHODynEqWithHolonomyLocal}, that is,
to the following system of $n$ equations
\begin{equation*}
(-1)^k\left(F^B - d_T\left(q_{2k-1}^B\right)\right)\derpars{\Lag}{q_k^B}{q_k^A}
+ \sum_{l=0}^k(-1)^l d_T^l\left(\derpar{\Lag}{q_l^A}\right) = 0 \, .
\end{equation*}
Hence, the condition $\restric{X_\Lag(q_j^A-s_j^A)}{\Im(s)} = 0$ gives the following equations
\begin{equation}\label{Chap04_eqn:GenLagHJLocal}
s_{j+1}^A - \sum_{i=0}^{k-2}q_{i+1}^B\derpar{s_j^A}{q_i^B} - s_{k}^B\derpar{s_j^A}{q_{k-1}^B} = 0 \quad ; \quad
\restric{F^A}{\Im(s)} - \sum_{i=0}^{k-2} q_{i+1}^B\derpar{s_{2k-1}^A}{q_i^B} - s_{k}^B\derpar{s_j^A}{q_{k-1}^B} = 0 \, .
\end{equation}
This is a system of $kn$ partial differential equations with $kn$ unknown functions $s_j^A$. Thus, a
section $s \in \Gamma(\rho^{2k-1}_{k-1})$ solution to the generalized $k$th-order Lagrangian Hamilton-Jacobi
problem must satisfy the local equations \eqref{Chap04_eqn:GenLagHJLocal}.

\subsection{The Lagrangian Hamilton-Jacobi problem}
\label{Chap04_sec:LagHJProb}

In general, to solve the generalized $k$th-order Lagrangian Hamilton-Jacobi problem can be a difficult
task, since it amounts to find $kn$-codimensional submanifolds of $\Tan^{2k-1}Q$ invariant by the
Euler-Lagrange vector field $X_\Lag$, or, equivalently, solutions to a large system of partial differential
equations with many unknown functions. Therefore, in order to simplify the problem, it is convenient to
impose some additional conditions to the section $s \in \Gamma(\rho^{2k-1}_{k-1})$, thus considering a
less general problem.

\begin{definition}\label{Chap04_def:LagHJProb}
The \textnormal{$k$th-order Lagrangian Hamilton-Jacobi problem} consists in the search of solutions
$s \in \Gamma(\rho^{2k-1}_{k-1})$ to the generalized $k$th-order Lagrangian Hamilton-Jacobi
problem satisfying $s^*\omega_\Lag = 0$. Such a section is called a
\textnormal{solution to the $k$th-order Lagrangian Hamilton-Jacobi problem}.
\end{definition}

With the new assumption in Definition \ref{Chap04_def:LagHJProb}, a straightforward consequence of
Proposition \ref{Chap04_prop:GenLagHJEquiv} is the following result.

\begin{proposition}\label{Chap04_prop:LagHJEquiv}
The following conditions on a section $s \in \Gamma(\rho^{2k-1}_{k-1})$ satisfying $s^*\omega_\Lag = 0$
are equivalent.
\begin{enumerate}
\item The section $s$ is a solution to the $k$th-order Lagrangian Hamilton-Jacobi problem.
\item $\d(s^*E_\Lag) = 0$.
\item $\Im(s)$ is a Lagrangian submanifold of $\Tan^{2k-1}Q$ invariant by $X_\Lag$.
\item The integral curves of $X_\Lag$ with initial conditions in $\Im(s)$ project
onto the integral curves of the vector field $X = \Tan\rho^{2k-1}_{k-1} \circ X_\Lag \circ s$.
\end{enumerate}
\end{proposition}

Let us compute in coordinates the equations for a section $s \in \Gamma(\rho_{k-1}^{2k-1})$ to be a
solution to the $k$th-order Lagrangian Hamilton-Jacobi equation. From Proposition
\ref{Chap04_prop:LagHJEquiv} we know that this is equivalent to $\d(s^*E_\Lag) = 0$, which in turn is
equivalent to $s^*(\d E_\Lag) = 0$, since the pull-back and the exterior derivative commute. Then,
since $E_\Lag \in \Cinfty(\Tan^{2k-1}Q)$, its exterior derivative is given by
\begin{equation*}
\d E_\Lag = \derpar{E_\Lag}{q_i^A}\d q_i^A + \derpar{E_\Lag}{q_j^A}\d q_j^A
\, , \quad 0 \leqslant i \leqslant k-1,\ k \leqslant j \leqslant 2k-1 \, .
\end{equation*}
Then, taking the pull-back of this $1$-form by the section $s(q_i^A) = (q_i^A,s_j^A)$, we obtain
\begin{equation*}
s^*(\d E_\Lag) = \left( \derpar{E_\Lag}{q_i^A} + \derpar{E_\Lag}{q_j^B}\derpar{s_j^B}{q_i^A} \right) \d q_i^A \, .
\end{equation*}
Hence, the condition $\d(s^*E_\Lag) = 0$ in Proposition \ref{Chap04_prop:LagHJEquiv} is locally equivalent
to the following $kn$ partial differential equations (on $\Im(s)$)
\begin{equation}\label{Chap04_eqn:LagHJLocal}
\derpar{E_\Lag}{q_i^A} + \derpar{E_\Lag}{q_j^B}\derpar{s_j^B}{q_i^A} = 0 \, .
\end{equation}
Therefore, a section $s \in \Gamma(\rho^{2k-1}_{k-1})$ given locally by $s(q_i^A) = (q_i^A,s_j^A(q_i^A))$
is a solution to the $k$th-order Lagrangian Hamilton-Jacobi problem if, and only if, the local functions
$s_j^A$ satisfy the system of $2kn$ partial differential equations given by \eqref{Chap04_eqn:GenLagHJLocal}
and \eqref{Chap04_eqn:LagHJLocal}. Note that these $2kn$ partial differential equations may not be
$\Cinfty(U)$-linearly independent.

In addition to the local equations for the section $s \in \Gamma(\rho^{2k-1}_{k-1})$, we can state the
equations for the characteristic Hamilton-Jacobi function. These equations are a generalization to higher-order
systems of the classical Lagrangian Hamilon-Jacobi equations \eqref{Chap02_eqn:HJLagFOHamiltonJacobiEq}.

Since $\omega_\Lag = -\d\theta_\Lag$, it is clear that $s^*\omega_\Lag = 0$ if, and only if,
$s^*(\d\theta_\Lag) = \d(s^*\theta_\Lag) = 0$; that is, the form $s^*\theta_\Lag \in \df^{1}(\Tan^{k-1}Q)$ is
closed. Then, using Poincar\'{e}'s Lemma, $s^*\theta_\Lag$ is locally exact, and thus there exists
$W \in \Cinfty(U)$, with $U \subseteq \Tan^{k-1}Q$ an open set, such that $s^*\theta_\Lag\vert_U = \d W$.
In coordinates, bearing in mind the coordinate expression \eqref{Chap02_eqn:LagHO1FormLocal} of the
Poincar\'{e}-Cartan $1$-form, the form $s^*\theta_\Lag$ is given locally by
\begin{equation*}
s^*\theta_\Lag = \sum_{i=0}^{k-1} \sum_{l=0}^{k-i-1} \restric{(-1)^ld_T^l\left( \derpar{\Lag}{q_{i+1+l}^A} \right)}{\Im(s)} \d q_i^A \, ,
\end{equation*}
since $\theta_\Lag$ is $\rho^{2k-1}_{k-1}$-semibasic in $\Tan^{2k-1}Q$. Hence, from the identity
$s^*\theta_\Lag = \d W$ we obtain
\begin{equation}\label{Chap04_eqn:LagHJCharFunctLocal}
\derpar{W}{q_i^A} =
\sum_{l=0}^{k-i-1} \restric{(-1)^ld_T^l\left( \derpar{\Lag}{q_{i+1+l}^A} \right)}{\Im(s)} \, ,
\end{equation}
which is a system of $kn$ partial differential equations for $W$ that clearly generalizes equations
\eqref{Chap02_eqn:HJLagFOHamiltonJacobiEq}.

\subsection{Complete solutions}
\label{Chap04_sec:LagHJCompleteSolutions}

In the above Sections we have stated the $k$th-order Hamilton-Jacobi problem in the Lagrangian formalism,
and a section $s \in \Gamma(\rho^{2k-1}_{k-1})$ solution to this problem gives a particular solution
to the dynamical equation \eqref{Chap02_eqn:LagHODynEq}. Nevertheless, this is not a complete solution
to the system, since only the integral curves of $X_\Lag$ with initial conditions in $\Im(s)$ can be
recovered from the solution to the Hamilton-Jacobi problem. Hence, in order to obtain a complete solution
to the problem, we need to foliate the phase space $\Tan^{2k-1}Q$ in such a way that every leaf is the image
set of a section solution to the $k$th-order Lagrangian Hamilton-Jacobi problem. The precise definition is
the following.

\begin{definition}
A \textnormal{complete solution to the $k$th-order Lagrangian Hamilton-Jacobi problem} is a local
diffeomorphism $\Phi \colon U \times \Tan^{k-1}Q \to \Tan^{2k-1}Q$, with $U \subseteq \R^{kn}$ an
open set, such that for every $\lambda \in U$, the map
$s_\lambda(\bullet) \equiv \Phi(\lambda,\bullet) \colon \Tan^{k-1}Q \to \Tan^{2k-1}Q$ is a solution
to the $k$th-order Lagrangian Hamilton-Jacobi problem.
\end{definition}

\begin{remark}
Usually, it is the set of maps $\{ s_\lambda \mid \lambda \in U\}$ which is called a complete solution
of the $k$th-order Lagrangian Hamilton-Jacobi problem, instead of the map $\Phi$. Both definitions are
clearly equivalent.
\end{remark}

It follows from this last definition that a complete solution provides $\Tan^{2k-1}Q$ with a foliation
transverse to the fibers, and that every leaf of this foliation has dimension $kn$ and is invariant by
the Euler-Lagrange vector field $X_\Lag$.

Let $\Phi$ be a complete solution, and we consider the family of vector fields
\begin{equation*}
\left\{ X_\lambda = \Tan\rho^{2k-1}_{k-1} \circ X_\Lag \circ s_\lambda \in \vf(\Tan^{k-1}Q)
\ \mid \ \lambda \in U \subseteq \R^{kn} \right\} \, ,
\end{equation*}
where $s_\lambda \equiv \Phi(\lambda,\bullet)$. Then, the integral curves of $X_\lambda$, for different
$\lambda \in U$, will provide all the integral curves of the Euler-Lagrange vector field $X_\Lag$. That is,
if $\bar{y} \in \Tan^{2k-1}Q$, then there exists $\lambda_o \in U$ such that if
$p_o = \rho^{2k-1}_{k-1}(\bar{y})$, then $s_{\lambda_o}(p_o) = \bar{y}$, and the integral curve of
$X_{\lambda_o}$ through $p_o$, lifted to $\Tan^{2k-1}Q$ by $s_{\lambda_o}$, gives the integral curve
of $X_\Lag$ through $\bar{y}$.

\section{The Hamiltonian formalism}
\label{Chap04_sec:HamiltonianFormalism}

Recall that, in the Hamiltonian formalism for a $k$th-order dynamical system, all the geometric
structures are the canonical Liouville forms of the cotangent bundle $\Tan^*(\Tan^{k-1}Q)$,
namely $\theta_{k-1} \in \df^{1}(\Tan^*(\Tan^{k-1}Q))$ and
$\omega_{k-1} = -\d\theta_{k-1} \in \df^{2}(\Tan^*(\Tan^{k-1}Q))$, and the dynamics of the system
are given by a Hamiltonian function $h \in \Cinfty(\Tan^*(\Tan^{k-1}Q))$. With these elements
we can state the dynamical equation \eqref{Chap02_eqn:HamHODynEq} for this Hamiltonian system, which is
\begin{equation*}
\inn(X_h)\,\omega_{k-1} = \d h \, .
\end{equation*}
Since $\omega_{k-1}$ is symplectic regardless of the Hamiltonian function provided, the above equation
has always a unique solution $X_h \in \vf(\Tan^*(\Tan^{k-1}Q))$. In addition, since the Lagrangian function
$\Lag \in \Cinfty(\Tan^{k}Q)$ is regular, the Legendre-Ostrogradsky map defined in
\eqref{Chap02_eqn:HamHOLegendreMapDef} is a local diffeomorphism, and hence we have locally a one-to-one
correspondence between the vector field solution to the Lagrangian dynamical equation and the vector
field solution to the Hamiltonian dynamical equation.

Observe that, as the formalism is developed in the cotangent bundle of a manifold, $\Tan^*(\Tan^{k-1}Q)$,
the statement of the Hamiltonian Hamilton-Jacobi problem for higher-order systems follows the same patterns
as in the first-order case described in Section \ref{Chap02_sec:HamiltonJacobiHam}.

\subsection{The generalized Hamiltonian Hamilton-Jacobi problem}
\label{Chap04_sec:GenHamHJProb}

As in Section \ref{Chap02_sec:HamiltonJacobiHam} and the Lagrangian formalism stated in the previous
Section, we first consider the generalized Hamilton-Jacobi problem in the Hamiltonian formalism.
Recall that for first-order dynamical system, this problem consists in finding $1$-forms
$\alpha \in \df^{1}(Q)$ and vector fields $X \in \vf(Q)$ such that the lifting of every integral
curve of $X$ to $\Tan^*Q$ by $\alpha$ is an integral curve of the Hamiltonian vector field.
For higher-order systems, the statement of the problem is almost the same.

\begin{definition}\label{Chap04_def:GenHamHJProblem}
The \textnormal{generalized $k$th-order Hamiltonian Hamilton-Jacobi problem} consists in finding a
$1$-form $\alpha \in \df^{1}(\Tan^{k-1}Q)$ and a vector field $X \in \vf(\Tan^{k-1}Q)$ such that,
if $\gamma \colon \R \to \Tan^{k-1}Q$ is an integral curve of $X$, then
$\alpha \circ \gamma \colon \R \to \Tan^*(\Tan^{k-1}Q)$ is an integral curve of $X_h$; that is,
\begin{equation}\label{Chap04_eqn:GenHamHJDef}
X \circ \gamma = \dot{\gamma} \Longrightarrow
X_h \circ (\alpha \circ \gamma) = \dot{\overline{\alpha \circ \gamma}} \, .
\end{equation}
\end{definition}

As in the Lagrangian formalism, it is clear from Definition \ref{Chap04_def:GenHamHJProblem} that the
vector field $X \in \vf(\Tan^*(\Tan^{k-1}Q))$ and the $1$-form $\alpha \in \df^{1}(\Tan^{k-1}Q)$
must be related. In particular, we have the following result.

\begin{proposition}\label{Chap04_prop:GenHamHJRelatedVF}
The pair $(\alpha,X) \in \df^{1}(\Tan^{k-1}Q) \times \vf(\Tan^{k-1}Q)$ safisfies the condition
\eqref{Chap04_eqn:GenHamHJDef} if, and only if, $X$ and $X_h$ are $\alpha$-related, that is,
$X_h \circ \alpha = \Tan\alpha \circ X$.
\end{proposition}
\begin{proof}
The proof of this result follows the same patterns that the proof of Proposition
\ref{Chap04_prop:GenLagHJRelatedVF}. In particular, if $(\alpha,X)$ satisfies the condition
\eqref{Chap04_eqn:GenHamHJDef}, then for every integral curve $\gamma$ of $X$, we have
\begin{equation*}
X_h \circ ( \alpha \circ \gamma) = \dot{\overline{\alpha \circ \gamma}} = \Tan \alpha \circ \dot\gamma
= \Tan \alpha \circ X \circ \gamma \, ,
\end{equation*}
but, since $X$ has integral curves through every point $\bar{y} \in \Tan^{k-1}Q$, this is equivalent to
$X_h \circ \alpha = \Tan \alpha \circ X$.

Conversely, if $X_{h}$ and $X$ are $\alpha$-related and $\gamma \colon \R \to \Tan^{k-1}Q$ is an
integral curve of $X$, we have
\begin{equation*}
X_h \circ \alpha \circ \gamma = \Tan \alpha \circ X \circ \gamma = \Tan \alpha \circ \dot{\gamma}
= \dot{\overline{\alpha \circ \gamma}} \, . \qedhere
\end{equation*}
\end{proof}

That is, the vector field $X \in \vf(\Tan^{k-1}Q)$ is related to the Hamiltonian vector field $X_h$
and the $1$-form $\alpha \in \df^{1}(\Tan^*(\Tan^{k-1}Q))$. Moreover, from Proposition
\ref{Chap04_prop:GenHamHJRelatedVF}, composing both sides of the $\alpha$-relation equality
$X_h \circ \alpha = \Tan\alpha \circ X$ with $\Tan\pi_{\Tan^{k-1}Q}$, and bearing in mind
that $\alpha \in \df^{1}(\Tan^{k-1}Q) = \Gamma(\pi_{\Tan^{k-1}Q})$, we obtain the following result.

\begin{corollary}
If $(\alpha,X)$ satisfies condition \eqref{Chap04_eqn:GenHamHJDef}, then
$X = \Tan\pi_{\Tan^{k-1}Q} \circ X_h \circ \alpha$.
\end{corollary}

That is, the vector field $X \in \vf(\Tan^{k-1}Q)$ is completely determined by the $1$-form $\alpha$,
and it is called the \textsl{vector field associated to $\alpha$}. The following diagram illustrates
the situation
\begin{equation*}
\xymatrix{
\Tan(\Tan^{k-1}Q) \ar@/_1.5pc/[rrr]_{\Tan \alpha} & \ & \ &
 \ar[lll]_-{\Tan\pi_{\Tan^{k-1}Q}} \Tan(\Tan^*(\Tan^{k-1}Q)) \\
\ & \ & \ & \ \\
\Tan^{k-1}Q \ar[uu]^{X} \ar@/_1.5pc/[rrr]_{\alpha} & \ & \ &
 \ar[lll]_{\pi_{\Tan^{k-1}Q}} \Tan^*(\Tan^{k-1}Q) \ar[uu]_{X_h}
}
\end{equation*}

Since the vector field $X$ is completely determined by the $1$-form $\alpha$, the problem of finding
a pair $(\alpha,X) \in \df^{1}(\Tan^{k-1}Q) \times \vf(\Tan^{k-1}Q)$ that satisfies the condition
\eqref{Chap04_eqn:GenHamHJDef} is equivalent to the problem of finding a $1$-form
$\alpha \in \df^{1}(\Tan^{k-1}Q)$ satisfying the same condition with the associated vector field
$\Tan\pi_{\Tan^{k-1}Q} \circ X_h \circ \alpha$. Hence, we can give the following definition.

\begin{definition}
A \textnormal{solution to the generalized $k$th-order Hamiltonian Hamilton-Jacobi problem} for $X_h$
is a $1$-form $\alpha \in \df^{1}(\Tan^{k-1}Q)$ such that if $\gamma \colon \R \to \Tan^{k-1}Q$ is an
integral curve of $X = \Tan\pi_{\Tan^{k-1}Q} \circ X_h \circ \alpha$, then
$\alpha \circ \gamma \colon \R \to \Tan^*(\Tan^{k-1}Q)$ is an integral curve of $X_h$; that is,
\begin{equation*}
\Tan\pi_{\Tan^{k-1}Q} \circ X_h \circ \alpha \circ \gamma = \dot \gamma
\Longrightarrow X_h \circ (\alpha \circ \gamma) = \dot{\overline{\alpha \circ \gamma}} \, .
\end{equation*}
\end{definition}

Finally, we have the analogous result to Proposition \ref{Chap04_prop:GenLagHJEquiv} in the Hamiltonian
formalism, and also analogous to Theorem \ref{Chap02_thm:HJHamFOEquivalenceGeneralized} in the higher-order
setting, that gives several equivalent conditions for a $1$-form $\alpha$ to be a solution to the
generalized $k$th-order Hamiltonian Hamilton-Jacobi problem.

\begin{proposition}\label{Chap04_prop:GenHamHJEquiv}
The following conditions on a $1$-form $\alpha \in \df^{1}(\Tan^{k-1}Q)$ are equivalent.
\begin{enumerate}
\item The form $\alpha$ is a solution to the generalized $k$th-order Hamiltonian Hamilton-Jacobi problem.
\item The submanifold $\Im(\alpha) \hookrightarrow \Tan^*(\Tan^{k-1}Q)$ is invariant under the flow of the
Hamiltonian vector field $X_h$ (that is, $X_h$ is tangent to the submanifold $\Im(\alpha)$).
\item The form $\alpha$ satisfies the equation
\begin{equation*}
\inn(X)\d\alpha = -\d(\alpha^*h) \, ,
\end{equation*}
where $X = \Tan\pi_{\Tan^{k-1}Q} \circ X_h \circ \alpha$ is the vector field associated to $\alpha$.
\end{enumerate}
\end{proposition}
\begin{proof}
This proof follows exactly the same patterns as the proof of Proposition \ref{Chap04_prop:GenLagHJEquiv},
taking into account the properties of the tautological form $\theta_{k-1} \in \df^{1}(\Tan^*(\Tan^{k-1}Q))$
of the cotangent bundle, that is, we have $\alpha^*\theta_{k-1} = \alpha$ for every $\alpha \in \df^{1}(\Tan^{k-1}Q)$.
Hence, taking the pull-back of the dynamical equation \eqref{Chap02_eqn:HamHODynEq} by $\alpha$ we obtain
\begin{equation*}
\inn(X)\d\alpha = -\d(\alpha^*h) \, ,
\end{equation*}
because we have
\begin{equation}\label{Chap04_eqn:PullBackSymplecticFormByAlpha}
\alpha^*\omega_{k-1} = \alpha^*(-\d\theta_{k-1}) = -\d(\alpha^*\theta_{k-1}) = -\d\alpha \, .
\end{equation}
In particular, we have:
\begin{description}
\item[\textnormal{($1 \, \Longleftrightarrow \, 2$)}]
Let $\alpha$ be a solution to the generalized $k$th-order Hamiltonian Hamilton-Jacobi problem. Then by
Proposition \ref{Chap04_prop:GenHamHJRelatedVF} the Hamiltonian vector field $X_{h} \in \vf(\Tan^*(\Tan^{k-1}Q))$
is $\alpha$-related to the vector field $X = \Tan\pi_{\Tan^{k-1}Q} \circ X_h \circ \alpha \in \vf(\Tan^{k-1}Q)$
associated to $\alpha$, and thus for every $\bar{y} \in \Tan^{k-1}Q$ we have
\begin{equation*}
X_h(\alpha(\bar{y})) = (X_h\circ \alpha)(\bar{y}) = (\Tan \alpha \circ X)(\bar{y}) = \Tan \alpha(X(\bar{y})) \, .
\end{equation*}
Hence, $X_h(\alpha(\bar{y})) = \Tan \alpha (X(\bar{y}))$ and therefore $X_h$ is tangent to the submanifold
$\Im(\alpha) \hookrightarrow \Tan^*(\Tan^{k-1}Q)$.

Conversely, if the submanifold $\Im(\alpha)$ is invariant under the flow of $X_h$, then
$X_h(\alpha(\bar{y})) \in \Tan_{\alpha(\bar{y})}\Im(\alpha)$, for every $\bar{y} \in \Tan^{k-1}Q$;
that is, there exists an element $u_{\bar{y}} \in \Tan_{\bar{y}}\Tan^{k-1}Q$ such that
$X_h(\alpha(\bar{y})) = \Tan_{\bar{y}}\alpha(u_{\bar{y}})$. If we define $X \in \vf(\Tan^{k-1}Q)$
as the vector field that satisfies $\Tan_{\bar{y}}\alpha(X_{\bar{y}}) = X_{h}(\alpha(\bar{y}))$,
then $X$ is a vector field in $\Tan^{k-1}Q$, since $X = \Tan\pi_{\Tan^{k-1}Q} \circ X_h \circ \alpha$,
and it is $\alpha$-related with $X_h$. Therefore, by Proposition \ref{Chap04_prop:GenHamHJRelatedVF},
$\alpha$ is a solution to the generalized $k$th-order Lagrangian Hamilton-Jacobi problem.

\item[\textnormal{($1 \, \Longleftrightarrow \, 3$)}]
Let $\alpha$ be a solution to the generalized $k$th-order Hamiltonian Hamilton-Jacobi problem. Taking the
pull-back of the Hamiltonian dynamical equation \eqref{Chap02_eqn:HamHODynEq} by the $1$-form $\alpha$ we have
\begin{equation*}
\alpha^*\inn(X_h)\omega_{k-1} = \alpha^*\d h = \d(\alpha^*h) \, ,
\end{equation*}
but since $X$ and $X_h$ are $\alpha$-related by Proposition \ref{Chap04_prop:GenHamHJRelatedVF}, we have
that $\alpha^*\inn(X_h)\omega_{k-1} = \inn(X)\alpha^*\omega_{k-1}$. Then, using relation
\eqref{Chap04_eqn:PullBackSymplecticFormByAlpha}, we obtain
\begin{equation*}
-\inn(X)\d\alpha = \d(\alpha^*h) \, .
\end{equation*}

Conversely, consider the following vector field along the $1$-form $\alpha \in \df^{1}(\Tan^{k-1}Q)$
\begin{equation*}
D_h = X_h \circ \alpha - \Tan \alpha \circ X \colon \Tan^{k-1}Q \to \Tan(\Tan^*(\Tan^{k-1}Q)) \, .
\end{equation*}
We want to prove that $D_h = 0$, or equivalently, since the Liouville $2$-form $\omega_{k-1}$ is symplectic,
$(\omega_{k-1})_{\alpha(\bar{y})}(D_h(\bar{y}),Z_{\alpha(\bar{y})}) = 0$ for every tangent
vector $Z_{\alpha(\bar{y})} \in \Tan_{\alpha(\bar{y})}\Tan^*(\Tan^{k-1}Q)$. Taking the pull-back of the
Hamiltonian dynamical equation \eqref{Chap02_eqn:HamHODynEq}, and using the hypothesis, we have
\begin{equation*}
\alpha^{*}(\inn(X_h)\omega_{k-1}) = \alpha^{*}\d h = \d(\alpha^{*}h) = - \inn(X)\d\alpha \, ,
\end{equation*}
and then the form $\alpha^{*}(\inn(X_h)\omega_{k-1}) + \inn(X)\d\alpha = 
\alpha^{*}(\inn(X_h)\omega_{k-1}) - \inn(X)\alpha^*\omega_{k-1} \in \df^1(\Tan^{k-1}Q)$
vanishes. Therefore, for every $\bar{y} \in \Tan^{k-1}Q$ and $u_{\bar{y}}\in \Tan_{\bar{y}}\Tan^{k-1}Q$, we have
\begin{align*}
0 &=
(\alpha^*\inn(X_h)\omega_{k-1} - \inn(X)\alpha^*\omega_{k-1})_{\bar{y}}(u_{\bar{y}}) \\
&= (\omega_{k-1})_{\alpha(\bar{y})}(X_h(\alpha(\bar{y})),\Tan_{\bar{y}}\alpha(u_{\bar{y}}))
- (\omega_{k-1})_{\alpha(\bar{y})}(\Tan_{\bar{y}}\alpha(X_{\bar{y}}),\Tan_{\bar{y}}\alpha(u_{\bar{y}})) \\
&= (\omega_{k-1})_{\alpha(\bar{y})}(X_h(\alpha(\bar{y})) - \Tan_{\bar{y}}\alpha(X_{\bar{y}}),\Tan_{\bar{y}}\alpha(u_{\bar{y}})) \\
&= (\omega_{k-1})_{\alpha(\bar{y})}(D_h(\bar{y}),\Tan_{\bar{y}}\alpha(u_{\bar{y}})) \, .
\end{align*}
Therefore, $(\omega_{k-1})_{\alpha(\bar{y})}(D_h(\bar{y}),A_{\alpha(\bar{y})}) = 0$, for every
$A_{\alpha(\bar{y})} \in \Tan_{\alpha(\bar{y})}\Im(\alpha)$. Now recall that since
$\df^{1}(\Tan^{k-1}Q) = \Gamma(\pi_{\Tan^{k-1}Q})$, then every $1$-form defines a canonical
splitting of the tangent space of $\Tan^*(\Tan^{k-1}Q)$ at every point given by
\begin{equation*}
\Tan_{\alpha(\bar{y})}\Tan^*(\Tan^{k-1}Q) = \Tan_{\alpha(\bar{y})}\Im(\alpha) \oplus
V_{\alpha(\bar{y})}(\pi_{\Tan^{k-1}Q}) \, .
\end{equation*}
Thus, we only need to prove that $(\omega_{k-1})_{\alpha(\bar{y})}(D_h(\bar{y}),B_{\alpha(\bar{y})}) = 0$, for
every vertical tangent vector $B_{\alpha(\bar{y})} \in V_{\alpha(\bar{y})}(\pi_{\Tan^{k-1}Q})$. Equivalently, as
$\omega_{k-1}$ is annihilated by the contraction of two $\pi_{\Tan^{k-1}Q}$-vertical vectors, it suffices
to prove that $D_h$ is vertical with respect to that submersion. Indeed,
\begin{align*}
\Tan\pi_{\Tan^{k-1}Q} \circ D_h &=
\Tan\pi_{\Tan^{k-1}Q} \circ (X_h \circ \alpha - \Tan \alpha \circ X) \\
&= \Tan\pi_{\Tan^{k-1}Q} \circ X_h \circ \alpha - \Tan\pi_{\Tan^{k-1}Q} \circ \Tan \alpha \circ X \\
&= \Tan\pi_{\Tan^{k-1}Q} \circ X_h \circ \alpha - \Tan(\pi_{\Tan^{k-1}Q}\circ \alpha) \circ X \\
&= \Tan\pi_{\Tan^{k-1}Q} \circ X_h \circ \alpha - X = 0 \, .
\end{align*}
Therefore $(\omega_{k-1})_{\alpha(\bar{y})}(D_h(\bar{y}),Z_{\alpha(\bar{y})}) = 0$, for every
$Z_{\alpha(\bar{y})} \in \Tan_{\alpha(\bar{y})}\Tan^*(\Tan^{k-1}Q)$, and as $\omega_{k-1}$ is symplectic, we have
that $X_h$ and $X$ are $\alpha$-related, and by Proposition \ref{Chap04_prop:GenHamHJRelatedVF} $\alpha$
is a solution to the generalized $k$th-order Lagrangian Hamilton-Jacobi problem. \qedhere
\end{description}
\end{proof}

Now we give in coordinates the condition for a $1$-form $\alpha \in \df^{1}(\Tan^{k-1}Q)$ to be a solution
to the generalized $k$th-order Hamiltonian Hamilton-Jacobi problem. Let $(q_0^A)$ be local coordinates in
$Q$ and $(q_i^A)$, $0 \leqslant i \leqslant k-1$, the induced natural coordinates in $\Tan^{k-1}Q$.
Then, $(q_i^A,p_A^i)$ are natural coordinates in $\Tan^*(\Tan^{k-1}Q)$, which are also the adapted
coordinates to the $\pi_{\Tan^{k-1}Q}$-bundle structure. Hence, a $1$-form $\alpha \in \df^{1}(\Tan^{k-1}Q)$
is given locally by $\alpha(q_i^A) = \alpha_A^i \d q_i^A$, where $\alpha_A^i$ are local
smooth functions in $\Tan^{k-1}Q$.

Then, if $\alpha \in \df^{1}(\Tan^{k-1}Q)$ is a solution to the generalized $k$th-order Hamiltonian
Hamilton-Jacobi problem, then by Proposition \ref{Chap04_prop:GenHamHJEquiv} this is equivalent to
require the Hamiltonian vector field $X_h \in \vf(\Tan^*(\Tan^{k-1}Q))$ to be tangent to the submanifold
$\Im(\alpha) \hookrightarrow \Tan^*(\Tan^{k-1}Q)$. This submanifold is locally defined by the $kn$ constraints
$p_A^i - \alpha_A^i = 0$. Thus, we must require $\Lie(X_h)(p_A^i-\alpha_A^i) \equiv X_h(p_A^i-\alpha_A^i) = 0$
(on $\Im(\alpha)$). From the geometric description of the Hamiltonian formalism for higher-order systems
given in Section \ref{Chap02_sec:HamiltonianAutonomousHigherOrder}, we know that the Hamiltonian vector field
solution to equation \eqref{Chap02_eqn:HamHODynEq} is locally given by
\begin{equation*}
X_h = \derpar{h}{p_A^i}\,\derpar{}{q_i^A} - \derpar{h}{q_i^A}\,\derpar{}{p_A^i} \, .
\end{equation*}
Hence, the conditions $\restric{X_h(p_A^i-\alpha_A^i)}{\Im(\alpha)} = 0$ give the equations
\begin{equation}\label{Chap04_eqn:GenHamHJLocal}
\derpar{h}{q_i^A} + \derpar{h}{p_B^j}\derpar{\alpha_A^i}{q_j^B} = 0  \, , \ \mbox{(on $\Im(\alpha)$)} \, .
\end{equation}
This is a system of $kn$ partial differential equations with $kn$ unknown functions $\alpha_A^i$ which
must be verified by every $1$-form $\alpha \in \df^{1}(\Tan^{k-1}Q)$ solution to the generalized $k$th-order
Hamiltonian Hamilton-Jacobi.

\subsection{The Hamiltonian Hamilton-Jacobi problem}
\label{Chap04_sec:HamHJProblem}

As in the Lagrangian setting, to solve the generalized $k$th-order Hamiltonian Hamilton-Jacobi problem
is, in general, a difficult task. Hence, it is convenient to consider a less general problem requiring some
additional conditions to the $1$-form $\alpha \in \df^{1}(\Tan^{k-1}Q)$. Observe that, from
\eqref{Chap04_eqn:PullBackSymplecticFormByAlpha}, the isotropic condition $\alpha^*\omega_{k-1} = 0$ is equivalent
to $\d\alpha = 0$, that is, $\alpha$ is a closed $1$-form in $\Tan^{k-1}Q$. Therefore, we have the
following definition.

\begin{definition}\label{Chap04_def:HamHJProblemClassic}
The \textnormal{$k$th-order Hamiltonian Hamilton-Jacobi problem} consists in finding closed $1$-forms
$\alpha \in \df^{1}(\Tan^{k-1}Q)$ solution to the generalized $k$th-order Hamiltonian Hamilton-Jacobi problem.
Such a form is called a \textnormal{solution to the $k$th-order Hamiltonian Hamilton-Jacobi problem}.
\end{definition}

Then, bearing in mind the additional assumption of being closed, a straightforward consequence of
Proposition \ref{Chap04_prop:GenHamHJEquiv} is the following result, which is the analogous to Proposition
\ref{Chap04_prop:LagHJEquiv} in this formalism.

\begin{proposition}\label{Chap04_prop:HamHJEquiv}
Let $\alpha \in \df^{1}(\Tan^{k-1}Q)$ be a closed $1$-form. The following assertions are equivalent.
\begin{enumerate}
\item The $1$-form $\alpha$ is a solution to the $k$th-order Hamiltonian Hamilton-Jacobi problem.
\item $\d(\alpha^*h) = 0$.
\item $\Im(\alpha)$ is a Lagrangian submanifold of $\Tan^*(\Tan^{k-1}Q)$ invariant by $X_h$.
\item The integral curves of $X_h$ with initial conditions in $\Im(\alpha)$ project
onto the integral curves of the vector field $X = \Tan\pi_{\Tan^{k-1}Q} \circ X_h \circ \alpha$.
\end{enumerate}
\end{proposition}

Let us compute in the natural coordinates of the cotangent bundle $\Tan^*(\Tan^{k-1}Q)$ the local
equations for a $1$-form $\alpha \in \df^{1}(\Tan^{k-1}Q)$ to be a solution to the $k$th-order
Hamiltonian Hamilton-Jacobi problem. From Definition \ref{Chap04_def:HamHJProblemClassic}, we must
require the form $\alpha$ to be closed, that is, $\d\alpha = 0$. Hence, if $\alpha = \alpha_A^i\d q_i^A$,
this condition gives the following $kn(kn-1)/2$ equations
\begin{equation}\label{Chap04_eqn:HamHJLocalClosedForm}
\derpar{\alpha_A^i}{q_j^B} - \derpar{\alpha_B^j}{q_i^A} = 0 \, , \quad
\mbox{with } A \neq B \mbox{ or } i \neq j \, .
\end{equation}

Equivalently, from Proposition \ref{Chap04_prop:HamHJEquiv}, we know that this condition is equivalent
to $\d(\alpha^*h) = \alpha^*(\d h) = 0$. Then, bearing in mind that the exterior derivative of $h$
is given locally by
\begin{equation*}
\d h = \derpar{h}{q_i^A}\d q_i^A + \derpar{h}{p_A^i} \d p_A^i \, ,
\end{equation*}
and taking the pull-back of $\d h$ by the $1$-form $\alpha = \alpha_A^i \d q_i^A$, we have
\begin{equation*}
\alpha^*(\d h) = \left( \derpar{h}{q_i^A} +
\derpar{h}{p_B^j}\derpar{\alpha_B^j}{q_i^A} \right) \d q_i^A \, .
\end{equation*}
Hence, the condition $\d(\alpha^*h) = 0$ in Proposition \ref{Chap04_prop:HamHJEquiv} holds if, and only if,
the following $kn$ partial differential equations hold on $\Im(\alpha)$
\begin{equation}\label{Chap04_eqn:HamHJLocal}
\derpar{h}{q_i^A} + \derpar{h}{p_B^j}\derpar{\alpha_B^j}{q_i^A} = 0 \, .
\end{equation}

Therefore, a $1$-form $\alpha \in \df^{1}(\Tan^{k-1}Q)$ given locally by $\alpha = \alpha_A^i\d q_i^A$
is a solution to the $k$th-order Hamiltonian Hamilton-Jacobi problem if, and only if, the local functions
$\alpha_A^i$ satisfy the system of partial differential equations given by \eqref{Chap04_eqn:GenHamHJLocal} and
\eqref{Chap04_eqn:HamHJLocalClosedForm}, or equivalently \eqref{Chap04_eqn:GenHamHJLocal} and
\eqref{Chap04_eqn:HamHJLocal}. Observe that these systems of partial differential equations may not be
$\Cinfty(U)$-linearly independent.

In addition to the local equations for the $1$-form $\alpha \in \df^{1}(\Tan^{k-1}Q)$, we can give the
equation for the characteristic Hamilton-Jacobi function. This equation is a generalization to higher-order
systems of the classical Hamiltonian Hamilton-Jacobi equation \eqref{Chap02_eqn:HJHamFOHamiltonJacobiEq}.

Since $\alpha \in \df^{1}(\Tan^{k-1}Q)$ is closed, by Poincar{\'e}'s Lemma there exists a local function
$W \in \Cinfty(U)$, with $U \subseteq \Tan^{k-1}Q$ an open set, such that $\alpha = \d W$. In coordinates,
the condition $\alpha = \d W$ gives the following $kn$ partial differential equations for $W$
\begin{equation*}\label{Chap04_eqn:HamHJCharFunctLocal}
\derpar{W}{q_i^A} = \alpha_A^i \, .
\end{equation*}
Finally, as $\displaystyle \alpha^*h = h(q_i^A,\alpha_A^i) = h\left( q_i^A, \partial W / \partial q_i^A \right)$,
the condition $\d(\alpha^*h) = 0$ is equivalent to $\alpha^*h$ being locally constant, and hence we obtain
\begin{equation}\label{Chap04_eqn:HamHJEquation}
h\left( q_i^A, \derpar{W}{q_i^A} \right) = \mbox{const.}
\end{equation}
This equation clearly generalizes the equation \eqref{Chap02_eqn:HJHamFOHamiltonJacobiEq} to higher-order Hamiltonian systems.

\subsection{Complete solutions}
\label{Chap04_sec:HamHJCompleteSolutions}

As in the Lagrangian formalism, in the above Sections we have stated the $k$th-order Hamilton-Jacobi
problem in the Hamiltonian formalism, and a $1$-form $\alpha \in \df^{1}(\Tan^{k-1}Q)$ solution to this
problem gives a particular solution to the dynamical equation \eqref{Chap02_eqn:HamHODynEq}, but not a
complete solution, since only some integral curves of the vector field $X_h$ can be recovered from the
solution to the Hamilton-Jacobi problem. Hence, we want to define the concept of complete solution in
this formulation, and the way to do so is analogous to Section \ref{Chap04_sec:LagHJCompleteSolutions}.

\begin{definition}
A \textnormal{complete solution to the $k$th-order Hamiltonian Hamilton-Jacobi problem} is a local
diffeomorphism $\Phi \colon U \times \Tan^{k-1}Q \to \Tan^*(\Tan^{k-1}Q)$, where $U \subseteq \R^{kn}$
is an open set, such that, for every $\lambda \in U$, the map
$\alpha_\lambda(\bullet)\equiv\Phi(\lambda,\bullet) \colon \Tan^{k-1}Q \to \Tan^*(\Tan^{k-1}Q)$
is a solution to the $k$th-order Hamiltonian Hamilton-Jacobi problem.

\noindent Then, the set $\{ \alpha_\lambda \equiv \Phi(\lambda,\bullet) \in \df^{1}(\Tan^{k-1}Q) \mid \lambda \in U\}$
is also called a \textnormal{complete solution to the $k$th-order Hamiltonian Hamilton-Jacobi problem}.
\end{definition}

It follows from the definition that a complete solution endows $\Tan^*(\Tan^{k-1}Q)$
with a foliation transverse to the fibers, and that the Hamiltonian vector field $X_h$
is tangent to the leaves.

Let $\{ \alpha_\lambda \mid \lambda \in U \}$ be a complete solution, and we consider the
set of associated vector fields
\begin{equation*}
\left\{ X_\lambda = \Tan\pi_{\Tan^{k-1}Q} \circ X_h \circ \alpha_\lambda \in \vf(\Tan^{k-1}Q)
\ \mid \ \lambda \in U \subseteq \R^{kn} \right\} \, .
\end{equation*}
Then, the integral curves of $X_\lambda$, for different $\lambda \in U$, will provide all the integral
curves of the Hamiltonian vector field $X_h$. That is, if $\beta \in \Tan^*(\Tan^{k-1}Q)$, then there exists
$\lambda_o \in U$ such that if $\bar{y} = \pi_{\Tan^{k-1}Q}(\beta)$, then $\alpha_{\lambda_o}(\bar{y}) = \beta$,
and the integral curve of $X_{\lambda_o}$ through $\bar{y}$, lifted  to $\Tan^*(\Tan^{k-1}Q)$ by
$\alpha_{\lambda_o}$, gives the integral curve of $X_h$ through $\beta$.

Let us assume that $\Phi \colon \R^{kn} \times \Tan^{k-1}Q \to \Tan^*(\Tan^{k-1}Q)$ is a global
diffeomorphism for simplicity. Then, given $\lambda = (\lambda_i^A) \in \R^{kn}$,
$0 \leqslant i \leqslant k-1$, $1 \leqslant A \leqslant n$, we consider the functions
$f_{j}^{B} \in \Cinfty(\Tan^*(\Tan^{k-1}Q))$, $0 \leqslant j \leqslant k-1$, $1 \leqslant B \leqslant n$, given by
\begin{equation*}
f_{j}^{B} = \pr_{j}^{B} \circ \p_1 \circ \Phi^{-1} \, ,
\end{equation*}
where $\p_1 \colon \R^{kn} \times \Tan^{k-1}Q \to \R^{kn}$ is the projection onto the first factor
and $\pr_{j}^{B} \colon \R^{kn} \to \R$ is given by $\pr_{j}^{B} = \pr^{B} \circ \pr_{j}$,
where $\pr^B$ and $\pr_j$ are the natural projections
\begin{equation*}
\begin{array}{rcl}
\pr_j \colon \R^{kn} & \longrightarrow & \R^{n} \\
(x_i^A) & \longmapsto & (x_{j}^1,\ldots,x_j^n)
\end{array} \quad ; \quad
\begin{array}{rcl}
\pr^{B} \colon \R^{n} & \longrightarrow & \R \\
(x^1,\ldots,x^n) & \longmapsto & x^{B}
\end{array} \, .
\end{equation*}
Therefore, $f_{j}^{B}(\alpha_{\lambda}(q_i^A)) = (\pr_{j}^{B} \circ \p_{1} \circ \Phi^{-1} \circ \Phi)(\lambda_i^A,q_i^A)
(\pr_j^B \circ \p_1)(\lambda_i^A,q_i^A)  = \lambda_{j}^{B}$.

\begin{proposition}\label{Chap04_prop:HamHJInvolution}
The functions $f_{j}^{B}$, $0 \leqslant j \leqslant k-1,$ $1 \leqslant B \leqslant n$ are in involution,
that is, $\{ f_{i}^{j},f_{a}^{b} \} = 0$ for every $a,b,i,j.$
\end{proposition}
\begin{proof}
Since $\Phi$ is a complete solution, for every $\beta \in \Tan^*(\Tan^{k-1}Q)$ there exists a unique
$\lambda \in \R^{kn}$ such that
$\alpha_\lambda(\pi_{\Tan^{k-1}Q}(\beta)) = \Phi(\lambda,\pi_{\Tan^{k-1}Q}(\beta)) = \beta$. Then we have
\begin{align*}
f_j^B(\beta) &= (f_j^B \circ \alpha_\lambda)(\pi_{\Tan^{k-1}Q}(\beta))
= (\pr_{j}^{B} \circ \p_1 \circ \Phi^{-1} \circ \Phi)(\lambda,\pi_{\Tan^{k-1}Q}(\beta)) \\
&= (\pr_j^B \circ \p_1)(\lambda,\pi_{\Tan^{k-1}Q}(\beta))
= \lambda_j^B \, ,
\end{align*}
that is, $f_j^B \circ \alpha_\lambda = (f_j^B \circ \Phi)(\lambda,\bullet) \colon \Tan^{k-1}Q \to \R$
is constant for every $\lambda \in \R^{kn}$. Therefore, we have
\begin{equation*}
\restric{\d f_{j}^{B}}{\Tan \Im(\alpha_{\lambda})} = 0 \, .
\end{equation*}

Now, since $\Phi$ is a complete solution, we have that $\alpha_{\lambda} = \Phi(\lambda,\bullet)$ is a
solution to the $k$th-order Hamiltonian Hamilton-Jacobi problem. Therefore, from Proposition
\ref{Chap04_prop:HamHJEquiv}, $\Im(\Phi_{\lambda})$ is a Lagrangian submanifold of
$(\Tan^{*}(\Tan^{k-1}Q),\omega_{k-1})$, and then
\begin{equation*}
\left(\Tan \Im(\alpha_{\lambda})\right)^{\bot} = \Tan \Im(\alpha_{\lambda}) \, ,
\end{equation*}
where $\left(\Tan \Im(\alpha_{\lambda})\right)^{\bot}$ denotes the $\omega_{k-1}$-orthogonal
of $\Tan \Im(\alpha_{\lambda})$ (see Definition \ref{Chap01_def:SymplecticOrthogonal} for details).

From this, the result follows from the definition of the induced Poisson bracket and the facts that the form
$\omega_{k-1}$ is symplectic, that $\d f_{j}^{B} \in (\Tan \Im(\alpha_{\lambda}))^{\bot} = \Tan \Im(\alpha_{\lambda})$,
and that there exists a unique vector field $X_{f_{j}^{B}} \in \vf(\Tan^*(\Tan^{k-1}Q))$ satisfying
$\inn(X_{f_{j}^{B}})\,\omega_{k-1} = \d f_{j}^{B}$ (see Definition \ref{Chap01_def:PoissonBracketDef}
for the definition of the induced Poisson bracket, Section \ref{Chap01_sec:CanonicalIsomorphismSymplectic}
for the properties of Hamiltonian vector fields associated to functions, and Section
\ref{Chap01_sec:LagrangianSubmanifolds} for the properties of Lagrangian submanifolds).
\end{proof}

\subsection{Relation with the Lagrangian formulation}

Up to this point we have stated both the Lagrangian and Hamiltonian Hamilton-Jacobi problems for a
$k$th-order autonomous system. Now, we establish a relation between the solutions of the Hamilton-Jacobi
problem in both formulations. In particular, we show that there exists a bijection between the set of
solutions of the (generalized) $k$th-order Lagrangian Hamilton-Jacobi problem and the set of solutions of
the (generalized) $k$th-order Hamiltonian Hamilton-Jacobi problem, given by the Legendre-Ostrogradsky map
defined in \eqref{Chap02_eqn:HamHOLegendreMapDef}.

Since we assumed from the beginning that the $k$th-order Lagrangian function $\Lag \in \Cinfty(\Tan^{k}Q)$
is regular, the Legendre-Ostrogradsky map $\Leg \colon \Tan^{2k-1}Q \to \Tan^*(\Tan^{k-1}Q)$ is a local
diffeomorphism. For the sake of simplicity, in this Section we assume that the $k$th-order Lagrangian
function is hyperregular, and therefore the Legendre-Ostrogradsky map is a global diffeomorphism.
Obviously, for regular but not hyperregular Lagrangian functions, all these results hold only in the
open sets where $\Leg$ is a local diffeomorphism.

\begin{remark}
Observe that if $\Lag$ is hyperregular, then $\Leg$ is a symplectomorphism and therefore the symplectic
structures are in correspondence. Therefore, the induced Poisson brackets also are in correspondence and
we have the analogous to Proposition \ref{Chap04_prop:HamHJInvolution} in the Lagrangian formalism,
where the Poisson bracket is determined by the Poincar\'e-Cartan $2$-form $\omega_{\Lag}$ as
$\{f,g\} = \omega_{\Lag}(X_f,X_g)$, $X_f,X_g \in \vf(\Tan^{2k-1}Q)$ being the Hamiltonian vector fields
of $f$ and $g$, respectively, determined by the symplectic form $\omega_\Lag$.
\end{remark}

In order to establish the relation between the Lagrangian and Hamiltonian formalism, we first need the
following technical result.

\begin{lemma}\label{Chap04_lemma:TechLemma}
Let $E_1 \stackrel{\pi_1}{\longrightarrow} M$ and $E_2 \stackrel{\pi_2}{\longrightarrow} M$ be two fiber
bundles, $F \colon E_1 \to E_2$ a fiber bundle morphism, and two $F$-related vector fields
$X_1 \in \vf(E_1)$ and $X_2 \in \vf(E_2)$.  If $s_1 \in \Gamma(\pi_1)$ is a section of $\pi_1$ and we define
a section of $\pi_2$ as $s_2 = F \circ s_1 \in \Gamma(\pi_2)$, then
\begin{equation*}
\Tan\pi_1 \circ X_1 \circ s_1 = \Tan\pi_2 \circ X_2 \circ s_2 \in \vf(M) \, .
\end{equation*}
\end{lemma}
\begin{proof}
As $F \colon E_1 \to E_2$ is a fiber bundle morphism (that is, $\pi_1 = \pi_2 \circ F$), and $X_1$ and $X_2$
are $F$-related (that is, $\Tan F \circ X_1 = X_2 \circ F$), we have the following commutative diagram
\begin{equation*}
\xymatrix{
\Tan E_1 \ar[rr]^{\Tan F} &  \ & \Tan E_2 \\
E_1 \ar[dr]^{\pi_1} \ar[u]^{X_1} \ar[rr]^{F} & \ & E_2 \ar[dl]_{\pi_2} \ar[u]_{X_2} \\
\ & M \ar@/^1.1pc/[lu]^{s_1} \ar@/_1.1pc/@{-->}[ru]_<(.6){s_2 = F \circ s_1} & \
}
\end{equation*}
Then we have
\begin{equation*}
\Tan\pi_1 \circ X_1 \circ s_1 = \Tan(\pi_2 \circ F) \circ X_1 \circ s_1
= \Tan\pi_2 \circ \Tan F \circ X_1 \circ s_1
= \Tan\pi_2 \circ X_2 \circ F \circ s_1
= \Tan\pi_2 \circ X_2 \circ s_2 \, . \qedhere
\end{equation*}
\end{proof}

Now we can state the equivalence theorem. 

\begin{theorem}\label{Chap04_thm:EquivalenceSolutionsLag&Ham}
Let $(\Tan^{2k-1}Q,\Lag)$ be a hyperregular Lagrangian system, and $(\Tan^*(\Tan^{k-1}Q),\omega_{k-1},h)$
its associated Hamiltonian system.
Then, if $s \in \Gamma(\rho^{2k-1}_{k-1})$ is a solution to the (generalized) $k$th-order Lagrangian
Hamilton-Jacobi  problem, then the $1$-form $\alpha = \Leg \circ s  \in \df^{1}(\Tan^{k-1}Q)$
is a solution to the (generalized) $k$th-order Hamiltonian Hamilton-Jacobi problem.

\noindent Conversely, if $\alpha \in \df^{1}(\Tan^{k-1}Q)$ is a solution to the (generalized) $k$th-order Hamiltonian
Hamilton-Jacobi problem, then the section $s  = \Leg^{-1} \circ \alpha \in \Gamma(\rho^{2k-1}_{k-1})$
is a solution to the (generalized) $k$th-order Lagrangian Hamilton-Jacobi problem.
\end{theorem}
\begin{proof}
First, let us prove that $\alpha = \Leg \circ s$ is indeed a $1$-form, that is, a section of the projection
$\pi_{\Tan^{k-1}Q}$. Computing, we have
\begin{equation*}
\pi_{\Tan^{k-1}Q} \circ \alpha = \pi_{\Tan^{k-1}Q} \circ \Leg \circ s = \rho^{2k-1}_{k-1} \circ s = \Id_{\Tan^{k-1}Q} \, ,
\end{equation*}
since $\Leg$ is a bundle morphism over $\Tan^{k-1}Q$.

Next, let $X = \Tan\rho^{2k-1}_{k-1} \circ X_\Lag \circ s,
\bar{X} = \Tan\pi_{\Tan^{k-1}Q} \circ X_h \circ \alpha \in \vf(\Tan^{k-1}Q)$ be the vector fields associated
to $s$ and $\alpha = \Leg \circ s$, respectively. Using Lemma \ref{Chap04_lemma:TechLemma} we have
$X = \bar{X}$, and hence both vector fields are denoted by $X$.

Suppose that $s$ is a solution to the generalized $k$th-order Lagrangian Hamilton-Jacobi problem, and let
$\gamma \colon \R \to \Tan^{k-1}Q$ be an integral curve of $X$. Then, using Theorem
\ref{Chap02_thm:HamHORelationLagHamRegular} and Proposition \ref{Chap04_prop:GenLagHJRelatedVF}, we have
\begin{align*}
X_h \circ (\alpha \circ \gamma) &= X_h \circ \Leg \circ s \circ \gamma
= \Tan\Leg \circ X_\Lag \circ s \circ \gamma \\
&= \Tan\Leg \circ \Tan s \circ X \circ \gamma
= \Tan(\Leg \circ s) \circ \dot\gamma \\
&= \Tan \alpha \circ \dot\gamma = \dot{\overline{\alpha \circ \gamma}} \, .
\end{align*}
That is, $\alpha \circ \gamma \colon \R \to \Tan^*(\Tan^{k-1}Q)$ is an integral curve of $X_h$, and
thus $\alpha$ is a solution to the generalized $k$th-order Hamiltonian Hamilton-Jacobi problem.

Now, in addition, we require $s^*\omega_\Lag = 0$; that is, $s$ is a solution to the $k$th-order
Lagrangian Hamilton-Jacobi problem. Then, using \eqref{Chap04_eqn:PullBackSymplecticFormByAlpha}
and the properties of the Legendre-Ostrogradsky map, we have
\begin{equation*}
\d\alpha = -\alpha^*\omega_{k-1} = -(\Leg \circ s)^*\omega_{k-1} = -s^*(\Leg^*\omega_{k-1})
= -s^*\omega_\Lag = 0 \, ,
\end{equation*}
and hence $\alpha$ is a solution to the $k$th-order Hamiltonian Hamilton-Jacobi problem.

The converse is proved in an analogous way, but using $\Leg^{-1}$ instead of $\Leg$. In particular,
let us first prove that $s = \Leg^{-1} \circ \alpha$ is a section of the projection $\rho^{2k-1}_{k-1}$.
Computing, we obtain
\begin{equation*}
\rho^{2k-1}_{k-1} \circ s = \rho^{2k-1}_{k-1} \circ \Leg^{-1} \circ \alpha = \pi_{\Tan^{k-1}\pi} \circ \alpha =
\Id_{\Tan^{k-1}Q} \, ,
\end{equation*}
since $\rho^{2k-1}_{k-1} = \pi_{\Tan^{k-1}Q} \circ \Leg$, and $\Leg$ is a diffeomorphism.

Next, let $X = \Tan\rho^{2k-1}_{k-1} \circ X_\Lag \circ s,
\bar{X} = \Tan\pi_{\Tan^{k-1}Q} \circ X_h \circ \alpha \in \vf(\Tan^{k-1}Q)$ be the vector fields associated
to $s = \Leg^{-1} \circ \alpha$ and $\alpha$, respectively. From Lemma \ref{Chap04_lemma:TechLemma} we have
$X = \bar{X}$, and hence both vector fields are denoted by $X$.

Assume that $\alpha$ is a solution to the generalized $k$th-order Hamiltonian Hamilton-Jacobi problem.
If $\gamma \colon \R \to \Tan^{k-1}Q$ is an integral curve of $X$, then, using Theorem
\ref{Chap02_thm:HamHORelationLagHamRegular} and condition \eqref{Chap04_eqn:GenHamHJDef}, we have
\begin{align*}
X_\Lag \circ (s \circ \gamma) &= X_\Lag \circ \Leg^{-1} \circ \alpha \circ \gamma
= \Tan\Leg^{-1} \circ X_h \circ \alpha \circ \gamma \\
&= \Tan\Leg^{-1} \circ \dot{\overline{\alpha \circ \gamma}}
= \Tan\Leg^{-1} \circ \Tan \alpha \circ \dot\gamma \\
&= \Tan(\Leg^{-1} \circ \alpha) \circ \dot\gamma
= \Tan s \circ \dot\gamma
= \dot{\overline{s \circ \gamma}} \, .
\end{align*}
That is, $s \circ \gamma$ is an integral curve of $X_\Lag$, and hence $s$ is a solution to the generalized
$k$th-order Lagrangian Hamilton-Jacobi problem.

Now, in addition, we require $\alpha$ to be closed; that is, $\alpha$ is a solution to the $k$th-order
Hamiltonian Hamilton-Jacobi problem. Then, using the relation \eqref{Chap04_eqn:PullBackSymplecticFormByAlpha}
and the properties of the Legendre-Ostrogradsky map, we have
\begin{equation*}
s^*\omega_\Lag = (\Leg^{-1} \circ \alpha)^*\omega_\Lag = \alpha^*((\Leg^{-1})^*\omega_\Lag)
= \alpha^*\omega_{k-1} = -\d\alpha = 0 \, ,
\end{equation*}
and hence $s$ is a solution to the $k$th-order Lagrangian Hamilton-Jacobi problem.
\end{proof}

\begin{remark}
This result can be extended to complete solutions in a natural way, applying it to every particular solution
given by the local diffeomorphism $\Phi$.
\end{remark}

Theorem \ref{Chap04_thm:EquivalenceSolutionsLag&Ham} allows us to show that the vector field associated to
a section solution to the (generalized) Hamilton-Jacobi problem is a semispray of type $1$.
First, we need the following technical result.

\begin{lemma}\label{Chap04_lemma:TechLemma2}
Let $X \in \vf(\Tan^rQ)$ be a semispray of type $1$ on $\Tan^rQ$, and $Y \in \vf(\Tan^sQ)$ ($s \leqslant r$)
which is $\rho^r_s$-related with $X$. Then $Y$ is a semispray of type $1$ on $\Tan^sQ$.
\end{lemma}
\begin{proof}
Let $\gamma \colon \R \to \Tan^rQ$ be an integral curve of $X$. Then, as $X$ is a semispray of type $1$,
there exists a curve $\phi \colon \R \to Q$ such that $j^{r}_0\phi = \gamma$. Furthermore, as $X$ and $Y$
are $\rho^r_s$-related, the curve $\rho^r_s \circ \gamma \colon \R \to \Tan^sQ$ is an integral curve of $Y$.
Hence, $\rho^r_s (j^{r}\phi) = j^s\phi$ is an integral curve of $Y$.

It remains to show that every integral curve of $Y$ is the projection to $\Tan^sQ$ via $\rho^r_s$ of an
integral curve of $X$, but this holds due to the fact that the vector fields are $\rho^r_s$-related and
$\rho^r_s$ is a surjective submersion. Therefore, $Y$ is a semispray of type $1$ in $\Tan^sQ$.
\end{proof}

\begin{proposition}\label{Chap04_prop:XSemispray1}
Let $(\Tan^{2k-1}Q,\Lag)$ be a hyperregular Lagrangian system, and $(\Tan^*(\Tan^{k-1}Q),\omega_{k-1},h)$
the associated Hamiltonian system. Then, if $\alpha \in \df^{1}(\Tan^{k-1}Q)$ is a solution to the
$k$th-order Hamiltonian Hamilton-Jacobi problem, the vector field
$X = \Tan\pi_{\Tan^{k-1}Q} \circ X_h \circ \alpha$ is a semispray of type $1$ on $\Tan^{k-1}Q$.
\end{proposition}
\begin{proof}
Let $s = \Leg^{-1} \circ \alpha \in \Gamma(\rho^{2k-1}_{k-1})$ be the section associated to $\alpha$,
which is a solution to the $k$th-order Lagrangian Hamilton-Jacobi problem by Theorem
\ref{Chap04_thm:EquivalenceSolutionsLag&Ham}. Then, by Lemma \ref{Chap04_lemma:TechLemma}, if
$X = \Tan\pi_{\Tan^{k-1}Q} \circ X_h \circ \alpha$ and $\bar{X} = \Tan\rho^{2k-1}_{k-1} \circ X_\Lag \circ s$
are the vector fields associated to $\alpha$ and $s = \Leg^{-1} \circ \alpha$ respectively, then
$X = \bar{X} = \Tan\rho^{2k-1}_{k-1} \circ X_\Lag \circ s$. Hence, as $X_\Lag \in \vf(\Tan^{2k-1}Q)$
is the Euler-Lagrange vector field solution to the equation \eqref{Chap02_eqn:LagHODynEq} and
$\Lag \in \Cinfty(\Tan^{k}Q)$ is a hyperregular Lagrangian function, we have that $X_\Lag$ is a semispray
of type $1$ on $\Tan^{2k-1}Q$. In particular, $X_\Lag \circ s$ is a semispray of type $1$ along $\rho^{2k-1}_{k-1}$
and, by Lemma \ref{Chap04_lemma:TechLemma2}, $X$ is a semispray of type $1$ on $\Tan^{k-1}Q$.
\end{proof}

As a consequence of Proposition \ref{Chap04_prop:XSemispray1}, the generalized $k$th-order Hamilton-Jacobi
problem can be stated in the following way.

\begin{definition}
The \textnormal{generalized $k$th-order Lagrangian} (resp., \textnormal{Hamiltonian})
\textnormal{Hamilton-Jacobi problem} consists in finding a section $s \in \Gamma(\rho^{2k-1}_{k-1})$
(resp., a $1$-form $\alpha \in \df^{1}(\Tan^{k-1}Q)$) such that, if $\gamma \colon \R \to Q$ satisfies
that $j^{k-1}_0\gamma$ is an integral curve of $X = \Tan\rho^{2k-1}_{k-1} \circ X_\Lag \circ s$
(resp., $X = \Tan\pi_{\Tan^{k-1}Q} \circ X_h \circ \alpha$), then $s \circ j^{k-1}_0\gamma \colon \R \to \Tan^{2k-1}Q$
(resp., $\alpha \circ j^{k-1}_0\gamma \colon \R \to \Tan^*(\Tan^{k-1}Q)$) is an integral curve of $X_\Lag$ (resp., $X_h$).
\end{definition}

\section{The Lagrangian-Hamiltonian formalism}
\label{Chap04_sec:UnifiedFormalism}

In this Section we state the geometric Hamilton-Jacobi problem for a $k$th-order dynamical system in the
Lagrangian-Hamiltonian unified formalism described in Chapter \ref{Chap:HOAutonomousDynamicalSystems}.
As we have pointed out in the Remark at the beginning of Section \ref{Chap02_sec:HamiltonJacobi},
the geometric Hamilton-Jacobi problem has been stated in the unified formalism in a recent work
\cite{art:DeLeon_Martin_Vaquero12} to study the Hamilton-Jacobi theory in dynamical systems
given in terms of singular Lagrangian functions. In this Section we do not follow the patterns of the
referred work, since our goal is not to generalize the results given there, but to give a geometric
description of the Hamilton-Jacobi problem in the Lagrangian-Hamiltonian formalism, combining the
results of Sections \ref{Chap04_sec:LagrangianFormalism} and \ref{Chap04_sec:HamiltonianFormalism}
with the results of Chapter \ref{Chap:HOAutonomousDynamicalSystems}. Therefore, we stick to the general
setting described at the beginning of this Chapter, and in particular $\Lag \in \Cinfty(\Tan^{k}Q)$
denotes a $k$th-order regular Lagrangian function, although for simplicity we will assume throughout
this Section that $\Lag$ is hyperregular.

Recall that, in the Lagrangian-Hamiltonian formalism for a $k$th-order dynamical system, we consider
the bundle $\W = \Tan^{2k-1}Q \times_{\Tan^{k-1}Q} \Tan^*(\Tan^{k-1}Q)$ with the canonical projections
$\rho_1 \colon \W \to \Tan^{2k-1}Q$ and $\rho_2 \colon \W \to \Tan^*(\Tan^{k-1}Q)$. It is clear from the
definition that the bundle $\W$ fibers over $\Tan^{k-1}Q$. Let $\rho_{\Tan^{k-1}Q} \colon \W \to \Tan^{k-1}Q$ be the
canonical projection. Obviously, we have $\rho_{\Tan^{k-1}Q} = \rho^{2k-1}_{k-1} \circ \rho_1 = \pi_{\Tan^{k-1}Q} \circ \rho_2$.
Hence, we have the following diagram
\begin{equation*}
\xymatrix{
\ & \W \ar[dl]_{\rho_1} \ar[dr]^{\rho_2} \ar[dd]_{\rho_{\Tan^{k-1}Q}} & \ \\
\Tan^{2k-1}Q \ar[dr]_{\rho^{2k-1}_{k-1}} & \ & \Tan^*(\Tan^{k-1}Q) \ar[dl]^-{\pi_{\Tan^{k-1}Q}} \\
\ & \Tan^{k-1}Q & \
}
\end{equation*}

We consider in $\W$ the presymplectic form $\Omega = \rho_2^*\,\omega_{k-1} \in \df^{2}(\W)$, where
$\omega_{k-1} \in \df^{2}(\Tan^*(\Tan^{k-1}Q))$ is the canonical symplectic form. In addition, from the
$k$th-order Lagrangian function $\Lag$, and using the canonical coupling function $\C \in \Cinfty(\W)$, we
construct a Hamiltonian function $H \in \Cinfty(\W)$ as $H = \C - \Lag$. Thus, the dynamical equation for
the system is \eqref{Chap03_eqn:UnifDynEqVF}, that is,
\begin{equation*}
\inn(X_{LH})\Omega = \d H \ , \quad X_{LH} \in \vf(\W) \, .
\end{equation*}
Following the constraint algorithm described in Section \ref{Chap01_sec:ConstraintAlgorithm}, a solution to
the equation \eqref{Chap03_eqn:UnifDynEqVF} exists on the points of a submanifold
$j_\Lag \colon \W_\Lag \hookrightarrow \W$ which can be identified with the graph of the Legendre-Ostrogradsky
map $\Leg \colon \Tan^{2k-1}Q \to \Tan^*(\Tan^{k-1}Q)$ associated to $\Lag$. Since the Lagrangian function
is regular, there exists a unique vector field $X_{LH}$ solution to \eqref{Chap03_eqn:UnifDynEqVF} and
tangent to $\W_\Lag$ (see Chapter \ref{Chap:HOAutonomousDynamicalSystems} for details).

\subsection{The generalized Lagrangian-Hamiltonian Hamilton-Jacobi problem}
\label{Chap04_sec:GenLagHamHJProblem}

We first state the generalized version of the Hamilton-Jacobi problem. Following the same patterns as in
previous Sections and \cite{art:DeLeon_Martin_Vaquero12} (see also an approach to the problem for higher-order
field theories in \cite{art:Vitagliano12}), the natural definition for the generalized Hamilton-Jacobi problem
in the unified setting is the following.

\begin{definition}\label{Chap04_def:GenLagHamHJDef}
The \textnormal{generalized $k$th-order Lagrangian-Hamiltonian Hamilton-Jacobi problem} (or
\textnormal{generalized $k$th-order unified Hamilton-Jacobi problem}) consists in finding a section
$s \in \Gamma(\rho_{\Tan^{k-1}Q})$ and a vector field $X \in \vf(\Tan^{k-1}Q)$ such that the following
conditions are satisfied:
\begin{enumerate}
\item The submanifold $\Im(s) \hookrightarrow \W$ is contained in $\W_\Lag$.
\item If $\gamma \colon \R \to \Tan^{k-1}Q$ is an integral curve of $X$, then its lifting to $\W$ by $s$,
$s \circ \gamma \colon \R \to \W$, is an integral curve of $X_{LH}$, that is,
\begin{equation}\label{Chap04_eqn:GenLagHamHJDef}
X \circ \gamma = \dot{\gamma} \Longrightarrow X_{LH} \circ (s \circ \gamma) = \dot{\overline{s \circ \gamma}} \, .
\end{equation}
\end{enumerate}
\end{definition}

It is clear from Definition \ref{Chap04_def:GenLagHamHJDef} that the vector field $X \in \vf(\Tan^{k-1}Q)$
cannot be chosen independently from the section $s \in \Gamma(\rho_{\Tan^{k-1}Q})$. Indeed, we can prove the following result.

\begin{proposition}\label{Chap04_prop:GenLagHamHJRelatedVF}
The pair $(s,X) \in \Gamma(\rho_{\Tan^{k-1}Q}) \times \vf(\Tan^{k-1}Q)$ satisfies the two conditions in Definition
\ref{Chap04_def:GenLagHamHJDef} if, and only if, $X_{LH}$ and $X$ are $s$-related.
\end{proposition}
\begin{proof}
The proof of this result follows the same patterns that the proof of Propositions
\ref{Chap04_prop:GenLagHJRelatedVF} and \ref{Chap04_prop:GenHamHJRelatedVF}. In particular, if the pair $(s,X)$
satisfies the two conditions in Definition \ref{Chap04_def:GenLagHamHJDef}, then for every integral curve
$\gamma$ of $X$, we have
\begin{equation*}
X_{LH} \circ (s \circ \gamma) = \dot{\overline{s \circ \gamma}} = \Tan s \circ \dot\gamma
= \Tan s \circ X \circ \gamma \, ,
\end{equation*}
but, since $X$ has integral curves through every point $\bar{y} \in \Tan^{k-1}Q$, this is equivalent to
$X_{LH} \circ s = \Tan s \circ X$.

Conversely, if $X_{LH}$ and $X$ are $s$-related, and $\gamma \colon \R \to \Tan^{k-1}Q$ is an integral
curve of $X$, we have
\begin{equation*}
X_{LH} \circ s \circ \gamma = \Tan s \circ X \circ \gamma = \Tan s \circ \dot{\gamma}
= \dot{\overline{s \circ \gamma}} \, ,
\end{equation*}
which proves condition \eqref{Chap04_eqn:GenLagHamHJDef}. The first condition in Definition
\ref{Chap04_def:GenLagHamHJDef} is then satisfied automatically, since every integral curve of $X_{LH}$
must lie in the submanifold $\W_\Lag \hookrightarrow \W$, as the vector field $X_{LH} \in \vf(\W)$
is tangent to $\W_\Lag$.
\end{proof}

That is, the vector field $X \in \vf(\Tan^{k-1}Q)$ is related to the vector field $X_{LH} \in \vf(\W)$
and the section $s \in \Gamma(\rho_{\Tan^{k-1}Q})$. Moreover, from Proposition \ref{Chap04_prop:GenLagHamHJRelatedVF},
composing both sides of the equality $X_{LH} \circ s = \Tan s \circ X$ with $\Tan\rho_{\Tan^{k-1}Q}$, and bearing
in mind that $s \in \Gamma(\rho_{\Tan^{k-1}Q})$, we obtain the following result.

\begin{corollary}\label{corol:GenLagHamHJRelatedVF}
If $s \in \Gamma(\rho_{\Tan^{k-1}Q})$ and $X \in \vf(\Tan^{k-1}Q)$ satisfy the two conditions in Definition
\ref{Chap04_def:GenLagHamHJDef}, then $X = \Tan\rho_{\Tan^{k-1}Q} \circ X_{LH} \circ s$.
\end{corollary}
\begin{proof}
If $(s,X)$ satisfy the two conditions in Definition \ref{Chap04_def:GenLagHamHJDef}, then from Proposition
\ref{Chap04_prop:GenLagHamHJRelatedVF} $X$ and $X_{LH}$ are $s$-related, that is, we have
$\Tan s \circ X = X_{LH} \circ s$. Then, composing both sides of the equality with $\Tan\rho_{\Tan^{k-1}Q}$
and bearing in mind that $\rho_{\Tan^{k-1}Q} \circ s = \Id_{\Tan^{k-1}Q}$, we have
$X = \Tan\rho_{\Tan^{k-1}Q} \circ X_{LH} \circ s$.
\end{proof}

That is, the vector field $X \in \vf(\Tan^{k-1}Q)$ is completely determined by the section
$s \in \Gamma(\rho_{\Tan^{k-1}Q})$, and it is called the \textsl{vector field associated to $s$}.
The following diagram illustrates the situation
\begin{equation*}
\xymatrix{
\Tan(\Tan^{k-1}Q) \ar@/_1.5pc/[rr]_{\Tan s}  & \ & \ar[ll]_-{\Tan\rho_{\Tan^{k-1}Q}} \Tan\W \\
\ & \ & \ \\
\Tan^{k-1}Q \ar[uu]^{X} \ar@/_1.5pc/[rr]_{s} & \ & \ar[ll]_-{\rho_{\Tan^{k-1}Q}} \W \ar[uu]_{X_{LH}}
}
\end{equation*}

Therefore, the search of a pair $(s,X) \in \Gamma(\rho_{\Tan^{k-1}Q}) \times \vf(\Tan^{k-1}Q)$
satisfying the two conditions in Definition \ref{Chap04_def:GenLagHamHJDef} is equivalent to the
search of a section $s \in \Gamma(\rho_{\Tan^{k-1}Q})$ satisfying the same conditions with the
associated vector field $\Tan\rho_{\Tan^{k-1}Q} \circ X_{LH} \circ s$. Thus, we can give
the following definition.

\begin{definition}\label{Chap04_def:GenLagHamHJSolution}
A \textnormal{solution to the generalized $k$th-order unified Hamilton-Jacobi problem} for $X_{LH}$
consists in finding a section $s \in \Gamma(\rho_{\Tan^{k-1}Q})$ satisfying the following conditions:
\begin{enumerate}
\item The submanifold $\Im(s) \hookrightarrow \W$ is contained in $\W_\Lag$.
\item If $\gamma \colon \R \to \Tan^{k-1}Q$ is an integral curve of $\Tan\rho_{\Tan^{k-1}Q} \circ X_{LH} \circ s \in \vf(\Tan^{k-1}Q)$,
then $s \circ \gamma \colon \R \to \W$ is an integral curve of $X_{LH}$, that is
\begin{equation*}
\Tan\rho_{\Tan^{k-1}Q} \circ X_{LH} \circ s \circ \gamma = \dot{\gamma} \Longrightarrow
X_{LH} \circ (s \circ \gamma) = \dot{\overline{s \circ \gamma}} \, .
\end{equation*}
\end{enumerate}
\end{definition}

Now we can state the following result, which is the analogous to Propositions
\ref{Chap04_prop:GenLagHJEquiv} and \ref{Chap04_prop:GenHamHJEquiv} in the unified formalism.

\begin{proposition}\label{Chap04_prop:GenLagHamHJEquiv}
The following assertions on a section $s \in \Gamma(\rho_{\Tan^{k-1}Q})$ are equivalent.
\begin{enumerate}
\item $s$ is a solution to the generalized $k$th-order unified Hamilton-Jacobi problem.
\item The submanifold $\Im(s) \hookrightarrow \W$ is invariant under the flow of the vector field $X_{LH}$
solution to equation \eqref{Chap03_eqn:UnifDynEqVF} (that is, $X_{LH}$ is tangent to the submanifold $\Im(s)$).
\item The section $s$ satisfies the equation
\begin{equation*}
\inn(X)(s^*\Omega) = \d(s^*H) \, ,
\end{equation*}
where $X = \Tan\rho_{\Tan^{k-1}Q} \circ X_{LH} \circ s$ is the vector field associated to $s$.
\end{enumerate}
\end{proposition}
\begin{proof}
This proof follows the same patterns as the proofs of Propositions \ref{Chap04_prop:GenLagHJEquiv}
and \ref{Chap04_prop:GenHamHJEquiv}. In particular,
\begin{description}
\item[\textnormal{($1 \, \Longleftrightarrow \, 2$)}]
Let $s$ be a solution to the generalized $k$th-order unified Hamilton-Jacobi problem. Then by
Proposition \ref{Chap04_prop:GenLagHamHJRelatedVF} the vector field $X_{LH} \in \vf(\W)$ solution to
equation \eqref{Chap03_eqn:UnifDynEqVF} is $s$-related to the vector field
$X = \Tan\rho_{\Tan^{k-1}Q} \circ X_{LH} \circ s \in \vf(\Tan^{k-1}Q)$ associated to $s$, and thus for every
$\bar{y} \in \Tan^{k-1}Q$ we have
\begin{equation*}
X_{LH}(s(\bar{y})) = (X_{LH}\circ s)(\bar{y}) = (\Tan s \circ X)(\bar{y}) = \Tan s(X(\bar{y})) \, .
\end{equation*}
Hence, $X_{LH}(s(\bar{y})) = \Tan s(X(\bar{y}))$, and therefore $X_{LH}$ is tangent to the
submanifold $\Im(s) \hookrightarrow \W_\Lag$.

Conversely, suppose that the submanifold $\Im(s) \hookrightarrow \W$ is invariant under the flow of
$X_{LH}$. Then, $X_{LH}(s(\bar{y})) \in \Tan_{s(\bar{y})}\Im(s)$, for every $\bar{y} \in \Tan^{k-1}Q$;
that is, there exists an element $u_{\bar{y}} \in \Tan_{\bar{y}}\Tan^{k-1}Q$ such that
$X_{LH}(s(\bar{y})) = \Tan_{\bar{y}}s(u_{\bar{y}})$. If we define $X \in \vf(\Tan^{k-1}Q)$
as the vector field that satisfies $\Tan_{\bar{y}}s(X_{\bar{y}}) = X_{LH}(s(\bar{y}))$,
then $X$ is a vector field in $\Tan^{k-1}Q$, since $X = \Tan\rho_{\Tan^{k-1}Q} \circ X_{LH} \circ s$,
and it is $s$-related with $X_{LH}$. Therefore, by Proposition \ref{Chap04_prop:GenLagHamHJRelatedVF},
$s$ is a solution to the generalized $k$th-order unified Hamilton-Jacobi problem.

\item[\textnormal{($1 \, \Longleftrightarrow \, 3$)}]
Let $s$ be a solution to the generalized $k$th-order unified Hamilton-Jacobi problem. Taking the
pull-back of the dynamical equation \eqref{Chap03_eqn:UnifDynEqVF} by the section $s$ we have
\begin{equation*}
s^*\inn(X_{LH})\Omega = s^*\d H = \d(s^*H) \, ,
\end{equation*}
but since $X$ and $X_{LH}$ are $s$-related by Proposition \ref{Chap04_prop:GenLagHamHJRelatedVF}, we have
that $s^*\inn(X_{LH})\Omega = \inn(X)s^*\Omega$, and hence we obtain
\begin{equation*}
\inn(X)s^*\Omega = \d(s^*H) \, .
\end{equation*}

Conversely, consider the following vector field along the section $s\in\Gamma(\rho_{\Tan^{k-1}Q})$
\begin{equation*}
D_{LH} = X_{LH} \circ s - \Tan s \circ X \colon \Tan^{k-1}Q \to \Tan\W \, .
\end{equation*}
We want to prove that $D_{LH} = 0$. Equivalently, we can prove that
$\Omega_{s(\bar{y})}(D_{LH}(\bar{y}),Z_{s(\bar{y})}) = 0$ for every tangent vector
$Z_{s(\bar{y})} \in \Tan_{s(\bar{y})}W$, thus implying that $D_{LH}(\bar{y}) \in \ker\Omega_{\bar{y}}$,
and then prove that $D_{LH} = 0$ in $\ker\Omega$ (recall that, in the unified
formalism, the $2$-form $\Omega \in \df^{2}(\W)$ is presymplectic). Taking the pull-back of the
dynamical equation \eqref{Chap03_eqn:UnifDynEqVF}, and using the hypothesis, we have
\begin{equation*}
s^{*}(\inn(X_{LH})\Omega) = s^{*}\d H = \d(s^{*}H) = \inn(X)(s^{*}\Omega) \, ,
\end{equation*}
and then the form $s^{*}(\inn(X_{LH})\Omega) - \inn(X)(s^{*}\Omega) \in \df^1(\Tan^{k-1}Q)$
vanishes. Therefore, for every $\bar{y} \in \Tan^{k-1}Q$ and $u_{\bar{y}}\in \Tan_{\bar{y}}\Tan^{k-1}Q$, we have
\begin{align*}
0 &=
(s^*\inn(X_{LH})\Omega - \inn(X)s^*\Omega)_{\bar{y}}(u_{\bar{y}}) \\
&= \Omega_{s(\bar{y})}(X_{LH}(s(\bar{y})),\Tan_{\bar{y}}s(u_{\bar{y}}))
- \Omega_{s(\bar{y})}(\Tan_{\bar{y}}s(X_{\bar{y}}),\Tan_{\bar{y}}s(u_{\bar{y}})) \\
&= \Omega_{s(\bar{y})}(X_{LH}(s(\bar{y})) - \Tan_{\bar{y}}s(X_{\bar{y}}),\Tan_{\bar{y}}s(u_{\bar{y}})) \\
&= \Omega_{s(\bar{y})}(D_{LH}(\bar{y}),\Tan_{\bar{y}}s(u_{\bar{y}})) \, .
\end{align*}
Therefore, $\Omega_{s(\bar{y})}(D_{LH}(\bar{y}),A_{s(\bar{y})}) = 0$, for every
$A_{s(\bar{y})} \in \Tan_{s(\bar{y})}\Im(s)$. Now recall that every section defines a canonical
splitting of the tangent space of $\W$ at every point given by
\begin{equation*}
\Tan_{s(\bar{y})}\W = \Tan_{s(\bar{y})}\Im(s) \oplus V_{s(\bar{y})}(\rho_{\Tan^{k-1}Q}) \, .
\end{equation*}
Hence, we only need to prove that $\Omega_{s(\bar{y})}(D_{LH}(\bar{y}),B_{s(\bar{y})}) = 0$, for
every vertical tangent vector $B_{s(\bar{y})} \in V_{s(\bar{y})}(\rho_{\Tan^{k-1}Q})$. Equivalently,
since the canonical symplectic form $\omega_{k-1} \in \df^{2}(\Tan^*(\Tan^{k-1}Q))$ vanishes by the
contraction of two $\pi_{\Tan^{k-1}Q}$-vertical vectors, then $\Omega$ is annihilated
by the contraction of two $\rho_{\Tan^{k-1}Q}$-vertical vectors, and therefore it suffices to prove that
$D_{LH}$ is vertical with respect to $\rho_{\Tan^{k-1}Q}$. Indeed,
\begin{align*}
\Tan\rho_{\Tan^{k-1}Q} \circ D_{LH} &=
\Tan\rho_{\Tan^{k-1}Q} \circ (X_{LH} \circ s - \Tan s \circ X)
= \Tan\rho_{\Tan^{k-1}Q} \circ X_{LH} \circ s - \Tan\rho_{\Tan^{k-1}Q} \circ \Tan s \circ X \\
&= \Tan\rho_{\Tan^{k-1}Q} \circ X_{LH} \circ s - \Tan(\rho_{\Tan^{k-1}Q}\circ s) \circ X
= \Tan\rho_{\Tan^{k-1}Q} \circ X_{LH} \circ s - X = 0 \, .
\end{align*}
Therefore
$\Omega_{s(\bar{y})}(D_{LH}(\bar{y}),Z_{s(\bar{y})}) = 0$, for every $Z_{s(\bar{y})} \in \Tan_{s(\bar{y})}\W$.
Therefore, we have proved that $D_{LH}(\bar{y}) \in \ker\Omega_{\bar{y}}$, and it remains to prove that
$D_{LH} = 0$ in this vector space. Recall that in Section \ref{Chap03_sec:GeometricalSettingStructures}
we proved that $\ker\Omega = \vf^{V(\rho_2)}(\W)$, and hence what we have just proved is that $D_{LH}$ is
$\rho_2$-vertical, that is,
\begin{equation*}
\Tan\rho_2 \circ D_{LH} = \Tan\rho_2 \circ (X_{LH} \circ s - \Tan s \circ X) = 0 \, ,
\end{equation*}
and, in particular, we have
\begin{equation*}
\Tan\rho_2 \circ X_{LH} \circ s = \Tan\rho_2 \circ \Tan s \circ X \, .
\end{equation*}
Now, from Lemma \ref{Chap03_lemma:HamRegHamiltonianVF} and Theorem \ref{Chap03_thm:EquivUnifHamVFReg},
the vector field $X_{LH} \in \vf(\W)$ solution to the dynamical equation \eqref{Chap03_eqn:UnifDynEqVF}
is $\rho_2$-related to the Hamiltonian vector field solution to equation \eqref{Chap02_eqn:HamHODynEq},
that is, $\Tan \rho_2 \circ X_{LH} = X_h \circ \rho_2$. Hence, the previous relation becomes
\begin{equation*}
X_h \circ \rho_2 \circ s = \Tan\rho_2 \circ \Tan s \circ X \, .
\end{equation*}
Then, from Theorem \ref{Chap02_thm:HamHORelationLagHamRegular}, the Hamiltonian vector field solution
to equation \eqref{Chap02_eqn:HamHODynEq} and the Lagrangian vector field solution to equation
\eqref{Chap02_eqn:LagHODynEq} are $\Leg$-related, that is, $\Tan\Leg \circ X_\Lag = X_h \circ \Leg$.
In particular, since $\Leg$ is a diffeomorphism, we have $X_h = \Tan\Leg \circ X_\Lag \circ \Leg^{-1}$.
Replacing $X_h$ by $\Tan\Leg \circ X_\Lag \circ \Leg^{-1}$ in the previous equation, we have
\begin{equation*}
\Tan\Leg \circ X_\Lag \circ \Leg^{-1} \circ \rho_2 \circ s = \Tan\rho_2 \circ \Tan s \circ X 
\Longleftrightarrow
X_\Lag \circ \Leg^{-1} \circ \rho_2 \circ s = \Tan\Leg^{-1} \circ \Tan\rho_2 \circ \Tan s \circ X 
\end{equation*}
Then, bearing in mind that $\Leg \circ \rho_1 = \rho_2 \Longrightarrow \rho_1 = \Leg^{-1} \circ \rho_2$,
we obtain
\begin{equation*}
X_\Lag \circ \rho_1 \circ s = \Tan\rho_1 \circ \Tan s \circ X 
\end{equation*}
Finally, using Lemma \ref{Chap03_lemma:LagLagrangianVF} and Theorem \ref{Chap03_thm:EquivUnifLagVF},
the vector field $X_{LH} \in \vf(\W)$ solution to the dynamical equation \eqref{Chap03_eqn:UnifDynEqVF}
is $\rho_1$-related to the Lagrangian vector field solution to equation \eqref{Chap02_eqn:LagHODynEq},
and hence
\begin{equation*}
\Tan\rho_1 \circ X_{LH} \circ s = \Tan\rho_1 \circ \Tan s \circ X \, ,
\end{equation*}
which is equivalent to $\Tan\rho_1 \circ (X_{LH} \circ s - \Tan s \circ X) = \Tan\rho_1 \circ D_{LH} = 0$.
That is, we have proved that if $D_{LH}$ is $\rho_2$-vertical, then it is also $\rho_1$-vertical. And,
reversing the reasoning, the converse is obvious. In particular, this implies that
$D_{LH}(\bar{y}) \in V_{s(\bar{y})}(\rho_1) \cap V_{s(\bar{y})}(\rho_2)$ but, since
$V_w(\rho_1) \cap V_w(\rho_2) = \{ 0 \}$ for every $w \in \W$, we proved that $D_{LH} = 0$, that is,
$X_{LH}$ and $X$ are $s$-related, and by Proposition \ref{Chap04_prop:GenLagHamHJRelatedVF}
$s$ is a solution to the generalized $k$th-order unified Hamilton-Jacobi problem. \qedhere
\end{description}
\end{proof}

Let us compute in coordinates the condition for a section $s \in \Gamma(\rho_{\Tan^{k-1}Q})$ to be a solution to the
generalized $k$th-order unified Hamilton-Jacobi problem. Let $(q_0^A)$ be a set of local coordinates
in $Q$, with $1 \leqslant A \leqslant n$, and $(q_i^A,q_{j}^A,p_A^{i})$, $0 \leqslant i \leqslant k-1$,
$k \leqslant j \leqslant 2k-1$ the induced local coordinates in $\W$ (see Section
\ref{Chap03_sec:GeometricalSettingPhaseSpace} for details). These coordinates are adapted to the
$\rho_{\Tan^{k-1}Q}$-bundle structure, and hence a section $s \in \Gamma(\rho_{\Tan^{k-1}Q})$ is given by
$s(q_i^A) = (q_i^A,s_j^A,\alpha_A^i)$, where $s_j^A,\alpha_A^i$ are local functions in $\Tan^{k-1}Q$.

From Proposition \ref{Chap04_prop:GenLagHamHJEquiv}, an equivalent condition for a section
$s \in \Gamma(\rho_{\Tan^{k-1}Q})$ to be a solution to the generalized $k$th-order unified Hamilton-Jacobi problem is
that the dynamical vector field $X_{LH}$ is tangent to the submanifold $\Im(s) \hookrightarrow \W$,
which is defined locally by the $2kn$ constraints $q_j^A - s_j^A = 0$  and $p_A^i - \alpha_A^i = 0$.
From the results on Section \ref{Chap03_sec:DynamicalEquationsVF} we know that the vector field
$X_{LH} \in \vf(\W)$ solution to the dynamical equation \eqref{Chap03_eqn:UnifDynEqVF} is given
locally by \eqref{Chap03_eqn:UnifDynEqVFSolutionWithHolonomy}, that is,
\begin{equation*}
X_{LH} = \sum_{l=0}^{2k-2} q_{l+1}^A\derpar{}{q_l^A} + F^A\derpar{}{q_{2k-1}^A} + \derpar{\Lag}{q_0^A}\derpar{}{p_A^0}
+ \left( \derpar{\Lag}{q_i^A} - p^{i-1}_A \right)\derpar{}{p^i_A} \, ,
\end{equation*}
where the functions $F^A$ are the solutions to equations \eqref{Chap03_eqn:UnifDynEqVFTangencyConditionWithHolonomy}.
Hence, requiring $X_{LH}(q_j^A-s_j^A) = 0$ and $X_{LH}(p_A^i - \alpha_A^i) = 0$ on $\Im(s)$, we obtain
the following system of $2kn$ partial differential equations
\begin{equation}\label{Chap04_eqn:GenLagHamHJLocal}
\begin{array}{l}
\displaystyle
s_{j+1}^A - \sum_{l=0}^{k-2} q_{l+1}^B\derpar{s_j^A}{q_l^B} - s_{k}^B\derpar{s_j^A}{q_{k-1}^B} = 0 \quad ; \quad
F^A - \sum_{l=0}^{k-2} q_{l+1}^B\derpar{s_{2k-1}^A}{q_l^B} - s_{k}^B\derpar{s_{2k-1}^A}{q_{k-1}^B} = 0 \, , \\[15pt]
\displaystyle
\derpar{\Lag}{q_A^0} - \sum_{l=0}^{k-2} q_{l+1}^B\derpar{\alpha^0_A}{q_l^B} - s_{k}^B\derpar{\alpha_A^0}{q_{k-1}^B} = 0 \quad ; \quad
\derpar{\Lag}{q_i^A} - \alpha_A^{i-1} - \sum_{l=0}^{k-2} q_{l+1}^B\derpar{\alpha^i_A}{q_l^B} - s_{k}^B\derpar{\alpha^i_A}{q_{k-1}^B} = 0 \, .
\end{array}
\end{equation}
with $1 \leqslant i \leqslant k-1$, $k \leqslant j \leqslant 2k-2$.
This is a system of $2kn$ partial differential equations with $2kn$ unknown function $s_j^A$, $\alpha_A^i$.
Hence, a section $s \in \Gamma(\rho_{\Tan^{k-1}Q})$ is a solution to the generalized $k$th-order Lagrangian-Hamiltonian
Hamilton-Jacobi problem if, and only if, its component functions satisfy the local equations
\eqref{Chap04_eqn:GenLagHamHJLocal}.

\subsection{The Lagrangian-Hamiltonian Hamilton-Jacobi problem}
\label{Chap04_sec:LagHamHJProblem}

As in the Lagrangian and Hamiltonian formalisms described in previous Sections, to solve the generalized
$k$th-order unified Hamilton-Jacobi problem is a difficult task in general, since we must find
$kn$-dimensional submanifolds of $\W$ contained in the submanifold $\W_\Lag = \graph(\Leg)$ and invariant
by the dynamical vector field $X_{LH}$. Hence, it is convenient to consider a less general problem and
require some additional conditions to the section $s \in \Gamma(\rho_{\Tan^{k-1}Q})$.

\begin{definition}\label{Chap04_def:LagHamHJDef}
The \textnormal{$k$th-order Lagrangian-Hamiltonian Hamilton-Jacobi problem} consists in finding sections
$s \in \Gamma(\rho_{\Tan^{k-1}Q})$ solution to the generalized $k$th-order unified Hamilton-Jacobi problem
such that $s^*\Omega = 0$. Such a section is called a
\textnormal{solution to the $k$th-order unified Hamilton-Jacobi problem}.
\end{definition}

From the definition of $\Omega \in \df^{2}(\W)$ given in Section \ref{Chap03_sec:GeometricalSettingStructures} we have
\begin{equation*}
s^*\Omega = s^*(\rho_2^*\omega_{k-1}) = (\rho_2 \circ s)^*\omega_{k-1} \, .
\end{equation*}
Hence, we have that $s^*\Omega = 0$ if, and only if, $(\rho_2 \circ s)^*\omega_{k-1} = 0$. As
$\Gamma(\pi_{\Tan^{k-1}Q}) = \df^{1}(\Tan^{k-1}Q)$, the section
$\rho_2 \circ s \in \Gamma(\pi_{\Tan^{k-1}Q})$ is a $1$-form in $\Tan^{k-1}Q$, and from relation
\eqref{Chap04_eqn:PullBackSymplecticFormByAlpha} we have
\begin{equation*}
(\rho_2 \circ s)^*\omega_{k-1} = (\rho_2 \circ s)^*(-\d\theta_{k-1}) =
-\d((\rho_2 \circ s)^*\theta_{k-1}) = -\d(\rho_2 \circ s) \, .
\end{equation*}
Hence, the condition $s^*\Omega = 0$ is equivalent to $\rho_2 \circ s \in \df^{1}(\Tan^{k-1}Q)$ being a
closed $1$-form. Therefore, Definition \ref{Chap04_def:LagHamHJDef} can be rewritten as follows.

\begin{definition}\label{Chap04_def:LagHamHJDefClosedForm}
The \textnormal{$k$th-order Lagrangian-Hamiltonian Hamilton-Jacobi problem} consists in finding sections
$s \in \Gamma(\rho_{\Tan^{k-1}Q})$ solution to the generalized $k$th-order unified Hamilton-Jacobi problem
such that $\rho_2 \circ s$ is a closed $1$-form in $\Tan^{k-1}Q$.
\end{definition}

Taking into account the additional assumption on the section $s \in \Gamma(\rho_{\Tan^{k-1}Q})$, a
straightforward consequence of Proposition \ref{Chap04_prop:GenLagHamHJEquiv} is the following result, which is
the analogous to Propositions \ref{Chap04_prop:LagHJEquiv} and \ref{Chap04_prop:HamHJEquiv} in the unified formalism.

\begin{proposition}\label{Chap04_prop:LagHamHJEquiv}
The following assertions on a section $s \in \Gamma(\rho_{\Tan^{k-1}Q})$
such that $s^*\Omega = 0$ are equivalent.
\begin{enumerate}
\item $s$ is a solution to the $k$th-order Lagrangian-Hamiltonian Hamilton-Jacobi problem.
\item $\d(s^*H) = 0$.
\item $\Im(s)$ is an isotropic submanifold of $\W$
invariant by $X_{LH}$.
\item The integral curves of $X_{LH}$ with initial conditions in $\Im(s)$
project onto the integral curves of $X = \Tan\rho_{\Tan^{k-1}Q} \circ X_{LH} \circ s$.
\end{enumerate}
\end{proposition}

Let us compute in the natural coordinates of $\W$ the local equations for a section
$s \in \Gamma(\rho_{\Tan^{k-1}Q})$ to be a solution to the $k$th-order unified Hamilton-Jacobi
problem. From Definition \ref{Chap04_def:LagHamHJDef}, we must require
$s^*\Omega = 0$ or, equivalently, following Definition \ref{Chap04_def:LagHamHJDefClosedForm},
we can require the $1$-form $\rho_2 \circ s$ to be closed, that is, $\d (\rho_2 \circ s) = 0$.
Locally, if $s(q_i^A) = (q_i^A,s_j^A,\alpha_A^i)$, then $(\rho_2 \circ s)(q_i^A) = (q_i^A,\alpha_A^i)$,
and the condition $\d(\rho_2 \circ s) = 0$ gives equations \eqref{Chap04_eqn:HamHJLocalClosedForm}, that is,
\begin{equation*}
\derpar{\alpha_A^i}{q_j^B} - \derpar{\alpha_B^j}{q_i^A} = 0 \, , \quad
\mbox{with } A \neq B \mbox{ or } i \neq j \, .
\end{equation*}

Equivalently, from Proposition \ref{Chap04_prop:LagHamHJEquiv} we know that this condition is equivalent
to $\d(s^*H) = s ^*(\d H) = 0$. Then, bearing in mind the coordinate expression
\eqref{Chap03_eqn:UnifHamiltonianFunctionDifferentialLocal} of the $1$-form $\d H$, the condition $\d(s^*H) = 0$
holds if, and only if, the following $kn$ partial differential equations are satisfied
\begin{equation}\label{Chap04_eqn:LagHamHJLocal}
\begin{array}{l}
\displaystyle q_{i+1}^B\derpar{\alpha^i_B}{q_0^A} + s_{k}^B\derpar{\alpha_B^{k-1}}{q_0^A} +
\alpha_B^{k-1}\derpar{s_{k}^B}{q_0^A} - \left( \derpar{\Lag}{q_0^A} + \derpar{\Lag}{q_{k}^B}\derpar{s_{k}^B}{q_0^A} \right) = 0 \, ,\\[15pt]
\displaystyle q_{i+1}^B\derpar{\alpha^i_B}{q_l^A} + s_{k}^B\derpar{\alpha_B^{k-1}}{q_l^A} + \alpha_A^{l-1} +
\alpha_B^{k-1}\derpar{s_{k}^B}{q_l^A} - \left( \derpar{\Lag}{q_l^A} + \derpar{\Lag}{q_{k}^B}\derpar{s_{k}^B}{q_l^A} \right) = 0 \, ,
\end{array}
\end{equation}
where $1 \leqslant l \leqslant k-1$.

Therefore, a section $s \in \Gamma(\rho_{\Tan^{k-1}Q})$ is a solution to the $k$th-order
Lagrangian-Hamiltonian Hamilton-Jacobi problem if, and only if, the local functions $s_j^A,\alpha_A^i$
satisfy the system of partial differential equations given by  \eqref{Chap04_eqn:GenLagHamHJLocal}
and \eqref{Chap04_eqn:HamHJLocalClosedForm}, or, equivalently, \eqref{Chap04_eqn:GenLagHamHJLocal}
and \eqref{Chap04_eqn:LagHamHJLocal}. Observe that the system of partial differential equations
may not be $\Cinfty(U)$-linearly independent.

\subsection{Equivalent formulation}
\label{Chap04_sec:UnifiedFormalismEquivalent}

Since the vector field $X_{LH} \in \vf(\W)$ solution to the dynamical equation \eqref{Chap03_eqn:UnifDynEqVF}
is tangent to the submanifold $\W_\Lag \hookrightarrow \W$, we can state the (generalized) $k$th-order unified
Hamilton-Jacobi problem directly in the submanifold $\W_\Lag$, which is the real phase space of the system.
Let $X_o \in \vf(\W_\Lag)$ be the unique vector field in $\W_\Lag$ which is $j_\Lag$-related to $X_{LH}$, and we
consider in $\W_\Lag$ the $2$-form $\Omega_o = j_\Lag^*\Omega \in \df^{2}(\W_\Lag)$ and the restricted
Hamiltonian function $H_o = j_\Lag^*H \in \Cinfty(\W_\Lag)$.

Observe that $(\W_\Lag,\Omega_o)$ is a symplectic manifold, since the map
$\rho_1^\Lag = \rho_1 \circ j_\Lag \colon \W_\Lag \to \Tan^{2k-1}Q$ is a diffeomorphism by Proposition
\ref{Chap03_prop:LagRho1LDiffeomorphism}, and in addition we have $(\rho_1^\Lag)^*\omega_\Lag = \Omega_o$.
That is, $\rho_1^\Lag$ is a symplectomorphism between the symplectic manifolds $(\Tan^{2k-1}Q,\omega_\Lag)$
and $(\W_\Lag,\Omega_o)$.

\begin{remark}
As the $k$th-order Lagrangian function is hyperregular, by Proposition
\ref{Chap03_prop:HamRegRho2LDiffeomorphism} we know that the map
$\rho_2^\Lag = \rho_2 \circ j_\Lag = \Leg \circ \rho_1^\Lag \colon \W_\Lag \to \Tan^*(\Tan^{k-1}Q)$
is also a diffeomorphism. In addition, since
\begin{equation*}
(\rho_2^\Lag)^*\omega_{k-1} = (\Leg \circ \rho_1^\Lag)^*\omega_{k-1} =
(\rho_1^\Lag)^*(\Leg^*\omega_{k-1}) = (\rho_1^\Lag)^*\omega_\Lag = \Omega_o \, ,
\end{equation*}
we deduce that the map $\rho_2^\Lag$ is also a symplectomorphism, in this case between the symplectic manifolds
$(W_\Lag,\Omega_o)$ and $(\Tan^*(\Tan^{k-1}Q),\omega_{k-1})$.
\end{remark}

Since $X_o$ and $X_{LH}$ are $j_\Lag$-related, $X_o$ is the unique vector field in $\W_\Lag$ solution
to the equation
\begin{equation}\label{Chap04_eqn:LagHamDynEqEquivalent}
\inn(X_o)\Omega_o = \d H_o \, .
\end{equation}
Let $\rho_{\Tan^{k-1}Q}^\Lag = \rho_{\Tan^{k-1}Q} \circ j_\Lag \colon \W_\Lag \to \Tan^{k-1}Q$ be the
canonical submersion, which is the restriction of $\rho_{\Tan^{k-1}Q}$ to $\W_\Lag$. Thus, Definition
\ref{Chap04_def:GenLagHamHJDef} can be reformulated as follows.

\begin{definition}
The \textnormal{generalized $k$th-order Lagrangian-Hamiltonian Hamilton-Jacobi problem in $\W_\Lag$}
(or \textnormal{generalized $k$th-order unified Hamilton-Jacobi problem in $\W_\Lag$}) consists in finding
a section $s_o \in \Gamma(\rho_{\Tan^{k-1}Q}^\Lag)$ and a vector field $X \in \vf(\Tan^{k-1}Q)$ such that
if $\gamma \colon \R \to \Tan^{k-1}Q$ is an integral curve of $X$, then $s_o \circ \gamma \colon \R \to \W_\Lag$
is an integral curve of $X_o$, that is,
\begin{equation}\label{Chap04_eqn:GenLagHamHJDefEquivalent}
X \circ \gamma = \dot{\gamma} \Longrightarrow
X_o \circ (s_o \circ \gamma) = \dot{\overline{s_o \circ \gamma}} \, .
\end{equation}
\end{definition}

In this formulation, Proposition \ref{Chap04_prop:GenLagHamHJRelatedVF} and Corollary
\ref{corol:GenLagHamHJRelatedVF} are stated as follows.

\begin{proposition}
The pair $(s_o,X) \in \Gamma(\rho_{\Tan^{k-1}Q}^\Lag) \times \vf(\Tan^{k-1}Q)$ satisfies condition
\eqref{Chap04_eqn:GenLagHamHJDefEquivalent} if, and only if, $X_o$ and $X$ are $s_o$-related.
\end{proposition}
\begin{proof}
This proof is analogous to the proofs of Propositions \ref{Chap04_prop:GenLagHJRelatedVF},
\ref{Chap04_prop:GenHamHJRelatedVF} and \ref{Chap04_prop:GenLagHamHJRelatedVF}.
\end{proof}

\begin{corollary}
If the pair $(s_o,X)$ satisfies condition \eqref{Chap04_eqn:GenLagHamHJDefEquivalent}, then
$X = \Tan\rho_{\Tan^{k-1}Q}^\Lag \circ X_o \circ s_o$. 
\end{corollary}

Hence, Definition \ref{Chap04_def:GenLagHamHJSolution} now reads as follows.

\begin{definition}
A \textnormal{solution to the generalized $k$th-order Lagrangian-Hamiltonian Hamilton-Jacobi problem
in $\W_\Lag$} is a section $s_o \in \Gamma(\rho_{\Tan^{k-1}Q}^\Lag)$ such that, if
$\gamma \colon \R \to \Tan^{k-1}Q$ is an integral curve of the vector field
$\Tan\rho_{\Tan^{k-1}Q}^\Lag \circ X_o \circ s_o \in \vf(\Tan^{k-1}Q)$, then
$s_o \circ \gamma \colon \R \to \W_\Lag$ is an integral curve of $X_o$,
that is
\begin{equation*}
\Tan\rho_{\Tan^{k-1}Q}^\Lag \circ X_o \circ s_o \circ \gamma = \dot{\gamma} \Longrightarrow
X_o \circ (s_o \circ \gamma) = \dot{\overline{s_o \circ \gamma}} \, .
\end{equation*}
\end{definition}

To close the generalized Hamilton-Jacobi problem, Proposition \ref{Chap04_prop:GenLagHamHJEquiv}
now is stated as follows.

\begin{proposition}\label{Chap04_prop:GenLagHamHJEquivalentFormulation}
The following assertions on a section $s_o \in \Gamma(\rho_{\Tan^{k-1}Q}^\Lag)$ are equivalent.
\begin{enumerate}
\item $s_o$ is a solution to the generalized $k$th-order Lagrangian-Hamiltonian Hamilton-Jacobi
problem in $\W_\Lag$.
\item The submanifold $\Im(s_o) \hookrightarrow \W_\Lag$ is invariant under the flow of the vector
field $X_o$ solution of the equation \eqref{Chap04_eqn:LagHamDynEqEquivalent} (that is, $X_{o}$ is
tangent to the submanifold $\Im(s_o)$).
\item The section $s_o$ satisfies the equation
\begin{equation}\label{Chap04_eqn:DynEqGenLagHamHJEquivalent}
\inn(X)(s_o^*\Omega_o) = \d(s_o^*H_o) \, ,
\end{equation}
where $X = \Tan\rho_{\Tan^{k-1}Q}^\Lag \circ X_o \circ s_o$ is the vector field associated to $s_o$.
\end{enumerate}
\end{proposition}
\begin{proof}
The proof of this result follows the patterns in the proofs of Propositions \ref{Chap04_prop:GenLagHJEquiv},
\ref{Chap04_prop:GenHamHJEquiv} and \ref{Chap04_prop:GenLagHamHJEquiv}.
\end{proof}

For the $k$th-order Lagrangian-Hamiltonian Hamilon-Jacobi problem stated in Section
\ref{Chap04_sec:LagHamHJProblem}, Definition \ref{Chap04_def:LagHamHJDef} in this formulation
is stated as follows.

\begin{definition}
The \textnormal{$k$th-order Lagrangian-Hamiltonian Hamilton-Jacobi problem in $\W_\Lag$} 
(or \textnormal{$k$th-order unified Hamilton-Jacobi problem in $\W_\Lag$}) consists in finding sections
$s_o \in \Gamma(\rho_{\Tan^{k-1}Q}^\Lag)$ solution to the generalized $k$th-order Lagrangian-Hamiltonian
Hamilton-Jacobi problem on $\W_\Lag$ such that $s_o^*\Omega_o = 0$.
\end{definition}

And Proposition \ref{Chap04_prop:LagHamHJEquiv} now reads:

\begin{proposition}
Let $s_o \in \Gamma(\rho_{\Tan^{k-1}Q}^\Lag)$ be a section such that $s_o^*\Omega_o = 0$.
Then, the following assertions are equivalent.
\begin{enumerate}
\item $s_o$ is a solution to the $k$th-order Lagrangian-Hamiltonian Hamilton-Jacobi problem in $\W_\Lag$.
\item $\d(s_o^*H_o) = 0$.
\item $\Im(s_o)$ is a Lagrangian submanifold of $\W_\Lag$ invariant by $X_{o}$.
\item The integral curves of $X_{o}$ with initial conditions in $\Im(s_o)$ project onto the integral
curves of the associated vector field $X = \Tan\rho_{\Tan^{k-1}Q}^\Lag \circ X_{o} \circ s_o$.
\end{enumerate}
\end{proposition}

Finally, the following result ensures the equivalence between the formulation given in the manifold
$\W$ in Sections \ref{Chap04_sec:GenLagHamHJProblem} and \ref{Chap04_sec:LagHamHJProblem}, and the one
given in this Section for the submanifold $\W_\Lag$.

\begin{proposition}\label{Chap04_prop:LagHamHJEquivalentFormulation}
Let $s_o \in \Gamma(\rho_{\Tan^{k-1}Q}^\Lag)$ be a solution to the (generalized) $k$th-order unified
Hamilton-Jacobi problem in $W_\Lag$. Then, the section $s = j_\Lag \circ s_o \in \Gamma(\rho_{\Tan^{k-1}Q})$
is a solution to the (generalized) $k$th-order Lagrangian-Hamiltonian Hamilton-Jacobi problem in $\W$.

\noindent Conversely, if $s \in \Gamma(\rho_{\Tan^{k-1}Q})$ is a solution to the (generalized) $k$th-order
unified Hamilton-Jacobi problem in $\W$, then there exists a section
$s_o \in \Gamma(\rho_{\Tan^{k-1}Q}^\Lag)$ which is a solution to the (generalized) $k$th-order
Lagrangian-Hamiltonian Hamilton-Jacobi problem in $\W_\Lag$.
\end{proposition}
\begin{proof}
Let $s_o \in \Gamma(\rho_{\Tan^{k-1}Q}^\Lag)$ be a solution to the generalized $k$th-order unified
Hamilton-Jacobi problem in $\W_\Lag$. First, let us prove that $s = j_\Lag \circ s_o \in \Gamma(\rho_{\Tan^{k-1}Q})$,
that is, $s = j_\Lag \circ s_o$ is a section of the projection $\rho_{\Tan^{k-1}Q}$. In fact,
\begin{equation*}
\rho_{\Tan^{k-1}Q} \circ s = \rho_{\Tan^{k-1}Q} \circ j_\Lag \circ s_o = \rho_{\Tan^{k-1}Q}^\Lag \circ s_o
= \Id_{\Tan^{k-1}Q} \, ,
\end{equation*}
since $s_o \in \Gamma(\rho_{\Tan^{k-1}Q}^\Lag)$.

Next, from Lemma \ref{Chap04_lemma:TechLemma} we have that if $X,\bar{X} \in \vf(\Tan^{k-1}Q)$ are the
associated vector fields to $s$ and $s_o$, respectively, then $X = \bar{X}$. Thus, we denote both vector
fields by $X$. Then, let $\gamma \colon \R \to \Tan^{k-1}Q$ be an integral curve of $X$, and we want to
prove that $s \circ \gamma \colon \R  \to \W$ is an integral curve of $X_{LH}$. Computing,
\begin{align*}
X_{LH} \circ (s \circ \gamma) &= X_{LH} \circ j_\Lag \circ s_o \circ \gamma
= \Tan j_\Lag \circ X_o \circ s_o \circ \gamma
= \Tan j_\Lag \circ \dot{\overline{s_o \circ \gamma}} \\
&= \Tan j_\Lag \circ \Tan s_o \circ \dot{\gamma}
= \Tan s \circ \dot{\gamma} = \dot{\overline{s \circ \gamma}} \, ,
\end{align*}
where we have used that $X_{LH}$ and $X_o$ are $j_\Lag$-related, and that $s_o$ fulfills condition
\eqref{Chap04_eqn:GenLagHamHJDefEquivalent} with the associated vector field. Thus, $s \circ \gamma$
is an integral curve of $X_{LH}$, that is, $s = j_\Lag \circ s_o$ is a solution to the generalized
$k$th-order Lagrangian-Hamiltonian Hamilton-Jacobi problem.

Now we suppose, in addition, that $s_o^*\Omega_o=0$. Then
\begin{equation*}
s^*\Omega = (j_\Lag \circ s_o)^*\Omega = s_o^*(j_\Lag^*\Omega) = s_o^*\Omega_o = 0 \, .
\end{equation*}
Hence, if $s_o$ is a solution to the $k$th-order unified Hamilton-Jacobi problem in $\W_\Lag$, then
$s = j_\Lag \circ s_o$ is a solution to the $k$th-order unified Hamilton-Jacobi problem in $\W$.

Conversely, let $s \in \Gamma(\rho_{\Tan^{k-1}Q})$ be a solution to the generalized $k$th-order
Lagrangian-Hamiltonian Hamilton-Jacobi problem in $\W$. Then, by the first condition in Definition
\ref{Chap04_def:GenLagHamHJDef}, we have $\Im(s) \subseteq \W_\Lag$, and thus there exists a map
$s_o \colon \Tan^{k-1}Q \to \W_\Lag$ such that $j_\Lag \circ s_o = s$. In addition, composing this
last equality with $\rho_{\Tan^{k-1}Q}^\Lag$, we have
\begin{equation*}
\Id_{\Tan^{k-1}Q} = \rho_{\Tan^{k-1}Q} \circ s = \rho_{\Tan^{k-1}Q} \circ j_o \circ s_o
= \rho_{\Tan^{k-1}Q}^\Lag \circ s_o \, ,
\end{equation*}
and thus $s_o$ is a section of the projection $\rho_{\Tan^{k-1}Q}^\Lag$. We will prove that this section
$s_o \in \Gamma(\rho_{\Tan^{k-1}Q}^\Lag)$ satisfies condition \eqref{Chap04_eqn:GenLagHamHJDefEquivalent}.
Again, from Lemma \ref{Chap04_lemma:TechLemma} we have that if $X,\bar{X} \in \vf(\Tan^{k-1}Q)$ are the
associated vector fields to $s$ and $s_o$, respectively, then $X = \bar{X}$. Hence we denote both vector
fields by $X$.

From Proposition \ref{Chap04_prop:GenLagHamHJEquivalentFormulation}, the section
$s_o \in \Gamma(\rho_{\Tan^{k-1}Q}^\Lag)$ defined in the previous paragraph is a solution to the generalized
$k$th-order Lagrangian-Hamiltonian Hamilton-Jacobi problem in $\W_\Lag$ if, and only if, the following
equation holds
\begin{equation*}
\inn(X)(s_o^*\Omega_o) - \d(s_o^*H_o) = 0 \, .
\end{equation*}
Computing, we have
\begin{align*}
\inn(X)(s_o^*\Omega_o) - \d(s_o^*H_o) &= \inn(X)(s_o^*(j_\Lag^*\Omega)) - \d(s_o^*(j_\Lag^*H))
= \inn(X)((j_\Lag \circ s_o)^*\Omega - \d((j_\Lag \circ s_o)^*H) \\
&= \inn(X)(s^*\Omega) - \d(s^*H) = 0 \, ,
\end{align*}
since by Proposition \ref{Chap04_prop:GenLagHamHJEquiv} the last equation holds whenever $s$ is a solution
to the generalized $k$th-order Lagrangian-Hamiltonian Hamilton-Jacobi problem.

Finally, let $s$ be a solution to the $k$th-order Lagrangian-Hamiltonian Hamilton-Jacobi problem, that is,
we suppose in addition that $s^*\Omega = 0$. Then,
\begin{equation*}
s_o^*\Omega_o = s_o^*(j_\Lag^*\Omega) = (j_\Lag \circ s_o)^*\Omega = s^*\Omega = 0 \, .
\end{equation*}
Therefore, if $s$ is a solution to the $k$th-order unified Hamilton-Jacobi problem in $\W$, then the
induced section $s_o \in \Gamma(\rho_{\Tan^{k-1}Q}^\Lag)$ satisfying $s = j_\Lag \circ s_o$ is a solution
to the $k$th-order unified Hamilton-Jacobi problem in $\W_\Lag$.
\end{proof}

\begin{remark}
The main drawback of this equivalent formulation of the $k$th-order Hamilton-Jacobi problem in the
Skinner-Rusk setting is that $\W_\Lag$ has not a natural set of coordinates. This is due to
the identification $\W_\Lag = \graph(\Leg)$, which implies that the coordinates in $\W_\Lag$ depend
on the $k$th-order Lagrangian function provided. Hence, it is easier to consider the problem in $\W$, where
we do have a set of natural induced coordinates, and drag the condition $\Im(s) \subset \W_\Lag$ at every step.
\end{remark}

\subsection{Complete solutions}

As in the Lagrangian and Hamiltonian formalisms, we are interested in finding not only a particular
solution to the Hamilton-Jacobi problem, but a complete solution. In order to do so, we generalize
the concept of complete solutions given in Sections \ref{Chap04_sec:LagHJCompleteSolutions} and
\ref{Chap04_sec:HamHJCompleteSolutions} to the unified formalism, bearing in mind the definition of
a particular solution given in Definition \ref{Chap04_def:LagHamHJDef}.

\begin{definition}
A \textnormal{complete solution to the $k$th-order unified Hamilton-Jacobi problem} is an embedding
$\Phi \colon U \times \Tan^{k-1}Q \to \W$, with $U \subseteq \R^{kn}$ an open set, such that the
following conditions are satisfied:
\begin{enumerate}
\item $\Im(\Phi) \subset \W_\Lag$ is an open subset, that is, $\Phi \colon U \times \Tan^{k-1}Q \to \W_\Lag$
is a local diffeomorphism.
\item For every $\lambda \in U$, the map $s_\lambda(\bullet) = \Phi(\lambda,\bullet) \colon \Tan^{k-1}Q \to \W$
is a solution to the $k$th-order Lagrangian-Hamiltonian Hamilton-Jacobi problem.
\end{enumerate}
\end{definition}

It is clear from this last definition that a complete solution to the $k$th-order unified Hamilton-Jacobi
problem endows the submanifold $\W_\Lag \hookrightarrow \W$ with a foliation transverse to the fibers,
and that every leaf of this foliation is invariant by the vector field $X_{LH}$ solution to the dynamical
equation \eqref{Chap03_eqn:UnifDynEqVF}.

\begin{remark}
It is important to point out that it is the submanifold $\W_\Lag \hookrightarrow \W$ which is endowed with
a foliation by a complete solution, rather than $\W$. This is due to the fact that the real phase space of
the problem is $\W_\Lag$, and not $\W$.
\end{remark}

Let $\Phi$ be a complete solution, and we consider the family of vector fields
\begin{equation*}
\left\{ X_\lambda = \Tan\rho_{\Tan^{k-1}Q} \circ X_{LH} \circ s_\lambda \in \vf(\Tan^{k-1}Q)
\ \mid \ \lambda \in U \subseteq \R^{kn} \right\} \, ,
\end{equation*}
where $s_\lambda \equiv \Phi(\lambda,\bullet)$. Then, the integral curves of $X_\lambda$, for different
$\lambda \in U$, provide all the integral curves of the vector field $X_{LH}$ solution to the dynamical
equation \eqref{Chap03_eqn:UnifDynEqVF}. That is, if $w \in \W_\Lag$, then there exists $\lambda_o \in U$
such that if $\bar{y} = (\rho_{\Tan^{k-1}Q} \circ j_\Lag)(w)$ then $s_{\lambda_o}(\bar{y}) = w$, and the
integral curve of $X_{\lambda_o}$ through $\bar{y}$, lifted by $s_{\lambda_o}$ to $\W_\Lag \hookrightarrow \W$,
gives the integral curve of $X_{LH}$ through $w$.

\subsection{Relation with the Lagrangian and Hamiltonian formalisms}

Finally, we state the relation between the solutions of the Hamilton-Jacobi problem in the unified
formalism and the solutions of the problem in the Lagrangian and Hamiltonian settings given in
Sections \ref{Chap04_sec:LagrangianFormalism} and \ref{Chap04_sec:HamiltonianFormalism}.
Observe that, since the $k$th-order Lagrangian function is hyperregular, the submanifold
$\W_\Lag \hookrightarrow \W$ is diffeomorphic (actually, symplectomorphic) to both $\Tan^{2k-1}Q$
and $\Tan^*(\Tan^{k-1}Q)$ via the maps $\rho_1^\Lag$ and $\rho_2^\Lag$, respectively. It is clear then
that this fact enables us to establish a one-to-one correspondence between the solutions of the
Hamilton-Jacobi problem in the three formalisms.

\begin{theorem}
If $s \in \Gamma(\rho_{\Tan^{k-1}Q})$ is a solution to the (generalized) $k$th-order
Lagrangian-Hamiltonian Hamilton-Jacobi problem, then the sections
$s_\Lag = \rho_1 \circ s \in \Gamma(\rho^{2k-1}_{k-1})$ and
$\alpha = \rho_2 \circ s \in \df^{1}(\Tan^{k-1}Q)$ are solutions to the (generalized)
$k$th-order Lagrangian and Hamiltonian Hamilton-Jacobi problems, respectively.

\noindent Conversely, if $s_\Lag \in \Gamma(\rho^{2k-1}_{k-1})$
(resp., $\alpha \in \df^{1}(\Tan^{k-1}Q)$) is a solution to the (generalized)
$k$th-order Lagrangian (resp., Hamiltonian) Hamilton-Jacobi problem, then 
$s = j_\Lag \circ (\rho_1^\Lag)^{-1} \circ s_\Lag \in \Gamma(\rho_{\Tan^{k-1}Q})$
(resp., $s = j_\Lag \circ (\rho_2^\Lag)^{-1} \circ \alpha \in \Gamma(\rho_{\Tan^{k-1}Q})$)
is a solution to the (generalized) $k$th-order Lagrangian-Hamiltonian Hamilton-Jacobi problem.
\end{theorem}
\begin{proof}
Let us first prove that if $s \in \Gamma(\rho_{\Tan^{k-1}Q})$, then the maps $s_\Lag = \rho_1 \circ s$
and $\alpha = \rho_2 \circ s$ are sections of the projections $\rho^{2k-1}_{k-1}$ and $\pi_{\Tan^{k-1}Q}$,
respectively. Computing, we have for $s_\Lag$
\begin{equation*}
\rho^{2k-1}_{k-1} \circ s_\Lag = \rho^{2k-1}_{k-1} \circ \rho_1 \circ s = \rho_{\Tan^{k-1}Q} \circ s = \Id_{\Tan^{k-1}Q} \, ,
\end{equation*}
and for $\alpha$
\begin{equation*}
\pi_{\Tan^{k-1}Q} \circ \alpha = \pi_{\Tan^{k-1}Q} \circ \rho_2 \circ s = \rho_{\Tan^{k-1}Q} \circ s = \Id_{\Tan^{k-1}Q} \, .
\end{equation*}
Therefore, we have $s_\Lag = \rho_1 \circ s \in \Gamma(\rho^{2k-1}_{k-1})$ and
$\alpha = \rho_2 \circ s \in \df^{1}(\Tan^{k-1}Q)$.

Next, from Lemmas \ref{Chap03_lemma:LagLagrangianVF} and \ref{Chap03_lemma:HamRegHamiltonianVF}, and Theorems
\ref{Chap03_thm:EquivUnifLagVF} and \ref{Chap03_thm:EquivUnifHamVFReg}, the vector field $X_{LH} \in\vf(\W)$
tangent to $\W_\Lag$ and solution to the dynamical equation \eqref{Chap03_eqn:UnifDynEqVF} is
$\rho_1$-related to the Lagrangian vector field $X_\Lag \in \vf(\Tan^{2k-1}Q)$ solution to equation
\eqref{Chap02_eqn:LagHODynEq}, and also $\rho_2$-related to the Hamiltonian vector field
$X_h \in \vf(\Tan^*(\Tan^{k-1}Q))$ solution to equation \eqref{Chap02_eqn:HamHODynEq}. Thus, using Lemma
\ref{Chap04_lemma:TechLemma}, we have that if $X,\bar{X},\tilde{X} \in \vf(\Tan^{k-1}Q)$ are the vector
fields associated to $s$, $s_\Lag = \rho_1 \circ s$ and $\alpha = \rho_2 \circ s$, respectively, then
$X = \bar{X} = \tilde{X}$. Hence, all of them are denoted by $X$.

Now, let $s \in \Gamma(\rho_{\Tan^{k-1}Q})$ be a solution to the generalized $k$th-order unified
Hamilton-Jacobi problem. We want to prove that $s_\Lag = \rho_1 \circ s \in \Gamma(\rho^{2k-1}_{k-1})$ and
$\alpha = \rho_2 \circ s \in \df^{1}(\Tan^{k-1}Q)$ are solutions to the generalized $k$th-order Lagrangian
and Hamiltonian Hamilton-Jacobi problems, respectively. Let $\gamma \colon \R \to \Tan^{k-1}Q$ be an integral
curve of $X$. Then, for the Lagrangian section,
\begin{align*}
X_\Lag \circ (s_\Lag \circ \gamma) &= X_\Lag \circ \rho_1 \circ s \circ \gamma
= \Tan\rho_1 \circ X_{LH} \circ s \circ \gamma \\
&= \Tan\rho_1 \circ \dot{\overline{s \circ \gamma}}
= \Tan(\rho_1 \circ s) \circ \dot{\gamma}
= \Tan s_\Lag \circ \dot{\gamma}
= \dot{\overline{s_\Lag \circ \gamma}} \, ,
\end{align*}
and, for the Hamiltonian $1$-form,
\begin{align*}
X_h \circ (\alpha \circ \gamma) &= X_h \circ \rho_2 \circ s \circ \gamma
= \Tan\rho_2 \circ X_{LH} \circ s \circ \gamma \\
&= \Tan\rho_2 \circ \dot{\overline{s \circ \gamma}}
= \Tan(\rho_2 \circ s) \circ \dot{\gamma}
= \Tan \alpha \circ \dot{\gamma}
= \dot{\overline{\alpha \circ \gamma}} \, .
\end{align*}
Now we suppose in addition that $s^*\Omega = 0$. Then, for the Lagrangian section $s_\Lag$ we have
\begin{equation*}
s_\Lag^*\omega_\Lag = (\rho_1 \circ s)^*\omega_\Lag = s^*(\rho_1^*\omega_\Lag) = s^*\Omega = 0 \, ,
\end{equation*}
since $\Omega = \rho_1^*\omega_\Lag$ by Lemma \ref{Chap03_lemma:LagCartanForms}.
Now, for the Hamiltonian $1$-form $\alpha$ we have
\begin{equation*}
\d\alpha = -\alpha^*\omega_{k-1} = -(\rho_2 \circ s)^*\omega_{k-1} = -s^*(\rho_2^*\omega_{k-1}) = -s^*\Omega = 0 \, ,
\end{equation*}
from the definition of $\Omega$ given in Section \ref{Chap03_sec:GeometricalSettingStructures}
and relation \eqref{Chap04_eqn:PullBackSymplecticFormByAlpha}.

Therefore, if $s \in \Gamma(\rho_{\Tan^{k-1}Q})$ is a solution to the (generalized) $k$th-order
Lagrangian-Hamiltonian Hamilton-Jacobi problem, then $s_\Lag = \rho_1 \circ s \in \Gamma(\rho^{2k-1}_{k-1})$
and $\alpha = \rho_2 \circ s \in \df^{1}(\Tan^{k-1}Q)$ are solutions to the (generalized)
$k$th-order Lagrangian and Hamiltonian Hamilton-Jacobi problems, respectively.

For the converse, let $s_\Lag \in \Gamma(\rho^{2k-1}_{k-1})$ be a solution to the generalized
$k$th-order Lagrangian Hamilton-Jacobi problem. We will first prove that the section
$s_o = (\rho_1^\Lag)^{-1} \circ s_\Lag \in \Gamma(\rho_{\Tan^{k-1}Q}^\Lag)$ satisfies condition
\eqref{Chap04_eqn:GenLagHamHJDefEquivalent}. First, let us prove that the map
$s_o = (\rho_1^\Lag)^{-1}\circ s_\Lag \colon \Tan^{k-1}Q \to \W_\Lag$ is a section of the projection
$\rho_{\Tan^{k-1}Q}^\Lag$. In fact,
\begin{equation*}
\rho_{\Tan^{k-1}Q}^\Lag \circ s_o = \rho_{\Tan^{k-1}Q}^\Lag \circ (\rho_1^\Lag)^{-1} \circ s_\Lag
= \rho^{2k-1}_{k-1} \circ s_\Lag = \Id_{\Tan^{k-1}Q} \, ,
\end{equation*}
since $\rho_{\Tan^{k-1}Q} \circ j_\Lag = \rho^{2k-1}_{k-1} \circ \rho_1 \circ j_\Lag$ implies
$\rho^{2k-1}_{k-1} = \rho_{\Tan^{k-1}Q}^\Lag \circ (\rho_1^\Lag)^{-1}$.

Next, note that since $\rho_1^\Lag$ is a diffeomorphism, it is clear that the Lagrangian vector field
$X_\Lag$ and the vector field $X_o \in \vf(\W_\Lag)$ solution to the equation
\eqref{Chap04_eqn:LagHamDynEqEquivalent} are $\rho_1^\Lag$-related (or, equivalently,
$(\rho_1^\Lag)^{-1}$-related). Thus, from Lemma \ref{Chap04_lemma:TechLemma} we have that if
$X,\bar{X} \in \vf(\Tan^{k-1}Q)$ are the associated vector fields to $s_\Lag$ and
$s_o = (\rho_1^\Lag)^{-1} \circ s_\Lag$, then $X = \bar{X}$.

Then, let $\gamma \colon \R \to \Tan^{k-1}Q$ be an integral curve of $X$. Computing, we have
\begin{align*}
X_o \circ (s_o \circ \gamma) &= X_o \circ (\rho_1^\Lag)^{-1} \circ s_\Lag \circ \gamma
= \Tan(\rho_1^\Lag)^{-1} \circ X_\Lag \circ s_\Lag \circ \gamma \\
&= \Tan(\rho_1^\Lag)^{-1} \circ \dot{\overline{s_\Lag \circ \gamma}}
= \Tan(\rho_1^\Lag)^{-1} \circ \Tan s_\Lag \circ \dot{\gamma} \\
&= \Tan((\rho_1^\Lag)^{-1} \circ s_\Lag) \circ \dot{\gamma}
= \Tan s_o \circ \dot \gamma
= \dot{\overline{s_o \circ \gamma}} \, .
\end{align*}

Now, let us suppose, in addition, that $s_\Lag^*\omega_\Lag = 0$. Then, since
$\rho_1^\Lag \colon (\W_\Lag,\Omega_o) \to (\Tan^{2k-1}Q,\omega_\Lag)$ is a symplectomorphism, we have
\begin{equation*}
s_o^*\Omega_o = ((\rho_1^\Lag)^{-1} \circ s_\Lag)^*\Omega_o
= s_\Lag^*(((\rho_1^\Lag)^{-1})^*\Omega_o) = s_\Lag^*\omega_\Lag = 0 \, .
\end{equation*}

Therefore, if $s_\Lag \in \Gamma(\rho^{2k-1}_{k-1})$ us a solution to the (generalized) $k$th-order
Lagrangian Hamilton-Jacobi problem, then the section
$s_o = (\rho_1^\Lag)^{-1} \circ s_\Lag \in \Gamma(\rho_{\Tan^{k-1}Q})$ is a solution to the (generalized)
$k$th-order Lagrangian-Hamiltonian Hamilton-Jacobi problem in $\W_\Lag$. Then, using Proposition
\ref{Chap04_prop:LagHamHJEquivalentFormulation}, the section
$s = j_\Lag \circ (\rho_1^\Lag)^{-1} \circ s_\Lag \in \Gamma(\rho_{\Tan^{k-1}Q}^\Lag)$ is a solution to the
(generalized) $k$th-oder Lagrangian-Hamiltonian Hamilton-Jacobi problem in $\W$.

As we have pointed out in the remark at the beginning of Section \ref{Chap04_sec:UnifiedFormalismEquivalent},
the map $\rho_2^\Lag \colon \W_\Lag \to \Tan^*(\Tan^{k-1}Q)$ is also a symplectomorphism,
between the symplectic manifolds $(\W_\Lag,\Omega_o)$ and $(\Tan^*(\Tan^{k-1}Q),\omega_{k-1})$. Therefore,
the same proof applies for the Hamiltonian $1$-form $\alpha \in \df^{1}(\Tan^{k-1}Q)$ solution to the
(generalized) $k$th-order Hamiltonian Hamilton-Jacobi problem.
\end{proof}

\section{Examples}
\label{Chap04_sec:Examples}

In this last Section of the Chapter, two physical models are analyzed as examples to show the application
of the formalism. Contrary to Chapter \ref{Chap:HOAutonomousDynamicalSystems}, in this Chapter both examples
are regular systems. The first example is the \textsl{end of a thrown javelin}, and the second one is the
\textsl{shape of a homogeneous deformed elastic cylindrical beam with fixed ends}.

\subsection{The end of a thrown javelin}
\label{Chap04_sec:Example1}

Let us consider the dynamical system that describes the motion of the end of a thrown javelin.
This gives rise to a $3$-dimensional second-order dynamical system, which is a particular case of the problem
of determining the trajectory of a particle rotating about a translating center \cite{art:Constantelos84}.
Let $Q = \R^3$ be the manifold modeling the configuration space for this system with coordinates
$(q_0^1,q_0^2,q_0^3) = (q_0^A)$. Using the induced coordinates in $\Tan^2\R^3$, the Lagrangian
function for this system is
\begin{equation*}
\Lag(q_0^A,q_1^A,q_2^A) = \frac{1}{2} \sum_{A=1}^{3} \left((q_1^A)^2 - (q_2^A)^2\right) \, .
\end{equation*}
This is a regular Lagrangian function since the Hessian matrix of $\Lag$ with respect to the second-order
velocities is
\begin{equation*}
\left( \derpars{\Lag}{q_2^B}{q_2^A} \right) = \begin{pmatrix} 1 & 0 & 0 \\ 0 & 1 & 0 \\ 0 & 0 & 1 \end{pmatrix} \, ,
\end{equation*}
which is a $3 \times 3$ invertible matrix.
 
\subsubsection{Lagrangian formulation of the Hamilton-Jacobi problem}

The Poincar\'{e}-Cartan forms $\theta_\Lag$ and $\omega_\Lag$, and the Lagrangian energy are
locally given by
\begin{equation*}
\begin{array}{c}
\displaystyle \theta_\Lag = \sum_{A=1}^{3} \left( (q_1^A + q_3^A)\d q_0^A - q_2^A \d q_1^A\right) \quad ; \quad
\omega_\Lag = \sum_{A=1}^{3} \left( \d q_0^A \wedge \d q_1^A + \d q_0^A \wedge \d q_3^A - \d q_1^A \wedge \d q_2^A \right) \, , \\
\displaystyle E_\Lag = \frac{1}{2} \sum_{A=1}^{3} \left( (q_1^A)^2 + 2q_1^Aq_3^A - (q_2^A)^2 \right).
\end{array}
\end{equation*}
Thus, the semispray of type $1$, $X_\Lag \in \vf(\Tan^3\R^3)$, solution to the dynamical equation
\eqref{Chap02_eqn:LagHODynEq} is
\begin{equation*}
X_\Lag = q_1^A\derpar{}{q_0^A} + q_2^A\derpar{}{q_1^A} + q_3^A\derpar{}{q_2^A} - q_2^A\derpar{}{q_3^A} \, .
\end{equation*}

Consider the projection $\rho^3_1 \colon \Tan^3\R^3 \to \Tan\R^3$. From Proposition \ref{Chap04_prop:GenLagHJEquiv}
we know that the generalized second-order Lagrangian Hamilton-Jacobi problem consists in finding sections
$s \in \Gamma(\rho^3_1)$ such that the Lagrangian vector field $X_\Lag$ is tangent to the submanifold
$\Im(s) \hookrightarrow \Tan^3\R^3$. Suppose that the section $s$ is given locally
by $s(q_0^A,q_1^A) = (q_0^A,q_1^A,s_2^A,s_3^A)$. As the submanifold $\Im(s)$
is defined locally by the constraint functions $q_2^A - s_2^A$ and $q_3^A - s_3^A$, then the tangency
condition gives the following system of $6$ partial differential equations for the component functions
of the section
\begin{equation*}
s_3^A - q_1^B\derpar{s_2^A}{q_0^B} - s_2^B\derpar{s_2^A}{q_1^B} = 0 \quad ; \quad
s_2^A + q_1^B\derpar{s_3^A}{q_0^B} + s_2^B\derpar{s_3^A}{q_1^B} = 0 \, ,
\end{equation*}
with $1 \leqslant A \leqslant 3$.

In order to obtain the equations of the second-order Lagrangian Hamilton-Jacobi problem, we require
in addition the section $s \in \Gamma(\rho^3_1)$ to satisfy the condition $\d(s^*E_\Lag) = 0$, or,
equivalently, $s^*\omega_\Lag = 0$. From the local expression of the Cartan $2$-form
$\omega_\Lag \in \df^{2}(\Tan^3\R^3)$ given above, taking the pull-back by the section
$s(q_0^A,q_1^A) = (q_0^A,q_1^A,s_2^A,s_3^A)$ we obtain
\begin{equation*}
s^*\omega_\Lag = \sum_{A=1}^{3} \left[ \d q_0^A \wedge \d q_1^A + \derpar{s_3^A}{q_0^B} \d q_0^A \wedge \d q_0^B
+ \left( \derpar{s_3^A}{q_1^B} + \derpar{s_2^B}{q_0^A} \right) \d q_0^A \wedge \d q_1^B
- \derpar{s_2^A}{q_1^B} \d q_1^A \wedge \d q_1^B \right].
\end{equation*}
Hence, the condition $s^*\omega_\Lag = 0$ gives the following partial differential equations
\begin{equation*}
\derpar{s_3^A}{q_0^B} = \derpar{s_2^A}{q_1^B} = \derpar{s_3^A}{q_1^B} + \derpar{s_2^B}{q_0^A} = 0 \, , \ \mbox{if } A \neq B \quad ;
\quad \derpar{s_3^A}{q_1^A} + \derpar{s_2^A}{q_0^A} + 1 = 0 \, .
\end{equation*}
Hence, the section $s \in \Gamma(\rho^3_1)$ is a solution to the second-order Lagrangian
Hamilton-Jacobi problem if the following system of partial differential equations holds
\begin{equation*}
\begin{array}{c}
\displaystyle s_3^A = q_1^B\derpar{s_2^A}{q_0^B} + s_2^A\derpar{s_2^A}{q_1^A} \quad ; \quad
q_1^A\derpar{s_3^A}{q_0^A} + s_2^B\derpar{s_3^A}{q_1^B} + s_2^A = 0 \, , \\[10pt]
\displaystyle \derpar{s_3^A}{q_1^B} + \derpar{s_2^B}{q_0^A} = 0 \quad ; \quad
\derpar{s_3^A}{q_1^A} + \derpar{s_2^A}{q_0^A} + 1 = 0 \, .
\end{array}
\end{equation*}

Finally, we compute the equations for the generating function $W$. The pull-back of the Cartan $1$-form
$\theta_\Lag$ by the section $s$ gives in coordinates
\begin{equation*}
s^*\theta_\Lag = \sum_{A=1}^{3} \left( (q_1^A + s_3^A)\d q_0^A - s_2^A \d q_1^A\right) \, .
\end{equation*}
Hence, requiring $s^*\theta_\Lag = \d W$ for a local function $W$ defined in $\Tan Q$ we obtain
\begin{equation*}
\derpar{W}{q_0^A} = q_1^A + s_3^A \quad ; \quad \derpar{W}{q_1^A} = -s_2^A \, ,
\end{equation*}
and thus from $\d(s^*E_\Lag) = 0$, we have $s^*E_\Lag = \mbox{const.}$, that is,
\begin{equation*}
\sum_{A=1}^{3} \left( q_1^A\derpar{W}{q_0^A} - \frac{1}{2}\left((q_1^A)^2 + \left( \derpar{W}{q_1^A}\right)^2 \right)\right) = \mbox{const.} 
\end{equation*}

\begin{remark}
This equation cannot be stated in the general case, since we need to clear the higher-order velocities
from the previous equations. This calculation is easy for this particular example, but it depends on
the Lagrangian function provided in every system.
\end{remark}

\subsubsection{Hamiltonian formulation of the Hamilton-Jacobi problem}

Now, to establish the Hamiltonian formalism for the Hamilton-Jacobi problem, we consider natural
coordinates $(q_0^A,q_1^A,p_A^0,p_A^1)$ on the cotangent bundle $\Tan^*(\Tan\R^3)$. Then the
Legendre-Ostrogradsky map $\Leg \colon \Tan^3\R^3 \to \Tan^*(\Tan\R^3)$ associated to the Lagrangian
function $\Lag$ is
\begin{equation*}
\Leg^*q_0^A = q_0^A \quad ; \quad \Leg^*q_1^A = q_1^A \quad ; \quad
\Leg^*p_A^0 = q_1^A + q_3^A \quad ; \quad \Leg^*p_A^1 = -q_2^A \, ,
\end{equation*}
and the inverse map $\Leg^{-1} \colon \Tan^*(\Tan\R^3) \to \Tan^3\R^3$ is given by
\begin{equation*}
(\Leg^{-1})^*q_0^A = q_0^A \quad ; \quad (\Leg^{-1})^*q_1^A = q_1^A \quad ; \quad
(\Leg^{-1})^*q_2^A = -p_A^1 \quad ; \quad (\Leg^{-1})^*q_3^A = p_A^0 - q_1^A \, .
\end{equation*}
From these coordinate expressions it is clear that the Legendre-Ostrogradsky map
is a global diffeomorphism, that is, $\Lag$ is a hyperregular Lagrangian function.

The Hamiltonian function $h \in \Cinfty(\Tan^*(\Tan\R^3))$ is
\begin{equation*}
h = (\Leg^{-1})^*E_\Lag = \sum_{A=1}^{3} \left[p_A^0q_1^A - \frac{1}{2} \left( (q_1^A)^2 + (p_A^1)^2 \right)\right].
\end{equation*}
Thus, the Hamiltonian vector field $X_h \in \vf(\Tan^*(\Tan Q))$ solution to equation
\eqref{Chap02_eqn:HamHODynEq} is
\begin{equation*}
X_h = q_1^A \derpar{}{q_0^A} - p_A^1 \derpar{}{q_1^A} + (q_1^A - p_A^0) \derpar{}{p_A^1} \, .
\end{equation*}

Consider the projection $\pi_{\Tan Q} \colon \Tan^*(\Tan \R^3) \to \Tan\R^3$. From Proposition
\ref{Chap04_prop:GenHamHJEquiv} we know that the generalized second-order Hamiltonian Hamilton-Jacobi
problem consists in finding $1$-forms $\alpha \in \df^{1}(\Tan \R^3)$ such that the Hamiltonian vector
field $X_h$ is tangent to the submanifold $\Im(\alpha) \hookrightarrow \Tan^*(\Tan \R^3)$. Suppose that
the $1$-form $\alpha$ is given locally by $\alpha = \alpha_A^0 \d q_0^A + \alpha_A^1 \d q_1^A$.
As the submanifold $\Im(\alpha)$ is defined locally by the constraint functions $p_A^0 - \alpha_A^0$ and
$p_A^1 - \alpha_A^1$, then the tangency condition gives the following system of $6$ partial differential
equations for the component functions of the form
\begin{equation*}
-q_1^B\derpar{\alpha_A^0}{q_0^B} + \alpha_B^1 \derpar{\alpha_A^0}{q_1^B} = 0 \quad ; \quad
q_1^A - \alpha_A^0 - q_1^B\derpar{\alpha_A^1}{q_0^B} + \alpha_B^1 \derpar{\alpha_A^1}{q_1^B} = 0 \, ,
\end{equation*}
with $1 \leqslant A \leqslant 3$.

In order to obtain the equations of the second-order Hamiltonian Hamilton-Jacobi problem, we require
in addition the $1$-form $\alpha \in \df^{1}(\Tan \R^3)$ to be closed. In coordinates, this condition is
\begin{equation*}
\derpar{\alpha_A^1}{q_0^B} - \derpar{\alpha_B^0}{q_1^A} = 0 \quad ; \quad
\derpar{\alpha_A^0}{q_0^B} - \derpar{\alpha_B^0}{q_0^A} = 0  \, , \ \mbox{if } A \neq B \quad ; \quad
\derpar{\alpha_A^1}{q_1^B} - \derpar{\alpha_B^1}{q_1^A} = 0 \, , \ \mbox{if } A \neq B \, .
\end{equation*}
Hence, the $1$-form $\alpha \in \df^{1}(\Tan \R^3)$ is a solution to the second-order Hamiltonian
Hamilton-Jacobi problem if the following system of partial differential equations holds
\begin{equation*}
\begin{array}{c}
\displaystyle-q_1^B\derpar{\alpha_A^0}{q_0^B} + \alpha_B^1 \derpar{\alpha_A^0}{q_1^B} = 0 \quad ; \quad
q_1^A - \alpha_A^0 - q_1^B\derpar{\alpha_A^1}{q_0^B} + \alpha_B^1 \derpar{\alpha_A^1}{q_1^B} = 0 \, , \\[10pt]
\displaystyle\derpar{\alpha_A^1}{q_0^B} - \derpar{\alpha_B^0}{q_1^A} = 0 \quad ; \quad
\derpar{\alpha_A^0}{q_0^B} - \derpar{\alpha_B^0}{q_0^A} = 0  \, , \ \mbox{if } A \neq B \quad ; \quad
\derpar{\alpha_A^1}{q_1^B} - \derpar{\alpha_B^1}{q_1^A} = 0 \, , \ \mbox{if } A \neq B \, .
\end{array}
\end{equation*}

Finally, we compute the equations for the generating function $W$. Requiring $\alpha = \d W$
for a local function $W$ defined in $\Tan Q$, we obtain
\begin{equation*}
\alpha_A^0 = \derpar{W}{q_0^A} \quad ; \quad \alpha_A^1 = \derpar{W}{q_1^A} \, ,
\end{equation*}
and thus from $\d(\alpha^*h) = 0$, we have $\alpha^*h = \mbox{const.}$, that is,
\begin{equation*}
\sum_{A=1}^{3} \left( q_1^A\derpar{W}{q_0^A} - \frac{1}{2}\left((q_1^A)^2 + \left( \derpar{W}{q_1^A}\right)^2 \right)\right) = \mbox{const.} \, .
\end{equation*}
Observe that this equations coincides with the Hamilton-Jacobi equation given previously in the Lagrangian
problem.

A particular solution of this Hamilton-Jacobi equation in dimension $1$  has been obtained
in \cite{art:Constantelos84}. This particular solution is
\begin{equation*}
W(q_0,q_1) = \sqrt{2}\int \d q_{1}\sqrt{-\frac{1}{2}q_{1}^{2}+c_{2}q_{1}-c_{1}}+c_{2}q_{0}
\quad , \quad (c_1, c_2 \in \R) \, .
\end{equation*}

\subsubsection{Lagrangian-Hamiltonian formulation of the Hamilton-Jacobi problem}

In the induced natural coordinates $(q_0^A,q_1^A,q_2^A,q_3^A,p^0_A,p^1_A)$ of $\W,$ the coordinate expressions
of the presymplectic form $\Omega = \rho_2^*\omega_1 \in \df^{2}(\W)$ and the Hamiltonian function
$H = \C - \Lag \in \Cinfty(\W)$ are
\begin{equation*}
\Omega = \d q_0^A \wedge \d p^0_A + \d q_1^A \wedge \d p^1_A \quad ; \quad
H = q_1^Ap^0_A + q_2^Ap^1_A - \frac{1}{2}\left( (q_1^A)^2 - (q_2^A)^2 \right) \, .
\end{equation*}
Thus, the semispray of type $1$ $X_{LH} \in \vf(\W)$ solution to the dynamical equation
\eqref{Chap03_eqn:UnifDynEqVF} and tangent to the submanifold $\W_\Lag = \graph(\Leg) \hookrightarrow \W$
has the following coordinate expression
\begin{equation*}
X_{LH} = q_1^A\derpar{}{q_0^A} + q_2^A\derpar{}{q_1^A} + q_3^A\derpar{}{q_2^A}
- q_2^A\derpar{}{q_3^A} + (q_1^A - p_A^1) \derpar{}{p^1_A} \, .
\end{equation*}

In the generalized second-order Lagrangian-Hamiltonian Hamilton-Jacobi problem we look for sections
$s \in \Gamma(\rho_{\Tan\R^3})$, given locally by $s(q_0^A,q_1^A) = (q_0^A,q_1^A,s_2^A,s_3^A,\alpha^0_A,\alpha^1_A)$,
such that the submanifold $\Im(s) \hookrightarrow \W$ is invariant under the flow of $X_{LH} \in \vf(\W)$.
Since the constraints defining locally $\Im(s)$ are $q_2^A - s_2^A = 0$, $q_3^A - s_3^A = 0$,
$p^0_A - \alpha^0_A = 0$, $p^1_A - \alpha^1_A = 0$, then the equations for the section $s$ are
\begin{equation*}
\begin{array}{c}
\displaystyle  s_3^A - q_1^B\derpar{s_2^A}{q_0^B} - s_2^B\derpar{s_2^A}{q_1^B} = 0 \quad ; \quad
s_2^A + q_1^B\derpar{s_3^A}{q_0^B} + s_2^B\derpar{s_3^A}{q_1^B} = 0 \, , \\[10pt]
\displaystyle -q_1^B\derpar{\alpha_A^0}{q_0^B} + s_2^B \derpar{\alpha_A^0}{q_1^B} = 0 \quad ; \quad
q_1^A - \alpha_A^0 - q_1^B\derpar{\alpha_A^1}{q_0^B} - s_2^B \derpar{\alpha_A^1}{q_1^B} = 0 \, .
\end{array}
\end{equation*}

For the second-order Lagrangian-Hamiltonian Hamilton-Jacobi problem, we require the form obtained by
projecting the section, $\rho_2 \circ s \in \df^{1}(\Tan\R^3)$, to be closed. In coordinates, if
$s = (q_0^A,q_1^A,s_2^A,s_3^A,\alpha^0_A,\alpha^1_A)$, then the $1$-form $\rho_2 \circ s$ is given by
$\rho_2\circ s = \alpha^0_A \d q_0^A + \alpha^1_A \d q_1^A$. Hence, a section $s \in \Gamma(\rho_{\Tan\R^3})$
solution to the second-order Lagrangian-Hamiltonian Hamilton-Jacobi problem for this system must satisfy the following system of
partial differential equations
\begin{equation*}
\begin{array}{c}
\displaystyle  s_3^A - q_1^B\derpar{s_2^A}{q_0^B} - s_2^B\derpar{s_2^A}{q_1^B} = 0 \quad ; \quad
s_2^A + q_1^B\derpar{s_3^A}{q_0^B} + s_2^B\derpar{s_3^A}{q_1^B} = 0 \, , \\[10pt]
\displaystyle -q_1^B\derpar{\alpha_A^0}{q_0^B} + s_2^B \derpar{\alpha_A^0}{q_1^B} = 0 \quad ; \quad
q_1^A - \alpha_A^0 - q_1^B\derpar{\alpha_A^1}{q_0^B} - s_2^B \derpar{\alpha_A^1}{q_1^B} = 0 \, , \\[10pt]
\displaystyle \derpar{\alpha_A^1}{q_0^B} - \derpar{\alpha_B^0}{q_1^A} = 0 \quad ; \quad
\derpar{\alpha_A^0}{q_0^B} - \derpar{\alpha_B^0}{q_0^A} = 0  \, , \ \mbox{if } A \neq B \quad ; \quad
\derpar{\alpha_A^1}{q_1^B} - \derpar{\alpha_B^1}{q_1^A} = 0 \, , \ \mbox{if } A \neq B \, .
\end{array}
\end{equation*}

\subsection{The shape of a homogeneous deformed elastic cylindrical beam with fixed ends}
\label{Chap04_sec:Example2}

Let us consider a deformed elastic cylindrical beam with both ends fixed. The problem is to determinate its
shape, that is, the bending of the axis of the beam. This system has been studied on many occasions,
such as \cite{book:Benson06}, where it is applied to the study of xylophones and tubular bells (Chapter 3, \S 3.9),
and \cite{book:Elsgoltz83}, where the Euler-Lagrange equations are derived from a variational principle
(Chapter VI, \S 4).

Let $Q$ be the $1$-dimensional smooth manifold modeling the configuration space of the system with
local coordinate $(q_0)$. Then, in the natural coordinates of $\Tan^2Q$, the second-order Lagrangian
function $\Lag \in \Cinfty(\Tan^2Q)$ for this system is
\begin{equation*}
\Lag(q_0,q_1,q_2) = \frac{1}{2}\mu q_2^2 + \rho q_0 \, ,
\end{equation*}
where $\mu, \rho \in \R$ are constants that represent physical parameters of the beam: $\rho$ is
the linear density and $\mu$ is a non-zero constant involving Young's modulus of the material, the radius of curvature
and the sectional moment of the beam (see \cite{book:Benson06} for a detailed description).
This is a regular second-order Lagrangian function, since the Hessian matrix of $\Lag$ with respect to
$q_2$ is
\begin{equation*}
\left( \derpars{\Lag}{q_2}{q_2} \right) = \mu \, ,
\end{equation*}
and has maximum rank equal to $1$ as $\mu \neq 0$.

\subsubsection{Lagrangian formulation of the Hamilton-Jacobi problem}

The local expressions for the Poincar\'{e}-Cartan forms $\theta_\Lag \in \df^1(\Tan^3Q)$ and
$\omega_\Lag\in\df^2(\Tan^3Q)$, and the Lagrangian energy $E_\Lag \in \Cinfty(\Tan^3Q)$ are
\begin{align*}
\theta_\Lag = \mu ( -q_3 \d q_0 + \mu q_2 \d q_1) \quad ; \quad
\omega_\Lag = \mu (-\d q_0 \wedge \d q_3 + \d q_1 \wedge \d q_2) \quad ; \quad
E_\Lag = -\rho q_0 + \frac{1}{2}\mu q_2^2 - \mu q_1q_3 \, .
\end{align*}
Thus, the semispray of type $1$, $X_\Lag \in \vf(\Tan^3Q)$, solution to equation
\eqref{Chap02_eqn:LagHODynEq} is given locally by
\begin{equation*}
X_\Lag = q_1\derpar{}{q_0} + q_2\derpar{}{q_1} + q_3\derpar{}{q_2} - \frac{\rho}{\mu}\derpar{}{q_3} \, .
\end{equation*}
Observe that the Euler-Lagrange equation for this $1$-dimensional system is
\begin{equation*}
\frac{d^4\gamma}{dt^4} = -\frac{\rho}{\mu} \, ,
\end{equation*}
where $\gamma \colon \R \to Q$ is a curve. Therefore, it is straightforward to obtain the general solution,
which is a polynomial with degree $4$ on the variable $t$ given by
\begin{equation*}
\gamma(t) = -\frac{\rho}{\mu}t^4 + c_3t^3 + c_2 t^2 + c_1 t + c_0 \, ,
\end{equation*}
where $c_0,c_1,c_2,c_3 \in \R$ are constants depending on the initial conditions given.

Now, we state the equations of the Lagrangian Hamilton-Jacobi problem for this system. Consider the
projection $\rho^3_1\colon\Tan^3Q \to \Tan Q$. By Proposition \ref{Chap04_prop:GenLagHJEquiv},
the generalized second-order Lagrangian Hamilton-Jacobi problem consists in finding sections
$s \in \Gamma(\rho^3_1)$, given locally by $s(q_0,q_1) = (q_0,q_1,s_2,s_3)$, such that the submanifold
$\Im(s) \hookrightarrow \Tan^3Q$ is invariant by the Lagrangian vector field $X_\Lag \in \vf(\Tan^3Q)$.
Since the constraints defining locally $\Im(s)$ are $q_2-s_2 = 0$, $q_3-s_3=0$, then the equations for
the section $s$ are
\begin{equation*}
s_3 - q_1\derpar{s_2}{q_0} - s_2\derpar{s_2}{q_1} = 0 \quad ; \quad
-\frac{\rho}{\mu} - q_1\derpar{s_3}{q_0} - s_2\derpar{s_3}{q_1} = 0 \, .
\end{equation*}

For the second-order Lagrangian Hamilton-Jacobi problem, we must require, in addition, that the
section $s \in \Gamma(\rho^3_1)$ satisfies $\d (s^*E_\Lag) = s^*\d E_\Lag = 0$. From the local
expression of the Lagrangian energy $E_\Lag \in \Cinfty(\Tan^3Q)$ given above, we have
\begin{equation*}
\d E_\Lag = -\rho \d q_0 - \mu q_3 \d q_1 + \mu q_2 \d q_2 - \mu q_1 \d q_3 \, .
\end{equation*}
Thus, taking the pull-back of $\d E_\Lag$ by the section $s(q_0,q_1) = (q_0,q_1,s_2,s_3)$, we obtain
\begin{equation*}
s^*\d E_\Lag = \mu \left( -\frac{\rho}{\mu} + s_2\derpar{s_2}{q_0} - q_1 \derpar{s_3}{q_0} \right) \d q_0
+ \mu \left( - s_3 + s_2 \derpar{s_2}{q_1} - q_1\derpar{s_3}{q_1} \right) \d q_1 \, .
\end{equation*}
Hence, the section $s \in \Gamma(\rho^3_1)$ is a solution to the second-order Lagrangian Hamilton-Jacobi
problem if its component functions satisfy the following system of partial differential equations
\begin{align*}
s_3 - q_1\derpar{s_2}{q_0} - s_2\derpar{s_2}{q_1} = 0 \quad ; \quad
-\frac{\rho}{\mu} - q_1\derpar{s_3}{q_0} - s_2\derpar{s_3}{q_1} = 0 \, , \\
-\frac{\rho}{\mu} - q_1 \derpar{s_3}{q_0} + s_2\derpar{s_2}{q_0} = 0 \quad ; \quad
- s_3 - q_1\derpar{s_3}{q_1} + s_2 \derpar{s_2}{q_1} = 0 \, .
\end{align*}
These $4$ partial differential equations are not linearly independent. In particular, the equations
obtained requiring $\d(s^*E_\Lag) = 0$ can be reduced to a single one by computing the pull-back of
the Poincar\'{e}-Cartan $2$-form by the section $s$,
\begin{equation*}
s^*\omega_\Lag = -\mu \left( \derpar{s_3}{q_1} + \derpar{s_2}{q_0} \right) \d q_0 \wedge \d q_1 \, .
\end{equation*}
Therefore, requiring $s^*\omega_\Lag = 0$ instead of the equivalent condition $\d(s^*E_\Lag) = 0$,
we have that $s$ is a solution to the second-order Lagrangian Hamilton-Jacobi problem if its component
functions satisfy the following equivalent system of partial differential equations
\begin{equation*}
s_3 - q_1\derpar{s_2}{q_0} - s_2\derpar{s_2}{q_1} = 0 \quad ; \quad
-\frac{\rho}{\mu} - q_1\derpar{s_3}{q_0} - s_2\derpar{s_3}{q_1} = 0 \quad ; \quad
\derpar{s_3}{q_1} + \derpar{s_2}{q_0} = 0 \, ,
\end{equation*}
where these $3$ equations are now linearly independent.

Finally, we compute the equations for the generating function $W$. The pull-back of $\theta_\Lag$ by
$s$ gives, in coordinates
\begin{equation*}
s^*\theta_\Lag = -\mu s_3 \d q_0 + \mu s_2 \d q_1 \, .
\end{equation*}
Thus, requiring $s^*\theta_\Lag = \d W$, for a local function $W$ in $\Tan Q$, we obtain
\begin{equation*}
\derpar{W}{q_0} = - \mu s_3 \quad ; \quad \derpar{W}{q_1} = \mu s_2 \, .
\end{equation*}
and thus from $\d(s^*E_\Lag) = 0$, we have $s^*E_\Lag = \mbox{const.}$, that is,
\begin{equation*}
-\rho q_0 + \frac{1}{2\mu}\left( \derpar{W}{q_1} \right)^2 + q_1\derpar{W}{q_0} = \mbox{const.} \, ,
\end{equation*}
which is the Hamilton-Jacobi equation for this problem.

\begin{remark}
Observe that, in this particular example, the Hamilton-Jacobi equation is clearly more difficult to
solve than the Euler-Lagrange equation. Therefore, this example shows that it is important to be careful
when applying the Hamilton-Jacobi theory to a system, since the Hamilton-Jacobi equations obtained can be
harder to solve than the usual Euler-Lagrange (or Hamilton's) equations of the system. Nevertheless,
observe that a solution of the system can be obtained from a solution $\gamma \colon \R \to Q$ of the
Euler-Lagrange equations as (see \cite{art:Vitagliano12})
\begin{equation*}
W(q_0,q_1) = \int_{t_0}^{t_1} \Lag (j^2_0\gamma(t)) \, \d t \, .
\end{equation*}
\end{remark}

\subsubsection{Hamiltonian formulation of the Hamilton-Jacobi problem}

Now, to establish the Hamiltonian formalism for the Hamilton-Jacobi problem, we consider natural
coordinates on $\Tan^*\Tan Q$. In these coordinates the Legendre-Ostrogradsky map
$\Leg \colon \Tan^3Q \to \Tan^*\Tan Q$ associated to the Lagrangian function $\Lag$ is locally given by
\begin{equation*}
\Leg^*q_0 = q_0 \quad ; \quad \Leg^*q_1 = q_1 \quad ; \quad
\Leg^*p^0 = -\mu q_3 \quad ; \quad \Leg^*p^1 = \mu q_2 \, .
\end{equation*}
Moreover, the inverse map $\Leg^{-1} \colon \Tan^*\Tan Q \to \Tan^3Q$ is
\begin{equation*}
(\Leg^{-1})^*q_0 = q_0 \quad ; \quad (\Leg^{-1})^*q_1 = q_1 \quad ; \quad
(\Leg^{-1})^*q_2 = \frac{p^1}{\mu} \quad ; \quad (\Leg^{-1})^*q_3 = -\frac{p^0}{\mu} \, .
\end{equation*}
From these coordinate expressions it is clear that $\Lag$ is a hyperregular Lagrangian function,
since the Legendre-Ostrogradsky map is a global diffeomorphism.

The Hamiltonian function $h \in \Cinfty(\Tan^*(\Tan Q))$ is
\begin{equation*}
h = (\Leg^{-1})^*E_\Lag = -\rho q_0 + \frac{(p^1)^2}{2\mu} + q_1p^0 \, .
\end{equation*}
Thus, the Hamiltonian vector field $X_h \in \vf(\Tan^*(\Tan Q))$ solution to equation
\eqref{Chap02_eqn:HamHODynEq} is
\begin{equation*}
X_h = q_1 \derpar{}{q_0} + \frac{p^1}{\mu} \derpar{}{q_1} + \rho\derpar{}{p^0} - p^0 \derpar{}{p^1} \, .
\end{equation*}

Consider the projection $\pi_{\Tan Q} \colon \Tan^*(\Tan Q) \to \Tan Q$. From Proposition
\ref{Chap04_prop:GenHamHJEquiv} we know that the generalized second-order Hamiltonian Hamilton-Jacobi
problem consists in finding $1$-forms $\alpha \in \df^{1}(\Tan Q)$ such that the Hamiltonian vector
field $X_h$ is tangent to the submanifold $\Im(\alpha) \hookrightarrow \Tan^*(\Tan Q)$. Suppose that
the $1$-form $\alpha$ is given locally by $\alpha = \alpha^0 \d q_0 + \alpha^1 \d q_1$.
As the submanifold $\Im(\alpha)$ is defined locally by the constraint functions $p^0 - \alpha^0$ and
$p^1 - \alpha^1$, then the tangency condition gives the following system of $2$ partial differential
equations for the component functions of the form
\begin{equation*}
\rho - q_1\derpar{\alpha^0}{q_0} - \frac{\alpha^1}{\mu}\derpar{\alpha^0}{q_1} = 0 \quad ; \quad
-\alpha^0 - q_1\derpar{\alpha^1}{q_0} - \frac{\alpha^1}{\mu}\derpar{\alpha^1}{q_1} = 0 \, .
\end{equation*}

In order to obtain the equations of the second-order Hamiltonian Hamilton-Jacobi problem, we require
in addition the $1$-form $\alpha \in \df^{1}(\Tan Q)$ to be closed. In coordinates, this condition is
\begin{equation*}
\derpar{\alpha^1}{q_0} - \derpar{\alpha^0}{q_1} = 0 \, .
\end{equation*}
Hence, the $1$-form $\alpha \in \df^{1}(\Tan Q)$ is a solution to the second-order Hamiltonian
Hamilton-Jacobi problem if the following system of $3$ partial differential equations holds
\begin{equation*}
\rho - q_1\derpar{\alpha^0}{q_0} - \frac{\alpha^1}{\mu}\derpar{\alpha^0}{q_1} = 0 \quad ; \quad
-\alpha^0 - q_1\derpar{\alpha^1}{q_0} - \frac{\alpha^1}{\mu}\derpar{\alpha^1}{q_1} = 0 \quad ; \quad
\derpar{\alpha^1}{q_0} - \derpar{\alpha^0}{q_1} = 0 \, .
\end{equation*}

Finally, we compute the equations for the generating function $W$. Requiring $\alpha = \d W$
for a local function $W$ defined in $\Tan Q$, we obtain
\begin{equation*}
\alpha^0 = \derpar{W}{q_0} \quad ; \quad \alpha^1 = \derpar{W}{q_1} \, ,
\end{equation*}
and thus from $\d(\alpha^*h) = 0$, we have $\alpha^*h = \mbox{const.}$, that is,
\begin{equation*}
-\rho q_0 + \frac{1}{2\mu}\left( \derpar{W}{q_1} \right)^2 + q_1\derpar{W}{q_0} = \mbox{const.} \, ,
\end{equation*}
which coincides with the Hamilton-Jacobi equation given previously in the Lagrangian
problem.

\subsubsection{Lagrangian-Hamiltonian formulation of the Hamilton-Jacobi problem}

In the induced natural coordinates $(q_0,q_1,q_2,q_3,p^0,p^1)$ of $\W,$ the coordinate expressions
of the presymplectic form $\Omega = \rho_2^*\omega_1 \in \df^{2}(\W)$ and the Hamiltonian function
$H = \C - \Lag \in \Cinfty(\W)$ are
\begin{equation*}
\Omega = \d q_0 \wedge \d p^0 + \d q_1 \wedge \d p^1 \quad ; \quad
H = q_1p^0 + q_2p^1 - \frac{1}{2}\mu q_2^2 - \rho q_0 \, .
\end{equation*}
Thus, the semispray of type $1$ $X_{LH} \in \vf(\W)$ solution to the dynamical equation
\eqref{Chap03_eqn:UnifDynEqVF} and tangent to the submanifold $\W_\Lag = \graph(\Leg) \hookrightarrow \W$
has the following coordinate expression
\begin{equation*}
X_{LH} = q_1\derpar{}{q_0} + q_2\derpar{}{q_1} + q_3\derpar{}{q_2}
- \frac{\rho}{\mu}\derpar{}{q_3} + \rho\derpar{}{p^0} - p^0 \derpar{}{p^1} \, .
\end{equation*}

In the following we state the equations for the (generalized) Lagrangian-Hamilonian Hamilton-Jacobi
problem for this dynamical system.

In the generalized second-order Lagrangian-Hamiltonian Hamilton-Jacobi problem we look for sections
$s \in \Gamma(\rho_{\Tan Q})$, given locally by $s(q_0,q_1) = (q_0,q_1,s_2,s_3,\alpha^0,\alpha^1)$,
such that the submanifold $\Im(s) \hookrightarrow \W$ is invariant under the flow of $X_{LH} \in \vf(\W)$.
Since the constraints defining locally $\Im(s)$ are $q_2-s_2=0$, $q_3-s_3=0$, $p^0-\alpha^0=0$,
$p^1-\alpha^1=0$, then the $4$ equations for the section $s$ are
\begin{equation*}
\begin{array}{l}
\displaystyle s_3 - q_1\derpar{s_2}{q_0} - s_2\derpar{s_2}{q_1} = 0 \quad ; \quad
-\frac{\rho}{\mu} - q_1\derpar{s_3}{q_0} - s_2\derpar{s_3}{q_1} = 0 \, , \\[10pt]
\displaystyle \rho - q_1\derpar{\alpha^0}{q_0} - s_2\derpar{\alpha^0}{q_1} = 0 \quad ; \quad
-\alpha^0 - q_1\derpar{\alpha^1}{q_0} - s_2\derpar{\alpha^1}{q_1} = 0 \, .
\end{array}
\end{equation*}

For the unified second-order Hamilton-Jacobi problem, we require in addition the section
$s \in \Gamma(\rho_{\Tan Q})$ to satisfy $s^*\Omega = 0$ or, equivalently, the form
$\rho_2 \circ s \in \df^{1}(\Tan Q)$ to be closed. In coordinates, if
$s = (q_0,q_1,s_2,s_3,\alpha^0,\alpha^1)$, then the $1$-form $\rho_2 \circ s$ is given by
$\rho_2\circ s = \alpha^0 \d q_0 + \alpha^1 \d q_1$. Hence, a section $s \in \Gamma(\rho_{\Tan Q})$
solution to the unified Hamilton-Jacobi problem for this system must satisfy the following system of
$5$ partial differential equations
\begin{equation*}
\begin{array}{c}
\displaystyle s_3 - q_1\derpar{s_2}{q_0} - s_2\derpar{s_2}{q_1} = 0 \quad ; \quad
-\frac{\rho}{\mu} - q_1\derpar{s_3}{q_0} - s_2\derpar{s_3}{q_1} = 0 \quad ; \quad
\displaystyle \derpar{\alpha^1}{q_0} - \derpar{\alpha^0}{q_1} = 0 \, , \\[10pt]
\displaystyle \rho - q_1\derpar{\alpha^0}{q_0} - s_2\derpar{\alpha^0}{q_1} = 0 \quad ; \quad
-\alpha^0 - q_1\derpar{\alpha^1}{q_0} - s_2\derpar{\alpha^1}{q_1} = 0 \, .
\end{array}
\end{equation*}


\clearpage
\chapter{Higher-order non-autonomous dynamical systems}
\label{Chap:HONonAutonomousDynamicalSystems}


Our aim in this Chapter is to introduce the geometric formulation of higher-order
non-autonomous systems, thus generalizing the results of Section \ref{Chap02_sec:NonAutonomous}
to the higher-order case, and the results of Section \ref{Chap02_sec:AutonomousHigherOrder}
and Chapter \ref{Chap:HOAutonomousDynamicalSystems} to the non-autonomous setting.

Observe that, unlike the autonomous case, for higher-order non-autonomous systems we do not
have a complete description of the Lagrangian and Hamiltonian formalisms
(partial studies on this subject can be found in
\cite{art:Crasmareanu00,proc:DeLeon_Marrero92,art:deLeon_Martin94_2,proc:DeLeon_Rodrigues87,art:Krupkova96}).
Therefore, instead of describing the unified formalism starting from the Lagrangian and Hamiltonian formulations,
in this Chapter we proceed backwards: we first describe the Skinner-Rusk formalism for higher-order non-autonomous
systems, and, from this setting, we derive both the Lagrangian and Hamiltonian formalisms for this kind of systems.

Taking into account these comments, the structure of the Chapter is the following.
In Section \ref{Chap05_sec:UnifiedFormalism} we describe the Lagrangian-Hamiltonian formalism
for higher-order non-autonomous systems: phase space, canonical structures and dynamical equations.
Then we describe the Lagrangian and Hamiltonian formalisms in Sections \ref{Chap05_sec:UnifiedToLagrangian}
and \ref{Chap05_sec:UnifiedToHamiltonian}, respectively.
Finally, two physical examples are studied in Section \ref{Chap05_sec:Examples}: the shape of a
non-homogeneous deformed elastic cylindrical beam with fixed ends, and a second-order relativistic
particle subjected to a time-depending potential.

Along this Chapter, we consider a $k$th-order non-autonomous Lagrangian dynamical system
with $n$ degrees of freedom. As in the first-order setting described in Section
\ref{Chap02_sec:NonAutonomous}, the configuration space for this system is a bundle
$\pi \colon E \to \R$, with $\dim E = n+1$. The dynamical information is given in terms
of a Lagrangian density which, by analogy with Sections \ref{Chap02_sec:AutonomousHigherOrder} and
\ref{Chap02_sec:NonAutonomous} is a $\bar{\pi}^k$-semibasic $1$-form,
$\Lag \in \df^{1}(J^{k}\pi)$. As in the first-order case, we write $\Lag = L\cdot(\bar{\pi}^k)^*\eta$,
where $\eta \in \df^{1}(\R)$ is the canonical volume form in $\R$ and $L \in \Cinfty(J^k\pi)$
is the Lagrangian function associated to $\Lag$ and $\eta$.

\section{Lagrangian-Hamiltonian unified formalism}
\label{Chap05_sec:UnifiedFormalism}

\subsection{Geometrical setting}

\subsubsection{Unified phase space and bundle structures. Local coordinates}

According to Sections \ref{Chap02_sec:AutonomousHigherOrder} and \ref{Chap02_sec:NonAutonomous},
let us consider the following bundles
\begin{equation*}
\W = J^{2k-1}\pi \times_{J^{k-1}\pi} \Tan^*(J^{k-1}\pi) \quad ; \quad
\W_r = J^{2k-1}\pi \times_{J^{k-1}\pi} \, J^{k-1}\pi^* \, ,
\end{equation*}
where $\Tan^*(J^{k-1}\pi)$ and $J^{k-1}\pi^*$ are the $k$th-order extended and reduced dual
jet bundles defined in Section \ref{Chap01_sec:HOJetBundlesDualBundles}, respectively.
The bundles $\W$ and $\W_r$ are called the \textsl{$k$th-order extended jet-momentum bundle} and the
\textsl{$k$th-order restricted jet-momentum bundle}, respectively.

\begin{remark}
The reason for taking these bundles is that we want to describe the Lagrangian and Hamiltonian formalisms
from this unified framework, and as we see in Sections \ref{Chap05_sec:UnifiedToLagrangian} and
\ref{Chap05_sec:UnifiedToHamiltonian}, those formalisms take place in the bundles $J^{2k-1}\pi$ and $J^{k-1}\pi^*$,
respectively.
\end{remark}

The bundles $\W$ and $\W_r$ are endowed with the canonical projections
\begin{equation*}
\rho_1 \colon \W \to J^{2k-1}\pi \quad ; \quad
\rho_2 \colon \W \to \Tan^*(J^{k-1}\pi) \quad ; \quad
\rho_{J^{k-1}\pi} \colon \W \to J^{k-1}\pi \quad ; \quad
\rho_\R \colon \W \to \R \, ,
\end{equation*}
\begin{equation*}
\rho_1^r \colon \W_r \to J^{2k-1}\pi \quad ; \quad
\rho_2^r \colon \W_r \to J^{k-1}\pi^* \quad ; \quad
\rho_{J^{k-1}\pi}^r \colon \W_r \to J^{k-1}\pi \quad ; \quad
\rho_\R^r \colon \W_r \to \R \, .
\end{equation*}

In addition, the natural quotient map $\mu \colon \Tan^*(J^{k-1}\pi) \to J^{k-1}\pi^*$ induces
a surjective submersion $\mu_\W \colon \W \to \W_r$. Thus, we have the following commutative diagram
\begin{equation*}
\xymatrix{
\ & \ & \W \ar@/_1.3pc/[llddd]_{\rho_1} \ar[d]^-{\mu_\W} \ar@/^1.3pc/[rrdd]^{\rho_2} & \ & \ \\
\ & \ & \W_r \ar[lldd]_{\rho_1^r} \ar[rrdd]^{\rho_2^r} & \ & \ \\
\ & \ & \ & \ & \Tan^*(J^{k-1}\pi) \ar[d]^-{\mu} \ar[lldd]_{\pi_{J^{k-1}\pi}}|(.25){\hole} \\
J^{2k-1}\pi \ar[rrd]^{\pi^{2k-1}_{k-1}} & \ & \ & \ & J^{k-1}\pi^* \ar[dll]^{\pi_{J^{k-1}\pi}^r} \\
\ & \ & J^{k-1}\pi \ar[d]^{\bar{\pi}^{k-1}} & \ & \ \\
\ & \ & \R & \ & \
}
\end{equation*}
where $\pi_{J^{k-1}\pi} \colon \Tan^*(J^{k-1}\pi) \to J^{k-1}\pi$ is the canonical
submersion and $\pi_{J^{k-1}\pi}^r \colon J^{k-1}\pi^* \to J^{k-1}\pi$ is the map
satisfying $\pi_{J^{k-1}\pi} = \pi_{J^{k-1}\pi}^r \circ \mu$.

Local coordinates in $\W$ and $\W_r$ are constructed as follows. Let $t$ be the global
coordinate in $\R$ such that the canonical volume form $\eta \in \df^{1}(\R)$ is given locally by
$\eta = \d t$, and $(U;(t,q^A))$, $1 \leqslant A \leqslant n$, a local chart of coordinates in $E$
adapted to the bundle structure. Then, the induced natural coordinates in the suitable open sets of
$J^{2k-1}\pi$, $\Tan^*(J^{k-1}\pi)$ and $J^{k-1}\pi^*$ are $(t,q_i^A,q_j^A)$, $(t,q_i^A,p,p_A^i)$
and $(t,q_i^A,p_A^i)$, respectively, where $1 \leqslant A \leqslant n$, $0 \leqslant i \leqslant k-1$,
$k \leqslant j \leqslant 2k-1$. Therefore, the natural coordinates in $\W$ and $\W_r$ are
$(t,q_i^A,q_j^A,p,p_A^i)$ and $(t,q_i^A,q_j^A,p_A^i)$, respectively. Note that
$\dim\W = 3kn + 2$ and $\dim\W_r = \dim\W - 1 = 3kn + 1$.

In these coordinates, the above projections have the following coordinate expressions
\begin{equation*}
\rho_1(t,q_i^A,q_j^A,p,p^i_A) = (t,q_i^A,q_j^A) \ ; \ \rho_2(t,q_i^A,q_j^A,p,p^i_A) = (t,q_i^A,p,p^i_A) \ ; \
\rho_{J^{k-1}\pi}(t,q_i^A,q_j^A,p,p^i_A) = (t,q_i^A) \, ,
\end{equation*}
\begin{equation*}
\rho_1^r(t,q_i^A,q_j^A,p^i_A) = (t,q_i^A,q_j^A) \ ; \ \rho_2^r(t,q_i^A,q_j^A,p^i_A) = (t,q_i^A,p^i_A) \ ; \
\rho_{J^{k-1}\pi}^r(t,q_i^A,q_j^A,p^i_A) = (t,q_i^A) \, ,
\end{equation*}
\begin{equation*}
\rho_\R(t,q_i^A,q_j^A,p,p^i_A) = t \ ; \ \rho_\R^r(t,q_i^A,q_j^A,p^i_A) = t \, .
\end{equation*}

\subsubsection{Canonical geometric structures}

The extended jet-momentum bundle $\W$ is endowed with some canonical geometric structures,
which are the generalization to the higher-order setting of the canonical structures introduced
in Section \ref{Chap02_sec:NonAutonomousUnified}.

Let $\Theta_{k-1} \in \df^{1}(\Tan^*(J^{k-1}\pi))$ and $\Omega_{k-1} = -\d\Theta_{k-1} \in \df^{2}(\Tan^*(J^{k-1}\pi))$
be the canonical forms of the cotangent bundle. Then, we define the following forms in $\W$
\begin{equation*}
\Theta = \rho_2^*\Theta_{k-1} \in \df^{1}(\W) \quad ; \quad
\Omega = \rho_2^*\Omega_{k-1} = -\d\Theta \in \df^{2}(\W) \, .
\end{equation*}
It is clear from the definition that $\Omega$ is a closed $2$-form. Nevertheless, this form has not maximal
rank in $\W$. Indeed, let $X \in \vf^{V(\rho_{2})}(\W)$. Then we have
\begin{equation*}
\inn(X)\Omega = \inn(X)\rho_{2}^*\,\Omega_{k-1} = \rho_{2}^*(\inn(Y)\Omega_{k-1}) \, ,
\end{equation*}
where $Y \in \vf(\Tan^{*}(J^{k-1}\pi))$ is a vector field $\rho_{2}$-related with $X$.
However, since $X$ is vertical with respect to $\rho_{2}$, we have $Y = 0$, and therefore
\begin{equation*}
\rho_{2}^{*}(\inn(Y)\Omega_{k-1}) = \rho_{2}^{*}(\inn(0)\Omega_{k-1}) = 0 \, .
\end{equation*}
In particular, $\{ 0 \} \varsubsetneq \vf^{V(\rho_2)}(\W) \subseteq \ker\Omega$,
and thus $\Omega$ has not maximal rank.

Bearing in mind the coordinate expressions of the Liouville forms of the cotangent bundle
given in Example \ref{Chap01_exa:CotangentBundle}, which in this case are
\begin{equation*}
\Theta_{k-1} = p_A^i\d q_i^A + p\d t \quad ; \quad
\Omega_{k-1} = \d q_i^A \wedge \d p_A^i - \d p \wedge \d t \, ,
\end{equation*}
and the local expression of the projection $\rho_2$ given above,
the forms $\Theta$ and $\Omega$ are given locally by
\begin{equation}\label{Chap05_eqn:UnifCanonicalFormsLocal}
\Theta = \rho_2^*(p_A^i\d q_i^A + p\d t) = p_A^i\d q_i^A + p\d t \quad ; \quad
\Omega = \rho_2^*(\d q_i^A \wedge \d p^i_A - \d p \wedge \d t) = \d q_i^A \wedge \d p^i_A - \d p \wedge \d t \, .
\end{equation}
It is clear from these coordinate expressions that $\Omega$ is closed.
Moreover, a local basis for $\ker\Omega$ is
\begin{equation}\label{Chap05_eqn:UnifCanonicalFormsKernelLocal}
\ker\Omega = \left\langle \derpar{}{q_k^A},\ldots,\derpar{}{q_{2k-1}^A} \right\rangle = \vf^{V(\rho_2)}(\W) \, .
\end{equation}
Thus, the pair $(\Omega,\rho_\R^*\eta)$ is a precosymplectic structure in $\W$.

The second canonical structure in $\W$ is the following.

\begin{definition}
The \textnormal{$k$th-order coupling $1$-form} in $\W$ is the $\rho_\R$-semibasic $1$-form
$\hat{\C} \in \df^{1}(\W)$ defined as follows: for every $w = (\bar{u},\alpha_q) \in \W$
(that is, $\bar{u} = j^{2k-1}_t\phi \in J^{2k-1}\pi$ and $\alpha_q \in \Tan^*_q (J^{k-1}\pi)$,
where $q = \pi^{2k-1}_{k-1}(\bar{u})$ is the projection of $\bar{u}$ to $J^{k-1}\pi$) and $v \in \Tan_w\W$, then
\begin{equation}\label{Chap05_eqn:UnifCouplingFormDef}
\langle \hat{\C}(w) , v \rangle = \langle \alpha_q , (\Tan_w(j^{k-1}\phi \circ \rho_\R))(v) \rangle \, .
\end{equation}
\end{definition}

$\hat{\C}$ being a $\rho_\R$-semibasic form, there exists a function $\hat{C} \in \Cinfty(\W)$
such that $\hat{\C} = \hat{C}\rho_\R^*\eta = \hat{C}\d t$. An easy computation in coordinates
gives the following local expression for the coupling $1$-form
\begin{equation}\label{Chap05_eqn:UnifCouplingFormLocal}
\hat{\C} = \left( p + p_A^iq_{i+1}^A \right) \d t \, .
\end{equation}

Let us denote $\hat{\Lag} = (\pi^{2k-1}_{k} \circ \rho_1)^*\Lag \in \df^{1}(\W)$. Since the Lagrangian
density is a $\bar{\pi}^{k}$-semibasic form, we have that $\hat{\Lag}$ is a $\rho_\R$-semibasic $1$-form,
and thus we can write $\hat{\Lag} = \hat{L}\rho_\R^*\eta = \hat{L}\d t$, where the function
$\hat{L} = (\pi^{2k-1}_{k} \circ \rho_1)^*L \in \Cinfty(\W)$ is the pull-back of the Lagrangian function
associated with $\Lag$ and $\eta$. Then, we define a \textsl{Hamiltonian submanifold}
\begin{equation*}
\W_o = \left\{ w \in \W \mid \hat{\Lag}(w) = \hat{\C}(w) \right\} \stackrel{j_o}{\hookrightarrow} \W \, .
\end{equation*}
Since both $\hat{\C}$ and $\hat{\Lag}$ are $\rho_\R$-semibasic $1$-forms, the submanifold
$\W_o$ is defined by the regular constraint $\hat{C} - \hat{L} = 0$. In the natural
coordinates of $\W$, bearing in mind the local expression \eqref{Chap05_eqn:UnifCouplingFormLocal}
of the coupling form, the constraint function is locally given by
\begin{equation}\label{Chap05_eqn:UnifConstraintFunctionW0}
\hat{C} - \hat{L} = p + p_A^iq_{i+1}^A - \hat{L} = 0 \, .
\end{equation}

\begin{proposition}\label{Chap05_prop:UnifW0DiffeomorphicWr}
The submanifold $\W_o \hookrightarrow \W$ is $1$-codimensional, $\mu_\W$-transverse
and diffeomorphic to $\W_r$. This diffeomorphism is given by the map $\mu_\W \circ j_o \colon \W_o \to \W_r$.
\end{proposition}
\begin{proof}
$\W_o$ is obviously $1$-codimensional, since it is defined by a single constraint function.

To see that $\W_o$ is diffeomorphic to $\W_r$, we show that the smooth map
$\mu_\W \circ j_o \colon \W_o \to \W_r$ is one-to-one. First, observe that for every
$(\bar{u},\alpha) \in \W_o$, we have that
$L(\pi^{2k-1}_{k}(\bar{u})) = \hat{L}(\bar{u},\alpha) = \hat{C}(\bar{u},\alpha)$,
and, on the other hand, that
\begin{equation}\label{Chap05_eqn:UnifW0DiffeomorphicWrProof1}
(\mu_\W \circ j_o)(\bar{u},\alpha) = \mu_\W(\bar{u},\alpha) = (\bar{u},\mu(\alpha)) = (\bar{u},[\alpha]) \, .
\end{equation}

Now, we first prove that $\mu_\W \circ j_o$ is injective. That is, if
$(\bar{u}_1,\alpha_1), (\bar{u}_2,\alpha_2) \in \W_o$ are two arbitrary points in $\W_o$,
then we want to prove that
\begin{equation*}
(\mu_\W \circ j_o)(\bar{u}_1,\alpha_1) = (\mu_\W \circ j_o)(\bar{u}_2,\alpha_2) \Longleftrightarrow
(\bar{u}_1,\alpha_1) = (\bar{u}_2,\alpha_2) \Longleftrightarrow
\bar{u}_1 = \bar{u}_2 \mbox{ and } \alpha_1 = \alpha_2 \, .
\end{equation*}
Using the expression \eqref{Chap05_eqn:UnifW0DiffeomorphicWrProof1} for $(\mu_\W \circ j_o)(\bar{u},\alpha)$, we have
\begin{equation*}
(\mu_\W \circ j_o)(\bar{u}_1,\alpha_1) = (\mu_\W \circ j_o)(\bar{u}_2,\alpha_2) \Longleftrightarrow
(\bar{u}_1,[\alpha_1]) = (\bar{u}_2,[\alpha_2]) \Longleftrightarrow
\bar{u}_1 = \bar{u}_2 \mbox{ and } [\alpha_1] = [\alpha_2] \, .
\end{equation*}
Hence, by definition of $\W_o$, we have $L(\pi^{2k-1}_{k}(\bar{u}_1)) = L(\pi^{2k-1}_{k}(\bar{u}_2))
= \hat{C}(\bar{u}_1,\alpha_1) = \hat{C}(\bar{u}_2,\alpha_2)$. Locally, from
the third equality we obtain
\begin{equation*}
p(\alpha_1) + p_A^i(\alpha_1)q_{i+1}^A(\bar{u}_1) = p(\alpha_2) + p_A^i(\alpha_2)q_{i+1}^A(\bar{u}_2) \, ,
\end{equation*}
but $[\alpha_1] = [\alpha_2]\, \Longrightarrow\,p^i_A(\alpha_1) = p_A^i([\alpha_1]) = p_A^i([\alpha_2]) = p^i_A(\alpha_2)$.
Then $p(\alpha_1) = p(\alpha_2)$, and $\alpha_1 = \alpha_2$; that is, the map $\mu_\W \circ j_o$ is injective.

Now, let us prove that $\mu_\W \circ j_o$ is surjective. That is, if
$(\bar{u},[\alpha]) \in \W_r$, we want to find $(\bar{u},\beta) \in j_o(\W_o)$
such that $[\beta] = [\alpha]$. It suffices to take $[\beta]$ such that,
in local coordinates of $\W$, it satisfies
\begin{equation*}
p_A^i(\beta) = p_A^i([\beta]) \quad , \quad
p(\beta) = L(\pi^{2k-1}_{k}(\bar{u})) - p_A^i([\alpha])q_{i+1}^A(\bar{u}) \, .
\end{equation*}
This $\beta$ exists as a consequence of the definition of $\W_o$, and therefore
the map $\mu_\W \circ j_o$ is surjective.

Hence, since $\mu_\W \circ j_o$ is a one-to-one submersion, then, by equality
on the dimensions of $\W_o$ and $\W_r$, it is a one-to-one local diffeomorphism,
and thus a global diffeomorphism.

Finally, in order to prove that $\W_o$ is $\mu_\W$-transverse, it is necessary to check if
$\Lie(Y)(\xi) \equiv Y(\xi) \neq 0$, for every $Y \in \ker(\Tan\mu_\W)$ and every constraint
function $\xi$ defining $\W_o$.
Since $\W_o$ is defined by the constraint function $\hat{C} - \hat{L} = 0$ and
$\ker(\Tan\mu_\W) = \{\partial/\partial p \}$, we have
\begin{equation*}
\derpar{}{p}(\hat{C} - \hat{L}) = \derpar{}{p}(p + p_A^iq_{i+1}^A - \hat{L}) = 1 \, ,
\end{equation*}
and thus $\W_o$ is $\mu_\W$-transverse.
\end{proof}

As a consequence of Proposition \ref{Chap05_prop:UnifW0DiffeomorphicWr},
the submanifold $\W_o$ induces a section $\hat{h} \in \Gamma(\mu_\W)$ defined as
$\hat{h} = j_o \circ (\mu_\W \circ j_o)^{-1} \colon \W_r \to \W$. This section
is specified by giving the \textsl{local Hamiltonian function}
\begin{equation}\label{Chap05_eqn:UnifHamiltonianFunctionLocal}
\hat{H} = -\hat{L} + p_A^iq_{i+1}^A \, ,
\end{equation}
that is, $\hat{h}(t,q_i^A,q_j^A,p^i_A) = (t,q_i^A,q_j^A,-\hat{H},p^i_A)$. The section $\hat{h}$ is
called a \textsl{Hamiltonian section} of $\mu_\W$, or a \textsl{Hamiltonian $\mu_\W$-section}.

Using this Hamiltonian $\mu_\W$-section, we can define the forms
\begin{equation*}
\Theta_r = \hat{h}^*\Theta_\W \in \df^{1}(\W_r) \quad ; \quad
\Omega_r = \hat{h}^*\Omega_\W \in \df^{2}(\W_r) \, ,
\end{equation*}
with local expressions
\begin{equation}\label{Chap05_eqn:UnifPrecosymplecticFormsLocal}
\Theta_r = p^i_A\d q_i^A + (\hat{L} - p^i_Aq_{i+1}^A)\d t \quad ; \quad
\Omega_r = \d q_i^A \wedge \d p^i_A + \d(p^i_A q_{i+1}^A - \hat{L}) \wedge \d t \, .
\end{equation}
Then, the triple $(\W_r,\Omega_r,(\rho_\R^r)^*\eta)$ is a precosymplectic Hamiltonian system.

Finally, as in the autonomous setting, it is necessary to introduce the following concepts in order
to give a complete description of higher-order Lagrangian systems in terms of the unified formalism.

\begin{definition}
A section $\psi \in \Gamma(\rho_\R^r)$ is \textnormal{holonomic of type $s$} in $\W_r$, $1 \leqslant s \leqslant 2k-1$,
if the section $\psi_1 = \rho_1^r \circ \psi \in \Gamma(\bar{\pi}^{2k-1})$ is holonomic of type $s$ in $J^{2k-1}\pi$.
\end{definition}

\begin{definition}
A vector field $X \in \vf(\W_r)$ is said to be a \textnormal{holonomic of type $s$} in $\W_r$,
$1 \leqslant s \leqslant 2k-1$, if every integral section $\psi$ of $X$ is holonomic of type $s$ in $\W_r$.
\end{definition}

In the natural coordinates of $\W_r$, the local expression of a semispray of type $s$ in $\W_r$ is
\begin{equation*}
X = f\derpar{}{t} + \sum_{i=0}^{2k-1-s}q_{i+1}^A\derpar{}{q_i^A} + \sum_{i=2k-s}^{2k-1}X_i^A\derpar{}{q_i^A}
+\sum_{i=0}^{k-1}G^i_A\derpar{}{p^i_A} \, ,
\end{equation*}
and, in particular, for a semispray of type $1$ in $\W_r$ we have
\begin{equation*}
X = f\derpar{}{t} + \sum_{i=0}^{2k-2}q_{i+1}^A\derpar{}{q_i^A} + X_{2k-1}^A\derpar{}{q_{2k-1}^A}
+\sum_{i=0}^{k-1}G^i_A\derpar{}{p^i_A} \, .
\end{equation*}

\subsection{Dynamical equations}
\label{Chap05_sec:UnifDynamicalEquations}

In this Section we analyze the dynamical equations for a $k$th-order non-autonomous dynamical system
in the unified formalism. First, we state the variational principle from which the equations of
motion can be derived. Then, we state the geometric equation for a $k$th-order non-autonomous
dynamical system in two different ways: in terms of sections and vector fields. Finally, we
prove that the variational principle is, in fact, equivalent to these geometric equations.

\subsubsection{Variational principle}

Let $\Gamma(\rho_\R^r)$ be the set of sections of $\rho_\R^r$, and consider the functional
\begin{equation*}\label{Chap05_eqn:UnifVariationalFunctionalDef}
\begin{array}{rcl}
\mathbf{LH} \colon \Gamma(\rho_\R^r) & \longrightarrow & \R \\
\psi & \longmapsto & \displaystyle \int_\R \psi^*\Theta_r
\end{array} \, ,
\end{equation*}
where the convergence of the integral is assumed.

\begin{definition} 
The \textnormal{$k$th-order Lagrangian-Hamiltonian variational problem} associated to the system
$(\W_r,\Omega_r,(\rho_\R^r)^*\eta)$ is the search for the critical (or stationary) holonomic
sections of the functional $\mathbf{LH}$ with respect to the variations of $\psi$ given by
$\psi_s = \sigma_s \circ \psi$, where $\left\{ \sigma_s \right\}$ is a local one-parameter group
of any compact-supported $\rho_\R^r$-vertical vector field $Z$ in $\W_r$, that is,
\begin{equation*}\label{Chap05_eqn:UnifDynEqVar}
\restric{\frac{d}{ds}}{s=0}\int_\R \psi_s^*\Theta_r = 0 \, .
\end{equation*}
\end{definition}

In the following Subsections we analyze the geometric dynamical equations in terms of sections
and vector fields. Then, we prove in Theorem \ref{Chap05_thm:UnifEquivalenceTheorem} that the critical
sections of the variational problem stated above are exactly the sections solution to the geometric
equations analyzed in the following.

\subsubsection{Dynamical equation for sections}

The \textsl{$k$th-order Lagrangian-Hamiltonian problem for sections} associated with the system
$(\W_r,\Omega_r,(\rho_\R^r)^*\eta)$ consists in finding holonomic sections $\psi \in \Gamma(\rho_\R^r)$
satisfying
\begin{equation}\label{Chap05_eqn:UnifDynEqSect}
\psi^*\inn(Y)\Omega_r = 0 \, , \quad \mbox{for every } Y \in \vf(\W_r) \, .
\end{equation}
In the natural coordinates of $\W_r$, let $Y \in \vf(\W_r)$ be a generic vector field given by
\begin{equation}\label{Chap05_eqn:UnifGenericVectorField}
Y = f \derpar{}{t} + f_i^A\derpar{}{q_i^A} + F_j^A\derpar{}{q_j^A} + G_A^i\derpar{}{p_A^i}\, .
\end{equation}
Then, bearing in mind the coordinate expression \eqref{Chap05_eqn:UnifPrecosymplecticFormsLocal} of $\Omega_r$,
the contraction $\inn(Y)\Omega_r$ gives the following $1$-form on $\W_r$
\begin{align*}
\inn(Y)\Omega_r
&= f\left( \derpar{\hat{L}}{q_r^A} \d q_r^A - q_{i+1}^A \d p_A^i - p_A^i \d q_{i+1}^A \right)
+ f_0^A\left( \d p_A^0 - \derpar{\hat{L}}{q_0^A} \d t \right)  \\
&\qquad{} + f_i^A\left( \d p_A^i - \derpar{\hat{L}}{q_i^A}\d t + p_A^{i-1} \d t \right)
+ F_k^A\left(p_A^{k-1} - \derpar{\hat{L}}{q_k^A}\right)\d t + G_A^i\left( q_{i+1}^A \d t - \d q_i^A \right) \, .
\end{align*}
Thus, taking the pull-back by the section $\psi = (t,q_i^A(t),q_j^A(t),p_A^i(t))$, we obtain
\begin{align*}
\psi^*\inn(Y)\Omega_r
&= f\left( \derpar{\hat{L}}{q_r^A} \dot{q}_r^A - q_{i+1}^A \dot{p}_A^i - p_A^i \dot{q}_{i+1}^A \right)\d t
+ f_0^A\left( \dot{p}_A^0 - \derpar{\hat{L}}{q_0^A} \right)\d t \\
&\qquad + f_i^A\left( \dot{p}_A^i - \derpar{\hat{L}}{q_i^A} + p_A^{i-1} \right)\d t
+ F_k^A\left(p_A^{k-1} - \derpar{\hat{L}}{q_k^A}\right)\d t + G_A^i\left( q_{i+1}^A - \dot{q}_i^A \right)\d t \, .
\end{align*}
Finally, requiring this last expression to vanish, and taking into account that the equation must hold for every
vector field $Y \in \vf(\W_r)$ (that is, it must hold for every function $f,f_i^A,F_j^A,G_A^i \in \Cinfty(\W_r)$)
we obtain the following system of $(2k+1)n+1$ equations
\begin{align}
\derpar{\hat{L}}{q_r^A} \dot{q}_r^A - q_{i+1}^A \dot{p}_A^i - p_A^i \dot{q}_{i+1}^A = 0 \, ,
\label{Chap05_eqn:UnifDynEqSectLocalRedundantEq} \\
\dot{q}_i^A = q_{i+1}^A \, , \label{Chap05_eqn:UnifDynEqSectHolonomyLocalPart} \\
\dot{p}_A^0 = \derpar{\hat{L}}{q_0^A} \, , \label{Chap05_eqn:UnifDynEqSectLocal1} \\
\dot{p}_A^i = \derpar{\hat{L}}{q_i^A} - p_A^{i-1} \, ,\label{Chap05_eqn:UnifDynEqSectLocal2} \\
p_A^{k-1} = \derpar{\hat{L}}{q_k^A} \, . \label{Chap05_eqn:UnifDynEqSectLegendreLocal}
\end{align}
It is easy to check that equation \eqref{Chap05_eqn:UnifDynEqSectLocalRedundantEq} is redundant,
since it is a consequence of the others. Equations \eqref{Chap05_eqn:UnifDynEqSectHolonomyLocalPart},
\eqref{Chap05_eqn:UnifDynEqSectLocal1} and \eqref{Chap05_eqn:UnifDynEqSectLocal2} are differential equations
whose solutions are the functions defining the section $\psi$. In particular, equations
\eqref{Chap05_eqn:UnifDynEqSectHolonomyLocalPart} are part of the system of differential equations that the section
$\psi$ must satisfy to be holonomic, and are automatically satisfied since we assumed the section $\psi$ to be
holonomic from the beginning. On the other hand, equations \eqref{Chap05_eqn:UnifDynEqSectLocal1} and
\eqref{Chap05_eqn:UnifDynEqSectLocal2} are the dynamical equations of the system. Finally, equations
\eqref{Chap05_eqn:UnifDynEqSectLegendreLocal} do not involve any derivative of $\psi$: they are
pointwise algebraic conditions. These equations arise from the $\rho_2^r$-vertical part of the
vector fields $Y$. Moreover, we have the following result.

\begin{lemma}\label{Chap05_lemma:UnifTechLemma1}
If $Y \in \vf^{V(\rho_2^r)}(\W_r)$, then $\inn(Y)\Omega_r \in \df^{1}(\W_r)$ is $\rho_\R^r$-semibasic.
\end{lemma}
\begin{proof}
A direct calculation in coordinates leads to this result. Bearing in mind that a local basis for
the $\rho_2^r$-vertical vector fields is given by \eqref{Chap05_eqn:UnifCanonicalFormsKernelLocal},
and the local expression \eqref{Chap05_eqn:UnifPrecosymplecticFormsLocal} of $\Omega_r$, we have
\begin{equation*}
\inn\left( \derpar{}{q_j^A} \right) \Omega_r =
\begin{cases}
\displaystyle \left(p_A^{k-1} - \derpar{\hat{L}}{q_k^A}\right)\d t \, , & \mbox{for } j = k \, , \\[10pt]
0 = 0 \cdot \d t \, , & \mbox{for } j = k+1,\ldots,2k-1.
\end{cases}
\end{equation*}
Thus, in both cases we obtain a $\rho_\R^r$-semibasic form.
\end{proof}

As a consequence of Lemma \ref{Chap05_lemma:UnifTechLemma1}, we can define the submanifold
\begin{equation}\label{Chap05_eqn:UnifFirstConstraintSubmanifoldDefSect}
\W_c = \left\{ [w] \in \W_r \mid (\inn(Y)\Omega_r)([w])=0 \mbox{ for every } Y \in \vf^{V(\rho_2^r)}(\W_r)\right\}
\stackrel{j_c}{\hookrightarrow} \W_r \, ,
\end{equation}
where every section $\psi$ solution to equation \eqref{Chap05_eqn:UnifDynEqSect} must take values.
It is called the \textsl{first constraint submanifold} of the Hamiltonian precosymplectic system
$(\W_r,\Omega_r,(\rho_\R^r)^*\eta)$.

Locally, $\W_c$ is defined in $\W_r$ by the constraints $p_A^{k-1} - \partial \hat{L} / \partial q_k^A = 0$,
as we have seen in \eqref{Chap05_eqn:UnifDynEqSectLegendreLocal} and in the proof of Lemma
\ref{Chap05_lemma:UnifTechLemma1}. In combination with equations \eqref{Chap05_eqn:UnifDynEqSectLocal2},
we have the following result, which is the analogous to Proposition
\ref{Chap03_prop:UnifGraphLegendreOstrogradskyMap} in the non-autonomous setting.

\begin{proposition}\label{Chap05_prop:UnifGraphLegendreOstrogradskyMap}
The submanifold $\W_c \hookrightarrow \W$ contains a submanifold $\W_\Lag \hookrightarrow \W_c$
which can be identified with the graph of a map $\Leg \colon J^{2k-1}\pi \to J^{k-1}\pi^*$ defined locally by
\begin{equation*}
\Leg^*t = t \quad ; \quad \Leg^*q_r^A = q_r^A \quad ; \quad
\Leg^*p_A^{r-1} = \sum_{i=0}^{k-r}(-1)^i \frac{d^i}{dt^{i}} \left( \derpar{\hat{L}}{q_{r+i}^A} \right) \, .
\end{equation*}
\end{proposition}
\begin{proof}
Since $\W_c$ is defined locally by the constraints \eqref{Chap05_eqn:UnifDynEqSectLegendreLocal},
it suffices to prove that these constraints, in combination with equations \eqref{Chap05_eqn:UnifDynEqSectLocal2}
give rise to the functions defining the map given in the statement, and thus to the submanifold $\W_\Lag$.
We proceed in coordinates.

The constraint functions defining $\W_c$, in combination with equations
\eqref{Chap05_eqn:UnifDynEqSectLocal2}, give rise to the following $n$ new constraint functions
\begin{equation*}
p^{k-2}_A - \left( \derpar{\hat{L}}{q_{k-1}^A} - \frac{d}{dt}\,p_A^{k-1} \right) =
p_A^{k-2} - \sum_{i=0}^{1}(-1)^i \frac{d^i}{dt^{i}}\left(\derpar{\hat{L}}{q_{k-1+i}^A}\right) = 0 \, ,
\end{equation*}
which define a new submanifold of $\W_c$. Combining these constraint functions again
with equations \eqref{Chap05_eqn:UnifDynEqSectLocal2}, we obtain the following
$n$ additional constraints
\begin{equation*}
p_A^{k-3} - \left( \derpar{\hat{L}}{q_{k-2}^A} - \frac{d}{dt}\,p_A^{k-2} \right) =
p_A^{k-3} - \sum_{i=0}^{2}(-1)^i \frac{d^i}{dt^{i}}\left(\derpar{\hat{L}}{q_{k-2+i}^A}\right) = 0 \, .
\end{equation*}
Iterating this process $k-3$ more times, we obtain a $kn$-codimensional submanifold
$\W_\Lag \hookrightarrow \W_r$ defined locally by the following constraints
\begin{equation*}
p_A^{r-1} = \sum_{i=0}^{k-r}(-1)^i \frac{d^i}{dt^{i}} \left( \derpar{\hat{L}}{q_{r+i}^A} \right) \, ,
\end{equation*}
with $1 \leqslant A \leqslant n$ and $1 \leqslant r \leqslant k$. Therefore, we may consider that
$\W_\Lag$ is the graph of a bundle morphism $\Leg \colon J^{2k-1}\pi \to J^{k-1}\pi^*$ over $J^{k-1}\pi$
locally given by
\begin{equation*}
\Leg^*t = t \quad ; \quad \Leg^*q_r^A = q_r^A \quad ; \quad
\Leg^*p_A^{r-1} = \sum_{i=0}^{k-r}(-1)^i \frac{d^i}{dt^{i}}\left( \derpar{\hat{L}}{q_{r+i}^A} \right) \, .
\qedhere
\end{equation*}
\end{proof}

Bearing in mind that the submanifold $\W_o \hookrightarrow \W$ is defined locally by the constraints
\eqref{Chap05_eqn:UnifConstraintFunctionW0}, that $\W_r$ is diffeomorphic to $\W_o$, and that $\W_\Lag$ is a
submanifold of $\W_c$, and thus a sumbanifold of $\W_r$, from the above Proposition we can state the following
result, which is a straightforward consequence of Proposition \ref{Chap05_prop:UnifGraphLegendreOstrogradskyMap}.

\begin{corollary}\label{Chap05_corol:UnifGraphExtendedLegendreOstrogradskyMapSect}
The submanifold $\W_\Lag \hookrightarrow \W$ is the graph of a bundle morphism
$\widetilde{\Leg} \colon J^{2k-1}\pi \to \Tan^*(J^{k-1}\pi)$ over $J^{k-1}\pi$ defined locally by
\begin{equation*}
\widetilde{\Leg}^*t = t \quad ; \quad
\widetilde{\Leg}^*q_r^A = q_r^A \, ,
\end{equation*}
\begin{equation*}
\widetilde{\Leg}^*p = \hat{L} - \sum_{r=1}^kq_r^A\sum_{i=0}^{k-r}(-1)^i\frac{d^i}{dt^{i}}\left( \derpar{\hat{L}}{q_{r+i}^A} \right) \quad ; \quad
\widetilde{\Leg}^*p_A^{r-1} = \sum_{i=0}^{k-r}(-1)^i \frac{d^i}{dt^{i}}\left( \derpar{\hat{L}}{q_{r+i}^A} \right) \, .
\end{equation*}
\end{corollary}

\begin{remark}
As in the autonomous setting described in Chapter \ref{Chap:HOAutonomousDynamicalSystems}, the
submanifold $\W_\Lag$ can be obtained from $\W_c$ using a constraint algorithm.
Hence, $\W_\Lag$ acts as the initial phase space of the system, as we see in the following,
and in next Section.
\end{remark}

\begin{definition}
The maps $\widetilde{\Leg} \colon J^{2k-1}\pi \to \Tan^*(J^{k-1}\pi)$ and $\Leg \colon J^{2k-1}\pi \to J^{k-1}\pi^*$
given by the above results are called the \textnormal{extended Legendre-Ostrogradsky map} and the
\textnormal{restricted Legendre-Ostrogradsky map} associated to the $k$th-order Lagrangian density $\Lag$, respectively.
\end{definition}

A justification of this terminology will be given in Section \ref{Chap05_sec:UnifiedToHamiltonian}.
An important result concerning both Legendre-Ostrogradsky maps is the following.

\begin{proposition}\label{Chap05_prop:UnifRankBothLegendreMaps}
For every $j^{2k-1}_t\phi \in J^{2k-1}\pi$ we have that
$\rank(\widetilde{\Leg}(j^{2k-1}_t\phi)) = \rank(\Leg(j^{2k-1}_t\phi))$.
\end{proposition}

We do not prove this result. Following the patterns in \cite{art:deLeon_Marin_Marrero96},
the idea is to compute in natural coordinates the local expressions of the Jacobian matrices
of $\Leg$ and $\widetilde{\Leg}$. Then, observe that the ranks of both maps depend on the rank of the
Hessian matrix of $L$ with respect to $q_k^A$ at the point $j^{2k-1}_t\phi$, and that the additional row in the
Jacobian matrix of $\widetilde{\Leg}$ is a linear combination of the others. See \cite{art:deLeon_Marin_Marrero96}
for details in the first-order case.

Now we can give the following definition.

\begin{definition}\label{Chap05_def:UnifRegularLagrangian}
A $k$th-order Lagrangian density $\Lag \in \df^{1}(J^{k}\pi)$ is \textnormal{regular} if the restricted Legendre-Ostrogradsky
map $\Leg$ is a local diffeomorphism. If the map $\Leg$ is a global diffeomorphism, then $\Lag$ is said to be
\textnormal{hyperregular}. Otherwise, $\Lag$ is said to be \textnormal{singular}.
\end{definition}

As a consequence of Proposition \ref{Chap05_prop:UnifRankBothLegendreMaps}, a $k$th-order Lagrangian density
$\Lag \in \df^{1}(J^{k}\pi)$ is regular if, and only if, the extended Legendre-Ostrogradsky $\widetilde{\Leg}$
is an immersion on $\Tan^*(J^{k-1}\pi)$. Moreover, computing in natural coordinates the local expression of the
tangent map to $\Leg$, the regularity condition for $\Lag$ is equivalent to
\begin{equation*}
\det\left( \derpars{L}{q_k^B}{q_k^A} \right)(j^{k}_t\phi) \neq 0 \, , \quad \mbox{for every } j^{k}_t\phi \in J^{k}\pi \, .
\end{equation*}

Equivalently, if we denote $\hat{p}^{r-1}_A = \Leg^*p_A^{r-1}$, then the Lagrangian density
$\Lag$ is regular if, and only if, the set $(t,q_i^A,\hat{p}_A^i)$, $0 \leqslant i \leqslant k-1$,
is a set of local coordinates in $J^{2k-1}\pi$. As in the Hamiltonian formalism for higher-order
autonomous systems described in Section \ref{Chap02_sec:HamiltonianAutonomousHigherOrder}, the local
functions $\hat{p}_A^i$ are called the \textsl{Jacobi-Ostrogradsky momentum coordinates}, and they satisfy
the relation
\begin{equation}\label{Chap05_eqn:UnifMomentumCoordRelation}
\hat{p}_A^{r-1} = \derpar{L}{q_r^A} - \frac{d}{dt}\,\hat{p}_A^r \quad ,
\end{equation}
which are exactly the relations given by \eqref{Chap05_eqn:UnifDynEqSectLocal2}.

Finally, observe that since the section $\psi \in \Gamma(\rho_\R^r)$ must take values in the submanifold
$\W_\Lag \hookrightarrow \W_r$, it is natural to consider the restriction of equation \eqref{Chap05_eqn:UnifDynEqSect}
to the submanifold $\W_\Lag$; that is, to restrict the set of vector fields to those tangent to $\W_\Lag$. Nevertheless,
the new equation may not be equivalent to the former. The following result gives a sufficient condition for these
two equations to be equivalent.

\begin{proposition}\label{Chap05_prop:UnifDynEqSectTangent}
If $\psi \in \Gamma(\rho_M^r)$ is holonomic in $\W_r$, then the equation \eqref{Chap05_eqn:UnifDynEqSect} is equivalent to
\begin{equation}\label{Chap05_eqn:UnifDynEqSectTangent}
\psi^*\inn(Y)\Omega_r = 0 \, , \quad \mbox{for every } Y \in \vf(\W_r) \mbox{ tangent to } \W_\Lag \, .
\end{equation}
\end{proposition}
\begin{proof}
We prove this result in coordinates. First of all, let us compute the coordinate expression of a vector
field $X \in \vf(\W_r)$ tangent to $\W_\Lag$. Let $X$ be a generic vector field locally given by
\eqref{Chap05_eqn:UnifGenericVectorField}, that is,
\begin{equation*}
X = f \derpar{}{t} + f_i^A\derpar{}{q_i^A} + F_j^A\derpar{}{q_j^A} + G_A^i\derpar{}{p_A^i}\, .
\end{equation*}
Thus, since $\W_\Lag$ is the submanifold of $\W_r$ defined locally by the $kn$ constraint functions
\begin{equation*}
\xi^{r-1}_A = p_A^{r-1} - \sum_{i=0}^{k-r}(-1)^i \frac{d^i}{dt^{i}} \left( \derpar{\hat{L}}{q_{r+i}^A} \right) \, ,
\end{equation*}
then the tangency condition of $X$ along $\W_\Lag$, which is $\Lie(X)(\xi^r_A) = 0$ (on $\W_\Lag$), gives
the following relations on the component functions of $X$
\begin{align*}
G_A^{k-1} &= f \derpars{\hat{L}}{t}{q_k^A} + f_i^B \derpars{\hat{L}}{q_i^B}{q_k^A}
+ F_k^B \derpars{\hat{L}}{q_k^B}{q_k^A} \, , \\
G_A^{k-2} &= f \left( \derpars{\hat{L}}{t}{q_{k-1}^A} - \frac{d}{dt} \, \derpars{\hat{L}}{t}{q_k^A} \right)
+ f_i^B \left( \derpars{\hat{L}}{q_i^B}{q_{k-1}^A} - \frac{d}{dt} \, \derpars{\hat{L}}{q_i^B}{q_k^A} \right)
+ F_k^B \left( \derpars{\hat{L}}{q_k^B}{q_{k-1}^A} - \frac{d}{dt} \, \derpars{\hat{L}}{q_k^B}{q_k^A} \right) \\
&\quad {} - \left( f_{i+1}^B \derpars{\hat{L}}{q_i^B}{q_k^A}
+ F_k^B \derpars{\hat{L}}{q_i^B}{q_k^A} + F_{k+1}^B \derpars{\hat{L}}{q_k^B}{q_k^A} \right) \, , \\
&\hspace{.1\textwidth} \vdots \hspace{.35\textwidth} \vdots \hspace{.35\textwidth} \vdots
\end{align*}
where the remaining calculations are omitted for simplicity.
That is, the tangency condition enables us to write the component functions $G_A^i$ as functions
$\widetilde{G}_A^i$ depending on the rest of the components $f,f_i^A,F_j^A$.

Now, from the previous calculations in this Section, recall that if $\psi(t) = (t,q_i^A(t),q_j^A(t),p_A^i(t))$,
then equation \eqref{Chap05_eqn:UnifDynEqSect} gives in coordinates
\begin{align*}
\psi^*\inn(X)\Omega_r
&= f\left( \derpar{\hat{L}}{q_r^A} \dot{q}_r^A - q_{i+1}^A \dot{p}_A^i - p_A^i \dot{q}_{i+1}^A \right)\d t
+ f_0^A\left( \dot{p}_A^0 - \derpar{\hat{L}}{q_0^A} \right)\d t \\
&\qquad + f_i^A\left( \dot{p}_A^i - \derpar{\hat{L}}{q_i^A} + p_A^{i-1} \right)\d t
+ F_k^A\left(p_A^{k-1} - \derpar{\hat{L}}{q_k^A}\right)\d t + G_A^i\left( q_{i+1}^A - \dot{q}_i^A \right)\d t \, .
\end{align*}
On the other hand, if we take a vector field $Y$ tangent to $\W_\Lag$, then we must replace the component functions
$G_A^i$ by $\widetilde{G}_A^i$ in the previous equation, thus obtaining
\begin{align*}
\psi^*\inn(Y)\Omega_r
&= f\left( \derpar{\hat{L}}{q_r^A} \dot{q}_r^A - q_{i+1}^A \dot{p}_A^i - p_A^i \dot{q}_{i+1}^A \right)\d t
+ f_0^A\left( \dot{p}_A^0 - \derpar{\hat{L}}{q_0^A} \right)\d t \\
&\qquad + f_i^A\left( \dot{p}_A^i - \derpar{\hat{L}}{q_i^A} + p_A^{i-1} \right)\d t
+ F_k^A\left(p_A^{k-1} - \derpar{\hat{L}}{q_k^A}\right)\d t + \widetilde{G}_A^i\left( q_{i+1}^A - \dot{q}_i^A \right)\d t \, .
\end{align*}
Finally, if $\psi$ is holonomic, then equations \eqref{Chap05_eqn:UnifDynEqSectHolonomyLocalPart} are satisfied,
and the last two terms of both $\inn(X)\Omega_r$ and $\inn(Y)\Omega_r$ vanish, thus obtaining
\begin{align*}
\psi^*\inn(X)\Omega_r
&= f\left( \derpar{\hat{L}}{q_r^A} \dot{q}_r^A - q_{i+1}^A \dot{p}_A^i - p_A^i \dot{q}_{i+1}^A \right)\d t
+ f_0^A\left( \dot{p}_A^0 - \derpar{\hat{L}}{q_0^A} \right)\d t \\
&\qquad + f_i^A\left( \dot{p}_A^i - \derpar{\hat{L}}{q_i^A} + p_A^{i-1} \right)\d t
+ F_k^A\left(p_A^{k-1} - \derpar{\hat{L}}{q_k^A}\right)\d t = \psi^*\inn(Y)\Omega_r \, .
\end{align*}
Hence, we have $\psi^*\inn(X)\Omega_r = 0$ if and only if $\psi^*\inn(Y)\Omega_r = 0$.
\end{proof}

\begin{remarks} \
\begin{itemize}
\item
Note that if the holonomy condition is not required in the statement of the problem, then the local
component functions $q_j^A(t)$, $k \leqslant j \leqslant 2k-1$, of the section $\psi$ remain undetermined, and
the $2kn$ equations \eqref{Chap05_eqn:UnifDynEqSectHolonomyLocalPart}, \eqref{Chap05_eqn:UnifDynEqSectLocal1}
and \eqref{Chap05_eqn:UnifDynEqSectLocal2} do not allow us to determinate them at first sight. Hence, the
holonomy condition can not be dropped from the statement of the problem, contrary to what happens
in the unified formalism for first-order non-autonomous systems (see Section
\ref{Chap02_sec:NonAutonomousUnified}).

In fact, the local functions $q_j^A(t)$ can be determined by the equations
\eqref{Chap05_eqn:UnifDynEqSectLocal1} and \eqref{Chap05_eqn:UnifDynEqSectLocal2}, bearing in mind
that the section $\psi$ must lie in the submanifold $\W_\Lag \hookrightarrow \W_r$.
It is easy to see that, by replacing the local expression of the restricted Legendre-Ostrogradsky map in the
equations \eqref{Chap05_eqn:UnifDynEqSectLocal1} and \eqref{Chap05_eqn:UnifDynEqSectLocal2}, these equations
lead to the Euler-Lagrange equations and to the remaining $(k-1)n$ equations that give the full holonomy condition
\begin{align}
\left(\dot{q}_j^B - q_{j+1}^B \right) \restric{\derpars{\hat{L}}{q_k^B}{q_k^A}}{\psi} - \sum_{i=k}^{j-1}\left(\dot{q}_i^B-q_{i+1}^B\right)(\cdots\cdots) = 0 \qquad (k \leqslant j \leqslant 2k-2) \, , \label{Chap05_eqn:UnifDynEqSectHolonomyLocalPart2} \\
\restric{\derpar{\hat{L}}{q_0^A}}{\psi} - \restric{\frac{d}{dt}\derpar{\hat{L}}{q_1^A}}{\psi}
+ \restric{\frac{d^2}{dt^2}\derpar{\hat{L}}{q_2^A}}{\psi} + \ldots +
(-1)^k \restric{\frac{d^k}{dt^k}\derpar{\hat{L}}{q_k^A}}{\psi} = 0 \, , \label{Chap05_eqn:UnifEulerLagrange}
\end{align}
where the terms in brackets $(\cdots)$ contain terms involving partial derivatives of the Lagrangian function
and iterated total time derivatives, and the first sum (for $j=k$) is empty. However, observe that these equations
may or may not be compatible, and a sufficient condition to ensure compatibility is the regularity of the Lagrangian
density. Thus, for singular Lagrangian densities, the holonomy condition for the section $\psi$ is required.\hfill$\lozenge$

\item The requirement of the section $\psi$ to be holonomic is a relevant difference from the first-order case
described in Section \ref{Chap02_sec:NonAutonomousUnified}, where the holonomy condition is deduced straightforwardly
from the dynamical equations when written in local coordinates.
Nevertheless, in the higher-order case, the equations allow us to recover only the holonomy of type $k$, as seen in
\eqref{Chap05_eqn:UnifDynEqSectHolonomyLocalPart} and in the autonomous case, and the highest-order holonomy
condition can only be recovered from the equations if the Lagrangian density is regular. Hence, this condition
is required ``ad hoc'' in the statement of the problem. \hfill$\lozenge$

\item The regularity of the Lagrangian density has no relevant role at first sight. However, as we have seen in the first remark,
equations \eqref{Chap05_eqn:UnifDynEqSectLocal1} and \eqref{Chap05_eqn:UnifDynEqSectLocal2} give the higher-order
Euler-Lagrange equations, which have a unique solution if the Lagrangian density is regular. For singular Lagrangians,
these equations may give rise to new constraints, and an adaptation of the constraint algorithm described in Section
\ref{Chap01_sec:ConstraintAlgorithm} should be used for finding a submanifold where the equations can be solved. \hfill$\lozenge$
\end{itemize}
\end{remarks}

\subsubsection{Dynamical equation for vector fields}

As in the first-order case described in Section \ref{Chap02_sec:NonAutonomousUnified}, if we assume that
the sections $\psi \in \Gamma(\rho_\R^r)$ solutions to equation \eqref{Chap05_eqn:UnifDynEqSect} are the
integral curves of some vector fields in $\W_r$, then we can state the problem in terms of vector fields.
The \textsl{$k$th-order Lagrangian-Hamiltonian problem for vector fields} associated with the system
$(\W_r,\Omega_r,(\rho_\R^r)^*\eta)$ consists in finding holonomic vector fields $X \in \vf(\W_r)$ such that
\begin{equation}\label{Chap05_eqn:UnifDynEqVF}
\inn(X)\Omega_r = 0 \quad ; \quad \inn(X)(\rho_\R^r)^*\eta \neq 0 \, .
\end{equation}

\begin{remark}
As in the first-order case described in Section \ref{Chap02_sec:NonAutonomousUnified}, the second equation in
\eqref{Chap05_eqn:UnifDynEqVF} is just a transverse condition for the vector field $X$ with respect to the
projection onto $\R$, and the non-zero value is usually fixed to $1$, thus giving the following equations
\begin{equation*}
\inn(X)\Omega_r = 0 \quad ; \quad \inn(X)(\rho_\R^r)^*\eta = 1 \, .
\end{equation*}
\end{remark}

Recall that $(\W_r,\Omega_r,(\rho_\R^r)^*\eta)$ is a precosymplectic manifold. Hence, equations
\eqref{Chap05_eqn:UnifDynEqVF} may not admit a solution $X \in \vf(\W_r)$ defined in the whole manifold,
but only on some submanifold of $\W_r$. Using an adapted version of the constraint algorithm described in
Section \ref{Chap01_sec:ConstraintAlgorithm} to precosymplectic manifolds, we can state the following result.

\begin{proposition}\label{Chap05_prop:UnifFirstConstraintSubmanifold}
A solution $X \in \vf(\W_r)$ to equations \eqref{Chap05_eqn:UnifDynEqVF} exists only
on the points of the submanifold $\mathcal{S}_c$ defined by
\begin{equation}\label{Chap05_eqn:UnifFirstConstraintSubmanifoldDefVF}
\mathcal{S}_c = \left\{ [w] \in \W_r \mid (\inn(Z)\d\hat{H})([w]) = 0 \mbox{ for every } Z \in \ker\Omega \right\}
\hookrightarrow \W_r \, .
\end{equation}
\end{proposition}

In the natural coordinates of $\W_r$, let us compute the constraint functions defining locally the submanifold
$\mathcal{S}_c$. Taking into account the coordinate expression \eqref{Chap05_eqn:UnifHamiltonianFunctionLocal}
of the local Hamiltonian function $\hat{H} \in \Cinfty(\W_r)$, then its differential is locally given by
\begin{equation}\label{Chap05_eqn:UnifHamiltonianFunctionDifferentialLocal}
\d\hat{H} = \sum_{i=0}^{k-1}(q_{i+1}^A\d p^i_A + p^i_A\d q_{i+1}^A) - \sum_{i=0}^{k} \derpar{\hat{L}}{q_i^A}\d q_i^A \, .
\end{equation}
Then, using the local basis for $\ker\Omega$ given in \eqref{Chap05_eqn:UnifCanonicalFormsKernelLocal}, we obtain
\begin{equation*}
\inn(Z)\d\hat{H} =
\begin{cases}
\displaystyle p_A^{k-1} - \derpar{\hat{L}}{q_k^A} \, , & \displaystyle \mbox{if } Z = \derpar{}{q_k^A} \, , \\[10pt]
0 \, , & \displaystyle \mbox{if } Z = \derpar{}{q_j^A} \, , \, j=k+1,\ldots,2k-1 \, .
\end{cases}
\end{equation*}
Therefore, $\mathcal{S}_c \hookrightarrow \W_r$ is a $n$-codimensional submanifold of $\W_r$ defined locally by the
constraints
\begin{equation}\label{Chap05_eqn:UnifFirstConstraintSubmanifoldLocal}
p_A^{k-1} - \derpar{\hat{L}}{q_k^A} = 0 \, .
\end{equation}
In particular, we have $\mathcal{S}_c = \W_c$, where $\W_c$ is the first constraint submanifold obtained for sections
in the previous Section.

As in the unified formalism for higher-order autonomous systems described in Chapter \ref{Chap:HOAutonomousDynamicalSystems},
in this setting we do not have a characterization of the submanifold $\W_c$ as the graph of a bundle morphism.

Let us compute in coordinates the local expression of the equations \eqref{Chap05_eqn:UnifDynEqVF}.
Let $X \in \vf(\W_r)$ be a generic vector field given locally by \eqref{Chap05_eqn:UnifGenericVectorField}.
Then, bearing in mind the local expression \eqref{Chap05_eqn:UnifPrecosymplecticFormsLocal} of the $2$-form
$\Omega_r$, the contraction $\inn(X)\Omega_r$ gives the following $1$-form on $\W_r$
\begin{align*}
\inn(X)\Omega_r
&= \left[ -f_0^A\derpar{\hat{L}}{q_0^A} + f_i^A\left( p_A^{i-1} - \derpar{\hat{L}}{q_i^A} \right)
+ F_k^A\left(p_A^{k-1} - \derpar{\hat{L}}{q_k^A}\right) +G_A^iq_{i+1}^A \right] \d t \\
&\quad{} + \left( f \derpar{\hat{L}}{q_0^A} - G_A^0 \right) \d q_0^A + \left( f\derpar{\hat{L}}{q_i^A} - fp_A^{i-1} - G_A^i \right) \d q_i^A
+ f\left( \derpar{\hat{L}}{q_k^A} - p_A^{k-1} \right) \d q_k^A \\
&\quad{} + \left( f_i^A - fq_{i+1}^A \right) \d p_A^i \, .
\end{align*}
Then, requiring this $1$-form to vanish, we obtain the following system of equations
\begin{align}
-f_0^A\derpar{\hat{L}}{q_0^A} + f_i^A\left( p_A^{i-1} - \derpar{\hat{L}}{q_i^A} \right)
+ F_k^A\left(p_A^{k-1} - \derpar{\hat{L}}{q_k^A}\right) + G_A^iq_{i+1}^A = 0 \, ,
\label{Chap05_eqn:UnifDynEqVFLocalRedundantEq1} \\
f_i^A = fq_{i+1}^A \, , \label{Chap05_eqn:UnifDynEqVFHolonomyLocalPart1} \\
G_A^0 = f\derpar{\hat{L}}{q_0^A} \quad ; \quad G_A^i = f\left( \derpar{\hat{L}}{q_i^A} - p_A^{i-1} \right) \, ,
\label{Chap05_eqn:UnifDynEqVFLocal1} \\
f\left(p_A^{k-1} - \derpar{\hat{L}}{q_k^A}\right) = 0 \, , \label{Chap05_eqn:UnifDynEqVFLegendreLocal1} \\
f \neq 0 \, , \label{Chap05_eqn:UnifDynEqVFLocalGaugeNonZero}
\end{align}
where $0 \leqslant i \leqslant k-1$ in \eqref{Chap05_eqn:UnifDynEqVFHolonomyLocalPart1},
and $1 \leqslant i \leqslant k-1$ in \eqref{Chap05_eqn:UnifDynEqVFLocal1}, and equation
\eqref{Chap05_eqn:UnifDynEqVFLocalGaugeNonZero} arises from the second equation in \eqref{Chap05_eqn:UnifDynEqVF}.
Fixing the non-zero value of the local function $f$ to $1$, the above equations become
\begin{align}
-f_0^A\derpar{\hat{L}}{q_0^A} + f_i^A\left( p_A^{i-1} - \derpar{\hat{L}}{q_i^A} \right)
+ F_k^A\left(p_A^{k-1} - \derpar{\hat{L}}{q_k^A}\right) + G_A^iq_{i+1}^A = 0 \, ,
\label{Chap05_eqn:UnifDynEqVFLocalRedundantEq2} \\
f_i^A = q_{i+1}^A \, , \label{Chap05_eqn:UnifDynEqVFHolonomyLocalPart2} \\
G_A^0 = \derpar{\hat{L}}{q_0^A} \quad ; \quad G_A^i = \derpar{\hat{L}}{q_i^A} - p_A^{i-1} \, ,
\label{Chap05_eqn:UnifDynEqVFLocal2} \\
p_A^{k-1} - \derpar{\hat{L}}{q_k^A} = 0 \, , \label{Chap05_eqn:UnifDynEqVFLegendreLocal2} \\
f = 1 \, . \label{Chap05_eqn:UnifDynEqVFLocalGaugeOne}
\end{align}
A simple calculation shows that equation \eqref{Chap05_eqn:UnifDynEqVFLocalRedundantEq2} is redundant,
since it is a combination of the others. Note that equations \eqref{Chap05_eqn:UnifDynEqVFHolonomyLocalPart2}
are part of the system of equations that the vector field $X$ must satisfy to be holonomic. In particular,
from these equations we deduce that $X$ is holonomic of type $k$ in $\W_r$. However, since the holonomy condition
is required in the statement of the problem, these equations are an identity. On the other hand, equations
\eqref{Chap05_eqn:UnifDynEqVFLegendreLocal2} are a compatibility condition stating that the vector fields $X$
solution to equations \eqref{Chap05_eqn:UnifDynEqVF} exist only with support on the submanifold $\W_c$ given by
Proposition \ref{Chap05_prop:UnifFirstConstraintSubmanifold}. Finally, equations \eqref{Chap05_eqn:UnifDynEqVFLocal2}
are the dynamical equations of the system.

Therefore, a vector field solution to equations \eqref{Chap05_eqn:UnifDynEqVF} is given in coordinates by
\begin{equation}\label{Chap05_eqn:UnifDynEqVFSolution}
X = \derpar{}{t} + q_{i+1}^A\derpar{}{q_i^A} + F_j^A\derpar{}{q_j^A} +
\derpar{\hat{L}}{q_0^A}\derpar{}{p_A^0} + \left( \derpar{\hat{L}}{q_i^A} - p_A^{i-1} \right) \derpar{}{p_A^i}\, .
\end{equation}
Moreover, since the holonomy condition is required from the beginning, the coordinate expression of
a holonomic vector field $X \in \vf(\W_r)$ solution to equations \eqref{Chap05_eqn:UnifDynEqVF} is
\begin{equation}\label{Chap05_eqn:UnifHolonomicVFSolution}
X = \derpar{}{t} + \sum_{i=0}^{2k-2} q_{i+1}^A\derpar{}{q_i^A} + F_{2k-1}^A\derpar{}{q_{2k-1}^A} +
\derpar{\hat{L}}{q_0^A}\derpar{}{p_A^0} + \left( \derpar{\hat{L}}{q_i^A} - p_A^{i-1} \right) \derpar{}{p_A^i}\, .
\end{equation}

Observe that the component function $F_{2k-1}^A$, $1 \leqslant A \leqslant n$, remain undetermined.
Nevertheless, since the vector field $X$ is defined at support on the submanifold $\W_c$, we must study
the tangency of $X$ along the submanifold $\W_c$. That is, we must require $\restric{\Lie(X)\xi}{\W_c} = 0$
for every constraint function $\xi$ defining locally $\W_c$. Hence, bearing in mind that the submanifold
$\W_c$ is defined locally by the $n$ constraints \eqref{Chap05_eqn:UnifFirstConstraintSubmanifoldLocal},
we must require
\begin{equation*}
\left(\derpar{}{t} + \sum_{i=0}^{2k-2} q_{i+1}^A\derpar{}{q_i^A} + F_{2k-1}^A\derpar{}{q_{2k-1}^A} +
\derpar{\hat{L}}{q_0^A}\derpar{}{p_A^0} + \left( \derpar{\hat{L}}{q_i^A} -p^{i-1}_A \right)\derpar{}{p^i_A}\right)
\left( p_A^{k-1} - \derpar{\hat{L}}{q_k^A} \right) = 0 \, ,
\end{equation*}
then we obtain the following $n$ equations
\begin{equation*}
p^{k-2}_A - \derpar{\hat{L}}{q_{k-1}^A} + \derpars{\hat{L}}{t}{q_k^A} + \sum_{i=0}^{k} q_{i+1}^B \derpars{\hat{L}}{q_i^B}{q_k^A} = 0 \, ,
\end{equation*}
which may be rewritten as
\begin{equation*}
p^{k-2}_A - \sum_{i=0}^{1}(-1)^i \frac{d^i}{dt^i} \left(\derpar{\hat{L}}{q_{k-1+i}^A}\right) = 0 \, .
\end{equation*}
These equations define a new submanifold $\W_1 \hookrightarrow \W_c$. Then, requiring $X$ to be tangent
to this new submanifold $\W_1$, we obtain the following $n$ additional constraints
\begin{equation*}
p^{k-3}_A - \sum_{i=0}^{2}(-1)^i \frac{d^i}{dt^i} \left(\derpar{\hat{L}}{q_{k-2+i}^A}\right) = 0 \, ,
\end{equation*}
which define a new submanifold $\W_2 \hookrightarrow \W_1 \hookrightarrow \W_c$. Iterating this process $k-3$
more times, the constraint algorithm delivers a $kn$-codimensional submanifold $\W_\Lag \hookrightarrow \W_r$
defined locally by the constraints
\begin{equation*}
p_A^{r-1} = \sum_{i=0}^{k-r}(-1)^i \frac{d^i}{dt^{i}} \left( \derpar{\hat{L}}{q_{r+i}^A} \right) \, ,
\end{equation*}
with $1 \leqslant A \leqslant n$ and $1 \leqslant r \leqslant k$. Moreover, the submanifold
$\W_\Lag \hookrightarrow \W_r$ can be identified with the graph of a bundle morphism
$\Leg \colon J^{2k-1}\pi \to J^{k-1}\pi^*$ over $J^{k-1}\pi$ defined locally by
\begin{equation*}
\Leg^*t = t \quad ; \quad \Leg^*q_r^A = q_r^A \quad ; \quad
\Leg^*p_A^{r-1} = \sum_{i=0}^{k-r}(-1)^i \frac{d^i}{dt^{i}} \left( \derpar{\hat{L}}{q_{r+i}^A} \right) \, .
\end{equation*}
Therefore, we recover Proposition \ref{Chap05_prop:UnifGraphLegendreOstrogradskyMap} and Corollary
\ref{Chap05_corol:UnifGraphExtendedLegendreOstrogradskyMapSect} from the constraint algorithm.
Finally, requiring $X$ to be tangent along the submanifold $\W_\Lag \hookrightarrow \W_r$,
we obtain the following $n$ equations
\begin{equation}\label{Chap05_eqn:UnifDynEqVFTangencyFinalLocal}
(-1)^k\left(F_{2k-1}^B - \frac{d}{dt}\,q_{2k-1}^B\right) \derpars{\hat{L}}{q_k^B}{q_k^A} + 
\sum_{i=0}^{k} (-1)^i\frac{d^i}{dt^i}\left( \derpar{\hat{L}}{q_i^A} \right) = 0 \ .
\end{equation}
These are the \textsl{$k$th-order Euler-Lagrange equations} for a vector field.
These equations may be compatible or not, and a sufficient condition to ensure compatibility
is the regularity of the Lagrangian density. In particular, we have the following result.

\begin{proposition}\label{Chap05_prop:UnifRegLagUniqueVF}
If $\Lag \in \df^1(J^k\pi)$ is a regular Lagrangian density, then there exists a unique vector
field $X \in \vf(\W_r)$ solution to equation \eqref{Chap05_eqn:UnifDynEqVF} and tangent to $\W_\Lag$.
\end{proposition}
\begin{proof}
Since the $k$th-order Lagrangian density $\Lag \in \df^{1}(J^{k}\pi)$ is regular, then equations
\eqref{Chap05_eqn:UnifDynEqVFTangencyFinalLocal} have a unique solution for $F_{2k-1}^A$, and
therefore the vector field $X \in \vf(\W_r)$ solution to equation \eqref{Chap05_eqn:UnifDynEqVF}
is unequivocally determined. In addition, since equations \eqref{Chap05_eqn:UnifDynEqVFTangencyFinalLocal}
are the necessary and sufficient condition for $X$ to be tangent along the submanifold $\W_\Lag$,
this condition holds whenever they the referred equations are compatible, as it is the case when the
$k$th-order Lagrangian density is regular.
\end{proof}

However, if the $k$th-order Lagrangian density $\Lag$ is not regular, then equations
\eqref{Chap05_eqn:UnifDynEqVFTangencyFinalLocal} may or may not be compatible, and the
compatibility condition may give rise to new constraints. In the most favorable cases,
there exists a submanifold $\W_f \hookrightarrow \W_\Lag$ (it could be $\W_f = \W_\Lag$)
such that there exist vector fields $X \in \vf(\W_r)$, tangent to $\W_f$, which are solutions
to the equations
\begin{equation}\label{Chap05_eqn:UnifDynEqVFSingular}
\restric{\inn(X)\Omega_r}{\W_f} = 0 \quad ; \quad \restric{\inn(X)(\rho_\R^r)^*\eta}{\W_f} = 1 \, .
\end{equation}
In this case, the equations \eqref{Chap05_eqn:UnifDynEqVFTangencyFinalLocal} are not compatible,
and the compatibility condition gives rise to new constraints, and the constraint algorithm continues.

\subsubsection{Equivalence of the dynamical equations in the unified formalism}

In the previous Sections we have stated the dynamical equations in the unified formalism in several ways.
First, we have stated the problem using a variational principle. Then we have stated a geometric equation
for sections of the bundle $\rho_\R^r \colon \W_r \to \R$, and we have analyzed it in coordinates. Finally,
we have stated a geometric equation for vector fields defined in $\W_r$, and we have studied the equation
and the tangency condition in coordinates. In this Section we prove that all these equations are equivalent.

\begin{theorem}\label{Chap05_thm:UnifEquivalenceTheorem}
The following assertions on a holonomic section $\psi \in \Gamma(\rho_\R^r)$ are equivalent.
\begin{enumerate}
\item $\psi$ is a solution to the Lagrangian-Hamiltonian variational problem.
\item $\psi$ is a solution to equation \eqref{Chap05_eqn:UnifDynEqSect}, that is,
\begin{equation*}
\psi^*\inn(Y)\Omega_r = 0 \, , \quad \text{for every }Y \in \vf(\W_r) \ (\mbox{tangent to } \W_\Lag) \, .
\end{equation*}
\item If $\psi$ is given locally by
\begin{equation*}
\psi(t) = (t,q_i^A(t),q_j^A(t),p_A^i(t)) \, ,
\end{equation*}
with $0 \leqslant i \leqslant k-1$, $k \leqslant j \leqslant 2k-1$, then the components of $\psi$ satisfy equations \eqref{Chap05_eqn:UnifDynEqSectLocal1} and \eqref{Chap05_eqn:UnifDynEqSectLocal2}, that is,
the following system of $kn$ differential equations
\begin{equation}\label{Chap05_eqn:UnifEquivalenceTheoremLocalDiffEq}
\dot{p}_A^0 = \derpar{\hat{L}}{q_0^A} \quad ; \quad \dot{p}_A^i = \derpar{\hat{L}}{q_i^A} - p_A^{i-1} \, .
\end{equation}
\item $\psi$ is a solution to equation
\begin{equation}\label{Chap05_eqn:UnifDynEqIC}
\inn(\dot{\psi})(\Omega_r \circ \psi) = 0 \, ,
\end{equation}
where $\dot{\psi} \colon \R \to \Tan\W_r$ is the canonical lifting of $\psi$ to $\Tan\W_r$.
\item $\psi$ is an integral curve of a vector field contained in a class of $\rho_\R^r$-transverse and
holonomic vector fields, $\left\{ X \right\} \subset \vf(\W_r)$, satisfying the first equation in
\eqref{Chap05_eqn:UnifDynEqVF}, that is,
\begin{equation*}
\inn(X)\Omega_r = 0 \, .
\end{equation*}
\end{enumerate}
\end{theorem}
\begin{proof}
\begin{description}
\item[\textnormal{($1 \, \Longleftrightarrow \, 2$)}]
We prove this equivalence following the patterns in \cite{art:Echeverria_DeLeon_Munoz_Roman07}.

Let $Z \in \vf^{V(\rho_\R^r)}(\W_r)$ be a compact-supported vector field, and $V \subseteq \R$ an open
set such that $\partial V$ is a $0$-dimensional manifold and $\rho_\R^r(\supp(Z)) \subseteq V$. Then,
\begin{align*}
\restric{\frac{d}{ds}}{s=0} \int_\R \psi^*_s\Theta_r
&= \restric{\frac{d}{ds}}{s=0} \int_V \psi^*_s\Theta_r
= \restric{\frac{d}{ds}}{s=0} \int_V \psi^*\sigma_s^*\Theta_r\\
&= \int_V\psi^*\left( \lim_{s\to0} \frac{\sigma_s^*\Theta_r - \Theta_r}{s} \right)
= \int_V\psi^*\Lie(Z)\Theta_r \\
&= \int_V \psi^*(\inn(Z)\d \Theta_r + \d\inn(Z)\Theta_r)\\
&= \int_V \psi^*(-\inn(Z)\Omega_r + \d\inn(Z)\Theta_r)\\
&= - \int_V \psi^*\inn(Z)\Omega_r + \int_V \d(\psi^*\inn(Z)\Theta_r) \\
&= - \int_V \psi^*\inn(Z)\Omega_r + \int_{\partial V}\psi^*\inn(Z)\Theta_r
= - \int_V\psi^*\inn(Z)\Omega_r \, ,
\end{align*}
as a consequence of Stoke's theorem and the assumptions made on the supports of the vertical vector
fields. Thus, by the fundamental Theorem of the variational calculus, we conclude
\begin{equation*}
\restric{\frac{d}{ds}}{s=0} \int_\R \psi_s^*\Theta_r = 0 \quad \Longleftrightarrow \quad \psi^*\inn(Z)\Omega_r = 0 \, ,
\end{equation*}
for every compact-supported $Z \in \vf^{V(\rho_\R^r)}(\W_r)$. However, since the compact-supported
vector fields generate locally the $\Cinfty(\W_r)$-module of vector fields in $\W_r$, it follows
that the last equality holds for every $\rho_\R^r$-vertical vector field $Z$ in $\W_r$.

Now, recall that for every point $w \in \Im(\psi)$, we have a canonical splitting of the tangent
space of $\W_r$ at $w$ in a $\rho_\R^r$-vertical subspace and a $\rho_\R^r$-horizontal subspace, that is,
\begin{equation*}
\Tan_w\W_r = V_w(\rho_\R^r) \oplus \Tan_w(\Im(\psi)) \, .
\end{equation*}
Thus, if $Y \in \vf(\W_r)$, then
\begin{equation*}
Y_w = (Y_w - \Tan_w(\psi \circ \rho_\R^r)(Y_w)) + \Tan_w(\psi \circ \rho_\R^r)(Y_w) \equiv Y_w^V + Y_w^{\psi} \, ,
\end{equation*}
with $Y_w^V \in V_w(\rho_\R^r)$ and $Y_w^{\psi} \in \Tan_w(\Im(\psi))$. Therefore
\begin{equation*}
\psi^*\inn(Y)\Omega_r= \psi^*\inn(Y^V)\Omega_r + \psi^*\inn(Y^{\psi})\Omega_r = \psi^*\inn(Y^{\psi})\Omega_r \, ,
\end{equation*}
since $\psi^*\inn(Y^V)\Omega_r = 0$, by the conclusion in the above paragraph. Now, as
$Y^{\psi}_w \in \Tan_w(\Im(\psi))$ for every $w \in \Im(\psi)$, then the vector field $Y^{\psi}$
is tangent to $\Im(\psi)$, and hence there exists a vector field $X \in \vf(\R)$ such that $X$ is
$\psi$-related with $Y^{\psi}$; that is, $\psi_*X = \restric{Y^{\psi}}{\Im(\psi)}$. Then
$\psi^*\inn(Y^{\psi})\Omega_r = \inn(X)\psi^*\Omega_r$. However, as $\dim\Im(\psi) = \dim \R = 1$ and
$\Omega_r$ is a $2$-form, we obtain that $\psi^*\inn(Y^{\psi})\Omega_r = 0$. Hence, we conclude that
$\psi^*\inn(Y)\Omega_r = 0$ for every $Y \in \vf(\W_r)$.

Taking into account the reasoning of the first paragraph, the converse is obvious since the condition
$\psi^*\inn(Y)\Omega_r = 0$, for every $Y \in \vf(\W_r)$, holds, in particular, for every
$Z \in \vf^{V(\rho_\R^r)}(\W_r)$.

\item[\textnormal{($2 \, \Longleftrightarrow \, 3$)}]
As we have seen in previous Sections, in the natural coordinates of $\W_r$ equation
\eqref{Chap05_eqn:UnifDynEqSect} gives locally the equations \eqref{Chap05_eqn:UnifDynEqSectLocalRedundantEq},
\eqref{Chap05_eqn:UnifDynEqSectHolonomyLocalPart}, \eqref{Chap05_eqn:UnifDynEqSectLocal1},
\eqref{Chap05_eqn:UnifDynEqSectLocal2} and \eqref{Chap05_eqn:UnifDynEqSectLegendreLocal}.
As stated previously, equation \eqref{Chap05_eqn:UnifDynEqSectLocalRedundantEq} is redundant,
since it is a combination of the others, and from equations \eqref{Chap05_eqn:UnifDynEqSectLegendreLocal}
we deduce that the section $\psi \in \Gamma(\rho_\R^r)$ lies in the submanifold $\W_c \hookrightarrow \W_r$,
and in particular in the submanifold $\W_\Lag = \graph(\Leg)$ when we combine these constraints with equations
\eqref{Chap05_eqn:UnifDynEqSectLocal2}. Hence, equation \eqref{Chap05_eqn:UnifDynEqSect} is locally equivalent
to equations \eqref{Chap05_eqn:UnifDynEqSectHolonomyLocalPart}, \eqref{Chap05_eqn:UnifDynEqSectLocal1} and
\eqref{Chap05_eqn:UnifDynEqSectLocal2}. However, since $\psi$ is assumed to be holonomic, equations
\eqref{Chap05_eqn:UnifDynEqSectHolonomyLocalPart} hold identically, and thus equation \eqref{Chap05_eqn:UnifDynEqSect}
is locally equivalent to equations \eqref{Chap05_eqn:UnifDynEqSectLocal1} and \eqref{Chap05_eqn:UnifDynEqSectLocal2},
that is, to equations \eqref{Chap05_eqn:UnifEquivalenceTheoremLocalDiffEq}.

\item[\textnormal{($3 \, \Longleftrightarrow \, 4$)}]
If $\psi \in \Gamma(\rho_\R^r)$ is locally given by
\begin{equation*}
\psi(t) = (t,q_i^A(t),q_j^A(t),p_A^i(t)) \, ,
\end{equation*}
then its canonical lifting to the tangent bundle of $\W_r$, $\dot{\psi} \colon \R \to \Tan\W_r$,
is locally given by
\begin{equation*}
\dot{\psi}(t) = (1,\dot{q}_i^A(t),\dot{q}_j^A(t),\dot{p}_A^i(t)) \, ,
\end{equation*}
and the inner product $\inn(\dot{\psi})(\Omega_r \circ \psi)$ gives, in coordinates,
\begin{align*}
\inn(\dot{\psi})(\Omega_r \circ \psi) &=
\left( p_A^i\dot{q}_{i+1}^A - \dot{q}_r^A\derpar{\hat{L}}{q_r^A} + \dot{p}_A^iq_{i+1}^A \right)\d t
+ \left( \derpar{\hat{L}}{q_0^A} - \dot{p}_A^0 \right) \d q_0^A \\
&\quad{} + \left( \derpar{\hat{L}}{q_i^A} - p_A^{i-1} - \dot{p}_A^i \right)\d q_i^A
+ \left( p_A^{k-1} - \derpar{\hat{L}}{q_k^A} \right)\d q_k^A
+ \left( \dot{q}_i^A - q_{i+1}^A \right)\d p_A^i \, .
\end{align*}
Now, requiring this last expression to vanish, we obtain
the system of $(2k+1)n+1$ equations
\begin{align*}
p_A^i\dot{q}_{i+1}^A - \dot{q}_r^A\derpar{\hat{L}}{q_r^A} + \dot{p}_A^iq_{i+1}^A = 0 \quad &; \quad
\dot{p}_A^0 = \derpar{\hat{L}}{q_0^A} \quad ; \quad
\dot{p}_A^i = \derpar{\hat{L}}{q_i^A} - p_A^{i-1} \, , \\
p_A^{k-1} = \derpar{\hat{L}}{q_k^A} &\quad ; \quad
\dot{q}_i^A = q_{i+1}^A \, .
\end{align*}
Observe that this system of equations is the same given by \eqref{Chap05_eqn:UnifDynEqSectLocalRedundantEq},
\eqref{Chap05_eqn:UnifDynEqSectLocal1},  \eqref{Chap05_eqn:UnifDynEqSectHolonomyLocalPart},
\eqref{Chap05_eqn:UnifDynEqSectLocal2}, \eqref{Chap05_eqn:UnifDynEqSectLegendreLocal}.
The same remarks given in the proof of ($2\,\Longleftrightarrow\,3$) apply in this case.
Thus, bearing in mind the above item, we have proved that equation \eqref{Chap05_eqn:UnifDynEqIC}
is locally equivalent to the $kn$ differential equations \eqref{Chap05_eqn:UnifEquivalenceTheoremLocalDiffEq}.

\item[\textnormal{($3 \, \Longleftrightarrow \, 5$)}]
As we have seen in the previous Section, taking $f = 1$ as a representative of the class of holonomic and
$\rho_\R^r$-transverse vector fields $\{ X \} \subseteq \vf(\W_r)$, a vector field solution to the first equation in
\eqref{Chap05_eqn:UnifDynEqVF} is given locally by \eqref{Chap05_eqn:UnifHolonomicVFSolution}, that is,
\begin{equation*}
X = \derpar{}{t} + \sum_{i=0}^{2k-2} q_{i+1}^A\derpar{}{q_i^A} + F_{2k-1}^A\derpar{}{q_{2k-1}^A} +
\derpar{\hat{L}}{q_0^A}\derpar{}{p_A^0} + \left( \derpar{\hat{L}}{q_i^A} - p_A^{i-1} \right) \derpar{}{p_A^i}\, ,
\end{equation*}
where the functions $F_{2k-1}^A$ are the solutions to equations \eqref{Chap05_eqn:UnifDynEqVFTangencyFinalLocal}.

Now, let $\psi \in \Gamma(\rho_\R^r)$ be an integral curve of $X$, that is, $\dot{\psi} = X \circ \psi$.
If $\psi$ is given locally by $\psi(t) = (t,q_i^A(t),q_j^A(t),p_A^i(t))$, then
$\dot{\psi}(t) = (1,\dot{q}_i^A(t),\dot{q}_j^A(t),\dot{p}_A^i(t))$, and the condition for $\psi$ to
be an integral curve is locally equivalent to the equations
\begin{equation*}
\dot{q}_i^A = q_{i+1}^A \quad ; \quad
\dot{q}_{2k-1}^A = F_{2k-1}^A \circ \psi \quad ; \quad
\dot{p}_A^0 = \derpar{\hat{L}}{q_0^A} \quad ; \quad
\dot{p}_A^i = \derpar{\hat{L}}{q_i^A} - p_A^{i-1} \, .
\end{equation*}
Since the curve is assumed to be holonomic, the first $2kn$ equations hold identically.
Hence, the condition of $\psi$ to be an integral curve of a $\rho_\R^r$-transverse and holonomic
vector field, $X \in \vf(\W_r)$, satisfying the first equation in \eqref{Chap05_eqn:UnifDynEqVF}
is locally equivalent to equations \eqref{Chap05_eqn:UnifEquivalenceTheoremLocalDiffEq}.\qedhere
\end{description}
\end{proof}


\section{The Lagrangian formalism}
\label{Chap05_sec:UnifiedToLagrangian}

In this Section we state the Lagrangian formalism for higher-order non-autonomous dynamical systems.
Since we have already described the Lagrangian-Hamiltonian unified formalism for this kinds of
systems, we proceed in an analogous way to Section \ref{Chap03_sec:UnifiedToLagrangian}: we ``recover''
the Lagrangian structures, equations and solutions from the ones in the unified formalism.

We do not distinguish between the regular and singular cases, since the results remain the same
in either case, but a few comments on the singular case are given.

\subsection{Geometric and dynamical structures}

As in the autonomous case described in Section \ref{Chap03_sec:UnifiedToLagrangian}, the first step
to give a Lagrangian formalism for higher-order non-autonomous dynamical systems is to define the usual geometric
structures of the Lagrangian formalism, namely the Poincar\'{e}-Cartan forms, in order to state the dynamical equations.

The first fundamental result is the following.

\begin{proposition}\label{Chap05_prop:LagRho1LDiffeomorphism}
The map $\rho_1^\Lag = \rho_1^r \circ j_\Lag \colon \W_\Lag \to J^{2k-1}\pi$ is a diffeomorphism.
\end{proposition}
\begin{proof}
Since $\W_\Lag = \graph(\Leg)$, it is clear that $J^{2k-1}\pi$ is diffeomorphic to $\W_\Lag$.
On the other hand, since $\rho_1$ is a surjective submersion by definition, its restriction
to the submanifold $\W_\Lag$ is also a surjective submersion and, due to the fact that
$\dim\W_\Lag = \dim J^{2k-1}\pi = 2kn + 1$, the map $\rho_1^\Lag$ is a bijective local diffeomorphism.
In particular, the map $\rho_1^\Lag$ is a global diffeomorphism.
\end{proof}

This result enables us to state a one-to-one correspondence between the set of solutions to the dynamical
equations in the unified formalism and the set of solutions to the dynamical equations in the Lagrangian formalism.
Now, let us define the geometric structures in the Lagrangian formalism. Using the extended Legendre-Ostrogradsky map
obtained in Corollary \ref{Chap05_corol:UnifGraphExtendedLegendreOstrogradskyMapSect}, we give the following definition.

\begin{definition}\label{Chap05_def:LagCartanForms}
The \textnormal{Poincar\'{e}-Cartan forms} are the forms $\Theta_\Lag \in \df^{1}(J^{2k-1}\pi)$
and $\Omega_\Lag \in \df^{2}(J^{2k-1}\pi)$ defined as
\begin{equation*}
\Theta_{\Lag} = \widetilde{\Leg}^*\Theta_{k-1} \quad ; \quad
\Omega_{\Lag} = \widetilde{\Leg}^*\Omega_{k-1} = -\d\Theta_\Lag \, ,
\end{equation*}
where $\Theta_{k-1} \in \df^{1}(\Tan^*(J^{k-1}\pi))$ and $\Omega_{k-1} = -\d\Theta_{k-1} \in \df^{2}(\Tan^*(J^{k-1}\pi))$
are the canonical Liouville forms in $\Tan^*(J^{k-1}\pi)$. The pair $(J^{2k-1}\pi,\Lag)$ is the
\textnormal{$k$th-order non-autonomous Lagrangian system} associated with $(\W_r,\Omega_r,(\rho_\R^r)^*\eta)$.
\end{definition}

Observe that the Poincar\'{e}-Cartan forms are defined from the canonical forms in the cotangent bundle
$\Tan^*(J^{k-1}\pi)$, rather than using the forms $\Theta$ and $\Omega$ in the unified setting.
Nevertheless, we have the following result.

\begin{lemma}\label{Chap05_lemma:LagCartanForms}
The Poincar\'{e}-Cartan $1$-form $\Theta_\Lag \in \df^{1}(J^{2k-1}\pi)$
satisfies $\Theta = \rho_1^*\Theta_{\Lag}$ and $\Theta_r = (\rho_1^r)^*\Theta_\Lag$.
\end{lemma}
\begin{proof}
A straightforward calculation leads to this result. Bearing in mind that $\rho_2 = \widetilde{\Leg} \circ \rho_1$,
for the first statement we have
\begin{equation*}
\rho_1^*\Theta_\Lag = \rho_1^*(\widetilde{\Leg}^*\Theta_{k-1})
= (\widetilde{\Leg} \circ \rho_1)^*\Theta_{k-1} = \rho_2^*\Theta_{k-1}
= \Theta \, .
\end{equation*}
On the other hand, taking into account that $\rho_1^r = \rho_1 \circ \hat{h}$, for the second
statement we have
\begin{equation*}
(\rho_1^r)^*\Theta_\Lag = (\rho_1 \circ \hat{h})^*\Theta_{\Lag})
= \hat{h}^*(\rho_1^*\Theta_\Lag) = \hat{h}^*\Theta = \Theta_r \, . \qedhere
\end{equation*}
\end{proof}

\begin{remark}
Since the pull-back of a form and the exterior derivative commute, Lemma \ref{Chap05_lemma:LagCartanForms}
also holds for the Poincar\'{e}-Cartan $2$-form, $\Omega_\Lag$.
\end{remark}

In the natural coordinates of $J^{2k-1}\pi$, bearing in mind the coordinate expression of the extended
Legendre-Ostrogradsky map $\widetilde{\Leg} \colon J^{2k-1}\pi \to \Tan^*(J^{k-1}\pi)$ given in Corollary
\ref{Chap05_corol:UnifGraphExtendedLegendreOstrogradskyMapSect}, the local expression of the Poincar\'e-Cartan $1$-form is
\begin{equation}\label{Chap05_eqn:LagCartan1FormLocal}
\Theta_\Lag = \sum_{r=1}^{k} \sum_{i=0}^{k-r}(-1)^i \frac{d^i}{dt^i}\left( \derpar{L}{q^A_{r+i}} \right)
(\d q^A_{r-1} - q_r^A\d t) + L\d t \, ,
\end{equation}
which coincides with the coordinate expression of the Poincar\'{e}-Cartan $1$-form obtained by D.J. Saunders
in \cite{art:Saunders87} and \cite{book:Saunders89} when the base manifold $M$ of the bundle $\pi\colon E \to M$
is $1$-dimensional. It is clear from this coordinate expression that $\Theta_\Lag$ is $\pi^{2k-1}_{k-1}$-semibasic.
Now, taking its exterior derivative and changing the sign, we obtain the local expression of the Poincar\'{e}-Cartan
$2$-form, which is
\begin{equation}\label{Chap05_eqn:LagCartan2FormLocal}
\begin{array}{l}
\displaystyle \Omega_\Lag = \sum_{r=1}^{k} \sum_{i=0}^{k-r}(-1)^{i+1} \left( \frac{d^i}{dt^i} \left( \derpars{L}{t}{q_{r+i}^A}\d t
+ \derpars{L}{q_j^B}{q_{r+i}^A}\d q_j^B \right) \wedge (\d q_{r-1}^A - q_r^A\d t) \right. \\[15pt]
\displaystyle \qquad\quad{} \left. - \frac{d^i}{dt^i}\left(\derpar{L}{q^A_{r+i}} \right) \d q_r^A \wedge \d t \right)
- \derpar{L}{q_j^B}\d q_j^B \wedge \d t \, .
\end{array}
\end{equation}

\begin{remark}
The Poincar\'{e}-Cartan $1$-form can be defined alternatively using the canonical structures of the higher-order
jet bundles. In particular, according to \cite{art:Saunders87} and \cite{book:Saunders89} (see also
\cite{art:Aldaya_Azcarraga78_2}, \cite{proc:Garcia_Munoz83}), we have
\begin{equation}\label{Chap05_eqn:LagCartan1FormAlternativeDef}
\Theta_\Lag = S_\eta^{(k)}(\d L) + (\pi^{2k-1}_{k})^*\Lag \in \df^{1}(J^{2k-1}\pi) \, ,
\end{equation}
where $S_\eta^{(k)}$ is the generalization to higher-order jet bundles of the operator used in the classical
Hamilton-Cartan formalism for problems in the calculus of variations which involve time explicitly
(see \cite{art:Saunders87} and \cite{book:Saunders89}, $\S 6.5$, for details).
\end{remark}

From the Poincar\'{e}-Cartan forms, the concept of regularity for a $k$th-order Lagrangian density is a
straightforward generalization of Definition \ref{Chap02_def:LagNAFORegularLagrangian}
for first-order non-autonomous dynamical systems.

\begin{definition}
A $k$th-order Lagrangian density $\Lag \in \df^{1}(J^{k}\pi)$ is \textnormal{regular} if the pair
$(\Omega_\Lag,(\bar{\pi}^{2k-1})^*\eta)$ is a cosymplectic structure in $J^{2k-1}\pi$.
Otherwise, $\Lag$ is \textnormal{singular}.
\end{definition}

Observe that, taking into account Section \ref{Chap01_sec:CosymplecticGeom}, a $k$th-order Lagrangian density
is regular if, and only if, the Poincar\'e-Cartan $2$-form $\Omega_\Lag \in \df^{2}(J^{2k-1}\pi)$ has maximal
rank $2n$. Moreover, bearing in mind the coordinate expression \eqref{Chap05_eqn:LagCartan2FormLocal}
of the Poincar\'{e}-Cartan $2$-form, the regularity condition for $\Lag$ is locally equivalent to
\begin{equation*}
\det\left( \derpars{L}{q_k^B}{q_k^A} \right)(j^{k}_t\phi) \neq 0, \quad \mbox{for every } j^{k}_t\phi \in J^{k}\pi \, .
\end{equation*}
Thus, this notion of regularity is equivalent to the one given in Definition \ref{Chap05_def:UnifRegularLagrangian}.
Moreover, bearing in mind Proposition \ref{Chap05_prop:UnifRankBothLegendreMaps}, we have the following result.

\begin{proposition}\label{Chap05_prop:CharacterizationRegularLagrangian}
Given a $k$th-order Lagrangian density $\Lag \in \df^{1}(J^{k}\pi)$, the following
statements are equivalent.
\begin{enumerate}
\item In every local chart of coordinates $(q_0^A,\ldots,q_k^A)$ of $J^{k}\pi$, we have
\begin{equation*}
\det\left( \derpars{L}{q_k^B}{q_k^A} \right)(j^{k}_t\phi) \neq 0, \quad \mbox{for every } j^{k}_t\phi \in J^{k}\pi \, .
\end{equation*}
\item $\Omega_\Lag$ has maximal rank on $J^{2k-1}\pi$.
\item The pair $(\Omega_\Lag,(\bar{\pi}^{2k-1})^*\eta)$ is a cosymplectic structure on $J^{2k-1}\pi$.
\item $\Leg \colon J^{2k-1}\pi \to J^{k-1}\pi^*$ is a local diffeomorphism.
\item $\widetilde{\Leg} \colon J^{2k-1}\pi \to \Tan^*(J^{k-1}\pi)$ is an immersion.
\end{enumerate}
\end{proposition}

\subsection{Dynamical equations}

Using the results stated in the previous Section, we can state the dynamical equations in
the Lagrangian formalism, and recover the solutions to these equations from the solutions
to the dynamical equations in the unified formalism.

\subsubsection{Variational principle}

First of all, let us state the variational problem from which the Lagrangian dynamical equations
are derived. Let $\Gamma(\pi)$ be the set of sections of $\pi$, and let us consider the functional
\begin{equation*}\label{Chap05_eqn:LagVariationalFunctionalDef}
\begin{array}{rcl}
\mathbf{L} \colon \Gamma(\pi) & \longrightarrow & \R \\
\phi & \longmapsto & \displaystyle \int_\R (j^{2k-1}\phi)^*\Theta_\Lag
\end{array} \, ,
\end{equation*}
where the convergence of the integral is assumed.

\begin{definition}
The \textnormal{$k$th-order Lagrangian variational problem} (also called \textnormal{generalized Hamilton variational problem})
for the $k$th-order Lagrangian system $(J^{2k-1}\pi,\Lag)$ is the search
for the critical (or stationary) sections of the functional $\mathbf{L}$ with respect to the variations
of $\phi$ given by $\phi_s = \sigma_s \circ \phi$, where $\left\{ \sigma_s \right\}$ is a local one-parameter
group of any compact-supported $Z \in \vf^{V(\pi)}(E)$; that is,
\begin{equation*}\label{Chap05_eqn:LagDynEqVar}
\restric{\frac{d}{ds}}{s=0}\int_\R (j^{2k-1}\phi_s)^*\Theta_\Lag = 0 \, .
\end{equation*}
\end{definition}

\begin{theorem}\label{Chap05_thm:UnifiedToLagrangianVar}
Let $\psi \in \Gamma(\rho_\R^r)$ be a holonomic section which is a solution to the Lagrangian-Hamiltonian
variational problem given by the functional $\mathbf{LH}$. Then, the section
$\psi_\Lag = \rho_1^r \circ \psi \in \Gamma(\bar{\pi}^{2k-1})$ is holonomic, and its projection
$\phi = \pi^{2k-1} \circ \psi_\Lag \in \Gamma(\pi)$ is a solution to the Lagrangian variational problem
given by the functional $\mathbf{L}$.

\noindent Conversely, given a section $\phi \in \Gamma(\pi)$ which is a solution to the Lagrangian variational
problem, the section $\psi = j_\Lag \circ (\rho_1^\Lag)^{-1} \circ j^{2k-1}\phi \in \Gamma(\rho_\R^r)$ is a
solution to the Lagrangian-Hamiltonian variational problem.
\end{theorem}
\begin{proof}
As $\psi \in \Gamma(\rho_\R^r)$ is holonomic,
then $\psi_\Lag = \rho_1^r \circ \psi\in \Gamma(\bar{\pi}^{2k-1})$
is a holonomic section, by definition.

Now, $\rho_1^r$ being a submersion, for every compact-supported vector field
$X \in \vf^{V(\bar{\pi}^{2k-1})}(J^{2k-1}\pi)$ there exist compact-supported vector fields
$Y \in \vf^{V(\rho_\R^r)}(\W_r)$ such that $X$ and $Y$ are $\rho_1^r$-related. In particular, this holds
if $X$ is the $(2k-1)$-jet lifting of a compact-supported $\pi$-vertical vector field in $E$; that is,
if we have $X = j^{2k-1}Z$, with $Z \in \vf^{V(\pi)}(E)$. We also denote by $\left\{ \sigma_s \right\}$ a local
one-parameter group for the compact-supported vector fields $Y \in \vf^{V(\rho_\R^r)}(\W_r)$.
Then, using this particular choice of $\rho_1^r$-related vector fields, we have
\begin{align*}
\restric{\frac{d}{ds}}{s=0}\int_\R (j^{2k-1}\phi_s)^*\Theta_\Lag
&= \restric{\frac{d}{ds}}{s=0}\int_\R(j^{2k-1}(\sigma_s \circ\phi))^*\Theta_\Lag
= \restric{\frac{d}{ds}}{s=0}\int_\R(j^{2k-1}\phi)^*(j^{2k-1}\sigma_s)^*\Theta_\Lag \\
&= \int_\R\psi_\Lag^*\Lie(j^{2k-1}Z)\Theta_\Lag 
= \int_\R\psi_\Lag^*(\inn(j^{2k-1}Z)\d\Theta_\Lag + \d\inn(j^{2k-1}Z)\Theta_\Lag) \\
&= \int_\R \psi^*(\rho_1^r)^*(\inn(j^{2k-1}Z)\d\Theta_\Lag + \d\inn(j^{2k-1}Z)\Theta_\Lag)
= \int_\R \psi^*(\inn(Y)\d\Theta_r + \d\inn(Y)\Theta_r) \\
&= \int_\R \psi^*\Lie(Y)\Theta_r 
= \restric{\frac{d}{ds}}{s=0}\int_\R\psi^*\sigma_s^*\Theta_r
= \restric{\frac{d}{ds}}{s=0} \int_\R \psi_s^*\Theta_r = 0 \, ,
\end{align*}
since $\psi$ is a critical section for the Lagrangian-Hamiltonian variational problem.

Conversely, if we have a section $\phi$ which is a solution to the Lagrangian variational problem,
then we can construct a section $\psi = j_\Lag \circ (\rho_1^\Lag)^{-1} \circ j^{2k-1}\phi$
of the projection $\rho_\R^r$ (remember that, in the unified formalism,
the dynamical equations have solutions only on the points of $\W_\Lag$, or in a subset of it).
Then, the above reasoning also shows that if $\phi$ is a solution to the Lagrangian variational problem,
then $\psi$ is a solution to the Lagrangian-Hamiltonian variational problem.
\end{proof}

\subsubsection{Dynamical equations for sections}

Using the previous results, we can state the Lagrangian equations for sections, and
recover the Lagrangian sections in $J^{2k-1}\pi$ from the sections in the unified formalism.

First, the \textsl{$k$th-order Lagrangian problem for sections} associated with the system
$(J^{2k-1}\pi,\Lag)$ consists in finding sections $\phi \in \Gamma(\pi)$ satisfying
\begin{equation}\label{Chap05_eqn:LagDynEqSect}
(j^{2k-1}\phi)^*\inn(Y)\Omega_\Lag = 0 \, , \quad \mbox{for every } Y \in \vf(J^{2k-1}\pi) \, .
\end{equation}

\begin{proposition}\label{Chap05_prop:UnifiedToLagrangianSect}
Let $\psi \in \Gamma(\rho_\R^r)$ be a holonomic section solution to equation \eqref{Chap05_eqn:UnifDynEqSect}.
Then the section $\psi_\Lag = \rho_1^r \circ \psi \in \Gamma(\bar{\pi}^{2k-1})$ is holonomic, and is a solution
to equation \eqref{Chap05_eqn:LagDynEqSect}.
\end{proposition}
\begin{proof}
Since, by definition, $\psi \in \Gamma(\rho_\R^r)$ is holonomic if $\rho_1^r \circ \psi \in \Gamma(\bar{\pi}^{2k-1})$
is holonomic, it is obvious that $\psi_\Lag = \rho_1^r \circ \psi$ is a holonomic section.

Now, recall that, as $\rho_1^r$ is a submersion, for every $Y \in \vf(J^{2k-1}\pi)$ there exist some
$Z \in \vf(\W_r)$ such that $Y$ and $Z$ are $\rho_1^r$-related. Note that this vector field is not unique,
since $Z + Z_o$, with $Z_o \in \ker\Tan\rho_1^r$, is also $\rho_1^r$-related with $Y$. Thus, using this
particular choice of $\rho_1^r$-related vector fields, we have
\begin{equation*}
\psi_\Lag^*\inn(Y)\Omega_\Lag = (\rho_1^r \circ \psi)^*\inn(Y)\Omega_\Lag =
\psi^*((\rho_1^r)^*\inn(Y)\Omega_\Lag) = \psi^*(\inn(Z)(\rho_1^r)^*\Omega_\Lag) =
\psi^*i(Z)\Omega_r \, .
\end{equation*}
Since the equality $\psi^*\inn(Z)\Omega_r = 0$ holds for every $Z \in \vf(\W_r)$, in particular it holds
for every $Z \in \vf(\W_r)$ which is $\rho_1^r$-related with $Y \in \vf(J^{2k-1}\pi)$. Hence, we obtain
\begin{equation*}
\psi_\Lag^*\inn(Y)\Omega_\Lag = \psi^*\inn(Z)\Omega_r = 0 \, . \qedhere
\end{equation*}
\end{proof}

The diagram for this situation is the following:
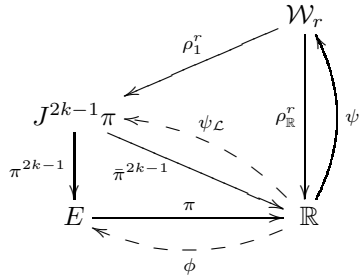
\begin{equation*}
\xymatrix{
\ & \ & \W_r \ar[dd]_-{\rho_\R^r} \ar[dll]_-{\rho_1^r}  \\
J^{2k-1}\pi \ar[d]_{\pi^{2k-1}} \ar[drr]_<(0.25){\bar{\pi}^{2k-1}} & \ & \ \\
E \ar[rr]^{\pi} & \ & \  \R \ar@/_1pc/[uu]_{\psi} \ar@/_1pc/@{-->}[ull]_{\psi_\Lag} \ar@/^1pc/@{-->}[ll]^{\phi} \\
}
\end{equation*}

Observe that, from this result, we do not have a one-to-one correspondence between sections $\psi \in \Gamma(\rho_\R^r)$
solutions to equation \eqref{Chap05_eqn:UnifDynEqSect} and sections $\psi_\Lag \in \Gamma(\bar{\pi}^{2k-1})$
solutions to equation \eqref{Chap05_eqn:LagDynEqSect}, but only that every holonomic section $\psi$ solution
to the dynamical equations in the unified formalism can be projected to a holonomic section $\psi_\Lag$
solution to the Lagrangian equations. Nevertheless, recall that sections $\psi$ solution to equation
\eqref{Chap05_eqn:UnifDynEqSect} take values in the submanifold $\W_\Lag$, which is diffeomorphic to
$J^{2k-1}\pi$, and thus it is possible to establish an equivalence using the diffeomorphism $\rho_1^\Lag$.
In order to establish this equivalence, we first need the following technical result.

\begin{lemma}\label{Chap05_lemma:LagCartanFormsTechLemma}
The Poincar\'{e}-Cartan $1$-form satisfies $(\rho_1^\Lag)^*\Theta_\Lag = j_\Lag^*\Theta_r$.
\end{lemma}
\begin{proof}
A simple calculation proves this result:
\begin{align*}
(\rho_1^\Lag)^*\Theta_\Lag &= (\rho_1^r\circ j_\Lag)^*\Theta_\Lag
= (\rho_1 \circ \hat{h} \circ j_\Lag)^*\Theta_\Lag 
= (\rho_1 \circ \hat{h} \circ j_\Lag)^*(\widetilde{\Leg}^*\Theta_{k-1}) \\
&= (\widetilde{\Leg} \circ \rho_1 \circ \hat{h} \circ j_\Lag)^*\Theta_{k-1}
= (\rho_2 \circ \hat{h} \circ j_\Lag)^*\Theta_{k-1}
= (\hat{h} \circ j_\Lag)^*\Theta
= j_\Lag^*\Theta_r \, . \qedhere
\end{align*}
\end{proof}

\begin{remark}
Since the exterior derivative and the pull-back commute, Lemma \ref{Chap05_lemma:LagCartanFormsTechLemma}
also holds for the Poincar\'{e}-Cartan $2$-form.
\end{remark}

Now we can state the remaining part of the equivalence between the solutions of the Lagrangian and unified formalisms.

\begin{proposition}\label{Chap05_prop:LagrangianToUnifiedSect}
Let $\psi_\Lag \in \Gamma(\bar{\pi}^{2k-1})$ be a holonomic section solution to the dynamical equation
\eqref{Chap05_eqn:LagDynEqSect}. Then the section
$\psi = j_\Lag \circ (\rho_1^\Lag)^{-1} \circ \psi_\Lag \in \Gamma(\rho_\R^r)$ is holonomic and it is a
solution to the equation \eqref{Chap05_eqn:UnifDynEqSect}.
\end{proposition}
\begin{proof}
By definition, a section $\psi \in \Gamma(\rho_\R^r)$ is holonomic if the section
$\rho_1^r \circ \psi \in \Gamma(\bar{\pi}^{2k-1})$ is holonomic. Computing, we have
\begin{equation*}
\rho_1^r \circ \psi = \rho_1^r \circ j_\Lag \circ (\rho_1^\Lag)^{-1} \circ \psi_\Lag
= \rho_1^\Lag \circ (\rho_1^\Lag)^{-1} \circ \psi_\Lag = \psi_\Lag \, .
\end{equation*}
Hence, since $\psi_\Lag$ is holonomic, the section $\psi = j_\Lag \circ (\rho_1^\Lag)^{-1} \circ \psi_\Lag$
is holonomic in $\W_r$.

Now, since $j_\Lag \colon \W_\Lag \to \W_r$ is an embedding, for every vector field $X \in \vf(\W_r)$
tangent to $\W_\Lag$ there exists a unique vector field $Y \in \vf(\W_\Lag)$ which is $j_\Lag$-related
with $X$. Hence, let us assume that $X \in \vf(\W_r)$ is tangent to $\W_\Lag$. Then we have
\begin{equation*}
\psi^*\inn(X)\Omega_r = (j_\Lag \circ (\rho_1^\Lag)^{-1} \circ \psi_\Lag)^*\inn(X)\Omega_r
= ((\rho_1^\Lag)^{-1} \circ \psi_\Lag)^*\inn(Y)j_\Lag^*\Omega_r \, .
\end{equation*}
Applying Lemma \ref{Chap05_lemma:LagCartanFormsTechLemma} we obtain
\begin{equation*}
((\rho_1^\Lag)^{-1} \circ \psi_\Lag)^*\inn(Y)j_\Lag^*\Omega_r
= ((\rho_1^\Lag)^{-1} \circ \psi_\Lag)^*\inn(Y)(\rho_1^\Lag)^*\Omega_\Lag
= (\rho_1^\Lag \circ (\rho_1^\Lag)^{-1} \circ \psi_\Lag)^*\inn(Z)\Omega_\Lag = \psi_\Lag^*\inn(Z)\Omega_\Lag \, ,
\end{equation*}
where $Z \in \vf(J^{2k-1}\pi)$ is the unique vector field related with $Y$ by the diffeomorphism $\rho_1^\Lag$.
Hence, since $\psi_\Lag^*\inn(Z)\Omega_\Lag = 0$ for every $Z \in \vf(J^{2k-1}\pi)$ by hypothesis, we have
proved that the section $\psi = j_\Lag \circ (\rho_1^\Lag)^{-1} \circ \psi_\Lag \in \Gamma(\rho_\R^r)$
satisfies the equation
\begin{equation*}
\psi^*\inn(X)\Omega_r = 0 \, , \quad \mbox{for every } X \in \vf(\W_r) \mbox{ tangent to }\W_\Lag \, .
\end{equation*}
However, from Proposition \ref{Chap05_prop:UnifDynEqSectTangent} we know that if $\psi \in \Gamma(\rho_\R^r)$
is holonomic, then this last equation is equivalent to equation \eqref{Chap05_eqn:UnifDynEqSect}, that is,
\begin{equation*}
\psi^*\inn(X)\Omega_r = 0 \, , \quad \mbox{for every } X \in \vf(\W_r) \, . \qedhere
\end{equation*}
\end{proof}

Finally, let us compute the local expression of the equation for the section $\psi_\Lag \in \Gamma(\bar{\pi}^{2k-1})$.
Suppose that $\psi \in \Gamma(\rho_\R^r)$ is given locally by $\psi(t) = (t,q_i^A(t),q_j^A(t),p_A^i(t))$,
$0 \leqslant i \leqslant k-1$, $k \leqslant j \leqslant 2k-1$. Since $\psi$ is holonomic and a solution
to equation \eqref{Chap05_eqn:UnifDynEqSect}, it must satisfy equations \eqref{Chap05_eqn:UnifDynEqSectHolonomyLocalPart},
\eqref{Chap05_eqn:UnifDynEqSectLocal1} and \eqref{Chap05_eqn:UnifDynEqSectLocal2}. The first group of
equations is automatically satisfied because of the holonomy assumption. Now, bearing in mind that the section
$\psi$ takes values in the submanifold $\W_\Lag = \graph(\Leg)$, equations \eqref{Chap05_eqn:UnifDynEqSectLocal1} and
\eqref{Chap05_eqn:UnifDynEqSectLocal2} can be $\rho_1^r$-projected to $J^{2k-1}\pi$, thus giving the following equations
for the section $\psi_\Lag = \rho_1^r \circ \psi$
\begin{equation*}
\restric{\derpar{L}{q_0^A}}{\psi_\Lag} - \restric{\frac{d}{dt}\derpar{L}{q_1^A}}{\psi_\Lag}
+ \restric{\frac{d^2}{dt^2}\derpar{L}{q_2^A}}{\psi_\Lag} + \ldots +
(-1)^k \restric{\frac{d^k}{dt^k}\derpar{L}{q_k^A}}{\psi_\Lag} = 0 \, .
\end{equation*}
Finally, bearing in mind that $\psi_\Lag$ is holonomic in $J^{2k-1}\pi$, there exists a section
$\phi \in \Gamma(\pi)$, whose local expression is $\phi(t) = (t,q_0^A(t))$, such that $j^{2k-1}\phi = \psi_\Lag$, and
thus the above equations can be rewritten in the following form
\begin{equation}\label{Chap05_eqn:EulerLagrangeEquations}
\restric{\derpar{L}{q_0^A}}{j^{2k-1}\phi} - \restric{\frac{d}{dt}\derpar{L}{q_1^A}}{j^{2k-1}\phi}
+ \restric{\frac{d^2}{dt^2}\derpar{L}{q_2^A}}{j^{2k-1}\phi} + \ldots +
(-1)^k \restric{\frac{d^k}{dt^k}\derpar{L}{q_k^A}}{j^{2k-1}\phi} = 0 \, .
\end{equation}
Therefore, we obtain the Euler-Lagrange equations for a $k$th-order non-autonomous system. As stated before,
equations \eqref{Chap05_eqn:EulerLagrangeEquations} may or may not be compatible, and in this last case a
constraint algorithm must be used in order to obtain a submanifold $S_f \hookrightarrow J^{2k-1}\pi$
(if such a submanifold exists) where the equations can be solved.

\subsubsection{Dynamical equations for vector fields}

Now, using the results stated at the beginning of the Section, we can state the Lagrangian dynamical
equations for vector fields, and recover a vector field solution to the Lagrangian equations starting
from a vector field solution to the equations in the unified formalism.

The \textsl{$k$th-order Lagrangian problem for vector fields} associated with the system
$(J^{2k-1}\pi,\Lag)$ consists in finding holonomic vector fields $X_\Lag \in \vf(J^{2k-1}\pi)$ such that
\begin{equation}\label{Chap05_eqn:LagDynEqVF}
\inn(X_\Lag)\Omega_\Lag = 0 \quad ; \quad \inn(X)(\bar{\pi}^{2k-1})^*\eta \neq 0 \, .
\end{equation}

\begin{remark}
As in the first-order case described in Section \ref{Chap02_sec:LagrangianNonAutonomousFirstOrder},
the second equation in \eqref{Chap05_eqn:LagDynEqVF} is a transversality condition for the vector field $X_\Lag$
with respect to the projection onto $\R$, and the non-zero value is usually fixed to $1$, thus giving the
following equations
\begin{equation*}
\inn(X)\Omega_\Lag = 0 \quad ; \quad \inn(X)(\bar{\pi}^{2k-1})^*\eta = 1 \, .
\end{equation*}
\end{remark}

First we need to state a correspondence between the set of vector fields in $\W_r$ tangent to $\W_\Lag$ and
the set of vector fields in $J^{2k-1}\pi$.

\begin{lemma}\label{Chap05_lemma:LagCorrespondenceVF}
Let $X \in \vf(\W_r)$ be a vector field tangent to $\W_\Lag$. Then there exists a unique vector field
$X_\Lag \in \vf(J^{2k-1}\pi)$ such that $X_\Lag \circ \rho_1^r \circ j_\Lag = \Tan\rho_1^r \circ X \circ j_\Lag$.
\end{lemma}
\begin{proof}
Since $X$ is tangent to $\W_\Lag$, there exists a unique $X_o \in \vf(\W_\Lag)$ such that
$\Tan j_\Lag \circ X_o = X \circ j_\Lag$. Furthermore, since $\rho_1^\Lag$ is a diffeomorphism, there
is a unique vector field $X_\Lag \in \vf(J^{2k-1}\pi)$ which is $\rho_1^\Lag$-related with $X_o$;
that is, $X_\Lag \circ \rho_1^\Lag = \Tan\rho_1^\Lag \circ X_o$. Then we have
\begin{equation*}
X_\Lag \circ \rho_1^r \circ j_\Lag = X_\Lag \circ \rho_1^\Lag = \Tan\rho_1^\Lag \circ X_o
= \Tan\rho_1^r \circ \Tan j_\Lag \circ X_o = \Tan \rho_1^r \circ X \circ j_\Lag \, . \qedhere
\end{equation*}
\end{proof}

The above result states that for every vector field $X \in \vf(\W_r)$ tangent to $\W_\Lag$ there exists
a unique vector field $X_\Lag \in \vf(J^{2k-1}\pi)$ such that the following diagram commutes
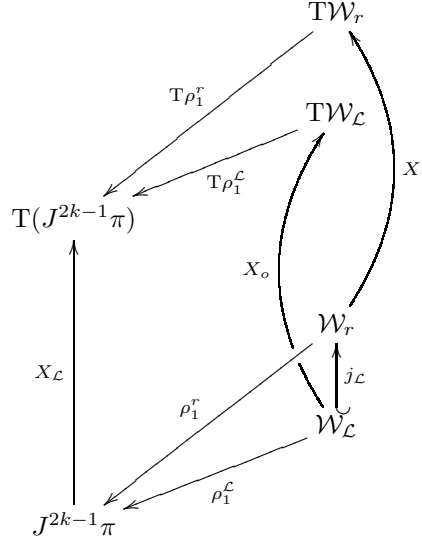
\begin{equation*}
\xymatrix{
\ & \ & \Tan\W_r \ar[ddll]_-{\Tan\rho_1^r} \\
\ & \ & \Tan\W_\Lag \ar[dll]^-{\Tan\rho_1^\Lag} \\
\Tan(J^{2k-1}\pi) & \ & \ \\
\ & \ & \W_r \ar[ddll]_{\rho_1^r} \ar@/_1.8pc/[uuu]_-{X} \\
\ & \ & \W_\Lag \ar[dll]^-{\rho_1^\Lag} \ar@/^1.8pc/[uuu]^-{X_o}|(.225){\hole} \ar@{_{(}->}[u]_{j_\Lag} \\
J^{2k-1}\pi \ar[uuu]^-{X_\Lag} & \ & \
}
\end{equation*}

As a consequence of Lemma \ref{Chap05_lemma:LagCorrespondenceVF}, we can establish a bijective correspondence
between the set of vector fields in $\W_r$ tangent to $\W_\Lag$ solution to the dynamical equations in the
unified formalism and the set of vector fields in $J^{2k-1}\pi$ solution to the Lagrangian dynamical equations
stated above. In particular, we have the following result.

\begin{theorem}\label{Chap05_thm:UnifiedToLagrangianVF}
Let $X \in \vf(\W_r)$ be a holonomic vector field solution to equations \eqref{Chap05_eqn:LagDynEqVF} and tangent to
$\W_\Lag$ (at least on the points of a submanifold $\W_f \hookrightarrow \W_\Lag$). Then there exists a
unique holonomic vector field $X_\Lag \in \vf(J^{2k-1}\pi)$ which is a solution to the equations
\eqref{Chap05_eqn:LagDynEqVF} (at least on the points of $S_f = \rho_1^r(\W_f)$).

\noindent Conversely, if $X_\Lag \in \vf(J^{2k-1}\pi)$ is a holonomic vector field, which is a solution to equations
\eqref{Chap05_eqn:LagDynEqVF} (at least on the points of a submanifold $S_f \hookrightarrow J^{2k-1}\pi$),
then there exists a unique holonomic vector field $X \in \vf(\W_r)$ tangent to $\W_\Lag$ which is a solution to
equations \eqref{Chap05_eqn:UnifDynEqVF} (at least on the points of $\W_f = (\rho_1^r)^{-1}(S_f) \hookrightarrow \W_r$).
\end{theorem}
\begin{proof}
Applying Lemma \ref{Chap05_lemma:LagCartanForms}, and taking $X_\Lag \in \vf(J^{2k-1}\pi)$ as the unique
vector field given by Lemma \ref{Chap05_lemma:LagCorrespondenceVF}, we have:
\begin{equation*}
\inn(X)\Omega_r = \inn(X)(\rho_1^r)^*\Omega_\Lag = (\rho_1^r)^*\inn(X_\Lag)\Omega_\Lag \, .
\end{equation*}
However, as $\rho_1^r$ is a surjective submersion, this is equivalent to
\begin{equation*}
\restric{\inn(X_\Lag)\Omega_\Lag}{\rho_1^r(\W_r)} = \restric{\inn(X_\Lag)\Omega_\Lag}{J^{2k-1}\pi} = \inn(X_\Lag)\Omega_\Lag \, ,
\end{equation*}
since $\rho_1^r(\W_r) = J^{2k-1}\pi$. Hence, we have proved that $\inn(X)\Omega_r = 0$ if, and only if,
$\inn(X_\Lag)\Omega_\Lag = 0$. The same reasoning proves that $\inn(X)(\rho_\R^r)^*\eta \neq 0$
if, and only if $\inn(X_\Lag)(\bar{\pi}^{2k-1})^*\eta \neq 0$.

In order to prove that $X_\Lag$ is holonomic, we compute its local expression in coordinates. From the
local expression \eqref{Chap05_eqn:UnifHolonomicVFSolution} for the vector field $X$ (where the functions
$F_{2k-1}^A$ are the solutions of equations \eqref{Chap05_eqn:UnifDynEqVFTangencyFinalLocal}), and using Lemma
\ref{Chap05_lemma:LagCorrespondenceVF}, we obtain that the local expression of the vector field
$X_\Lag \in \vf(J^{2k-1}\pi)$ is
\begin{equation*}
X_\Lag = \derpar{}{t} + \sum_{i=0}^{2k-2}q_{i+1}^A\derpar{}{q_i^A} + F_{2k-1}^A\derpar{}{q_{2k-1}^A} \, ,
\end{equation*}
which is the local expression for a holonomic vector field in $J^{2k-1}\pi$. Reversing this reasoning
we prove that if $X_\Lag$ is holonomic, then the vector field $X \in \vf(\W_r)$ is holonomic.
\end{proof}

\begin{remark}
It is important to point out that, if $\Lag$ is not a regular $k$th-order Lagrangian density, then
$X$ is holonomic of type $k$ in $\W_r$, but not necessarily holonomic of type $1$, as it is shown in
equations \eqref{Chap05_eqn:UnifDynEqVFHolonomyLocalPart2}. When translated to the Lagrangian setting,
this means that $X_\Lag$ may be a solution to the Lagrangian equations for vector fields, but the trajectories
given by its integral sections are not solutions to the dynamical system (the sections solution to the dynamical
problem must be holonomic, but the integral sections of $X_\Lag$ are only holonomic of type $k$). Therefore,
the holonomy condition can not be dropped from the statement of the problem, since for singular Lagrangians
this must be imposed as an additional condition. This constitutes a relevant difference from the case of
first-order dynamical systems, where this condition ($X_\Lag$ being holonomic) is obtained straightforwardly
in the unified formalism.

For singular $k$th-order Lagrangian densities, only in the most interesting cases can we assure the existence
of a submanifold $\W_f \hookrightarrow \W_\Lag$ and vector fields $X \in \vf(\W_r)$ tangent to $\W_f$ which
are solutions to equations \eqref{Chap05_eqn:UnifDynEqVFSingular}. Then, considering the submanifold
$S_f = \rho_1^\Lag(\W_f) \hookrightarrow J^{2k-1}\pi$, in the best cases we have that those holonomic vector fields
$X_\Lag$ exist, perhaps on another submanifold $S^h_f \hookrightarrow S_f$ where they are tangent, and are solutions
to equations
\begin{equation*}
\restric{\inn(X_\Lag)\Omega_\Lag}{S^h_f} = 0 \quad ; \quad \restric{\inn(X_\Lag)(\bar{\pi}^{2k-1})^*\eta}{S^h_f} = 1 \, .
\end{equation*}
\end{remark}

Notice that Theorem \ref{Chap05_thm:UnifiedToLagrangianVF} states that there is a one-to-one correspondence
between vector fields $X \in \vf(\W_r)$ solutions to equations \eqref{Chap05_eqn:UnifDynEqSect} and vector fields
$X_\Lag \in \vf(J^{2k-1}\pi)$ solutions to \eqref{Chap05_eqn:LagDynEqVF}, but not uniqueness. In fact, we cannot
assure uniqueness of the vector field $X_\Lag$ unless the Lagrangian density is regular, as we can see in the following result.

\begin{corollary}
If the $k$th-order Lagrangian density $\Lag\in\df^{1}(J^{k}\pi)$ is regular, then there is a unique
holonomic vector field, $X_\Lag \in \vf(J^{2k-1}\pi)$, which is a solution to equations \eqref{Chap05_eqn:LagDynEqVF}.
\end{corollary}
\begin{proof}
If the Lagrangian density $\Lag \in \df^{1}(J^{k}\pi)$ is regular, using Proposition
\ref{Chap05_prop:UnifRegLagUniqueVF}, there exists a unique holonomic vector field, $X \in \vf(\W_r)$,
solution to equations \eqref{Chap05_eqn:UnifDynEqVF} and tangent to $\W_\Lag$. Then, using Theorem
\ref{Chap05_thm:UnifiedToLagrangianVF}, related to this unique vector field in $\W_r$ there is a unique vector
field $X_\Lag \in \vf(J^{2k-1}\pi)$, which is holonomic and is a solution to equations \eqref{Chap05_eqn:LagDynEqVF}.
\end{proof}

In other words, uniqueness of the vector field $X_\Lag$ is a consequence of uniqueness of $X$.

\subsubsection{Equivalence of the dynamical equations in the Lagrangian formalism}

Finally, we state the equivalence Theorem in the Lagrangian formalism, which is the analogous to
Theorem \ref{Chap05_thm:UnifEquivalenceTheorem}. This result is a straightforward consequence of
Theorems \ref{Chap05_thm:UnifEquivalenceTheorem}, \ref{Chap05_thm:UnifiedToLagrangianVar} and
\ref{Chap05_thm:UnifiedToLagrangianVF}, and of Propositions \ref{Chap05_prop:UnifiedToLagrangianSect}
and \ref{Chap05_prop:LagrangianToUnifiedSect}, and hence we omit the proof.

\begin{theorem}\label{Chap05_thm:LagEquivalenceTheorem}
The following assertions on a section $\phi \in \Gamma(\pi)$ are equivalent.
\begin{enumerate}
\item $\phi$ is a solution to the Lagrangian variational problem.
\item $j^{2k-1}\phi$ is a solution to equation \eqref{Chap05_eqn:LagDynEqSect}, that is,
\begin{equation*}
(j^{2k-1}\phi)^*\inn(Y)\Omega_\Lag = 0 \, , \quad \mbox{for every } Y \in \vf(J^{2k-1}\pi) \, .
\end{equation*}
\item In natural coordinates, if $\phi = (t,q_0^A(t))$, then $j^{2k-1}\phi = (t,q_0^A(t),q_1^A(t),\ldots,q_{2k-1}^A(t))$
is a solution to the $k$th-order Euler-Lagrange equations given by \eqref{Chap05_eqn:EulerLagrangeEquations}, that is,
\begin{equation*}
\restric{\derpar{L}{q_0^A}}{j^{2k-1}\phi} - \restric{\frac{d}{dt}\derpar{L}{q_1^A}}{j^{2k-1}\phi}
+ \restric{\frac{d^2}{dt^2}\derpar{L}{q_2^A}}{j^{2k-1}\phi} + \ldots +
(-1)^k \restric{\frac{d^k}{dt^k}\derpar{L}{q_k^A}}{j^{2k-1}\phi} = 0 \, .
\end{equation*}
\item Denoting $\psi_\Lag = j^{2k-1}\phi$, then $\psi_\Lag$ is a solution to the equation
\begin{equation*}
\inn(\dot{\psi}_\Lag)(\Omega_\Lag \circ \psi_\Lag) = 0 \, ,
\end{equation*}
where $\dot{\psi}_\Lag \colon \R \to \Tan(J^{2k-1}\pi)$ is the canonical lifting of $\psi_\Lag$ to the tangent bundle.
\item $j^{2k-1}\phi$ is an integral curve of a vector field contained in a class of $\bar{\pi}^{2k-1}$-transverse
holonomic vector fields, $\left\{ X_\Lag \right\} \subset \vf(J^{2k-1}\pi)$, satisfying the first equation in
\eqref{Chap05_eqn:LagDynEqVF}, that is,
\begin{equation*}
\inn(X_\Lag)\Omega_\Lag = 0 \, .
\end{equation*}
\end{enumerate}
\end{theorem}


\section{The Hamiltonian formalism}
\label{Chap05_sec:UnifiedToHamiltonian}

In order to describe the Hamiltonian formalism on the basis of the unified one, we must distinguish
between the regular and non-regular cases. In fact, the only ``non-regular'' case that we consider is the
almost-regular one, so we first need to generalize the concept of \textsl{almost-regular Lagrangian density}
to the higher-order non-autonomous setting. Moreover, we must introduce the dynamical information in the
Hamiltonian formalism from the dynamics in the unified setting, that is, from the Hamiltonian $\mu_\W$-section
$\hat{h} \in \Gamma(\mu_\W)$.

\subsection{Geometrical setting}

Let $\widetilde{\Leg} \colon J^{2k-1}\pi \to \Tan^*(J^{k-1}\pi)$ and $\Leg \colon J^{2k-1}\pi \to J^{k-1}\pi^*$ be
the extended and restricted Legendre-Ostrogradsky maps given by Proposition \ref{Chap05_prop:UnifGraphLegendreOstrogradskyMap}
and Corollary \ref{Chap05_corol:UnifGraphExtendedLegendreOstrogradskyMapSect}, respectively.
Then, let us denote by $\widetilde{\P} = \Im(\widetilde{\Leg}) = \widetilde{\Leg}(J^{2k-1}\pi)
\stackrel{\tilde{\jmath}}{\hookrightarrow} \Tan^*(J^{k-1}\pi)$ the image of the extended Legendre-Ostrogradsky map;
and by $\P = \Im(\Leg) = \Leg(J^{2k-1}\pi) \stackrel{\jmath}{\hookrightarrow} J^{k-1}\pi^*$ the image of the
restricted Legendre-Ostrogradsky map. Let $\bar{\pi}_\P = \bar{\pi}_{J^{k-1}\pi}^r \circ \jmath \colon \P \to \R$
be the natural projection, and $\Leg_o$ the map defined by $\Leg = \jmath \circ \Leg_o$.

\begin{remark}
If the Poincar\'{e}-Cartan $1$-form is defined without using the Legendre-Ostrogradsky map, as we have seen
in \eqref{Chap05_eqn:LagCartan1FormAlternativeDef}, then we can define the extended Legendre-Ostrogradsky
map in an alternative, but equivalent, way. In particular, since $\Theta_\Lag \in \df^{1}(J^{2k-1}\pi)$
is a $\pi^{2k-1}_{k-1}$-semibasic form, we can define a bundle morphism
$\widetilde{\Leg} \colon J^{2k-1}\pi \to \Tan^*(J^{k-1}\pi)$ over $J^{k-1}\pi$ as follows: for every
$v \in \Tan(J^{2k-1}\pi)$,
\begin{equation*}
\langle \Theta_\Lag, v \rangle = \left\langle \Tan\pi^{2k-1}_{k-1}(v) , \widetilde{\Leg}(\tau_{J^{2k-1}\pi}(v)) \right\rangle \, ,
\end{equation*}
where $\tau_{J^{2k-1}\pi} \colon \Tan(J^{2k-1}\pi) \to J^{2k-1}\pi$ is the canonical submersion. The map
$\widetilde{\Leg}$ is the extended Legendre-Ostrogradsky map. From this, the restricted Legendre-Ostrogradsky
map $\Leg \colon J^{2k-1}\pi \to J^{k-1}\pi^*$ is defined by composition with the canonical quotient map
$\mu \colon \Tan^*(J^{k-1}\pi) \to J^{k-1}\pi^*$, that is, $\Leg = \mu \circ \widetilde{\Leg}$.
All the remarks and properties for both Legendre-Ostrogradsky maps stated in Section \ref{Chap05_sec:UnifDynamicalEquations}
hold, and in addition we have $\Theta_\Lag = \widetilde{\Leg}^*\Theta_{k-1}$ and $\Omega_\Lag = \widetilde{\Leg}^*\Omega_{k-1}$.

Observe that these are the usual definitions of the Legendre-Ostrogradsky maps, as we have seen in Sections
\ref{Chap02_sec:HamiltonianAutonomousFirstOrder}, \ref{Chap02_sec:HamiltonianAutonomousHigherOrder} and
\ref{Chap02_sec:NonAutonomousHamiltonian}, thus justifying the notation and terminology adopted
in Section \ref{Chap05_sec:UnifDynamicalEquations}.
\end{remark}

With the previous notations, we can give the following definition.

\begin{definition}
A $k$th-order Lagrangian density $\Lag \in \df^{1}(J^{k}\pi)$ is \textnormal{almost-regular} if
\begin{enumerate}
\item $\P$ is a closed submanifold of $J^{k-1}\pi^*$.
\item $\Leg$ is a submersion onto its image.
\item For every $j^{2k-1}_t\phi \in J^{2k-1}\pi$, the fibers $\Leg^{-1}(\Leg(j^{2k-1}_t\phi))$ are connected
submanifolds of $J^{2k-1}\pi$.
\end{enumerate}
\end{definition}

Observe that, in particular, if $\Lag$ is a $k$th-order almost-regular Lagrangian density, then $\Leg_o$ is a
surjective submersion, and thus admits global sections on $\P$, that is, maps $\gamma \colon \P \to J^{2k-1}\pi$ satisfying
$\Leg_o \circ \gamma = \Id_{\P}$. Let $\Gamma(\Leg_o)$ be the set of sections of $\Leg_o$.

As a consequence of Proposition \ref{Chap05_prop:UnifRankBothLegendreMaps}, we have that $\widetilde{\P}$
is diffeomorphic to $\P$. This diffeomorphism is $\widetilde{\mu} = \mu \circ \tilde{\jmath} \colon \widetilde{\P} \to \P$.
This enables us to state the following result.

\begin{lemma}\label{Chap05_lemma:HamProjectedHamiltonianSection}
If the $k$th-order Lagrangian density $\Lag \in \df^{1}(J^{k}\pi)$ is, at least, almost-regular, the Hamiltonian
section $\hat{h} \in \Gamma(\mu_\W)$ induces a Hamiltonian section $h \in \Gamma(\mu)$ defined by
\begin{equation*}
h([\omega]) = (\rho_2 \circ \hat{h})([(\rho_2^r)^{-1}(\jmath([\omega]))]),\quad \mbox{for every } [\omega] \in \P.
\end{equation*}
\end{lemma}
\begin{proof}
It is clear that, given $[\omega] \in \P$, the section $\hat{h}$ maps every point
$(j^{2k-1}_t\phi,[\omega]) \in (\rho^r_2)^{-1}([\omega])$ into $\rho_2^{-1}[\rho_2(\hat{h}(j^{2k-1}_t\phi,[\omega]))]$.
So we have the diagram
\begin{equation*}
\xymatrix{
\widetilde{\mathcal{P}} \ar[rr]^{\tilde{\jmath}} \ar[d]^-{\tilde{\mu}} & \ & \Tan^*(J^{k-1}\pi) \ar[d]^-{\mu} & \ & \W \ar[d]_-{\mu_\W} \ar[ll]_-{\rho_2} \\
\mathcal{P} \ar[rr]^-{\jmath} \ar@{-->}[urr]^-{h}& \ & J^{k-1}\pi^* & \ & \W_r \ar@/_0.7pc/[u]_{\hat{h}} \ar[ll]_-{\rho_2^r}
}
\end{equation*}
Thus, the crucial point is the $\rho_2$-projectability of the local function $\hat{H}$. However, since a
local base for $\ker\Tan\rho_2$ is given by \eqref{Chap05_eqn:UnifCanonicalFormsKernelLocal}, that is,
\begin{equation*}
\ker\Tan\rho_2 = \left\langle \derpar{}{q_k^A},\ldots,\derpar{}{q_{2k-1}^A} \right\rangle \, ,
\end{equation*}
we have that $\hat{H}$ is $\rho_2$-projectable if and only if
\begin{equation*}
p_{A}^{k-1} = \derpar{L}{q_k^A} \, .
\end{equation*}
This condition is fulfilled whenever $[\omega] \in \mathcal{P}$, which implies that
$\rho_2(\hat{h}((\rho_2^r)^{-1}([\omega]))) \in \widetilde{\mathcal{P}}$.
\end{proof}

As in the unified setting, the Hamiltonian $\mu$-section is specified by the local Hamiltonian function
$H \in \Cinfty(\P)$, that is,
\begin{equation*}
h(t,q_i^A,p^i_A) = (t,q_i^A,-H,p_A^i) \, .
\end{equation*}
Using the Hamiltonian $\mu$-section we define the \textsl{Hamilton-Cartan forms}
$\Theta_h = h^*\Theta_{k-1} \in \df^{1}(\P)$ and $\Omega_h = h^*\Omega_{k-1} \in \df^{2}(\P)$.
Observe that $\Leg_o^*\Theta_h = \Theta_\Lag$ and $\Leg_o^*\Omega_h = \Omega_\Lag$. Then, the triple
$(\P,\Omega_h,\bar{\pi}_\P^*\eta)$ is the \textsl{$k$th-order non-autonomous Hamiltonian system}
associated with $(\W_r,\Omega_r,(\rho_\R^r)^*\eta)$.

\begin{remark}
The Hamiltonian $\mu$-section can be defined in an equivalent way, without passing
through the unified formalism, as follows: $h = \tilde{\jmath} \circ \widetilde{\mu}^{-1}$.
\end{remark}

\subsection{Regular and hyperregular Lagrangian densities}
\label{Chap05_sec:HamiltonianRegularCase}

Now we analyze the case when $\Lag$ is a $k$th-order regular Lagrangian density, although by simplicity
we focus on the hyperregular case (the regular case is recovered from this by restriction on the corresponding
open sets where the restricted Legendre-Ostrogradsky map is a local diffeomorphism). This means that the phase
space of the system is $J^{k-1}\pi^*$ (or the corresponding open sets).

In this case we have $\P = J^{k-1}\pi^*$, $\Leg_o = \Leg$ and $\bar{\pi}_\P = \bar{\pi}_{J^{k-1}\pi}$,
and the $k$th-order non-autonomous Hamiltonian system is now described by the triple
$(J^{k-1}\pi^*,\Omega_h,(\bar{\pi}_{J^{k-1}\pi}^r)^*\eta)$.
In addition, the Hamiltonian $\mu$-section $h$ can be defined as $h = \widetilde{\Leg} \circ \Leg^{-1}$,
and we can give the explicit coordinate expression for the local Hamiltonian function $H$, which is
\begin{equation}\label{Chap05_eqn:HamRegHamiltonianFunctionLocal}
H(t,q_i^A,p_A^i) = \sum_{i=0}^{k-2}p_A^iq_{i+1}^A + p_A^{k-1}(\Leg^{-1})^*q_k^A - (\pi_{k}^{2k-1} \circ \Leg^{-1})^*L(t,q_i^A) \, .
\end{equation}
Moreover, in this case we can also give the coordinate expression of the Hamilton-Cartan forms:
bearing in mind the local expression of $\Theta_{k-1}$ and $\Omega_{k-1}$ given in Example
\ref{Chap01_exa:CotangentBundle}, the coordinate expression of the forms $\Theta_{h}$ and $\Omega_{h}$ is
\begin{equation}\label{Chap05_eqn:HamHamiltonCartanForms}
\Theta_h = p_A^i\d q_i^A - H\d t \quad ; \quad
\Omega_h = \d q_i^A \wedge \d p_A^i + \d H \wedge \d t \, .
\end{equation}

In this setting, the fundamental result is the following, which is the analogous to Proposition
\ref{Chap05_prop:LagRho1LDiffeomorphism} in the Hamiltonian formalism.

\begin{proposition}\label{Chap05_prop:HamRegRho2LDiffeomorphism}
If the $k$th-order Lagrangian density $\Lag \in \df^{1}(J^{k}\pi)$ is hyperregular, then the map
$\rho_2^\Lag = \rho_2^r \circ j_\Lag \colon \W_\Lag \to J^{k-1}\pi^*$ is a diffeomorphism.
\end{proposition}
\begin{proof}
The following diagram is commutative
\begin{equation*}
\xymatrix{
\ & \ & \W_r \ar@/_1.3pc/[ddll]_-{\rho_1^r} \ar@/^1.3pc/[ddrr]^-{\rho_2^r} \ & \ \\
\ & \ & \W_\Lag \ar[dll]_-{\rho_1^\Lag} \ar[drr]^-{\hat{\rho}_2^\Lag} \ar@{^{(}->}[u]^{j_\Lag} & \ & \ \\
J^{2k-1}\pi \ar[rrrr]^-{\Leg} & \ & \ & \ & J^{k-1}\pi^*
}
\end{equation*}
that is, we have $\rho_2^\Lag = \rho_2^r \circ j_\Lag = \Leg \circ \rho_1^\Lag$. Now, by Proposition
\ref{Chap05_prop:LagRho1LDiffeomorphism}, the map $\rho_1^\Lag$ is a diffeomorphism. In addition, as $\Lag$
is hyperregular, the map $\Leg$ is also a diffeomorphism, and thus $\rho_2^\Lag$ is a composition of
diffeomorphisms, and hence a diffeomorphism itself.
\end{proof}

This last result allows us to recover the Hamiltonian formalism in the same way we recovered the Lagrangian one
in the previus Section: using the diffeomorphism to define a correspondence between the solutions of both equations.

\subsubsection{Variational principle}

Given the Hamiltonian system $(J^{k-1}\pi^*,\Omega_h,(\bar{\pi}_{J^{k-1}\pi}^r)^*\eta)$,
let $\Gamma(\bar{\pi}_{J^{k-1}\pi}^r)$ be the set of sections of $\bar{\pi}_{J^{k-1}\pi}^r$.
Consider the functional
\begin{equation*}\label{Chap05_eqn:HamRegVariationalFunctionalDef}
\begin{array}{rcl}
\mathbf{H} \colon \Gamma(\bar{\pi}_{J^{k-1}\pi}^r) & \longrightarrow & \R \\
\psi_h & \longmapsto & \displaystyle \int_\R \psi_h^*\Theta_h
\end{array} \, ,
\end{equation*}
where the convergence of the integral is assumed.

\begin{definition}
The \textnormal{$k$th-order Hamiltonian variational problem}
(or \textnormal{generalized Hamilton-Jacobi variational problem}) for the $k$th-order Hamiltonian system
$(J^{k-1}\pi^*,\Omega_h,(\bar{\pi}_{J^{k-1}\pi}^r)^*\eta)$ is the search for the critical (or stationary)
sections of the functional $\mathbf{H}$ with respect to the variations of $\psi_h$ given by
$(\psi_h)_s = \sigma_s \circ \psi_h$, where $\left\{ \sigma_s \right\}$ is a local one-parameter group of
any compact-supported $Z \in \vf^{V(\bar{\pi}_{J^{k-1}\pi}^r)}(J^{k-1}\pi^*)$; that is
\begin{equation}\label{Chap05_eqn:HamRegDynEqVar}
\restric{\frac{d}{ds}}{s=0}\int_\R (\psi_h)^*_s\Theta_h = 0 \, .
\end{equation}
\end{definition}

\begin{theorem}\label{Chap05_thm:UnifiedToHamiltonianRegVar}
Let $\psi \in \Gamma(\rho_\R^r)$ be a critical section for the Lagrangian-Hamiltonian variational problem
given by the functional $\mathbf{LH}$. Then, the section $\psi_h = \rho_2^r \circ \psi \in \Gamma(\bar{\pi}_{J^{k-1}\pi}^r)$
is a critical section for the Hamiltonian variational problem given by the functional $\mathbf{H}$.

\noindent Conversely, given a section $\psi_h \in \Gamma(\bar{\pi}_{J^{k-1}\pi}^r)$ solution to the Hamiltonian
variational problem, the section $\psi = j_\Lag \circ (\rho_2^\Lag)^{-1} \circ \psi_h \in \Gamma(\rho_\R^r)$ is a solution
to the Lagrangian-Hamiltonian variational problem.
\end{theorem}
\begin{proof}
The proof of this result is analogous to the proof given for Theorem \ref{Chap05_thm:UnifiedToLagrangianVar}.

Since $\rho_2^r$ is a submersion, for every compact-supported vector field
$X \in \vf^{V(\bar{\pi}_{J^{k-1}\pi}^r)}(J^{k-1}\pi^*)$ there exist compact-supported vector fields
$Y \in \vf^{V(\rho_\R^r)}(\W_r)$ such that $X$ and $Y$ are $\rho_2^r$-related.
We also denote by $\left\{ \sigma_s \right\}$ a local one-parameter group for the compact-supported vector fields
$Y \in \vf^{V(\rho_\R^r)}(\W_r)$. Then, using this particular choice of $\rho_2^r$-related vector fields, we have
\begin{align*}
\restric{\frac{d}{ds}}{s=0}\int_\R (\psi_h)_s^*\Theta_h
&= \restric{\frac{d}{ds}}{s=0}\int_\R(\sigma_s \circ \psi_h)^*\Theta_h
= \restric{\frac{d}{ds}}{s=0}\int_\R\psi_h^*\sigma_s^*\Theta_h \\
&= \int_\R\psi_h^*\Lie(X)\Theta_h
= \int_\R\psi_h^*(\inn(X)\d\Theta_h + \d\inn(X)\Theta_h) \\
&= \int_\R \psi^*(\rho_2^r)^*(\inn(X)\d\Theta_h + \d\inn(X)\Theta_h)
= \int_\R \psi^*(\inn(Y)\d\Theta_r + \d\inn(Y)\Theta_r) \\
&= \int_\R \psi^*\Lie(Y)\Theta_r
= \restric{\frac{d}{ds}}{s=0}\int_\R\psi^*\sigma_s^*\Theta_r
= \restric{\frac{d}{ds}}{s=0} \int_\R \psi_s^*\Theta_r = 0 \, ,
\end{align*}
since $\psi$ is a critical section for the Lagrangian-Hamiltonian variational problem.

Conversely, if we have a section $\psi_h$ which is a solution to the Hamiltonian variational problem,
then we can construct a section $\psi = j_\Lag \circ (\rho_2^\Lag)^{-1} \circ \psi_h \in \Gamma(\rho_\R^r)$.
Then, the above reasoning also shows that if $\psi_h$ is a solution to
the Hamiltonian variational problem, then $\psi$ is a solution to the Lagrangian-Hamiltonian variational problem.
\end{proof}

\subsubsection{Dynamical equations for sections}

As in Section \ref{Chap05_sec:UnifiedToLagrangian}, using the results given in previous Sections,
we can now state the Hamiltonian equations for sections in the hyperregular case, and recover the
Hamiltonian solutions in $J^{k-1}\pi^*$ from the solutions in the unified setting.

The \textsl{$k$th-order (hyperregular) Hamiltonian problem for sections} associated with the cosymplectic
Hamiltonian system $(J^{k-1}\pi^*,\Omega_h,(\bar{\pi}_{J^{k-1}\pi}^r)^*\eta)$
consists in finding sections $\psi_h \in \Gamma(\bar{\pi}_{J^{k-1}\pi}^r)$ characterized by the equation
\begin{equation}\label{Chap05_eqn:HamRegDynEqSect}
\psi_h^*\inn(Y)\Omega_h = 0 \, , \quad \mbox{for every } Y \in \vf(J^{k-1}\pi^*) \, .
\end{equation}

\begin{proposition}\label{Chap05_prop:UnifiedToHamiltonianRegSect}
Let $\Lag \in \df^{1}(J^{k}\pi)$ be a hyperregular $k$th-order Lagrangian density, and
$\psi \in \Gamma(\rho_\R^r)$ a section solution to equation \eqref{Chap05_eqn:UnifDynEqSect}. Then
$\psi_h = \rho_2^r \circ \psi \in \Gamma(\bar{\pi}_{J^{k-1}\pi}^r)$ is a solution to equation
\eqref{Chap05_eqn:HamRegDynEqSect}.
\end{proposition}
\begin{proof}
The proof of this result is analogous to the proof given for Proposition \ref{Chap05_prop:UnifiedToLagrangianSect}.

As $\rho_2^r$ is a submersion, for every $Y \in \vf(J^{k-1}\pi^*)$ there exist some $Z \in \vf(\W_r)$ which is
$\rho_2^r$-related with $Y$. Observe that this vector field $Z$ is not unique, since $Z + Z_o$, with $Z_o$ any
$\rho_2^r$-vertical vector field, is also $\rho_2^r$-related with $Y$. Thus, using this particular choice of
$\rho_2^r$-related vector fields, we have
\begin{equation*}
\psi_h^*\inn(Y)\Omega_h = (\rho_2^r \circ \psi)^*\inn(Y)\Omega_h =
\psi^*((\rho_2^r)^*\inn(Y)\Omega_h) = \psi^*(\inn(Z)(\rho_2^r)^*\Omega_h) =
\psi^*i(Z)\Omega_r \, .
\end{equation*}
Since the equality $\psi^*\inn(Z)\Omega_r = 0$ holds for every $Z \in \vf(\W_r)$, in particular it holds
for every $Z \in \vf(\W_r)$ which is $\rho_2^r$-related with $Y \in \vf(J^{k-1}\pi^*)$. Hence, we obtain
\begin{equation*}
\psi_h^*\inn(Y)\Omega_h = \psi^*\inn(Z)\Omega_r = 0 \, . \qedhere
\end{equation*}
\end{proof}

The diagram illustrating this situation is the following:
\begin{equation*}
\xymatrix{
\W_r \ar[dd]^-{\rho_\R^r} \ar[drr]^-{\rho_2^r} & \ & \  \\
\ & \ & J^{k-1}\pi^* \ar[dll]_{\bar{\pi}_{J^{k-1}\pi}^r} \\
\R \ar@/^1pc/[uu]^{\psi} \ar@/_1pc/@{-->}[urr]_{\psi_h = \rho_2^r \circ \psi} & \ & \ \\
}
\end{equation*}

\begin{remark}
Observe that, for the Hamiltonian sections, the condition of holonomy on the section $\psi$ is not required.
\end{remark}

As for the Lagrangian sections given by Proposition \ref{Chap05_prop:UnifiedToLagrangianSect}, this last
result does not give an equivalence between sections $\psi \in \Gamma(\rho_\R^r)$, which are solutions to
equation \eqref{Chap05_eqn:UnifDynEqSect}, and sections $\psi_h \in \Gamma(\bar{\pi}_{J^{k-1}\pi}^r)$,
which are solutions to equation \eqref{Chap05_eqn:HamRegDynEqSect}, but only that a section solution
to the former equation can be projected to a section solution to the latter. However, recall that sections
$\psi$ which are solutions to the dynamical equations in the unified formalism  take values in $\W_\Lag$,
and hence we are able to establish the equivalence using the diffeomorphism $\rho_2^\Lag$. As in the Lagrangian
formalism, we first need the following technical result to state the full equivalence, which is the analogous
to Lemma \ref{Chap05_lemma:LagCartanFormsTechLemma} in the Hamiltonian formalism.

\begin{lemma}\label{Chap05_lemma:HamHamiltonCartanFormsTechLemma}
The Hamilton-Cartan $1$-form satisfies $(\rho_2^\Lag)^*\Theta_h = j_\Lag^*\Theta_r$.
\end{lemma}
\begin{proof}
An easy computation proves this result:
\begin{equation*}
(\rho_2^\Lag)^*\Theta_h = (\rho_2^r\circ j_\Lag)^*\Theta_h = j_\Lag^*\Theta_r \, . \qedhere
\end{equation*}
\end{proof}

\begin{remark}
Since the exterior derivative and the pull-back commute, Lemma \ref{Chap05_lemma:HamHamiltonCartanFormsTechLemma}
also holds for the Hamilton-Cartan $2$-form.
\end{remark}

Now we can state the remaining part of the equivalence between the solutions of the Hamiltonian and unified formalisms.

\begin{proposition}\label{Chap05_prop:HamiltonianRegToUnifiedSect}
Let $\Lag \in \df^{1}(J^{k}\pi)$ be a hyperregular Lagrangian density, and
$\psi_h \in \Gamma(\bar{\pi}_{J^{k-1}}^r)$ a section solution to the dynamical equation
\eqref{Chap05_eqn:HamRegDynEqSect}. Then the section
$\psi = j_\Lag \circ (\rho_2^\Lag)^{-1} \circ \psi_h \in \Gamma(\rho_\R^r)$ is holonomic and a solution to the equation
\eqref{Chap05_eqn:UnifDynEqSect}.
\end{proposition}
\begin{proof}
The proof of this result is analogous to the proof given for Proposition \ref{Chap05_prop:LagrangianToUnifiedSect}.

Since $j_\Lag \colon \W_\Lag \to \W_r$ is an embedding, for every vector field $X \in \vf(\W_r)$
tangent to $\W_\Lag$ there exists a unique vector field $Y \in \vf(\W_\Lag)$ which is $j_\Lag$-related
with $X$. Hence, let us assume that $X \in \vf(\W_r)$ is tangent to $\W_\Lag$. Then we have
\begin{equation*}
\psi^*\inn(X)\Omega_r = (j_\Lag \circ (\rho_2^\Lag)^{-1} \circ \psi_h)^*\inn(X)\Omega_r
= ((\rho_2^\Lag)^{-1} \circ \psi_h)^*\inn(Y)j_\Lag^*\Omega_r \, .
\end{equation*}
Applying Lemma \ref{Chap05_lemma:HamHamiltonCartanFormsTechLemma} we obtain
\begin{equation*}
((\rho_2^\Lag)^{-1} \circ \psi_h)^*\inn(Y)j_\Lag^*\Omega_r
= ((\rho_2^\Lag)^{-1} \circ \psi_h)^*\inn(Y)(\rho_2^\Lag)^*\Omega_h
= (\rho_2^\Lag \circ (\rho_2^\Lag)^{-1} \circ \psi_h)^*\inn(Z)\Omega_h = \psi_h^*\inn(Z)\Omega_h \, ,
\end{equation*}
where $Z \in \vf(J^{k-1}\pi^*)$ is the unique vector field related with $Y$ by the diffeomorphism $\rho_2^\Lag$.
Hence, since $\psi_h^*\inn(Z)\Omega_h = 0$ for every $Z \in \vf(J^{k-1}\pi^*)$ by hypothesis, we have
proved that the section $\psi = j_\Lag \circ (\rho_2^\Lag)^{-1} \circ \psi_h \in \Gamma(\rho_\R^r)$
satisfies the equation
\begin{equation*}
\psi^*\inn(X)\Omega_r = 0 \, , \quad \mbox{for every } X \in \vf(\W_r) \mbox{ tangent to }\W_\Lag \, .
\end{equation*}
However, from Proposition \ref{Chap05_prop:UnifDynEqSectTangent} we know that if $\psi \in \Gamma(\rho_\R^r)$
is holonomic, then this last equation is equivalent to equation \eqref{Chap05_eqn:UnifDynEqSect}, that is,
\begin{equation*}
\psi^*\inn(X)\Omega_r = 0 \, , \quad \mbox{for every } X \in \vf(\W_r) \, .
\end{equation*}

It remains to prove that $\psi$ is holonomic in $\W_r$. By definition, a section $\psi \in \Gamma(\rho_\R^r)$
is holonomic if its projection $\rho_1^r \circ \psi \in \Gamma(\bar{\pi}^{2k-1})$ is holonomic in $J^{2k-1}\pi$.
We prove it in coordinates: if $\psi_h(t) = (t,q_i^A(t),p_A^i(t))$ is a solution to equation
\eqref{Chap05_eqn:HamRegDynEqSect}, then we have just proved that $\psi(t) = (t,q_i^A(t),q_j^A(t),p_A^i(t))$
is a solution to equation \eqref{Chap05_eqn:UnifDynEqSect} which, in coordinates, gives the equations
\eqref{Chap05_eqn:UnifDynEqSectHolonomyLocalPart}, \eqref{Chap05_eqn:UnifDynEqSectLocal1} and
\eqref{Chap05_eqn:UnifDynEqSectLocal2}, along with the equations defining the submanifold $\W_c \hookrightarrow \W_r$.
Then, bearing in mind that the section $\psi$ lies in the submanifold $\W_\Lag = \graph(\Leg)$, its projection
$\rho_1^r \circ \psi$ must satisfy the equations \eqref{Chap05_eqn:UnifDynEqSectHolonomyLocalPart}, 
\eqref{Chap05_eqn:UnifDynEqSectHolonomyLocalPart2} and \eqref{Chap05_eqn:UnifEulerLagrange}. Then, since
the $k$th-order Lagrangian density $\Lag$ is hyperregular, equations \eqref{Chap05_eqn:UnifDynEqSectHolonomyLocalPart2}
have a unique solution, thus obtaining the following equations for the section $\rho_1^r \circ \psi$
\begin{equation*}
\dot{q}_i^B - q_{i+1}^B = 0 \, , \qquad 0 \leqslant i \leqslant 2k-2 \, ,
\end{equation*}
in addition to the Euler-Lagrange equations. In particular, these equations are the local equations giving
the holonomy condition for the section $\rho_1^r \circ \psi$. Hence, $\rho_1^r \circ \psi$ is holonomic in
$J^{2k-1}\pi$, and therefore $\psi$ is holonomic in $\W_r$.
\end{proof}

Finally, let us compute the local expression of equation \eqref{Chap05_eqn:HamRegDynEqSect}.
If $\psi(t) = (t,q_i^A(t),q_j^A(t),p_A^i(t)) \in \Gamma(\rho_\R^r)$ is a solution to equation
\eqref{Chap05_eqn:UnifDynEqSect}, then equations \eqref{Chap05_eqn:UnifDynEqSectHolonomyLocalPart},
\eqref{Chap05_eqn:UnifDynEqSectLocal1} and \eqref{Chap05_eqn:UnifDynEqSectLocal2} hold. Now, bearing
in mind the local expression \eqref{Chap05_eqn:HamRegHamiltonianFunctionLocal} for the local Hamiltonian function
$H$, we obtain the following $2kn$ equations for the section $\psi_h = \rho_2^r \circ \psi = (t,q_i^A(t),p_A^i(t))$
\begin{equation}\label{Chap05_eqn:HamiltonEquations}
\dot{q}_i^A = \restric{\derpar{H}{p_A^i}}{\psi_h} \quad ; \quad \dot{p}_A^i = - \restric{\derpar{H}{q_i^A}}{\psi_h} \, .
\end{equation}
So we obtain the Hamilton equations for a $k$th-order non-autonomous system.

\subsubsection{Dynamical equations for vector fields}

Next, using the results stated at the beginning of this Section, we can now state the Hamiltonian dynamical
equations for vector fields, and recover the Hamiltonian vector field from the unique vector field solution
to the dynamical equations \eqref{Chap05_eqn:UnifDynEqVF} in the unified formalism.

The \textsl{$k$th-order (hyperregular) Hamiltonian problem for vector fields} associated with the
Hamiltonian system $(J^{k-1}\pi^*,\Omega_r,(\bar{\pi}_{J^{k-1}\pi}^r)^*\eta)$ consists in finding
vector fields $X_h \in \vf(J^{k-1}\pi^*)$ such that
\begin{equation}\label{Chap05_eqn:HamRegDynEqVF}
\inn(X_h)\Omega_h = 0 \quad ; \quad \inn(X_h)(\bar{\pi}_{J^{k-1}\pi}^r)^*\eta \neq 0 \, .
\end{equation}

\begin{remark}
As in the first-order case described in Section \ref{Chap02_sec:NonAutonomousHamiltonian}, the second equation in
\eqref{Chap05_eqn:HamRegDynEqVF} is a transversality condition for the vector field $X_h$ with respect to the
projection onto $\R$, and the non-zero value is usually fixed to $1$, thus giving the following equations
\begin{equation*}
\inn(X_h)\Omega_h = 0 \quad ; \quad \inn(X_h)(\bar{\pi}_{J^{k-1}\pi}^r)^*\eta = 1 \, .
\end{equation*}
\end{remark}

Now that the problem is stated, we recover a vector field solution to equations \eqref{Chap05_eqn:HamRegDynEqVF}
from a vector field solution to equations \eqref{Chap05_eqn:UnifDynEqVF}. Since $\rho_2^\Lag$ is a diffeomorphism
by Proposition \ref{Chap05_prop:HamRegRho2LDiffeomorphism}, the reasoning we follow is the same as that for the
Lagrangian formalism.

\begin{lemma}\label{Chap05_lemma:HamRegCorrespondenceVF}
Let $\Lag \in \df^{1}(J^{k}\pi)$ be a $k$th-order hyperregular Lagrangian density, and $X \in \vf(\W_r)$ a
vector field tangent to $\W_\Lag$. Then there exists a unique vector field $X_h \in \vf(J^{k-1}\pi^*)$
such that $X_h \circ \rho_2^r \circ j_\Lag = \Tan\rho_2^r \circ X \circ j_\Lag$.
\end{lemma}
\begin{proof}
The proof of this result is similar to the proof given for Lemma \ref{Chap05_lemma:LagCorrespondenceVF}.

Since $X$ is tangent to $\W_\Lag$, there exists a unique $X_o \in \vf(\W_\Lag)$ such that
$\Tan j_\Lag \circ X_o = X \circ j_\Lag$. Furthermore, since $\rho_2^r$ is a diffeomorphism, there is a
unique vector field $X_h \in \vf(J^{k-1}\pi^*)$ which is $\rho_2^\Lag$-related with $X_o$; that is,
$X_h \circ \rho_2^\Lag = \Tan\rho_2^\Lag \circ X_o$.
Then we have
\begin{equation*}
X_h \circ \rho_2^r \circ j_\Lag = X_h \circ \rho_2^\Lag = \Tan\rho_2^\Lag \circ X_o
= \Tan\rho_2^r \circ \Tan j_\Lag \circ X_o = \Tan\rho_2^r \circ X \circ j_\Lag \, . \qedhere
\end{equation*}
\end{proof}

This result states that, for every $X \in \vf(\W_r)$ tangent to $\W_\Lag$ we can define implicitly a unique
vector field $X_h \in \vf(J^{k-1}\pi^*)$ such that the following diagram commutes
\begin{equation*}
\xymatrix{
\Tan\W_r \ar[ddrr]^-{\Tan\rho_2^r} & \ & \ \\
\Tan\W_\Lag \ar[drr]_-{\Tan\rho_2^\Lag} & \ & \ \\
\ & \ & \Tan(J^{k-1}\pi^*) \\
\W_r \ar[ddrr]^-{\rho_2^r} \ar@/^1.8pc/[uuu]^-{X} & \ & \ \\
\W_\Lag \ar[drr]_-{\rho_2^\Lag} \ar@{^{(}->}[u]^-{j_\Lag} \ar@/_1.8pc/[uuu]_-{X_o}|(.23){\hole} & \ & \ \\
\ & \ & J^{k-1}\pi^* \ar[uuu]_-{X_h} \\
}
\end{equation*}

As a consequence of Lemma \ref{Chap05_lemma:HamRegCorrespondenceVF} we can give the following result, which
states a one-to-one correspondence between the set of vector fields solution to the dynamical equation in the
unified formalism and the set of vector fields solution to the dynamical equation in the Hamiltonian formalism.

\begin{theorem}\label{Chap05_thm:UnifiedToHamiltonianRegVF}
Let $\Lag \in \df^{1}(J^{k}\pi)$ be a $k$th-order hyperregular Lagrangian density, and $X \in \vf(\W_r)$
the vector field solution to equations \eqref{Chap05_eqn:UnifDynEqVF} and tangent to $\W_\Lag$. Then, there
exists a unique vector field $X_h \in \vf(J^{k-1}\pi^*)$, which is a solution to the equations
\eqref{Chap05_eqn:HamRegDynEqVF}.

\noindent Conversely, if $X_h \in \vf(J^{k-1}\pi^*)$ is a solution to equations \eqref{Chap05_eqn:HamRegDynEqVF},
then there exists a unique holonomic vector field $X \in \vf(\W_r)$, tangent to $\W_\Lag$, which is a solution
to equations \eqref{Chap05_eqn:UnifDynEqVF}.
\end{theorem}
\begin{proof}
The proof of this result is analogous to the proof given for Theorem \ref{Chap05_thm:UnifiedToLagrangianVF},
Lemma \ref{Chap05_lemma:HamRegCorrespondenceVF} now being used to obtain the vector field $X_h \in \vf(J^{k-1}\pi^*)$
instead of Lemma \ref{Chap05_lemma:LagCorrespondenceVF}.

Taking $X_h \in \vf(J^{k-1}\pi^*)$ as the unique vector field given by Lemma \ref{Chap05_lemma:HamRegCorrespondenceVF}
we have
\begin{equation*}
\inn(X)\Omega_r = \inn(X)(\rho_2^r)^*\Omega_h = (\rho_2^r)^*\inn(X_h)\Omega_h \, ,
\end{equation*}
but, as $\rho_2^r$ is a surjective submersion, this is equivalent to
\begin{equation*}
\restric{\inn(X_h)\Omega_h}{\rho_2^r(\W_r)} = \restric{\inn(X_h)\Omega_h}{J^{k-1}\pi^*} \, ,
\end{equation*}
since $\rho_2^r(\W_r) = J^{k-1}\pi^*$.
The converse is immediate, reversing this reasoning. Hence, we have proved that $\inn(X)\Omega_r = 0$ if, and only if,
$\inn(X_h)\Omega_h = 0$. The same reasoning proves that $\inn(X)(\rho_\R^r)^*\eta \neq 0$
if, and only if $\inn(X_h)(\bar{\pi}_{J^{k-1}\pi^*}^r)^*\eta \neq 0$.

It remains to prove that if $X_h$ is a solution to equation \eqref{Chap05_eqn:HamRegDynEqVF}, then $X$ is holonomic.
In order to prove this, we compute the coordinate expression of $X$. In particular, a generic vector
field $X \in \vf(\W_r)$ is given locally by \eqref{Chap05_eqn:UnifGenericVectorField}, that is,
\begin{equation*}
X = f \derpar{}{t} + f_i^A\derpar{}{q_i^A} + F_j^A\derpar{}{q_j^A} + G_A^i\derpar{}{p_A^i}\, .
\end{equation*}
Since we have just proved that $X$ is a solution to equation \eqref{Chap05_eqn:UnifDynEqVF}, its component functions
must satisfy equations \eqref{Chap05_eqn:UnifDynEqVFHolonomyLocalPart2} and \eqref{Chap05_eqn:UnifDynEqVFLocal2}
(taking $f = 1$ as a representative of the equivalence class). Hence, the vector field $X$ is given locally by
\eqref{Chap05_eqn:UnifDynEqVFSolution}, that is,
\begin{equation*}
X = \derpar{}{t} + q_{i+1}^A\derpar{}{q_i^A} + F_j^A\derpar{}{q_j^A} +
\derpar{\hat{L}}{q_0^A}\derpar{}{p_A^0} + \left( \derpar{\hat{L}}{q_i^A} - p_A^{i-1} \right) \derpar{}{p_A^i}\, .
\end{equation*}
In addition, by Lemma \ref{Chap05_lemma:HamRegCorrespondenceVF}, the vector field $X$ is tangent to the submanifold
$\W_\Lag \hookrightarrow \W_r$, which is given locally by the $kn$ constraints
\begin{equation*}
p_A^{r-1} = \sum_{i=0}^{k-r}(-1)^i \frac{d^i}{dt^{i}} \left( \derpar{\hat{L}}{q_{r+i}^A} \right) \, ,
\end{equation*}
Therefore, requiring $X$ to be tangent to this submanifold, we obtain the following system of $kn$ equations
\begin{align*}
\left(F_k^B-q_{k+1}^B\right)\derpars{\hat{L}}{q_k^B}{q_k^A} = 0 \, , \\
\left(F_{k+1}^B - q_{k+2}^B\right)\derpars{\hat{L}}{q_k^B}{q_k^A} -
\left(F_k^B-q_{k+1}^B \right) \frac{d}{dt}\left(\derpars{\hat{L}}{q_k^B}{q_k^A}\right) = 0 \, , \\
\vdots \qquad \qquad \qquad \qquad \\
\left(F_{2k-2}^B - q_{2k-1}^B\right)\derpars{\hat{L}}{q_k^B}{q_k^A} -
 \sum_{i=0}^{k-3} \left(F_{k+i}^B-q_{k+i+1}^B \right) (\cdots) = 0 \, , \\
(-1)^k\left(F_{2k-1}^B - \frac{d}{dt}q_{2k-1}^B\right) \derpars{\hat{L}}{q_k^B}{q_k^A} + 
\sum_{i=0}^{k} (-1)^i\frac{d^i}{dt^i}\left( \derpar{\hat{L}}{q_i^A} \right) - \sum_{i=0}^{k-2} \left(F_{k+i}^B-q_{k+i+1}^B \right)
 (\cdots) = 0 \, ,
\end{align*}
and, since the $k$th-order Lagrangian density is hyperregular, the Hessian matrix of $\hat{L}$ with respect
to $q_k^A$ is invertible, and these equations reduce to
\begin{align*}
F_i^A = q_{i+1}^A \quad , \quad  (k \leqslant i \leqslant 2k-2) \, , \\
(-1)^k\left(F_{2k-1}^B - \frac{d}{dt} q_{2k-1}^B \right) \derpars{\hat{L}}{q_k^B}{q_k^A} +
\sum_{i=0}^{k} (-1)^i\frac{d^i}{dt^i}\left( \derpar{\hat{L}}{q_i^A} \right) = 0 \, ,
\end{align*}
from where we deduce the coordinate expression of $X$, which is
\begin{equation*}
X = \derpar{}{t} + \sum_{i=0}^{2k-2} q_{i+1}^A\derpar{}{q_i^A} + F_{2k-1}^A\derpar{}{q_{2k-1}^A} +
\derpar{\hat{L}}{q_0^A}\derpar{}{p_A^0} + \left( \derpar{\hat{L}}{q_i^A} - p_A^{i-1} \right) \derpar{}{p_A^i}\, ,
\end{equation*}
where $F_{2k-1}^A$ are the unique solutions to the previous system of equations.
It is clear from this local expression that the vector field $X$ is holonomic in $\W_r$.
\end{proof}

Observe that, in this setting, the vector field $X_h \in \vf(J^{k-1}\pi^*)$ solution to equations
\eqref{Chap05_eqn:HamRegDynEqVF} is unique, since the $k$th-order Lagrangian density is hyperregular,
and hence the vector field $X \in \vf(\W_r)$ solution to equation \eqref{Chap05_eqn:UnifDynEqVF}
is unique by Proposition \ref{Chap05_prop:UnifRegLagUniqueVF}.

In local coordinates, bearing in mind the coordinate expression \eqref{Chap05_eqn:HamHamiltonCartanForms}
of the Hamilton-Cartan $2$-form $\Omega_h$, the coordinate expression of the vector field $X_h \in \vf(J^{k-1}\pi)$
solution to equations \eqref{Chap05_eqn:HamRegDynEqVF} is
\begin{equation*}
X_h = \derpar{H}{p_A^i} \, \derpar{}{q_i^A} - \derpar{H}{q_i^A}\,\derpar{}{p_A^i} \, .
\end{equation*}

\subsubsection{Equivalence of the Hamiltonian dynamical equations in the hyperregular case}

To close the hyperregular case, we state the equivalence Theorem in the Hamiltonian formalism,
which is the analogous to Theorems \ref{Chap05_thm:UnifEquivalenceTheorem} and \ref{Chap05_thm:LagEquivalenceTheorem}.
This result is a direct consequence of Theorems \ref{Chap05_thm:UnifEquivalenceTheorem},
\ref{Chap05_thm:UnifiedToHamiltonianRegVar} and \ref{Chap05_thm:UnifiedToHamiltonianRegVF}, and of Propositions
\ref{Chap05_prop:UnifiedToHamiltonianRegSect} and \ref{Chap05_prop:HamiltonianRegToUnifiedSect}, and thus
we do not prove it.

\begin{theorem}\label{Chap05_thm:HamEquivalenceTheoremRegular}
The following assertions on a section $\psi_h \in \Gamma(\bar{\pi}_{J^{k-1}\pi}^r)$ are equivalent.
\begin{enumerate}
\item $\psi_h$ is a solution to the Hamiltonian variational problem.
\item $\psi_h$ is a solution to equation \eqref{Chap05_eqn:HamRegDynEqSect}, that is,
\begin{equation*}
\psi_h^*\inn(Y)\Omega_h = 0, \quad \mbox{for every } Y \in \vf(J^{k-1}\pi^*) \, .
\end{equation*}
\item In natural coordinates, if $\psi_h(t) = (t,q_i^A(t),p_A^i(t))$, $0 \leqslant i \leqslant k-1$,
then the components of $\psi_h$ satisfy the $k$th-order non-autonomous Hamilton equations given by
\eqref{Chap05_eqn:HamiltonEquations}, that is,
\begin{equation*}
\dot{q}_i^A = \restric{\derpar{H}{p_A^i}}{\psi_h} \quad ; \quad \dot{p}_A^i = - \restric{\derpar{H}{q_i^A}}{\psi_h} \, .
\end{equation*}
\item $\psi_h$ is a solution to the equation
\begin{equation*}
\inn(\dot{\psi}_h)(\Omega_h \circ \psi_h) = 0 \, ,
\end{equation*}
where $\dot{\psi}_h \colon \R \to \Tan(J^{k-1}\pi^*)$ is the canonical lifting of $\psi_h$ to the tangent bundle.
\item $\psi_h$ is an integral curve of a vector field contained in a class of $\bar{\pi}_{J^{k-1}\pi}^r$-transverse
vector fields, $\left\{ X_h \right\} \subset \vf(J^{k-1}\pi^*)$, satisfying the first equation in
\eqref{Chap05_eqn:HamRegDynEqVF}, that is,
\begin{equation*}
\inn(X_h)\Omega_h = 0 \, .
\end{equation*}
\end{enumerate}
\end{theorem}

\subsection{Singular (almost-regular) Lagrangian densities}
\label{Chap05_sec:HamiltonianSingularCase}

To close the Hamiltonian formalism of higher-order non-autonomous systems, we analyze the case of
non-regular Lagrangian densities. Nevertheless, in order to give a general framework for singular systems
we must require some minimal regularity conditions to the $k$th-order Lagrangian density. Therefore,
throughout this Section we assume that the $k$th-order Lagrangian density is, at least, almost-regular.

In this case, Proposition \ref{Chap05_prop:HamRegRho2LDiffeomorphism} no longer holds, since the Hamiltonian
phase space is $\P = \Im(\Leg)$, and $\dim\P \leqslant \dim\W_\Lag$. This fact implies that we are not
able to recover the Hamiltonian solutions directly from the unified setting, and we are forced to pass
through the Lagrangian formalism and use the Legendre-Ostrogradsky map to obtain the Hamiltonian solutions
to the dynamical equations. Moreover, in this case the correspondence is not one-to-one, but for every solution
in the Hamiltonian formalism there are several solutions in both the unified and Lagrangian formalisms
that project to the given Hamiltonian solution.

\subsubsection{Variational principle}

Given the Hamiltonian system $(\P,\Omega_h,\bar{\pi}_\P^*\eta)$, let $\Gamma(\bar{\pi}_\P)$ be the set
of sections of $\bar{\pi}_\P$. Consider the functional
\begin{equation*}\label{Chap05_eqn:HamSingVariationalFunctionalDef}
\begin{array}{rcl}
\mathbf{H} \colon \Gamma(\bar{\pi}_\P) & \longrightarrow & \R \\
\psi_h & \longmapsto & \displaystyle \int_\R \psi_h^*\Theta_h
\end{array} \, ,
\end{equation*}
where the convergence of the integral is assumed.

\begin{definition} 
The \textnormal{$k$th-order Hamiltonian variational problem} (or \textnormal{generalized Hamilton-Jacobi
variational problem}) for the $k$th-order Hamiltonian system $(\P,\Omega_h,\bar{\pi}_\P^*\eta)$ is the
search for the critical (or stationary) sections of the functional $\mathbf{H}$ with respect to the variations
of $\psi_h$ given by $(\psi_h)_s = \sigma_s \circ \psi_h$, where $\left\{ \sigma_s \right\}$ is a local
one-parameter group of any compact-supported $Z \in \vf^{V(\bar{\pi}_\P)}(\P)$; that is
\begin{equation*}\label{Chap05_eqn:HamSingDynEqVar}
\restric{\frac{d}{ds}}{s=0}\int_\R (\psi_h)_s^*\Theta_h = 0 \, .
\end{equation*}
\end{definition}

\begin{theorem}\label{Chap05_thm:UnifiedToHamiltonianSingVar}
Let $\psi \in \Gamma(\rho_\R^r)$ be a critical section for the Lagrangian-Hamiltonian variational problem
given by the functional $\mathbf{LH}$. Then, the section
$\psi_h = \Leg_o \circ \rho_1^r \circ \psi = \Leg_o \circ \psi_\Lag \in \Gamma(\bar{\pi}_\P)$
is a critical section for the Hamiltonian variational problem given by the functional $\mathbf{H}$.

\noindent Conversely, if $\psi_h \in \Gamma(\bar{\pi}_\P)$ is a section solution to the Hamiltonian variational problem,
then the section $\psi = j_\Lag \circ (\rho_1^\Lag)^{-1} \circ \gamma \circ \psi_h \in \Gamma(\rho_\R^r)$
is a solution to the Lagrangian-Hamiltonian variational problem, where $\gamma \in \Gamma(\Leg_o)$ is some
section of $\Leg_o$.
\end{theorem}
\begin{proof}
From Theorem \ref{Chap05_thm:UnifiedToLagrangianVar} we know that if $\psi \in \Gamma(\rho_\R^r)$ is a
solution to the Lagrangian-Hamiltonian variational problem, then the section
$\phi = \bar{\pi}^{2k-1} \circ \rho_1^r \circ \psi \in \Gamma(\pi)$ is a solution to the Lagrangian
variational problem. And, conversely, if $\phi \in \Gamma(\pi)$ is a solution to the Lagrangian variational
problem, then the section $\psi = j_\Lag \circ (\rho_1^\Lag)^{-1} \circ j^{2k-1}\phi$ is a solution
to the Lagrangian variational problem.

Hence, it suffices to prove that the restricted Legendre-Ostrogradsky map gives a correspondence between
the solution of the Lagrangian and Hamiltonian variational problems. That is, if $\phi \in \Gamma(\pi)$
is a solution to the Lagrangian variational problem, then the section
$\psi_h = \Leg_o \circ j^{2k-1}\phi \in \Gamma(\bar{\pi}_\P)$ is a solution to the Hamiltonian variational
problem. And, conversely, if if $\psi_h \in \Gamma(\bar{\pi}_\P)$ is a solution to the Hamiltonian variational
problem, then $\psi_\Lag = \gamma \circ \psi_\Lag$ is holonomic, and its projection is a solution to the
Lagrangian variational problem.

Since $\Leg_o \colon J^{2k-1}\pi \to \P$ is a submersion, for every compact-supported vector field
$Z \in \vf^{V(\bar{\pi}_\P)}(\P)$, there exist compact-supported vector fields
$Y \in \vf^{V(\bar{\pi}^{2k-1})}(J^{2k-1}\pi)$ such that $Z$ and $Y$ are $\Leg_o$-related. In particular,
some of these vector fields in $J^{2k-1}\pi$ are the $(2k-1)$-jet lifting of compact-supported $\pi$-vertical
vector fields $X \in \vf^{V(\pi)}(E)$, that is, we have $Y = j^{2k-1}X$. We denote by $\{\sigma_s\}$ a local one-parameter
group for the compact-supported vector fields $X \in \vf^{V(\pi)}(E)$. Then, using this particular choice of
$\Leg_o$-related vector fields, we have
\begin{align*}
\restric{\frac{d}{ds}}{s=0} \int_\R(\psi_h)_s^*\Theta_{h}
&= \restric{\frac{d}{ds}}{s=0} \int_\R(\sigma_s \circ \psi_h)^*\Theta_{h}
= \restric{\frac{d}{ds}}{s=0} \int_\R\psi_h^*(\sigma_s^*\Theta_{h})
= \int_\R \psi_h^*\Lie(Z)\Theta_{h} \\
&= \int_\R \psi_h^*(\inn(Z)\d\Theta_h + \d\inn(Z)\Theta_h)
= \int_\R \psi_\Lag^*(\Leg_o^*(\inn(Z)\d\Theta_h + \d\inn(Z)\Theta_h)) \\
&= \int_\R \psi_\Lag^*(\inn(Y)\d\Theta_\Lag + \d\inn(Y)\Theta_\Lag)
= \int_\R\psi_\Lag^*\Lie(Y)\Theta_\Lag
= \int_\R(j^{2k-1}\phi)^*\Lie(j^{2k-1}X)\Theta_\Lag \\
&= \restric{\frac{d}{ds}}{s=0}\int_\R(j^{2k-1}\phi)^*(j^{2k-1}\sigma_s)^*\Theta_\Lag
= \restric{\frac{d}{ds}}{s=0}\int_\R(j^{2k-1}(\sigma_s \circ\phi))^*\Theta_\Lag \\
&= \restric{\frac{d}{ds}}{s=0}\int_\R (j^{2k-1}\phi_s)^*\Theta_\Lag = 0 \, ,
\end{align*}
since $\phi$ is a critical section for the Lagrangian variational problem.

For the converse, if suffices to reverse this reasoning with $\psi_\Lag = \gamma \circ \psi_h$, and bearing
in mind that $\Leg_o \circ \gamma = \Id_\P$. It is important to point out that, in general, not every
section $\gamma \circ \psi_h \in \Gamma(\bar{\pi}^{2k-1})$ is holonomic in $J^{2k-1}\pi$. Nevertheless,
following the same patterns as in the theory of singular non-autonomous first-order mechanical systems
\cite{art:DeLeon_Marin_Marrero_Munoz_Roman02}, it can be proved that some of the sections $\psi_\Lag = \gamma \circ \psi_h$
are holonomic.
\end{proof}

\begin{remark}
Observe that the sections $\psi \in \Gamma(\rho_\R^r)$ and $\phi \in \Gamma(\pi)$ obtained from a given
$\psi_h \in \Gamma(\bar{\pi}_\P)$ are not necessarily unique.
\end{remark}

\subsubsection{Dynamical equations for sections}

As in the previous Section, we now state the Hamiltonian equations for sections in the almost-regular case, and we recover the
Hamiltonian solutions in $\P$ from the solutions in the unified setting.

The \textsl{$k$th-order (almost-regular) Hamiltonian problem for sections} associated with the
Hamiltonian system $(\P,\Omega_h,\bar{\pi}_\P^*\eta)$ consists in finding sections
$\psi_h \in \Gamma(\bar{\pi}_{\P})$ characterized by the equation
\begin{equation}\label{Chap05_eqn:HamSingDynEqSect}
\psi_h^*\inn(Y)\Omega_h = 0 \, , \quad \mbox{for every } Y \in \vf(\P) \, .
\end{equation}

\begin{proposition}\label{Chap05_prop:UnifiedToHamiltonianSingSect}
Let $\Lag \in \df^{1}(J^{k}\pi)$ be a $k$th-order almost-regular Lagrangian density, and
$\psi \in \Gamma(\rho_\R^r)$ a section solution to equation \eqref{Chap05_eqn:UnifDynEqSect}. Then, the section
$\psi_h = \Leg_o \circ \rho_1^r \circ \psi = \Leg_o \circ \psi_\Lag \in \Gamma(\bar{\pi}_\P)$ is a solution
to equation \eqref{Chap05_eqn:HamSingDynEqSect}.

\noindent Conversely, let $\psi_h \in \Gamma(\bar{\pi}_\P)$ be a solution to equation
\eqref{Chap05_eqn:HamSingDynEqSect}. Then 
$\psi = j_\Lag \circ (\rho_1^\Lag)^{-1} \circ \gamma \circ \psi_h \in \Gamma(\rho_\R^r)$ is holonomic and a solution
to the equation \eqref{Chap05_eqn:UnifDynEqSect} for some $\gamma \in \Gamma(\Leg_o)$.
\end{proposition}
\begin{proof}
Since the Lagrangian density is almost-regular, the map $\Leg_o$ is a submersion onto its image, $\P$.
Hence, for every $Y \in \vf(\P)$ there exist some $Z \in \vf(J^{2k-1}\pi)$ such that $Z$ is $\Leg_o$-related
with $Y$. Note that this vector field $Z$ is not unique in general, since $\ker\Tan\Leg_o \neq \{ 0 \}$,
and the vector field $Z + Z_o$, with $Z_o \in \vf^{V(\Leg_o)}(J^{2k-1}\pi)$, is also $\Leg_o$-related with $Y$.
Using this particular choice of $\Leg_o$-related vector fields, we have
\begin{equation*}
\psi_h^*\inn(Y)\Omega_h = (\Leg_o \circ \psi_\Lag)^*\inn(Y)\Omega_h = \psi_\Lag^*(\Leg_o^*\inn(Y)\Omega_h)
= \psi_\Lag^*\inn(Z)\Leg_o^*\Omega_h = \psi_\Lag^*\inn(Z)\Omega_\Lag \, ,
\end{equation*}
where $\psi_\Lag = \rho_1^r \circ \psi$. Then, using Proposition \ref{Chap05_prop:UnifiedToLagrangianSect}, we have proved
\begin{equation*}
\psi_h^*\inn(Y)\Omega_h = \psi_\Lag^*\inn(Z)\Omega_\Lag = 0 \, .
\end{equation*}

The converse is clear reversing the reasoning and using Proposition \ref{Chap05_prop:LagrangianToUnifiedSect},
since $\Leg_o \circ \gamma = \Id_\P$ and, in particular, we have $\gamma^*\Theta_\Lag = \Theta_h$.
As in the proof of Theorem \ref{Chap05_thm:UnifiedToHamiltonianSingVar}, in general not every section
$\gamma \circ \psi_h \in \Gamma(\bar{\pi}^{2k-1})$ is holonomic, but it can be proved
that some of the sections $\psi_\Lag = \gamma \circ \psi_h$ are so.
\end{proof}

The diagram for this situation is the following:
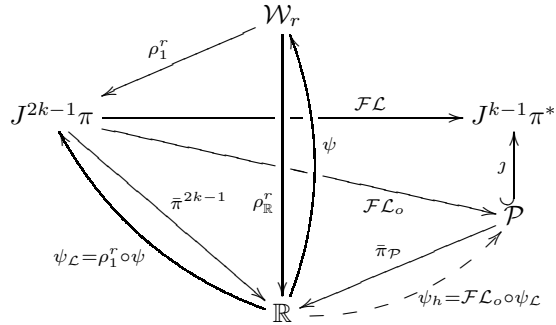
\begin{equation*}
\xymatrix{
\ & \ & \W_r \ar[ddd]_<(0.6){\rho_\R^r} \ar[dll]_{\rho_1^r} & \ & \ \\
J^{2k-1}\pi \ar[ddrr]^{\bar{\pi}^{2k-1}} \ar[rrrr]^<(0.65){\Leg}|(.5){\hole}|(.56){\hole} \ar[drrrr]_<(0.7){\Leg_o}|(.5){\hole}|(.57){\hole} & \ & \ & \ & J^{k-1}\pi^* \\
\ & \ & \ & \ & \mathcal{P} \ar@{^{(}->}[u]^{\jmath} \ar[dll]_{\bar{\pi}_\P} \\
\ & \ & \R \ar@/_1pc/[uuu]_<(0.535){\psi} \ar@/^1pc/[uull]^{\psi_\Lag = \rho_1^r \circ \psi} \ar@/_1pc/@{-->}[urr]_<(.6){\psi_h = \Leg_o \circ \psi_\Lag} & \ & \ \\
}
\end{equation*}

\subsubsection{Dynamical equations for vector fields}

Now we state the Hamiltonian dynamical equations for vector fields in the almost-regular case, and we recover
a Hamiltonian vector field from a vector field solution to the dynamical equations \eqref{Chap05_eqn:UnifDynEqVFSingular}
in the unified formalism.

The \textsl{$k$th-order (almost-regular) Hamiltonian problem for vector fields} associated with
the Hamiltonian system $(\P,\Omega_r,\bar{\pi}_{\P}^*\eta)$ consists in finding vector fields
$X_h \in \vf(\P)$ such that
\begin{equation}\label{Chap05_eqn:HamSingDynEqVF}
\inn(X_h)\Omega_h = 0 \quad ; \quad \inn(X_h)\bar{\pi}_{\P}^*\eta \neq 0 \, .
\end{equation}

Now that the problem is stated, we recover a vector field solution to equations \eqref{Chap05_eqn:HamSingDynEqVF}
from a vector field solution to equations \eqref{Chap05_eqn:UnifDynEqVF}.

\begin{theorem}\label{Chap05_thm:UnifiedToHamiltonianSingVF}
Let $\Lag \in \df^{1}(J^{k}\pi)$ be a $k$th-order almost-regular Lagrangian density, and $X \in \vf(\W_r)$
the vector field solution to equations \eqref{Chap05_eqn:UnifDynEqVF} and tangent to $\W_\Lag$
(at least on the points of a submanifold $\W_f \hookrightarrow \W_\Lag$). Then, there exists a vector field
$X_h \in \vf(\P)$, which is a solution to the equations \eqref{Chap05_eqn:HamSingDynEqVF} (at least on the
points of $\P_f = \rho_2^r(\W_f) \hookrightarrow \P$).

\noindent Conversely, if $X_h \in \vf(\P)$ is a solution to equations \eqref{Chap05_eqn:HamSingDynEqVF}
(at least on the points of a submanifold $\P_f \hookrightarrow \P$), then there exist some holonomic
vector fields $X \in \vf(\W_r)$, tangent to $\W_\Lag$, which are solutions to equations \eqref{Chap05_eqn:UnifDynEqVF}
(at least on the points of $\W_f = (\rho_2^r)^{-1}(\P_f)$).
\end{theorem}
\begin{proof}
From Theorem \ref{Chap05_thm:UnifiedToLagrangianVF} we have a bijective correspondence between holonomic
vector fields $X_\Lag \in \vf(J^{2k-1}\pi)$ solution to equations \eqref{Chap05_eqn:LagDynEqVF} (at least
on the points of a submanifold $S_f \hookrightarrow J^{2k-1}\pi$) and holonomic vector fields $X \in \vf(\W_r)$,
tangent to $\W_\Lag$, solution to equations \eqref{Chap05_eqn:UnifDynEqVF} (at least on the points of a
submanifold $\W_f \hookrightarrow \W_r)$. Hence, it suffices to prove that we can establish a correspondence
between the set of vector fields solution to the Lagrangian dynamical equations, and the set of vector fields
solution to the Hamiltonian dynamical equations.

Since the Lagrangian density is almost-regular, the map $\Leg_o$ is a surjective submersion on $\P$. Hence,
for every $X_h \in \vf(\P)$ there exist some $X_\Lag \in \vf(J^{2k-1}\pi)$ (not necessarily unique) such
that $X_h$ and $X_\Lag$ are $\Leg_o$-related, that is, $X_h \circ \Leg_o = \Tan\Leg_o \circ X_\Lag$. And,
conversely, for every vector field $X_\Lag \in \vf(J^{2k-1}\pi)$, there exists a vector field $X_h \in \vf(\P)$
$\Leg_o$-related with $X_\Lag$. Using this particular choice of $\Leg_o$-related vector fields, we have
\begin{equation*}
\inn(X_\Lag)\Omega_\Lag = \inn(X_\Lag)\Leg_o^*\Omega_h = \Leg_o^*\inn(X_h)\Omega_h =
\restric{\inn(X_h)\Omega_h}{\Leg_o(J^{2k-1}\pi)} = \restric{\inn(X_h)\Omega_h}{\P} \, ,
\end{equation*}
since $\Leg_o$ is a surjective submersion on $\P$.
The converse is immediate, reversing this reasoning. Hence, we have proved that $\inn(X_\Lag)\Omega_\Lag = 0$
is equivalent to $\inn(X_h)\Omega_h = 0$ whenever $X_\Lag$ and $X_h$ are $\Leg_o$-related. The same reasoning
proves that $\inn(X_\Lag)(\bar{\pi}^{2k-1})^*\eta \neq 0$ is equivalent to $\inn(X_h)\bar{\pi}_{\P}^*\eta \neq 0$.
Observe that the reasoning remains the same replacing $J^{2k-1}\pi$ by $S_f$ and $\P$ by $\P_f$.

As in the proof of Proposition \ref{Chap05_prop:UnifiedToHamiltonianSingSect}, not every vector field
$X_\Lag \in \vf(J^{2k-1}\pi)$ which is $\Leg_o$-related with a vector field $X_h \in \vf(\P)$ is holonomic.
Nevertheless, it can be proved following the patterns in \cite{art:DeLeon_Marin_Marrero_Munoz_Roman02} that
those holonomic vector fields $\Leg_o$-related with $X_h$ exist, maybe on another submanifold
$S^h_f \hookrightarrow S_f$.
\end{proof}

The diagram illustrating the statement and proof of Theorem \ref{Chap05_thm:UnifiedToHamiltonianSingVF}
is the following:
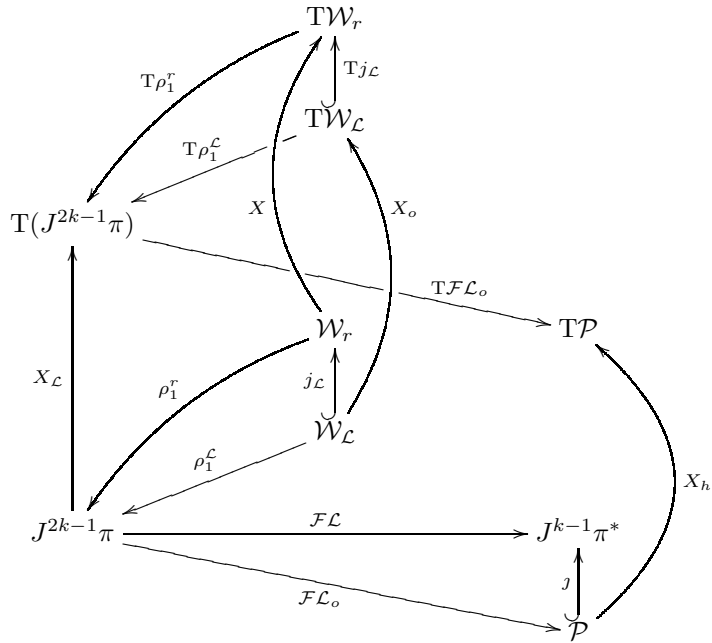
\begin{equation*}
\xymatrix{
\ & \ & \Tan\W_r \ar@/_1pc/[ddll]_-{\Tan\rho_1^r} & \ & \  \\
\ & \ & \Tan\W_\Lag \ar[dll]_-{\Tan\rho^\Lag_1}|<(.095){\hole} \ar@{^{(}->}[u]_-{\Tan j_\Lag} & \ & \  \\
\Tan(J^{2k-1}\pi) \ar[rrrrd]^(.75){\Tan\Leg_o}|<(.355){\hole}|<(.57){\hole} & \ & \ & \ & \ \\
\ & \ & \W_r \ar@/_1pc/[ddll]_-{\rho_1^r} \ar@/^1.95pc/[uuu]^(.4){X} & \ & \Tan\P \\
\ & \ & \W_\Lag \ar[dll]_-{\rho^\Lag_1} \ar@{^{(}->}[u]^-{j_\Lag} \ar@/_1.75pc/[uuu]_(.7){X_o} & \ & \ \\
J^{2k-1}\pi \ar[rrrr]^{\Leg} \ar[drrrr]_{\Leg_o} \ar[uuu]^-{X_\Lag} & \ & \ & \ & J^{k-1}\pi^* \\
\ & \ & \  & \ & \P \ar@{^{(}->}[u]^{\jmath} \ar@/_3pc/[uuu]_-{X_{h}}
}
\end{equation*}

\subsubsection{Equivalence of the Hamiltonian equations in the almost-regular case}

To close the Hamiltonian formalism for $k$th-order non-autonomous dynamical systems, we state the equivalence Theorem
in the almost-regular case, which is the analogous to Theorems \ref{Chap05_thm:UnifEquivalenceTheorem},
\ref{Chap05_thm:LagEquivalenceTheorem} and \ref{Chap05_thm:HamEquivalenceTheoremRegular}
This result is a direct consequence of Theorems \ref{Chap05_thm:UnifEquivalenceTheorem},
\ref{Chap05_thm:UnifiedToHamiltonianSingVar} and \ref{Chap05_thm:UnifiedToHamiltonianSingVF}, and of Proposition
\ref{Chap05_prop:UnifiedToHamiltonianSingSect}. Hence, the proof is omitted.

\begin{theorem}\label{Chap05_thm:HamEquivalenceTheoremSingular}
The following assertions on a section $\psi_h \in \Gamma(\bar{\pi}_\P)$ are equivalent.
\begin{enumerate}
\item $\psi_h$ is a solution to the Hamiltonian variational problem.
\item $\psi_h$ is a solution to the equation \eqref{Chap05_eqn:HamSingDynEqSect}, that is,
\begin{equation*}
\psi_h^*\inn(Y)\Omega_h = 0 \, , \quad \text{for every } Y \in \vf(\P) \, .
\end{equation*}
\item $\psi_h$ is a solution to the equation
\begin{equation*}
\inn(\dot{\psi}_h)(\Omega_h \circ \psi_h) = 0\ ,
\end{equation*}
where $\dot{\psi}_h \colon \R \to \Tan\P$ is the canonical lifting of $\psi_h$ to $\Tan\P$.
\item $\psi_h$ is an integral curve of a vector field contained in a class of $\bar{\pi}_\P$-transverse vector fields,
$\left\{ X_h \right\} \subset \vf(\P)$, satisfying the first equation in \eqref{Chap05_eqn:HamSingDynEqVF}, that is,
\begin{equation*}
\inn(X_h)\Omega_h = 0 \, .
\end{equation*}
\end{enumerate}
\end{theorem}


\section{Examples}
\label{Chap05_sec:Examples}

In this last Section of the Chapter, two physical models are analyzed as examples to show the application
of the formalism. The first example is a regular system, the \textsl{shape of a deformed elastic cylindrical
beam with fixed ends}, while the second is a singular one, the \textsl{second-order relativistic particle}
analyzed in Section \ref{Chap03_exa:2ndOrderRelativisticParticle}, now subjected to a potential depending
on time and position.

\subsection{The shape of a deformed elastic cylindrical beam with fixed ends}
\label{Chap05_exa:CylindricalBeam}

Let us consider a deformed elastic cylindrical beam with both ends fixed, as in Section \ref{Chap04_sec:Example2},
and let us consider the same problem: to determinate the axis of the beam. Strictly speaking, this is not a
time-dependent mechanical system, but if the beam is not homogeneous, it can be modeled using a configuration
bundle over a compact subset of $\R$, where every point in the base manifold represents the position of a
transverse section of the beam with respect to one of the fixed ends, thus allowing us to show an application
of our formalism. For simplicity, instead of a compact subset, we take the whole real line as the base manifold.

The configuration bundle for this system is $\pi \colon E \to \R$, where $E$ is a $2$-dimensional smooth
manifold. Let us denote by $x$ the global coordinate in $\R$, and the canonical volume form in $\R$ by
$\eta \in \df^{1}(\R)$, with local expression $\eta = \d x$. Natural coordinates in $E$ adapted to the
bundle structure are $(x,q_0)$, where $q_0$ represents the bending of the beam. Now, taking natural coordinates
in the second-order jet bundle of $\pi$, the second-order Lagrangian density for this system,
$\Lag \in \df^{1}(J^{2}\pi)$, is locally given by
\begin{equation*}\label{Chap05_eqn:ExampleRegular_LagrangianFunction}
\Lag(x,q_0,q_1,q_2) = L\cdot(\bar{\pi}^{2})^*\eta = \left( \frac{1}{2}\mu(x) q_2^2 + \rho(x) q_0 \right) \d x \, ,
\end{equation*}
where $\mu, \rho \in \Cinfty(J^{2}\pi)$ are functions that only depend on the coordinate $x$ and represent
physical parameters of the beam: $\rho$ is the linear density and $\mu$ is a non-vanishing function involving
Young's modulus of the material, the radius of curvature and the sectional moment of the cross-section considered
(see \cite{book:Benson06} for a detailed description). This is a regular Lagrangian density, since the Hessian
matrix of the  Lagrangian function $L \in \Cinfty(J^{2}\pi)$ associated with $\Lag$ with respect to $q_2$ is
\begin{equation*}
\left( \derpars{L}{q_2}{q_2} \right) = \mu(x) \, ,
\end{equation*}
and this $1 \times 1$ matrix has maximum rank, since $\mu$ is a non-vanishing function.

\begin{remark}
If the beam is homogeneous, $\mu$ and $\rho$ are constants (with $\mu \neq 0$), and thus the Lagrangian
density is ``autonomous'', that is, it does not depend explicitly on the coordinate of the base manifold.
This case is analyzed in \cite{book:Elsgoltz83}, and in Section \ref{Chap04_sec:Example1} to obtain the
Hamilton-Jacobi equations.
\end{remark}

As this is a second-order system, in the unified setting we consider the bundles $\W = J^{3}\pi \times_{J^{1}\pi} \Tan^*(J^{1}\pi)$
and $\W_r = J^{3}\pi \times_{J^{1}\pi} J^{1}\pi^*$, that is, the following diagram
\begin{equation*}
\xymatrix{
\ & \ & J^{3}\pi \times_{J^1\pi} \Tan^*(J^1\pi) \ar@/_1.3pc/[llddd]_{\rho_1} \ar[d]^-{\mu_\W} \ar@/^1.3pc/[rrdd]^{\rho_2} & \ & \ \\
\ & \ & J^3\pi \times_{J^1\pi} J^1\pi^* \ar[lldd]_{\rho_1^r} \ar[rrdd]^{\rho_2^r} & \ & \ \\
\ & \ & \ & \ & \Tan^*(J^{1}\pi) \ar[d]^-{\mu} \ar[lldd]_{\pi_{J^{1}\pi}}|(.25){\hole} \\
J^{3}\pi \ar[rrd]^{\pi^{3}_{1}} & \ & \ & \ & J^{1}\pi^* \ar[dll]^{\pi_{J^{1}\pi}^r} \\
\ & \ & J^{1}\pi \ar[d]^{\bar{\pi}^{1}} & \ & \ \\
\ & \ & \R & \ & \
}
\end{equation*}
Natural coordinates in $\W$ and $\W_r$ are $(x,q_0,q_1,q_2,q_3,p,p^0,p^1)$
and $(x,q_0,q_1,q_2,q_3,p^0,p^1)$, respectively.

Now, using the notation and terminology introduced throughout this Chapter,
if $\Theta_{1} \in \df^{1}(\Tan^*(J^{1}\pi))$ and $\Omega_{1} \in \df^{2}(\Tan^*(J^{1}\pi))$ are
the canonical Liouville forms of the cotangent bundle $\Tan^*(J^{1}\pi)$, we define the forms
$\Theta = \rho_2^*\Theta_{1} \in \df^{1}(\W)$ and $\Omega = \rho_2^*\Omega_{1} \in \df^{2}(\W)$
whose local expressions are
\begin{equation*}
\Theta = p^0\d q_0 + p^1\d q_1 + p\d x \quad ; \quad
\Omega = \d q_0 \wedge \d p^0 + \d q_1 \wedge \d p^1 - \d p \wedge \d x \, .
\end{equation*}
The coupling $1$-form $\hat{\C} \in \df^{1}(\W)$, whose local expression is given by
\eqref{Chap05_eqn:UnifCouplingFormLocal}, has the following local expression in this case
\begin{equation*}
\hat{\C} = \hat{C}\cdot\rho_\R^*\eta = (p + p^0q_1 + p^1q_2)\d x \, .
\end{equation*}
Then, denoting $\hat{\Lag} = (\pi^3_2 \circ \rho_1)^*\Lag \in \df^{1}(\W)$, we introduce the Hamiltonian
submanifold
\begin{equation*}
\W_o = \left\{ w \in \W \mid \hat{\Lag}(w) = \hat{\C}(w) \right\} \stackrel{j_o}{\hookrightarrow} \W \, ,
\end{equation*}
which is locally defined by the constraint function $\hat{C} - \hat{L} = 0$, whose coordinate expression is
\begin{equation*}
\hat{C} - \hat{L} = p + p^0 q_1 + p^1 q_2 - \frac{1}{2} \mu(x) q_2^2 - \rho(x) q_0 = 0 \, .
\end{equation*}
Finally, we construct the Hamiltonian $\mu_\W$-section $\hat{h} \in \Gamma(\mu_\W)$, which is specified
by giving the local Hamiltonian function $\hat{H}$, whose local expression is
\begin{equation*}\label{Chap05_eqn:ExampleRegular_UnifiedHamiltonianFunction}
\hat{H}(x,q_0,q_1,q_2,q_3,p^0,p^1) = p^0q_1 + p^1q_2 - \frac{1}{2}\mu(x) q_2^2 - \rho(x) q_0 \, ,
\end{equation*}
that is, we have $\hat{h}(x,q_0,q_1,q_2,q_3,p^0,p^1) = (x,q_0,q_1,q_2,q_3,-\hat{H},p^0,p^1)$.
Using this Hamiltonian section, we define the forms $\Theta_r = \hat{h}^*\Theta \in \df^{1}(\W_r)$ and
$\Omega_r = \hat{h}^*\Omega \in \df^{2}(\W_r)$, with local expressions
\begin{align*}
&\Theta_r = p^0\d q_0 + p^1 \d q_1 + \left(\frac{1}{2}\mu(x) q_2^2 + \rho(x) q_0 - p^0q_1 - p^1q_2 \right)\d x \, , \\
&\Omega_r = \d q_0 \wedge \d p^0 + \d q_1 \wedge \d p^1 + \left( -\rho(x)\d q_0 + p^0 \d q_1 + (p^1 - \mu(x)q_2)\d q_2 + q_1 \d p^0 + q_2 \d p^1\right) \wedge \d x \, .
\end{align*}
Notice that we omit the derivative of $H$ with respect to $x$ in the last summand of $\Omega_r$,
since $\d x \wedge \d x = 0$, and thus it is not relevant for the final result.

Now we can derive the dynamical equations of the system. First, let us describe the dynamics of this system
in terms of sections $\psi \in \Gamma(\rho_\R^r)$. In order to do so, let $Y \in \vf(\W_r)$ be a generic vector field
locally given by
\begin{equation*}
Y = f\derpar{}{x} + f_0\derpar{}{q_0} + f_1\derpar{}{q_1} + F_2\derpar{}{q_2} + F_3\derpar{}{q_3} + G^0\derpar{}{p^0}
+ G^1\derpar{}{p^1} \, .
\end{equation*}
Then, if $\psi(x) = (x,q_0(x),q_1(x),q_2(x),q_3(x),p^0(x),p^1(x))$ is a holonomic section of the projection
$\rho_\R^r$, equation \eqref{Chap05_eqn:UnifDynEqSect} leads to the following $5$ equations (the redundant
equation \eqref{Chap05_eqn:UnifDynEqSectLocalRedundantEq} is omitted)
\begin{align}
\dot{q}_0 = q_1 \quad ; \quad \dot{q}_1 = q_2 \, , \label{Chap05_eqn:ExampleRegular_Holonomy} \\
\dot{p}^0 = \rho(x) \quad ; \quad \dot{p}^1 = -p^0 \, , \label{Chap05_eqn:ExampleRegular_DiffEquations} \\
p^1 = q_2\mu(x) \, . \label{Chap05_eqn:ExampleRegular_LastMomentumCoord1}
\end{align}
Equations \eqref{Chap05_eqn:ExampleRegular_Holonomy} give us the condition of holonomy of type $2$ for the
section, which are also redundant since we assume that $\psi$ is holonomic. Equation
\eqref{Chap05_eqn:ExampleRegular_LastMomentumCoord1} is a pointwise algebraic condition, which, in combination
with the second equation in \eqref{Chap05_eqn:ExampleRegular_DiffEquations}, state that the section $\psi$ must
lie in a submanifold $\W_\Lag$ defined locally by the constraints
\begin{equation*}
p^0 = -q_2 \derpar{\mu}{x} - q_3\mu \quad ; \quad p^1 = q_2\mu \, .
\end{equation*}

Now we compute the local expression of the map $\Leg \colon J^{3}\pi \to J^{1}\pi^*$; from Proposition
\ref{Chap05_prop:UnifGraphLegendreOstrogradskyMap} we know the general expression for this map, and we obtain
\begin{equation}\label{Chap05_eqn:ExampleRegular_RestrictedLegendreMap}
\Leg^*p^0 = -q_2 \derpar{\mu}{x} - q_3\mu \quad ; \quad \Leg^*p^1 = q_2\mu \, .
\end{equation}
From this, the coordinate expression of the extended Legendre-Ostrogradsky map given by Corollary
\ref{Chap05_corol:UnifGraphExtendedLegendreOstrogradskyMapSect} in this example is
\begin{equation*}\label{Chap05_eqn:ExampleRegular_ExtendedLegendreMap}
\widetilde{\Leg}^*p^0 = -q_2 \derpar{\mu}{x} - q_3\mu \quad ; \quad \widetilde{\Leg}^*p^1 = q_2\mu \quad ; \quad
\widetilde{\Leg}^*p = -\frac{1}{2}\mu q_2^2 + q_1q_2 \derpar{\mu}{x} + q_1q_3\mu + q_0\rho \, .
\end{equation*}

Therefore, the section $\psi \in \Gamma(\rho_\R^r)$ solution to the dynamical equations is a holonomic section
of the projection $\rho_\R^r$, which lies in the submanifold $\W_\Lag \hookrightarrow \W_r$ defined by the above
constraint functions, and whose last component functions satisfy the differential equations
\begin{equation*}
\dot{p}^0 = \rho(x) \quad ; \quad \dot{p}^1 = -p^0 \, .
\end{equation*}

Now we state the Lagrangian-Hamiltonian problem for vector fields: we wish to find holonomic vector fields
$X \in \vf(\W_r)$ solution to equations \eqref{Chap05_eqn:UnifDynEqVF}. If $X$ is locally given by
\begin{equation*}
X = f\derpar{}{x} + f_0\derpar{}{q_0} + f_1\derpar{}{q_1} + F_2\derpar{}{q_2} + F_3\derpar{}{q_3} + G^0\derpar{}{p^0} + 
G^1\derpar{}{p^1} \, ,
\end{equation*}
then equations \eqref{Chap05_eqn:UnifDynEqVF} lead to the following system of $6$ equations
(again, the redundant equation \eqref{Chap05_eqn:UnifDynEqVFLocalRedundantEq1} is omitted)
\begin{align}
f_0 = f\cdot q_1 \quad ; \quad f_1 = f\cdot q_2 \, , \label{Chap05_eqn:ExampleRegular_Semispray2} \\
G^0 = f\cdot\rho(x) \quad ; \quad G^1 = -f\cdot p^0 \, , \label{Chap05_eqn:ExampleRegular_DynamicalEquations} \\
f \neq 0 \, , \label{Chap05_eqn:ExampleRegular_FixingGauge} \\
f\cdot \left(p^1 - q_2\mu(x)\right) = 0 \, . \label{Chap05_eqn:ExampleRegular_LastMomentumCoord2}
\end{align}
Equations \eqref{Chap05_eqn:ExampleRegular_Semispray2} give us the condition of holonomic of type $2$
in $\W_r$ for the vector field $X$. In addition, equation \eqref{Chap05_eqn:ExampleRegular_LastMomentumCoord2}
is an algebraic relation from which we obtain, in coordinates, the result stated in Proposition
\ref{Chap05_prop:UnifFirstConstraintSubmanifold}, that is, the vector field $X$ is defined along a
submanifold $\W_c \hookrightarrow \W_r$ defined by
\begin{equation*}
\W_c = \left\{ [w] \in \W_r \mid \xi([w]) = 0 \right\} \, ,
\end{equation*}
where $\xi = p^1 - q_2\mu(x)$. Thus, using \eqref{Chap05_eqn:ExampleRegular_Semispray2} and
\eqref{Chap05_eqn:ExampleRegular_DynamicalEquations}, and taking $f = 1$ as a representative of the
equivalence class, we have that $X$ is given locally by
\begin{equation}\label{Chap05_eqn:ExampleRegular_VectorFieldBeforeTangency}
X = \derpar{}{x} + q_1\derpar{}{q_0} + q_2\derpar{}{q_1} + F_2\derpar{}{q_2} + F_3\derpar{}{q_3} + \rho\derpar{}{p^0}
- p^0\derpar{}{p^1} \, .
\end{equation}
Notice that the functions $F_2$ and $F_3$ in \eqref{Chap05_eqn:ExampleRegular_VectorFieldBeforeTangency}
are not determined yet. However, recall that the statement of the problem requires the vector field $X$
to be holonomic, from where we can determinate the component function $F_2$ as $F_2 = q_3$, and then the
vector field given by \eqref{Chap05_eqn:ExampleRegular_VectorFieldBeforeTangency} becomes
\begin{equation}\label{Chap05_eqn:ExampleRegular_HolonomicVectorFieldBeforeTangency}
X = \derpar{}{x} + q_1\derpar{}{q_0} + q_2\derpar{}{q_1} + q_3\derpar{}{q_2} + F_3\derpar{}{q_3} + \rho\derpar{}{p^0}
- p^0\derpar{}{p^1} \, .
\end{equation}
Moreover, since the vector field $X$ is defined along the submanifold $\W_c$, we must require $X$
to be tangent to $\W_c$. This condition is locally equivalent to checking if the following identity holds
\begin{equation*}
\restric{\Lie(X)\xi}{\W_c} = 0 \, .
\end{equation*}
As we have seen in Section \ref{Chap05_sec:UnifDynamicalEquations}, this equation leads to
\begin{equation*}
\restric{\Lie(X)\xi}{\W_c} = -p^0 + q_2\derpar{\mu}{x} + q_3\mu = 0 \, .
\end{equation*}
This is a new constraint defining a submanifold $\W_\Lag \hookrightarrow \W_r$ which can be identified
with the graph of the restricted Legendre-Ostrogradsky map, as we have seen in the coordinate expression
\eqref{Chap05_eqn:ExampleRegular_RestrictedLegendreMap} of $\Leg$. Then, requiring $X$ to be tangent to this
new submanifold, we obtain
\begin{equation}\label{Chap05_eqn:ExampleRegular_EulerLagrangeVectFieldEq}
\Lie(X)\left(p^0 - q_2\derpar{\mu}{x} - q_3\mu \right) = \rho + q_2 \frac{\partial^2\mu}{\partial x^2} + 2q_3\derpar{\mu}{x} + F_3\mu = 0 \, .
\end{equation}
Equation \eqref{Chap05_eqn:ExampleRegular_EulerLagrangeVectFieldEq} is the Euler-Lagrange equation for $X$.
Observe that, since $\mu$ is a non-vanishing function, this equation has a unique solution for $F_3$.
Hence, there is a unique vector field $X \in \vf(\W_r)$ solution to the dynamical equations
\eqref{Chap05_eqn:UnifDynEqVF}, which is tangent to the submanifold $\W_\Lag \hookrightarrow \W_r$,
and is given locally by
\begin{equation*}
X = \derpar{}{x} + q_1\derpar{}{q_0} + q_2\derpar{}{q_1} + q_3\derpar{}{q_2}
- \frac{1}{\mu}\left( \rho + q_2 \frac{\partial^2\mu}{\partial x^2} + 2q_3\derpar{\mu}{x} \right) \derpar{}{q_3}
+ \rho\derpar{}{p^0} - p^0\derpar{}{p^1} \, .
\end{equation*}

Finally, we recover the Lagrangian and Hamiltonian formalisms. For the Lagrangian solutions,
by Proposition \ref{Chap05_prop:UnifiedToLagrangianSect}, from the holonomic section
$\psi \in \Gamma(\rho_\R^r)$ solution to equation \eqref{Chap05_eqn:UnifDynEqSect} we can recover
a holonomic section $\psi_\Lag = \rho_1^r \circ \psi \in \Gamma(\bar{\pi}^3)$ solution to equation
\eqref{Chap05_eqn:LagDynEqSect}. In particular, if we have $\psi(x) = (x,q_0(x),q_1(x),q_2(x),q_3(x),p^0(x),p^1(x))$,
then $\psi_\Lag(x) = (x,q_0(x),q_1(x),q_2(x),q_3(x))$ is a holonomic section solution to equations
\eqref{Chap05_eqn:ExampleRegular_DiffEquations}, which, bearing in mind the local expression
\eqref{Chap05_eqn:ExampleRegular_RestrictedLegendreMap} of the restricted Legendre-Ostrogradsky map
in this example, can be written locally as
\begin{align}
\rho + \dot{q}_2\derpar{\mu}{x} + q_2\frac{\partial^2\mu}{\partial x^2} + \dot{q}_3\mu + q_3\derpar{\mu}{x} = 0 \, ,
\label{Chap05_eqn:ExampleRegular_EulerLagrangeSections} \\
(\dot{q}_2 - q_3)\mu = 0 \, . \label{Chap05_eqn:ExampleRegular_FullHolonomySections}
\end{align}
Equation \eqref{Chap05_eqn:ExampleRegular_FullHolonomySections} gives the condition for the section
$\psi_\Lag$ to be holonomic, and it is redundant since we required this condition to be fulfilled at
the beginning. Now, if $\phi(x) = (x,y(x))$ is a section of $\pi$ such that $j^{3}\phi = \psi_\Lag$,
then the Euler-Lagrange equation can be written locally
\begin{equation*}
\frac{d^2}{d x^2}(\mu\ddot{y}) + \rho = 0 \ .
\end{equation*}
In the case of an homogeneous beam, the Euler-Lagrange equation reduces to $\mu y^{(iv)} + \rho = 0$.

For the Lagrangian vector field, from Lemma \ref{Chap05_lemma:LagCorrespondenceVF} and Theorem
\ref{Chap05_thm:UnifiedToLagrangianVF}, we can recover, from the holonomic vector field $X \in \vf(\W_r)$,
a holonomic vector field $X_\Lag \in \vf(J^{3}\pi)$ which is a solution to equations \eqref{Chap05_eqn:LagDynEqVF},
and is locally given by
\begin{equation*}
X_\Lag = \derpar{}{x} + q_1\derpar{}{q_0} + q_2\derpar{}{q_1} + q_3\derpar{}{q_2}
- \frac{1}{\mu}\left( \rho + q_2 \frac{\partial^2\mu}{\partial x^2} + 2q_3\derpar{\mu}{x} \right) \derpar{}{q_3} \, .
\end{equation*}

For the Hamiltonian solutions, since $\Lag$ is a second-order regular Lagrangian density we can use the results stated in
Section \ref{Chap05_sec:HamiltonianRegularCase} and recover the Hamiltonian solutions directly from the
unified formalism. For the Hamiltonian sections, using Proposition \ref{Chap05_prop:UnifiedToHamiltonianRegSect},
from a section $\psi \in \Gamma(\rho_\R^r)$ fulfilling equation \eqref{Chap05_eqn:UnifDynEqSect} we can recover a section
$\psi_h = \rho_2^r \circ \psi \in \Gamma(\bar{\pi}_{J^1\pi}^r)$ solution to equation \eqref{Chap05_eqn:HamRegDynEqSect}.
In particular, if we have $\psi(x) = (x,q_0(x),q_1(x),q_2(x),q_3(x),p^0(x),p^1(x))$, then
$\psi_h(x) = (x,q_0(x),q_1(x),p^0(x),p^1(x))$ is a section solution to equations \eqref{Chap05_eqn:ExampleRegular_Holonomy}
and \eqref{Chap05_eqn:ExampleRegular_DiffEquations}, which can be written locally as
\begin{equation*}
\dot{q}_0 = \restric{\derpar{H}{p^0}}{\psi_h} \quad ; \quad
\dot{q}_1 = \restric{\derpar{H}{p^1}}{\psi_h} \quad ; \quad
\dot{p}^0 = -\restric{\derpar{H}{q_0}}{\psi_h} \quad ; \quad
\dot{p}^1 = -\restric{\derpar{H}{q_1}}{\psi_h} \, ,
\end{equation*}
where $H \in \Cinfty(J^{1}\pi^*)$ is the local Hamiltonian function with local expression
\begin{equation*}
H(x,q_0,q_1,p^0,p^1) = p^0q_1 + \frac{(p^1)^2}{2\mu} - \rho q_0 \, .
\end{equation*}

For the Hamiltonian vector field, from Lemma \ref{Chap05_lemma:HamRegCorrespondenceVF} and Theorem
\ref{Chap05_thm:UnifiedToHamiltonianRegVF}, the vector field $X \in \vf(\W_r)$ gives a vector field
$X_h \in \vf(J^1\pi^*)$ solution to equations \eqref{Chap05_eqn:HamRegDynEqVF}, which is locally given by
\begin{equation*}
X_h = \derpar{}{x} + q_1\derpar{}{q_0} + \frac{p^1}{\mu}\derpar{}{q_1} + \rho\derpar{}{p^0} - p^0\derpar{}{p^1} \, .
\end{equation*}

\subsection{The second-order relativistic particle subjected to a potential}
\label{Chap05_exa:2ndOrderRelativisticParticlePotential}

Let us consider a relativistic particle whose action is proportional to its extrinsic curvature.
This system has been analyzed in several works
\cite{art:Batlle_Gomis_Pons_Roman88,art:Nesterenko89,art:Pisarski86,art:Plyushchay88}, and in Section
\ref{Chap03_exa:2ndOrderRelativisticParticle} using the Lagrangian-Hamiltonian formalism. Now assume
that this system is subjected to the action of a generic potential $V$ depending only on the time
and the position of the particle, thus obtaining a time-dependent dynamical system.

The configuration bundle for this system is $E \stackrel{\pi}{\to} \R$, where $E$ is a $(n+1)$-dimensional
smooth manifold. Let $t$ be the global coordinate in $\R$, and $\eta \in \df^{1}(\R)$ the volume form
in $\R$ with local expression $\eta = \d t$. Natural coordinates in $E$ adapted to the bundle structure
are denoted by $(t,q_0^i)$, $1 \leqslant i \leqslant n$. Now, bearing in mind the natural coordinates in the
second-order jet bundle of $\pi$, the second-order Lagrangian density for this system, $\Lag \in \df^{1}(J^{2}\pi)$,
is locally given by
\begin{equation}\label{Chap05_eqn:ExampleSingular_Lagrangian}
\Lag(t,q_0^i,q_1^i,q_2^i) = \left( \frac{\alpha}{(q_1^i)^2} \left[ (q_1^i)^2(q_2^i)^2 - (q_1^iq_2^i)^2 \right]^{1/2} + V(t,q_0^i) \right)\d t \equiv \left(\frac{\alpha}{(q_1^i)^2} \sqrt{g} + V(t,q_0^i)\right) \d t  \, ,
\end{equation}
where $\alpha$ is some nonzero constant and $V \in \Cinfty(J^{2}\pi)$ is a function depending only on $t$ and $q_0^i$.
As we have seen in Section \ref{Chap03_exa:2ndOrderRelativisticParticle}, this is a singular Lagrangian density,
since the Hessian matrix of the Lagrangian function $L \in \Cinfty(J^{2}\pi)$ associated with $\Lag$ with respect
to $q_2^A$ is
\begin{equation*}
\derpars{L}{q_2^B}{q_2^A} = \begin{cases}
\displaystyle \frac{\alpha}{2(q_1^i)^2\sqrt{g^3}} \left[ \left((q_1^iq_2^i)^2 - 2(q_1^i)^2(q_2^i)^2 \right)q_1^Bq_1^A \right.   & 
 \\
\displaystyle \qquad\qquad\qquad\quad \left. + (q_1^i)^2(q_1^iq_2^i)(q_2^Bq_1^A-q_1^Bq_2^A) - (q_1^i)^2(q_2^i)^2q_2^Bq_2^A \right] \, ,
& \mbox{ if } B \neq A \, , \\[10pt]
\displaystyle \frac{\alpha}{\sqrt{g^3}}\left[ g - (q_2^i)^2q_1^Aq_1^A + 2(q_1^iq_2^i)q_1^Aq_2^A - (q_1^i)^2q_2^Aq_2^A \right] \, , & \mbox{ if } B = A \, ,
\end{cases}
\end{equation*}
and a long calculation shows that
\begin{equation*}
\det \left( \derpars{L}{q_2^B}{q_2^A} \right) = 0 \, .
\end{equation*}

As in the previous example, this is a second-order system, and therefore we consider the bundles
$\W = J^3\pi \times_{J^{1}\pi} \Tan^*(J^{1}\pi)$ and $\W_r = J^{3}\pi \times_{J^{1}\pi} J^{1}\pi^*$.
Hence, we obtain the following diagram
\begin{equation*}
\xymatrix{
\ & \ & J^{3}\pi \times_{J^1\pi} \Tan^*(J^1\pi) \ar@/_1.3pc/[llddd]_{\rho_1} \ar[d]^-{\mu_\W} \ar@/^1.3pc/[rrdd]^{\rho_2} & \ & \ \\
\ & \ & J^3\pi \times_{J^1\pi} J^1\pi^* \ar[lldd]_{\rho_1^r} \ar[rrdd]^{\rho_2^r} & \ & \ \\
\ & \ & \ & \ & \Tan^*(J^{1}\pi) \ar[d]^-{\mu} \ar[lldd]_{\pi_{J^{1}\pi}}|(.25){\hole} \\
J^{3}\pi \ar[rrd]^{\pi^{3}_{1}} & \ & \ & \ & J^{1}\pi^* \ar[dll]^{\pi_{J^{1}\pi}^r} \\
\ & \ & J^{1}\pi \ar[d]^{\bar{\pi}^{1}} & \ & \ \\
\ & \ & \R & \ & \
}
\end{equation*}
Natural coordinates in $\W$ and $\W_r$ are denoted $(t,q_0^i,q_1^i,q_2^i,q_3^i,p,p_i^0,p_i^1)$ and
$(t,q_0^i,q_1^i,q_2^i,q_3^i,p_i^0,p_i^1)$, respectively. Now, if $\Theta_1 \in \df^{1}(\Tan^*(J^{1}\pi))$ and
$\Omega_{1} \in \df^2(\Tan^*(J^{1}\pi))$ are the canonical forms of the cotangent bundle of $J^{1}\pi$, we define
\begin{equation*}
\Theta = \rho_2^*\Theta_1=p_i^0 \d q_0^i + p_i^1 \d q_1^i + p\d t\in \df^{1}(\W)\quad ; \quad
\Omega = \rho_2^*\Omega_1=dq_0^i \wedge dp_i^0 + dq_1^i \wedge dp_i^1 - \d p \wedge \d t
\in \df^{2}(\W) \ .
\end{equation*}
The coupling $1$-form $\hat{\C} \in \df^{1}(\W)$, whose coordinate expression is
\eqref{Chap05_eqn:UnifCouplingFormLocal}, in this case is given locally by
\begin{equation*}
\hat{\C} = \hat{C}\cdot \rho_\R^*\eta = (p + p_i^0q_1^i + p_i^1q_2^i) \d t \, .
\end{equation*}
From this, and denoting $\hat{\Lag} = (\pi^3_2 \circ \rho_1)^*\Lag \in \df^{1}(\W)$ we can introduce the Hamiltonian
submanifold $\W_o \stackrel{j_o}{\hookrightarrow} \W$, which is locally defined by the constraint function
$\hat{C} - \hat{L} = 0$, whose coordinate expression is
\begin{equation*}
\hat{C} - \hat{L} = p + p_i^0q_1^i + p_i^1q_2^i - \frac{\alpha}{(q_1^i)^2} \sqrt{g} - V(t,q_0^i) \, .
\end{equation*}
This allows us to construct the Hamiltonian $\mu_\W$-section $\hat{h} \in \Gamma(\mu_\W)$,
which is specified by giving the local Hamiltonian function $\hat{H}$, whose local expression is
\begin{equation*}
\hat{H}(t,q_0^i,q_1^i,q_2^i,q_3^i,p_i^0,p_i^1) = p_i^0q_1^i + p_i^1q_2^i - \frac{\alpha}{(q_1^i)^2} \sqrt{g} - V(t,q_0^i) \, ,
\end{equation*}
that is, we have $\hat{h}(t,q_0^i,q_1^i,q_2^i,q_3^i,p_i^0,p_i^1) = (t,q_0^i,q_1^i,q_2^i,q_3^i,-\hat{H},p_i^0,p_i^1)$.
Using this Hamiltonian section, we define the forms $\Theta_r = \hat{h}^*\Theta \in \df^{1}(\W_r)$ and
$\Omega_r = \hat{h}^*\Omega \in \df^{2}(\W_r)$, with local expressions
\begin{align*}
&\Theta_r = p_i^0 \d q_0^i + p_i^1 \d q_1^i
+ \left( \frac{\alpha}{(q_1^i)^2} \sqrt{g} + V(t,q_0^i) - p_i^0q_1^i - p_i^1q_2^i \right)\d t \, , \\
&\Omega_r = \d q_0^i \wedge \d p_i^0 + \d q_1^i \wedge \d p_i^1 + \left( q_1^A\d p_A^0 + q_2^A\d p_A^1 - \derpar{V}{q_0^i}\d q_0^i \right. \\
&\qquad \qquad + \left[ p^0_A + \frac{\alpha}{((q_1^i)^2)^2\sqrt{g}}\left[ \left((q_1^i)^2(q_2^i)^2 - 2(q_1^iq_2^i)^2\right)q_1^A + (q_1^iq_2^i)(q_1^i)^2q_2^A \right] \right]\d q_1^A \\
&\qquad \qquad + \left. \left[ p_A^1 - \frac{\alpha}{(q_1^i)^2\sqrt{g}}\left( (q^i_1)^2 q_2^A - (q_1^iq_2^i)q_1^A \right) \right]\d q_2^A \right) \wedge \d t \, .
\end{align*}

Now we derive the dynamical equations of the system. In order to state the Lagrangian-Hamiltonian problem for
sections, let $Y \in \vf(\W_r)$ be a generic vector field locally given by
\begin{equation*}
Y = f\derpar{}{t} + f_0^i \derpar{}{q_0^i} + f_1^i \derpar{}{q_1^i} + F_2^i \derpar{}{q_2^i} + 
F_3^i \derpar{}{q_3^i} + G_i^0 \derpar{}{p_i^0} + G_i^1 \derpar{}{p_i^1} \ .
\end{equation*}
Now, if $\psi(t) = (t,q_0^i(t),q_1^i(t),q_2^i(t),q_3^i(t),p_i^0(t),p_i^1(t))$ is a holonomic section of the
projection $\rho_\R^r$, then equation \eqref{Chap05_eqn:UnifDynEqSect} leads to the following $5n$ equations
(the redundant equation \eqref{Chap05_eqn:UnifDynEqSectLocalRedundantEq} is omitted)
\begin{align}
\dot{q}_0^A = q_1^A \quad , \quad \dot{q}_1^A = q_2^A \, , \label{Chap05_eqn:ExampleSingular_Holonomy} \\
\dot{p}_A^0 = \derpar{V}{q_0^A} \quad ; \quad \dot{p}_A^1 = - p^0_A - 
\frac{\alpha}{((q_1^i)^2)^2\sqrt{g}}\left[ \left((q_1^i)^2(q_2^i)^2 - 2(q_1^iq_2^i)^2\right)q_1^A + 
(q_1^iq_2^i)(q_1^i)^2q_2^A \right] \, , \label{Chap05_eqn:ExampleSingular_DiffEquations} \\
p_A^1 - \frac{\alpha}{(q_1^i)^2\sqrt{g}}\left( (q^i_1)^2 q_2^A - (q_1^iq_2^i)q_1^A \right) = 0 \, .
\label{Chap05_eqn:ExampleSingular_LastMomentumCoord1}
\end{align}
Equations \eqref{Chap05_eqn:ExampleSingular_Holonomy} give the condition of holonomy of type $2$ for the
section $\psi$, which are also redundant since the holonomy of $\psi$ is required from the beginning. Equations
\eqref{Chap05_eqn:ExampleSingular_LastMomentumCoord1} are an algebraic condition, from which, in combination with
the second group of equations \eqref{Chap05_eqn:ExampleSingular_DiffEquations}, we conclude that the section $\psi$
lies in a submanifold $\W_\Lag \hookrightarrow \W_r$ defined locally by the constraints
\begin{align*}
&p^0_A = \frac{\alpha}{(q_1^i)^2\sqrt{g^3}} \left[ \left( (q_2^i)^2g + (q_1^i)^2(q_2^i)^2(q_1^iq_3^i) -
(q_1^i)^2(q_1^iq_2^i)(q_ 2^iq_3^i) \right)q_1^A\right] \\
&\qquad\qquad\quad + \frac{\alpha}{(q_1^i)^2\sqrt{g^3}} \left[ \left( ((q_1^i)^2)^2(q_2^iq_3^i) - (q_1^i)^2(q_1^iq_2^i)(q_1^iq_3^i) - 
(q_1^iq_2^i)g \right)q_2^A - (q_1^i)^2gq_3^A\right] \, , \\
&p^1_A = \frac{\alpha}{(q^i_1)^2\sqrt{g}} \left[ (q^i_1)^2q_2^A - (q^i_1q^i_2)q^A_1 \right] \, .
\end{align*}

Let us compute the local expression of the restricted Legendre-Ostrogradsky map, $\Leg \colon J^{3}\pi \to J^{1}\pi^*$,
which is obtained from Proposition \ref{Chap05_prop:UnifGraphLegendreOstrogradskyMap}. In this example,
the coordinate expression of $\Leg$ is
\begin{align}
\nonumber
&\Leg^*p^0_A = \frac{\alpha}{(q_1^i)^2\sqrt{g^3}} \left[ \left( (q_2^i)^2g + (q_1^i)^2(q_2^i)^2(q_1^iq_3^i) -
(q_1^i)^2(q_1^iq_2^i)(q_ 2^iq_3^i) \right)q_1^A\right] \\
\label{Chap05_eqn:ExampleSingular_RestrictedLegendreMap}
&\qquad\qquad\quad + \frac{\alpha}{(q_1^i)^2\sqrt{g^3}} \left[ \left( ((q_1^i)^2)^2(q_2^iq_3^i) - (q_1^i)^2(q_1^iq_2^i)(q_1^iq_3^i) - 
(q_1^iq_2^i)g \right)q_2^A - (q_1^i)^2gq_3^A\right] \, , \\
&\Leg^*p^1_A = \frac{\alpha}{(q^i_1)^2\sqrt{g}} \left[ (q^i_1)^2q_2^A - (q^i_1q^i_2)q^A_1 \right] \, . \nonumber
\end{align}
From this, the local expression of the extended Legendre-Ostrogradsky map
$\widetilde{\Leg} \colon J^{3}\pi \to \Tan^*(J^1\pi)$ in this example is
\begin{align}
\nonumber
&\widetilde{\Leg}^*p^0_A = \frac{\alpha}{(q_1^i)^2\sqrt{g^3}} \left[ \left( (q_2^i)^2g + (q_1^i)^2(q_2^i)^2(q_1^iq_3^i) -
(q_1^i)^2(q_1^iq_2^i)(q_ 2^iq_3^i) \right)q_1^A\right] \\
\nonumber
&\qquad\qquad\quad + \frac{\alpha}{(q_1^i)^2\sqrt{g^3}} \left[ \left( ((q_1^i)^2)^2(q_2^iq_3^i) - (q_1^i)^2(q_1^iq_2^i)(q_1^iq_3^i) - 
(q_1^iq_2^i)g \right)q_2^A - (q_1^i)^2gq_3^A\right] \, , \\ 
\label{Chap05_eqn:ExampleSingular_ExtendedLegendreMap}
&\widetilde{\Leg}^*p^1_A = \frac{\alpha}{(q^i_1)^2\sqrt{g}} \left[ (q^i_1)^2q_2^A - (q^i_1q^i_2)q^A_1 \right] \, , \\
\nonumber
&\widetilde{\Leg}^*p = \frac{\alpha}{(q_1^i)^2} \sqrt{g} + V(t,q_0^i)
- \left(\frac{\alpha}{(q_1^i)^2\sqrt{g^3}} \left[ \left( (q_2^i)^2g + (q_1^i)^2(q_2^i)^2(q_1^iq_3^i) -
(q_1^i)^2(q_1^iq_2^i)(q_ 2^iq_3^i) \right)q_1^A\right] \right. \\
\nonumber
&\qquad\qquad\quad \left .+ \frac{\alpha}{(q_1^i)^2\sqrt{g^3}} \left[ \left( ((q_1^i)^2)^2(q_2^iq_3^i) - (q_1^i)^2(q_1^iq_2^i)(q_1^iq_3^i) - (q_1^iq_2^i)g \right)q_2^A - (q_1^i)^2gq_3^A\right] \right)q_1^A \\
\nonumber
&\qquad \qquad \quad - \frac{\alpha}{(q^i_1)^2\sqrt{g}} \left[ (q^i_1)^2q_2^A - (q^i_1q^i_2)q^A_1 \right]q_2^A \, .
\end{align}
Hence, the section $\psi \in \Gamma(\rho_\R^r)$ is holonomic and lies in the submanifold $\W_\Lag \hookrightarrow \W_r$
defined by the constraint functions given by \eqref{Chap05_eqn:ExampleSingular_RestrictedLegendreMap}, and its last
component functions satisfy the $2n$ differential equations
\begin{equation*}
\dot{p}_A^0 = \derpar{V}{q_0^A} \quad ; \quad \dot{p}_A^1 = - p^0_A - 
\frac{\alpha}{((q_1^i)^2)^2\sqrt{g}}\left[ \left((q_1^i)^2(q_2^i)^2 - 2(q_1^iq_2^i)^2\right)q_1^A + 
(q_1^iq_2^i)(q_1^i)^2q_2^A \right] \, .
\end{equation*}

Now we state the Lagrangian-Hamiltonian problem for vector fields, that is, we wish to find a vector field
$X \in \vf(\W_r)$ solution to equations \eqref{Chap05_eqn:UnifDynEqVF}. If the vector field $X$ is locally
given by
\begin{equation*}
X = f\derpar{}{t} + f_0^A \derpar{}{q_0^A} + f_1^A \derpar{}{q_1^A} + F_2^A \derpar{}{q_2^A} + 
F_3^A \derpar{}{q_3^A} + G_A^0 \derpar{}{p_A^0} + G_A^1 \derpar{}{p_A^1} \, ,
\end{equation*}
then equations \eqref{Chap05_eqn:UnifDynEqVF} lead to the following $5n+1$ equations (the redundant equation
\eqref{Chap05_eqn:UnifDynEqVFLocalRedundantEq2} is omitted)
\begin{align}
f_0^A = f\cdot q_1^A \quad ; \quad f_1^A = f\cdot q_2^A \, , \label{Chap05_eqn:ExampleSingular_Semispray1} \\
G_A^0 = f\cdot\derpar{V}{q_0^A} \ ; \ G_A^1 = - f\cdot\left(p^0_A + \frac{\alpha}{((q_1^i)^2)^2\sqrt{g}}
\left[ \left((q_1^i)^2(q_2^i)^2 - 2(q_1^iq_2^i)^2\right)q_1^A + (q_1^iq_2^i)(q_1^i)^2q_2^A \right] \right) \, ,
\label{Chap05_eqn:ExampleSingular_DynamicalEquations} \\
f \neq 0 \, , \label{Chap05_eqn:ExampleSingular_FixingGauge} \\
f\cdot \left( p_A^1 - \frac{\alpha}{(q_1^i)^2\sqrt{g}}\left( (q^i_1)^2 q_2^A - (q_1^iq_2^i)q_1^A \right) \right) = 0
\label{Chap05_eqn:ExampleSingular_LastMomentumCoord2} \, .
\end{align}
From equations \eqref{Chap05_eqn:ExampleSingular_Semispray1} we obtain the condition of semispray of type $2$
for the vector field $X$. In addition, equations \eqref{Chap05_eqn:ExampleSingular_LastMomentumCoord2} are
algebraic relations between the coordinates of $\W_r$ which give, in coordinates, the result stated in
Proposition \ref{Chap05_prop:UnifFirstConstraintSubmanifold}. Thus, using \eqref{Chap05_eqn:ExampleSingular_Semispray1},
\eqref{Chap05_eqn:ExampleSingular_DynamicalEquations} and \eqref{Chap05_eqn:ExampleSingular_FixingGauge},
and taking $f = 1$ as a representative of the equivalence class, the vector field $X$ is given locally by
\begin{equation}\label{Chap05_eqn:ExampleSingular_VectorFieldBeforeTangency}
X = \derpar{}{t} + q_1^A \derpar{}{q_0^A} + q_2^A \derpar{}{q_1^A} + F_2^A \derpar{}{q_2^A} + 
F_3^A \derpar{}{q_3^A} + \derpar{V}{q_0^A} \derpar{}{p_A^0} + G_A^1 \derpar{}{p_A^1} \, ,
\end{equation}
where the functions $G_A^1$ are determined by \eqref{Chap05_eqn:ExampleSingular_DynamicalEquations}. If, in
addition, we require the vector field $X$ to be holonomic, this condition reduces the set of vector fields
$X \in \vf(\W_r)$ given by \eqref{Chap05_eqn:ExampleSingular_VectorFieldBeforeTangency} to the following ones
\begin{equation}
\label{Chap05_eqn:ExampleSingular_SemisprayType1dBeforeTangency}
X = \derpar{}{t} + q_1^A \derpar{}{q_0^A} + q_2^A \derpar{}{q_1^A} + q_3^A \derpar{}{q_2^A} + 
F_3^A \derpar{}{q_3^A} + \derpar{V}{q_0^A} \derpar{}{p_A^0} + G_A^1 \derpar{}{p_A^1} \, .
\end{equation}
Notice that the functions $F_3^A$ are not determinated until the tangency of the vector field $X$ along
$\W_c$ is required. Since this example has a Lagrangian density far more complicated than the previous
example, in this case we study directly the tangency of the vector field along the submanifold
$\W_\Lag = \graph(\Leg)$. From the expression in local coordinates
\eqref{Chap05_eqn:ExampleSingular_RestrictedLegendreMap} of the map $\Leg$, we obtain the primary
constraints defining the closed submanifold $\P = \Im(\Leg) \stackrel{\jmath}{\hookrightarrow} J^{1}\pi^*$, which are
\begin{equation}\label{Chap05_eqn:ExampleSingular_Constraints0}
\phi^{(0)}_1 \equiv p^1_iq_1^i = 0 \quad ; \quad
\phi^{(0)}_2 \equiv (p_i^1)^2 - \frac{\alpha^2}{(q^i_1)^2} = 0 \, .
\end{equation}
Let $\Leg_o \colon J^{3}\pi \to \P$ be the map defined by $\Leg = \jmath \circ \Leg_o$. Then, the submanifold
$\W_\Lag = \graph(\Leg_o)$ is defined by
\begin{equation*}
\W_\Lag = \left\{ [w] \in \W_r \mid \xi_0^A([w]) = \xi_1^A([w]) = \phi_1^{(0)}([w]) = \phi_2^{(1)}([w]) = 0 \right\} \, ,
\end{equation*}
where $\xi_r^A = p_A^r - \Leg^*p_A^r$.

Next, we compute the tangency condition for the vector field $X \in \vf(\W_r)$ given locally by
\eqref{Chap05_eqn:ExampleSingular_SemisprayType1dBeforeTangency} along the submanifold $\W_\Lag \hookrightarrow \W_r$,
by checking if the following identities hold
\begin{align}
\restric{\Lie(X)\xi^A_0}{\W_\Lag} = 0 \quad &; \quad \restric{\Lie(X)\xi^A_1}{\W_\Lag} = 0 \, ,
\label{Chap05_eqn:ExampleSingular_LagEquations1} \\
\restric{\Lie(X)\phi^{(0)}_1}{\W_\Lag} = 0  \quad &; \quad \restric{\Lie(X)\phi^{(0)}_2}{\W_\Lag} = 0 \, .
\label{Chap05_eqn:ExampleSingular_LieDerConstraints0}
\end{align}
As we have seen in Section \ref{Chap05_sec:UnifDynamicalEquations}, equations
\eqref{Chap05_eqn:ExampleSingular_LagEquations1} give us the Lagrangian equations for the vector field $X$.
However, equations \eqref{Chap05_eqn:ExampleSingular_LieDerConstraints0} do not hold, since
\begin{equation*}
\Lie(X)\phi^{(0)}_1 = \Lie(X)(p^1_iq_1^i) = - p_i^0q_1^i \quad ; \quad
\Lie(X)\phi^{(0)}_2 = \Lie(X)((p^1_i)^2 - \alpha^2 / (q_1^i)^2) = - 2p_i^0p_i^1 \, ,
\end{equation*}
and hence we obtain two first-generation secondary constraints
\begin{equation}\label{Chap05_eqn:ExampleSingular_Constraints1}
\phi^{(1)}_1 \equiv p_i^0 q_1^i = 0 \quad ; \quad \phi^{(1)}_2 \equiv p_i^0p_i^1 = 0 \, ,
\end{equation}
that define a new submanifold $\W_1 \hookrightarrow \W_\Lag$. Now, by checking the tangency of the vector
field $X$ along this new submanifold, we obtain
\begin{equation*}
\Lie(X)\phi^{(1)}_1 = \Lie(X)(p_i^0q_1^i) = 0 \quad ; \quad \Lie(X)\phi^{(1)}_2 = \Lie(X)(p_i^0p_i^1) = -(p^0_i)^2 \, ,
\end{equation*}
and a second-generation secondary constraint appears,
\begin{equation}\label{Chap05_eqn:ExampleSingular_Constraints2}
\phi^{(2)} \equiv (p^0_i)^2 = 0 \, ,
\end{equation}
which defines a new submanifold $\W_2 \hookrightarrow \W_1$. Finally, the tangency of the vector field
$X$ along this submanifold gives no new constraints, since
\begin{equation*}
\Lie(X)\phi^{(2)} = \Lie(X)((p^0_i)^2) = 0 \, .
\end{equation*}
So we have two primary constraints \eqref{Chap05_eqn:ExampleSingular_Constraints0}, two first-generation
secondary constraints \eqref{Chap05_eqn:ExampleSingular_Constraints1}, and a single second-generation
secondary constraint \eqref{Chap05_eqn:ExampleSingular_Constraints2}. Notice that these five constraints
only depend on $q_1^A$, $p^0_A$ and $p^1_A$, and so they are $\rho_2^r$-projectable.

Notice that we still have to check \eqref{Chap05_eqn:ExampleSingular_LagEquations1}. As we have seen in
Section \ref{Chap05_sec:UnifDynamicalEquations}, we obtain the following equations
\begin{align}
\left(F_3^B - \frac{d}{dt} q_3^B \right)\derpars{\hat{L}}{q_2^B}{q_2^A} +  \derpar{\hat{L}}{q_0^A} -
\frac{d}{dt}\left(\derpar{\hat{L}}{q_1^A}\right) +  \frac{d^2}{dt^2}\left(\derpar{\hat{L}}{q_2^A}\right)
+  \left(F_2^B - q_3^B\right)\frac{d}{dt}\left(\derpars{\hat{L}}{q_2^B}{q_2^A}\right) = 0 \, ,
\label{Chap05_eqn:ExampleSingular_EulerLagrangeInitial} \\
\left(F_2^B - q_3^B\right)\derpars{\hat{L}}{q_2^B}{q_2^A} = 0 \, . \label{Chap05_eqn:ExampleSingular_SemisprayType1Condition}
\end{align}
Since we have already required the vector field $X$ to be holonomic in $\W_r$, equations
\eqref{Chap05_eqn:ExampleSingular_SemisprayType1Condition} are satisfied identically and equations
\eqref{Chap05_eqn:ExampleSingular_EulerLagrangeInitial} become
\begin{equation}\label{Chap05_eqn:ExampleSingular_EulerLagrangeFinal}
\left(F_3^B - \frac{d}{dt} q_3^B \right) \derpars{\hat{L}}{q_2^B}{q_2^A} + \derpar{\hat{L}}{q_0^A}
- \frac{d}{dt}\left(\derpar{\hat{L}}{q_1^A}\right) + \frac{d^2}{dt^2}\left(\derpar{\hat{L}}{q_2^A}\right) = 0 \, .
\end{equation}
Using the results of Section \ref{Chap03_exa:2ndOrderRelativisticParticle}, and bearing in mind the
coordinate expressions \eqref{Chap03_eqn:ExampleSingular_Lagrangian} and \eqref{Chap05_eqn:ExampleSingular_Lagrangian}
of the autonomous and non-autonomous Lagrangians, respectively, we deduce that this equation is compatible
if, and only if,
\begin{equation*}
\derpar{V}{q_0^A} = 0 \ , \ \mbox{for }1 \leqslant A \leqslant n \, .
\end{equation*}
That is, we have $n$ first-generation secondary constraints arising from the tangency condition of $X$ along
$\W_\Lag$, thus defining a new submanifold $\W_3 \hookrightarrow \W_2$ with constraint functions
\begin{equation*}
\phi^{(1)}_{3,A} \equiv \derpar{V}{q_0^A} = 0 \quad \mbox{for } 1 \leqslant A \leqslant n \, .
\end{equation*}
Observe that, since $V$ is a function that depends only on $t$ and $q_0^A$, these new constraints also
depend only on the coordinates $t$ and $q_0^A$, and thus they are $\rho_2^r$-projectable. From a physical
viewpoint, these constraints mean that the dynamics of the particle can take place on every level set of
the potential with respect to the position coordinates.

Finally, we recover the Lagrangian and Hamiltonian dynamics from the unified formalism. For the Lagrangian
solutions, using Proposition \ref{Chap05_prop:UnifiedToLagrangianSect}, we know that from the holonomic
section $\psi \in \Gamma(\rho_\R^r)$ solution to equation \eqref{Chap05_eqn:UnifDynEqSect} we can recover
a holonomic section $\psi_\Lag = \rho_1^r \circ \psi \in \Gamma(\bar{\pi}^3)$ solution to equation
\eqref{Chap05_eqn:LagDynEqSect}. In particular, if
$\psi(t) = (t,q_0^i(t),q_1^i(t),q_2^i(t),q_3^i(t),p_i^0(t),p_i^1(t))$, then
$\psi_\Lag(t) = (t,q_0^i(t),q_1^i(t),q_2^i(t),q_3^i(t))$ is a holonomic section
solution to equations \eqref{Chap05_eqn:ExampleSingular_DiffEquations}. Now, bearing in mind the
local expression \eqref{Chap05_eqn:ExampleSingular_RestrictedLegendreMap} of the restricted Legendre-Ostrogradsky
map, equations \eqref{Chap05_eqn:ExampleSingular_DiffEquations} give the last $n$ equations of the holonomy
condition for $\psi_\Lag$, which are identically satisfied since the holonomy condition has been already required,
and the classical second-order Euler-Lagrange equations
\begin{equation*}
\restric{\derpar{L}{q_0^A}}{\psi_\Lag} - \restric{\frac{d}{dt}\derpar{L}{q_1^A}}{\psi_\Lag}
+ \restric{\frac{d^2}{d t^2}\derpar{L}{q_2^A}}{\psi_\Lag} = 0 \, .
\end{equation*}

For the Lagrangian vector field, from Lemma \ref{Chap05_lemma:LagCorrespondenceVF} and Theorem
\ref{Chap05_thm:UnifiedToLagrangianVF}, we can recover from a holonomic vector field $X \in \vf(\W_r)$
solution to equation \eqref{Chap05_eqn:UnifDynEqVF} a holonomic vector field, $X_\Lag \in \vf(J^3\pi)$,
solution to equations \eqref{Chap05_eqn:LagDynEqVF} on the points of a submanifold $S_f \hookrightarrow J^3\pi$
given by $S_f = \rho_1^r(\W_3)$), and this vector field is locally given by
\begin{equation*}
X_\Lag = \derpar{}{t} + q_1^A \derpar{}{q_0^A} + q_2^A \derpar{}{q_1^A} + q_3^A \derpar{}{q_2^A} + F_3^A \derpar{}{q_3^A} \, ,
\end{equation*}
where $F_3^A$ are the solutions of equations \eqref{Chap05_eqn:ExampleSingular_EulerLagrangeFinal}, and we
have taken $f = 1$ as a representative of the equivalence class.

One can check that if the holonomy condition is not required at the beginning and we perform all this procedure
with the vector field given by \eqref{Chap05_eqn:ExampleSingular_VectorFieldBeforeTangency}, the final result
is the same. This means that, in this case, the holonomy condition does not give any additional constraint.

Now, since $\Lag$ is an almost-regular Lagrangian density, for the Hamiltonian dynamics we must use the results
stated in Section \ref{Chap05_sec:HamiltonianSingularCase} and recover the Hamiltonian solutions passing through
the Lagrangian formalism. For the Hamiltonian sections, by Proposition \ref{Chap05_prop:UnifiedToHamiltonianSingSect},
from a section $\psi \in \Gamma(\rho_\R^r)$ solution to equation \eqref{Chap05_eqn:UnifDynEqSect}, we can recover
a section $\psi_h = \Leg \circ \rho_1^r \circ \psi \in \Gamma(\bar{\pi}_\P)$ solution to the equation
\eqref{Chap05_eqn:HamSingDynEqSect}.

For the Hamiltonian vector fields, we know that there are holonomic vector fields $X_\Lag \in \vf(J^{3}\pi)$,
solutions to equations \eqref{Chap05_eqn:LagDynEqVF} at support on the submanifold $S_f = \rho_1^r(\W_3)$
which are $\Leg_o$-projectable on $\P_f = \Leg_o(S_f) = \rho_2^r(\W_3)$, tangent to $\P_f$ and solutions to
the Hamilton equations.


\clearpage
\thispagestyle{empty}


\clearpage
\chapter{Second-order classical field theories}
\label{Chap:HOClassicalFieldTheories}


The aim of this Chapter is to state the geometric formulation of second-order field theories,
thus generalizing the results of Section \ref{Chap02_sec:FieldTheories} to the second-order case.
Observe that, unlike in previous Chapters, in which we give geometric formulations for dynamical
systems of arbitrary order, in this Chapter we focus on the second-order case. The reason to do so
is that our formulation can not be straightforwardly generalized to field theories of order greater
than $2$. Some comments on this subject, and higher-order field theories in general, are given in
Section \ref{Chap06_sec:HigherOrderCase}.

Observe also that, as in the geometric description of higher-order non-autonomous dynamical systems
given in Chapter \ref{Chap:HONonAutonomousDynamicalSystems}, for second-order field theories we do
not have a complete description of the Lagrangian and Hamiltonian formalisms. This is due, mainly,
to the following two reasons. First, given a $k$th-order Lagrangian density for a field theory, the
definition of the Poincar\'{e}-Cartan forms is not unique, and while for the second-order case
it is proved that all these forms are equivalent \cite{book:Saunders89,art:Saunders_Crampin90},
this is not true for the general higher-order case. The second reason, which is closely related
with the first one, is the choice of the Hamiltonian phase space. Since the Poincar\'{e}-Cartan forms
are not unique, neither is the Legendre map in higher-order field theories, and therefore we have several
different options for the Hamiltonian phase space of the theory
\cite{art:Aldaya_Azcarraga80,art:Francaviglia_Krupka82,art:Kolar84,proc:Krupka84}.

Taking into account the above comments, in this Chapter we proceed in an analogous way to Chapter
\ref{Chap:HONonAutonomousDynamicalSystems}: we first describe the Lagrangian-Hamiltonian unified
formalism for second-order field theories, and, from this setting, we derive both the Lagrangian
and Hamiltonian formalisms for this kind of systems. Therefore, the structure of the Chapter is the
following. In Section \ref{Chap06_sec:SymmetricMultimomenta} we introduce the space of 2-symmetric
multimomenta, which is the Hamiltonian phase space that we choose to set up our formulation. Using
this Hamiltonian phase space, in Section \ref{Chap06_sec:UnifiedFormalism} we describe the unified
formalism for second-order field theories: phase space, canonical structures and field equations,
written in terms of sections and multivector fields. Then we describe the Lagrangian and Hamiltonian
formalisms in Sections \ref{Chap06_sec:LagrangianFormalism} and \ref{Chap06_sec:HamiltonianFormalism},
respectively. Next, two physical examples are studied in Section \ref{Chap06_sec:Examples} to
illustrate the application of the formalism: the bending of a clamped plate under a uniform load,
and the classic Korteweg--de Vries equation. Finally, Section \ref{Chap06_sec:HigherOrderCase} is
devoted to give some comments on field theories of order greater than $2$, and the main issues that
prevent us from generalizing this formulation to field theories of arbitrary order.

Along this Chapter, we consider a second-order Lagrangian field theory with $n$ fields depending on
$m$ independent variables. As in Section \ref{Chap02_sec:FieldTheories}, the configuration space
for this theory is a bundle $\pi \colon E \to M$, where $M$ is a $m$-dimensional orientable smooth
manifold with fixed volume form $\eta \in \df^{m}(M)$, and $\dim E = m + n$. The physical information
is given in terms of a second-order Lagrangian density $\Lag \in \df^{m}(J^2\pi)$, which is a
$\bar{\pi}^2$-semibasic $m$-form. Because of this, we can write $\Lag = L \cdot (\bar{\pi}^2)^*\eta$,
where $L \in \Cinfty(J^2\pi)$ is the second-order Lagrangian function associated to $\Lag$ and $\eta$.
Multi-index notation introduced in Section \ref{Chap01_sec:HOJetBundlesMultiIndex} is used.

\section{The space of 2-symmetric multimomenta}
\label{Chap06_sec:SymmetricMultimomenta}

According to \cite{proc:Carinena_Crampin_Ibort91,art:Echeverria_DeLeon_Munoz_Roman07,
art:Echeverria_Munoz_Roman00_JMP}, and the results stated in Section \ref{Chap02_sec:FieldTheories}
and Chapter \ref{Chap:HONonAutonomousDynamicalSystems}, the appropriate choices of Hamiltonian phase
spaces seem to be the extended and restricted dual jet bundles introduced in Section
\ref{Chap01_sec:HOJetBundlesDualBundles}, namely $\Lambda_2^m(J^1\pi)$ and $J^1\pi^*$. Nevertheless,
these bundles have too many multimomentum coordinates in order to establish a correspondence between
``velocities'' and multimomenta in terms of derivatives of the second-order Lagrangian function $L$.

In particular, from the results in Sections \ref{Chap01_sec:HOJetBundlesDef&Coord} and
\ref{Chap01_sec:HOJetBundlesDualBundles} we know that if $(x^i,u^\alpha)$, $1 \leqslant i \leqslant m$,
$1 \leqslant \alpha \leqslant n$ are local coordinates in $E$ adapted to the bundle structure, then
the induced natural coordinates in $J^2\pi$, $\Lambda_2^m(J^{1}\pi)$ and $J^1\pi^*$ are
$(x^i,u^\alpha,u_i^\alpha,u_I^\alpha)$, $(x^i,u^\alpha,u_i^\alpha,p,p_\alpha^i,p_\alpha^{ij})$ and
$(x^i,u^\alpha,u_i^\alpha,p_\alpha^i,p_\alpha^{ij})$, respectively, with
$1 \leqslant i,j \leqslant m$, $1 \leqslant \alpha \leqslant n$, $|I| = 2$. Hence, the dual jet
bundles of $\pi$ have $nm + nm^2$ multimomentum coordinates (and an additional one in the extended
bundle, which is identified with the local Hamiltonian function), while the second-order jet bundle
$J^2\pi$ has $nm + nm(m+1)/2$ ``velocity'' coordinates. That is, although we have the same number
of first-order velocities and multimomentum coordinates ($nm$), there are $nm(m-1)/2$ more
second-order multimomenta than second-order ``velocities''. Therefore, we should consider a
Hamiltonian phase space with less second-order multimomentum coordinates.

A way to do so is introducing some relations among the second-order multimomentum coordinates, thus
defining submanifolds of the aforementioned bundles with less second-order multimomenta.
According to \cite{phd:Campos} and \cite{art:Saunders_Crampin90}, let us consider the submanifold
$J^2\pi^\dagger \hookrightarrow \Lambda_2^m(J^1\pi)$ defined by
\begin{equation*}
J^2\pi^\dagger = \left\{ \omega \in \Lambda_2^m(J^1\pi) \, \mid \, p_\alpha^{ij} = p_\alpha^{ji} \
\mbox{ for every } 1 \leqslant i,j\leqslant m \, , \, 1 \leqslant \alpha \leqslant n \right\} \, .
\end{equation*}
By definition, it is clear that this submanifold is $\pi_{J^{1}\pi}$-transverse. Therefore,
$J^2\pi^\dagger$ fibers over $J^{1}\pi$, and we have the canonical projections
\begin{equation*}
\pi_{J^1\pi}^\dagger \colon J^2\pi^\dagger \to J^1\pi \quad ; \quad
\bar{\pi}_{J^1\pi}^\dagger = \bar{\pi}^{1} \circ \pi_{J^1\pi}^\dagger \colon J^2\pi^\dagger \to M \, ,
\end{equation*}
which are the natural restrictions of the canonical projections of $\Lambda^m_2(J^1\pi)$ to the
submanifold $J^2\pi^\dagger$. From the induced coordinates
$(x^i,u^\alpha,u_i^\alpha,p,p_\alpha^i,p_\alpha^{ij})$ of $\Lambda^m_2(J^1\pi)$ we obtain the natural
coordinates in $J^2\pi^\dagger$ adapted to the bundle structure, which are
$(x^i,u^\alpha,u^\alpha_i,p,p_\alpha^{i},p_\alpha^{I})$, where $|I| = 2$. Then, the natural embedding
$j_s \colon J^2\pi^\dagger \hookrightarrow \Lambda_2^m(J^1\pi)$ is given locally by
\begin{equation}\label{Chap06_eqn:SymmetricMultimomentaEmbeddingLocal}
\begin{array}{c}
\displaystyle j_s^*x^i = x^i \quad ; \quad j_s^*u^\alpha = u^\alpha \quad ; \quad j_s^*u_i^\alpha = u_i^\alpha
\quad ; \quad j_s^*p_\alpha^i = p_\alpha^i \, , \\[10pt]
\displaystyle j_s^*p_\alpha^{ij} = \frac{1}{n(ij)} \, p_\alpha^{1_i+1_j} \, , \quad \mbox{where }
n(ij) = \begin{cases} 1 \, , & \mbox{ if } i=j \\ 2 \, , & \mbox{ if } i \neq j \end{cases} \, .
\end{array}
\end{equation}
The submanifold $J^2\pi^\dagger \hookrightarrow \Lambda_2^m(J^1\pi)$ is called the
\textsl{extended $2$-symmetric multimomentum bundle}, and its dimension is given by
\begin{equation*}
\dim J^{2}\pi^\dagger = \dim\Lambda_2^m(J^1\pi) - \frac{nm(m-1)}{2} = m+n+2mn+\frac{nm(m+1)}{2} + 1 \, .
\end{equation*}
In particular, this submanifold has $nm(m+1)/2$ second-order multimomentum coordinates, as we want.

All the geometric structures defined in Section \ref{Chap01_sec:HOJetBundlesDualBundles} for
$\Lambda_2^m(J^1\pi)$ restrict to $J^2\pi^\dagger$. In particular, let us denote
$\Theta_1^s = j_s^*\Theta_1 \in \df^{m}(J^2\pi^\dagger)$ and
$\Omega_1^s = j_s^*\Omega_1 = -\d\Theta_1^s \in \df^{m+1}(J^2\pi^\dagger)$ the pull-back of the
Liouville $m$ and $(m+1)$-forms to $J^2\pi^\dagger$, which we call the
\textsl{symmetrized Liouville $m$ and $(m+1)$-forms}. In addition, from the canonical pairing
$\C \colon J^2\pi \times_{J^1\pi} \Lambda_2^m(J^1\pi) \to \Lambda_1^m(J^1\pi)$, we can define
a pairing $\C^s \colon J^2\pi \times_{J^1\pi} J^2\pi^\dagger \to \Lambda_1^m(J^1\pi)$ as
\begin{equation*}
\C^s(j^{2}_x\phi,\omega) = \C(j^{2}_x\phi,j_s(\omega)) = (j^1\phi)_{j^1_x\phi}^* \ j_s(\omega) \, ,
\end{equation*}
which we call the \textsl{symmetrized canonical pairing}. As in Section
\ref{Chap01_sec:HOJetBundlesDualBundles}, since $\C^s$ takes values in $\Lambda_1^m(J^1\pi)$, there
exists a function $C^s \in \Cinfty(J^2\pi \times_{J^1\pi} J^2\pi^\dagger)$ such that
$C^s(j^2_x\phi,\omega) \cdot (\bar{\pi}_{J^1\pi}^\dagger)^*\eta = (j^1\phi)_{j^1_x\phi}^* \ j_s(\omega)$.

Let us compute in coordinates the local expressions of the symmetrized Liouville forms and of the
symmetrized canonical pairing. Recall that, in the induced natural coordinates of
$\Lambda_2^m(J^1\pi)$, the coordinate expressions of the Liouville forms are given by
\eqref{Chap01_eqn:HOJetBundlesDualBundlesCanonicalFormsLocal}, and the canonical pairing $\C$ is
given by \eqref{Chap01_eqn:HOJetBundlesDualBundlesCanonicalPairingLocal}. In the second-order case,
the aforementioned local expressions are
\begin{align*}
&\Theta_{1} = p\d^mx + p^i_\alpha \d u^\alpha \wedge \d^{m-1}x_i + p^{ij}_\alpha \d u_i^\alpha \wedge \d^{m-1}x_j \, ,\\
&\Omega_{1} = -\d p \wedge \d^mx - \d p^i_\alpha \wedge \d u^\alpha \wedge \d^{m-1}x_i - 
\d p^{ij}_\alpha \wedge \d u_i^\alpha \wedge \d^{m-1}x_j \, , \\
& \C(x^i,u^\alpha,u_i^\alpha,p,p_\alpha^i,p_\alpha^{ij}) = (p + p_\alpha^iu_i^\alpha + 
p_\alpha^{ij}u_{1_i+1_j}^\alpha)\d^mx \, .
\end{align*}
Then, bearing in mind the local expression \eqref{Chap06_eqn:SymmetricMultimomentaEmbeddingLocal}
of the canonical embedding $j_s \colon J^2\pi^\dagger \hookrightarrow \Lambda_2^m(J^1\pi)$, the
coordinate expressions of $\Theta_1^s$ and $\Omega_1^s$ become
\begin{equation}\label{Chap06_eqn:SymmetricMultimomentaLiouvilleFormsLocal}
\begin{array}{l}
\displaystyle \Theta_{1}^s = p \d^mx + p^i_\alpha \d u^\alpha \wedge \d^{m-1}x_i +
 \frac{1}{n(ij)} \, p^{1_i+1_j}_\alpha \d u_i^\alpha \wedge \d^{m-1}x_j \, , \\[10pt]
\displaystyle \Omega_{1}^s = -\d p \wedge \d^mx - 
\d p^i_\alpha \wedge \d u^\alpha \wedge \d^{m-1}x_i -
 \frac{1}{n(ij)} \, \d p^{1_i+1_j}_\alpha \wedge \d u_i^\alpha \wedge \d^{m-1}x_j \, ,
\end{array}
\end{equation}
while the local expression of the symmetrized canonical pairing is
\begin{equation}\label{Chap06_eqn:SymmetricMultimomentaCanonicalPairingLocal}
\C^s(x^i,u^\alpha,u^\alpha_i,u^\alpha_I,p,p_\alpha^i,p_\alpha^I) = 
(p + p_\alpha^iu_i^\alpha + p_\alpha^{I}u_{I}^\alpha) \d^mx \, .
\end{equation}

An important fact concerning the pull-back of the multisymplectic $(m+1)$-form $\Omega_1$ to
$J^2\pi^\dagger$ is that it is multisymplectic in $J^2\pi^\dagger$. Since $\Omega_1^s = -\d\Theta_1^s$
is obviously closed, it suffices to show that this form is $1$-nondegenerate, that is,
$\inn(X)\Omega_1^s = 0$ if, and only if, $X = 0$. In coordinates, let $X \in \vf(J^2\pi^\dagger)$
be a generic vector field locally given by
\begin{equation*}
X = f^i\derpar{}{x^i} + F^\alpha\derpar{}{u^\alpha} + F_i^\alpha\derpar{}{u_i^\alpha}
+ g \derpar{}{p} + G_\alpha^i\derpar{}{p_\alpha^i} + G_\alpha^I\derpar{}{p_\alpha^I} \, .
\end{equation*}
Then, taking into account the coordinate expression \eqref{Chap06_eqn:SymmetricMultimomentaLiouvilleFormsLocal}
of the $(m+1)$-form $\Omega_1^s$, the $m$-form $\inn(X)\Omega_1^s$ is locally given by
\begin{align*}
\inn(X)\Omega_1^s &=
f^k\left( \d p \wedge \d^{m-1}x_k - \d p_\alpha^i \wedge \d u^\alpha \wedge 
\d^{m-2}x_{ik} - \frac{1}{n(ij)}\,\d p^{1_i+1_j}_\alpha \wedge \d u_i^\alpha \wedge \d^{m-2}x_{jk} \right) \\
&\qquad {} + 
F^\alpha \d p_\alpha^i \wedge \d^{m-1}x_i + F_i^\alpha \frac{1}{n(ij)} \, \d p^{1_i+1_j} \wedge \d^{m-1}x_j
- g\d^mx - G_\alpha^i\d u^\alpha \wedge \d^{m-1}x_i \\
 &\qquad {} - G_\alpha^I\sum_{1_i+1_j=I}\frac{1}{n(ij)} \d u_i^\alpha \wedge \d^{m-1}x_j \, ,
\end{align*}
where $\d^{m-2}x_{jk} = \inn(\partial / \partial x^k) \d^{m-1}x_j$.
From this coordinate expression it is clear that $\inn(X)\Omega_1^s = 0$ if, and only if, $X = 0$.
Hence $\Omega_1^s$ is multisymplectic.

Finally, recall that in the extended Hamiltonian formalism described in Section
\ref{Chap02_sec:FieldTheories} we define the restricted multimomentum bundle as the quotient of
the extended multimomentum bundle by constant affine transformations along the fibers of $\pi^1$.
Analogously, we define the \textsl{restricted $2$-symmetric multimomentum bundle} as the quotient
bundle
\begin{equation*}
J^2\pi^\ddagger = J^2\pi^\dagger \big/ \Lambda^m_1(J^1\pi) \, .
\end{equation*}
This bundle is endowed with some canonical projections: the natural quotient map,
$\mu \colon J^2\pi^\dagger \to J^2\pi^\ddagger$, and the canonical projections
$\pi_{J^1\pi}^\ddagger \colon J^2\pi^\ddagger \to J^1\pi$ and
$\bar{\pi}_{J^1\pi}^\ddagger \colon J^2\pi^\ddagger \to M$, which satisfy
$\pi_{J^1\pi}^\dagger = \pi_{J^1\pi}^\ddagger \circ \mu$ and
$\bar{\pi}_{J^1\pi}^\dagger = \bar{\pi}_{J^1\pi}^\ddagger \circ \mu$.

Since the quotient $J^2\pi^\ddagger$ can be defined alternatively as the submanifold of $J^1\pi^*$
defined by the $nm(m-1)/2$ local constraints $p_\alpha^{ij} - p_\alpha^{ji} = 0$, natural coordinates
$(x^i,u^\alpha,u^\alpha_i,p_\alpha^i,p_\alpha^{ij})$ in $J^1\pi^*$ induce local coordinates
$(x^i,u^\alpha,u^\alpha_i,p_\alpha^i,p_\alpha^I)$ in $J^2\pi^\ddagger$. Observe that
\begin{equation*}
\dim J^{2}\pi^\ddagger = \dim J^{2}\pi^\dagger - 1 = m+n+2mn+\frac{nm(m+1)}{2} \, .
\end{equation*}


\section{Lagrangian-Hamiltonian unified formalism}
\label{Chap06_sec:UnifiedFormalism}

\subsection{Geometrical setting}
\label{Chap06_sec:UnifiedFormalismSetting}

\subsubsection{Unified phase space and bundle structures. Local coordinates}

According to the results in Section \ref{Chap02_sec:FieldTheoriesUnified}, let us consider the following
fiber bundles
\begin{equation*}
\W = J^3\pi \times_{J^1\pi} J^{2}\pi^\dagger \quad ; \quad
\W_r = J^3\pi \times_{J^1\pi} J^{2}\pi^\ddagger \, ,
\end{equation*}
where $J^2\pi^\dagger$ and $J^2\pi^\ddagger$ are the extended and restricted $2$-symmetric
multimomentum bundles defined in the previous Section, respectively. The bundles $\W$ and $\W_r$
are called the \textsl{extended $2$-symmetric jet-multimomentum bundle} and the
\textsl{restricted $2$-symmetric jet-multimomentum bundle}, respectively.

The bundles $\W$ and $\W_r$ are endowed with the canonical projections
\begin{equation*}
\rho_1 \colon \W \to J^{3}\pi \quad ; \quad \rho_2 \colon \W \to J^{2}\pi^\dagger \quad ; \quad
\rho_{J^{1}\pi} \colon \W \to J^{1}\pi \quad ; \quad \rho_M \colon \W \to M \, ,
\end{equation*}
\begin{equation*}
\rho_1^r \colon \W_r \to J^{3}\pi \quad ; \quad \rho^r_2 \colon \W_r \to J^{2}\pi^\ddagger
\quad ; \quad \rho_{J^{1}\pi}^r \colon \W_r \to J^{1}\pi \quad ; \quad \rho_M^r \colon \W_r \to M \, .
\end{equation*}

In addition, the natural quotient map $\mu \colon J^{2}\pi^\dagger \to J^{2}\pi^\ddagger$ induces
a surjective submersion $\mu_\W \colon \W \to \W_r$. Thus, we have the following commutative diagram
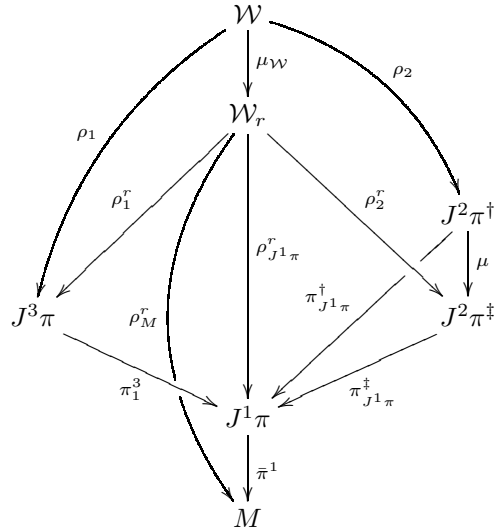
\begin{equation*}
\xymatrix{
\ & \ & \W \ar@/_1.3pc/[llddd]_{\rho_1} \ar[d]^-{\mu_\W} \ar@/^1.3pc/[rrdd]^{\rho_2} & \ & \ \\
\ & \ & \W_r \ar[lldd]_{\rho_1^r} \ar[rrdd]^{\rho_2^r} \ar[ddd]^<(0.4){\rho_{J^{1}\pi}^r} \ar@/_2.5pc/[dddd]_-{\rho_M^r}|(.675){\hole} & \ & \ \\
\ & \ & \ & \ & J^{2}\pi^\dagger \ar[d]^-{\mu} \ar[lldd]_{\pi_{J^{1}\pi}^\dagger}|(.25){\hole} \\
J^{3}\pi \ar[rrd]_{\pi^{3}_{1}} & \ & \ & \ & J^{2}\pi^\ddagger \ar[dll]^{\pi_{J^{1}\pi}^\ddagger} \\
\ & \ & J^{1}\pi \ar[d]^{\bar{\pi}^{1}} & \ & \ \\
\ & \ & M & \ & \
}
\end{equation*}

Let $(x^i,u^\alpha)$ be a set of local coordinates in $E$ adapted to the bundle structure and such
that $\eta = \d x^1 \wedge \ldots \wedge \d x^m \equiv \d^mx$. Then, we denote by
$(x^i,u^\alpha,u^\alpha_{i},u^\alpha_{I},u^\alpha_{J})$ and
$(x^i,u^\alpha,u^\alpha_i,p,p_\alpha^i,p_\alpha^{I})$
the induced local coordinates in $J^3\pi$ and $J^{2}\pi^\dagger$, respectively, with $|I| = 2$
and $|J|=3$. Thus, $(x^i,u^\alpha,u^\alpha_i,p_\alpha^i,p_\alpha^{I})$ are the natural
coordinates in $J^{2}\pi^\ddagger$, and the coordinates in $\W$ and $\W_r$ are
$(x^i,u^\alpha,u^\alpha_i,u^\alpha_{I},u^\alpha_{J},p,p_\alpha^{i},p_\alpha^{I})$ and
$(x^i,u^\alpha,u^\alpha_i,u^\alpha_{I},u^\alpha_{J},p_\alpha^{i},p_\alpha^{I})$, respectively.
Observe that
\begin{equation*}
\dim\W = m + n + 2nm + nm(m+1) + \frac{nm(m+1)(m+2)}{6} + 1 \quad ; \quad
\dim\W_r = \dim\W - 1 \, .
\end{equation*}
In these coordinates, the above projections have the following coordinate expressions
\begin{align*}
&\rho_1(x^i,u^\alpha,u^\alpha_i,u^\alpha_{I},u^\alpha_{J},p,p_\alpha^{i},p_\alpha^{I}) = 
\rho_1^r(x^i,u^\alpha,u^\alpha_i,u^\alpha_{I},u^\alpha_{J},p_\alpha^{i},p_\alpha^{I}) =
(x^i,u^\alpha,u^\alpha_i,u^\alpha_{I},u^\alpha_{J}) \, , \\
&\rho_2(x^i,u^\alpha,u^\alpha_i,u^\alpha_{I},u^\alpha_{J},p,p_\alpha^{i},p_\alpha^{I}) =
(x^i,u^\alpha,u^\alpha_i,p,p_\alpha^{i},p_\alpha^{I}) \, , \\
&\rho_2^r(x^i,u^\alpha,u^\alpha_i,u^\alpha_{I},u^\alpha_{J},p_\alpha^{i},p_\alpha^{I}) =
(x^i,u^\alpha,u^\alpha_i,p_\alpha^{i},p_\alpha^{I}) \, , \\
&\rho_{J^{k-1}\pi}(x^i,u^\alpha,u^\alpha_i,u^\alpha_{I},u^\alpha_{J},p,p_\alpha^{i},p_\alpha^{I}) =
\rho_{J^{k-1}\pi}^r(x^i,u^\alpha,u^\alpha_i,u^\alpha_{I},u^\alpha_{J},p_\alpha^{i},p_\alpha^{I}) =
(x^i,u^\alpha,u^\alpha_i) \, , \\
&\rho_M(x^i,u^\alpha,u^\alpha_i,u^\alpha_{I},u^\alpha_{J},p,p_\alpha^{i},p_\alpha^{I}) =
\rho_M^r(x^i,u^\alpha,u^\alpha_i,u^\alpha_{I},u^\alpha_{J},p,p_\alpha^{i},p_\alpha^{I}) = (x^i) \, .
\end{align*}

\subsubsection{Canonical geometric structures}

The extended $2$-symmetric jet-multimomentum bundle $\W$ is endowed with some canonical geometric
structures, which are the generalization to the second-order setting of the canonical structures
introduced in Section \ref{Chap02_sec:FieldTheoriesUnified}.

Let $\Theta_{1}^s \in \df^{m}(J^{2}\pi^\dagger)$ and $\Omega_{1}^s \in \df^{m+1}(J^{2}\pi^\dagger)$
be the symmetrized Liouville forms. Then we define the following forms in $\W$
\begin{equation*}\label{Chap06_eqn:UnifiedCanonicalFormDef}
\Theta = \rho_2^*\Theta_1^s \in \df^{m}(\W) \quad ; \quad
\Omega = \rho_2^*\Omega_1^s \in \df^{m+1}(\W) \, .
\end{equation*}

Bearing in mind the local expressions \eqref{Chap06_eqn:SymmetricMultimomentaLiouvilleFormsLocal}
of the forms $\Theta_1^s$ and $\Omega_1^s$, and the coordinate expression of the projection $\rho_2$
given above, we obtain the coordinate expression of the these forms, which are
\begin{equation*}\label{Chap06_eqn:UnifUnifiedCanonicalFormsLocal}
\begin{array}{l}
\displaystyle
\Theta = p\d^mx + p_\alpha^i \d u^\alpha \wedge \d^{m-1}x_i + \frac{1}{n(ij)} \, p_\alpha^{1_i+1_j} \d u_i^\alpha \wedge \d^{m-1}x_j \, , \\[10pt]
\displaystyle
\Omega = - \d p \wedge \d^mx - \d p_\alpha^i \wedge \d u^\alpha \wedge \d^{m-1}x_i -
\frac{1}{n(ij)} \, \d p_\alpha^{1_i+1_j} \wedge \d u_i^\alpha \wedge \d^{m-1}x_j \, .
\end{array}
\end{equation*}
Observe that, although $\Omega_1^s$ is multisymplectic, the $(m+1)$-form $\Omega$ is
premultisymplectic, since it is closed and $1$-degenerate. Indeed, let $X \in \vf^{V(\rho_{2})}(\W)$.
Then we have
\begin{equation*}
\inn(X)\Omega = \inn(X)\rho_{2}^*\,\Omega_{1}^s = \rho_{2}^*(\inn(Y)\Omega_{1}^s) \, ,
\end{equation*}
where $Y \in \vf(J^2\pi^\dagger)$ is a vector field $\rho_{2}$-related with $X$. However, since
$X$ is vertical with respect to $\rho_{2}$, we have $Y = 0$, and therefore
\begin{equation*}
\rho_{2}^{*}(\inn(Y)\Omega_{1}) = \rho_{2}^{*}(\inn(0)\Omega_{1}^s) = 0 \, .
\end{equation*}
In particular, $\{ 0 \} \varsubsetneq \vf^{V(\rho_2)}(\W) \subseteq \ker\Omega$, and thus $\Omega$
is $1$-degenerate. In coordinates, the $\Cinfty(\W)$-module $\vf^{V(\rho_2)}(\W)$ is locally given by
\begin{equation}\label{Chap06_eqn:PremultisymplecticKernelLocal}
\vf^{V(\rho_2)}(\W) = \left\langle \derpar{}{u^\alpha_{I}},\derpar{}{u^\alpha_{J}} \right\rangle \, ,
\end{equation}
with $|I| = 2$ and $|J| = 3$.

The second canonical structure in $\W$ is the following.

\begin{definition}
The \textnormal{second-order coupling $m$-form} in $\W$ is the $\rho_M$-semibasic $m$-form
$\hat{\C} \in \df^{m}(\W)$ defined as follows: for every $(j^3_x\phi,\omega) \in \W$ we have
\begin{equation*}\label{Chap06_eqn:UnifCouplingFormDef}
\hat{\C}(j^3_x\phi,\omega) = \C^s(\pi^3_2(j^3_x\phi),\omega)  \, .
\end{equation*}
\end{definition}

As $\hat{\C}$ is a $\rho_M$-semibasic $m$-form, there exists a function $\hat{C} \in \Cinfty(\W)$
such that $\hat{\C} = \hat{C}\cdot\rho_M^*\eta$. Bearing in mind the local expression
\eqref{Chap06_eqn:SymmetricMultimomentaCanonicalPairingLocal} of $C^s$, the coordinate expression
of the second-order coupling form is
\begin{equation}\label{Chap06_eqn:UnifCouplingFormLocal}
\hat{\C} = \left( p + p_\alpha^iu_i^\alpha + p_\alpha^{I}u_{I}^\alpha \right)\d^mx \, .
\end{equation}

Let us denote $\hat{\Lag} = (\pi^3_2 \circ \rho_1)^*\Lag \in \df^{m}(\W)$. Since the second-order
Lagrangian density is a $\bar{\pi}^2$-semibasic form, we have that $\hat{\Lag}$ is a
$\rho_M$-semibasic $m$-form, and thus we can write $\hat{\Lag} = \hat{L} \cdot \rho_M^*\eta$, where
$\hat{L} = (\pi^3_2 \circ \rho_1)^*L \in \Cinfty(\W)$ is the pull-back of the Lagrangian
function associated with $\Lag$ and $\eta$. Then, we define a \textsl{Hamiltonian submanifold}
\begin{equation*}
\W_o = \left\{ w \in \W \mid \hat{\Lag}(w) = \hat{\C}(w) \right\} \stackrel{j_o}{\hookrightarrow} \W \, .
\end{equation*}
Since both $\hat{\Lag}$ and $\hat{\C}$ are $\rho_M$-semibasic $m$-forms, the submanifold $\W_o$
is defined by the constraint $\hat{C} - \hat{L} = 0$. In local coordinates, bearing in mind the local
expression \eqref{Chap06_eqn:UnifCouplingFormLocal} of $\hat{\C}$, the constraint function is
\begin{equation*}
p + p_\alpha^iu_i^\alpha + p_\alpha^{I}u_{I}^\alpha - \hat{L} = 0 \, , \quad (|I|=2) \, .
\end{equation*}

\begin{proposition}\label{Chap06_prop:UnifW0DiffeomorphicWr}
The submanifold $\W_o \hookrightarrow \W$ is $1$-codimensional, $\mu_\W$-transverse, and the map
$\Phi = \mu_\W \circ j_o \colon \W_o \to \W_r$ is a diffeomorphism.
\end{proposition}
\begin{proof}
This proof follows the same patterns as the proof of Proposition \ref{Chap05_prop:UnifW0DiffeomorphicWr}.

$\W_o$ is obviously $1$-codimensional, since it is defined by a single constraint function.

To see that $\Phi = \mu_\W \circ j_o \colon \W_o \to \W$ is a diffeomorphism, we show that it is
one-to-one. First, for every $(j^3_x\phi,\omega) \in \W_o$, we have
$L(\pi^{3}_{2}(j^3_x\phi)) = \hat{L}(j^3_x\phi,\omega) = \hat{C}(j^3_x\phi,\omega)$, and
\begin{equation*}
(\mu_\W \circ j_o)(j^3_x\phi,\omega) = \mu_\W(j^3_x\phi,\omega) = (j^3_x\phi,\mu(\omega)) = (j^3_x\phi,[\omega]) \, .
\end{equation*}

First, $\mu_\W \circ j_o$ is injective; in fact, let
$(j^3_x\phi_1,\omega_1), (j^3_x\phi_2,\omega_2) \in \W_o$, then we wish to prove that
\begin{align*}
(\mu_\W \circ j_o)(j^3_x\phi_1,\omega_1) = (\mu_\W \circ j_o)(j^3_x\phi_2,\omega_2) &\Longleftrightarrow
(j^3_x\phi_1,\omega_1) = (j^3_x\phi_2,\omega_2) \\ &\Longleftrightarrow
j^3_x\phi_1 = j^3_x\phi_2 \mbox{ and } \omega_1 = \omega_2 \, .
\end{align*}
Now, using the previous expression for $(\mu_\W \circ j_o)(j^3_x\phi,\omega)$, we have
\begin{align*}
(\mu_\W \circ j_o)(j^3_x\phi_1,\omega_1) = (\mu_\W \circ j_o)(j^3_x\phi_2,\omega_2) &\Longleftrightarrow
(j^3_x\phi_1,[\omega_1]) = (j^3_x\phi_2,[\omega_2]) \\ &\Longleftrightarrow
j^3_x\phi_1 = j^3_x\phi_2 \mbox{ and } [\omega_1] = [\omega_2] \, .
\end{align*}
Hence, by definition of $\W_o$, we have $L(\pi^{3}_{2}(j^3_x\phi_1)) = L(\pi^{3}_{2}(j^3_x\phi_2))
= \hat{C}(j^3_x\phi_1,\omega_1) = \hat{C}(j^3_x\phi_2,\omega_2)$. Locally, from the third equality we obtain
\begin{equation*}
p(\omega_1) + p_\alpha^i(\omega_1)u_{i}^\alpha(j^3_x\phi_1) + p_\alpha^I(\omega_1)u_I^\alpha(j^3_x\phi_1)
= p(\omega_2) + p_\alpha^i(\omega_2)u_{i}^\alpha(j^3_x\phi_2) + p_\alpha^I(\omega_2)u_I^\alpha(j^3_x\phi_2) \, ,
\end{equation*}
but $[\omega_1] = [\omega_2]$ implies
$p^i_\alpha(\omega_1) = p_\alpha^i([\omega_1]) = p_\alpha^i([\omega_2]) = p^i_\alpha(\omega_2)$
and $p^I_\alpha(\omega_1) = p_\alpha^I([\omega_1]) = p_\alpha^I([\omega_2]) = p^I_\alpha(\omega_2)$.
Then $p(\omega_1) = p(\omega_2)$, and $\omega_1 = \omega_2$.

Furthermore, $\mu_\W \circ j_o$ is surjective. In fact, given $(j^3_x\phi,[\omega]) \in \W_r$, we
wish to find $(j^3_x\phi,\zeta) \in j_o(\W_o)$ such that $[\zeta] = [\omega]$. It suffices to take
$[\zeta]$ such that, in local coordinates of $\W$,
\begin{equation*}
p_\alpha^i(\zeta) = p_\alpha^i([\zeta]) \quad , \quad
p_\alpha^I(\zeta) = p_\alpha^I([\zeta]) \quad , \quad
p(\zeta) = L(\pi^{3}_{2}(j^3_x\phi)) - p_\alpha^i([\omega])u_{i}^\alpha(j^3_x\phi) - p_\alpha^I([\omega])u_I^\alpha(j^3_x\phi) \, .
\end{equation*}
This $\zeta$ exists as a consequence of the definition of $\W_o$. Now, since $\mu_\W \circ j_o$
is a one-to-one submersion, then, by equality on the dimensions of $\W_o$ and $\W_r$, it is a
one-to-one local diffeomorphism, and thus a global diffeomorphism.

Finally, in order to prove that $\W_o$ is $\mu_\W$-transversal, it is necessary to check if
$\Lie(X)(\xi) \equiv X(\xi) \neq 0$, for every $X \in \ker\Tan\mu_\W$ and every constraint function
$\xi$ defining $\W_o$. Since $\W_o$ is defined by the constraint $\hat{C} - \hat{L} = 0$ and
$\ker\Tan\mu_\W = \langle \partial/\partial p \rangle$, we have
\begin{equation*}
\derpar{}{p}(\hat{C} - \hat{L}) = \derpar{}{p}(p + p_\alpha^iu_{i}^\alpha + p_\alpha^Iu_I^\alpha - \hat{L}) = 1 \neq 0 \, ,
\end{equation*}
then $\W_o$ is $\mu_\W$-transverse.
\end{proof}

As a consequence of Proposition \ref{Chap06_prop:UnifW0DiffeomorphicWr}, the submanifold $\W_o$
induces a section $\hat{h} \in \Gamma(\mu_\W)$ defined as
$\hat{h} = j_o \circ \Phi^{-1} \colon \W_r \to \W$,
which is called a \textsl{Hamiltonian section of $\mu_\W$}, or a \textsl{Hamiltonian $\mu_\W$-section}.
This section is specified by giving the \textsl{local Hamiltonian function}
\begin{equation}\label{Chap06_eqn:UnifHamiltonianFunctionLocal}
\hat{H}(x^i,u^\alpha,u^\alpha_i,u^\alpha_{I},u^\alpha_{J},p_\alpha^i,p_\alpha^{I}) =
p_\alpha^iu_i^\alpha + p_\alpha^{I}u_{I}^\alpha - \hat{L}(x^i,u^\alpha,u^\alpha_i,u^\alpha_{I}) \, ,
\end{equation}
that is, $\hat{h}(x^i,u^\alpha,u^\alpha_i,u^\alpha_{I},u^\alpha_{J},p_\alpha^i,p_\alpha^{I}) =
(x^i,u^\alpha,u^\alpha_i,u^\alpha_{I},u^\alpha_{J},-\hat{H},p_\alpha^i,p_\alpha^{I})$.
Observe that $\hat{h}$ satisfies $\rho_1^r = \rho_1 \circ \hat{h}$ and
$\rho_2^r = \mu \circ \rho_2 \circ \hat{h}$.

Using this Hamiltonian $\mu_\W$-section, we define the following forms in $\W_r$:
$\Theta_r = \hat{h}^*\Theta \in \df^{m}(\W_r)$ and
$\Omega_r = \hat{h}^*\Omega = -\d\Theta_r \in \df^{m+1}(\W_r)$ with local expressions
\begin{equation}\label{Chap06_eqn:UnifHamiltonCartanFormsLocal}
\begin{array}{l}
\displaystyle
\Theta_r = - \hat{H} \d^mx + p_\alpha^i\d u^\alpha \wedge \d^{m-1}x_i + \frac{1}{n(ij)} \, p_\alpha^{1_i+1_j}\d u_i^\alpha \wedge \d^{m-1}x_j \, , \\[10pt]
\displaystyle
\Omega_r = \d \hat{H} \wedge \d^mx - \d p_\alpha^i \wedge \d u^\alpha \wedge \d^{m-1}x_i - \frac{1}{n(ij)} \, \d p_\alpha^{1_i+1_j} \wedge \d u_i^\alpha \wedge \d^{m-1}x_j \, .
\end{array}
\end{equation}
Then, the pair $(\W_r,\Omega_r)$ is a premultisymplectic Hamiltonian system.

Finally, as we have done in Chapters \ref{Chap:HOAutonomousDynamicalSystems} and
\ref{Chap:HONonAutonomousDynamicalSystems} for higher-order dynamical systems, we generalize the
definition of holonomic sections and multivector fields to the unified setting in order to give a
complete description of second-order Lagrangian field theories in terms of the Lagrangian-Hamiltonian
formalism.

\begin{definition}
A section $\psi \in \Gamma(\rho_M^r)$ is \textnormal{holonomic of type $s$ in $\W_r$},
$1 \leqslant s \leqslant 3$, if the projected section $\rho_1^r \circ \psi \in \Gamma(\bar{\pi}^3)$
is holonomic of type $s$ in $J^3\pi$.
\end{definition}

In coordinates, if $\psi(x^i) = (x^i,u^\alpha,u^\alpha_i,u^\alpha_{I},u^\alpha_{J},p_\alpha^i,p_\alpha^{I})$,
then the condition for $\psi$ to be holonomic of type $s$ in $\W_r$ gives the partial
differential equations \eqref{Chap01_eqn:HOJetBundlesHolonomyConditionSect1} with $k = 3$ (or, equivalently,
\eqref{Chap01_eqn:HOJetBundlesHolonomyConditionSect2} with $k = 3$).

\begin{definition}
A multivector field $\X \in \vf^{m}(\W_r)$ is \textnormal{holonomic of type $s$ in $\W_r$},
$1 \leqslant s \leqslant 3$, if
\begin{enumerate}
\item $\X$ is integrable.
\item $\X$ is $\rho_M^r$-transverse.
\item The integral sections $\psi \in \Gamma(\rho_M^r)$ of $\X$ are holonomic of type $s$ in $\W_r$.
\end{enumerate}
\end{definition}

In natural coordinates, let $\X \in \vf^{m}(\W_r)$ be a locally decomposable and
$(\rho_M^r)$-transverse multivector field locally given by
\begin{equation*}
\X = \bigwedge_{j=1}^{m}
f_j \left(  \derpar{}{x^j} + F_j^\alpha\derpar{}{u^\alpha} + F_{i,j}^\alpha\derpar{}{u_i^\alpha} + F_{I,j}^\alpha\derpar{}{u_{I}^\alpha}
+ F_{J,j}^\alpha\derpar{}{u_{J}^\alpha} + G_{\alpha,j}^i\derpar{}{p_\alpha^i} + G_{\alpha,j}^{I}\derpar{}{p_\alpha^{I}} \right) \, ,
\end{equation*}
with $f_j$ non-vanishing local functions. Then, the condition for $\X$ to be holonomic of type
$s$ in $\W_r$ gives equations \eqref{Chap01_eqn:MultiVFHolonomyLocal} with $k=3$. In particular,
the local expression for a locally decomposable holonomic of type $1$ multivector field
$\X \in \vf^{m}(\W_r)$ is
\begin{equation*}
\X = \bigwedge_{j=1}^{m}
f_j \left(  \derpar{}{x^j} + u_j^\alpha\derpar{}{u^\alpha} + u_{1_i+1_j}^\alpha\derpar{}{u_i^\alpha} + u_{I+1_j}^\alpha\derpar{}{u_{I}^\alpha}
+ F_{J,j}^\alpha\derpar{}{u_{J}^\alpha} + G_{\alpha,j}^i\derpar{}{p_\alpha^i} + G_{\alpha,j}^{I}\derpar{}{p_\alpha^{I}} \right) \, .
\end{equation*}

\subsection{Field equations}
\label{Chap06_sec:UnifFieldEquations}

In this Section we state the field equations for a second-order classical field theory in the
unified formalism. The equations are given in two different ways: first we state the geometric
equation for sections, and then the geometric equation for multivector fields. Both equations are
analyzed locally in-depth. Finally, we prove that these two ways of obtaining the field equations are
equivalent.

\subsubsection{Field equations for sections}

The \textsl{second-order Lagrangian-Hamiltonian problem for sections} associated with the
premultisymplectic system $(\W_r,\Omega_r)$ consists in finding holonomic sections
$\psi \in \Gamma(\rho_M^r)$ satisfying the following condition
\begin{equation}\label{Chap06_eqn:UnifFieldEqSect}
\psi^*\inn(X)\Omega_r = 0 \, , \quad \mbox{for every } X \in \vf(\W_r) \, .
\end{equation}
In the natural coordinates of $\W_r$, let $X \in \vf(\W_r)$ be a generic vector field given by
\begin{equation}\label{Chap06_eqn:UnifGenericVectorField}
X = f^i\derpar{}{x^i} + F^\alpha\derpar{}{u^\alpha} + F_i^\alpha\derpar{}{u_i^\alpha}
+ F_I^\alpha\derpar{}{u_I^\alpha} + F_J^\alpha\derpar{}{u_J^\alpha}
+ G_\alpha^i\derpar{}{p_\alpha^i} + G_\alpha^I\derpar{}{p_\alpha^I} \, .
\end{equation}
Then, bearing in mind the local expression \eqref{Chap06_eqn:SymmetricMultimomentaLiouvilleFormsLocal}
of $\Omega_r$, the $m$-form $\inn(X)\Omega_r$ is given locally by
\begin{align*}
\inn(X)\Omega_r
&= f^k\left(-\d p_\alpha^i \wedge \d u^\alpha \wedge \d^{m-2}x_{ik} - \frac{1}{n(ij)} \, \d p_\alpha^{1_i+1_j} \wedge \d u_i^\alpha \wedge \d^{m-2}x_{jk} - u_i^\alpha\d p_\alpha^i \wedge \d^{m-1}x_k \right. \\
&\qquad{} - p_\alpha^i \d u_i^\alpha \wedge \d^{m-1}x_k - u_{I}^\alpha \d p_\alpha^{I} \wedge \d^{m-1}x_k - p_\alpha^{I} \d u_{I}^\alpha \wedge \d^{m-1}x_k + \derpar{\hat{L}}{u^\alpha}\d u^\alpha \wedge \d^{m-1}x_k \\ 
&\qquad{} \left. + \derpar{\hat{L}}{u_i^\alpha} \d u_i^\alpha \wedge \d^{m-1}x_k + \derpar{\hat{L}}{u_{I}^\alpha} \d u_{I}^\alpha \wedge \d^{m-1}x_k \right)
+ F^\alpha\left( \d p_\alpha^i \wedge \d^{m-1}x_i - \derpar{\hat{L}}{u^\alpha} \d^mx \right) \\
&\quad{} + F_i^\alpha\left( \frac{1}{n(ij)} \, \d p_\alpha^{1_i+1_j} \wedge \d^{m-1}x_j + p_\alpha^i \d^mx - \derpar{\hat{L}}{u_i^\alpha}\d^mx \right)
+ F_{I}^\alpha \left( p_\alpha^{I} - \derpar{\hat{L}}{u_{I}^\alpha} \right) \d^mx \\
& \quad{} + G_\alpha^i\left( u_i^\alpha \d^mx -\d u^\alpha \wedge \d^{m-1}x_i \right)
+ G_\alpha^I \left( u_{I}^\alpha \d^mx - \sum_{1_i+1_j=I} \frac{1}{n(ij)} \, \d u_i^\alpha \wedge \d^{m-1}x_j \right) \, .
\end{align*}
Thus, taking the pull-back of this last expression by a section $\psi \in \Gamma(\rho^r_M)$ with
local expression
\begin{equation*}
\psi(x^i) = (x^i,u^\alpha(x^i),u^\alpha_i(x^i),u^\alpha_{I}(x^i),u^\alpha_{J}(x^i),p_\alpha^i(x^i),p_\alpha^{I}(x^i)) \, ,
\end{equation*}
we obtain the following $\rho_M^r$-semibasic $m$-form
\begin{align*}
\psi^*\inn(X)\Omega_r
&= \left[ f^k\left( \derpar{p_\alpha^i}{x^k}\derpar{u^\alpha}{x^i} + \derpar{p_\alpha^i}{x^i}\derpar{u^\alpha}{x^k} + \frac{1}{n(ij)} \derpar{p_\alpha^{1_i+1_j}}{x^k}\derpar{u_i^\alpha}{x^j}
+ \frac{1}{n(ij)} \derpar{p_\alpha^{1_i+1_j}}{x^j}\derpar{u_i^\alpha}{x^k} - u_i^\alpha \derpar{p_\alpha^i}{x^k} - p_\alpha^i \derpar{u_i^\alpha}{x^k} \right. \right. \\
&\qquad{} \left. - u_{I}^\alpha\derpar{p_\alpha^{I}}{x^k} - p_\alpha^{I} \derpar{u_{I}^\alpha}{x^k}
+ \derpar{\hat{L}}{u^\alpha}\derpar{u^\alpha}{x^k} + \derpar{\hat{L}}{u_i^\alpha} \derpar{u_i^\alpha}{x^k}
+ \derpar{\hat{L}}{u_{I}^\alpha} \derpar{u_{I}^\alpha}{x^k} \right)
+ F^\alpha\left( \sum_{i=1}^{m} \derpar{p_\alpha^i}{x^i} - \derpar{\hat{L}}{u^\alpha} \right) \\
&\quad{} + F_i^\alpha\left( \sum_{j=1}^{m} \frac{1}{n(ij)} \,\derpar{p_\alpha^{1_i+1_j}}{x^j} + p_\alpha^i - \derpar{\hat{L}}{u_i^\alpha} \right)
+ F_{I}^\alpha \left( p_\alpha^{I} - \derpar{\hat{L}}{u_{I}^\alpha} \right)
+ G_\alpha^i\left( u_i^\alpha - \derpar{u^\alpha}{x^i} \right) \\
&\quad{} \left. + G_\alpha^{I} \left( u_{I}^\alpha - \sum_{1_i+1_j=I} \frac{1}{n(ij)} \, \derpar{u_i^\alpha}{x^j} \right) \right] \d^mx \, .
\end{align*}
Finally, requiring this last expression to vanish for every vector field $X \in \vf(\W_r)$ (that is,
the equality must hold for every local function $f^i,F^\alpha,F_i^\alpha,F_{I}^\alpha,G_\alpha^i,G_\alpha^{I}$)
we obtain the following system of equations
\begin{align}
\sum_{i=1}^{m}\derpar{p_\alpha^i}{x^i} - \derpar{\hat{L}}{u^\alpha} = 0 \, , \label{Chap06_eqn:UnifFieldEqSectLocal} \\
\sum_{j=1}^{m} \frac{1}{n(ij)} \, \derpar{p_\alpha^{1_i+1_j}}{x^j} + p_\alpha^i - \derpar{\hat{L}}{u_i^\alpha} = 0 \, ,
\label{Chap06_eqn:UnifFieldEqSectRelationMomenta} \\
p_\alpha^{I} - \derpar{\hat{L}}{u_{I}^\alpha} = 0 \, , \label{Chap06_eqn:UnifFieldEqSectLegendreLocal} \\
u_i^\alpha - \derpar{u^\alpha}{x^i} = 0 \quad ; \quad
u_{I}^\alpha - \sum_{1_i+1_j=I} \frac{1}{n(ij)} \, \derpar{u_i^\alpha}{x^j} = 0 \, .
\label{Chap06_eqn:UnifFieldEqSectHolonomy}
\end{align}
A long but straightforward calculation shows that the $m$ equations along the coefficients $f^k$ are a
combination of the others, and thus we omit them. Observe that equations
\eqref{Chap06_eqn:UnifFieldEqSectHolonomy}
give partially the holonomy condition for the section $\psi$, but since we required this condition
from the beginning, these equations are automatically satisfied. On the other hand, equations
\eqref{Chap06_eqn:UnifFieldEqSectLegendreLocal} do not involve any partial derivative of the
component functions of $\psi$: they are point-wise algebraic conditions that must be fulfilled for
every section $\psi \in \Gamma(\rho_M^r)$ solution to the field equation
\eqref{Chap06_eqn:UnifFieldEqSect}. These equations arise from the $\rho_2^r$-vertical part of the
vector fields $X \in \vf(\W_r)$, as it is shown in the following result.

\begin{lemma}\label{Chap06_lemma:UnifTechLemma1}
If $X \in \vf^{V(\rho_2^r)}(\W_r)$, then $\inn(X)\Omega_r \in \df^{m}(\W_r)$ is $\rho_M^r$-semibasic.
\end{lemma}
\begin{proof}
This result is easy to prove in coordinates. In the natural coordinates of $\W_r$, the
$\Cinfty(\W_r)$-module of $\rho_2^r$-vertical vector fields is given by
\eqref{Chap06_eqn:PremultisymplecticKernelLocal}, that is,
\begin{equation*}
\vf^{V(\rho_2^r)}(\W_r) = \left\langle \derpar{}{u_I^\alpha} \right\rangle \, ,
\end{equation*}
with $2 \leqslant |I| \leqslant 3$. Then, bearing in mind the local expression
\eqref{Chap06_eqn:UnifHamiltonCartanFormsLocal} of $\Omega_r$, we have
\begin{equation*}
\inn\left( \derpar{}{u_I^\alpha} \right) \Omega_r =
\begin{cases}
\displaystyle \left( p_\alpha^{I} - \derpar{\hat{L}}{u_I^\alpha} \right) \d^{m}x \, , & \mbox{for } |I| = 2 \, , \\[10pt]
0 = 0 \cdot \d^{m}x \, , & \mbox{for } |I| > 2 \, .
\end{cases}
\end{equation*}
Thus, in both cases we obtain a $\rho_M^r$-semibasic $m$-form.
\end{proof}

As a consequence of this result, we can define the submanifold
\begin{equation}\label{Chap06_eqn:CompSubmanifoldSect}
\W_c = \left\{ w \in \W_r \mid (\inn(X)\Omega_r)(w) = 0\ \mbox{ for every } X \in \vf^{V(\rho_2^r)}(\W_r) \right\}
\stackrel{j_c}{\hookrightarrow} \W_r \, ,
\end{equation}
where every section $\psi \in \Gamma(\rho_M^r)$ solution to the equation \eqref{Chap06_eqn:UnifFieldEqSect}
must take values. This submanifold is called the \textsl{first constraint submanifold}
of the premultisymplectic system $(\W_r,\Omega_r)$, and has codimension $nm(m+1)/2$.

As we have seen in the proof of Lemma \ref{Chap06_lemma:UnifTechLemma1}, the submanifold
$\W_c \hookrightarrow \W_r$ is locally defined by the constraints
\eqref{Chap06_eqn:UnifFieldEqSectLegendreLocal}. In combination with equations
\eqref{Chap06_eqn:UnifFieldEqSectRelationMomenta}, we have the following result, which is the
analogous to Propositions \ref{Chap03_prop:UnifGraphLegendreOstrogradskyMap} and
\ref{Chap05_prop:UnifGraphLegendreOstrogradskyMap} in second-order field theories.

\begin{proposition}\label{Chap06_prop:UnifGraphLegendreMap}
A solution $\psi \in \Gamma(\rho_M^r)$ to equation \eqref{Chap06_eqn:UnifFieldEqSect} takes values
in a $nm$-codimensional submanifold $\W_\Lag \hookrightarrow \W_c$ which is identified with the graph
of a bundle map $\Leg \colon J^3\pi \to J^{2}\pi^\ddagger$ over $J^1\pi$ defined in coordinates by
\begin{equation}\label{Chap06_eqn:UnifRestrictedLegendreMapLocal}
\Leg^*p^i_\alpha = \derpar{\hat{L}}{u_i^\alpha} - \sum_{j=1}^{m}\frac{1}{n(ij)} \, \frac{d}{dx^j} \, \derpar{\hat{L}}{u_{1_i+1_j}^\alpha} \quad ; \quad
\Leg^*p^I_\alpha = \derpar{\hat{L}}{u_I^\alpha} \, .
\end{equation}
\end{proposition}
\begin{proof}
Since $\W_c$ is defined locally by the constraints \eqref{Chap06_eqn:UnifFieldEqSectLegendreLocal},
it suffices to prove that these contraints, together with the remaining local equations for the
section $\psi \in \Gamma(\rho_M^r)$ to be a solution to the equation \eqref{Chap06_eqn:UnifFieldEqSect},
give rise to the local functions defining the map given above, and thus to the submanifold $\W_\Lag$.

Replacing $p^I_\alpha$ by $\partial \hat{L} / \partial u_I^\alpha$ in equations
\eqref{Chap06_eqn:UnifFieldEqSectRelationMomenta}, we obtain
\begin{equation*}
p_\alpha^i - \derpar{\hat{L}}{u_i^\alpha}
+ \sum_{j=1}^{m} \frac{1}{n(ij)} \, \frac{d}{dx^j} \, \derpar{\hat{L}}{u_{1_i+1_j}^\alpha} = 0 \, .
\end{equation*}
Therefore, these constraints define a submanifold $\W_\Lag \hookrightarrow \W_c$, which can be
identified with the graph of a map $\Leg \colon J^3\pi \to J^2\pi^\ddagger$ given by
\begin{equation*}
\Leg^*x^i = x^i \quad ; \quad \Leg^*u^\alpha = u^\alpha \quad ; \quad \Leg^*u_i^\alpha \, ,
\end{equation*}
\begin{equation*}
\Leg^*p^i_\alpha = \derpar{\hat{L}}{u_i^\alpha} - \sum_{j=1}^{m}\frac{1}{n(ij)} \, \frac{d}{dx^j} \, \derpar{\hat{L}}{u_{1_i+1_j}^\alpha} \quad ; \quad
\Leg^*p^I_\alpha = \derpar{\hat{L}}{u_I^\alpha} \, . \qedhere
\end{equation*}
\end{proof}

\begin{definition}
The bundle map $\Leg \colon J^3\pi \to J^2\pi^\ddagger$ over $J^1\pi$ is called the
\textnormal{restricted Legendre map} associated with the second-order Lagrangian density $\Lag$.
\end{definition}

Observe that $\dim\W_\Lag = \dim J^3\pi = m + n + mn + nm(m+1)/2 + nm(m+1)(m+2)/6$.

\begin{remark}
The terminology ``Legendre map'' is justified, since $\Leg$ is a fiber bundle morphism from the
Lagrangian phase space to the Hamiltonian phase space that identifies the multimomenta coordinates
with functions on partial derivatives of the Lagrangian function, and thus generalizes the Legendre
map in first-order field theories (see \cite{art:Echeverria_DeLeon_Munoz_Roman07,
art:Echeverria_Munoz_Roman00_RMP}), and first-order and higher-order mechanics (see
\cite{book:Abraham_Marsden78} for first-order mechanics and \cite{book:DeLeon_Rodrigues85} for the
higher-order setting).
\end{remark}

According to \cite{art:Saunders_Crampin90}, we can give the following definition.

\begin{definition}\label{Chap06_def:UnifRegularLagrangianDef}
A second-order Lagrangian density $\Lag \in \df^{m}(J^2\pi)$ is \textnormal{regular} if, for
every point $j^3_x\phi \in J^{3}\pi$, we have
\begin{equation*}
\rank(\Leg(j^3_x\phi)) = \dim J^{2}\pi + \dim J^{1}\pi - \dim E = \dim J^2\pi^\ddagger \, .
\end{equation*}
Otherwise, the Lagrangian density is said to be \textnormal{singular}.
\end{definition}

Hence, a second-order Lagrangian density $\Lag \in \df^{m}(J^2\pi)$ is regular if, and only if, the
restricted Legendre map $\Leg \colon J^3\pi \to J^{2}\pi^\ddagger$ associated to $\Lag$ is a
submersion onto $J^{2}\pi^\ddagger$. This implies that there exist local sections of $\Leg$, that is,
maps $\sigma \colon U \to J^3\pi$, with $U \subset J^{2}\pi^\ddagger$ an open set, such that
$\Leg \circ \sigma = \textnormal{Id}_U$. If $\Leg$ admits a global section
$\Upsilon \colon J^2\pi^\ddagger \to J^3\pi$, then the Lagrangian density is said to be
\textsl{hyperregular}.

Observe that $\dim J^3\pi \geqslant J^2\pi^\ddagger$, since
\begin{align*}
m + n + nm + \frac{nm(m+1)}{2} + \frac{nm(m+1)(m+2)}{6} \geqslant m + n + 2nm + \frac{nm(m+1)}{2} \, ,
\end{align*}
and the equality holds if, and only if, $m = 1$. Therefore, unlike in higher-order mechanics
(see Chapters \ref{Chap:HOAutonomousDynamicalSystems} and \ref{Chap:HONonAutonomousDynamicalSystems})
or first-order field theories (see Section \ref{Chap02_sec:FieldTheories}),
the Legendre map in second-order field theories cannot be a local diffeomorphism due to
dimension restrictions.

Computing the local expression of the tangent map to $\Leg$ in a natural chart of $J^{3}\pi$,
the regularity condition for the Lagrangian density $\Lag$ is equivalent to
\begin{equation*}
\det\left( \derpars{L}{u_I^\beta}{u_K^\alpha} \right)(j^3_x\phi) \neq 0
\, , \quad \mbox{for every } j^3_x\phi \in J^3\pi \, ,
\end{equation*}
where $|I|=|K|=2$. That is, the Hessian of the Lagrangian function associated with $\Lag$ and $\eta$
with respect to the highest order ``velocities'' is a regular matrix at every point, which is the
usual definition of regular Lagrangian densities.

Note that since $\W_r$ is diffeomorphic to the submanifold $\W_o \hookrightarrow \W$ by Proposition
\ref{Chap06_prop:UnifW0DiffeomorphicWr}, and $\W_o$ is defined locally by the constraint
$p + p_\alpha^iu_i^\alpha + p_\alpha^Iu_I^\alpha - \hat{L} = 0$, the restricted Legendre map
$\Leg \colon J^3\pi \to J^2\pi^\ddagger$ can be extended in a canonical way to a map
$\widetilde{\Leg} \colon J^3\pi \to J^2\pi^\dagger$, defining $\widetilde{\Leg}^*p$ as the pull-back
of the local Hamiltonian function $-\hat{H}$. This enables us to state the following result, which is
a straightforward consequence of Proposition \ref{Chap06_prop:UnifGraphLegendreMap}, and is the analogous
to Corollary \ref{Chap05_corol:UnifGraphExtendedLegendreOstrogradskyMapSect} on second-order field
theories.

\begin{corollary}\label{Chap06_corol:UnifGraphExtendedLegendreMap}
The submanifold $\W_\Lag \hookrightarrow \W$ is the graph of a bundle map
$\widetilde{\Leg} \colon J^3\pi \to J^2\pi^\dagger$ over $J^1\pi$ defined locally by
\begin{equation}\label{Chap06_eqn:UnifExtendedLegendreMapLocal}
\begin{array}{l}
\displaystyle \widetilde{\Leg}^*p^i_\alpha = \derpar{\hat{L}}{u_i^\alpha} - \sum_{j=1}^{m}\frac{1}{n(ij)} \,
\frac{d}{dx^j}\left( \derpar{\hat{L}}{u_{1_i+1_j}^\alpha} \right) \quad ; \quad
\widetilde{\Leg}^*p^I_\alpha = \derpar{\hat{L}}{u_I^\alpha} \, , \\[10pt]
\displaystyle\widetilde{\Leg}^*p = \hat{L} - u_i^\alpha\left(\derpar{\hat{L}}{u_i^\alpha} - \sum_{j=1}^{m}\frac{1}{n(ij)} \,
\frac{d}{dx^j}\left( \derpar{\hat{L}}{u_{1_i+1_j}^\alpha} \right) \right) - u_{I}^\alpha\derpar{\hat{L}}{u_I^\alpha} \, ,
\end{array}
\end{equation}
and satisfying $\Leg = \mu \circ \widetilde{\Leg}$.
\end{corollary}

The bundle map $\widetilde{\Leg} \colon J^3\pi \to J^2\pi^\dagger$ is the
\textsl{extended Legendre map} associated with the Lagrangian density $\Lag$. An important result
concerning both Legendre maps is the following, which is the analogous to Proposition
\ref{Chap05_prop:UnifRankBothLegendreMaps} for second-order field theories.

\begin{proposition}\label{Chap06_prop:UnifRankBothLegendreMaps}
For every $j^3_x\phi \in J^3\pi$ we have $\rank(\widetilde{\Leg}(j^3_x\phi)) = \rank(\Leg(j^3_x\phi))$.
\end{proposition}

Following the same patterns as in \cite{art:deLeon_Marin_Marrero96} for first-order mechanical
systems, the proof of this result consists in computing in a natural chart of coordinates the local
expressions of the Jacobian matrices of both maps, $\widetilde{\Leg}$ and $\Leg$. Then, observe that
the ranks of both maps depend on the rank of the Hessian matrix of the Lagrangian function with
respect to the highest order velocities, and that the additional row in the Jacobian matrix of
$\widetilde{\Leg}$ is a combination of the others. Since it is just a long calculation in coordinates,
we omit the proof of this result.

Notice that the component functions $u_J^\alpha$ with $|J| = 3$ of the section
$\psi \in \Gamma(\rho_M^r)$ are not yet determined, since the coordinate expression of the field
equation \eqref{Chap06_eqn:UnifFieldEqSect} does not give any condition on these functions. In fact,
these functions are determined by the equations \eqref{Chap06_eqn:UnifFieldEqSectLocal} and
\eqref{Chap06_eqn:UnifFieldEqSectRelationMomenta}. Indeed, since the section $\psi \in \Gamma(\rho_M^r)$
must take values in the submanifold $\W_\Lag$ given by Proposition \ref{Chap06_prop:UnifGraphLegendreMap},
then by replacing the local expression of the restricted Legendre map in equations
\eqref{Chap06_eqn:UnifFieldEqSectLocal} and \eqref{Chap06_eqn:UnifFieldEqSectRelationMomenta} we obtain
the second-order Euler-Lagrange equations for field theories
\begin{equation}\label{Chap06_eqn:UnifEulerLagrangeSect}
\restric{\derpar{\hat{L}}{u^\alpha}}{\psi} - \restric{\frac{d}{dx^i} \, \derpar{\hat{L}}{u_i^\alpha}}{\psi} + \restric{\sum_{|I|=2} \frac{d^2}{dx^{I}} \, \derpar{\hat{L}}{u_I^\alpha}}{\psi} = 0 \, , \quad (1 \leqslant \alpha \leqslant n) \, .
\end{equation}

Finally, observe that since the section $\psi \in \Gamma(\rho_M^r)$ must take values in the
submanifold $\W_\Lag \hookrightarrow \W_r$, it is natural to consider the restriction of equation
\eqref{Chap06_eqn:UnifFieldEqSect} to the submanifold $\W_\Lag$; that is, to restrict the set of
vector fields to those tangent to $\W_\Lag$. Nevertheless, the new equation may not be equivalent
to the former. The following result gives a sufficient condition for these two equations to be equivalent.

\begin{proposition}\label{Chap06_prop:UnifFieldEqSectTangent}
If $\psi \in \Gamma(\rho_M^r)$ is holonomic in $\W_r$, then the equation
\eqref{Chap06_eqn:UnifFieldEqSect} is equivalent to
\begin{equation}\label{Chap06_eqn:UnifFieldEqSectTangent}
\psi^*\inn(Y)\Omega_r = 0 \, , \quad \mbox{for every } Y \in \vf(\W_r) \mbox{ tangent to } \W_\Lag \, .
\end{equation}
\end{proposition}
\begin{proof}
We prove this result in coordinates. First of all, let us compute the coordinate expression of a
vector field $X \in \vf(\W_r)$ tangent to $\W_\Lag$. Let $X$ be a generic vector field locally given by
\eqref{Chap06_eqn:UnifGenericVectorField}, that is,
\begin{equation*}
X = f^i \derpar{}{x^i} + F^\alpha\derpar{}{u^\alpha} + F^\alpha_i\derpar{}{u_i^\alpha} + F_I^\alpha\derpar{}{u_I^\alpha}
+ F_J^\alpha\derpar{}{u_J^\alpha} + G_\alpha^i\derpar{}{p_\alpha^i} + G_\alpha^I\derpar{}{p_\alpha^I} \, .
\end{equation*}
Then, since $\W_\Lag$ is the submanifold of $\W_r$ defined locally by the $nm + nm(m+1)/2$
constraint functions $\xi^i_\alpha$, $\xi^I_\alpha$ with coordinate expressions
\begin{equation*}
\xi^i_\alpha = p_\alpha^i - \derpar{\hat{L}}{u_i^\alpha}
+ \sum_{j=1}^{m} \frac{1}{n(ij)} \, \frac{d}{dx^j} \, \derpar{\hat{L}}{u_{1_i+1_j}^\alpha}
\quad ; \quad \xi^I_\alpha = p_\alpha^I - \derpar{\hat{L}}{u_I^\alpha} \, ,
\end{equation*}
then the tangency condition of $X$ along $\W_\Lag$, which is
$\Lie(X)\xi^i_\alpha = \Lie(X)\xi^I_\alpha = 0$ (on $\W_\Lag$), gives the following relation
on the component functions of $X$
\begin{align*}
G_\alpha^i &=
f^k\left( \derpars{\hat{L}}{x^k}{u_i^\alpha} - \frac{1}{n(ij)} \, \frac{d}{dx^j} \, \derpars{\hat{L}}{x^k}{u_{1_i+1_j}^\alpha} \right)
+ F^\beta\left( \derpars{\hat{L}}{u^\beta}{u_i^\alpha} - \frac{1}{n(ij)} \, \frac{d}{dx^j} \, \derpars{\hat{L}}{u^\beta}{u_{1_i+1_j}^\alpha} \right) \\
&\quad {} + F_k^\beta\left( \derpars{\hat{L}}{u_k^\beta}{u_i^\alpha} - \frac{1}{n(ij)} \, \frac{d}{dx^j} \, \derpars{\hat{L}}{u_k^\beta}{u_{1_i+1_j}^\alpha} \right)
+ F^\beta_I\left( \derpars{\hat{L}}{u^\beta_I}{u_i^\alpha} - \frac{1}{n(ij)} \, \frac{d}{dx^j} \, \derpars{\hat{L}}{u^\beta_I}{u_{1_i+1_j}^\alpha} \right) \\
&\quad {} - \frac{1}{n(ij)} \left( F_j^\beta \derpars{\hat{L}}{u^\beta}{u_{1_i+1_j}^\alpha}
+ F_{1_k+1_j}^\beta\derpars{\hat{L}}{u_k^\beta}{u_{1_i+1_j}^\alpha}
+ F_{I+1_j}^\beta\derpars{\hat{L}}{u_I^\beta}{u_{1_i+1_j}^\alpha} \right) \, , \\
G_\alpha^I &= f^i\derpars{\hat{L}}{x^i}{u_I^\alpha} + F^\beta\derpars{\hat{L}}{u^\beta}{u_I^\alpha}
+ F_i^\beta\derpars{\hat{L}}{u_i^\beta}{u_I^\alpha} + F_J^\beta\derpars{\hat{L}}{u_J^\beta}{u_I^\alpha} \, .
\end{align*}
Hence, the tangency condition enables us to write the component functions $G_\alpha^i$, $G_\alpha^I$
as functions $\widetilde{G}_\alpha^i$, $\widetilde{G}_\alpha^I$ depending on the rest of the components
$f^i,F^\alpha,F^\alpha_i,F^\alpha_I,F^\alpha_J$.

Now, if $\psi(x^i) = (x^i,u^\alpha,u^\alpha_i,u^\alpha_I,u^\alpha_J,p_\alpha^i,p_\alpha^I)$,
then the equation \eqref{Chap06_eqn:UnifFieldEqSect} gives in coordinates
\begin{align*}
\psi^*\inn(X)\Omega_r
&= \left[ f^k ( \cdots )
+ F^\alpha\left( \sum_{i=1}^{m} \derpar{p_\alpha^i}{x^i} - \derpar{\hat{L}}{u^\alpha} \right) 
+ F_i^\alpha\left( \sum_{j=1}^{m} \frac{1}{n(ij)} \,\derpar{p_\alpha^{1_i+1_j}}{x^j} + p_\alpha^i - \derpar{\hat{L}}{u_i^\alpha} \right)\right. \\
&\quad {} \left. + F_{I}^\alpha \left( p_\alpha^{I} - \derpar{\hat{L}}{u_{I}^\alpha} \right) 
+ G_\alpha^i\left( -\derpar{u^\alpha}{x^i} + u_i^\alpha \right)
+ G_\alpha^{I} \left( u_{I}^\alpha - \sum_{1_i+1_j=I} \frac{1}{n(ij)} \, \derpar{u_i^\alpha}{x^j} \right) \right] \d^mx \, .
\end{align*}
where the terms $(\cdots)$ contain a long expression with several partial derivatives of the component
functions and the Lagrangian function, which is not relevant in this proof. On the other hand, if we
take a vector field $Y$ tangent to $\W_\Lag$, then we must replace the component functions $G_\alpha^i$
and $G_\alpha^I$ by $\widetilde{G}_\alpha^i$ and $\widetilde{G}_\alpha^I$ in the previous equation, thus obtaining
\begin{align*}
\psi^*\inn(Y)\Omega_r
&= \left[ f^k ( \cdots )
+ F^\alpha\left( \sum_{i=1}^{m} \derpar{p_\alpha^i}{x^i} - \derpar{\hat{L}}{u^\alpha} \right) 
+ F_i^\alpha\left( \sum_{j=1}^{m} \frac{1}{n(ij)} \,\derpar{p_\alpha^{1_i+1_j}}{x^j} + p_\alpha^i - \derpar{\hat{L}}{u_i^\alpha} \right)\right. \\
&\quad {} \left. + F_{I}^\alpha \left( p_\alpha^{I} - \derpar{\hat{L}}{u_{I}^\alpha} \right) 
+ \widetilde{G}_\alpha^i\left( -\derpar{u^\alpha}{x^i} + u_i^\alpha \right)
+ \widetilde{G}_\alpha^{I} \left( u_{I}^\alpha - \sum_{1_i+1_j=I} \frac{1}{n(ij)} \, \derpar{u_i^\alpha}{x^j} \right) \right] \d^mx \, .
\end{align*}
Finally, if $\psi$ is holonomic, then equations \eqref{Chap06_eqn:UnifFieldEqSectHolonomy} are satisfied,
and the last two terms of both $\inn(X)\Omega_r$ and $\inn(Y)\Omega_r$ vanish, thus obtaining
\begin{align*}
\psi^*\inn(X)\Omega_r
&= \left[ f^k ( \cdots )
+ F^\alpha\left( \sum_{i=1}^{m} \derpar{p_\alpha^i}{x^i} - \derpar{\hat{L}}{u^\alpha} \right) 
+ F_i^\alpha\left( \sum_{j=1}^{m} \frac{1}{n(ij)} \,\derpar{p_\alpha^{1_i+1_j}}{x^j} + p_\alpha^i - \derpar{\hat{L}}{u_i^\alpha} \right)\right. \\
&\quad {} \left. + F_{I}^\alpha \left( p_\alpha^{I} - \derpar{\hat{L}}{u_{I}^\alpha} \right) \right] \d^mx = \psi^*\inn(Y)\Omega_r \, .
\end{align*}
Hence, we have $\inn(X)\Omega_r = 0$ if, and only if, $\inn(Y)\Omega_r = 0$.
\end{proof}

\begin{remarks} \
\begin{itemize}
\item
Observe that, contrary to first-order field theories the holonomy condition is not recovered from
the coordinate expression of the field equations (see Section \ref{Chap02_sec:FieldTheoriesUnified}).
Moreover, in this case, unlike in higher-order time-dependent mechanical systems, not even a condition
for the holonomy of type $2$ can be obtained (see Sections \ref{Chap03_sec:DynamicalEquations} and
\ref{Chap05_sec:UnifiedFormalism}). This is due to the constraints $p_\alpha^{ij} - p_\alpha^{ji} = 0$
introduced in Section \ref{Chap06_sec:SymmetricMultimomenta} to define both the extended and restricted
$2$-symmetric multimomentum bundles. Hence, the full holonomy condition is necessarily required in
this formalism.

It is important to point out that, although the holonomy condition cannot be obtained from the
field equation, a holonomic section $\psi \in \Gamma(\rho_M^r)$ satisfies the local equations
\eqref{Chap06_eqn:UnifFieldEqSectHolonomy}. Hence, a holonomic section can be a solution to the equation
\eqref{Chap06_eqn:UnifFieldEqSect}. \hfill$\lozenge$

\item
The regularity of the Lagrangian density seems to play a secondary role in this formulation, because
the holonomy of the section solution to the equation \eqref{Chap06_eqn:UnifFieldEqSect} is necessarily
required, regardless of the regularity of the Lagrangian density given. Nevertheless, recall that the
Euler-Lagrange equations \eqref{Chap06_eqn:UnifEulerLagrangeSect} may not be compatible if the 
second-order Lagrangian density is singular, and thus the regularity of $\Lag$ still determines if
the section $\psi \in \Gamma(\rho_M^r)$ solution to the equation \eqref{Chap06_eqn:UnifFieldEqSect}
lies in $\W_\Lag$ or in a submanifold of $\W_\Lag$. If $\Lag$ is singular, in the most favorable cases,
there exists a submanifold $\W_f \hookrightarrow \W_\Lag$ where the section $\psi$ takes values.
\hfill$\lozenge$
\end{itemize}
\end{remarks}

\subsubsection{Field equations for multivector fields}

The \textsl{second-order Lagrangian-Hamiltonian problem for multivector fields} associated with the
premultisymplectic manifold $(\W_r,\Omega_r)$ consists in finding a class of locally decomposable
holonomic multivector fields $\{\X\} \subset \vf^m(\W_r)$ satisfying the following field equation
\begin{equation}\label{Chap06_eqn:UnifFieldEqMultiVF}
\inn(\X)\Omega_r = 0 \ , \quad \mbox{for every } \X \in \{\X\} \subseteq \vf^{m}(\W_r) \, .
\end{equation}

Since the $(m+1)$-form $\Omega_r$ is premultisymplectic, equation \eqref{Chap06_eqn:UnifFieldEqMultiVF}
may not admit a global solution $\X \in \vf^{m}(\W_r)$, but only defined on some submanifold of $\W_r$.
Using an adapted version of the constraint algorithm described in Section
\ref{Chap01_sec:ConstraintAlgorithm} for premultisymplectic manidolds
\cite{art:DeLeon_Marin_Marrero_Munoz_Roman05}, we have the following result.

\begin{proposition}\label{Chap06_prop:UnifFirstConstraintSubmanifold}
A solution $\X \in \vf^{m}(\W_r)$ to equation \eqref{Chap06_eqn:UnifFieldEqMultiVF} exists only on the points
of the submanifold $\W_c \hookrightarrow \W_r$ defined by
\begin{align*}
\W_c &= \left\{ w \in \W_r \mid (\inn(Z)\d\hat{H})(w) = 0 \, , \mbox{ for every }
Z \in \ker\Omega \right\} \\
&= \left\{ w \in \W_r \mid (\inn(Y)\Omega_r)(w) = 0 \, , \mbox { for every }
Y \in \vf^{V(\rho_2^r)}(\W_r)\right\} \, .
\end{align*}
\end{proposition}

The submanifold $\W_c \hookrightarrow \W_r$ is the so-called \textsl{compatibility submanifold} for
the premultisymplectic system $(\W_r,\Omega_r)$. Observe that we denoted this submanifold by $\W_c$,
which is the notation used for the first constraint submanifold defined in
\eqref{Chap06_eqn:CompSubmanifoldSect}. Indeed, both submanifolds are equal. In order to prove this,
recall that the first constraint submanifold is defined locally by the constraints
$p^I_\alpha - \partial\hat{L}/u_I^\alpha = 0$. Hence, it suffices to prove that the compatibility
submanifold given by Proposition \ref{Chap06_prop:UnifFirstConstraintSubmanifold} is defined locally
by the same constraints.

In fact, in natural coordinates, the coordinate expression for the local Hamiltonian function
$\hat{H}$ is given by \eqref{Chap06_eqn:UnifHamiltonianFunctionLocal}, and thus we have
\begin{align*}
\d\hat{H} &= -\derpar{\hat{L}}{u^\alpha}\d u^\alpha + \left( p_\alpha^i - \derpar{\hat{L}}{u_i^\alpha} \right)\d u_i^\alpha
+ \left( p_\alpha^I - \derpar{\hat{L}}{u_I^\alpha} \right) \d u_I^\alpha + u_i^\alpha \d p_\alpha^i + u_I^\alpha \d p_\alpha^I \, .
\end{align*}
Now, bearing in mind that $\ker\Omega$ is the $(nm(m+1)/2 + nm(m+1)(m+2)/6)$-dimensional
$\Cinfty(\W)$-module locally given by \eqref{Chap06_eqn:PremultisymplecticKernelLocal},
the functions $\inn(Z)\d\hat{H}$ for $Z \in \ker\Omega$ have the following coordinate expressions
\begin{equation*}
\inn\left( \derpar{}{u_I^\alpha} \right) \d\hat{H} = p_\alpha^I - \derpar{\hat{L}}{u_I^\alpha} \ \mbox{ for }|I| = 2 \quad ; \quad
\inn\left( \derpar{}{u_J^\alpha} \right) \d\hat{H} = 0 \ \mbox{ for } |J| = 3 \, .
\end{equation*}
Therefore, the submanifold $\W_c \hookrightarrow \W_r$ is locally defined by the $nm(m+1)/2$
constraints $p_\alpha^I - \partial\hat{L} / \partial u_I^\alpha = 0$. In particular, it is equal
to the submanifold defined in \eqref{Chap06_eqn:CompSubmanifoldSect}, and we have 
\begin{equation*}
\dim\W_c = \dim\W_r - nm(m+1)/2 = m + n + 2mn + \frac{nm(m+1)}{2} + \frac{nm(m+1)(m+2)}{6} \, .
\end{equation*}

Now we compute the coordinate expression of the equation \eqref{Chap06_eqn:UnifFieldEqMultiVF} in a
local chart of $\W_r$. From the results in Section \ref{Chap01_sec:MultivectorFields} and
\cite{art:Echeverria_Munoz_Roman98}, a representative $\X$ of a class of locally decomposable,
integrable and $\rho_M^r$-transverse $m$-vector fields $\{\X\} \subset \vf^m(\W_r)$ can be written
in coordinates
\begin{equation}\label{Chap06_eqn:UnifGenericMultiVFLocal}
\X = f \bigwedge_{j=1}^{m}
\left(  \derpar{}{x^j} + F_j^\alpha\derpar{}{u^\alpha} + F_{i,j}^\alpha\derpar{}{u_i^\alpha} + F_{I,j}^\alpha\derpar{}{u_{I}^\alpha}
+ F_{J,j}^\alpha\derpar{}{u_{J}^\alpha} + G_{\alpha,j}^i\derpar{}{p_\alpha^i} + G_{\alpha,j}^{I}\derpar{}{p_\alpha^{I}} \right) \, ,
\end{equation}
where $f$ is a non-vanishing local function. Taking $f = 1$ as a representative of the equivalence
class, the contraction $\inn(\X)\Omega_r$ gives locally the following $1$-form
\begin{align*}
\inn(\X)\Omega_r &=
\left( G_{\alpha,k}^i F_i^\alpha + G_{\alpha,i}^iF_k^\alpha + \frac{1}{n(ij)} \, G_{\alpha,k}^{1_i+1_j}F_{i,j}^\alpha
+ \frac{1}{n(ij)} \, G_{\alpha,j}^{1_i+1_j}F_{i,k}^\alpha - u_i^\alpha G_{\alpha,k}^i - p_{\alpha}^iF_{i,k}^\alpha \right. \\
&\quad{} \left. - u_{I}^\alpha G_{\alpha,k}^I - p_\alpha^I F_{I,k}^\alpha + \derpar{\hat{L}}{u^\alpha} \, F_k^\alpha
+ \derpar{\hat{L}}{u_i^\alpha} \, F_{i,k}^\alpha + \derpar{\hat{L}}{u_I^\alpha} \, F_{I,k}^\alpha \right) \d x^k
+ \left( \sum_{i=1}^{m} G_{\alpha,i}^i - \derpar{\hat{L}}{u^\alpha} \right) \d u^\alpha \\
&\quad{} + \left( \sum_{j=1}^{m} \frac{1}{n(ij)} \, G_{\alpha,j}^{1_i+1_j} + p_\alpha^i - \derpar{\hat{L}}{u_i^\alpha} \right) \d u_i^\alpha
+ \left( p_\alpha^I - \derpar{\hat{L}}{u_I^\alpha} \right) \d u_I^\alpha + \left( u_j^\alpha - F_j^\alpha \right) \d p_\alpha^j \\
&\quad{} + \left( u_I^\alpha - \sum_{1_i+1_j=I}\frac{1}{n(ij)} \, F_{i,j}^\alpha \right) \d p_\alpha^I \, .
\end{align*}
Then, requiring this $1$-form to vanish, we obtain the coordinate expression of equation
\eqref{Chap06_eqn:UnifFieldEqMultiVF}, which is the following system of equations
\begin{align}
F_j^\alpha = u_j^\alpha \quad ; \quad \sum_{1_i+1_j=I} \frac{1}{n(ij)} \, F_{i,j}^\alpha = u_I^\alpha \, ,
\label{Chap06_eqn:UnifFieldEqMultiVFHolonomy} \\
\sum_{i=1}^{m} G_{\alpha,i}^{i} = \derpar{\hat L}{u^\alpha} \, , \label{Chap06_eqn:UnifFieldEqMultiVFLocal1} \\
\sum_{j=1}^{m} \frac{1}{n(ij)} \, G_{\alpha,j}^{1_i+1_j} = \derpar{\hat L}{u_i^\alpha} - p_\alpha^i \, ,
\label{Chap06_eqn:UnifFieldEqMultiVFLocal2} \\
p_\alpha^K - \derpar{\hat L}{u^\alpha_K} = 0 \, , \quad |K| = 2 \, . \label{Chap06_eqn:HOMomentaMultiVF}
\end{align}
The $m$ additional equations alongside the $1$-forms $\d x^i$ are a straightforward consequence of
the others and the tangency condition that follows, and thus we omit them. Therefore, a locally
decomposable and $\rho_M^r$-transverse multivector field solution to the field equation
\eqref{Chap06_eqn:UnifFieldEqMultiVF} is given in coordinates by
\begin{equation*}
\X = \bigwedge_{j=1}^{m}
\left(  \derpar{}{x^j} + u_j^\alpha\derpar{}{u^\alpha} + F_{i,j}^\alpha\derpar{}{u_i^\alpha} + F_{I,j}^\alpha\derpar{}{u_{I}^\alpha}
+ F_{J,j}^\alpha\derpar{}{u_{J}^\alpha} + G_{\alpha,j}^i\derpar{}{p_\alpha^i} + G_{\alpha,j}^{I}\derpar{}{p_\alpha^{I}} \right) \, ,
\end{equation*}
where the functions $F_{i,j}^\alpha$, $G_{\alpha,j}^i$ and $G_{\alpha,j}^I$ must satisfy the
equations \eqref{Chap06_eqn:UnifFieldEqMultiVFHolonomy}, \eqref{Chap06_eqn:UnifFieldEqMultiVFLocal1} and
\eqref{Chap06_eqn:UnifFieldEqMultiVFLocal2}. Note that most of the component functions remain undetermined,
and that there can be several different functions satisfying the aforementioned equations. However, recall
that the statement of the problem requires the class of multivector fields to be holonomic.
In coordinates, this implies that equations \eqref{Chap01_eqn:MultiVFHolonomyLocal} are satisfied with
$k = 3$ and $r=1$, and, more particularly, equations \eqref{Chap06_eqn:UnifFieldEqMultiVFHolonomy} are
satisfied. Thus, a locally decomposable and holonomic multivector field $\X$ solution to the
field equation \eqref{Chap06_eqn:UnifFieldEqMultiVF} has the following coordinate expression
\begin{equation*}
\X = \bigwedge_{j=1}^{m}
\left(  \derpar{}{x^j} + u_j^\alpha\derpar{}{u^\alpha} + u_{1_i+1_j}^\alpha\derpar{}{u_i^\alpha} + u_{I+1_j}^\alpha\derpar{}{u_{I}^\alpha}
+ F_{J,j}^\alpha\derpar{}{u_{J}^\alpha} + G_{\alpha,j}^i\derpar{}{p_\alpha^i} + G_{\alpha,j}^{I}\derpar{}{p_\alpha^{I}} \right) \, ,
\end{equation*}
with $G_{\alpha,j}^i$ and $G_{\alpha,j}^{I}$ satisfying \eqref{Chap06_eqn:UnifFieldEqMultiVFLocal1}
and \eqref{Chap06_eqn:UnifFieldEqMultiVFLocal2}.

Observe that equations \eqref{Chap06_eqn:HOMomentaMultiVF} are a compatibility condition for
the multivector field $\X$, which state that the multivector field solution to the field equation
\eqref{Chap06_eqn:UnifFieldEqMultiVF} exists only at support on the submanifold $\W_c$. Hence, we
recover in coordinates the result stated in Proposition \ref{Chap06_prop:UnifFirstConstraintSubmanifold}.

Let us analyze the tangency of the multivector field $\X$ along the submanifold
$\W_c \hookrightarrow \W_r$. Recall that, since $\X$ is locally decomposable, that is, we have
$\X = X_1 \wedge \ldots \wedge X_m$ on an open neighborhood around every point, the tangency of
$\X$ along the submanifold $\W_c$ is equivalent to the tangency of every $X_i$ along $\W_c$. That
is, we must require that $\restric{\Lie(X_k)\xi}{\W_c} = 0$ for every constraint function $\xi$
defining $\W_c$ and for every $1 \leqslant k \leqslant m$. 

Therefore, since the submanifold $\W_c \hookrightarrow \W_r$ is locally defined by the
$nm(m+1)/2$ constraint functions
$\xi^{K}_{\alpha} = p_\alpha^{K} - \partial\hat{L} / \partial u_{K}^\alpha$, we must
check if the condition $\Lie(X_j)\xi_\alpha^{K} \equiv X_j(\xi_{\alpha}^{K}) = 0$ holds on $\W_c$
for every $1 \leqslant j \leqslant m$, $1 \leqslant \alpha \leqslant n$, $|K| = 2$. Computing, we obtain
\begin{align*}
&\left( \derpar{}{x^j} + u_j^\alpha\derpar{}{u^\alpha} + u_{1_i+1_j}^\alpha\derpar{}{u_i^\alpha} + u_{I+1_j}^\alpha\derpar{}{u_{I}^\alpha} + F_{J,j}^\alpha\derpar{}{u_{J}^\alpha} + G_{\alpha,j}^i\derpar{}{p_\alpha^i} + G_{\alpha,j}^{I}\derpar{}{p_\alpha^{I}}\right)\left( p_\alpha^K - \derpar{\hat{L}}{u_K^\alpha} \right) = 0 \\
&\qquad \Longleftrightarrow
G_{\alpha,j}^K - \derpars{\hat{L}}{x^j}{u_K^\alpha} - u_j^\beta\derpars{\hat{L}}{u^\beta}{u_K^\alpha}
- u_{1_i+1_j}^\beta \derpars{\hat{L}}{u_i^\beta}{u_K^\alpha} - u_{I+1_j}^\beta\derpars{\hat{L}}{u_I^\beta}{u_K^\alpha} = 0 \\
&\qquad \Longleftrightarrow
G_{\alpha,j}^K - \frac{d}{dx^j}\, \derpar{\hat{L}}{u_K^\alpha} = 0 \, .
\end{align*}
Hence, the tangency condition enables us to determinate all the coefficient functions $G_{\alpha,j}^K$,
since we obtain $nm^2(m+1)/2$ equations, one for each function. Now, taking into account equations
\eqref{Chap06_eqn:UnifFieldEqMultiVFLocal2} and the coefficients $G_{\alpha,j}^K$ that we have determined, we obtain
\begin{equation*}
\sum_{j=1}^{m} \frac{1}{n(ij)} \, G_{\alpha,j}^{1_i+1_j} - \derpar{\hat L}{u_i^\alpha} + p_\alpha^i = 0
\ \Longleftrightarrow \
p_\alpha^i - \derpar{\hat{L}}{u_i^\alpha} + \sum_{j=1}^{m} \frac{1}{n(ij)} \, \frac{d}{dx^j} \, \derpar{\hat{L}}{u_{1_i+1_j}^\alpha} = 0 \, .
\end{equation*}
Hence, the tangency condition for the multivector field $\X$ along $\W_{c}$ gives rise to $mn$ new
constraints defining a submanifold of $\W_{c}$ that coincides with the submanifold $\W_\Lag$
introduced in Proposition \ref{Chap06_prop:UnifGraphLegendreMap}. Now we must study the tangency of $\X$ along
the new submanifold $\W_\Lag$. After a long but straightforward calculation, we obtain
\begin{align*}
G_{\alpha,k}^{i} =
\frac{d}{dx^k} \, \derpar{\hat{L}}{u_i^\alpha}
- \frac{d}{dx^k} \sum_{j=1}^{m} \frac{1}{n(ij)} \, \frac{d}{dx^j} \, \derpar{\hat{L}}{u_{1_i+1_j}^\alpha}
- \sum_{j=1}^{m}\frac{1}{n(ij)}\left( F_{I+1_j,k}^\beta - \frac{d}{dx^k} \, u_{I+1_j}^\beta \right) \derpars{\hat{L}}{u_I^\beta}{u_{1_i+1_j}^\alpha} \, .
\end{align*}
Therefore, the tangency condition along the submanifold $\W_\Lag$ enables us to determinate all the functions
$G_{\alpha,k}^{i}$. Now, taking into account equations \eqref{Chap06_eqn:UnifFieldEqMultiVFLocal1}, we have
\begin{align*}
&\sum_{i=1}^{m} G_{\alpha,i}^{i} - \derpar{\hat{L}}{u^\alpha} = 0 \Longleftrightarrow \\
& \derpar{\hat{L}}{u^\alpha}
- \frac{d}{dx^i} \, \derpar{\hat{L}}{u_i^\alpha}
+ \sum_{|I|=2} \frac{d^{2}}{dx^{I}} \, \derpar{\hat{L}}{u_{I}^\alpha}
+ \sum_{i=1}^{m}\sum_{j=1}^{m}\frac{1}{n(ij)}\left( F_{I+1_j,i}^\beta - \frac{d}{dx^i} \, u_{I+1_j}^\beta \right) \derpars{\hat{L}}{u_I^\beta}{u_{1_i+1_j}^\alpha} = 0 \, .
\end{align*}
These $n$ equations are the \textsl{second-order Euler-Lagrange equations} for a locally decomposable
holonomic multivector field. Observe that if $\hat{\Lag}$ is a second-order regular Lagrangian density,
then the Hessian of $\hat{L}$ with respect to the second-order velocities is regular, and we can assure
the existence of a local multivector field $\X$ solution to the equation
\eqref{Chap06_eqn:UnifFieldEqMultiVF}, defined at support on $\W_\Lag \hookrightarrow \W_r$, and tangent
to $\W_\Lag$. A global solution is then obtained using partitions of the unity.

If the Lagrangian density is not regular, then the above equations may or may not be compatible,
and may give rise to new constraints. In the most favorable cases there exists a submanifold
$\W_f \hookrightarrow \W_\Lag$ (where we admit $\W_f = \W_\Lag)$ where we have a well-defined holonomic
multivector field at support on $\W_f$, and tangent to $\W_f$, solution to the equation
\begin{equation}\label{Chap06_eqn:UnifFieldEqMultiVFSing}
\restric{\inn(\X)\Omega_r}{\W_f} = 0 \, .
\end{equation}

\subsubsection{Equivalence of the field equations in the unified formalism}

In the previous Sections we have stated the field equations in the unified formalism in several ways.
First, we have stated a geometric equation for sections of the bundle $\rho_M^r \colon \W_r \to M$,
and we have analyzed it in coordinates. Then, we have stated a geometric equation for multivector fields
defined in $\W_r$, and we have studied the equation and the tangency condition in coordinates.
In this Section we prove that all these equations are equivalent.

\begin{theorem}\label{Chap06_thm:UnifEquivalenceTheorem}
The following assertions on a holonomic section $\psi \in \Gamma(\rho_M^r)$ are equivalent.
\begin{enumerate}
\item $\psi$ is a solution to the equation \eqref{Chap06_eqn:UnifFieldEqSect}, that is,
\begin{equation*}
\psi^*\inn(X)\Omega_r = 0 \, , \quad \mbox{for every } X \in \vf(\W_r) \, .
\end{equation*}
\item If $\psi$ is given locally by
$\psi(x^i) = (x^i,u^\alpha(x^i),u^\alpha_j(x^i),u^\alpha_I(x^i),u^\alpha_J(x^i),p_\alpha^{j}(x^i),p_\alpha^{I}(x^i))$,
then the component functions of $\psi$ satisfy equations \eqref{Chap06_eqn:UnifFieldEqSectLocal}
and \eqref{Chap06_eqn:UnifFieldEqSectRelationMomenta}, that is, the following system of $n + nm$
partial differential equations
\begin{equation}\label{Chap06_eqn:EquivalenceTheoremUnifiedLocal}
\sum_{i=1}^{m}\derpar{p_\alpha^i}{x^i} = \derpar{\hat{L}}{u^\alpha} \quad ; \quad
\sum_{j=1}^{m} \frac{1}{n(ij)} \, \derpar{p_\alpha^{1_i+1_j}}{x^j} = \derpar{\hat{L}}{u_i^\alpha} - p_\alpha^i \, .
\end{equation}
\item $\psi$ is a solution to the equation
\begin{equation}\label{Chap06_eqn:UnifFieldEqIntSect}
\inn(\Lambda^m\psi^\prime)(\Omega_r \circ \psi) = 0 \, ,
\end{equation}
where $\Lambda^m\psi^\prime \colon M \to \Lambda^m\Tan\W_r$ is the canonical lifting of $\psi$.
\item $\psi$ is an integral section of a multivector field contained in a class of locally
decomposable holonomic multivector fields $\{ \X \} \subset \vf^{m}(\W_r)$, tangent to $\W_\Lag$,
and satisfying the equation \eqref{Chap06_eqn:UnifFieldEqMultiVF}, that is,
\begin{equation*}
\inn(\X)\Omega_r = 0 \, .
\end{equation*}
\end{enumerate}
\end{theorem}
\begin{proof}
We prove this result following the same patterns as the proof of Theorem
\ref{Chap05_thm:UnifEquivalenceTheorem}.

\begin{description}
\item[\textnormal{($1 \, \Longleftrightarrow \, 2$)}]
From the results in the previous Sections, the field equation \eqref{Chap06_eqn:UnifFieldEqSect}
gives in coordinates the equations \eqref{Chap06_eqn:UnifFieldEqSectLocal},
\eqref{Chap06_eqn:UnifFieldEqSectRelationMomenta}, \eqref{Chap06_eqn:UnifFieldEqSectLegendreLocal}
and \eqref{Chap06_eqn:UnifFieldEqSectHolonomy}.
As stated there, the equations \eqref{Chap06_eqn:UnifFieldEqSectLegendreLocal} are the local
constraints defining the first constraint submanifold $\W_c \hookrightarrow \W_r$. In addition,
since we assume that the section $\psi \in \Gamma(\rho_M^r)$ is holonomic, the equations
\eqref{Chap06_eqn:UnifFieldEqSectHolonomy} are satisfied. Therefore, the equation \eqref{Chap06_eqn:UnifFieldEqSect}
is locally equivalent to equations \eqref{Chap06_eqn:UnifFieldEqSectLocal} and
\eqref{Chap06_eqn:UnifFieldEqSectRelationMomenta}, that is, to equations \eqref{Chap06_eqn:EquivalenceTheoremUnifiedLocal}.

\item[\textnormal{($2 \, \Longleftrightarrow \, 3$)}]
If $\psi \in \Gamma(\rho_M^r)$ is locally given by
\begin{equation*}
\psi(x^i) = (x^i,u^\alpha(x^i),u^\alpha_i(x^i),u^\alpha_I(x^i),u^\alpha_J(x^i),p_\alpha^j(x^i),p_\alpha^I(x^i)) \, ,
\end{equation*}
then its canonical lifting to $\Lambda^m\Tan\W_r$ is locally given by
$\Lambda^m\psi^\prime = \psi^\prime_1 \wedge \ldots \wedge \psi_m^\prime$, with
\begin{equation*}
\psi^\prime_j = \left(0,\ldots,0,1,0,\ldots,0,\frac{d}{dx^j}u^\alpha,\frac{d}{dx^j}u^\alpha_i,
\frac{d}{dx^j}u^\alpha_I,\frac{d}{dx^j}u^\alpha_J,\frac{d}{dx^j}p_\alpha^i,\frac{d}{dx^j}p_\alpha^I\right) \, ,
\end{equation*}
where $d/dx^j$ is the $j$th coordinate total derivative, and the $1$ is at the $j$th position. Then,
the inner product $\inn(\Lambda^m\psi^\prime)(\Omega_r \circ \psi)$ gives, in coordinates,
\begin{align*}
\inn(\Lambda^m\psi^\prime)(\Omega_r \circ \psi) &=
\sum_{i=1}^{m}\left( \cdots \right)\d x^i
+ \left( \derpar{\hat{L}}{u^\alpha} - \frac{dp_\alpha^i}{dx^i} \right) \d u^\alpha 
+ \left( \derpar{\hat{L}}{u_i^\alpha} - p_\alpha^{i} - \sum_{j=1}^{m} \frac{1}{n(ij)} \, \frac{dp_\alpha^{1_i+1_j}}{dx^j} \right)\d u_i^\alpha \\
&\quad{} + \left( p_\alpha^I - \derpar{\hat{L}}{u_I^\alpha} \right)\d u_I^\alpha
+ \left( \frac{du^\alpha}{dx^i} - u_i^\alpha \right)\d p_\alpha^i +
\left( \sum_{1_i+1_j=I} \frac{1}{n(ij)} \, \frac{du^\alpha_i}{dx^j} - u_I^\alpha \right) \d p_\alpha^I \, ,
\end{align*}
where the terms $(\cdots)$ along the forms $\d x^i$ involve of partial derivatives of the Lagrangian
function and of the rest of component functions. Now, requiring this last expression to vanish, we obtain
equations \eqref{Chap06_eqn:UnifFieldEqSectLocal}, \eqref{Chap06_eqn:UnifFieldEqSectRelationMomenta}, \eqref{Chap06_eqn:UnifFieldEqSectLegendreLocal}
and \eqref{Chap06_eqn:UnifFieldEqSectHolonomy}, along with $m$ additional equations which are a combination of those.
Same comments as in the proof of the previous item apply. In particular,
equations \eqref{Chap06_eqn:UnifFieldEqSectLegendreLocal} are the local constraints defining the first constraint submanifold
$\W_c \hookrightarrow \W_r$, and equations \eqref{Chap06_eqn:UnifFieldEqSectHolonomy} are automatically satisfied
because of the holonomy assumption. Therefore, the equation \eqref{Chap06_eqn:UnifFieldEqIntSect}
is locally equivalent to equations \eqref{Chap06_eqn:UnifFieldEqSectLocal} and
\eqref{Chap06_eqn:UnifFieldEqSectRelationMomenta}, that is, to equations \eqref{Chap06_eqn:EquivalenceTheoremUnifiedLocal}.

\item[\textnormal{($2\, \Longleftrightarrow \, 4$)}]
From the results in the previous Section, if $\X \in \vf^{m}(\W_r)$ is a generic locally decomposable
multivector field locally given by \eqref{Chap06_eqn:UnifGenericMultiVFLocal}, then, taking $f = 1$
as a representative of the equivalence class, the field equation \eqref{Chap06_eqn:UnifFieldEqMultiVF}
is locally equivalent to the equations \eqref{Chap06_eqn:UnifFieldEqMultiVFHolonomy},
\eqref{Chap06_eqn:UnifFieldEqMultiVFLocal1}, \eqref{Chap06_eqn:UnifFieldEqMultiVFLocal2} and \eqref{Chap06_eqn:HOMomentaMultiVF}.
As already stated, equations \eqref{Chap06_eqn:HOMomentaMultiVF} give, in coordinates, the compatibility
submanifold $\W_c$ obtained using the premultisymplectic version of the constraint algorithm in
\cite{art:DeLeon_Marin_Marrero_Munoz_Roman05}.
On the other hand, since the multivector field $\X$ is assumed to be holonomic, then equations
\eqref{Chap06_eqn:UnifFieldEqMultiVFHolonomy} are satisfied. Hence, the field equation
\eqref{Chap06_eqn:UnifFieldEqMultiVF} is locally equivalent to equations \eqref{Chap06_eqn:UnifFieldEqMultiVFLocal1}
and \eqref{Chap06_eqn:UnifFieldEqMultiVFLocal2}.

Let $\psi \in \Gamma(\rho_M^r)$ be an integral section of $\X$ locally given by
$\psi(x^i) = (x^i,u^\alpha,u^\alpha_i,u^\alpha_I,u^\alpha_J,p_\alpha^{i},p_\alpha^{I})$. Then,
the condition of integral section is locally equivalent to the following system of equations
\begin{equation*}
\derpar{u^\alpha}{x^i} = F_i^\alpha \circ \psi \quad ; \quad
\derpar{u^\alpha_i}{x^j} = F_{i,j}^\alpha \circ \psi \quad ; \quad
\derpar{u^\alpha_I}{x^j} = F_{I,j}^\alpha \circ \psi \quad ; \quad
\derpar{u^\alpha_J}{x^j} = F_{J,j}^\alpha \circ \psi \, ,
\end{equation*}
\begin{equation*}
\derpar{p_\alpha^i}{x^j} = G_{\alpha,j}^i \circ \psi \quad ; \quad
\derpar{p_\alpha^I}{x^j} = G_{\alpha,j}^I \circ \psi \, .
\end{equation*}
Replacing these equations in \eqref{Chap06_eqn:UnifFieldEqMultiVFHolonomy},
\eqref{Chap06_eqn:UnifFieldEqMultiVFLocal1} and \eqref{Chap06_eqn:UnifFieldEqMultiVFLocal2},
we obtain the following system of partial differential equations for the component functions
of $\psi$
\begin{equation*}
\derpar{u^\alpha}{x^i} = u_i^\alpha \quad ; \quad
\derpar{u^\alpha_i}{x^j} = u_{1_i+1_j}^\alpha \quad ; \quad
\derpar{u^\alpha_I}{x^j} = u_{I+1_j}^\alpha \quad ; \quad
\derpar{u^\alpha_J}{x^j} = F_{J,j}^\alpha \, ,
\end{equation*}
\begin{equation*}
\sum_{i=1}^{m}\derpar{p_\alpha^i}{x^i} = \derpar{\hat{L}}{u^\alpha} \quad ; \quad
\sum_{j=1}^{m} \frac{1}{n(ij)} \, \derpar{p_\alpha^{1_i+1_j}}{x^j} = \derpar{\hat{L}}{u_i^\alpha} - p_\alpha^i \, .
\end{equation*}
Since the multivector field $\X$ is holonomic and tangent to $\W_\Lag$, the first equations are
identically satisfied. Thus, the condition of $\psi$ to be an integral section of a locally
decomposable holonomic multivector field $\X \in \vf^{m}(\W_r)$, tangent to $\W_\Lag$, and
satisfying the equation \eqref{Chap06_eqn:UnifFieldEqMultiVF} is locally equivalent to equations
\eqref{Chap06_eqn:EquivalenceTheoremUnifiedLocal}. \qedhere
\end{description}
\end{proof}


\section{Lagrangian formalism}
\label{Chap06_sec:LagrangianFormalism}

In this Section we state the Lagrangian formalism for second-order field theories. As in Section
\ref{Chap05_sec:UnifiedToLagrangian}, we have already stated the unified formalism for
second-order field theories, and thus we will ``recover'' the Lagrangian structures and solutions
from the unified setting.

The results remain the same for both regular and singular second-order Lagrangian densities. Thus,
no distinction will be made in this matter.

\subsection{Geometrical setting}
\label{Chap06_sec:LagrangianFormalismGeometricalSetting}

In order to establish the field equations in the Lagrangian formalism, we must define the
Poincar\'{e}-Cartan $m$ and $(m+1)$-forms in $J^3\pi$. Since the constraint algorithm delivers a
unique extended Legendre map in the unified framework (see Proposition
\ref{Chap06_prop:UnifGraphLegendreMap} and Corollary \ref{Chap06_corol:UnifGraphExtendedLegendreMap}),
we can give the following definition.

\begin{definition}\label{Chap06_def:PoincareCartanForms}
Let $\Theta_1^s \in \df^{m}(J^2\pi^\dagger)$ and $\Omega_1^s \in \df^{m+1}(J^2\pi^\dagger)$ be the
symmetrized Liouville forms in $J^2\pi^\dagger$. The \textnormal{Poincar\'{e}-Cartan forms} in
$J^3\pi$ are defined as
\begin{equation*}
\Theta_\Lag = \widetilde{\Leg}^*\Theta_1^s \in \df^{m}(J^3\pi) \quad ; \quad
\Omega_\Lag = \widetilde{\Leg}^*\Omega_1^s = -\d\Theta_\Lag \in \df^{m+1}(J^3\pi) \, .
\end{equation*}
\end{definition}

The Poincar\'{e}-Cartan forms can also be recovered directly from the unified formalism. In fact:

\begin{lemma}\label{Chap06_lemma:LagRelatedForms}
Let $\Theta = \rho_2^*\Theta_1^s$ and $\Theta_r = \hat{h}^*\Theta$ be the canonical $m$-forms
defined in $\W$ and $\W_r$, respectively. Then, the Poincar\'{e}-Cartan $m$-form satisfies
$\Theta = \rho_1^*\Theta_\Lag$ and $\Theta_r = (\rho_1^r)^*\Theta_\Lag$.
\end{lemma}
\begin{proof}
A straightforward computation leads to this result. For the first statement we have
\begin{equation*}
\rho_1^*\Theta_\Lag = \rho_1^*(\widetilde{\Leg}^*\Theta_1^s)
= (\widetilde{\Leg} \circ \rho_1)^*\Theta_1^s = \rho_2^*\Theta_1^s = \Theta\, ,
\end{equation*}
and from this the second statement follows
\begin{equation*}
(\rho_1^r)^*\Theta_\Lag = (\rho_1 \circ \hat{h})^*\Theta_\Lag
= \hat{h}^*(\rho_1^*\Theta_\Lag) = \hat{h}^*\Theta = \Theta_r \, . \qedhere
\end{equation*}
\end{proof}

\begin{remark}
As the pull-back of a form by a function and the exterior derivative
commute, this result also holds for the Poincar\'{e}-Cartan $(m+1)$-form.
\end{remark}

In the natural coordinates $(x^i,u^\alpha,u_i^\alpha,u_I^\alpha,u_J^\alpha)$ of $J^3\pi$,
bearing in mind the local expression \eqref{Chap06_eqn:SymmetricMultimomentaLiouvilleFormsLocal}
of $\Theta_1^s$, and \eqref{Chap06_eqn:UnifExtendedLegendreMapLocal} of the extended Legendre map,
the local expression of the Poincar\'{e}-Cartan $m$-form is
\begin{align*}
\Theta_\Lag &= \left( \derpar{L}{u_i^\alpha} - \sum_{j=1}^{m}\frac{1}{n(ij)} \, \frac{d}{dx^j} \, \derpar{L}{u_{1_i+1_j}^\alpha} \right)(\d u^\alpha \wedge \d^{m-1}x_i - u_i^\alpha\d^mx) \\
&\qquad {} + \frac{1}{n(ij)} \, \derpar{L}{u_{1_i+1_j}^\alpha} \, ( \d u_i^\alpha \wedge \d^{m-1}x_j - u_{1_i+1_j}^\alpha\d^mx ) + L \d^mx \, .
\end{align*}

An important fact regarding the Poincar\'{e}-Cartan $(m+1)$-form $\Omega_\Lag$ is that it is
$1$-degenerate when $m > 1$, regardless of the regularity of the Lagrangian density. Indeed, since
the restricted Legendre map $\Leg \colon J^3\pi \to J^2\pi^\ddagger$ is a submersion with
$\dim J^3\pi > \dim J^2\pi^\ddagger$, and $\rank(\Leg) = \rank(\widetilde{\Leg})$,
there exists a non-zero vector field $X \in \vf(J^3\pi)$ which is $\widetilde{\Leg}$-related to
$\mathbf{0} \in \vf(J^2\pi^\dagger)$, that is, $\Tan\widetilde{\Leg} \circ X = \mathbf{0} \circ \widetilde{\Leg}$.
Then, we have
\begin{equation*}
\inn(X)\Omega_\Lag = \inn(X)\widetilde{\Leg}^*\Omega_1^s = \widetilde{\Leg}^*\inn(\mathbf{0}) \Omega_1^s = 0 \, .
\end{equation*}

Finally, the following result enables us to establish a one-to-one correspondence between the
solutions of the unified formalism and the solutions of the Lagrangian equations that we state
in the following Sections.

\begin{proposition}\label{Chap06_prop:LagDiffWL}
The map $\rho_1^\Lag = \rho_1^r \circ j_\Lag \colon \W_\Lag \to J^3\pi$ is a diffeomorphism.
\end{proposition}
\begin{proof}
Since $\W_\Lag = \graph(\Leg)$, it is clear that $J^{3}\pi$ is diffeomorphic to $\W_\Lag$.
On the other hand, since $\rho_1$ is a surjective submersion by definition, its restriction
to the submanifold $\W_\Lag$ is also a surjective submersion and, due to the fact that
$\dim\W_\Lag = \dim J^{3}\pi$, the map $\rho_1^\Lag$ is a bijective local diffeomorphism.
In particular, the map $\rho_1^\Lag$ is a global diffeomorphism.
\end{proof}

\subsection{Field equations}

Using the results stated in the previous Section, we can state the field equations in
the Lagrangian formalism, and recover the solutions to these equations from the solutions
to the field equations in the unified formalism.

\subsubsection{Field equations for sections}

Using the previous results, we can state the Lagrangian equations for sections, and
recover the Lagrangian sections in $J^{3}\pi$ from the sections in the unified formalism.

First, the \textsl{second-order Lagrangian problem for sections} associated with the system
$(J^{3}\pi,\Lag)$ consists in finding sections $\phi \in \Gamma(\pi)$ satisfying
\begin{equation}\label{Chap06_eqn:LagFieldEqSect}
(j^{3}\phi)^*\inn(X)\Omega_\Lag = 0 \, , \quad \mbox{for every } X \in \vf(J^{3}\pi) \, .
\end{equation}

\begin{proposition}\label{Chap06_prop:UnifToLagSect}
Let $\psi \in \Gamma(\rho_M^r)$ be a holonomic section solution to the equation
\eqref{Chap06_eqn:UnifFieldEqSect}. Then the section $\psi_\Lag = \rho_1^r \circ \psi \in \Gamma(\bar{\pi}^3)$
is holonomic, and its projection $\phi = \pi^3 \circ \psi_\Lag$ is a solution to equation
\eqref{Chap06_eqn:LagFieldEqSect}.
\end{proposition}
\begin{proof}
By definition, a section $\psi \in \Gamma(\rho_M^r)$ is holonomic if the section
$\rho_1^r \circ \psi \in \Gamma(\bar{\pi}^3)$ is holonomic. Hence, $\psi_\Lag = \rho_1^r \circ \psi$
is clearly a holonomic section.

Now, since $\rho_1^r \colon \W_r \to J^3\pi$ is a submersion, for every vector field
$X \in \vf(J^3\pi)$ there exist some vector fields $Y \in \vf(\W_r)$ such that $X$ and $Y$ are
$\rho_1^r$-related. Observe that this vector field $Y$ is not unique because the vector field
$Y + Y_o$, with $Y_o \in \ker\Tan\rho_1^r$ is also $\rho_1^r$-related with $X$. Thus, using this
particular choice of $\rho_1^r$-related vector fields, we have
\begin{equation*}
\psi_\Lag^*\inn(X)\Omega_\Lag = (\rho_1^r \circ \psi)^*\inn(X)\Omega_\Lag
= \psi^*((\rho_1^r)^*\inn(X)\Omega_\Lag) = \psi^*\inn(Y)(\rho_1^r)^*\Omega_\Lag
= \psi^*\inn(Y)\Omega_r \, .
\end{equation*}
Since the equality $\psi^*\inn(Y)\Omega_r = 0$ holds for every $Y \in \vf(\W_r)$, it holds, in
particular, for every $Y \in \vf(\W_r)$ which is $\rho_1^r$-related with $X \in \vf(J^3\pi)$.
Hence we obtain
\begin{equation*}
\psi_\Lag^*\inn(X)\Omega_\Lag = \psi^*\inn(Y)\Omega_r = 0 \, . \qedhere
\end{equation*}
\end{proof}

The following diagram illustrates the situation of the above Proposition
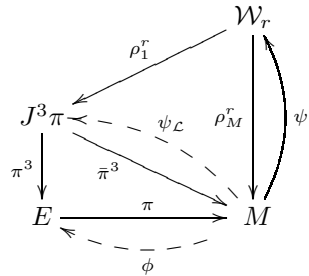
\begin{equation*}
\xymatrix{
\ & \ & \W_r \ar[dd]_-{\rho_M^r} \ar[dll]_-{\rho_1^r}  \\
J^{3}\pi \ar[d]_{\pi^{3}} \ar[drr]_<(0.25){\bar{\pi}^{3}} & \ & \ \\
E \ar[rr]^{\pi} & \ & \  M \ar@/_1pc/[uu]_{\psi} \ar@/_1pc/@{-->}[ull]_{\psi_\Lag} \ar@/^1pc/@{-->}[ll]^{\phi} \\
}
\end{equation*}

Observe that Proposition \ref{Chap06_prop:UnifToLagSect} states that every section solution to the field
equation in the unified formalism projects to a section solution to the field equation in the
Lagrangian formalism, but it does not establish an equivalence between the solutions. This
equivalence does exist, due to the facts that the map $\rho_1^\Lag \colon \W_\Lag \to J^3\pi$ is a
diffeomorphism, and that every section $\psi \in \Gamma(\rho_M^r)$ solution to equation
\eqref{Chap06_eqn:UnifFieldEqSect} takes values in the submanifold
$\W_\Lag = \graph(\Leg) \hookrightarrow \W_r$. In order to establish this equivalence, we first
need the following technical result.

\begin{lemma}\label{Chap06_lemma:LagCartanFormsTechLemma}
The Poincar\'{e}-Cartan forms satisfy $(\rho_1^\Lag)^*\Theta_\Lag = j_\Lag^*\Theta_r$
and $(\rho_1^\Lag)^*\Omega_\Lag = j_\Lag^*\Omega_r$.
\end{lemma}
\begin{proof}
Since the exterior derivative and the pull-back commute, it suffices to prove the statement for
the $m$-forms. We have
\begin{align*}
(\rho_1^\Lag)^*\Theta_\Lag &= (\rho_1^r\circ j_\Lag)^*\Theta_\Lag
= (\rho_1 \circ \hat{h} \circ j_\Lag)^*\Theta_\Lag 
= (\rho_1 \circ \hat{h} \circ j_\Lag)^*(\widetilde{\Leg}^*\Theta_1^s) \\
&= (\widetilde{\Leg} \circ \rho_1 \circ \hat{h} \circ j_\Lag)^*\Theta_1^s
= (\rho_2 \circ \hat{h} \circ j_\Lag)^*\Theta_1^s
= (\hat{h} \circ j_\Lag)^*\Theta
= j_\Lag^*\Theta_r \, . \qedhere
\end{align*}
\end{proof}

Now we can state the remaining part of the equivalence between the solutions of the Lagrangian and
unified formalisms.

\begin{proposition}\label{Chap06_prop:LagToUnifSect}
Let $\psi_\Lag \in \Gamma(\bar{\pi}^3)$ be a holonomic section solution to the field equation
\eqref{Chap06_eqn:LagFieldEqSect}. Then the section
$\psi = j_\Lag \circ (\rho_1^\Lag)^{-1} \circ \psi_\Lag \in \Gamma(\rho_M^r)$
is holonomic and it is a solution to the equation \eqref{Chap06_eqn:UnifFieldEqSect}.
\end{proposition}
\begin{proof}
By definition, a section $\psi \in \Gamma(\rho_M^r)$ is holonomic in $\W_r$ if the section
$\rho_1^r \circ \psi \in \Gamma(\bar{\pi}^3)$ is holonomic in $J^3\pi$. Computing, we have
\begin{equation*}
\rho_1^r \circ \psi = \rho_1^r \circ j_\Lag \circ (\rho_1^\Lag)^{-1} \circ \psi_\Lag = \psi_\Lag \, ,
\end{equation*}
since 
$\rho_1^r \circ j_\Lag = \rho_1^\Lag \Leftrightarrow \rho_1^r \circ j_\Lag \circ (\rho_1^\Lag)^{-1} =
\textnormal{Id}_{J^3\pi}$.
Hence, as $\psi_\Lag$ is holonomic, the section $\psi = j_\Lag \circ (\rho_1^\Lag)^{-1} \circ \psi_\Lag$
is holonomic in $\W_r$.

Now, since $j_\Lag \colon \W_\Lag \to \W_r$ is an embedding, for every vector field $X \in \vf(\W_r)$
tangent to $\W_\Lag$, there exists a unique vector field $Y \in \vf(\W_\Lag)$ which is $j_\Lag$-related
with $X$. Hence, let us assume that $X \in \vf(\W_r)$ is tangent to $\W_\Lag$. Then we have
\begin{equation*}
\psi^*\inn(X)\Omega_r = (j_\Lag \circ (\rho_1^\Lag)^{-1} \circ \psi_\Lag)^*\inn(X)\Omega_r
= ((\rho_1^\Lag)^{-1} \circ \psi_\Lag)^*\inn(Y)j_\Lag^*\Omega_r \, .
\end{equation*}
Applying Lemma \ref{Chap06_lemma:LagCartanFormsTechLemma} we obtain
\begin{equation*}
((\rho_1^\Lag)^{-1} \circ \psi_\Lag)^*\inn(Y)j_\Lag^*\Omega_r
= ((\rho_1^\Lag)^{-1} \circ \psi_\Lag)^*\inn(Y)(\rho_1^\Lag)^*\Omega_\Lag
= (\rho_1^\Lag \circ (\rho_1^\Lag)^{-1} \circ \psi_\Lag)^*\inn(Z)\Omega_\Lag = \psi_\Lag^*\inn(Z)\Omega_\Lag \, ,
\end{equation*}
where $Z \in \vf(J^3\pi)$ is the unique vector field related with $Y \in \vf(\W_\Lag)$ by the
diffeomorphism $\rho_1^\Lag$. Hence, as $\psi_\Lag^*\inn(Z)\Omega_\Lag = 0$, for every
$Z \in \vf(J^3\pi)$ by hypothesis, we just proved that the section
$\psi = j_\Lag \circ (\rho_1^\Lag)^{-1} \circ \psi_\Lag$ satisfies the equation
\begin{equation*}
\psi^*\inn(X)\Omega_r = 0 \, , \quad \mbox{for every } X \in \vf(\W_r) \mbox{ tangent to }\W_\Lag \, .
\end{equation*}
However, from Proposition \ref{Chap06_prop:UnifFieldEqSectTangent} we know that if $\psi \in \Gamma(\rho_M^r)$
is a holonomic section, then the last equation is equivalent to the equation
\eqref{Chap06_eqn:UnifFieldEqSect}, that is,
\begin{equation*}
\psi^*\inn(X)\Omega_r = 0 \, , \quad \mbox{for every } X \in \vf(\W_r) \, . \qedhere
\end{equation*}
\end{proof}

Let us compute the local equation for the section
$\psi_\Lag = \rho_1^r \circ \psi \in \Gamma(\bar{\pi}^3)$.
Assume that the section $\psi \in \Gamma(\rho_M^r)$ is given locally by
$\psi(x^i) = (x^i,u^\alpha,u^\alpha_i,u^\alpha_{I},u^\alpha_{J},p_\alpha^i,p_\alpha^{I})$.
Since $\psi$ is a holonomic section solution to equation \eqref{Chap06_eqn:UnifFieldEqSect},
it must satisfy the local equations \eqref{Chap06_eqn:UnifFieldEqSectLocal},
\eqref{Chap06_eqn:UnifFieldEqSectRelationMomenta} and \eqref{Chap06_eqn:UnifFieldEqSectHolonomy}.
The equations \eqref{Chap06_eqn:UnifFieldEqSectHolonomy} are automatically satisfied as a consequence of the
assumption of $\psi$ being holonomic. Now, taking into account that $\psi$ takes values in the
submanifold $\W_\Lag \cong \textnormal{graph}(\Leg)$, the equations \eqref{Chap06_eqn:UnifFieldEqSectLocal}
and \eqref{Chap06_eqn:UnifFieldEqSectRelationMomenta} can be $\rho_1^\Lag$-projected to $J^3\pi$, thus giving
the following system of $n$ partial differential equations for the component functions of the
section $\psi_\Lag = \rho_1^r \circ \psi$
\begin{equation*}
\restric{\derpar{L}{u^\alpha}}{\psi_\Lag} - \restric{\frac{d}{dx^i} \, \derpar{L}{u_i^\alpha}}{\psi_\Lag}
+ \sum_{|I|=2} \restric{\frac{d^2}{dx^{I}} \, \derpar{L}{u_I^\alpha}}{\psi_\Lag} = 0 \qquad
(1 \leqslant \alpha \leqslant n) \, ,
\end{equation*}
where the section $\psi_\Lag$ is locally given by
$\psi_\Lag(x^i) = (x^i,u^\alpha,u^\alpha_i,u^\alpha_{I},u^\alpha_{J})$. Finally, since $\psi_\Lag$
is holonomic in $J^3\pi$, there exists a section $\phi \in \Gamma(\pi)$ with local expression
$\phi(x^i) = (x^i,u^\alpha(x^i))$ satisfying $j^3\phi = \psi_\Lag$. Then, the above equations can
be rewritten as follows
\begin{equation} \label{Chap06_eqn:EulerLagrangeEquations}
\restric{\derpar{L}{u^\alpha}}{j^3\phi} - \restric{\frac{d}{dx^i} \, \derpar{L}{u_i^\alpha}}{j^3\phi}
+ \sum_{|I|=2} \restric{\frac{d^2}{dx^{I}} \, \derpar{L}{u_I^\alpha}}{j^3\phi} = 0 \qquad
(1 \leqslant \alpha \leqslant n) \, .
\end{equation}
Therefore, we obtain the Euler-Lagrange equations for a second-order field theory. As stated before,
equations \eqref{Chap06_eqn:EulerLagrangeEquations} may or may not be compatible, and in this last case a
constraint algorithm must be used to obtain a submanifold $S_f \hookrightarrow J^{3}\pi$
(if such a submanifold exists) where the equations can be solved.

\subsubsection{Field equations for multivector fields}

Now, using the results stated at the beginning of the Section, we can state the Lagrangian field
equation for multivector fields, and recover a solution to the Lagrangian equation starting
from a solution to the equation in the unified formalism.

The \textsl{Lagrangian problem for multivector fields} associated with the system $(J^3\pi,\Lag)$
consists in finding a class of locally decomposable holonomic multivector fields
$\{\X_\Lag\} \subset \vf^m(J^3\pi)$ satisfying the following field equation
\begin{equation}\label{Chap06_eqn:LagFieldEqMultiVF}
\inn(\X_\Lag)\Omega_\Lag = 0 \, , \quad \mbox{for every } \X_\Lag \in \{\X_\Lag\} \subseteq \vf^{m}(J^3\pi) \, .
\end{equation}

First we need to state a correspondence between the set of multivector fields in $\W_r$ tangent to
$\W_\Lag$ and the set of multivector fields in $J^{3}\pi$.

\begin{lemma}\label{Chap06_lemma:LagRelatedMultiVF}
Let $\X \in \vf^{m}(\W_r)$ be a multivector field tangent to $\W_\Lag \hookrightarrow \W_r$. Then
there exists a unique multivector field $\X_\Lag \in \vf^m(J^3\pi)$ such that
$\X_\Lag \circ \rho_1^r \circ j_\Lag = \Lambda^m\Tan\rho_1^r \circ \X \circ j_\Lag$.

\noindent Conversely, if $\X_\Lag \in \vf^m(J^3\pi)$, then there exists a unique multivector
field $\X \in \vf^m(\W_r)$ tangent to $\W_\Lag$ such that
$\X_\Lag \circ \rho_1^r \circ j_\Lag = \Lambda^m\Tan\rho_1^r \circ \X \circ j_\Lag$.
\end{lemma}
\begin{proof}
As the multivector field $\X$ is tangent to $\W_\Lag$, there exists a unique multivector field
$\X_o \in \vf^{m}(\W_\Lag)$ which is $j_\Lag$-related to $\X$, that is,
$\Lambda^m\Tan j_\Lag \circ \X_o = \X \circ j_\Lag$. Furthermore, since
$\rho_1^\Lag \colon \W_\Lag \to J^3\pi$ is a diffeomorphism, there is a unique multivector field
$\X_\Lag \in \vf^{m}(J^3\pi)$ which is $\rho_1^\Lag$-related to $\X_o$; that is,
$\X_\Lag \circ \rho_1^\Lag = \Lambda^m\Tan \rho_1^\Lag \circ \X_o$. Then we have
\begin{equation*}
\X_\Lag \circ \rho_1^r \circ j_\Lag = \X_\Lag \circ \rho_1^\Lag
= \Lambda^m\Tan\rho_1^\Lag \circ \X_o
= \Lambda^m\Tan\rho_1^r \circ \Lambda^m\Tan j_\Lag \circ \X_o
= \Lambda^m\Tan\rho_1^r \circ \X \circ j_\Lag \, .
\end{equation*}
The converse is proved reversing this reasoning.
\end{proof}

The above result states that there is a one-to-one correspondence between the set of multivector
fields $\X \in \vf^m(\W_r)$ tangent to $\W_\Lag$ and the set of multivector fields
$\X_\Lag \in \vf^{m}(J^3\pi)$, which makes the following diagram commutative
\begin{equation*}
\xymatrix{
\ & \ & \ & \Lambda^m\Tan\W_r \ar[dlll]_{\Lambda^m\Tan\rho_1^r} \\
\Lambda^m\Tan J^3\pi & \ & \ & \Lambda^m\Tan\W_\Lag \ar[lll]^{\Lambda^m\Tan\rho_1^\Lag} \ar@{^{(}->}[u]^{\Lambda^m\Tan j_\Lag} \\
 \ & \ & \ & \ \\
\ & \ & \ & \W_r \ar[dlll]_{\rho_1^r} \ar@/_2.25pc/[uuu]_{\X} \\
J^3\pi \ar[uuu]^{\X_\Lag} & \ & \ & \W_\Lag \ar[lll]^{\rho_1^\Lag} \ar@{_{(}->}[u]_{j_\Lag} \ar@/^1.5pc/[uuu]^{\X_o}|(.32)\hole
}
\end{equation*}

As a consequence of Lemma \ref{Chap06_lemma:LagRelatedMultiVF}, we can establish a bijective
correspondence between the set of multivector fields in $\W_r$ tangent to $\W_\Lag$ solution to the
field equation in the unified formalism and the set of multivector fields in $J^{3}\pi$ solution
to the Lagrangian field equation stated above. In particular, we have the following result.

\begin{theorem}\label{Chap06_thm:UnifToLagMultiVF}
Let $\X \in \vf^m(\W_r)$ be a locally decomposable holonomic multivector field solution to the
equation \eqref{Chap06_eqn:UnifFieldEqMultiVF} (at least on the points of a submanifold
$\W_f \hookrightarrow \W_\Lag$) and tangent to $\W_\Lag$ (resp. tangent to $\W_f$). Then there
exists a unique locally decomposable holonomic multivector field $\X_\Lag \in \vf^m(J^3\pi)$
solution to the equation \eqref{Chap06_eqn:LagFieldEqMultiVF} (at least on the points of
$S_f = \rho_1^\Lag(\W_f)$, and tangent to $S_f$).

\noindent Conversely, if $\X_\Lag \in \vf^{m}(J^3\pi)$ is a locally decomposable holonomic multivector field
solution to the equation \eqref{Chap06_eqn:LagFieldEqMultiVF} (at least on the points of a submanifold
$S_f \hookrightarrow J^3\pi$, and tangent to $S_f$), then there exists a unique locally decomposable
holonomic multivector field $\X \in \vf^{m}(\W_r)$ which is a solution to the equation
\eqref{Chap06_eqn:UnifFieldEqMultiVF} (at least on the points of $(\rho_1^\Lag)^{-1}(S_f) \hookrightarrow \W_\Lag$),
and tangent to $\W_\Lag$ (resp. tangent to $\W_f$).
\end{theorem}
\begin{proof}
Applying Lemmas \ref{Chap06_lemma:LagRelatedForms} and \ref{Chap06_lemma:LagRelatedMultiVF}, we have
\begin{equation*}
\restric{\inn(\X)\Omega_r}{\W_\Lag} = \restric{\inn(\X)(\rho_1^r)^*\Omega_\Lag}{\W_\Lag}
= \restric{(\rho_1^r)^*\inn(\X_\Lag)\Omega_\Lag}{\W_\Lag}
= \restric{\inn(\X_\Lag)\Omega_\Lag}{\rho_1^r(\W_\Lag)}
= \restric{\inn(\X_\Lag)\Omega_\Lag}{J^3\pi} \, .
\end{equation*}
Hence, $\X_\Lag$ is a solution to the equation $\inn(\X_\Lag)\Omega_\Lag = 0$ if, and only if,
$\X$ is a solution to the equation $\inn(\X)\Omega_r = 0$.

Now we must prove that $\X_\Lag$ is holonomic if, and only if, $\X$ is holonomic.
Observe that, following the same reasoning as above, we have
\begin{align*}
\restric{\inn(\X)(\rho_M^r)^*\eta}{\W_\Lag} &= \restric{\inn(\X)(\bar{\pi}^3 \circ \rho_1^r)^*\eta}{\W_\Lag}
= \restric{(\rho_1^r)^*\inn(\X_\Lag)(\bar{\pi}^3)^*\eta}{\W_\Lag} \\
&= \restric{\inn(\X_\Lag)(\bar{\pi}^3)^*\eta}{\rho_1^r(\W_\Lag)}
= \restric{\inn(\X_\Lag)(\bar{\pi}^3)^*\eta}{J^3\pi} \, .
\end{align*}
Hence, $\X_\Lag$ is $\bar{\pi}^3$-transverse if, and only if, $\X$ is $\rho_M^r$-transverse.

Now, let us assume that $\X \in \vf^m(\W_r)$ is holonomic, and let $\psi \in \Gamma(\rho_M^r)$ be
an integral section of $\X$. Then, the section $\psi_\Lag = \rho_1^r \circ \psi \in \Gamma(\bar{\pi}^3)$
is holonomic by definition, and we have
\begin{equation*}
\X_\Lag \circ \psi_\Lag = \X_\Lag \circ \rho_1^r \circ \psi
= \Lambda^m\Tan\rho_1^r \circ \X \circ \psi
= \Lambda^m\Tan\rho_1^r \circ \Lambda^m\psi^\prime
= \Lambda^m\psi_\Lag^\prime \, ,
\end{equation*}
where $\Lambda^m\psi^\prime \colon M \to \Lambda^m\Tan\W_r$ is the canonical lifting of $\psi$ to
$\Lambda^m\Tan\W_r$. That is, $\psi_\Lag$ is an integral section of $\X_\Lag$. Hence, if $\X$ is
holonomic, then $\X_\Lag$ is holonomic.

For the converse, let us assume that $\X_\Lag \in \vf^m(J^3\pi)$ is holonomic, and let
$\psi_\Lag \in \Gamma(\bar{\pi}^3)$ be an integral section of $\X_\Lag$. Then, the section
$\psi = j_\Lag \circ (\rho_1^\Lag)^{-1} \circ \psi_\Lag \in \Gamma(\rho_M^3)$ satisfies
\begin{equation*}
\rho_1^r \circ \psi = \rho_1^r \circ j_\Lag \circ (\rho_1^\Lag)^{-1} \circ \psi_\Lag = \psi_\Lag \, ,
\end{equation*}
since $\rho_1^r \circ j_\Lag = \rho_1^\Lag$ implies $\rho_1^r \circ j_\Lag \circ (\rho_1^\Lag)^{-1} = \Id_{J^3\pi}$.
Therefore, the section $\psi = j_\Lag \circ (\rho_1^\Lag)^{-1} \circ \psi_\Lag$ is holonomic. Finally,
since the multivector field $\X$ is tangent to $\W_\Lag$, there exists a unique multivector
field $\X_o \in \vf^{m}(\W_\Lag)$ satisfying $\Lambda^m\Tan j_\Lag \circ \X_o = \X \circ j_\Lag$.
In addition, since the map $\rho_1^\Lag$ is a diffeomorphism, $\X_\Lag$ and $\X_o$ are
$(\rho_1^\Lag)^{-1}$-related; that is,
$\X_o \circ (\rho_1^\Lag)^{-1} = (\Lambda^m\Tan\rho_1^\Lag)^{-1} \circ \X_\Lag$. Then we have
\begin{align*}
\X \circ \psi &= \X \circ j_\Lag \circ (\rho_1^\Lag)^{-1} \circ \psi_\Lag
= \Lambda^m\Tan j_\Lag \circ \X_o \circ (\rho_1^\Lag)^{-1} \circ \psi_\Lag
= \Lambda^m\Tan j_\Lag \circ (\Lambda^m\Tan\rho_1^\Lag)^{-1} \circ \X_\Lag \circ \psi_\Lag \\
&= \Lambda^m\Tan j_\Lag \circ (\Lambda^m\Tan\rho_1^\Lag)^{-1} \circ \Lambda^m\psi_\Lag^\prime 
= \Lambda^m(j_\Lag \circ (\rho_1^\Lag)^{-1} \circ \psi_\Lag)^\prime
= \Lambda^m\psi^\prime \, .
\end{align*}
Hence, $\psi$ is an integral section of $\X$. Therefore, $\X$ is holonomic if, and only if, $\X_\Lag$
is holonomic.
\end{proof}

Let $\X_\Lag \in \vf^m(J^3\pi)$ be a locally decomposable multivector field. From the results in
Section \ref{Chap01_sec:MultivectorFields} and in \cite{art:Echeverria_Munoz_Roman98} we know that
$\X_\Lag$ admits the following local expression
\begin{equation*}\label{Chap06_eqn:LagGenericMultiVFLocal}
\X_\Lag = f \bigwedge_{j=1}^{m}
\left(  \derpar{}{x^j} + F_j^\alpha\derpar{}{u^\alpha} + F_{i,j}^\alpha\derpar{}{u_i^\alpha} + F_{I,j}^\alpha\derpar{}{u_{I}^\alpha}
+ F_{J,j}^\alpha\derpar{}{u_{J}^\alpha} \right) \, .
\end{equation*}
Taking $f = 1$ as a representative of the equivalence class, since $\X_\Lag$ is required to be
holonomic, it must satisfy the equations \eqref{Chap01_eqn:MultiVFHolonomyLocal} with $k = 3$ and $r = 1$, that is,
\begin{equation*}
F_j^\alpha = u_j^\alpha \quad ; \quad
F_{i,j}^\alpha = u_{1_i+1_j}^\alpha \quad ; \quad
F_{I,j}^\alpha = u_{I + 1_j}^\alpha \, .
\end{equation*}
In addition, $\X_\Lag$ is a solution to the equation \eqref{Chap06_eqn:LagFieldEqMultiVF}. Bearing in mind the
local equations \eqref{Chap06_eqn:UnifFieldEqMultiVFLocal1} and \eqref{Chap06_eqn:UnifFieldEqMultiVFLocal2}
for the multivector field $\X$, and the fact that $\X$ is tangent to the submanifold $\W_\Lag = \graph(\Leg)$,
we obtain that the local equations for the component functions of $\X_\Lag$ are
\begin{equation*}
\derpar{L}{u^\alpha}
- \frac{d}{dx^i} \, \derpar{L}{u_i^\alpha}
+ \sum_{|I|=2} \frac{d^2}{dx^{I}} \, \derpar{L}{u_{I}^\alpha}
+ \sum_{i=1}^{m}\sum_{j=1}^{m}\frac{1}{n(ij)}\left( F_{I+1_j,i}^\beta - \frac{d}{dx^i} \, u_{I+1_j}^\beta \right) \derpars{L}{u_I^\beta}{u_{1_i+1_j}^\alpha} = 0 \, ,
\end{equation*}
which are the second-order Euler-Lagrange equations for a multivector field.

\subsubsection{Equivalence of the field equations in the Lagrangian formalism}

Finally, we state the equivalence Theorem in the Lagrangian formalism, which is the analogous to
Theorem \ref{Chap06_thm:UnifEquivalenceTheorem} in this formulation. This result is a straightforward
consequence of Theorems \ref{Chap06_thm:UnifEquivalenceTheorem} and \ref{Chap06_thm:UnifToLagMultiVF},
and of Propositions \ref{Chap06_prop:UnifToLagSect} and \ref{Chap06_prop:LagToUnifSect},
and hence we omit the proof.

\begin{theorem}\label{Chap06_thm:LagEquivalanceTheorem}
The following assertions on a section $\phi \in \Gamma(\pi)$ are equivalent.
\begin{enumerate}
\item $j^{3}\phi$ is a solution to equation \eqref{Chap06_eqn:LagFieldEqSect}, that is,
\begin{equation*}
(j^3\phi)^*\inn(X)\Omega_\Lag = 0 \, , \quad \mbox{for every } X \in \vf(J^3\pi) \, .
\end{equation*}
\item In natural coordinates, if $\phi$ is given by $\phi(x^i) = (x^i,u^\alpha)$, then
$j^{3}\phi(x^i) = (x^i,u^\alpha,u^\alpha_i,u^\alpha_I,u^\alpha_J)$ is a solution to the
second-order Euler-Lagrange equations given by \eqref{Chap06_eqn:EulerLagrangeEquations}, that is,
\begin{equation*}
\restric{\derpar{L}{u^\alpha}}{j^3\phi} - \restric{\frac{d}{dx^i} \, \derpar{L}{u^\alpha_i}}{j^3\phi}
+ \restric{\sum_{|I|=2}\frac{d^2}{dx^I} \, \derpar{L}{u^\alpha_I}}{j^3\phi} = 0 \, .
\end{equation*}
\item $\psi_\Lag = j^3\phi$ is a solution to the equation
\begin{equation*}
\inn(\Lambda^m\psi_\Lag^\prime)(\Omega_\Lag \circ \psi_\Lag) = 0 \, ,
\end{equation*}
where $\Lambda^m\psi_\Lag^\prime \colon M \to \Lambda^m\Tan(J^3\pi)$ is the canonical lifting of $\psi_\Lag$.
\item $j^3\phi$ is an integral section of a multivector field contained in a class of locally
decomposable holonomic multivector fields $\{ \X_\Lag \} \subset \vf^{m}(J^3\pi)$ satisfying
equation \eqref{Chap06_eqn:LagFieldEqMultiVF}, that is,
\begin{equation*}
\inn(\X_\Lag)\Omega_\Lag = 0 \, .
\end{equation*}
\end{enumerate}
\end{theorem}


\section{Hamiltonian formalism}
\label{Chap06_sec:HamiltonianFormalism}

In order to describe the Hamiltonian formalism on the basis of the unified one, we must distinguish
between the regular and non-regular cases. In fact, the only ``non-regular'' case that we consider
is the almost-regular one, so we first need to generalize the concept of
\textsl{almost-regular Lagrangian density} to second-order field theories. On the other hand, recall
that the geometric information of the theory is given by the canonical Liouville forms of the
extended $2$-symmetric multimomentum bundle. Hence, we only need to introduce the physical information in the
Hamiltonian formalism from the Hamiltonian $\mu_\W$-section $\hat{h} \in \Gamma(\mu_\W)$
defined in the unified setting.

\subsection{Geometrical setting}

Let $\widetilde{\Leg} \colon J^3\pi \to J^2\pi^\dagger$ be the extended Legendre map obtained in
\eqref{Chap06_eqn:UnifExtendedLegendreMapLocal} and $\Leg \colon J^3\pi \to J^2\pi^\ddagger$ the
restricted Legendre map obtained in \eqref{Chap06_eqn:UnifRestrictedLegendreMapLocal}. Let us denote by
$\widetilde{\P} = \Im(\widetilde{\Leg}) = \widetilde{\Leg}(J^3\pi) \stackrel{\tilde{\jmath}}{\hookrightarrow} J^2\pi^\dagger$
and $\P = \Im(\Leg) = \Leg(J^3\pi) \stackrel{\jmath}{\hookrightarrow} J^2\pi^\ddagger$
the image sets of the extended and restricted Legendre maps, respectively, which we assume to be submanifolds.
We denote $\bar{\pi}_\P \colon \P \to M$ the natural projection, and $\Leg_o$ the map defined by
$\Leg = \jmath \circ \Leg_o$. With these notations, we can give the following definition.

\begin{definition}
A second-order Lagrangian density $\Lag \in \df^{m}(J^2\pi)$ is said to be \textnormal{almost-regular} if
\begin{enumerate}
\item $\P$ is a closed submanifold of $J^2\pi^\ddagger$.
\item $\Leg$ is a submersion onto its image.
\item For every $j^3_x\phi \in J^3\pi$, the fibers $\Leg^{-1}(\Leg(j^3_x\phi))$ are connected submanifolds
of $J^3\pi$.
\end{enumerate}
\end{definition}

If the second-order Lagrangian density is almost-regular, the Legendre map is a submersion onto its
image, and therefore it admits local sections defined on the submanifold $\P \hookrightarrow J^2\pi^\ddagger$.
We denote by $\Gamma_\P(\Leg)$ the set of local sections of $\Leg$ defined on the submanifold $\P$.
Observe that if $\Lag$ is regular, then $\Gamma_\P(\Leg)$ is exactly the set of
local sections of $\Leg$.

As a consequence of Proposition \ref{Chap06_prop:UnifRankBothLegendreMaps}, we have that
$\widetilde{\P}$ is diffeomorphic to $\P$. This diffeomorphism is
$\widetilde{\mu} = \mu \circ \tilde{\jmath} \colon \widetilde{\P} \to \P$. This enables us to state
the following result.

\begin{lemma}\label{Chap06_lemma:HamHamiltonianSection}
If the second-order Lagrangian density $\Lag \in \df^{m}(J^2\pi)$ is, at least, almost-regular, then the Hamiltonian section
$\hat{h} \in \Gamma(\mu_\W)$ induces a Hamiltonian section $h \in \Gamma(\mu)$ defined by
\begin{equation*}
h([\omega]) = (\rho_2 \circ \hat{h}) ([(\rho_2^r)^{-1}(\jmath([\omega]))]) \, , \quad \mbox{for every } [\omega] \in \P \, .
\end{equation*}
\end{lemma}
\begin{proof}
It is clear that, given $[\omega] \in J^{2}\pi^\ddagger$, the section $\hat{h}$ maps every point
$(j^3_x\phi,[\omega]) \in (\rho^r_2)^{-1}([\omega])$ into $\rho_2^{-1}[\rho_2(\hat{h}(j^3_x\phi,[\omega]))]$.
So we have the diagram
\begin{equation*}
\xymatrix{
\widetilde{\P} \ar[rr]^-{\tilde{\jmath}} \ar[d]^-{\tilde{\mu}} & \ & J^{2}\pi^\dagger \ar[d]^-{\mu} & \ & \W \ar[d]_-{\mu_\W} \ar[ll]_-{\rho_2} \\
\P \ar[rr]^-{\jmath} \ar[urr]^-{h}& \ & J^{2}\pi^\ddagger & \ & \W_r \ar@/_0.7pc/[u]_{\hat{h}} \ar[ll]_-{\rho_2^r}
}
\end{equation*}
Thus, the crucial point is the $\rho_2$-projectability of the local function $\hat{H}$. However,
since a local base for $\ker\Tan\rho_2$ is given by \eqref{Chap06_eqn:PremultisymplecticKernelLocal},
then we have that $\hat{H}$ is $\rho_2$-projectable if, and only if,
\begin{equation*}
p_{\alpha}^{I} = \derpar{L}{u_I^\alpha} \, .
\end{equation*}
This condition is fulfilled when $[\omega] \in \P = \Im(\Leg)$, which implies that
$\rho_2[\hat{h}((\rho_2^r)^{-1}([\omega]))] \in \widetilde{\P}$.
\end{proof}

As in the unified setting, this Hamiltonian $\mu$-section is specified by a local Hamiltonian function
$H \in \Cinfty(\P)$, that is,
\begin{equation*}
h(x^i,u^\alpha,u_i^\alpha,p_\alpha^i,p_\alpha^I) = (x^i,u^\alpha,u_i^\alpha,-H,p_\alpha^i,p_\alpha^I) \, .
\end{equation*}
Using the Hamiltonian $\mu$-section we define the \textsl{Hamilton-Cartan forms}
$\Theta_h = h^*\Theta_1^s \in \df^{m}(\P)$ and $\Omega_h = h^*\Omega_1^s \in \df^{m+1}(\P)$.
Observe that $\Leg_o^*\Theta_h = \Theta_\Lag$ and $\Leg_o^*\Omega_h = \Omega_\Lag$. Then,
the pair $(\P,\Omega_h)$ is the \textsl{second-order Hamiltonian field theory} associated with
$(\W_r,\Omega_r)$.

\begin{remark}
The Hamiltonian $\mu$-section can be defined in some equivalent ways without passing through the
unified formalism. First, we can define it as $h = \tilde{\jmath} \circ \widetilde{\mu}^{-1}$.
From this, bearing in mind the definition of $\widetilde{\P}$ and $\P$ as the image sets of the
extended and restricted Legendre maps, respectively, we can also define the Hamiltonian $\mu$-section
as $h = \widetilde{\Leg} \circ \gamma$, where $\gamma \in \Gamma_\P(\Leg)$.
\end{remark}

\subsection{Hyperregular and regular Lagrangian densities}

For the sake of simplicity, we assume throughout this Section that the second-order Lagrangian
density $\Lag \in \df^{m}(J^2\pi)$ is hyperregular, and that $\Upsilon \colon J^2\pi^\ddagger \to J^3\pi$
is a global section of $\Leg$. All the results stated also hold for regular Lagrangians,
restricting to the corresponding open sets where the Legendre map admits local sections.

Observe that if the Lagrangian density is hyperregular, then we have $\P = J^2\pi^\ddagger$ and
$\Leg_o = \Leg$. In addition, the local Hamiltonian function associated to the Hamiltonian
$\mu$-section $h$ has the following coordinate expression
\begin{equation}\label{Chap06_eqn:HamRegHamiltonianFunctionLocal}
H(x^i,u^\alpha,u_i^\alpha,p_\alpha^i,p_\alpha^I) = p_\alpha^iu_i^\alpha + p_\alpha^If_I^\alpha
- (\pi_2^3 \circ \Upsilon)^*L \, ,
\end{equation}
where $f_I^\alpha(x^i,u^\alpha,u^\alpha_i,p_\alpha^i,p_\alpha^I) = \Upsilon^*u_I^\alpha$.
Therefore, the Hamilton-Cartan $m$ and $(m+1)$-forms have the following coordinate expression
\begin{align*}
&\Theta_h = - H \d^mx + p_\alpha^i\d u^\alpha \wedge \d^{m-1}x_i + \frac{1}{n(ij)} \, p_\alpha^{1_i+1_j}\d u_i^\alpha \wedge \d^{m-1}x_j \, , \\
&\Omega_h = \d H \wedge \d^mx - \d p_\alpha^i \wedge \d u^\alpha \wedge \d^{m-1}x_i
- \frac{1}{n(ij)} \, \d p_\alpha^{1_i+1_j} \wedge \d u_i^\alpha \wedge \d^{m-1}x_j \, .
\end{align*}
In addition, since $\Im(\Leg) = J^2\pi^\ddagger$, then the Hamiltonian sections $h$ and $\hat{h}$
satisfy $h \circ \rho_2^r = \rho_2 \circ \hat{h}$, that is, the following diagram commutes
\begin{equation*}
\xymatrix{
\W \ar[drr]^{\rho_2} & \ & \ \\
\W_r \ar[u]^{\hat{h}} \ar[drr]_{\rho_2^r} & \ & J^2\pi^\dagger \\
\ & \ & J^2\pi^\ddagger \ar[u]_{h}
}
\end{equation*}

In addition to the previous comments, in the hyperregular case we can give the following result
on the $1$-nondegeneracy of the Hamilton-Cartan $(m+1)$-form.

\begin{proposition}\label{Chap06_prop:HamRegHamiltonCartanFormMultisymplectic}
If the Lagrangian density is hyperregular, then the Hamilton-Cartan $(m+1)$-form
$\Omega_h = h^*\Omega_1^s \in \df^{m+1}(J^2\pi^\ddagger)$ is a multisymplectic form in $J^2\pi^\ddagger$.
\end{proposition}
\begin{proof}
A direct computation in coordinates leads to this result. Let $\Upsilon \in \Gamma(\Leg)$ be a
global section of the restricted Legendre map, and assume that the local Hamiltonian function $H$
is given locally by \eqref{Chap06_eqn:HamRegHamiltonianFunctionLocal}. Then we have the following
local expression for $\d H$
\begin{equation*}
\d H = - \derpar{L}{x^i}\,\d x^i - \derpar{L}{u^\alpha}\,\d u^\alpha
+ \left( p_\alpha^i - \derpar{L}{u_i^\alpha} \right)\d u_i^\alpha
+ \left( p_\alpha^I - \derpar{L}{u_I^\alpha} \right)\d f_I^\alpha
+ u_i^\alpha\d p_\alpha^i + f_I^\alpha \d p_\alpha^I \, ,
\end{equation*}
where
\begin{equation*}
\d f_I^\alpha = \derpar{f_I^\alpha}{x^j}\,\d x^j + \derpar{f_I^\alpha}{u^\beta}\,\d u^\beta
+ \derpar{f_I^\alpha}{u^\beta_j}\,\d u^\beta_j + \derpar{f_I^\alpha}{p_\beta^j}\,\d p^j_\beta
+ \derpar{f_I^\alpha}{p^K_\beta}\,\d p^K_\beta \, .
\end{equation*}
Observe that since $H$ takes values in $J^2\pi^\ddagger = \Im(\Leg)$, we have
$p_\alpha^I - \partial L / \partial u_I^\alpha = 0$. Thus, the expression of $\d H$ becomes
\begin{equation*}
\d H = - \derpar{L}{x^i}\,\d x^i - \derpar{L}{u^\alpha}\,\d u^\alpha
+ \left( p_\alpha^i - \derpar{L}{u_i^\alpha} \right)\d u_i^\alpha
+ u_i^\alpha\d p_\alpha^i + f_I^\alpha \d p_\alpha^I \, ,
\end{equation*}
and therefore the Hamilton-Cartan $(m+1)$-form is locally given by
\begin{align*}
\Omega_h &= \left[ - \derpar{L}{u^\alpha}\,\d u^\alpha
+ \left( p_\alpha^i - \derpar{L}{u_i^\alpha} \right)\d u_i^\alpha
+ u_i^\alpha\d p_\alpha^i + f_I^\alpha \d p_\alpha^I \right] \wedge \d^mx \\
&\qquad {}- \d p_\alpha^i \wedge \d u^\alpha \wedge \d^{m-1}x_i - \frac{1}{n(ij)} \, \d p_\alpha^{1_i+1_j} \wedge \d u_i^\alpha \wedge \d^{m-1}x_j \, .
\end{align*}
Now, since the $\Cinfty(J^2\pi^\ddagger)$-module of vector fields $\vf(J^2\pi^\ddagger)$ admits
the following local base
\begin{equation*}
\vf(J^2\pi^\ddagger) = \left\langle \derpar{}{x^i} \, , \,
\derpar{}{u^\alpha} \, , \, \derpar{}{u^\alpha_i} \, , \,
\derpar{}{p_\alpha^i} \, , \, \derpar{}{p_\alpha^I} \right\rangle \, ,
\end{equation*}
then the contraction of $\Omega_h$ with every vector field in the local base of $\vf(J^{2}\pi^\ddagger)$
gives the following $m$-forms
\begin{align*}
&\inn\left(\derpar{}{x^k}\right)\Omega_h =
- \d H \wedge \d^{m-1}x_k - \d p_\alpha^i \wedge \d u^\alpha \wedge \d^{m-2}x_{ik}
- \frac{1}{n(ij)}\d p_\alpha^{1_i+1_j} \wedge \d u_i^\alpha \wedge \d^{m-2}x_{jk} \, , \\
&\inn\left(\derpar{}{u^\alpha}\right)\Omega_h =
- \derpar{L}{u^\alpha} \d^{m}x + \d p_\alpha^i  \wedge \d^{m-1}x_{i} \, , \\
&\inn\left(\derpar{}{u^\alpha_i}\right)\Omega_h =
\left( p_\alpha^i - \derpar{L}{u_i^\alpha} \right) \d^{m}x + \frac{1}{n(ij)}\d p_\alpha^{1_i+1_j} \wedge \d^{m-1}x_{j} \, , \\
&\inn\left(\derpar{}{p_\alpha^i}\right)\Omega_h =
u_i^\alpha \d^{m}x - \d u^\alpha \wedge \d^{m-1}x_{i}\, , \\
&\inn\left(\derpar{}{p_\alpha^I}\right)\Omega_h =
f_I^\alpha \d^{m}x - \sum_{1_i+1_j=I} \frac{1}{n(ij)} \d u_i^\alpha \wedge \d^{m-1}x_{j} \, .
\end{align*}
From this it is clear that $\inn(X)\Omega_h = 0$ if, and only if, $X = 0$, that is, $\Omega_h$
is multisymplectic.
\end{proof}

As it has been pointed out in Section \ref{Chap06_sec:LagrangianFormalismGeometricalSetting},
the Poincar\'{e}-Cartan $(m+1)$-form can not be multisymplectic in $J^3\pi$, due to the fact
that the restricted Legendre map is, at the best, a submersion onto $J^2\pi^\ddagger$.
Nevertheless, the restriction of the form $\Omega_\Lag$ to some submanifold can be multisymplectic,
as we show in the following result, which is a direct consequence of Proposition
\ref{Chap06_prop:HamRegHamiltonCartanFormMultisymplectic}.

\begin{corollary}
Let $\Lag \in \df^{m}(J^2\pi)$ be a second-order hyperregular Lagrangian density,
$\Upsilon \in \Gamma(\Leg)$ a global section of $\Leg$, and $\Im(\Upsilon) \hookrightarrow J^3\pi$
the image set of $\Upsilon$, whose natural embedding is canonically identified with $\Upsilon$.
Then the $(m+1)$-form $\Upsilon^*\Omega_\Lag \in \df^{m+1}(\Im(\Upsilon))$ is  a multisymplectic
form in $\Im(\Upsilon)$.
\end{corollary}
\begin{proof}
From Definition \ref{Chap06_def:PoincareCartanForms} we have
\begin{equation*}
\Upsilon^*\Omega_\Lag = \Upsilon^*(\widetilde{\Leg}^*\Omega_1^s)
= (\widetilde{\Leg} \circ \Upsilon)^*\Omega_1^s = h^*\Omega_1^s = \Omega_h \, .
\end{equation*}
Then, since $\Lag$ is a second-order hyperregular Lagrangian density, from Proposition
\ref{Chap06_prop:HamRegHamiltonCartanFormMultisymplectic} we have that $\Omega_h$ is multisymplectic
in $J^2\pi^\ddagger$. Therefore, $\Upsilon^*\Omega_\Lag$ is multisymplectic in $\Im(\Upsilon)$.
\end{proof}

\subsubsection{Field equations for sections}

As in Section \ref{Chap06_sec:LagrangianFormalism}, using the results given in previous Sections,
we can now state the Hamiltonian field equation for sections in the hyperregular case, and recover the
Hamiltonian solutions in $J^{2}\pi^\ddagger$ from the solutions in the unified setting.

The \textsl{second-order (hyperregular) Hamiltonian problem for sections} associated with the
Hamiltonian system $(J^2\pi^\ddagger,\Omega_h)$ consists in finding sections
$\psi_h \in \Gamma(\bar{\pi}_{J^{1}\pi}^\ddagger)$ characterized by the equation
\begin{equation}\label{Chap06_eqn:HamRegFieldEqSect}
\psi_h^*\inn(X)\Omega_h = 0 \, , \quad \mbox{for every } X \in \vf(J^{2}\pi^\ddagger) \, .
\end{equation}

\begin{proposition}\label{Chap06_prop:UnifiedToHamiltonianRegSect}
Let $\psi \in \Gamma(\rho_M^r)$ be a section solution to the equation \eqref{Chap06_eqn:UnifFieldEqSect}.
Then the section
$\psi_h = \rho_2^r \circ \psi \in \Gamma(\bar{\pi}_{J^1\pi}^\ddagger)$ is a solution to the equation
\eqref{Chap06_eqn:HamRegFieldEqSect}.
\end{proposition}
\begin{proof}
The proof of this result is analogous to the proof of Proposition \ref{Chap06_prop:UnifToLagSect}.
In particular, since the map $\rho_2^r \colon \W_r \to J^2\pi^\ddagger$ is a submersion, for every vector field
$X \in \vf(J^2\pi^\ddagger)$ there exist some vector fields $Y \in \vf(\W_r)$ such that $X$ and $Y$
are $\rho_2^r$-related. Observe that this vector field $Y$ is not unique, as the vector field $Y + Y_o$,
with $Y_o \in \ker\Tan\rho_2^r$, is also $\rho_2^r$-related with $X$. Thus, using this particular
choice of $\rho_2^r$-related vector fields, we have
\begin{align*}
\psi_h^*\inn(X)\Omega_h &= (\rho_2^r \circ \psi)^*\inn(X)\Omega_h
= \psi^*((\rho_2^r)^*\inn(X)\Omega_h) = \psi^*\inn(Y)(\rho_2^r)^*\Omega_h \\
&= \psi^*\inn(Y)(h \circ \rho_2^r)^*\Omega_1^s = \psi^*\inn(Y)(\rho_2 \circ \hat{h})^*\Omega_1^s
= \psi^*\inn(Y)\Omega_r \, .
\end{align*}
Since the equality $\psi^*\inn(Y)\Omega_r = 0$ holds for every $Y \in \vf(\W_r)$, in particular
it holds for every $Y \in \vf(\W_r)$ which is $\rho_2^r$-related with $X \in \vf(J^2\pi^\ddagger)$.
Hence we obtain
\begin{equation*}
\psi_h^*\inn(X)\Omega_h = \psi^*\inn(Y)\Omega_r = 0 \, . \qedhere
\end{equation*}
\end{proof}

The diagram illustrating the situation of the above Proposition is the following
\begin{equation*}
\xymatrix{
\W_r \ar[drr]^{\rho^r_2} \ar[dd]^{\rho^r_M} & \ & \\
\ & \ & J^2\pi^\ddagger \ar[dll]_{\bar{\pi}_{J^1\pi}^\ddagger} \\
M \ar@/_1pc/@{-->}[urr]_{\psi_h = \rho_2^r \circ \psi} \ar@/^1pc/[uu]^{\psi} & \ & \
}
\end{equation*}

Observe that, as in the Lagrangian formalism described in Section \ref{Chap06_sec:LagrangianFormalism},
Proposition \ref{Chap06_prop:UnifiedToHamiltonianRegSect} states that every section solution to the field
equation in the unified formalism projects to a section solution to the field equation in the
Hamiltonian formalism, but it does not establish a correspondence between the solutions. As in the
Lagrangian setting, this correspondence does exist, but in this formulation it is not one-to-one.
This is due to the fact that the map $\rho_2^\Lag \colon \W_\Lag \to J^2\pi^\ddagger$ is a submersion,
and not a diffeomorphism, as in the Lagrangian setting. This implies that for every section $\psi_h$
solution to the Hamiltonian field equation there are several sections solution to the field equation
in the unified formalism that project to $\psi_h$.

\begin{proposition}\label{Chap06_prop:HamToUnifSectReg}
Let $\Lag \in \df^{m}(J^2\pi)$ be a second-order hyperregular Lagrangian density, and
$\psi_h \in \Gamma(\bar{\pi}_{J^1\pi}^\ddagger)$ a section solution to the field equation
\eqref{Chap06_eqn:HamRegFieldEqSect}. Then the section
$\psi = \sigma \circ \psi_h \in \Gamma(\rho_M^r)$
is a solution to the equation \eqref{Chap06_eqn:UnifFieldEqSect},
where $\sigma \in \Gamma(\rho_2^r)$ is a global section of the projection $\rho_2^r$.
\end{proposition}
\begin{proof}
First, let us prove that the global section $\sigma \in \Gamma(\rho_2^r)$ does exist.
As the second-order Lagrangian density is hyperregular, there exists a global section of $\Leg$,
which we denote by $\Upsilon \in \Gamma(\Leg)$.
Then, we define $\sigma = j_\Lag \circ (\rho_1^\Lag)^{-1} \circ \Upsilon$.
This map $\sigma \colon J^2\pi^\ddagger \to \W_r$ is a section of the projection $\rho_2^r$,
since we have
\begin{equation*}
\rho_2^r \circ \sigma = \rho_2^r \circ j_\Lag \circ (\rho_1^\Lag)^{-1} \circ \Upsilon
= \rho_2^\Lag \circ (\rho_1^\Lag)^{-1} \circ \Upsilon = \Leg \circ \Upsilon = \Id_{J^2\pi^\ddagger} \, .
\end{equation*}
Moreover, as $\Upsilon$ is a global section of $\Leg$, $\rho_1^\Lag$ is a diffeomorphism, and
$j_\Lag$ is an embedding, we deduce that $\sigma$ is a global section of $\rho_2^r$.

Now we prove that $\psi$ is a solution to equation \eqref{Chap06_eqn:UnifFieldEqSect}. Computing,
\begin{equation*}
\psi^*\inn(X)\Omega_r = (\sigma \circ \psi_h)^*\inn(X)\Omega_r
= \psi_h^*\inn(Y)\Omega_h \, ,
\end{equation*}
where $Y \in \vf(J^2\pi^\ddagger)$ is a vector field $\rho_2^r$-related with $X$, and we have used
that $\Omega_r = (\rho_2^r)^*\Omega_h$ implies $\sigma^*\Omega_r = \sigma^*((\rho_2^r)^*\Omega_h) =
(\rho_2^r \circ \sigma)^*\Omega_h = \Omega_h$.
\end{proof}

Let us compute the local equations for the section
$\psi_h = \rho_2^r \circ \psi \in \Gamma(\bar{\pi}_{J^1\pi}^\ddagger)$. If the section
$\psi \in \Gamma(\rho_M^r)$ is locally given by
$\psi(x^i) = (x^i,u^\alpha,u^\alpha_i,u^\alpha_I,u^\alpha_J,p_\alpha^i,p_\alpha^I)$,
then the section $\psi_h = \rho_2^r \circ \psi$ is given in coordinates by
$\psi(x^i) = (x^i,u^\alpha,u^\alpha_i,p_\alpha^i,p_\alpha^I)$. Now, bearing in mind that the
section $\psi$ solution to the equation \eqref{Chap06_eqn:UnifFieldEqSect} must satisfy the local equations
\eqref{Chap06_eqn:UnifFieldEqSectLocal}, \eqref{Chap06_eqn:UnifFieldEqSectRelationMomenta} and
\eqref{Chap06_eqn:UnifFieldEqSectHolonomy}, that the section $\psi$ takes values in the submanifold
$\W_\Lag \cong \graph(\Leg)$, and the local expression
\eqref{Chap06_eqn:HamRegHamiltonianFunctionLocal} of the Hamiltonian function $H$ in the hyperregular
case, we obtain the following system of partial differential equations for the section $\psi_h$
\begin{equation}\label{Chap06_eqn:HamiltonEquationsReg}
\derpar{u^\alpha}{x^i} = \derpar{H}{p_\alpha^i} \quad ; \quad
\sum_{1_i+1_j=I}\frac{1}{n(ij)}\,\derpar{u_i^\alpha}{x^j} = \derpar{H}{p_\alpha^I} \quad ; \quad
\sum_{i=1}^m\derpar{p_\alpha^i}{x^i} = -\derpar{H}{u^\alpha} \quad ; \quad
\sum_{j=1}^m\derpar{p_\alpha^{1_i+1_j}}{x^j} = -\derpar{H}{u_i^\alpha} \, .
\end{equation}

\subsubsection{Field equations for multivector fields}

Next, using the results stated at the beginning of this Section, we can now state the Hamiltonian
field equation for a multivector field, and recover the Hamiltonian solutions from the solutions
to the field equation \eqref{Chap06_eqn:UnifFieldEqMultiVF} in the unified formalism.

The \textsl{second-order (hyperregular) Hamiltonian problem for multivector fields} associated with
the system $(J^2\pi^\ddagger,\Omega_h)$ consists in finding a class of locally decomposable,
integrable and $(\bar{\pi}_{J^1\pi}^\ddagger)$-transverse multivector
fields $\{\X_h\} \subset \vf^m(J^2\pi^\ddagger)$ satisfying the following field equation
\begin{equation}\label{Chap06_eqn:HamRegFieldEqMultiVF}
\inn(\X_h)\Omega_h = 0 \ , \quad \mbox{for every } \X_h \in \{\X_h\} \subseteq \vf^{m}(J^2\pi^\ddagger) \, .
\end{equation}

In order to recover the solutions to the field equation for multivector fields, we first need the
following technical result, which is similar to Lemma \ref{Chap06_lemma:LagRelatedMultiVF}.

\begin{lemma}\label{Chap06_lemma:HamRelatedMultiVFReg}
Let $\X \in \vf^{m}(\W_r)$ be a multivector field tangent to $\W_\Lag \hookrightarrow \W_r$.
Then there exists a unique multivector field $\X_h \in \vf^m(J^2\pi^\ddagger)$ such that
$\X_h \circ \rho_2^r \circ j_\Lag = \Lambda^m\Tan\rho_2^r \circ \X \circ j_\Lag$.

\noindent Conversely, if $\X_h \in \vf^m(J^2\pi^\ddagger)$, then there exist multivector fields
$\X \in \vf^m(\W_r)$ tangent to $\W_\Lag$ such that
$\X_h \circ \rho_2^r \circ j_\Lag = \Lambda^m\Tan\rho_2^r \circ \X \circ j_\Lag$.
\end{lemma}
\begin{proof}
The proof of this result is analogous to the proof of Lemma \ref{Chap06_lemma:LagRelatedMultiVF},
bearing in mind that the map $\rho_2^\Lag = \rho_2^r \circ j_\Lag \colon \W_\Lag \to J^2\pi^\ddagger$
is a submersion onto $J^2\pi^\ddagger$. In particular, since the multivector field $\X$ is tangent to
$\W_\Lag$, there exists a unique multivector field $\X_o \in \vf^{m}(\W_\Lag)$ which is
$j_\Lag$-related to $\X$, that is, $\Lambda^m\Tan j_\Lag \circ \X_o = \X \circ j_\Lag$. On the
other hand, as $\rho_2^\Lag \colon \W_\Lag \to J^2\pi^\ddagger$ is a submersion, there is a unique
multivector field $\X_h \in \vf^{m}(J^2\pi^\ddagger)$ which is $\rho_2^\Lag$-related to $\X_o$; that
is, $\X_h \circ \rho_2^\Lag = \Lambda^m\Tan \rho_1^\Lag \circ \X_o$. Then, computing, we have
\begin{equation*}
\X_h \circ \rho_2^r \circ j_\Lag = \X_h \circ \rho_2^\Lag
= \Lambda^m\Tan\rho_2^\Lag \circ \X_o
= \Lambda^m\Tan\rho_2^r \circ \Lambda^m\Tan j_\Lag \circ \X_o
= \Lambda^m\Tan\rho_2^r \circ \X \circ j_\Lag \, .
\end{equation*}
The converse is proved reversing this reasoning, but now the multivector field
$\X_o \in \vf^m(\W_\Lag)$ which is $\rho_2^\Lag$-related with the given $\X_h \in \vf^m(J^2\pi^\ddagger)$
is not unique, since $\rho_2^\Lag$ is a submersion with $\ker\Tan\rho_2^\Lag \neq \{ 0 \}$.
\end{proof}

As in the Lagrangian formalism, the previous result gives a correspondence between the set of
multivector fields $\X \in \vf^m(\W_r)$ tangent to $\W_\Lag$ and the set of multivector fields
$\X_h \in \vf^{m}(J^2\pi^\ddagger)$ such that the following diagram is commutative
\begin{equation*}
\xymatrix{
\Lambda^m\Tan\W_r \ar[drrr]^{\Lambda^m\Tan\rho_2^r} & \ & \ & \ \\
\Lambda^m\Tan\W_\Lag \ar[rrr]_{\Lambda^m\Tan\rho_2^\Lag} \ar@{_{(}->}[u]_{\Lambda^m\Tan j_\Lag} & \ & \ & \Lambda^m\Tan(J^2\pi^\ddagger) \\
\ & \ & \ & \ \\
\W_r \ar[drrr]^{\rho_2^r} \ar@/^2.25pc/[uuu]^{\X} & \ & \ & \ \\
\W_\Lag \ar[rrr]_{\rho_2^\Lag} \ar@{^{(}->}[u]^{j_\Lag} \ar@/_1.5pc/[uuu]_{\X_o}|(.32)\hole & \ & \ & J^2\pi^\ddagger \ar[uuu]^{\X_h}
}
\end{equation*}
Nevertheless, observe that in the Hamiltonian formalism, the map
$\rho_2^\Lag = \rho_2^r \circ j_\Lag \colon \W_\Lag \to J^2\pi^\ddagger$ is a submersion (instead
of a diffeomorphism, as in the Lagrangian setting), and thus the correspondence is not one-to-one.
In particular, for every multivector field $\X \in \vf^m(\W_r)$ tangent to $\W_\Lag$
we can define a unique multivector field $\X_h \in \vf^m(J^2\pi^\ddagger)$ such that the previous
diagram commutes. But since $\rho_2^\Lag$ is a submersion, for every $\X_h \in \vf^m(J^2\pi^\ddagger)$
there are several multivector fields $\X \in \vf^m(\W_r)$, tangent to $\W_\Lag$, satisfying the same
property.

Using Lemma \ref{Chap06_lemma:HamRelatedMultiVFReg} we can now state the (non-bijective) correspondence
between the multivector fields in $J^2\pi^\ddagger$ solution to equation
\eqref{Chap06_eqn:HamRegFieldEqMultiVF} and the multivector fields in $\W_r$ solution to the
field equation \eqref{Chap06_eqn:UnifFieldEqMultiVF}.

\begin{theorem}\label{Chap06_thm:UnifToHamMultiVFReg}
Let $\X \in \vf^m(\W_r)$ be a locally decomposable, integrable and $\rho_M^r$-transverse
multivector field solution to the equation \eqref{Chap06_eqn:UnifFieldEqMultiVF} and tangent to
$\W_\Lag$. Then there exists a locally decomposable, integrable and
$(\bar{\pi}_{J^1\pi}^\ddagger)$-transverse multivector field $\X_h \in \vf^m(J^2\pi^\ddagger)$
solution to the equation \eqref{Chap06_eqn:HamRegFieldEqMultiVF}.

\noindent Conversely, if $\X_h \in \vf^{m}(J^2\pi^\ddagger)$ is a locally decomposable, integrable and
$(\bar{\pi}_{J^1\pi}^\ddagger)$-transverse multivector field solution to the equation
\eqref{Chap06_eqn:HamRegFieldEqMultiVF}, then there exist locally decomposable, integrable and
$\rho_M^r$-transverse multivector fields $\X \in \vf^{m}(\W_r)$ tangent to $\W_\Lag$ solution to
the equation \eqref{Chap06_eqn:UnifFieldEqMultiVF}.
\end{theorem}
\begin{proof}
The proof of this result is analogous to the proof of Theorem \ref{Chap06_thm:UnifToLagMultiVF}.
In particular, applying Lemma \ref{Chap06_lemma:HamRelatedMultiVFReg}, we have
\begin{equation*}
\restric{\inn(\X)\Omega_r}{\W_\Lag} = \restric{\inn(\X)(\rho_2^r)^*\Omega_h}{\W_\Lag}
= \restric{(\rho_2^r)^*\inn(\X_h)\Omega_h}{\W_\Lag}
= \restric{\inn(\X_h)\Omega_h}{\rho_2^r(\W_\Lag)}
= \restric{\inn(\X_h)\Omega_h}{J^2\pi^\ddagger} \, .
\end{equation*}
Hence, $\X_h$ is a solution to the equation $\inn(\X_h)\Omega_h = 0$ if, and only if,
$\X$ is a solution to the equation $\inn(\X)\Omega_r = 0$.

Observe that, following the same reasoning as above, we have
\begin{align*}
\restric{\inn(\X)(\rho_M^r)^*\eta}{\W_\Lag} &= \restric{\inn(\X)(\bar{\pi}_{J^1\pi}^\ddagger \circ \rho_2^r)^*\eta}{\W_\Lag}
= \restric{(\rho_2^r)^*\inn(\X_h)(\bar{\pi}_{J^1\pi}^\ddagger)^*\eta}{\W_\Lag} \\
&= \restric{\inn(\X_h)(\bar{\pi}^3)^*\eta}{\rho_1^r(\W_\Lag)}
= \restric{\inn(\X_h)(\bar{\pi}_{J^1\pi}^\ddagger)^*\eta}{J^2\pi^\ddagger} \, .
\end{align*}
Hence, $\X_h$ is $\bar{\pi}_{J^1\pi}^\ddagger$-transverse if, and only if, $\X$ is $\rho_M^r$-transverse.

Now, let us assume that $\X \in \vf^m(\W_r)$ is integrable, and let $\psi \in \Gamma(\rho_M^r)$ be
an integral section of $\X$. Then, the section $\psi_h = \rho_2^r \circ \psi \in \Gamma(\bar{\pi}_{J^1\pi}^\ddagger)$
satisfies
\begin{equation*}
\X_h \circ \psi_h = \X_h \circ \rho_2^r \circ \psi
= \Lambda^m\Tan\rho_2^r \circ \X \circ \psi
= \Lambda^m\Tan\rho_2^r \circ \Lambda^m\psi^\prime
= \Lambda^m\psi_h^\prime \, ,
\end{equation*}
where $\Lambda^m\psi^\prime \colon M \to \Lambda^m\Tan\W_r$ is the canonical lifting of $\psi$ to
$\Lambda^m\Tan\W_r$. That is, $\psi_h$ is an integral section of $\X_h$. Hence, if $\X$ is
integrable, then $\X_h$ is integrable.

For the converse, let us assume that $\X_h \in \vf^m(J^2\pi^\ddagger)$ is integrable, and let
$\psi_h \in \Gamma(\bar{\pi}_{J^1\pi}^\ddagger)$ be an integral section of $\X_h$. Then, the section
$\psi = \sigma \circ \psi_h \in \Gamma(\rho_M^r)$, defined as in Proposition \ref{Chap06_prop:HamToUnifSectReg},
satisfies
\begin{align*}
\X \circ \psi &= \X \circ \sigma \circ \psi_h
= \Lambda^m\Tan \sigma \circ \X_h \circ \psi_h
= \Lambda^m\Tan \sigma \circ \Lambda^m\psi_h^\prime 
= \Lambda^m\psi^\prime \, ,
\end{align*}
where we have used the fact that if the multivector fields $\X_h$ and $\X$ are $\rho_2^r$-related,
then they are also $\sigma$-related. Therefore, if $\X_h$ is integrable, so is $\X$.
\end{proof}

Let $\X_h \in \vf^m(J^2\pi^\ddagger)$ be a locally decomposable multivector field. From the results
given in Section \ref{Chap01_sec:MultivectorFields} and \cite{art:Echeverria_Munoz_Roman98}, we know
that $\X_h$ is given in the natural coordinates of $J^2\pi^\ddagger$ by
\begin{equation}\label{Chap06_eqn:HamGenericMultiVFLocalReg}
\X_h = f \bigwedge_{j=1}^{m}
\left(  \derpar{}{x^j} + F_j^\alpha\derpar{}{u^\alpha} + F_{i,j}^\alpha\derpar{}{u_i^\alpha}
+ G_{\alpha,j}^i\derpar{}{p_\alpha^i} + G_{\alpha,j}^{I}\derpar{}{p_\alpha^{I}} \right) \, .
\end{equation}
Taking $f = 1$ as a representative of the equivalence class, since $\X_h$ is a solution to the equation
\eqref{Chap06_eqn:HamRegFieldEqMultiVF}, we obtain that the local equations for the component functions
of $\X_h$ are
\begin{equation*}
F_j^\alpha = \derpar{H}{p_\alpha^j} \quad ; \quad
\sum_{1_i+1_j=I}\frac{1}{n(ij)}\, F_{i,j}^\alpha = \derpar{H}{p_\alpha^I} \quad ; \quad
\sum_{i=1}^m G^{i}_{\alpha,i} = -\derpar{H}{u^\alpha} \quad ; \quad
\sum_{j=1}^m G^{1_i+1_j}_{\alpha,j} = -\derpar{H}{u_i^\alpha} \, .
\end{equation*}

\subsubsection{Equivalence of the Hamiltonian field equations in the hyperregular case}

Finally, we state the equivalence Theorem for the Hamiltonian formalism in the hyperregular case.
This result is analogous to Theorems \ref{Chap06_thm:UnifEquivalenceTheorem} and
\ref{Chap06_thm:LagEquivalanceTheorem}, and is a direct consequence of Theorems
\ref{Chap06_thm:UnifEquivalenceTheorem} and \ref{Chap06_thm:UnifToHamMultiVFReg},
and of Propositions \ref{Chap06_prop:UnifiedToHamiltonianRegSect} and
\ref{Chap06_prop:HamToUnifSectReg}, and hence we omit the proof.

\begin{theorem}\label{Chap06_thm:HamRegEquivalenceTheorem}
The following assertions on a section $\psi_h \in \Gamma(\bar{\pi}_{J^1\pi}^\ddagger)$
are equivalent.
\begin{enumerate}
\item $\psi_h$ is a solution to equation \eqref{Chap06_eqn:HamRegFieldEqSect}, that is,
\begin{equation*}
\psi_h^*\inn(X)\Omega_h = 0 \, , \quad \mbox{for every } X \in \vf(J^2\pi^\ddagger) \, .
\end{equation*}
\item In natural coordinates, if $\psi_h$ is given by
$\psi_h(x^i) = (x^i,u^\alpha,u_i^\alpha,p_\alpha^i,p_\alpha^I)$, then
its component functions are a solution to the equations \eqref{Chap06_eqn:HamiltonEquationsReg}, that is,
\begin{equation*}
\derpar{u^\alpha}{x^i} = \derpar{H}{p_\alpha^i} \quad ; \quad
\sum_{1_i+1_j=I}\frac{1}{n(ij)}\,\derpar{u_i^\alpha}{x^j} = \derpar{H}{p_\alpha^I} \quad ; \quad
\sum_{i=1}^m\derpar{p_\alpha^i}{x^i} = -\derpar{H}{u^\alpha} \quad ; \quad
\sum_{j=1}^m\derpar{p_\alpha^{1_i+1_j}}{x^j} = -\derpar{H}{u_i^\alpha} \, .
\end{equation*}
\item $\psi_h$ is a solution to the equation
\begin{equation*}
\inn(\Lambda^m\psi_h^\prime)(\Omega_h \circ \psi_h) = 0 \, ,
\end{equation*}
where $\Lambda^m\psi_h^\prime \colon M \to \Lambda^m\Tan(J^2\pi^\ddagger)$ is the canonical lifting of $\psi_h$.
\item $\psi_h$ is an integral section of a multivector field contained in a class of locally
decomposable, integrable and $(\bar{\pi}_{J^1\pi}^\ddagger)$-transverse multivector fields
$\{ \X_h \} \subset \vf^{m}(J^2\pi^\ddagger)$ satisfying equation \eqref{Chap06_eqn:HamRegFieldEqMultiVF},
that is,
\begin{equation*}
\inn(\X_h)\Omega_h = 0 \, .
\end{equation*}
\end{enumerate}
\end{theorem}

\subsection{Singular (almost-regular) Lagrangian densities}

Finally, we study the case of second-order singular Lagrangian densities, although the only
non-regular case that we study is the almost-regular one, since some minimal regularity conditions
must be required to the second-order Lagrangian density in order to give a general framework.
Thus, throughout this Section we assume that the second-order Lagrangian density is, at least,
almost-regular.

Recall that, for almost-regular Lagrangian densities, only in the most favorable cases does there
exist a submanifold $\W_f \hookrightarrow \W_\Lag$ where the field equations can be solved. In this
situation, the solutions in the Hamiltonian formalism cannot be obtained directly from the projection
of the solutions in the unified setting, but rather by passing through the Lagrangian formalism and
using the Legendre map. Recall that, in this case, the phase space of the system is
$\P = \Im(\Leg) \hookrightarrow J^2\pi^\ddagger$.

\subsubsection{Field equations for sections}

As for the hyperregular case, we now state the Hamiltonian field equation for sections in the
almost-regular case, and we recover the Hamiltonian solutions in $\P$ from the solutions in the
unified formalism.

The \textsl{second-order (almost-regular) Hamiltonian problem for sections} associated with the
Hamiltonian system $(\P,\Omega_h)$ consists in finding sections $\psi_h \in \Gamma(\bar{\pi}_\P)$
characterized by the equation
\begin{equation}\label{Chap06_eqn:HamSingFieldEqSect}
\psi_h^*\inn(X)\Omega_h = 0 \, , \quad \mbox{for every } X \in \vf(\P) \, .
\end{equation}

\begin{proposition}\label{Chap06_prop:UnifiedToHamiltonianSingSect}
Let $\Lag \in \df^{m}(J^2\pi)$ be an almost-regular Lagrangian density. Let $\psi \in \Gamma(\rho_M^r)$
be a solution to the equation \eqref{Chap06_eqn:UnifFieldEqSect}. Then, the section
$\psi_h = \Leg_o \circ \rho_1^r \circ \psi = \Leg_o \circ \psi_\Lag \in \Gamma(\bar{\pi}_\P)$ is a solution
to the equation \eqref{Chap06_eqn:HamSingFieldEqSect}.

\noindent Conversely, let $\psi_h \in \Gamma(\bar{\pi}_\P)$ be a solution to equation
\eqref{Chap06_eqn:HamSingFieldEqSect}. Then 
$\psi = j_\Lag \circ (\rho_1^\Lag)^{-1} \circ \gamma \circ \psi_h \in \Gamma(\rho_\R^r)$ is a solution
to the equation \eqref{Chap06_eqn:UnifFieldEqSect}, where $\gamma \in \Gamma_\P(\Leg_o)$.
\end{proposition}
\begin{proof}
Since the Lagrangian density $\Lag$ is assumed to be almost-regular, then the map $\Leg_o$ is a
submersion onto its image, $\P$. Thus, for every vector field $X \in \vf(\P)$ there exist some vector
fields $Y \in \vf(J^3\pi)$ such that $X$ and $Y$ are $\Leg_o$-related. Using this particular choice of
$\Leg_o$-related vector fields, we have
\begin{equation*}
\psi_h^*\inn(X)\Omega_h = (\Leg_o \circ \psi_\Lag)^*\inn(X)\Omega_h
=\psi_\Lag^*(\Leg_o^*\inn(X)\Omega_h) = \psi_\Lag^*\inn(Y)\Leg_o^*\Omega_h
= \psi_\Lag^*\inn(Y)\Omega_\Lag \, .
\end{equation*}
Then we have proved
\begin{equation*}
\psi_h^*\inn(X)\Omega_h = \psi_\Lag^*\inn(Y)\Omega_\Lag = 0 \, ,
\end{equation*}
since the last equality holds for every $Y \in \vf(J^3\pi)$ and, in particular,
for every vector field $\Leg_o$-related to a vector field in $\P$. Therefore,
using Proposition \ref{Chap06_prop:UnifToLagSect}, the result follows.

The converse is proved reversing the reasoning and using Proposition \ref{Chap06_prop:LagToUnifSect},
since $\Leg_o \circ \gamma = \Id_\P$ and, in particular, we have $\gamma^*\Theta_\Lag = \Theta_h$.
\end{proof}

The diagram illustrating this last result is the following.
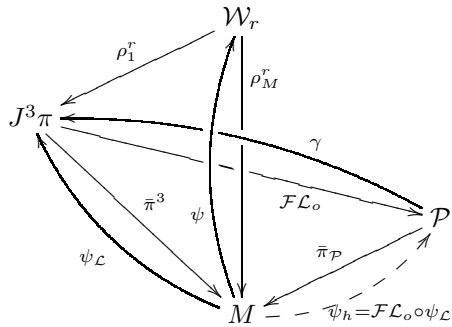
\begin{equation*}
\xymatrix{
\ & \ & \W_r \ar[dll]_{\rho^r_1} \ar[ddd]^<(0.15){\rho^r_M}|(.4){\hole} & \ & \\
J^3\pi \ar[ddrr]^{\bar{\pi}^3} \ar[rrrrd]_<(0.65){\Leg_o}|(.44){\hole}|(.515){\hole} & \ & \ & \ & \ \\
\ & \ & \ & \ & \P \ar[dll]_{\bar{\pi}_\P} \ar@/_1pc/[ullll]_<(0.3){\gamma}|(.58){\hole} \\
\ & \ & M \ar@/^1pc/[uull]^{\psi_\Lag} \ar@/_1pc/@{-->}[urr]_{\psi_h = \Leg_o \circ \psi_\Lag} \ar@/^1pc/[uuu]^<(0.3){\psi} & \ & \
}
\end{equation*}

\subsubsection{Field equations for multivector fields}

Next, we state the Hamiltonian field equation for a multivector field in the almost-regular case,
and recover the Hamiltonian solutions from the solutions to the field equation
\eqref{Chap06_eqn:UnifFieldEqMultiVF} in the unified formalism.

The \textsl{second-order (almost-regular) Hamiltonian problem for multivector fields} associated with
the system $(\P,\Omega_h)$ consists in finding a class of locally decomposable, integrable
and $\bar{\pi}_\P$-transverse multivector
fields $\{\X_h\} \subset \vf^m(\P)$ satisfying the following field equation
\begin{equation}\label{Chap06_eqn:HamSingFieldEqMultiVF}
\inn(\X_h)\Omega_h = 0 \ , \quad \mbox{for every } \X_h \in \{\X_h\} \subseteq \vf^{m}(\P) \, .
\end{equation}

Since the second-order Lagrangian density is almost-regular, assume that there exists a submanifold
$\W_f \hookrightarrow \W_\Lag$ and a multivector field $\X \in \vf^m(\W_r)$, defined at support on
$\W_f$ and tangent to $\W_f$, which is a solution to the equation
\eqref{Chap06_eqn:UnifFieldEqMultiVFSing}. Now consider the submanifolds
$S_f = \rho_1^\Lag(\W_f) \hookrightarrow J^{3}\pi$ and
$\P_f = \Leg(S_f) \hookrightarrow \P \hookrightarrow J^2\pi^\ddagger$.
With these notations,  we can state the following result, which is the analogous theorem to Theorem
\ref{Chap06_thm:UnifToHamMultiVFReg} in the case of almost-regular Lagrangian densities.

\begin{theorem}\label{Chap06_thm:UnifToHamMultiVFSing}
Let $\X \in \vf^m(\W_r)$ be a locally decomposable, integrable and $\rho_M^r$-transverse multivector
field, defined at support on $\W_f$ and tangent to $\W_f$, which is a solution to the equation
\eqref{Chap06_eqn:UnifFieldEqMultiVFSing}. Then there exists a locally decomposable, integrable and
$\bar{\pi}_{\P}$-transverse multivector field $\X_h \in \vf^m(\P)$, defined at support
on $\P_f$ and tangent to $\P_f$, which is a solution to the equation \eqref{Chap06_eqn:HamSingFieldEqMultiVF}.

\noindent Conversely, if $\X_h \in \vf^{m}(\P)$ is a locally decomposable, integrable and
$\bar{\pi}_{\P}$-transverse multivector field defined at support on $\P_f$ and tangent
to $\P_f$ which is a solution to the equation \eqref{Chap06_eqn:HamSingFieldEqMultiVF}, then there exist
locally decomposable, integrable and $\rho_M^r$-transverse multivector fields $\X \in \vf^{m}(\W_r)$,
defined at support on $\W_f$ and tangent to $\W_f$, which are solutions to the equation
\eqref{Chap06_eqn:UnifFieldEqMultiVFSing}.
\end{theorem}
\begin{proof}
Using Theorem \ref{Chap06_thm:UnifToLagMultiVF}, there is a one-to-one correspondence between
holonomic multivector fields $\X_\Lag \in \vf^m(J^{3}\pi)$ solution to the field equation
\eqref{Chap06_eqn:LagFieldEqMultiVF} (at least on the points of a submanifold $S_f \hookrightarrow J^{3}\pi$)
and holonomic multivector fields $\X \in \vf^m(\W_r)$, tangent to $\W_\Lag$, solution to equations
\eqref{Chap06_eqn:UnifFieldEqMultiVF} (at least on the points of a submanifold $\W_f \hookrightarrow \W_r)$.
Hence, it suffices to prove that we can establish a correspondence between multivector fields in $J^3\pi$
solution to the Lagrangian field equation, and multivector fields in $\P$ solution to the Hamiltonian field equation.

Since the Lagrangian density is almost-regular, the map $\Leg_o$ is a surjective submersion on $\P$. Hence,
for every $\X_h \in \vf^m(\P)$ there exist some $\X_\Lag \in \vf^m(J^{3}\pi)$ (not necessarily unique) such
that $\X_h$ and $\X_\Lag$ are $\Leg_o$-related, that is, $\X_h \circ \Leg_o = \Lambda^m\Tan\Leg_o \circ \X_\Lag$.
And, conversely, for every multivector field $\X_\Lag \in \vf^m(J^{3}\pi)$, there exists a multivector field
$\X_h \in \vf^m(\P)$ which is $\Leg_o$-related with $X_\Lag$. Using this particular choice of $\Leg_o$-related
multivector fields, we have
\begin{equation*}
\inn(\X_\Lag)\Omega_\Lag = \inn(\X_\Lag)\Leg_o^*\Omega_h = \Leg_o^*\inn(\X_h)\Omega_h =
\restric{\inn(\X_h)\Omega_h}{\Leg_o(J^{3}\pi)} = \restric{\inn(\X_h)\Omega_h}{\P} \, ,
\end{equation*}
since $\Leg_o$ is a surjective submersion on $\P$.
The converse is immediate, reversing this reasoning. Hence, we have proved that $\inn(\X_\Lag)\Omega_\Lag = 0$
is equivalent to $\inn(\X_h)\Omega_h = 0$ whenever $\X_\Lag$ and $\X_h$ are $\Leg_o$-related. The same reasoning
proves that $\inn(\X_\Lag)(\bar{\pi}^{3})^*\eta \neq 0$ is equivalent to $\inn(\X_h)\bar{\pi}_{\P}^*\eta \neq 0$.
Observe that the reasoning remains the same replacing $J^{3}\pi$ by $S_f$ and $\P$ by $\P_f$.

Now, let us assume that $\X_\Lag \in \vf^m(J^3\pi)$ is integrable, and let $\psi_\Lag \in \Gamma(\bar{\pi}^3)$ be
an integral section of $\X_\Lag$. Then, the section $\psi_h = \Leg_o \circ \psi_\Lag \in \Gamma(\bar{\pi}_\P)$
satisfies
\begin{equation*}
\X_h \circ \psi_h = \X_h \circ \Leg_o \circ \psi_\Lag
= \Lambda^m\Tan\Leg_o \circ \X_\Lag \circ \psi_\Lag
= \Lambda^m\Tan\Leg_o \circ \Lambda^m\psi^\prime_\Lag
= \Lambda^m\psi_h^\prime \, .
\end{equation*}
That is, $\psi_h$ is an integral section of $\X_h$. Hence, if $\X_\Lag$ is integrable, then $\X_h$ is
integrable.

For the converse, let us assume that $\X_h \in \vf^m(\P)$ is integrable, and let
$\psi_h \in \Gamma(\bar{\pi}_\P)$ be an integral section of $\X_h$. Then, the section
$\psi_\Lag = \gamma \circ \psi_h \in \Gamma(\rho_M^r)$, with $\gamma \in \Gamma_\P(\Leg_o)$ satisfies
\begin{align*}
\X_\Lag \circ \psi_\Lag &= \X_\Lag \circ \gamma \circ \psi_h
= \Lambda^m\Tan \gamma \circ \X_h \circ \psi_h
= \Lambda^m\Tan \gamma \circ \Lambda^m\psi_h^\prime 
= \Lambda^m\psi_\Lag^\prime \, ,
\end{align*}
where we have used the fact that if the multivector fields $\X_h$ and $\X_\Lag$ are $\Leg_o$-related,
then they are also $\gamma$-related. Therefore, if $\X_h$ is integrable, so is $\X_\Lag$.
\end{proof}

The diagram that illustrates the situation of the previous Theorem is the following
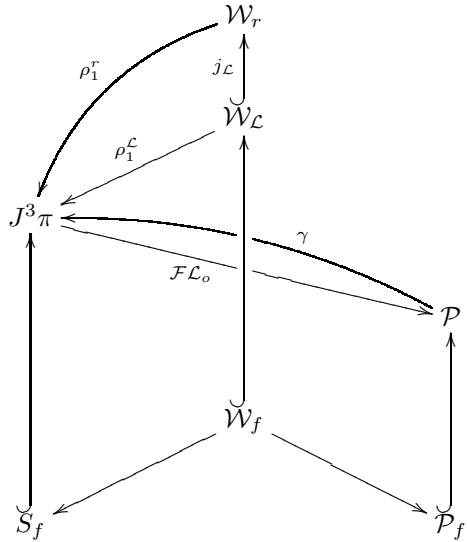
\begin{equation*}
\xymatrix{
\ & \ & \W_r \ar@/_1.3pc/[ddll]_{\rho_1^r} & \ & \ \\
\ & \ & \W_\Lag \ar[dll]_{\rho_1^\Lag} \ar@{^{(}->}[u]^{j_\Lag} & \ & \ \\
J^3\pi \ar[drrrr]_<(0.35){\Leg_o}|(.51){\hole} & \ & \ & \ & \ \\
\ & \ & \ & \ & \P \ar@/_1pc/[ullll]_<(0.35){\gamma}|(.51){\hole} \\
\ & \ & \W_f \ar@{^{(}->}[uuu] \ar[dll] \ar[drr] & \ & \ \\
S_f \ar@{^{(}->}[uuu] & \ & \ & \ & \P_f \ar@{^{(}->}[uu] \\
}
\end{equation*}

\subsubsection{Equivalence of the Hamiltonian field equations in the almost-regular case}

Finally, we state the equivalence Theorem for the Hamiltonian formalism in the almost-regular case,
which is the analogous to Theorems \ref{Chap06_thm:UnifEquivalenceTheorem},
\ref{Chap06_thm:LagEquivalanceTheorem} and \ref{Chap06_thm:HamRegEquivalenceTheorem} in the almost-regular
setting. Since this result is a straightforward consequence of Theorems
\ref{Chap06_thm:UnifEquivalenceTheorem} and \ref{Chap06_thm:UnifToHamMultiVFSing},
and of Proposition \ref{Chap06_prop:UnifiedToHamiltonianSingSect}, we omit the proof.

\begin{theorem}\label{Chap06_thm:HamSingEquivalenceTheorem}
The following assertions on a section $\psi_h \in \Gamma(\bar{\pi}_{\P})$ are equivalent.
\begin{enumerate}
\item $\psi_h$ is a solution to equation \eqref{Chap06_eqn:HamSingFieldEqSect}, that is,
\begin{equation*}
\psi_h^*\inn(X)\Omega_h = 0 \, , \quad \mbox{for every } X \in \vf(\P) \, .
\end{equation*}
\item $\psi_h$ is a solution to the equation
\begin{equation*}
\inn(\Lambda^m\psi_h^\prime)(\Omega_h \circ \psi_h) = 0 \, ,
\end{equation*}
where $\Lambda^m\psi_h^\prime \colon M \to \Lambda^m\Tan\P$ is the canonical lifting of $\psi_h$.
\item $\psi_h$ is an integral section of a multivector field contained in a class of locally
decomposable, integrable and $\bar{\pi}_\P$-transverse multivector fields
$\{ \X_h \} \subset \vf^{m}(\P)$ satisfying equation \eqref{Chap06_eqn:HamSingFieldEqMultiVF},
that is,
\begin{equation*}
\inn(\X_h)\Omega_h = 0 \, .
\end{equation*}
\end{enumerate}
\end{theorem}


\section{Examples}
\label{Chap06_sec:Examples}

In this Section, two physical models are analyzed as examples to show the application of the
formalisms. The first example is a regular field theory, the \textsl{bending of a loaded and clamped
plate}, while the second is a singular one, the well-known \textsl{Korteweg-de Vries equation}.

\subsection{Loaded and clamped plate}
\label{Chap06_exa:LoadedClampedPlate}

Let us consider a plate with clamped edges. We wish to determine the bending (or deflection)
perpendicular to the plane of the plate under the action of an external force given by a uniform
load. This system has been studied using a previous version of the unified formalism in
\cite{art:Campos_DeLeon_Martin_Vankerschaver09}, and can be modeled as a second-order field theory,
taking $M = \R^2$ as the base manifold (the plate) and the ``vertical'' bending as a fiber bundle
$E = \R^2 \times \R \stackrel{\pi}{\longrightarrow} \R^2$ (that is, the fibers are $1$-dimensional).

We consider in $M = \R^2$ the canonical coordinates $(x,y)$ of the Euclidean plane, and in $E = \R^3$
we take the global coordinates $(x,y,u)$ adapted to the bundle structure. Recall that $\R^2$ admits
a canonical volume form $\eta = \d x \wedge \d y \in \df^{2}(\R^2)$.

\begin{remark}
Note that this is the ``smaller'' higher-order field theory that can be considered: dimension $2$
in the base manifold, $1$-dimensional fibers and second-order.
\end{remark}

In the induced coordinates $(x,y,u,u_1,u_2,u_{(2,0)},u_{(1,1)},u_{(0,2)})$ of $J^2\pi$, the second-order
Lagrangian density $\Lag \in \df^{2}(J^2\pi)$ for this field theory is given by
\begin{equation*}
\Lag = \frac{1}{2}( u_{(2,0)}^2 + 2u_{(1,1)}^2 + u_{(0,2)}^2 - 2qu ) \, \d x \wedge \d y \, ,
\end{equation*}
where $q \in \R$ is a constant modeling the uniform load on the plate.

\subsubsection{Lagrangian-Hamiltonian formalism}

Following Section \ref{Chap06_sec:UnifiedFormalismSetting}, we consider the fiber bundles
\begin{equation*}
\W = J^3\pi \times_{J^1\pi} J^2\pi^\dagger \quad ; \quad
\W_r = J^3\pi \times_{J^1\pi} J^2\pi^\ddagger \, ,
\end{equation*}
with the natural coordinates introduced in the aforementioned Section, which are
\begin{equation}\label{Chap06_eqn:ExampleRegular_LocalCoordinatesW}
(x,y,u,u_1,u_2,u_{(2,0)},u_{(1,1)},u_{(0,2)},u_{(3,0)},u_{(2,1)},u_{(1,2)},u_{(0,3)},p,p^1,p^2,p^{(2,0)},p^{(1,1)},p^{(0,2)}) \, ,
\end{equation}
and
\begin{equation}\label{Chap06_eqn:ExampleRegular_LocalCoordinatesWr}
(x,y,u,u_1,u_2,u_{(2,0)},u_{(1,1)},u_{(0,2)},u_{(3,0)},u_{(2,1)},u_{(1,2)},u_{(0,3)},p^1,p^2,p^{(2,0)},p^{(1,1)},p^{(0,2)}) \, ,
\end{equation}
respectively. Observe that, in this example, we have $\dim J^3\pi = 12$ and $\dim J^2\pi^\ddagger = 10$,
and therefore $\dim\W = 18$ and $\dim\W_r = 17$.

The Hamiltonian $\mu_\W$-section $\hat{h} \in \Gamma(\mu_\W)$ is specified by the local Hamiltonian
function, whose coordinate expression \eqref{Chap06_eqn:UnifHamiltonianFunctionLocal} in this case is
\begin{equation}\label{Chap06_eqn:ExampleRegular_UnifHamiltonianLocal}
\hat{H} = p^1u_1 + p^2u_2 + p^{(2,0)}u_{(2,0)} + p^{(1,1)}u_{(1,1)} + p^{(0,2)}u_{(0,2)}
- \frac{1}{2}u_{(2,0)}^2 - u_{(1,1)}^2 - \frac{1}{2}u_{(0,2)}^2 + qu \, ,
\end{equation}
and the canonical forms in $\W_r$ are given by
\begin{equation}\label{Chap06_eqn:ExampleRegular_UnifiedCanonicalForms}
\begin{array}{l}
\displaystyle \Theta_r = -\hat{H}\d x \wedge \d y + p^1 \d u \wedge \d y - p^2 \d u \wedge \d x + p^{(2,0)}\d u_1 \wedge \d y 
- \frac{1}{2} \, p^{(1,1)} \d u_1 \wedge \d x \\[5pt]
\displaystyle \qquad{} + \frac{1}{2} \, p^{(1,1)} \d u_2 \wedge \d y - p^{(0,2)}\d u_2 \wedge \d x \, , \\[5pt]
\displaystyle \Omega_r = \d \hat{H} \wedge \d x \wedge \d y - \d p^1 \d u \wedge \d y + \d p^2 \d u \wedge \d x - \d p^{(2,0)}\d u_1 \wedge \d y 
+ \frac{1}{2} \, \d p^{(1,1)} \d u_1 \wedge \d x \\[5pt]
\displaystyle \qquad{} - \frac{1}{2} \, \d p^{(1,1)} \d u_2 \wedge \d y + \d p^{(0,2)}\d u_2 \wedge \d x \, .
\end{array}
\end{equation}

Let $\psi \in \Gamma(\rho_M^r)$ be a holonomic section. Then, taking into account the local expression
\eqref{Chap06_eqn:ExampleRegular_UnifHamiltonianLocal} of the local Hamiltonian function $\hat{H}$
and \eqref{Chap06_eqn:ExampleRegular_UnifiedCanonicalForms} of the canonical forms in $\W_r$,
the field equation \eqref{Chap06_eqn:UnifFieldEqSect} gives in coordinates the following system of $11$
equations
\begin{align}
\derpar{p^1}{x} + \derpar{p^2}{y} + q = 0 \, , \label{Chap06_eqn:ExampleRegular_DiffEquations} \\
\derpar{p^{(2,0)}}{x} + \frac{1}{2}\derpar{p^{(1,1)}}{y} + p^1 = 0 \quad ; \quad
\frac{1}{2} \derpar{p^{(1,1)}}{x} + \derpar{p^{(0,2)}}{y} + p^2 = 0 \, ,
\label{Chap06_eqn:ExampleRegular_RelationMomenta} \\[6pt]
p^{(2,0)} - u_{(2,0)} = 0 \quad ; \quad p^{(1,1)} - 2u_{(1,1)} = 0 \quad ; \quad p^{(0,2)} - u_{(0,2)} = 0 \, , 
\label{Chap06_eqn:ExampleRegular_LegendreMap} \\[6pt]
u_1 - \derpar{u}{x} = 0 \quad ; \quad u_2 - \derpar{u}{y} = 0 \, , \label{Chap06_eqn:ExampleRegular_Holonomy1} \\
u_{(2,0)} - \derpar{u_1}{x} = 0 \quad ; \quad
u_{(1,1)} - \frac{1}{2}\left( \derpar{u_1}{y} + \derpar{u_2}{x} \right) = 0 \quad ; \quad
u_{(0,2)} - \derpar{u_2}{y} = 0 \, . \label{Chap06_eqn:ExampleRegular_Holonomy2}
\end{align}
Equations \eqref{Chap06_eqn:ExampleRegular_Holonomy1} and \eqref{Chap06_eqn:ExampleRegular_Holonomy2}
are automatically satisfied, since we require the section $\psi$ to be holonomic at the beginning.
On the other hand, combining equations \eqref{Chap06_eqn:ExampleRegular_RelationMomenta} and
\eqref{Chap06_eqn:ExampleRegular_LegendreMap}, we obtain the constraints defining the submanifold
$\W_\Lag$, and hence the Legendre map associated to this Lagrangian density, which is the fiber bundle
map $\Leg \colon J^3\pi \to J^2\pi^\ddagger$ given locally by
\begin{equation}\label{Chap06_eqn:ExampleRegular_FullLegendreMap}
\begin{array}{c}
\displaystyle \Leg^*p^1 = -u_{(3,0)} - u_{(1,2)} \quad ; \quad
\Leg^*p^2 = -u_{(2,1)} - u_{(0,3)} \, , \\[10pt]
\displaystyle \Leg^*p^{(2,0)} = u_{(2,0)} \quad ; \quad
\Leg^*p^{(1,1)} = 2u_{(1,1)} \quad ; \quad
\Leg^*p^{(0,2)} = u_{(0,2)} \, .
\end{array}
\end{equation}
Observe that the tangent map of $\Leg$ at every point $j^3_x\phi \in J^3\pi$ is given in coordinates
by the $10 \times 12$ real matrix
\begin{equation*}
\Tan_{j^3_x\phi}\Leg = 
\left(
\begin{array}{cccccccccccc}
1 & 0 & 0 & 0 & 0 & 0 & 0 & 0 & 0 & 0 & 0 & 0 \\
0 & 1 & 0 & 0 & 0 & 0 & 0 & 0 & 0 & 0 & 0 & 0 \\
0 & 0 & 1 & 0 & 0 & 0 & 0 & 0 & 0 & 0 & 0 & 0 \\
0 & 0 & 0 & 1 & 0 & 0 & 0 & 0 & 0 & 0 & 0 & 0 \\
0 & 0 & 0 & 0 & 1 & 0 & 0 & 0 & 0 & 0 & 0 & 0 \\
0 & 0 & 0 & 0 & 0 & 0 & 0 & 0 & -1 & 0 & -1 & 0 \\
0 & 0 & 0 & 0 & 0 & 0 & 0 & 0 & 0 & -1 & 0 & -1 \\
0 & 0 & 0 & 0 & 0 & 1 & 0 & 0 & 0 & 0 & 0 & 0 \\
0 & 0 & 0 & 0 & 0 & 0 & 2 & 0 & 0 & 0 & 0 & 0 \\
0 & 0 & 0 & 0 & 0 & 0 & 0 & 1 & 0 & 0 & 0 & 0
\end{array}
\right) \, .
\end{equation*}
From this it is clear that $\rank(\Leg(j^3_x\phi)) = 10 = \dim J^2\pi^\ddagger$. Hence, the restricted
Legendre map is a submersion onto $J^2\pi^\ddagger$, and thus the second-order Lagrangian density
$\Lag \in \df^{2}(J^2\pi)$ is regular.

Finally, combining equations \eqref{Chap06_eqn:ExampleRegular_DiffEquations}, \eqref{Chap06_eqn:ExampleRegular_RelationMomenta}
and \eqref{Chap06_eqn:ExampleRegular_LegendreMap}, we obtain the second-order Euler-Lagrange equation
for this field theory
\begin{equation}\label{Chap06_eqn:ExampleRegular_EulerLagrangeEq}
u_{(4,0)} + 2u_{(2,2)} + u_{(0,4)} = q \Longleftrightarrow
\frac{\partial^4 u}{\partial x^4} + 2\frac{\partial^4 u}{\partial x^2 \partial y^2} + \frac{\partial^4 u}{\partial y^4} = q \, .
\end{equation}
This is the classical equation for the bending of a clamped plate under a uniform load.

Now, let $\X \in \vf^2(\W_r)$ be a locally decomposable bivector field given locally by
\eqref{Chap06_eqn:UnifGenericMultiVFLocal}, that is,
\begin{align*}
\X &= \left(  \derpar{}{x} + F_1\derpar{}{u} + F_{1,1}\derpar{}{u_1} + F_{2,1}\derpar{}{u_2}
+ F_{(2,0),1} \derpar{}{u_{(2,0)}} + F_{(1,1),1} \derpar{}{u_{(1,1)}} + F_{(0,2),1} \derpar{}{u_{(0,2)}} \right. \\
&\qquad{} + F_{(3,0),1} \derpar{}{u_{(3,0)}} + F_{(2,1),1} \derpar{}{u_{(2,1)}} + F_{(1,2),1} \derpar{}{u_{(1,2)}}
+ F_{(0,3),1} \derpar{}{u_{(0,3)}} \\
&\qquad{} + \left. G_{1}^1 \derpar{}{p^1} + G_{1}^2 \derpar{}{p^2} + G_{1}^{(2,0)} \derpar{}{p^{(2,0)}}
+ G_{1}^{(1,1)} \derpar{}{p^{(1,1)}} + G_{1}^{(0,2)} \derpar{}{p^{(0,2)}} \right) \\
&\quad{} \wedge \left(  \derpar{}{y} + F_2\derpar{}{u} + F_{1,2}\derpar{}{u_1} + F_{2,2}\derpar{}{u_2}
+ F_{(2,0),2} \derpar{}{u_{(2,0)}} + F_{(1,1),2} \derpar{}{u_{(1,1)}} + F_{(0,2),2} \derpar{}{u_{(0,2)}} \right. \\
&\qquad{} + F_{(3,0),2} \derpar{}{u_{(3,0)}} + F_{(2,1),2} \derpar{}{u_{(2,1)}} + F_{(1,2),2} \derpar{}{u_{(1,2)}}
+ F_{(0,3),2} \derpar{}{u_{(0,3)}} \\
&\qquad{} + \left. G_{2}^1 \derpar{}{p^1} + G_{2}^2 \derpar{}{p^2} + G_{2}^{(2,0)} \derpar{}{p^{(2,0)}}
+ G_{2}^{(1,1)} \derpar{}{p^{(1,1)}} + G_{2}^{(0,2)} \derpar{}{p^{(0,2)}} \right) \, .
\end{align*}
Then, taking into account the coordinate expressions \eqref{Chap06_eqn:ExampleRegular_UnifHamiltonianLocal}
of the local Hamiltonian function $\hat{H}$ and \eqref{Chap06_eqn:ExampleRegular_UnifiedCanonicalForms}
of the $3$-form $\Omega_r$, the equation \eqref{Chap06_eqn:UnifFieldEqMultiVF} gives in coordinates the
following system of $11$ equations for the component functions of the bivector field $\X$
\begin{align}
F_1 = u_1 \quad ; \quad F_2 = u_2 \, , \label{Chap06_eqn:ExampleRegular_Holonomy1MVF} \\
F_{1,1} = u_{(2,0)} \quad ; \quad
\frac{1}{2} \left( F_{1,2} + F_{2,1} \right) = u_{(1,1)} \quad ; \quad
F_{2,2} = u_{(0,2)} \, , \label{Chap06_eqn:ExampleRegular_Holonomy2MVF} \\
G_1^1 + G_2^2 = - q \, , \label{Chap06_eqn:ExampleRegular_DiffEquationsMVF} \\
G_1^{(2,0)} + \frac{1}{2} \, G_2^{(1,1)} = -p^1 \quad ; \quad
\frac{1}{2} \, G_1^{(1,1)} + G_2^{(0,2)} = -p^2 \, , \label{Chap06_eqn:ExampleRegular_RelationMomentaMVF} \\
p^{(2,0)} - u_{(2,0)} = 0 \quad ; \quad
p^{(1,1)} - 2u_{(1,1)} = 0 \quad ; \quad
p^{(0,2)} - u_{(0,2)} = 0 \, . \label{Chap06_eqn:ExampleRegular_LegendreMapMVF}
\end{align}
Moreover, if we assume that $\X$ is holonomic, then we have the following $8$ additional equations
\begin{equation}\label{Chap06_eqn:ExampleRegular_Holonomy3MVF}
\begin{array}{c}
F_{1,2} = u_{(1,1)} \quad ; \quad F_{2,1} = u_{(1,1)} \quad ; \quad
F_{(2,0),1} = u_{(3,0)} \quad ; \quad F_{(2,0),2} = u_{(2,1)} \, , \\[5pt]
F_{(1,1),1} = u_{(2,1)} \quad ; \quad F_{(1,1),2} = u_{(1,2)} \quad ; \quad
F_{(0,2),1} = u_{(1,2)} \quad ; \quad F_{(0,2),2} = u_{(0,3)} \, .
\end{array}
\end{equation}
Observe that equations \eqref{Chap06_eqn:ExampleRegular_LegendreMapMVF} are the equations defining
the first constraint submanifold $\W_c \hookrightarrow \W_r$. As we have seen in Section
\ref{Chap06_sec:UnifFieldEquations}, the tangency condition for the bivector field $\X$ along $\W_c$
enables us to determine all the coefficients $G_i^I$, with $i = 1,2$ and $|I| = 2$, in the following way
\begin{align*}
G_1^{(2,0)} = u_{(3,0)} \quad ; \quad G_1^{(1,1)} = 2u_{(2,1)} \quad ; \quad G_1^{(0,2)} = u_{(1,2)} \, , \\
G_2^{(2,0)} = u_{(2,1)} \quad ; \quad G_2^{(1,1)} = 2u_{(1,2)} \quad ; \quad G_2^{(0,2)} = u_{(0,3)} \, .
\end{align*}
Then, replacing these functions in equations \eqref{Chap06_eqn:ExampleRegular_RelationMomentaMVF},
we obtain the following $2$ additional constraints
\begin{equation*}
p^1 + u_{(3,0)} + u_{(1,2)} = 0 \quad ; \quad p^2 + u_{(2,1)} + u_{(0,3)} = 0 \, ,
\end{equation*}
which define a new submanifold $\W_\Lag \hookrightarrow \W_r$. Analyzing the tangency of the bivector
field $\X$ along this new submanifold $\W_\Lag$, we obtain the following $4$ equations, which
enable us to determinate the coefficients $G_i^j$ as follows
\begin{align*}
G_1^1 + F_{(3,0),1} + F_{(1,2),1} = 0 \quad ; \quad G_1^2 + F_{(2,1),1} + F_{(0,3),1} = 0 \, , \\
G_2^1 + F_{(3,0),2} + F_{(1,2),2} = 0 \quad ; \quad G_1^2 + F_{(2,1),2} + F_{(0,3),2} = 0 \, .
\end{align*}
Hence, replacing these expressions on equations \eqref{Chap06_eqn:ExampleRegular_DiffEquationsMVF},
we obtain the second-order Euler-Lagrange equation for a bivector field, which is
\begin{equation}\label{Chap06_eqn:ExampleRegular_EulerLagrangeEqMultiVF}
F_{(3,0),1} + F_{(1,2),1} + F_{(2,1),2} + F_{(0,3),2} = q \, .
\end{equation}
Observe that if $\psi \in \Gamma(\rho_M^r)$ is an integral section of $\X$, then its component functions
must satisfy the second-order Euler-Lagrange equation \eqref{Chap06_eqn:ExampleRegular_EulerLagrangeEq}.

\subsubsection{Lagrangian formalism}

Now we recover the Lagrangian structures and equations from the unified setting. In order to obtain
the Poincar\'{e}-Cartan $2$-form $\Theta_\Lag = \widetilde{\Leg}^*\Theta_1^s \in \df^{2}(J^3\pi)$,
we need the extended Legendre map $\widetilde{\Leg} \colon J^3\pi \to J^2\pi^\dagger$. From the results
in Section \ref{Chap06_sec:UnifFieldEquations}, and bearing in mind the coordinate expression
\eqref{Chap06_eqn:ExampleRegular_FullLegendreMap} of the restricted Legendre map in this example,
we have that the extended Legendre map is given locally by
\begin{equation*}
\begin{array}{c}
\displaystyle \widetilde{\Leg}^*p^1 = -u_{(3,0)} - u_{(1,2)} \quad ; \quad
\widetilde{\Leg}^*p^2 = -u_{(2,1)} - u_{(0,3)} \, , \\[10pt]
\displaystyle \widetilde{\Leg}^*p^{(2,0)} = u_{(2,0)} \quad ; \quad
\widetilde{\Leg}^*p^{(1,1)} = 2u_{(1,1)} \quad ; \quad
\widetilde{\Leg}^*p^{(0,2)} = u_{(0,2)} \, , \\[8pt]
\displaystyle \widetilde{\Leg}^*p = u_{(3,0)}u_1 + u_{(1,2)}u_1 + u_{(2,1)}u_2 + u_{(0,3)}u_2
- \frac{1}{2}u_{(2,0)}^2 - u_{(1,1)}^2 - \frac{1}{2}u_{(0,2)}^2 - qu \, .
\end{array}
\end{equation*}
Therefore, the Poincar\'{e}-Cartan $2$-form is given locally by
\begin{align*}
\Theta_\Lag &=
\left( \frac{1}{2}u_{(2,0)}^2 + u_{(1,1)}^2 + \frac{1}{2}u_{(0,2)}^2 + qu
- u_{(3,0)}u_1 - u_{(1,2)}u_1 - u_{(2,1)}u_2 - u_{(0,3)}u_2 \right) \d x \wedge \d y \\
&\quad{} - (u_{(3,0)} + u_{(1,2)}) \d u \wedge \d y + (u_{(2,1)} + u_{(0,3)}) \d u \wedge \d x
+ u_{(2,0)}\d u_1 \wedge \d y - u_{(1,1)}\d u_1 \wedge \d x \\[5pt]
&\quad{} + u_{(1,1)}\d u_2 \wedge \d y - u_{(0,2)}\d u_2 \wedge \d x \, .
\end{align*}

Now, if $\Omega_\Lag = -\d\Theta_\Lag$, we recover the Lagrangian solutions for the field equations
from the unified formalism. In particular, if $\psi \in \Gamma(\rho^r_M)$ is a holonomic section
solution to the field equation \eqref{Chap06_eqn:UnifFieldEqSect}, then the section
$\psi_\Lag = \rho_1^r \circ \psi \in \Gamma(\bar{\pi}^3)$ is holonomic and is a solution to the field
equation \eqref{Chap06_eqn:LagFieldEqSect} by Proposition \ref{Chap06_prop:UnifToLagSect}.
In coordinates, the component functions of the section
$\psi_\Lag = j^3\phi$, for some $\phi(x,y) = (x,y,u(x,y)) \in \Gamma(\pi)$, are a solution to the
second-order Euler-Lagrange equation \eqref{Chap06_eqn:ExampleRegular_EulerLagrangeEq}, that is,
\begin{equation*}
u_{(4,0)} + 2u_{(2,2)} + u_{(0,4)} = q \, .
\end{equation*}
Finally, if $\X \in \vf^{2}(\W_r)$ is a locally decomposable holonomic bivector field
solution to the field equation \eqref{Chap06_eqn:UnifFieldEqMultiVF}, then,
using Theorem \ref{Chap06_thm:UnifToLagMultiVF}, there exists a unique locally decomposable
holonomic bivector field $\X_\Lag \in \vf^{2}(J^3\pi)$ solution to the equation
\eqref{Chap06_eqn:LagFieldEqMultiVF}. In coordinates, a locally decomposable holonomic bivector field
$\X_\Lag$ is given by
\begin{align*}
\X_\Lag &= \left(  \derpar{}{x} + u_1\derpar{}{u} + u_{(2,0)}\derpar{}{u_1} + u_{(1,1)}\derpar{}{u_2}
+ u_{(3,0)} \derpar{}{u_{(2,0)}} + u_{(2,1)} \derpar{}{u_{(1,1)}} + u_{(1,2)} \derpar{}{u_{(0,2)}} \right. \\
&\qquad{} + \left. F_{(3,0),1} \derpar{}{u_{(3,0)}} + F_{(2,1),1} \derpar{}{u_{(2,1)}} + F_{(1,2),1} \derpar{}{u_{(1,2)}}
+ F_{(0,3),1} \derpar{}{u_{(0,3)}} \right) \\
&\quad{} \wedge \left(  \derpar{}{y} + u_2\derpar{}{u} + u_{(1,1)} \derpar{}{u_1} + u_{(0,2)} \derpar{}{u_2}
+ u_{(2,1)} \derpar{}{u_{(2,0)}} + u_{(1,2)} \derpar{}{u_{(1,1)}} + u_{(0,3)} \derpar{}{u_{(0,2)}} \right. \\
&\qquad{} + \left. F_{(3,0),2} \derpar{}{u_{(3,0)}} + F_{(2,1),2} \derpar{}{u_{(2,1)}} + F_{(1,2),2} \derpar{}{u_{(1,2)}}
+ F_{(0,3),2} \derpar{}{u_{(0,3)}} \right) \, .
\end{align*}
Then, the component functions of this bivector field must satisfy the equation
\eqref{Chap06_eqn:ExampleRegular_EulerLagrangeEqMultiVF}, that is,
\begin{equation*}
F_{(3,0),1} + F_{(1,2),1} + F_{(2,1),2} + F_{(0,3),2} = q \, .
\end{equation*}

\subsubsection{Hamiltonian formalism}

Since the Lagrangian density is regular, the Hamiltonian formalism takes place in an open set of
$J^2\pi^\ddagger$. In fact, $\Lag \in \df^{2}(J^2\pi)$ is a hyperregular Lagrangian density, since
the restricted Legendre map admits global sections. For instance, the map defined locally by
\begin{equation*}
\Upsilon = 
\left(x,y,u,u_1,u_2,p^{(2,0)},\frac{1}{2} \, p^{(1,1)},p^{(0,2)},-\frac{1}{2} \, p^1,-\frac{1}{2} \, p^2,-\frac{1}{2} \, p^1,-\frac{1}{2} \, p^2\right) \, ,
\end{equation*}
is a section of $\Leg$ defined everywhere in $J^2\pi^\ddagger$.

In the natural coordinates of $J^2\pi^\ddagger$, using Lemma \ref{Chap06_lemma:HamHamiltonianSection},
and bearing in mind the coordinate expression
\eqref{Chap06_eqn:ExampleRegular_UnifHamiltonianLocal} of the local Hamiltonian function $\hat{H}$,
the local Hamiltonian function $H$ that specifies the Hamiltonian $\mu$-section $h$ is given by
\begin{equation*}
H = p^1u_1 + p^2u_2 + \frac{1}{2}\left(p^{(2,0)}\right)^2 + \frac{1}{4}\left(p^{(1,1)}\right)^2 + \frac{1}{2}\left(p^{(0,2)}\right)^2 + qu \, .
\end{equation*}
Hence, the Hamilton-Cartan $2$-form $\Theta_h \in \df^{2}(J^2\pi^\ddagger)$ is given locally by
\begin{align*}
\Theta_h &= \left( - p^1u_1 - p^2u_2 - \frac{1}{2}\left(p^{(2,0)}\right)^2 - \frac{1}{4}\left(p^{(1,1)}\right)^2
- \frac{1}{2}\left(p^{(0,2)}\right)^2 - qu \right) \d x \wedge \d y + p^1 \d u \wedge \d y \\
&\quad{} - p^2 \d u \wedge \d x + p^{(2,0)}\d u_1 \wedge \d y - \frac{1}{2} \, p^{(1,1)} \d u_1 \wedge \d x
+ \frac{1}{2} \, p^{(1,1)} \d u_2 \wedge \d y - p^{(0,2)}\d u_2 \wedge \d x \, .
\end{align*}

Now we recover the Hamiltonian field equations and solutions from the unified setting. First, let
$\psi \in \Gamma(\rho_M^r)$ be a (holonomic) section solution to the field equation \eqref{Chap06_eqn:UnifFieldEqSect}.
Then, as the second-order Lagrangian density is hyperregular, using Proposition
\ref{Chap06_prop:UnifiedToHamiltonianRegSect} we know that the section
$\psi_h = \rho_2^r \circ \psi \in \Gamma(\bar{\pi}_{J^1\pi}^\ddagger)$ is a
solution to the equation \eqref{Chap06_eqn:HamRegFieldEqSect}. In coordinates, the component functions
of $\psi_h$ must satisfy the following system of $8$ partial differential equations
\begin{equation*}
\begin{array}{c}
\displaystyle \derpar{u}{x} = u_1 \quad ; \quad \derpar{u}{y} = u_2 \quad ; \quad
\derpar{u_1}{x} = p^{(2,0)} \quad ; \quad
\derpar{u_2}{x} + \derpar{u_1}{y} = p^{(1,1)} \quad ; \quad
\derpar{u_2}{y} = p^{(0,2)} \, , \\[15pt]
\displaystyle \derpar{p^1}{x} + \derpar{p^2}{y} = q \quad ; \quad
\derpar{p^{(2,0)}}{x} + \frac{1}{2} \, \derpar{p^{(1,1)}}{y} = - p^1 \quad ; \quad
\frac{1}{2} \, \derpar{p^{(1,1)}}{x} + \derpar{p^{(0,2)}}{y} = - p^2 \, .
\end{array}
\end{equation*}
Finally, if $\X \in \vf^2(\W_r)$ is a locally decomposable bivector field solution to the equation
\eqref{Chap06_eqn:UnifFieldEqMultiVF}, then, by Theorem \ref{Chap06_thm:UnifToHamMultiVFReg},
there exists a locally decomposable bivector field
$\X_h \in \vf^{2}(J^2\pi^\ddagger)$ solution to the equation \eqref{Chap06_eqn:HamRegFieldEqMultiVF}. If $\X_h$
is locally given by \eqref{Chap06_eqn:HamGenericMultiVFLocalReg}, that is,
\begin{align*}
\X_h &= \left( \derpar{}{x} + F_1\derpar{}{u} + F_{1,1}\derpar{}{u_1} + F_{2,1}\derpar{}{u_2}
+ G_{1}^1 \derpar{}{p^1} + G_{1}^2 \derpar{}{p^2} \right. \\
&\qquad{} + \left. G_{1}^{(2,0)} \derpar{}{p^{(2,0)}} + G_{1}^{(1,1)} \derpar{}{p^{(1,1)}} + G_{1}^{(0,2)} \derpar{}{p^{(0,2)}} \right) \\
&\quad{} \wedge \left( \derpar{}{y} + F_2\derpar{}{u} + F_{1,2}\derpar{}{u_1} + F_{2,2}\derpar{}{u_2}
+ G_{2}^1 \derpar{}{p^1} + G_{2}^2 \derpar{}{p^2} \right. \\
&\qquad{} + \left. G_{2}^{(2,0)} \derpar{}{p^{(2,0)}} + G_{2}^{(1,1)} \derpar{}{p^{(1,1)}} + G_{2}^{(0,2)} \derpar{}{p^{(0,2)}} \right) \, ,
\end{align*}
then its component functions must satisfy the following $8$ equations
\begin{equation*}
\begin{array}{c}
\displaystyle F_1 = u_1 \quad ; \quad F_2 = u_2 \quad ; \quad
F_{1,1} = p^{(2,0)} \quad ; \quad
F_{2,1} + F_{1,2} = p^{(1,1)} \quad ; \quad
F_{2,2} = p^{(0,2)} \, , \\[10pt]
\displaystyle G^{1}_{1} + G^{2}_2 = q \quad ; \quad
G^{(2,0)}_1 + \frac{1}{2} \, G^{(1,1)}_2 = - p^1 \quad ; \quad
\frac{1}{2} \, G^{(1,1)}_1 + G^{(0,2)}_2 = - p^2 \, .
\end{array}
\end{equation*}

\subsection{Korteweg--de Vries equation}
\label{Chap06_exa:KdVEquation}

In the following we derive the Korteweg--de Vries equation, usually denoted as the KdV equation
for short, using the geometric formalism introduced in this Chapter. The KdV equation is a
mathematical model of waves on shallow water surfaces, and has become the prototypical example
of a non-linear partial differential equation whose solutions can be specified exactly. Many papers
are devoted to analyzing this model and, in particular, some previous multisymplectic descriptions
of it are available, for instance in \cite{art:Ascher_McLachlan2005,proc:Gotay88,art:Zhao_Qin2000}.
A further analysis using a different version of the unified formalism is given in \cite{art:Vitagliano10}.

The usual form of the KdV equation is
\begin{equation}\label{Chap06_eqn:ExampleSingular_KdVEquation}
\derpar{y}{t} - 6y\derpar{y}{x} + \frac{\partial^3y}{\partial x^3} = 0 \, ,
\end{equation}
that is, a non-linear, dispersive partial differential equation for a real function $y$ depending
on two real variables, the space $x$ and the time $t$. It is known that the KdV equation can be
derived from a least action principle as the Euler-Lagrange equation of the Lagrangian density
\begin{equation*}
\Lag = \frac{1}{2}\derpar{u}{x}\derpar{u}{t} - \left(\derpar{u}{x}\right)^3 - \frac{1}{2} \left(\frac{\partial^2u}{\partial x^2}\right)^2 \, ,
\end{equation*}
where $y = \partial u / \partial x$. It is therefore clear that we can use our formulation to derive
the Korteweg--de Vries equation as the field equations of a second-order field theory with a
$2$-dimensional base manifold and a $1$-dimensional fiber over this base.

Hence, let us consider $M = \R^2$ with global coordinates $(x,t)$, and $E = \R^2 \times \R$ with
natural coordinates adapted to the bundle structure, $(x,t,u)$. In these coordinates, the canonical
volume form in $\R^2$ is given by $\eta = \d x \wedge \d t \in \df^2(\R^2)$.

In the induced coordinates $(x,t,u,u_1,u_2,u_{(2,0)},u_{(1,1)},u_{(0,2)})$ of $J^2\pi$,
the second-order Lagrangian density $\Lag \in \df^{2}(J^2\pi)$ given above may be written as
\begin{equation*}
\Lag = \frac{1}{2} \left( u_1u_2 - 2u_1^3 - u_{(2,0)}^2\right) \d x \wedge \d t \, .
\end{equation*}

\subsubsection{Lagrangian-Hamiltonian formalism}

Following Section \ref{Chap06_sec:UnifiedFormalismSetting}, consider the fiber bundles
\begin{equation*}
\W = J^3\pi \times_{J^1\pi} J^2\pi^\dagger \quad ; \quad
\W_r = J^3\pi \times_{J^1\pi} J^2\pi^\ddagger \, ,
\end{equation*}
with the natural coordinates \eqref{Chap06_eqn:ExampleRegular_LocalCoordinatesW}
and \eqref{Chap06_eqn:ExampleRegular_LocalCoordinatesWr}, respectively. Observe that,
as in the previous example, we have $\dim J^3\pi = 12$ and
$\dim J^2\pi^\ddagger = 10$, and therefore $\dim\W = 18$ and $\dim\W_r = 17$.

The Hamiltonian $\mu_\W$-section $\hat{h} \in \Gamma(\mu_\W)$ is specified
by the local Hamiltonian function, whose coordinate expression \eqref{Chap06_eqn:UnifHamiltonianFunctionLocal}
in this case is
\begin{equation}\label{Chap06_eqn:ExampleSingular_UnifHamiltonianLocal}
\hat{H} = p^1u_1 + p^2u_2 + p^{(2,0)}u_{(2,0)} + p^{(1,1)}u_{(1,1)} + p^{(0,2)}u_{(0,2)}
- \frac{1}{2}u_1u_2 + u_1^3 + \frac{1}{2}u_{(2,0)}^2 \, .
\end{equation}
Then, the canonical forms in $\W_r$ have the coordinate expressions
\eqref{Chap06_eqn:ExampleRegular_UnifiedCanonicalForms}, just replacing the local Hamiltonian function
\eqref{Chap06_eqn:ExampleRegular_UnifHamiltonianLocal} by \eqref{Chap06_eqn:ExampleSingular_UnifHamiltonianLocal}.

Let $\psi \in \Gamma(\rho_M^r)$ be a holonomic section. Then, bearing in mind the coordinate expression
\eqref{Chap06_eqn:ExampleSingular_UnifHamiltonianLocal} of the local Hamiltonian function $\hat{H}$ and
\eqref{Chap06_eqn:ExampleRegular_UnifiedCanonicalForms} of the canonical forms in $\W_r$, the field equation
\eqref{Chap06_eqn:UnifFieldEqSect} gives in coordinates the following system of $11$ equations
\begin{align}
\derpar{p^1}{x} + \derpar{p^2}{t} = 0 \, , \label{Chap06_eqn:ExampleSingular_DiffEquations} \\
\derpar{p^{(2,0)}}{x} + \frac{1}{2} \, \derpar{p^{(1,1)}}{t} + p^1 - \frac{1}{2}u_2 + 3u_1^2 = 0 \quad ; \quad
\frac{1}{2} \, \derpar{p^{(1,1)}}{x} + \derpar{p^{(0,2)}}{t} + p^2 - \frac{1}{2}u_1 = 0 \, ,
\label{Chap06_eqn:ExampleSingular_RelationMomenta} \\[6pt]
p^{(2,0)} + u_{(2,0)} = 0 \quad ; \quad p^{(1,1)} = 0 \quad ; \quad p^{(0,2)} = 0 \, ,
\label{Chap06_eqn:ExampleSingular_LegendreMap} \\[6pt]
u_1 - \derpar{u}{x} = 0 \quad ; \quad u_2 - \derpar{u}{t} = 0 \, , \label{Chap06_eqn:ExampleSingular_Holonomy1} \\
u_{(2,0)} - \derpar{u_1}{x} = 0 \quad ; \quad
u_{(1,1)} - \frac{1}{2}\left( \derpar{u_1}{t} + \derpar{u_2}{x} \right) = 0 \quad ; \quad
u_{(0,2)} - \derpar{u_2}{t} = 0 \, . \label{Chap06_eqn:ExampleSingular_Holonomy2}
\end{align}
As in the Example analyzed in the previous Section, equations \eqref{Chap06_eqn:ExampleSingular_Holonomy1}
and \eqref{Chap06_eqn:ExampleSingular_Holonomy2} are automatically satisfied because we require the
section to be holonomic from the beginning. On the other hand, combining equations
\eqref{Chap06_eqn:ExampleSingular_RelationMomenta} and \eqref{Chap06_eqn:ExampleSingular_LegendreMap}
we obtain the constraints defining the submanifold $\W_\Lag$, and in particular the coordinate
expression of the restricted Legendre map $\Leg \colon J^3\pi \to J^2\pi^\ddagger$ associated to this
second-order Lagrangian density, which is
\begin{equation}\label{Chap06_eqn:ExampleSingular_FullLegendreMap}
\begin{array}{c}
\displaystyle \Leg^*p^1 = \frac{1}{2}u_2 - 3u_1^2 + u_{(3,0)} \quad ; \quad
\Leg^*p^2 = \frac{1}{2} u_1 \, , \\[10pt]
\displaystyle \Leg^*p^{(2,0)} = - u_{(2,0)} \quad ; \quad
\Leg^*p^{(1,1)} = 0 \quad ; \quad
\Leg^*p^{(0,2)} = 0 \, .
\end{array}
\end{equation}
The tangent map of $\Leg$ at every point $j^3_x\phi \in J^3\pi$ is given in coordinates by
\begin{equation*}
\Tan_{j^3_x\phi}\Leg = 
\left(
\begin{array}{cccccccccccc}
1 & 0 & 0 & 0 & 0 & 0 & 0 & 0 & 0 & 0 & 0 & 0 \\
0 & 1 & 0 & 0 & 0 & 0 & 0 & 0 & 0 & 0 & 0 & 0 \\
0 & 0 & 1 & 0 & 0 & 0 & 0 & 0 & 0 & 0 & 0 & 0 \\
0 & 0 & 0 & 1 & 0 & 0 & 0 & 0 & 0 & 0 & 0 & 0 \\
0 & 0 & 0 & 0 & 1 & 0 & 0 & 0 & 0 & 0 & 0 & 0 \\
0 & 0 & 0 & -6u_1 & 1/2 & 0 & 0 & 0 & 1 & 0 & 0 & 0 \\
0 & 0 & 0 & 1/2 & 0 & 0 & 0 & 0 & 0 & 0 & 0 & 0 \\
0 & 0 & 0 & 0 & 0 & -1 & 0 & 0 & 0 & 0 & 0 & 0 \\
0 & 0 & 0 & 0 & 0 & 0 & 0 & 0 & 0 & 0 & 0 & 0 \\
0 & 0 & 0 & 0 & 0 & 0 & 0 & 0 & 0 & 0 & 0 & 0
\end{array}
\right) \, .
\end{equation*}
From this it is clear that $\rank(\Leg(j^3_x\phi)) = 7 < 10 = \dim J^2\pi^\ddagger$.
Hence, the second-order Lagrangian density $\Lag \in \df^{2}(J^2\pi)$ is singular.

Finally, combining equations \eqref{Chap06_eqn:ExampleSingular_DiffEquations},
\eqref{Chap06_eqn:ExampleSingular_RelationMomenta} and \eqref{Chap06_eqn:ExampleSingular_LegendreMap},
we obtain the second-order Euler-Lagrange equation for this field theory
\begin{equation}\label{Chap06_eqn:ExampleSingular_EulerLagrangeEq}
u_{(1,1)} - 6u_1u_{(2,0)} + u_{(4,0)} = 0 \Longleftrightarrow
\frac{\partial^2 u}{\partial t \, \partial x} - 6\,\derpar{u}{x} \, \frac{\partial^2 u}{\partial x^2} + \frac{\partial^4 u}{\partial x^4} = 0 \, ,
\end{equation}
which, taking $y = \partial u / \partial x$, is the usual Korteweg-de Vries equation
\eqref{Chap06_eqn:ExampleSingular_KdVEquation}.

Now, let $\X \in \vf^2(\W_r)$ be a locally decomposable $2$-vector field given locally by
\eqref{Chap06_eqn:UnifGenericMultiVFLocal}, that is,
\begin{align*}
\X &= \left(  \derpar{}{x} + F_1\derpar{}{u} + F_{1,1}\derpar{}{u_1} + F_{2,1}\derpar{}{u_2}
+ F_{(2,0),1} \derpar{}{u_{(2,0)}} + F_{(1,1),1} \derpar{}{u_{(1,1)}} + F_{(0,2),1} \derpar{}{u_{(0,2)}} \right. \\
&\qquad{} + F_{(3,0),1} \derpar{}{u_{(3,0)}} + F_{(2,1),1} \derpar{}{u_{(2,1)}} + F_{(1,2),1} \derpar{}{u_{(1,2)}}
+ F_{(0,3),1} \derpar{}{u_{(0,3)}} \\
&\qquad{} + \left. G_{1}^1 \derpar{}{p^1} + G_{1}^2 \derpar{}{p^2} + G_{1}^{(2,0)} \derpar{}{p^{(2,0)}}
+ G_{1}^{(1,1)} \derpar{}{p^{(1,1)}} + G_{1}^{(0,2)} \derpar{}{p^{(0,2)}} \right) \\
&\quad{} \wedge \left(  \derpar{}{t} + F_2\derpar{}{u} + F_{1,2}\derpar{}{u_1} + F_{2,2}\derpar{}{u_2}
+ F_{(2,0),2} \derpar{}{u_{(2,0)}} + F_{(1,1),2} \derpar{}{u_{(1,1)}} + F_{(0,2),2} \derpar{}{u_{(0,2)}} \right. \\
&\qquad{} + F_{(3,0),2} \derpar{}{u_{(3,0)}} + F_{(2,1),2} \derpar{}{u_{(2,1)}} + F_{(1,2),2} \derpar{}{u_{(1,2)}}
+ F_{(0,3),2} \derpar{}{u_{(0,3)}} \\
&\qquad{} + \left. G_{2}^1 \derpar{}{p^1} + G_{2}^2 \derpar{}{p^2} + G_{2}^{(2,0)} \derpar{}{p^{(2,0)}}
+ G_{2}^{(1,1)} \derpar{}{p^{(1,1)}} + G_{2}^{(0,2)} \derpar{}{p^{(0,2)}} \right) \, .
\end{align*}
Then the field equation \eqref{Chap06_eqn:UnifFieldEqMultiVF} gives in coordinates the following system
of $11$ equations
\begin{align}
F_1 = u_1 \quad ; \quad F_2 = u_2 \, , \label{Chap06_eqn:ExampleSingular_Holonomy1MVF} \\
F_{1,1} = u_{(2,0)} \quad ; \quad \frac{1}{2} \left( F_{1,2} + F_{2,1} \right) = u_{(1,1)} \quad ; \quad
F_{2,2} = u_{(0,2)} \, , \label{Chap06_eqn:ExampleSingular_Holonomy2MVF} \\
G_1^1 + G_2^2 = 0 \, , \label{Chap06_eqn:ExampleSingular_DiffEquationsMVF} \\
G_1^{(2,0)} + \frac{1}{2} \, G_2^{(1,1)} = \frac{1}{2} u_2 - 3u_1^2 - p^1 \quad ; \quad
\frac{1}{2}G_1^{(1,1)} + G_2^{(0,2)} = \frac{1}{2} u_1 - p^2 \, , \label{Chap06_eqn:ExampleSingular_RelationMomentaMVF} \\
p^{(2,0)} + u_{(2,0)} = 0 \quad ; \quad
p^{(1,1)} = 0 \quad ; \quad
p^{(0,2)} = 0 \, . \label{Chap06_eqn:ExampleSingular_LegendreMapMVF}
\end{align}
Moreover, if we assume that $\X$ is holonomic, then we have the following $8$ additional equations
\begin{equation}\label{Chap06_eqn:ExampleSingular_Holonomy3MVF}
\begin{array}{c}
F_{1,2} = u_{(1,1)} \quad ; \quad F_{2,1} = u_{(1,1)} \quad ; \quad
F_{(2,0),1} = u_{(3,0)} \quad ; \quad F_{(2,0),2} = u_{(2,1)} \, , \\[5pt]
F_{(1,1),1} = u_{(2,1)} \quad ; \quad F_{(1,1),2} = u_{(1,2)} \quad ; \quad
F_{(0,2),1} = u_{(1,2)} \quad ; \quad F_{(0,2),2} = u_{(0,3)} \, .
\end{array}
\end{equation}
Observe that equations \eqref{Chap06_eqn:ExampleSingular_LegendreMapMVF} are the equations defining
the first constraint submanifold $\W_c \hookrightarrow \W_r$. As we have seen in Section
\ref{Chap06_sec:UnifFieldEquations}, the tangency condition for the $2$-vector field $\X$ along $\W_c$
enables us to determine all the coefficients $G_i^I$, with $i = 1,2$ and $|I| = 2$, as follows
\begin{align*}
G_1^{(2,0)} = - u_{(3,0)} \quad ; \quad G_1^{(1,1)} = 0 \quad ; \quad G_1^{(0,2)} = 0 \, , \\
G_2^{(2,0)} = - u_{(2,1)} \quad ; \quad G_2^{(1,1)} = 0 \quad ; \quad G_2^{(0,2)} = 0 \, .
\end{align*}
Then, replacing these expressions in equations \eqref{Chap06_eqn:ExampleSingular_RelationMomentaMVF},
we obtain the following $2$ additional constraints
\begin{equation*}
p^1 - \frac{1}{2}u_2 + 3u_1^2 - u_{(3,0)} = 0 \quad ; \quad p^2 - \frac{1}{2} u_1 = 0 \, ,
\end{equation*}
which define a new submanifold $\W_\Lag \hookrightarrow \W_r$. Analyzing the tangency
of the $2$-vector field along this new submanifold $\W_\Lag$, we obtain the following $4$ equations,
which enable us to determinate the coefficient functions $G_i^j$ as follows
\begin{align*}
G_1^1 - \frac{1}{2}u_{(1,1)} + 6u_1u_{(2,0)} - F_{(3,0),1} = 0 \quad ; \quad
G_1^2 - \frac{1}{2}u_{(2,0)} = 0 \, , \\[2pt]
G_2^1 - \frac{1}{2}u_{(0,2)} + 6u_1u_{(1,1)} - F_{(3,0),2} = 0 \quad ; \quad
G_2^2 - \frac{1}{2}u_{(1,1)} = 0 \, .
\end{align*}
Hence, replacing these expressions on equations \eqref{Chap06_eqn:ExampleSingular_DiffEquationsMVF},
we obtain the second-order Euler-Lagrange equation for a $2$-vector field, which is
\begin{equation*}
u_{(1,1)} - 6u_1u_{(2,0)} + F_{(3,0),1} = 0 \, ,
\end{equation*}
from where we can determinate $F_{(3,0),1}$ as
\begin{equation}\label{Chap06_eqn:ExampleSingular_EulerLagrangeEqMultiVF}
F_{(3,0),1} = 6u_1u_{(2,0)} - u_{(1,1)}  \, .
\end{equation}
Observe that if $\psi \in \Gamma(\rho_M^r)$ is an integral section of $\X$, then its component functions
must satisfy the second-order Euler-Lagrange equation \eqref{Chap06_eqn:ExampleSingular_EulerLagrangeEq}.

\begin{remark}
Observe that, in this case, the Lagrangian density is singular, but there are no additional
constraints. This implies that the final constraint submanifold is the whole submanifold $\W_\Lag$
in the unified formalism.
\end{remark}

\subsubsection{Lagrangian formalism}

Now we recover the Lagrangian formalism from the unified setting. First, we need the coordinate
expression of the extended Legendre map $\widetilde{\Leg} \colon J^3\pi \to J^2\pi^\dagger$. From
the results in Section \ref{Chap06_sec:UnifFieldEquations}, the local expression of $\widetilde{\Leg}$ is
\begin{equation*}
\begin{array}{c}
\displaystyle \widetilde{\Leg}^*p^1 = \frac{1}{2}u_2 - 3u_1^2 + u_{(3,0)} \quad ; \quad
\widetilde{\Leg}^*p^2 = \frac{1}{2} u_1 \, , \\[10pt]
\displaystyle \widetilde{\Leg}^*p^{(2,0)} = - u_{(2,0)} \quad ; \quad
\widetilde{\Leg}^*p^{(1,1)} = 0 \quad ; \quad
\widetilde{\Leg}^*p^{(0,2)} = 0 \, , \\[10pt]
\displaystyle \widetilde{\Leg}^*p =
- \frac{1}{2}u_1u_2 + 2u_1^3 - u_{(3,0)}u_1 + \frac{1}{2}u_{(2,0)}^2 \, .
\end{array}
\end{equation*}
Therefore, the Poincar\'{e}-Cartan $2$-form $\Theta_\Lag = \widetilde{\Leg}^*\Theta_1^s \in \df^{2}(J^3\pi)$
is given locally by
\begin{align*}
\Theta_\Lag &=
\left( \frac{1}{2}u_1u_2 - 2u_1^3 + u_{(3,0)}u_1 - \frac{1}{2}u_{(2,0)}^2 \right) \d x \wedge \d y
+ \left( \frac{1}{2}u_2 - 3u_1^2 + u_{(3,0)} \right) \d u \wedge \d y \\
&\quad{} - \frac{1}{2}u_1 \d u \wedge \d x
- u_{(2,0)}\d u_1 \wedge \d y \, .
\end{align*}

Then, let $\psi \in \Gamma(\rho^r_M)$ be a holonomic section solution to the field equation
\eqref{Chap06_eqn:UnifFieldEqSect}. Then, from Proposition \ref{Chap06_prop:UnifToLagSect} we know that
the section $\psi_\Lag = \rho_1^r \circ \psi \in \Gamma(\bar{\pi}^3)$
is holonomic and is a solution to the Lagrangian field equation \eqref{Chap06_eqn:LagFieldEqSect}. In
coordinates, the component functions of the section $\psi_\Lag = j^3\phi$, for some
$\phi(x,t) = (x,t,u(x,t)) \in \Gamma(\pi)$, are a solution to the second-order Euler-Lagrange equation
\begin{equation*}
u_{(1,1)} - 6u_1u_{(2,0)} + u_{(4,0)} = 0 \, .
\end{equation*}
On the other hand, if $\X \in \vf^{2}(\W_r)$ is a locally decomposable holonomic $2$-vector field
solution to the field equation \eqref{Chap06_eqn:UnifFieldEqMultiVF}, then, by Theorem
\ref{Chap06_thm:UnifToLagMultiVF}, there exists a unique locally decomposable
holonomic $2$-vector field $\X_\Lag \in \vf^{2}(J^3\pi)$ solution to the equation
\eqref{Chap06_eqn:LagFieldEqMultiVF}. In coordinates, a locally decomposable holonomic $2$-vector
field in $J^3\pi$ is given by
\begin{align*}
\X_\Lag &= \left(  \derpar{}{x} + u_1\derpar{}{u} + u_{(2,0)}\derpar{}{u_1} + u_{(1,1)}\derpar{}{u_2}
+ u_{(3,0)} \derpar{}{u_{(2,0)}} + u_{(2,1)} \derpar{}{u_{(1,1)}} + u_{(1,2)} \derpar{}{u_{(0,2)}} \right. \\
&\qquad{} + \left. F_{(3,0),1} \derpar{}{u_{(3,0)}} + F_{(2,1),1} \derpar{}{u_{(2,1)}} + F_{(1,2),1} \derpar{}{u_{(1,2)}}
+ F_{(0,3),1} \derpar{}{u_{(0,3)}} \right) \\
&\quad{} \wedge \left(  \derpar{}{t} + u_2\derpar{}{u} + u_{(1,1)} \derpar{}{u_1} + u_{(0,2)} \derpar{}{u_2}
+ u_{(2,1)} \derpar{}{u_{(2,0)}} + u_{(1,2)} \derpar{}{u_{(1,1)}} + u_{(0,3)} \derpar{}{u_{(0,2)}} \right. \\
&\qquad{} + \left. F_{(3,0),2} \derpar{}{u_{(3,0)}} + F_{(2,1),2} \derpar{}{u_{(2,1)}} + F_{(1,2),2} \derpar{}{u_{(1,2)}}
+ F_{(0,3),2} \derpar{}{u_{(0,3)}} \right) \, .
\end{align*}
Then, the component functions of this $2$-vector field must satisfy the equation
\eqref{Chap06_eqn:ExampleRegular_EulerLagrangeEqMultiVF}, that is,
\begin{equation*}
F_{(3,0),1} = 6u_1u_{(2,0)} - u_{(1,1)}  \, .
\end{equation*}

\subsubsection{Hamiltonian formalism}

Since the Lagrangian density is singular, the Hamiltonian formalism takes place in the submanifold
$\P = \Im(\Leg) \hookrightarrow J^2\pi^\ddagger$. In this case, we can not recover the Hamiltonian
solutions directly from the unified setting, but rather passing through the Lagrangian formulation.
Bearing in mind the coordinate expression \eqref{Chap06_eqn:ExampleSingular_FullLegendreMap} of the
restricted Legendre map, the submanifold $\P$ is defined locally by the constraints
\begin{equation*}
p^2 - \frac{1}{2}u_1 = 0 \quad ; \quad
p^{(1,1)} = 0 \quad ; \quad
p^{(0,2)} = 0 \, .
\end{equation*}
Observe that $\dim\P = \rank(\Leg) = 7$.

A set of natural coordinates $(x,t,u,u_1,u_2,p^1,p^2,p^{(2,0)},p^{(1,1)},p^{(0,2)})$ in the restricted
$2$-symmetric multimomentum bundle $J^2\pi^\ddagger$
induces coordinates $(x,t,u,u_1,u_2,p^1,p^{(2,0)})$ in $\P$, with the natural embedding
$\jmath \colon \P \hookrightarrow J^2\pi^\ddagger$ given locally by
\begin{equation}\label{Chap06_eqn:ExampleSingular_EmbeddingHamiltonian}
\jmath^*p^2 = \frac{1}{2}u_1 \quad ; \quad
\jmath^*p^{(1,1)} = 0 \quad ; \quad \jmath^*p^{(0,2)} = 0 \, .
\end{equation}

In these coordinates, using Lemma \ref{Chap06_lemma:HamHamiltonianSection}, and bearing in mind the
coordinate expressions \eqref{Chap06_eqn:ExampleSingular_UnifHamiltonianLocal} of the local
Hamiltonian function $\hat{H}$ and \eqref{Chap06_eqn:ExampleSingular_EmbeddingHamiltonian} of the
natural embedding $\P \hookrightarrow J^2\pi^\ddagger$, the local Hamiltonian function specifying 
the Hamiltonian section $h \in \Gamma(\mu)$ is given by
\begin{equation*}
H = p^1u_1 + u_1^3 - \frac{1}{2}\left(p^{(2,0)}\right)^2 \, .
\end{equation*}

Therefore, the Hamilton-Cartan $2$-form $\Theta_h = h^*\Theta_1^s \in \df^{2}(\P)$ is given locally by
\begin{align*}
\Theta_h &= \left( \frac{1}{2}\left(p^{(2,0)}\right)^2 - p^1u_1 - u_1^3 \right) \d x \wedge \d t 
+ p^1 \d u \wedge \d t - \frac{1}{2}u_1 \d u \wedge \d x + p^{(2,0)}\d u_1 \wedge \d t  \, .
\end{align*}

Now we recover the Hamiltonian field equations. If $\psi \in \Gamma(\rho_M^r)$ is a (holonomic)
section solution to the field equation \eqref{Chap06_eqn:UnifFieldEqSect}, then, using Proposition
\ref{Chap06_prop:UnifiedToHamiltonianSingSect}, the section
$\psi_h = \Leg \circ \rho_1^r \circ \psi \in \Gamma(\bar{\pi}_\P)$ is a solution to the equation
\eqref{Chap06_eqn:HamSingFieldEqSect}. In coordinates, the component functions of $\psi_h$ must satisfy the
following system of $4$ partial differential equations
\begin{equation*}
\derpar{u}{x} = u_1 \quad ; \quad
\frac{1}{2}\derpar{u}{t} = p^1 + 3u_1^2 \quad ; \quad
\derpar{p^1}{x} + \frac{1}{2}\derpar{u_1}{t} = 0 \quad ; \quad
\derpar{u_1}{x} = -p^{(2,0)} \, .
\end{equation*}
Finally, if $\X \in \vf^2(\W_r)$ is a locally decomposable $2$-vector field solution to the equation
\eqref{Chap06_eqn:UnifFieldEqMultiVF}, then, using Theorem \ref{Chap06_thm:UnifToHamMultiVFSing},
there exists a locally decomposable $2$-vector field $\X_h \in \vf^{2}(\P)$ solution to the equation
\eqref{Chap06_eqn:HamSingFieldEqMultiVF}. If $\X_h$ is locally given by
\begin{align*}
\X_h &= \left(  \derpar{}{x} + F_1\derpar{}{u} + F_{1,1}\derpar{}{u_1} + F_{2,1}\derpar{}{u_2}
+ G_{1}^1 \derpar{}{p^1} + G_{1}^{(2,0)} \derpar{}{p^{(2,0)}} \right) \\
&\quad{} \wedge \left(  \derpar{}{t} + F_2\derpar{}{u} + F_{1,2}\derpar{}{u_1} + F_{2,2}\derpar{}{u_2}
+ G_{2}^1 \derpar{}{p^1} + G_{2}^{(2,0)} \derpar{}{p^{(2,0)}} \right) \, ,
\end{align*}
then its component functions must satisfy the following $4$ equations
\begin{equation*}
F_1 = u_1 \quad ; \quad \frac{1}{2} F_2 = p^1 + 3u_1^2 \quad ; \quad
G^1_1 + \frac{1}{2}F_{1,2} = 0 \quad ; \quad
F_{1,1} = -p^{(2,0)} \, .
\end{equation*}


\section{The higher-order case}
\label{Chap06_sec:HigherOrderCase}

As we have stated at the beginning of this Chapter, this formulation fails when we try to generalize
it to a classical field theory of order greater or equal than $3$. The main obstruction is also the
fundamental tool that we have used to obtain a unique Legendre map from the constraint algorithm in
the unified setting: the space of $2$-symmetric multimomenta.

In particular, the relation among the multimomentum coordinates that we have introduced in Section
\ref{Chap06_sec:SymmetricMultimomenta}, $p^{ij}_\alpha = p_{\alpha}^{ji}$ for every
$1 \leqslant i,j \leqslant m$ and every $1 \leqslant \alpha \leqslant n$, can indeed be generalized
to higher-order field theories. In particular, we have the following result, which has been proved in
\cite{phd:Campos}, \S $4.2.6$.

\begin{theorem*}
Let $(x^i,u^\alpha_I,p,p^{Ii}_\alpha)$ be an adapted system of coordinates on $\Lambda^m_2(J^{k-1}\pi)$. The relation
\begin{equation*}
I!\cdot p_\alpha^{Ii} = I^\prime!\cdot p_\alpha^{I^\prime i^\prime} \, , \
\mbox{whenever } I+1_i = I^\prime + 1_{i^\prime} \mbox{ and }
|I| = |I^\prime| = k-1 \, ,
\end{equation*}
is invariant under change of coordinates.
\end{theorem*}

From this result, a straightforward consequence is that the relation
$I!\cdot p_\alpha^{Ii} = I^\prime!\cdot p_\alpha^{I^\prime i^\prime}$ defines a submanifold
of $\Lambda^m_2(J^{k-1}\pi)$, that is, the following result, which is also stated
in \cite{phd:Campos}, \S $4.2.6$, holds.
 
\begin{corollary*}
The space of $k$-symmetric multimomenta
\begin{equation*}
J^{k}\pi^\dagger = \left\{ \omega \in \Lambda^m_2(J^{k-1}\pi) \mid
I!\cdot p_\alpha^{Ii} = I^\prime!\cdot p_\alpha^{I^\prime i^\prime} \, , \,
I+1_i = I^\prime + 1_{i^\prime} , \,
|I| = |I^\prime| = k-1 \right\} \, ,
\end{equation*}
is an embedded submanifold of $\Lambda^m_2(J^{k-1}\pi)$. A system of adapted coordinates $(x^i,u^\alpha)$
on $E$ induces coordinates $(x^i,u_I^\alpha,p,p_\alpha^{I^\prime i},p_\alpha^{K})$ on $J^{k}\pi^\dagger$,
where $0 \leqslant |I^\prime| < |I| \leqslant k-1$ and $|K| = k$. The natural embedding
is then given in coordinates by $j_s^*p^{Ii}_\alpha = p_\alpha^{I+1_i}/(I(i)+1)$, for $|I| = k-1$. This manifold is transverse to
$\pi_{J^{k-1}\pi} \colon \Lambda^m_2(J^{k-1}\pi) \to J^{k-1}\pi$, and therefore fibers over $J^{k-1}\pi$.
\end{corollary*}

That is, we can generalize both the extended and restricted $2$-symmetric multimomentum bundles
to higher-order field theories. The main issue, however, is that the previous Theorem only ensures
that the ``symmetric'' relation among the multimomentum coordinates holds for the highest-order
multimomenta. That is, this relation of symmetry on the multimomenta is not invariant under change
of coordinates for lower orders, and hence we do not obtain a submanifold of $\Lambda^m_2(J^{k-1}\pi)$.

When translated to the formulation, this implies that the field equations in the unified formalism
enable us to ``fix'' the highest-order multimomenta as usual, that is,
\begin{equation*}
p_\alpha^K - \derpar{\hat{L}}{u_K^\alpha} = 0 \, .
\end{equation*}
Nevertheless, when requiring the multivector field $\X$ to be tangent to the submanifold defined
by these constraints, there are many more coefficient functions $G^I_{\alpha,j}$ to be determined
than equations obtained by the tangency condition, preventing us to obtain a well-defined submanifold
of $\W_c$ and a univocally defined Legendre map. For more details and comments, as well as the explicit
calculations in the third-order case, we refer to \cite{phd:Campos}, \S $4.2.6$.

Observe, however, that the key point to obtain a unique restricted Legendre map has been to consider
a Hamiltonian phase space which has the same number of multimomentum coordinates than the number
of ``velocities'' in which the Lagrangian density depends, that is, we have considered a Hamiltonian
phase space where Definition \ref{Chap06_def:UnifRegularLagrangianDef} is equivalent to the
restricted Legendre map being a submersion. The aforementioned Definition is a particular case,
for $k=2$, of the following Definition, stated in \cite{art:Saunders_Crampin90}.

\begin{definition*}
A $k$th-order Lagrangian density $\Lag \in \df^{m}(J^k\pi)$ is \textnormal{regular} if the
restricted Legendre map associated to $\Lag$ satisfies
\begin{equation*}
\rank(\Leg) = \dim J^k\pi + \dim J^{k-1}\pi - \dim E 
= m + n + n\sum_{r=1}^{k-1} \binom{m+r-1}{r} + n\sum_{r=1}^{k} \binom{m+r-1}{r} \, .
\end{equation*}
\end{definition*}

Therefore, it may seem natural to consider a Hamiltonian phase space whose dimension coincides with
this number. Note that this Hamiltonian phase space would have $n\sum_{r=1}^{k} \binom{m+r-1}{r}$
multimomenta coordinates (plus one on the extended bundle), which is exactly the number of generalized
coordinates of ``velocities'' in which depends a $k$th-order Lagrangian density. For now, this is still
work in progress.



\cleardoublepage
\phantomsection
\addcontentsline{toc}{chapter}{Conclusions and further research}
\renewcommand{\chaptermark}[1]{\markboth{#1}{}}
\renewcommand{\sectionmark}[1]{\markright{#1}{}}
\chapter*{Conclusions and further research}
\chaptermark{Conclusions and further research}
\sectionmark{Conclusions and further research}

In this final Chapter of this dissertation we summarize the main contributions of the work.
A list of publications derived from this work, or related to it, is also given. Finally,
some further lines of research are pointed out at the end.

\section*{Summary of contributions}
\addcontentsline{toc}{section}{Summary of contributions}

The starting point of this work has been the geometric formulations for dynamical systems and
field theories, and, in particular, first-order dynamical systems and field theories.
Our work has been devoted to generalize these geometric formalisms to higher-order theories
using a Skinner-Rusk approach.

Among the results stated in this dissertation, I wish to point out the following ones.
\begin{itemize}
\item Starting from the Lagrangian and Hamiltonian formalisms for higher-order systems (Section
\ref{Chap02_sec:AutonomousHigherOrder}), we have generalized the unified formalism originally
stated by R. Skinner and R. Rusk in \cite{art:Skinner_Rusk83} (Section
\ref{Chap02_sec:SkinnerRuskAutonomousFirstOrder}) to higher-order systems. The dynamical equations
are stated both for vector fields and integral curves. These results can be found in
Sections \ref{Chap03_sec:GeometricalSetting} and \ref{Chap03_sec:DynamicalEquations}.

\item Following the patterns in \cite{art:Skinner_Rusk83}, which have been reviewed in Section
\ref{Chap02_sec:SkinnerRuskAutonomousFirstOrder}, we recover both the Lagrangian and Hamiltonian
dynamics for higher-order systems from the dynamics in the unified formalism. Regular and singular
cases are distinguished when necessary, and our results are consistent with the results in the
literature \cite{book:DeLeon_Rodrigues85}, which have been reviewed in Section
\ref{Chap02_sec:AutonomousHigherOrder}. These results are found in Sections
\ref{Chap03_sec:UnifiedToLagrangian} and \ref{Chap03_sec:UnifiedToHamiltonian}.

\item Starting from the geometric description of the Hamilton-Jacobi theory established in
\cite{art:Carinena_Gracia_Marmo_Martinez_Munoz_Roman06} (Section \ref{Chap02_sec:HamiltonJacobi}),
and using the geometric Lagrangian and Hamiltonian formulations for higher-order autonomous systems
(Section \ref{Chap02_sec:AutonomousHigherOrder}), we give the geometric formulation of the
Hamilton-Jacobi problem for regular higher-order autonomous dynamical systems. Following the
patterns in the aforementioned work, we distinguish between the generalized and standard versions
of the Hamilton-Jacobi problem. These results are given in Sections \ref{Chap04_sec:LagrangianFormalism}
and \ref{Chap04_sec:HamiltonianFormalism}.

\item Using the results obtained in Chapter \ref{Chap:HOAutonomousDynamicalSystems}, and following
the ideas in \cite{art:DeLeon_Martin_Vaquero12} (which is not reviewed in this dissertation), we
give the geometric description of the Hamilton-Jacobi problem for higher-order systems in the
Lagrangian-Hamiltonian formalism. Contrary to the aforementioned paper, we restrict ourselves to
the case of dynamical systems given in terms of a regular Lagrangian function, that is, the singular
case is not considered. In addition, both the Lagrangian and
Hamiltonian geometric formulations of the Hamilton-Jacobi problem for higher-order systems are
recovered from the unified setting. Section \ref{Chap04_sec:UnifiedFormalism} is devoted to give
these results.

\item We have stated the Lagrangian-Hamiltonian unified formalism for higher-order non-autonomous
dynamical systems, thus generalizing the results in \cite{art:Barbero_Echeverria_Martin_Munoz_Roman07}
(Section \ref{Chap02_sec:NonAutonomousUnified}) to the higher-order setting, or, equivalently,
generalizing the results in Chapter \ref{Chap:HOAutonomousDynamicalSystems} to the case of time
dependent dynamical systems. Equations are stated in terms of integral sections and vector fields.
Moreover, our approach uses a general fiber bundle $\pi \colon E \to \R$,
instead of the common trivial fiber bundle $E = Q \times \R$ found in the literature
\cite{book:Abraham_Marsden78,art:Chinea_deLeon_Marrero94,art:Echeverria_Munoz_Roman91,art:Ranada92}.
Observe that, although every fiber bundle over $\R$ is trivializable, by sticking to the general framework
we can give more easily the generalization to the case when the base manifold is not $1$-dimensional, that is,
to field theories. These results are found in Section \ref{Chap05_sec:UnifiedFormalism}.

\item Starting from the unified formalism obtained in Section \ref{Chap05_sec:UnifiedFormalism}, we
derived both the Lagrangian and Hamiltonian formalisms for higher-order non-autonomous dynamical
systems, thus completing previous partial studies on this subject
\cite{art:Crasmareanu00,proc:DeLeon_Marrero92,art:deLeon_Martin94_2,proc:DeLeon_Rodrigues87,art:Krupkova96}.
In particular, since the constraint algorithm in the unified formalism delivers the full
Legendre-Ostrogradsky map associated to the Lagrangian density, we are able to define the
Poincar\'{e}-Cartan forms, and state the Lagrangian equations. For the Hamiltonian formalism,
we can recover a local Hamiltonian function from the local Hamiltonian function of the unified
formalism, both in the regular and singular (almost-regular) cases. These results can be found
in Sections \ref{Chap05_sec:UnifiedToLagrangian} and \ref{Chap05_sec:UnifiedToHamiltonian}.

\item We have given an unambiguous geometric formulation for second-order classical field theories.
In particular, by introducing a relation of symmetry among the second-order multimomentum coordinates,
and using a Skinner-Rusk approach, we have been able to obtain a unique Legendre transformation
from the constraint algorithm, which coincides with the known well-defined second-order Legendre map
\cite{art:Saunders91,art:Saunders_Crampin90}
thus removing the ambiguities in the definition of this geometric
structure. Moreover, the field equation is stated in terms of multivector fields and their integral
sections. These results are stated in Sections \ref{Chap06_sec:SymmetricMultimomenta} and
\ref{Chap06_sec:UnifiedFormalism}.

\item Bearing in mind that a unique Legendre map is obtained as a consequence of the constraint
algorithm in the unified setting, we are able to define a unique Poincar\'{e}-Cartan $m$-form,
thus effectively removing any ambiguity in second-order field theories. Moreover, the
Poincar\'{e}-Cartan $m$-form that we obtain coincides with the one obtained in
\cite{art:Saunders87,book:Saunders89,art:Saunders_Crampin90}, which is derived in an alternative way.
This shows that our results are consistent with the results in the literature. Then, a well-defined
geometric Lagrangian formalism for second-order field theories is given following the patterns in
Section \ref{Chap02_sec:LagFieldTheories}. Moreover, the $1$-nondegeneracy of the Poincar\'{e}-Cartan
$(m+1)$-form is discussed, and we prove that this form can not be multisymplectic, regardless of the
second-order Lagrangian density provided. All these results are given in Section
\ref{Chap06_sec:LagrangianFormalism}.

\item Since we do have a uniquely well-defined Legendre map, we are able to give a geometric
Hamiltonian formalism for second-order field theories in the space of symmetric multimomenta,
and both the regular and singular (almost-regular) cases are analyzed in Section
\ref{Chap06_sec:HamiltonianFormalism}.

\item Our approach differs from \cite{art:Campos_DeLeon_Martin_Vankerschaver09} in that our
formulation enables us to obtain a unique Legendre map from the algorithm, and the tangency condition
does not give rise to ambiguities. In addition, we do state
the field equation in several equivalent ways, contrary to the aforementioned work, where the field
equation is stated only in terms of Ehresmann connections. Nevertheless, the field
equations obtained are identical, and the formalism given in \cite{art:Campos_DeLeon_Martin_Vankerschaver09}
allows the authors to recover the full holonomy condition from the field equation.

Furthermore, our approach also differs from \cite{art:Vitagliano10} in that our formulation
makes no use of infinite-order jet bundles. That is, although in \cite{art:Vitagliano10} all the
ambiguities in higher-order classical field theories are removed, the phase space of the system has
infinite dimension. In our work we do not use infinite-order jets, and therefore all the manifolds
have finite dimension.

\item The case of classical field theories of order greater than $2$ is discussed in Section
\ref{Chap06_sec:HigherOrderCase}. Some results are generalized to the higher-order setting,
and some further research is pointed out.

\item Several physical models, in both regular and non-regular cases, have been studied as examples
to show the applications of every geometric formulation given in this dissertation. These examples
are contained in Sections \ref{Chap03_sec:Examples}, \ref{Chap04_sec:Examples}, \ref{Chap05_sec:Examples}
and \ref{Chap06_sec:Examples}. In particular, the physical models studied are:
the Pais-Uhlenbeck oscillator, a second-order relativistic particle (both free and subjected to a
time-dependent potential), the end of a thrown javelin, the shape of a deformed elastic cylindrical
beam with fixed ends (both homogeneous and non-homogeneous cases), the bending of a clamped plate
under a uniform load, and the classic Korteweg--de Vries equation.
\end{itemize}

Since most of the features of the unified formalism for higher-order theories are common to every
theory analyzed in this dissertation, we do give the general conclusions and comments on this
formulation in the following.

\begin{itemize}
\item Although the forms in the unified phase spaces are defined straightforwardly from the canonical
forms in the Hamiltonian phase space, this is not the case of the coupling functions. In first-order
theories, the coupling functions are just the canonical pairing that arises naturally from the duality
between the Lagrangian and Hamiltonian phase spaces. For higher-order theories, these spaces are no
longer dual to each other, and thus the generalization is not straightforward. Nevertheless, the
Lagrangian phase space in all these theories can be canonically embedded into a bundle which
is naturally dual to the Hamiltonian phase space, and therefore the coupling function can be defined
as the restriction of the canonical pairing to the submanifold which we identify with the Lagrangian
phase space.

\item Contrary to the first-order case (see Sections \ref{Chap02_sec:SkinnerRuskAutonomousFirstOrder},
\ref{Chap02_sec:NonAutonomousUnified} and \ref{Chap02_sec:FieldTheoriesUnified}), the holonomy
condition is not recovered from the coordinate expression of the dynamical (resp., field) equations,
regardless of the regularity of the Lagrangian function which is considered. That is, this
condition must be required as an extra assumption for higher-order theories. Nevertheless,
some partial holonomy conditions are still recovered from the equations and, although the formulation
does not give the full condition, it is still useful when dealing with singular systems.
\begin{itemize}
\item For higher-order dynamical systems, the full holonomy condition can indeed be recovered
from the constraint algorithm when the Lagrangian function is regular. However, to do so we require
that the vector field solution to the dynamical equations is tangent to the submanifold $\graph(\Leg)$,
which is a condition required ``ad-hoc'', since no equation requires this condition to hold. When
sticking to the general framework, we must require the full holonomy condition to hold from
the beginning.
\item In the case of second-order field theories, only the first level of the holonomy condition
can be recovered, instead of the first and the second. This is due to the symmetry relation defined
among the multimomentum coordinates, which prevents us of obtaining separate equations for every
second-order partial derivative. Nevertheless, a holonomic section or multivector field still
satisfies these equations, and therefore can be solutions to the field equations.
\end{itemize}

\item Again, contrary to first-order theories, the full Legendre map (or Legendre-Ostrogradsky map)
is not obtained from the coordinate expression of the geometric equations or the compatibility
submanifold. In particular, only the highest-order momentum coordinates are fixed in both cases,
and the full transformation is obtained as a consequence of the constraint algorithm when the
solutions are required to lie or be tangent to the compatibility submanifold.

\item The regularity of the Lagrangian function seems to play a secondary role on higher-order
theories, since the holonomy condition must be required even in the regular case. Nevertheless,
as we have seen in Sections \ref{Chap03_sec:DynamicalEquations}, \ref{Chap05_sec:UnifDynamicalEquations}
and \ref{Chap06_sec:UnifFieldEquations}, after delivering the full Legendre transformation, the
constraint algorithm delivers the Euler-Lagrange equations, which must hold in order to have
well-defined solutions in the whole submanifold $\graph(\Leg)$. For regular Lagrangians, the Euler-Lagrange
equations are compatible, and thus this is the final step of the constraint algorithm: the solutions
to the equations are well-defined in all the points of $\graph(\Leg)$, and induce dynamics in this
submanifold. However, for singular Lagrangians, these equations may not be compatible. In this case,
new constraints may appear, and the algorithm continues.
\end{itemize}

\section*{List of publications}
\addcontentsline{toc}{section}{List of publications}

Publications (research papers and conference proceedings) derived from this work are
\cite{art:Colombo_deLeon_Prieto_Roman13_2,art:Colombo_deLeon_Prieto_Roman14_JPA,art:Prieto_Roman11,
proc:Prieto_Roman12,art:Prieto_Roman12,art:Prieto_Roman13,art:Prieto_Roman14}
in the Bibliography found on page \pageref{Chap:Bibliography}. In addition, there have been
$9$ contributions to national and international congresses and workshops derived from this work,
$6$ of them being talks and $3$ of them posters. In addition, $4$ talks based on this work have been
given in seminars.

The list of publications ordered by Chapters, but keeping the numeration in the
Bibliography, is the following.

\paragraph{Chapter 3. Unified formalism for higher-order autonomous dynamical systems}
\begin{description}
\item{\cite{art:Prieto_Roman11}}
P.D. {Prieto-Mart\'{\i}nez} and N.~{Rom\'{a}n-Roy},
``{Lagrangian}-{Hamiltonian} unified formalism for au\-to\-no\-mous higher-order dynamical systems'',
\textit{J. Phys. A: Math. Teor.} \textbf{44}(38) (2011) 385203.
\end{description}

\paragraph{Chapter 4. Geometric Hamilton-Jacobi theory for higher-order autonomous systems}
\begin{description}
\item{\cite{art:Colombo_deLeon_Prieto_Roman13_2}}
L.~{Colombo}, M.~{de Le\'{o}n}, P.D. {Prieto-Mart\'{\i}nez}, and N.~{Rom\'{a}n-Roy},
``Unified formalism for the generalized $k$th-order {H}amilton--{J}acobi problem'',
\textit{Int. J. Geom. Methods Mod. Phys.} \textbf{11}(9) (2014) 1460037.

\item{\cite{art:Colombo_deLeon_Prieto_Roman14_JPA}}
L.~{Colombo}, M.~{de Le\'{o}n}, P.D. {Prieto-Mart\'{\i}nez}, and N.~{Rom\'{a}n-Roy},
``Geometric {Hamilton}--{Jacobi} theory for higher--order autonomous systems'',
\textit{J. Phys. A: Math. Teor.} \textbf{47}(23) (2014) 235203.
\end{description}

\paragraph{Chapter 5. Higher-order non-autonomous dynamical systems}
\begin{description}
\item{\cite{proc:Prieto_Roman12}} P.D. {Prieto-Mart\'{\i}nez} and N.~{Rom\'an-Roy},
``{Skinner}-{Rusk} unified formalism for higher-order systems'',
Proceedings of the XX International Fall Workshop on Geometry and Physics. Madrid.
\textit{AIP Conference Proceedings} \textbf{1460} (2012) 216--220.

\item{\cite{art:Prieto_Roman12}} P.D. {Prieto-Mart\'{\i}nez} and N.~{Rom\'{a}n-Roy},
``Unified formalism for higher-order non-autonomous dynamical systems'',
\textit{J. Math. Phys.} \textbf{53}(3) (2012) 032901.

\item{\cite{art:Prieto_Roman13}} P.D. {Prieto-Mart\'{\i}nez} and N.~{Rom\'{a}n-Roy},
``Higher-order {Mechanics}: {Variational} {Principles} and other topics'',
\textit{J. Geom. Mech.} \textbf{5}(4) (2013) 493--510.
\end{description}

\paragraph{Chapter 6. Second-order classical field theories}
\begin{description}
\item{\cite{art:Prieto_Roman14}} P.D. {Prieto-Mart\'{\i}nez} and N.~{Rom\'{a}n-Roy},
``A multisymplectic unified formalism for second-order classical field theories'',
{arXiv}:1402.4087 [math-ph],  2014.

(Submitted to \textit{J. Geom. Mech.}).
\end{description}

In addition, there is another work in progress in collaboration with Leonardo Colombo (ICMAT, Madrid, Spain)
which is partially based on the results of Chapter \ref{Chap:HOAutonomousDynamicalSystems} and
\cite{master:Colombo,art:Colombo_Martin11}, and involves higher-order dynamical systems on
principal bundles. Some first results have already been published in \cite{art:Colombo_Prieto14}, but we do not
include them in this dissertation, since the study of higher-order systems on Lie groups would require
additional structures and mathematical background which are beyond the scope of this Ph.D. dissertation.

\section*{Further research}
\addcontentsline{toc}{section}{Further research}

Finally, I wish to point out some future lines of research that arise from the geometric
formulations stated in this dissertation.

\subsection*{Reduction by symmetries in higher-order theories}

The problem of reduction of dynamical systems with symmetries has deserved the interest of theoretical
physicists and mathematicians, with the purpose of reducing the number of evolution equations, by
finding integrals of motion. In particular, geometric treatment of this subject has been revealed as
a powerful tool in the study of this question. The pioneering and fundamental work on this topic is
\cite{art:Marsden_Weinstein74} (see also \cite{book:Abraham_Marsden78,book:Libermann_Marle87,
book:Weinstein77}).

The procedure in the aforementioned work has been generalized and extended to many different situations:
presymplectic autonomous Hamiltonian systems \cite{art:Echeverria_DeLeon_Munoz_Roman99},
non-autonomous mechanical systems \cite{art:DeLeon_Martin96}
nonholonomic systems \cite{art:Bates_Sniatycki93,art:Cantrijn_DeLeon_Marrero_Martin98,art:Marle95},
higher-order autonomous dynamical systems \cite{art:DeLeon_Martin94_1,art:deLeon_Martin94_2,
art:DeLeon_Martin95,art:DeLeon_Pitanga_Rodrigues94}, optimal control \cite{art:DeLeon_Cortes_Marin_Martinez04,
art:Echeverria_Marin_Munoz_Roman03,art:Martinez_Cortes_DeLeon01} and several formulations first-order
classical field theories \cite{art:DeLeon_Martin_Santamaria04_1,art:Echeverria_Munoz_Roman99,
art:Munteanu_Rey_Salgado04,art:Marrero_Roman_Salgado_Vilarino11,art:Roman_Salgado_Vilarino07},
including $k$-symplectic, $k$-cosymplectic and multisymplectic formulations, although
reduction and symmetries in first-order field theories is still an open research line.

Our geometric formulations for higher-order non-autonomous dynamical systems (Chapter
\ref{Chap:HONonAutonomousDynamicalSystems}) and second-order field theories (Chapter
\ref{Chap:HOClassicalFieldTheories}) may help in the generalization of the concepts of
\textsl{symmetry} and \textsl{conservation law} for these kinds of systems, and may also
prove useful on describing a geometric treatment of the reduction procedure.

\subsection*{Nonholonomic constraints in higher-order systems}

Constraints in first-order dynamical systems and field theories can be classified into two large
families: holonomic and nonholonomic constraints, and they are ``easily'' differentiated. A constraint
is \textsl{holonomic} if it can be written in the form $F = 0$, with $F$ being a function depending
only on the coordinates of position (or the fields) of the system, or the total derivative of such
a function. If a constraint can not be written in these forms (which mostly implies that it must
depend on the velocities and cannot be written as the total derivative of a function on the basis),
then we say that this constraint is \textsl{nonholonomic}.

Nonholonomic constraints appear naturally on a wide variety of systems, and the prototypical example
is the rolling motion of a disk on a horizontal plane \cite{art:Borisov_Mamaev_Kilin03,
art:Cushman_Hermans_Kemppainen95}. Because of this, many works are devoted to the study of theories
with nonholonomic constraints \cite{art:Cantrijn_DeLeon_Marrero_Martin98,art:Carinena_Gracia_Marmo_Martinez_Munoz_Roman10,
art:Cortes_DeLeon99,art:Cortes_deLeon_Martin_Martinez02,art:DeLeon_Martin97,art:DeLeon_Martin_Salgado_Vilarino08,
art:DeLeon_Martin_Santamaria04,art:Grabowski_DeLeon_Marrero_Martin09,art:Hussein_Bloch08,
art:Vankerschaver_Cantrijn_DeLeon_Martin05}.

Nevertheless, as we have seen in Section \ref{Chap01_sec:HOJetBundlesProlongationHolonomicSections},
there are several ``levels'' of holonomy in higher-order theories. Therefore, keeping
the above definitions, a constraint which is nonholonomic could be holonomic on some levels, in the
sense that it could the lift of a function in the basis up to a certain order, and thus the nonholonomy
would come only from the last orders of derivation. Adopting the terminology
of the aforementioned Section, we could refine the classification of constraints in higher-order
theories defining the concepts of \textsl{holonomic constraints of type $r$}, with $1 \leqslant r \leqslant k$,
where $k$ is the order of the tangent bundle or jet bundle being considered. In this terminology,
holonomic constraints would be \textsl{holonomic of type $1$}, and the concept of nonholonmic
constraint, as defined above, would correspond to ``not being holonomic of type $r$ for any
$1 \leqslant r \leqslant k$''.

We believe that the study of these constraints can be useful to simplify geometric models of
constrained systems in higher-order theories, and can serve as a first step to give a general
geometric formulation for the Hamilton-Jacobi problem of singular higher-order systems.

\subsection*{Hamilton-Jacobi theory}

As we have pointed out in the Introduction, the geometric formulation of
the Hamilton-Jacobi theory has been generalized to many different situations. Following this
line of research, the results in Chapter \ref{Chap:HOHamiltonJacobi} can be generalized
to higher-order non-autonomous dynamical systems and multisymplectic second-order classical
field theories using the results in Chapters \ref{Chap:HONonAutonomousDynamicalSystems} and
\ref{Chap:HOClassicalFieldTheories}, respectively.

Another line of research on Hamilton-Jacobi theory is the study of the Hamilton-Jacobi problem
for higher-order autonomous systems given in terms of a singular Lagrangian function, following
the patterns in \cite{art:DeLeon_Martin_Vaquero12} and using the results in Section
\ref{Chap04_sec:UnifiedFormalism}. Observe that, in order to achieve this goal, a better knowledge
of nonholonomic constraints in higher-order theories would prove useful.

\subsection*{Higher-order field theories}

There are several open problems on higher-order field theories. The first one, which is
the most interesting from a mathematical point of view (although not so interesting from the physical one),
has been pointed out in Section \ref{Chap06_sec:HigherOrderCase}: to obtain an unambiguous geometric
formulation for field theories of order greater or equal than $3$. Bearing in mind the results
in Chapter \ref{Chap:HOClassicalFieldTheories}, and the comments in Section
\ref{Chap06_sec:HigherOrderCase}, the natural way to solve this problem should consist in finding
the obstruction that prevents us to define the symmetry relation among every level of multimomentum
coordinates. At the present day, we are working on a generalization of the intrinsic definition given in
\cite{book:Saunders89,art:Saunders_Crampin90} of the $2$-symmetric multimomentum bundle. We do believe
that generalizing the results of D.J. Saunders and M. Crampin is the first step to obtain the
symmetry relation among all the multimomenta. Observe that, to the best of our knowledge, there
are no ``natural'' classical field theories of order greater than $2$, and therefore this open
problem is interesting only from a purely mathematical point of view.

A second open problem in field theories consists in giving a complete geometric description of
real second-order theories, which could facilitate the comprehension of the classical models.
More particularly, it would be very interesting to obtain a geometric model for general relativity,
using the Hilbert-Einstein second-order Lagrangian density.

Finally, a third open problem on second-order field theory consists on establishing the variational
principles from which the field equations can be derived, following the patterns in
\cite{art:Prieto_Roman13}, and to prove the equivalence between solutions to the variational
problem and solution to the field equations given in Chapter \ref{Chap:HOClassicalFieldTheories}.
This is work in progress at the present time.


\backmatter


\cleardoublepage
\phantomsection
\addcontentsline{toc}{chapter}{Bibliography}
\label{Chap:Bibliography}


\cleardoublepage
\phantomsection
\addcontentsline{toc}{chapter}{Index}

\begin{theindex}

  \item almost-regular Lagrangian function/density
    \subitem $1$st-order autonomous systems, \hyperpage{40}
    \subitem $1$st-order non-autonomous systems, \hyperpage{63}
    \subitem $2$nd-order field theories, \hyperpage{209}
    \subitem $k$th-order autonomous systems, \hyperpage{57}
    \subitem $k$th-order non-autonomous systems, \hyperpage{164}

  \indexspace

  \item canonical embedding
    \subitem iterated jet bundles, \hyperpage{19}
    \subitem iterated tangent bundles, \hyperpage{26}
  \item canonical isomorphism
    \subitem cosymplectic manifolds, \hyperpage{13}
    \subitem multisymplectic manifolds, \hyperpage{14}
    \subitem symplectic manifolds, \hyperpage{9}
  \item canonical multisymplectic form
    \subitem in $\Lambda^{m}_{2}(J^{k-1}\pi)$, \hyperpage{22}
    \subitem of the multicotangent bundle, \hyperpage{14}
  \item canonical pairing between $J^{k}\pi$ and $\Lambda^{m}_{2}(J^{k-1}\pi)$, 
		\hyperpage{22}
  \item canonical structure form of $J^k\pi$, \hyperpage{18}
  \item canonical symplectic form of $\Tan^*Q$, \hyperpage{8}
  \item canonical vector field of $\Tan^{k}M$, \hyperpage{27}
  \item Cartan distribution of order $k$, \hyperpage{18}
  \item coisotropic submanifold, \hyperpage{10}
  \item compatibility submanifold, \hyperpage{35}
  \item complete solution to Hamilton-Jacobi problem
    \subitem Hamiltonian formalism, $1$st-order, \hyperpage{52}
    \subitem Hamiltonian formalism, $k$th-order, \hyperpage{118}
    \subitem Lagrangian formalism, $1$st-order, \hyperpage{50}
    \subitem Lagrangian formalism, $k$th-order, \hyperpage{112}
    \subitem unified formalism, $k$th-order, \hyperpage{130}
  \item contact module of order $k$, \hyperpage{18}
  \item coordinate total derivative, \hyperpage{21}
  \item cosymplectic manifold, \hyperpage{12}
  \item cosymplectic structure, \hyperpage{12}
  \item coupling function/form
    \subitem $1$st-order autonomous systems, \hyperpage{43}
    \subitem $1$st-order field theories, \hyperpage{79}
    \subitem $1$st-order non-autonomous systems, \hyperpage{67}
    \subitem $2$nd-order field theories, \hyperpage{191}
    \subitem $k$th-order autonomous systems, \hyperpage{85}
    \subitem $k$th-order non-autonomous systems, \hyperpage{143}

  \indexspace

  \item Darboux coordinates
    \subitem cosymplectic manifolds, \hyperpage{12}
    \subitem symplectic manifolds, \hyperpage{8}
  \item decomposable multivector field, \hyperpage{31}

  \indexspace

  \item Euler-Lagrange equations
    \subitem $1$st-order autonomous systems, \hyperpage{40}
    \subitem $1$st-order field theories, \hyperpage{73}
    \subitem $1$st-order non-autonomous systems, \hyperpage{60}
    \subitem $2$nd-order field theories, \hyperpage{197}, 
		\hyperpage{206}
    \subitem $k$th-order autonomous systems, \hyperpage{55}
    \subitem $k$th-order non-autonomous systems, \hyperpage{150}, 
		\hyperpage{161}

  \indexspace

  \item first constraint submanifold, \hyperpage{35}
  \item fundamental exact sequences, \hyperpage{25}

  \indexspace

  \item generalized Hamilton-Jacobi problem
    \subitem Hamiltonian formalism, $1$st-order, \hyperpage{50}
    \subitem Hamiltonian formalism, $k$th-order, \hyperpage{113}
    \subitem Lagrangian formalism, $1$st-order, \hyperpage{48}
    \subitem Lagrangian formalism, $k$th-order, \hyperpage{108}
    \subitem unified formalism in $\W_\Lag$, $k$th-order, 
		\hyperpage{128}
    \subitem unified formalism, $k$th-order, \hyperpage{122}

  \indexspace

  \item Hamilton equations
    \subitem $1$st-order autonomous systems, \hyperpage{41}
    \subitem $1$st-order field theories, \hyperpage{76}
    \subitem $1$st-order non-autonomous systems, \hyperpage{64}
    \subitem $2$nd-order field theories, \hyperpage{213}
    \subitem $k$th-order autonomous systems, \hyperpage{58}
    \subitem $k$th-order non-autonomous systems, \hyperpage{169}
    \subitem symplectic manifolds, \hyperpage{9}
  \item Hamilton-Cartan forms
    \subitem $1$st-order field theories, \hyperpage{75}
    \subitem $1$st-order non-autonomous systems, \hyperpage{63}
    \subitem $2$nd-order field theories, \hyperpage{210}
    \subitem $k$th-order non-autonomous systems, \hyperpage{165}
  \item Hamilton-Jacobi equation(s)
    \subitem Hamiltonian formalism, $1$st-order, \hyperpage{52}
    \subitem Hamiltonian formalism, $k$th-order, \hyperpage{117}
    \subitem Lagrangian formalism, $1$st-order, \hyperpage{50}
    \subitem Lagrangian formalism, $k$th-order, \hyperpage{112}
  \item Hamilton-Jacobi problem
    \subitem Hamiltonian formalism, $1$st-order, \hyperpage{51}
    \subitem Hamiltonian formalism, $k$th-order, \hyperpage{116}
    \subitem Lagrangian formalism, $1$st-order, \hyperpage{49}
    \subitem Lagrangian formalism, $k$th-order, \hyperpage{111}
    \subitem unified formalism in $\W_\Lag$, $k$th-order, 
		\hyperpage{128}
    \subitem unified formalism, $k$th-order, \hyperpage{126}
  \item Hamiltonian (multi)vector field
    \subitem $1$st-order autonomous systems, \hyperpage{41}
    \subitem $1$st-order field theories, \hyperpage{77}
    \subitem $1$st-order non-autonomous systems, \hyperpage{64}
    \subitem $2$nd-order field theories (reg.), \hyperpage{214}
    \subitem $2$nd-order field theories (sing.), \hyperpage{217}
    \subitem $k$th-order autonomous systems, \hyperpage{57}
    \subitem $k$th-order non-autonomous systems (reg.), \hyperpage{169}
    \subitem $k$th-order non-autonomous systems (sing.), 
		\hyperpage{174}
    \subitem symplectic manifolds, \hyperpage{9}
  \item Hamiltonian equation (regular)
    \subitem $1$st-order autonomous systems (curves), \hyperpage{41}
    \subitem $1$st-order autonomous systems (v.f.), \hyperpage{41}
    \subitem $1$st-order field theories (m.v.f), \hyperpage{77}
    \subitem $1$st-order field theories (sect.), \hyperpage{76}
    \subitem $1$st-order non-autonomous systems (sect.), \hyperpage{64}
    \subitem $1$st-order non-autonomous systems (v.f.), \hyperpage{64}
    \subitem $2$nd-order field theories (m.v.f.), \hyperpage{214}
    \subitem $2$nd-order field theories (sect.), \hyperpage{212}
    \subitem $k$th-order autonomous systems (curves), \hyperpage{58}
    \subitem $k$th-order autonomous systems (v.f.), \hyperpage{57}
    \subitem $k$th-order non-autonomous systems (sect.), 
		\hyperpage{167}
    \subitem $k$th-order non-autonomous systems (v.f.), \hyperpage{169}
    \subitem $k$th-order non-autonomous systems (var.), \hyperpage{166}
  \item Hamiltonian equation (singular)
    \subitem $1$st-order autonomous systems (v.f.), \hyperpage{42}
    \subitem $1$st-order non-autonomous systems (sect.), \hyperpage{65}
    \subitem $1$st-order non-autonomous systems (v.f.), \hyperpage{65}
    \subitem $2$nd-order field theories (m.v.f.), \hyperpage{217}
    \subitem $2$nd-order field theories (sect.), \hyperpage{216}
    \subitem $k$th-order autonomous systems (v.f.), \hyperpage{58}
    \subitem $k$th-order non-autonomous systems (sect.), 
		\hyperpage{174}
    \subitem $k$th-order non-autonomous systems (v.f.), \hyperpage{174}
    \subitem $k$th-order non-autonomous systems (var.), \hyperpage{172}
  \item Hamiltonian function/section
    \subitem $1$st-order autonomous systems (unif.), \hyperpage{44}
    \subitem $1$st-order autonomous systems (reg.), \hyperpage{41}
    \subitem $1$st-order autonomous systems (sing.), \hyperpage{42}
    \subitem $1$st-order field theories, \hyperpage{75}
    \subitem $1$st-order field theories (unif.), \hyperpage{79}
    \subitem $1$st-order non-autonomous systems, \hyperpage{63}
    \subitem $1$st-order non-autonomous systems (unif.), \hyperpage{67}
    \subitem $2$nd-order field theories, \hyperpage{210}
    \subitem $2$nd-order field theories (unif.), \hyperpage{192}
    \subitem $k$th-order autonomous systems (reg.), \hyperpage{57}
    \subitem $k$th-order autonomous systems (sing.), \hyperpage{58}
    \subitem $k$th-order autonomous systems (unif.), \hyperpage{85}
    \subitem $k$th-order non-autonomous systems, \hyperpage{165}
    \subitem $k$th-order non-autonomous systems (unif.), 
		\hyperpage{144}
    \subitem symplectic manifolds, \hyperpage{9}
  \item holonomic (multi)vector field
    \subitem $2$nd-order field theories (unif.), \hyperpage{193}
    \subitem higher-order jet bundles (m.v.f.), \hyperpage{34}
    \subitem higher-order tangent bundles (v.f.), \hyperpage{30}
    \subitem $k$th-order autonomous systems (unif.), \hyperpage{85}
    \subitem $k$th-order non-autonomous systems (unif.), 
		\hyperpage{145}
  \item holonomic curves/sections
    \subitem $2$nd-order field theories (unif.), \hyperpage{193}
    \subitem higher-order jet bundles, \hyperpage{17}
    \subitem higher-order tangent bundles, \hyperpage{30}
    \subitem $k$th-order autonomous systems (unif.), \hyperpage{85}
    \subitem $k$th-order non-autonomous systems (unif.), 
		\hyperpage{145}
  \item hyperregular Lagrangian function/density
    \subitem $1$st-order autonomous systems, \hyperpage{40}
    \subitem $1$st-order field theories, \hyperpage{75}
    \subitem $1$st-order non-autonomous systems, \hyperpage{63}
    \subitem $2$nd-order field theories, \hyperpage{196}
    \subitem $k$th-order autonomous systems, \hyperpage{56}
    \subitem $k$th-order non-autonomous systems, \hyperpage{148}

  \indexspace

  \item integrable multivector field, \hyperpage{32}
  \item integral manifold of a multivector field, \hyperpage{32}
  \item isotropic submanifold, \hyperpage{10}

  \indexspace

  \item Jacobi-Ostrogradsky momentum coordinates
    \subitem $k$th-order autonomous systems, \hyperpage{56}
    \subitem $k$th-order non-autonomous systems, \hyperpage{148}
  \item jet-(multi)momentum bundle(s)
    \subitem $1$st-order autonomous systems, \hyperpage{42}
    \subitem $1$st-order field theories, \hyperpage{78}
    \subitem $1$st-order non-autonomous systems, \hyperpage{66}
    \subitem $2$nd-order field theories ($2$-symmetric), 
		\hyperpage{190}
    \subitem $k$th-order autonomous systems, \hyperpage{83}
    \subitem $k$th-order non-autonomous systems, \hyperpage{141}

  \indexspace

  \item $k$-jet manifold, \hyperpage{16}
  \item $k$-vector field, \hyperpage{31}
  \item $k$th holonomic lift
    \subitem of a tangent vector, \hyperpage{20}
    \subitem of a vector field, \hyperpage{21}
  \item $k$th prolongation of sections in jet bundles, \hyperpage{17}
  \item $k$th-order extended dual jet bundle, \hyperpage{21}
  \item $k$th-order lift of a curve of $\Tan^{k}M$, \hyperpage{24}
  \item $k$th-order reduced dual jet bundle, \hyperpage{22}
  \item $k$th-order tangent bundle, \hyperpage{23}

  \indexspace

  \item Lagrangian (multi)vector field
    \subitem $1$st-order autonomous systems, \hyperpage{39}
    \subitem $1$st-order field theories, \hyperpage{73}
    \subitem $1$st-order non-autonomous systems, \hyperpage{60}
    \subitem $2$nd-order field theories, \hyperpage{207}
    \subitem $k$th-order autonomous systems, \hyperpage{54}
    \subitem $k$th-order non-autonomous systems, \hyperpage{161}
  \item Lagrangian energy
    \subitem $1$st-order autonomous systems, \hyperpage{38}
    \subitem $k$th-order autonomous systems, \hyperpage{53}
  \item Lagrangian equation
    \subitem $1$st-order autonomous systems (curves), \hyperpage{40}
    \subitem $1$st-order autonomous systems (v.f.), \hyperpage{39}
    \subitem $1$st-order field theories (m.v.f.), \hyperpage{73}
    \subitem $1$st-order field theories (sect.), \hyperpage{73}
    \subitem $1$st-order non-autonomous systems (sect.), \hyperpage{60}
    \subitem $1$st-order non-autonomous systems (v.f.), \hyperpage{60}
    \subitem $2$nd-order field theories (m.v.f.), \hyperpage{207}
    \subitem $2$nd-order field theories (sect.), \hyperpage{205}
    \subitem $k$th-order autonomous systems (curves), \hyperpage{55}
    \subitem $k$th-order autonomous systems (v.f.), \hyperpage{54}
    \subitem $k$th-order non-autonomous systems (sect.), 
		\hyperpage{159}
    \subitem $k$th-order non-autonomous systems (v.f.), \hyperpage{161}
    \subitem $k$th-order non-autonomous systems (var.), \hyperpage{158}
  \item Lagrangian submanifold, \hyperpage{10}
  \item Lagrangian system/field theory
    \subitem $1$st-order autonomous systems, \hyperpage{38}
    \subitem $1$st-order field theories, \hyperpage{72}
    \subitem $1$st-order non-autonomous systems, \hyperpage{59}
    \subitem $k$th-order autonomous systems, \hyperpage{53}
    \subitem $k$th-order non-autonomous systems, \hyperpage{156}
  \item Lagrangian-Hamiltonian equation
    \subitem $1$st-order autonomous systems (curves), \hyperpage{47}
    \subitem $1$st-order autonomous systems (v.f.), \hyperpage{44}
    \subitem $1$st-order field theories (m.v.f.), \hyperpage{81}
    \subitem $1$st-order field theories (sect.), \hyperpage{80}
    \subitem $1$st-order non-autonomous systems (sect.), \hyperpage{68}
    \subitem $1$st-order non-autonomous systems (v.f.), \hyperpage{69}
    \subitem $2$nd-order field theories (m.v.f.), \hyperpage{199}
    \subitem $2$nd-order field theories (sect.), \hyperpage{193}
    \subitem $k$th-order autonomous systems (curves), \hyperpage{90}
    \subitem $k$th-order autonomous systems (v.f.), \hyperpage{86}
    \subitem $k$th-order non-autonomous systems (sect.), 
		\hyperpage{146}
    \subitem $k$th-order non-autonomous systems (v.f.), \hyperpage{150}
    \subitem $k$th-order non-autonomous systems (var.), \hyperpage{145}
  \item Legendre/Legendre-Ostrogradsky map
    \subitem $1$st-order autonomous systems, \hyperpage{40}
    \subitem $1$st-order field theories (ext.), \hyperpage{74}
    \subitem $1$st-order field theories (rest.), \hyperpage{74}
    \subitem $1$st-order non-autonomous systems (ext.), \hyperpage{62}
    \subitem $1$st-order non-autonomous systems (rest.), \hyperpage{62}
    \subitem $2$nd-order field theories (ext.), \hyperpage{197}
    \subitem $2$nd-order field theories (rest.), \hyperpage{196}
    \subitem $k$th-order autonomous systems, \hyperpage{56}
    \subitem $k$th-order non-autonomous systems (ext.), \hyperpage{148}
    \subitem $k$th-order non-autonomous systems (rest.), 
		\hyperpage{148}
  \item Liouville forms
    \subitem of $\Lambda^{m}_{2}(J^{k-1}\pi)$, \hyperpage{22}
    \subitem of the cotangent bundle, \hyperpage{8}
  \item Liouville vector field of $\Tan^{k}M$, \hyperpage{27}
  \item locally decomposable multivector field, \hyperpage{31}

  \indexspace

  \item multi-index, \hyperpage{15}
  \item multisymplectic form of order $k$, \hyperpage{13}
  \item multisymplectic manifold of order $k$, \hyperpage{13}
  \item multivector field of degree $k$, \hyperpage{31}

  \indexspace

  \item Poincar\'{e}-Cartan forms
    \subitem $1$st-order autonomous systems, \hyperpage{38}
    \subitem $1$st-order field theories, \hyperpage{71}
    \subitem $1$st-order non-autonomous systems, \hyperpage{59}
    \subitem $2$nd-order field theories, \hyperpage{203}
    \subitem $k$th-order autonomous systems, \hyperpage{53}
    \subitem $k$th-order non-autonomous systems, \hyperpage{156}
  \item Poisson bracket induced by a symplectic form, \hyperpage{11}
  \item precosymplectic manifold, \hyperpage{12}
  \item precosymplectic structure, \hyperpage{12}
  \item premultisymplectic form of order $k$, \hyperpage{13}
  \item presymplectic form, \hyperpage{7}
  \item presymplectic manifold, \hyperpage{7}

  \indexspace

  \item Reeb vector field of a cosymplectic manifold, \hyperpage{13}
  \item regular Lagrangian function/density
    \subitem $1$st-order autonomous systems, \hyperpage{38}, 
		\hyperpage{40}
    \subitem $1$st-order field theories, \hyperpage{72}, \hyperpage{75}
    \subitem $1$st-order non-autonomous systems, \hyperpage{60}, 
		\hyperpage{63}
    \subitem $2$nd-order field theories, \hyperpage{196}
    \subitem $k$th-order autonomous systems, \hyperpage{54}, 
		\hyperpage{56}
    \subitem $k$th-order non-autonomous systems, \hyperpage{148}, 
		\hyperpage{157}
  \item $\rho^k_r$-bundle structure of $\Tan^kM$, \hyperpage{24}

  \indexspace

  \item semispray
    \subitem higher-order tangent bundles, \hyperpage{30}
    \subitem $k$th-order autonomous systems (unif.), \hyperpage{85}
    \subitem $k$th-order non-autonomous systems (unif.), 
		\hyperpage{145}
  \item singular Lagrangian function/density
    \subitem $1$st-order autonomous systems, \hyperpage{38}
    \subitem $1$st-order field theories, \hyperpage{72}
    \subitem $1$st-order non-autonomous systems, \hyperpage{60}
    \subitem $2$nd-order field theories, \hyperpage{196}
    \subitem $k$th-order autonomous systems, \hyperpage{54}
    \subitem $k$th-order non-autonomous systems, \hyperpage{148}, 
		\hyperpage{157}
  \item solution to generalized Hamilton-Jacobi problem
    \subitem Hamiltonian formalism, $1$st-order, \hyperpage{51}
    \subitem Hamiltonian formalism, $k$th-order, \hyperpage{114}
    \subitem Lagrangian formalism, $1$st-order, \hyperpage{48}
    \subitem Lagrangian formalism, $k$th-order, \hyperpage{109}
    \subitem unified formalism in $\W_\Lag$, $k$th-order, 
		\hyperpage{128}
    \subitem unified formalism, $k$th-order, \hyperpage{123}
  \item solution to Hamilton-Jacobi problem
    \subitem Hamiltonian formalism, $1$st-order, \hyperpage{51}
    \subitem Hamiltonian formalism, $k$th-order, \hyperpage{116}
    \subitem Lagrangian formalism, $1$st-order, \hyperpage{49}
    \subitem Lagrangian formalism, $k$th-order, \hyperpage{111}
    \subitem unified formalism, $k$th-order, \hyperpage{126}
  \item symmetric multimomentum bundle
    \subitem $2$nd-order, extended, \hyperpage{188}
    \subitem $2$nd-order, restricted, \hyperpage{189}
  \item symmetrized canonical pairing, \hyperpage{188}
  \item symmetrized Liouville forms, \hyperpage{188}
  \item symplectic form, \hyperpage{7}
  \item symplectic manifold, \hyperpage{7}
  \item symplectic map, \hyperpage{8}
  \item symplectic orthogonal, \hyperpage{10}
  \item symplectomorphism, \hyperpage{8}

  \indexspace

  \item tautological form
    \subitem of $\Lambda^{m}_{2}(J^{k-1}\pi)$, \hyperpage{22}
    \subitem of the cotangent bundle, \hyperpage{8}
    \subitem of the multicotangent bundle, \hyperpage{14}
  \item total derivative, \hyperpage{21}
  \item total time derivative (autonomous), \hyperpage{29}
  \item transverse multivector, \hyperpage{33}
  \item Tulczyjew's derivation, \hyperpage{29}

  \indexspace

  \item vertical endomorphism
    \subitem first-order jet bundles, \hyperpage{19}
    \subitem higher-order tangent bundles, \hyperpage{28}

\end{theindex}

\end{document}